\documentclass[
11pt, % The default document font size, options: 10pt, 11pt, 12pt
english, % ngerman for German
singlespacing, % Single line spacing, alternatives: onehalfspacing or doublespacing
headsepline, % Uncomment to get a line under the header
]{MastersDoctoralThesis} % The class file specifying the document structure

\usepackage[utf8]{inputenc} % Required for inputting international characters
\usepackage[T1]{fontenc} % Output font encoding for international characters
\usepackage{slashed}
\usepackage{amsmath,amssymb,amsfonts,amssymb,amsthm}
\usepackage{wrapfig}
\usepackage{yfonts,color,xcolor}
\usepackage[caption = false]{subfig}
\usepackage{multirow}
\usepackage{mathpazo} % Use the Palatino font by default
\usepackage{pdfpages}
\usepackage[bottom]{footmisc} 
 
\usepackage[most]{tcolorbox}

\usepackage{stackengine}

\tcbset{width=\textwidth,colback=gray!3.5!white, colframe=black!25!gray, 
        highlight math style= {enhanced, %<-- needed for the ’remember’ options
            colframe=red,colback=red!10!white,boxsep=0pt}
        }

\newcommand{\diff}{\mathrm{d}}
\newcommand{\nn}{\nonumber}
\newcommand{\myeq}[1]{
\begin{tcolorbox}[ams align]
#1
\end{tcolorbox}
}
\definecolor{darkblue}{HTML}{004C93} 

\definecolor{colOr}{rgb}{1, 0.796, 0.619}
\definecolor{colGr}{rgb}{0.666, 0.850, 0.666}
\definecolor{colBl}{rgb}{0.647, 0.784, 0.882}

\usepackage[backend=bibtex,style=ieee,natbib=true,maxnames=10,dashed=false]{biblatex} % Use the bibtex backend with the authoryear citation style (which resembles APA)
\usepackage[autostyle=true]{csquotes} % Required to generate language-dependent quotes in the bibliography

\thesistitle{Producing and constraining self-interacting hidden sector dark matter}
\supervisor{Pr. Thomas \textsc{Hambye}} % Your supervisor's name, this is used in the title page, print it elsewhere with \supname
\examiner{Pr. Firstname \textsc{Lastname}} % Your examiner's name, this is not currently used anywhere in the template, print it elsewhere with \examname
\degree{Doctor of Philosophy in Physics} % Your degree name, this is used in the title page and abstract, print it elsewhere with \degreename
\author{Laurent \textsc{Vanderheyden}} % Your name, this is used in the title page and abstract, print it elsewhere with \authorname
\addresses{} % Your address, this is not currently used anywhere in the template, print it elsewhere with \addressname

\subject{Physical Sciences} % Your subject area, this is not currently used anywhere in the template, print it elsewhere with \subjectname
 % Keywords for your thesis, this is not currently used anywhere in the template, print it elsewhere with \keywordnames
\university{\href{https://www.ulb.be/}{Université Libre de Bruxelles}} % Your university's name and URL, this is used in the title page and abstract, print it elsewhere with \univname
\department{\href{https://sciences.ulb.be/departement-physique}{Département de Physique}} % Your department's name and URL, this is used in the title page and abstract, print it elsewhere with \deptname
\group{\href{https://www2.ulb.ac.be/sciences/physth/index.html}{Service de Physique Théorique}} % Your research group's name and URL, this is used in the title page, print it elsewhere with \groupname
\faculty{\href{https://sciences.ulb.be/}{Faculté des Sciences}} % Your faculty's name and URL, this is used in the title page and abstract, print it elsewhere with \facname

\AtBeginDocument{
\hypersetup{pdftitle=\ttitle} % Set the PDF's title to your title
\hypersetup{pdfauthor=\authorname} % Set the PDF's author to your name
\hypersetup{pdfkeywords=\keywordnames} % Set the PDF's keywords to your keywords
}

\begin{document}
\pagestyle{plain}
%\includepdf[page={1},noautoscale = true,scale=3]{Cover_page/DM_SM_v2_b5.pdf}
\includepdf[page={1}]{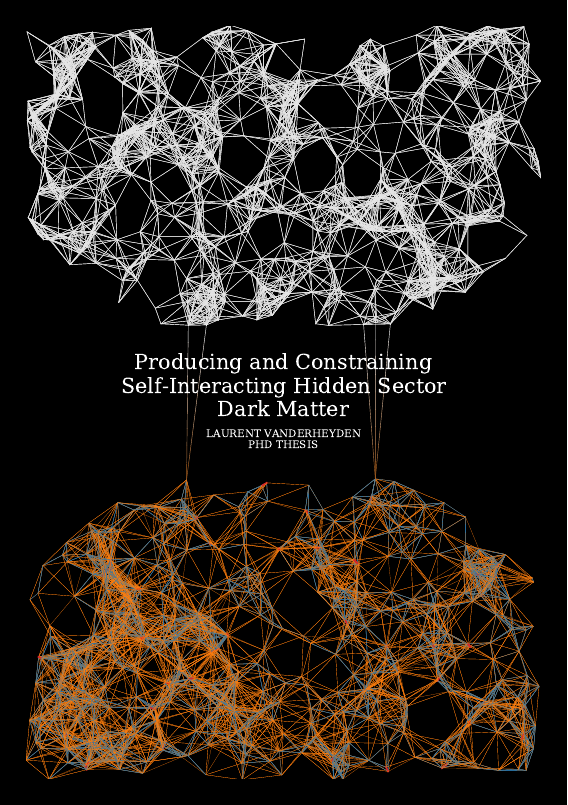}
\vfill
\newpage
\thispagestyle{empty}
\clearpage\null\newpage

\frontmatter % Use roman page numbering style (i, ii, iii, iv...) for the pre-content pages

\pagestyle{plain} % Default to the plain heading style until the thesis style is called for the body content

%----------------------------------------------------------------------------------------
%	TITLE PAGE
%----------------------------------------------------------------------------------------
%\includepdf[page={1}]{DM_SM.pdf}
%\clearpage\null\newpage
\includepdf[page={1}]{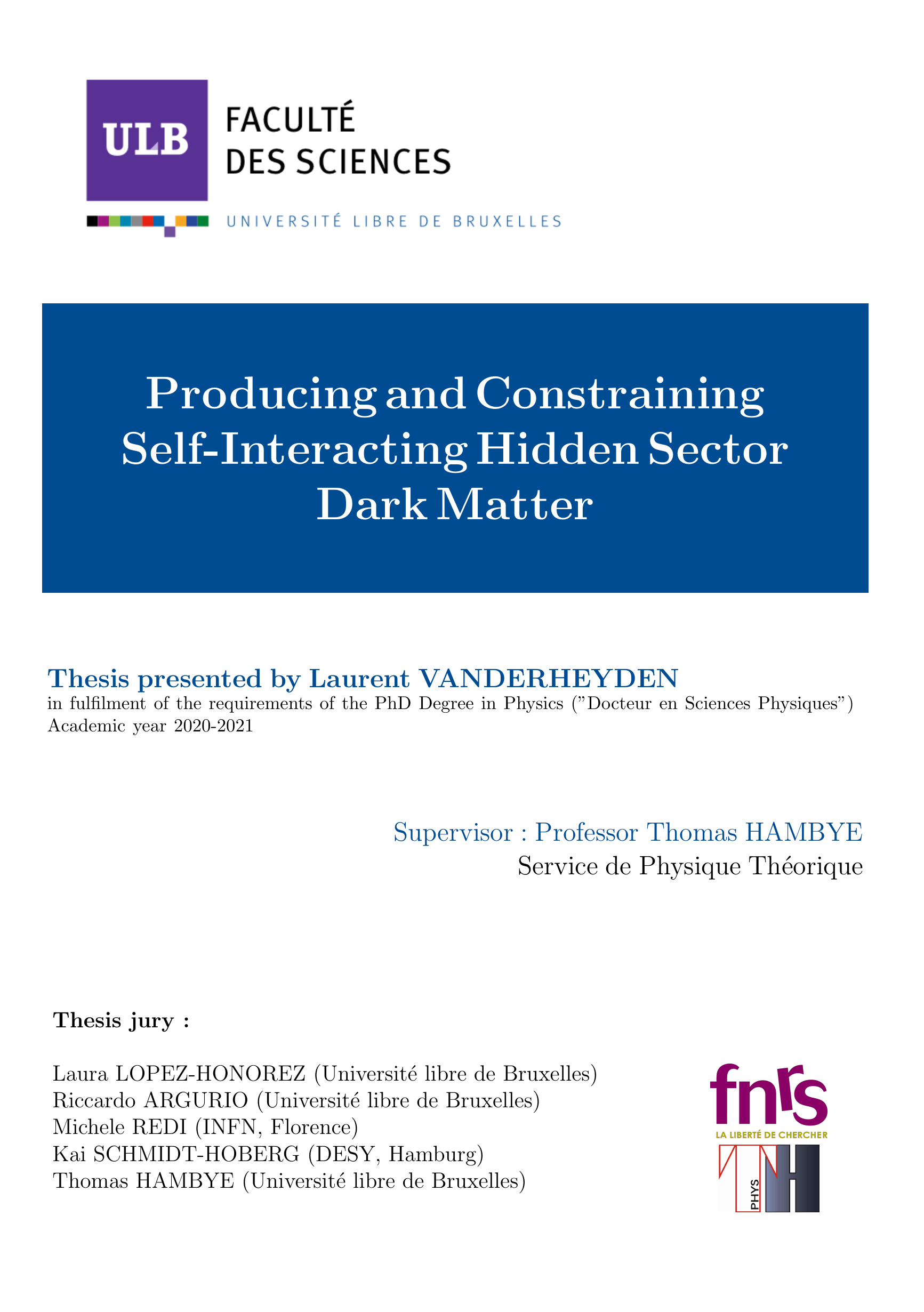}
\begin{center}
{\large 2017-2021}
\end{center}

\begin{abstract}
\addchaptertocentry{\abstractname} % Add the abstract to the table of contents
\yinipar{I}ncompatibilities between numerical simulations, theoretical predictions and cosmological observations at small and large scales suggest that the Dark Matter (DM) could strongly self-interact. It has been shown that introducing a light mediator would greatly help in accounting for large DM self-interactions. This mediator can allow portal interactions between the "hidden sector" containing the DM and the mediator and the "visible sector" containing the ordinary SM particles we know. However, more and more of the parameter space is excluded due to strong tensions between cosmological and astrophysical observations, on the one hand, and the fact that there is, until now, still no hint of DM in particle physics experiments, on the other hand.
\\

In this thesis, we start by reviewing all constraints applying on self-interacting DM with a light mediator considering well-known Higgs portal and Kinetic Mixing portal models as concrete examples. Then, we elaborate on the Sommerfeld effect, and we study its many implications on self-interacting DM models with light mediator. Next, we investigate how to account for the DM relic abundance in such portal models which present a generic structure with three populations (Standard Model, DM and mediator particles) connected to each other through three interactions resulting from two coupling strengths. It appears that this can be done through five different dynamical ways which give rise to nine regimes among which four are new. Those new production regimes are relevant for a situation where the hidden sector particles are not thermally connected to the visible vector particles.
\\

We further study the possibility of a thermally disconnected Hidden Sector and establish the general allowed parameter space for DM thermal candidates in the hidden-to-visible temperature ratio versus DM mass plane. We show that, in this framework, simple portal models such as the two we consider can alleviate tensions at small scale while offering suitable DM candidates. Other minimal self-interacting DM with light mediator frameworks which consistently fulfil all constraints are also proposed and explored.
\\

\end{abstract}

\begin{extraAbstract}
\addchaptertocentry{\abstractnameFR} % Add the abstract to the table of contents
\yinipar{L}es incompatibilités entre les simulations numériques, les prédictions théori-ques et les observations cosmologiques à courtes et larges échelles suggèrent que la Matière Noire (MN) auto-interagit fortement. Il a été démontré qu’introduire un médiateur léger peut permettre de rendre compte de telles auto-interactions pour la MN. Un tel médiateur permet, de plus, des interactions "portails" entre le "secteur caché" (contenant la MN et le médiateur léger) et le "secteur visible" (contenant les particules que l'on connaît déjà, c'est-à-dire du Modèle Standard). Toutefois, l’espace des paramètres se voit de plus en plus exclu aux vues d’une tension entre, d’une part, les observations cosmologiques et astrophysiques et, d’autre part, le fait que l'on ait à ce jour toujours pas mis en évidence des traces de MN dans aucune expérience de physique des particules.
\\

Nous commençons par détailler toutes les contraintes qui s’appliquent aux modèles pour lesquels la MN auto-interagit au moyen d’un médiateur léger. Pour cela, nous considérons les deux exemples de modèles connus du portail de Higgs et du portail de mélange cinétique. Nous discutons ensuite l’effet Sommerfeld et étudions ses nombreuses implications sur de tels modèles. Nous poursuivons en établissant comment rendre compte de la densité relique de MN observée aujourd’hui dans l'Univers dans de tels modèles, c'est-à-dire dans des modèles qui présentent une structure générique composée de trois populations de particules (le Modèle Standard, la MN et le médiateur) connectées les unes aux autres au travers de trois interactions résultantes de deux couplages fondamentaux entre particules. Il apparaît que cela peut être fait selon cinq manières dynamiques différentes qui donnent lieu à neuf régimes de production de la MN dont quatre sont nouveaux. Ces nouveaux régimes de production sont pertinents pour la situation où les particules du secteur caché ne sont pas en équilibre thermique avec les particules du secteur visible.
\\

Nous étudions ensuite plus en détail la possibilité d’un secteur caché qui n’est pas thermiquement connecté au secteur visible, en déterminant en particulier l’espace des paramètres autorisé pour des candidats de MN thermaux en termes du ratio des températures des deux secteurs et de la masse de la MN. Nous montrons ensuite que si les deux secteurs ne sont pas thermalement connectés, les deux modèles simples de portails que nous considérons permettent de diminuer les tensions à courtes échelles tout en offrant des candidats de MN viables. Enfin, nous proposons et explorons d’autres modèles minimaux de MN auto-interagissante qui satisfont aux nombreuses contraintes de manière consistante.
\\

MOTS-CLÉS: Matière noire, secteur caché, auto-interactions, médiateur léger, mécanisme de production, portail.
\end{extraAbstract}

%----------------------------------------------------------------------------------------
%	ACKNOWLEDGEMENTS
%----------------------------------------------------------------------------------------

%\begin{acknowledgements}
%\addchaptertocentry{\acknowledgementname} % Add the acknowledgements to the table of contents
%\yinipar{I} would like now to greatly thank every person which contributed directly or indirectly to this thesis. Let me start with the one who trusted in me and make it possible, Thomas Hambye my supervisor. I am much obliged to him as he advised me and taught me so much since I was still a first year master student. Now it has been almost six years that I am hanging out in the Service of Physique Théorique.
%\end{acknowledgements}
\begin{acknowledgements}
\addchaptertocentry{\acknowledgementname} % Add the acknowledgements to the table of contents
\yinipar{A}u travers de ces quelques lignes, je souhaiterais remercier chaque personne qui aura directement ou indirectement contribué à ce travail, en commençant, bien sur, par mon promoteur Thomas Hambye. Durant ces cinq dernières années, il m'aura énormément appris, tant d'un point de vue professionnel que personnel. Depuis le premier jour, il a pris son rôle à coeur en m'enseignant tout ce qu'il pouvait, en me guidant dans ce monde qui m'était alors totalement inconnu, sans jamais s'impatienter. Il m'a fait découvrir le monde de la recherche, ses obligations, ses opportunités, ses voyages et les incroyables rencontres qui y sont liées. Grâce à lui, j'ai le sentiment d'avoir vécu mes années de thèse à fond et n'ai aucun regret. I also warmly thank Laura Lopez-Honorez, Riccardo Argurio, Michele Redi and Kai Schmidt-Hoberg, the members of my jury for agreeing to read my manuscript and for the useful remarks and comments they have made during the private defense.
\\

Je tiens également à remercier l'ensemble des membres du Service de Physique Théorique qui m'a accueilli durant toute la durée de mon mémoire de fin d'étude et de mon doctorat. Je pense notamment à Michel Tytgat avec qui j'ai eu le plaisir de collaborer plusieurs fois, à Petr Tyniakov grâce à qui j'ai pu participer à l'encadre-ment du cours de relativité restreinte, à Laura Lopez-Honorez pour avoir apporté tant de bonne humeur et de dynamisme au quotidien et enfin à Jean-Marie Frère qui aura contribué à mon apprentissage. Je n'oublie évidement pas Isabelle Renders et Sarah Bouzerda qui m'ont aidées à maintes reprises au sujet de tâches administratives qui m'étaient parfois obscures. I thank everybody I met during my thesis in the group: Iason, Aritra, Deanna, Marco, Mikhail, Xunjie, Quentin, Sam, Baptiste, Raghuveer, Yoann, Mikaël, Daniele, Julian and many more. Let thank also Camillo, Chaïma, Rupert, Matteo and Jérôme, my collaborators, which all helped me to be the physicist I became. 
\\

Durant ma thèse et mes années passées à l'ULB, j'ai eu la chance de rencontrer et côtoyer bon nombre de personnes qui ont permis de transformer un quotidien parfois répétitif en des instants inoubliables. Je pense à toutes ces VUB et interminables pauses cafés passées avec notamment Martin, Adrien, Béa, Pawel, Louis, Daniel, Romain, Colin, Quentin, ... . Je retiendrai tout particulièrement Bilal, Antoine, Jérôme et Tanguy sans qui ces années à l'ULB n'auraient jamais pu être aussi extraordinaires. Je ne compte plus les heures passées aux salles d'étude, salles de repos, cafés, restaurants, billards à discuter de tout et de rien, à divaguer mais surtout à nous soutenir les uns les autres. Leur amitié durant mes études et ensuite durant ma thèse restera inoubliable.
\\

Ma passion pour la physique est née bien avant d'entrer à l'université, grâce notamment à ma rencontre avec Salahdin, lorsque nous avions dix ans environ. Nous nous sommes tout de suite parfaitement entendus et n'avons jamais cessé depuis de parler de physique et de sciences en général. Bien que nos chemins se soient quelques peu éloignés, il a toujours su éveiller ma curiosité et entretenir ma passion pour la physique, pour ça et pour notre longue amitié, je lui serai éternellement reconnaissant.
\\

Sans aucun doute, ma famille a joué un rôle primordial dans l'accomplissement de cette thèse. Que ce soit Bruno, Isabelle, Catherine, Papa ou Maman, vous avez tous toujours cru en moi. Jamais vous n'avez douté et ce même lorsque j'ai pu traverser des moments plus difficiles. Tantôt tour à tour, tantôt à l'unissons, vous avez tous su me remotiver et me montrer comment grandir, comment passer au-dessus des épreuves. Vous avez et êtes toujours des exemples pour moi, c'est donc pour moi une immense fierté de pouvoir enfin vous présenter ma thèse, mon travail. Le "petit" ne l'est peut-être plus tant que ça finalement ...
\\

Enfin, je finirai ces lignes en remerciant ma compagne (et presque épouse!) Élise qui, sans le moindre doute, m'aura été et m'est toujours d'un grand soutien. Depuis notre rencontre, tu es derrière moi, me motives, me forces à me dépasser et ce comme nul autre n'a pu le faire. Tout en suivant ton propre chemin, en menant de longues et passionnantes études, tu as su m'épauler en y mettant du coeur et de l'énergie. Il est certain que sans toi à mes côtés, je ne serais jamais arrivé là où j'en suis aujourd'hui. Grâce à toi, j'ai également rencontré ta famille, Alain, Evelyne et Christophe qui eux aussi m'auront soutenu et motivé, notamment en s'intéressant à ma thèse malgré le sujet plus qu'abstrait mais aussi en suivant mon parcours de près et en me conseillant. Pour ça, encore merci.
\end{acknowledgements}

%----------------------------------------------------------------------------------------
%	LIST OF CONTENTS/FIGURES/TABLES PAGES
%----------------------------------------------------------------------------------------

\tableofcontents % Prints the main table of contents

\listoffigures % Prints the list of figures

\listoftables % Prints the list of tables

%----------------------------------------------------------------------------------------
%	ABBREVIATIONS
%----------------------------------------------------------------------------------------

\begin{abbreviations}{ll} % Include a list of abbreviations (a table of two columns)

\textbf{BBN} & \textbf{B}ig \textbf{B}ang \textbf{N}ucleosynthesis\\
\textbf{CDM} & \textbf{C}old \textbf{D}ark \textbf{M}atter\\
\textbf{CMB} & \textbf{C}osmic \textbf{M}icrowave \textbf{B}ackground\\
\textbf{DG} & \textbf{D}warf \textbf{G}alaxy\\
\textbf{DM} & \textbf{D}ark \textbf{M}atter\\
\textbf{FLRW} & \textbf{F}ridemann - \textbf{L}emaître \textbf{R}obertson \textbf{W}alker\\
\textbf{FI} & \textbf{F}reeze-\textbf{I}n\\
\textbf{FO} & \textbf{F}reeze-\textbf{O}ut\\
\textbf{FS} & \textbf{F}ree-\textbf{S}treaming\\
\textbf{HP} & \textbf{H}iggs \textbf{P}ortal\\
\textbf{HS} & \textbf{H}idden \textbf{S}ector\\
\textbf{IR} & \textbf{I}nfra\textbf{r}ed\\
\textbf{KM} & \textbf{K}inetic \textbf{M}ixing\\
\textbf{LHC} & \textbf{L}arge \textbf{H}adron \textbf{C}ollider\\
\textbf{MW} & \textbf{M}ilky \textbf{W}ay\\
\textbf{MR} & \textbf{M}atter-\textbf{R}adiation\\
\textbf{NFW} & \textbf{N}avarro-\textbf{F}renk-\textbf{W}hite\\
\textbf{NR} & \textbf{N}on-\textbf{R}elativistic\\
\textbf{QCD} & \textbf{Q}uantum \textbf{C}hromo\textbf{D}ynamics\\
\textbf{SIDM} & \textbf{S}elf \textbf{I}nteracting \textbf{D}ark \textbf{M}atter\\
\textbf{SM} & \textbf{S}tandard \textbf{M}odel\\
\textbf{VEV} & \textbf{V}acuum \textbf{E}xpectation \textbf{V}alue\\
\textbf{VS} & \textbf{V}isible \textbf{S}ector\\

\end{abbreviations}

%----------------------------------------------------------------------------------------
%	PHYSICAL CONSTANTS/OTHER DEFINITIONS
%----------------------------------------------------------------------------------------

%\begin{constants}{lr@{${}={}$}l} % The list of physical constants is a three column table
%
%% The \SI{}{} command is provided by the siunitx package, see its documentation for instructions on how to use it
%
%Speed of Light & $c_{0}$ & \SI{2.99792458e8}{\meter\per\second} (exact)\\
%%Constant Name & $Symbol$ & $Constant Value$ with units\\
%
%\end{constants}

%----------------------------------------------------------------------------------------
%	SYMBOLS
%----------------------------------------------------------------------------------------

%\begin{symbols}{lll} % Include a list of Symbols (a three column table)
%
%$a$ & distance & \si{\meter} \\
%$P$ & power & \si{\watt} (\si{\joule\per\second}) \\
%%Symbol & Name & Unit \\
%
%\addlinespace % Gap to separate the Roman symbols from the Greek
%
%$\omega$ & angular frequency & \si{\radian} \\
%
%\end{symbols}

%----------------------------------------------------------------------------------------
%	DEDICATION
%----------------------------------------------------------------------------------------

%\dedicatory{For/Dedicated to/To my\ldots} 

%----------------------------------------------------------------------------------------
%	THESIS CONTENT - CHAPTERS
%----------------------------------------------------------------------------------------

\mainmatter % Begin numeric (1,2,3...) page numbering

\pagestyle{thesis} % Return the page headers back to the "thesis" style

% Include the chapters of the thesis as separate files from the Chapters folder
% Uncomment the lines as you write the chapters

%\include{chapter_1}
\chapter*{Introduction}\label{ch:intro}
\addcontentsline{toc}{chapter}{Introduction}
\markboth{Introduction}{Introduction}
\yinipar{T}o understand its environment is one of the oldest quests of humanity. In particular, the observations of the sky have been the matter of mystery and questions since the "dawn of time". It has been observed in particular since a very long time that globally, that is to say for large angular scales, the sky looks similar in all directions: the distribution of astrophysical objects looks almost the same. Since then, more and more refined observations did not stop to confirm this isotropy of the Universe. If, to the isotropy, one adds the Copernican principle which says that there is no privileged place in the Universe, one gets the cosmological principle. Hence, the statement that the observable Universe is homogeneous and isotropic at large scale is by today the fundamental principle in cosmology. Furthermore, it has been observed since very long too that the Universe develops more complex structures at small scales: The Earth belongs to our solar system which itself belongs to our galaxy, the Milky Way. In the Universe, many other spiral galaxies, like ours, exist and many of them have already been studied in detail.
\\

Those numerous observations have brought more questions than answers concerning the composition of the Universe. Indeed, Newton's law of gravitation has failed to explain several phenomena observed at various scales from the galactic scale (galaxy rotation curves) to galactic cluster scale (mass distribution in the Bullet cluster, distribution of galaxies) all the way to cosmological scales (Cosmic Microwave Background anisotropies). There are two general ways of solving these issues: either Newton's law is wrong at the relevant scales, or observations are missing "something". The first idea is not favoured because in practice it is very hard to modify Newton's law at these several scales in a consistent way that could account for all these observations at the same time. However, the second idea has been largely investigated this past century and people first named this missing piece of the Universe "Dark Matter" (DM) as J. Oort \cite{1932BAN.....6..249O}, F. Zwicky \cite{Zwicky:1933gu} and V. Rubin \cite{1970ApJ...159..379R} did. Later, it became clear that from cosmological scale observations one also needs "dark energy" on top of DM. According to observations, 26\% and 69\% of the energy content of the Universe should be due to DM and to dark energy, respectively. Only 5\% would then be due to the presence of the ordinary matter we know, as made of protons, neutrons, electrons in stars, and in galactic and inter-galactic gas.
\\

So far, only the gravitational effects that DM induces could be seen, as these effects are necessary to explain the various observations above. However, the microscopic nature of DM "as a particle" is still not established and there are many different possibilities of particle physics realisations. Alternatively, it is not fully ruled out that DM could be made of primordial black holes, massive compact halo objects, .... To establish what DM is made of is one of the most fundamental research topics in theoretical and experimental physics of the last decades even if the existence and characteristics of this new kind of matter intrigue physicists since almost a hundred years. Of course, as long as we will not establish what DM is made of, we will have no absolute proof for the existence of the DM. However, an extremely impressive series of experiments and a long series of theoretical consistency arguments do constrain already in many ways the properties of the DM particle, if it is made of a new particle. They establish in particular that these particles must be dark (i.e. feebly coupled to the photon) otherwise we would see it, stable (i.e. a lifetime larger than the age of the Universe) otherwise it would not be present today, be highly non-relativistic today and do not interact much with ordinary matter.
\\

In the following, we will assume that the DM is made of a new particle, which is by far the most plausible of all possibilities. The mass of the DM particle is still very poorly constrained. It could go all the way from a tiny fraction of eV to the Planck scale ($\sim 10^{19}$ GeV). However, its interactions are more constrained. For example, DM candidates which have electroweak interactions are already excluded or highly constrained \cite{Aprile:2017iyp,Akerib:2016vxi,Cui:2017nnn,Sirunyan:2018dub,Sirunyan:2018xlo,Sirunyan:2018fpy,Sirunyan:2018wcm,Sirunyan:2018gka,Aaboud:2018xdl,Sirunyan:2017leh,Aaboud:2017phn,Aaboud:2017rzf,Aaboud:2017bja,Hooper:2019xss} by many particle physics experiments (direct detection, indirect detection, colliders). These numerous constraints are such that the standard Weakly Interacting Massive Particle (or WIMP) scenario which used to be the main possibility for explaining the DM relic density becomes more and more disfavoured. Hence, the lack of proof of the existence of DM in particle physics experiments tends to suggest that it could be neutral under all SM gauge groups, in other words, it is a singlet of the $SU(3)_{c}\times SU(2)_{L}\times U(1)_{Y}$ SM fundamental interactions. Pushing this principle to its limit, DM particles would be insensitive to all SM particles except through gravity and would form a separate sector evolving on its own. One can then distinguish the "Visible Sector" (VS) made of all SM particles and the "Hidden Sector" (HS) made of DM particles and possibly other particles. Indeed, without any additional symmetry, there is nothing which forbids the DM to have other interactions, typically new gauge interactions in the new sector, the HS. Hence, the HS could be much more complex than if it was just composed of DM particles and could be ruled by new interactions and/or new gauge symmetries. All particles of the VS would be a singlet of all HS gauge groups as all HS particles are singlets of all SM gauge groups such that the only already established interaction that would couple the two sectors is gravity.
\\

If there is no additional connection other than gravity between the VS and the HS, this would be the end of the story as to see the DM particles through gravity is presumably almost impossible. Alternatively, the existence of a portal between the two sectors, i.e. a new renormalisable (or not) interaction which couples to a singlet combination of the particles of the two sectors at the same time, is perfectly possible. Hence, in such scenarios, DM particles are connected to the VS thanks to this additional interaction, the portal, and thus one can hope that through this portal one could constrain and/or detect DM particles. A DM candidate living in a HS and talking to the VS through a portal can be seen as an extension, a generalisation \cite{Chung:1998zb,Kolb:1998ki,Randall:2015xza,Berlin:2016vnh,Berlin:2016gtr,Harigaya:2016nlg,Tenkanen:2016jic,Berlin:2017ife,Bramante:2017obj,Kolb:2017jvz,Blanco:2017sbc,Cirelli:2018iax,Davoudiasl:2019xeb,Kim:2019udq,Heurtier:2019eou,Baker:2019ndr,Chway:2019kft,Heurtier:2019beu} of the traditional WIMP scenario, as we will see below. The idea of a DM candidate being part of a HS and coupling to the VS through a portal has already been suggested long ago in scenarios of the "mirror DM" type in which the HS is a copy of the VS. There are, in these specific mirror DM frameworks, many different ways to account for the DM relic abundance, see \cite{Kobzarev:1966qya,Blinnikov:1982eh,Foot:1991bp,Hodges:1993yb,Berezhiani:1995am} for more details.
\\

The idea of a new HS is not relevant only for the DM issue, it is a much broader concept which can be relevant for many other purposes. One can cite the Minimal Supersymmetric Standard Model (MSSM) in which a HS is often used for the purpose of breaking supersymmetry, \cite{ArkaniHamed:2005yv,Strassler:2006im,Hooper:2008im,Feng:2008ya}. More generally, HS also exists in string theories by constructions, see top-down string constructions \cite{Cvetic:2012kj}. On top of the supersymmetric models, one also needs or encounters HS in particle physics models aiming to solve other issues than DM, such as the hierarchy problem, composite Higgs models, see for example \cite{Kaul:1981uk,Montull:2012ik,Carmi:2012in,Barbieri:2015lqa}.
\\

Another major argument favouring the whole concept of a HS arises when looking at the Universe at small scale. Indeed, the DM hypothesis was introduced in order to explain different phenomena at large scale. However, there also exist inconsistencies observed at small scales (i.e. at galactic scales) which can be solved if the DM undergoes large self-interactions \cite{Kauffmann:1993gv,Zavala:2009ms,Zwaan:2009dz,Moore:1999nt,Klypin:1999uc,BoylanKolchin:2011de,BoylanKolchin:2011dk,Tollerud:2014zha,Garrison-Kimmel:2014vqa,Dubinski:1991bm,Navarro:1995iw,Navarro:1996gj,Flores:1994gz,Moore:1994yx,Moore:1999gc,Tulin:2017ara,Kaplinghat:2015aga,Kamada:2016euw,Burkert:1995yz,McGaugh:1998tq,vandenBosch:2000rza,Borriello:2000rv,deBlok:2001hbg,deBlok:2001rgg,Marchesini:2002vm,Gentile:2005de,Gentile:2006hv,KuziodeNaray:2006wh,KuziodeNaray:2007qi,Salucci:2007tm,Oh:2015xoa,Oman:2015xda,deNaray:2009xj,Navarro:1996gj,Bullock:1999he}. A theory of a strongly interacting DM is difficult to build if substantially connected to the VS. Indeed, such strong self-interactions between DM particles could have many imprints on different experiments such as direct detection experiments or could have an impact on astrophysical observations (e.g. supernovae). As a direct consequence, the very simple idea of a relatively strongly interacting HS evolving on its own and feebly coupled to the VS emerges naturally from all possible observations and it seems absolutely necessary to properly investigate this possibility.
\\

All the above brings a series of new questions concerning in particular the portal, "How strongly connected to the VS, the DM can be?", and on the structure of the HS itself, "How strongly self-interacting the HS has to be?". Those two fundamental questions, albeit not new, are still unclear now. These are the main questions we will ask in this thesis together with the question of knowing how we can account for the observed DM relic density in frameworks based on a HS. Thus, this thesis is all about the dynamics of such models. The aim is really to be able to establish the simple and minimal models of DM connected to the VS through a portal which could answer these questions or, at least, clarify the situation.
\\

The first two chapters, "\textbf{Part I}", are essentially introductory. To start this study, in \textbf{Chapter} \ref{ch:constr} we first present all relevant evidences favouring the hypothesis of DM as well as arguments pointing out the possibly self-interacting behaviour of this DM. We will also see how such scenarios are constrained in many ways explaining the physical principles these constraints are based on. To discuss this in a concrete way, we will consider two well-known HS portal models. One is based on a Higgs portal and one on a kinetic mixing portal. These models contain a light mediator in the HS, which is fully relevant for both the DM relic density and the DM self-interactions issues, as we will see. \textbf{Chapter} \ref{ch:som} will be dedicated to the analysis of the Sommerfeld effect, which is a physical phenomenon highly relevant for all these constraints especially for having large enough self-interactions thanks to the presence of the light mediator. The next two chapters, "\textbf{Part II}", concern essentially the DM relic density issue in HS frameworks. In \textbf{Chapter} \ref{ch:prod} we will see both qualitatively and quantitatively how this needed HS can be produced and along which dynamical production mechanisms. We investigate the full allowed parameter space of a generic thermal DM candidate living within a HS in \textbf{Chapter} \ref{ch:secl}. In the last two chapters, "\textbf{Part III}", we come back to the self-interactions issue proposing ways to fulfil all constraints discussed in Part I. In \textbf{Chapter} \ref{ch:TpT}, we propose a solution which is based on chapter \ref{ch:secl}: a HS which thermalises within itself, but not with the VS thermal bath. The last chapter, \textbf{Chapter} \ref{ch:other_minimal} will allow us to explore other minimal solutions which are also perfectly compatible with the idea of a HS. Finally, we draw our conclusions.
\\

Before going deeper in the subject of this work, we give here the list of preprints and published articles, sorted by chronological order, on which this thesis is based:

\begin{itemize}
\item T. Hambye, M. H. G. Tytgat, J. Vandecasteele, and \textbf{L. Vanderheyden}, “Dark matter direct detection is testing freeze-in”, \textit{Physical Review D}, vol. 98, no. 7, p. 75017, 2018. \\
DOI : 10.1103/PhysRevD.98.075017 - arXiv: 1807.05022.
\item T. Hambye, M. H. G. Tytgat, J. Vandecasteele, and \textbf{L. Vanderheyden}, “Dark matter from dark photons: a taxonomy of dark matter production”, \textit{Physical Review D}, vol. 100, no. 9, p. 95018, 2019. \\
DOI : 10.1103/PhysRevD.100.095018 - arXiv: 1908.09864.
\item  T. Hambye and \textbf{L. Vanderheyden}, "Minimal self-interacting dark matter models with light mediator", \textit{Journal of Cosmology and Astroparticle Physics}, vol. 05, 001, 2020. \\
DOI : 10.1088/1475-7516/2020/05/001 - arXiv : 1912.11708.
\item M. Lucca, T. Hambye and \textbf{L. Vanderheyden}, "Dark matter as a heavy thermal hot relic", \textit{Physics Letters B}, vol. 807, p. 135553\\
DOI : 10.1016/j.physletb.2020.135553 - arXiv : 2003.04936
\item R. Coy, T. Hambye, M. H. G. Tytgat, and \textbf{L. Vanderheyden}, “The domain of thermal dark matter candidates” \\
DOI : [Submitted in PRD] - arXiv: 2105.01263.
\item \textbf{L. Vanderheyden}, “Dark matter from dark photons”, [Contribution to the 2021 EW session of the 55th Rencontres de Moriond] \\
DOI : / - arXiv: 2105.07039.
\end{itemize}

There is another published article that we did not discuss in this thesis as it concerns a work did during my master thesis and is not directly related to HS nor to self-interacting DM scenarios. However, one can learn more about neutrino line produced by DM annihilation in:

\begin{itemize}
\item C. El Aisati, C. Garcia-Cely, T. Hambye, and \textbf{L. Vanderheyden}, "Prospects for discovering a neutrino line induced by dark matter annihilation", \textit{Journal of Cosmology and Astroparticle Physics}, vol. 10, 021, 2017.\\
DOI : 10.1088/1475-7516/2017/10/021 - arXiv : 1706.06600
\end{itemize}

\part{The Self-Interacting Dark Matter with light mediator problematic}
\chapter{SIDM with a light mediator: a highly constrained scenario}\label{ch:constr}
\yinipar{I}n this first introductory chapter, we will review the main constraints applying on self-interacting DM scenarios (SIDM) starting with historical evidences which brought the whole idea of DM and of SIDM. The following discussion is based on already well-established works as well as on personal contributions which have been published \cite{Hambye:2018dpi,Hambye_2020}.

\section{Gravitational evidences for DM}\label{sec:CST-relic_censity}
In the first half of the twentieth century, two physicists pointed out that we could have missed a large proportion of the gravitational mass of the observable Universe. The Dutch Jan Oort and the Swiss Fritz Zwicky studied the velocity of extra-galactic objects and found inconsistencies with Newton's law of gravity. While the first one studied the motion of stars in galaxies and the second the motion of galaxies in the Coma cluster\footnote{A cluster is a group of galaxies bounded by gravity.}, they both used the Doppler shift to determine the velocity of these luminous objects.

\subsection{Velocities in galaxy clusters}
The rotation velocity of galaxies in such clusters is not arbitrary. Starting from Newton's law and Virial's theorem, one gets that the time average of the kinetic energy of all particles in a system (the particles are galaxies in the study of Zwicky) is equal to the opposite of the time average of the potential energy of the same particles,

\myeq{
\left\langle\sum _{i}m_{i}\vert\vec{v}_{i}\vert ^{2}\right\rangle = -\left\langle\sum _{i}\vec{r}_{i}\cdot\vec{F}_{i}\right\rangle, \label{eq:virial}
}

\noindent where the brackets ``$\left\langle\right\rangle$'' indicate a time average quantity, $m_{i}, \vec{r}_{i}, \vec{v}_{i}$ are the mass, the position and the velocity of the particle (galaxy) i and $\vec{F}_{i}$ is the total force applying on the particle i. As the observable Universe is homogeneous and isotropic at large scale, such as galaxy cluster scales, and as the distribution of these galaxies in the cluster was presenting a global spherical symmetry, Fritz Zwicky could apply Eq. \ref{eq:virial} to the gravitational force (i.e. $\vec{F}_{i}=G\sum_{j}m_{i}m_{j}\vec{r}_{ij}/\vert\vec{r}_{ij}\vert ^{3}$)\footnote{Where we define $\vec{r}_{ij}\equiv \vec{r}_{i} - \vec{r}_{j}$ and where the gravitational constant is $G=6.674\times 10^{-11}$ m$^{3}$kg$^{-1}$s$^{-2}$.} and easily conclude that velocity of those galaxies within the Coma cluster were inversely proportional to the square root of their distance to the gravitational centre of the cluster. He got,

\myeq{
\vert \vec{v}_{i}\vert \propto \frac{1}{\sqrt{\vert\vec{r}_{i}\vert}}.
}

\noindent He compared velocities obtained by Doppler shift and those obtained with the help of Virial's theorem and found that velocities did not match at all. Even though his measures were not very accurate, he could qualitatively show that the velocity of galaxies was not decreasing with the distance to the centre of the cluster. He observed that galaxies sitting in the periphery of the cluster were moving with velocity up to three times the one predicted by Newton's law. He could not figure out how this matter could move so fast without being ejected from the galaxy. Indeed, these observed velocities are much bigger than the predicted ones. If we believe in the theory of Gravitation and if there is nothing we are missing, with such velocities, stars should have been ejected of galaxies since a long time ago! We have then two solutions: either there is a physical phenomenon, totally unknown, which we are missing, or gravitation's law have to be modified at these relevant scales but unchanged at larger scale (where predictions match with observations). On the one hand, changing gravitation's law at several scales in a consistent way that could account for all observations at the same time is a very challenging option. Until now, this was not successful and there is no clue for this being actually possible. On the other hand, one can easily imagine that, for the first option, if there were more massive objects inside the virtual sphere of radius $r_{\star}$\footnote{$r_{\star}$ being the distance to the cluster centre of a chosen galaxy which appears to move fast.}, the gravitational attraction would be much stronger and this could simply explain why stars are not ejected outside galaxies. Thus, Zwicky understood that a simple way for his observations to match with the predicted velocities was to assume that there was additional gravitational mass within the cluster. He named this missing mass \textit{Dunkle Materie} (Dark Matter or DM) \cite{Zwicky:1933gu} as it was invisible, see also \cite{1932BAN.....6..249O}. It was one of the very first time, the idea of DM was introduced. Since Oort and Zwicky many scientists worked on this topic and performed the same measurement on many more objects with more and more precision. This leads us to the very first quantitative evidence for DM, galaxy rotation curves.

%\yinipar{T}here exist several motivations favoring the hypothesis of DM: galaxy rotation curves, bullet cluster, CMB anisotropies, ... Even if the most stringent one comes from the study of CMB spectrum, we will review here the historical arguments (galaxy rotation curves and bullet cluster). 
%As CMB anisotropies provide more than a unique constraint on the DM abundance,it will be detailed in a separate section entirely dedicated to the CMB (see Section \ref{sec:CST-CMB}).

\subsection{Rotation curves}\label{subsec:rot}
\begin{figure}
\centering
\includegraphics[scale=0.4]{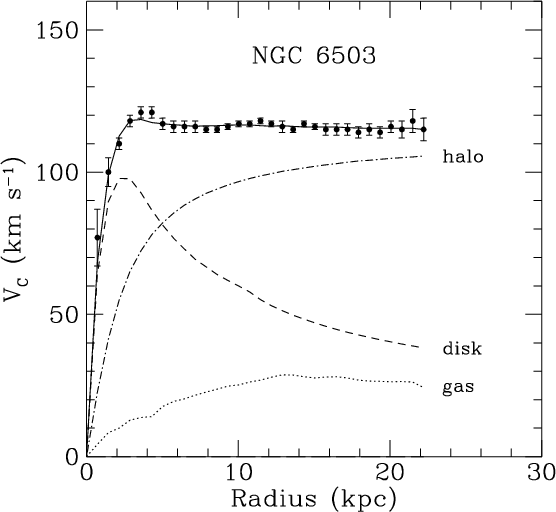}
\caption[Velocity of stars in NGC 6503 as a function of their distance to the centre]{Velocity of stars in NGC 6503 as a function of their distance to the centre, see \cite{1991MNRAS.249..523B}. The dots show data points extracted from observations while the lines indicate what should be expected for a galaxy made of a gas (dotted), a disk of matter (dashed) or a halo of DM (dot-dashed).}\label{fig:rotation-curve}
\end{figure}

Vera Rubin \cite{1970ApJ...159..379R}, an American physicist, was the first to perform a precise analysis of galaxy rotation curves and could quantify for the first time the discrepancies. Her measurements of the mass-to-luminosity ratio show that there is about five times more mass than expected from visible matter. As an example, Figure \ref{fig:rotation-curve} shows measured velocities (dots) of stars in NGC 6503 (taken from \cite{1991MNRAS.249..523B}), predicted velocities for visible matter, gas and a halo of DM are respectively shown in dashed, dotted and dot-dashed. One can see that, for this galaxy, introducing a halo of additional matter interacting only through gravity would explain the observed curve. At the time of those first quantitative measurements, this was only an example, for now, many more galaxies have been studied in great details and most of them are suggesting additional invisible matter (see e.g. \cite{Trimble:1987ee,Salucci:1996bf} and papers therein).

\subsection{Bullet cluster}\label{subsec:bullet}
\begin{figure}
\centering
\includegraphics[scale=0.42]{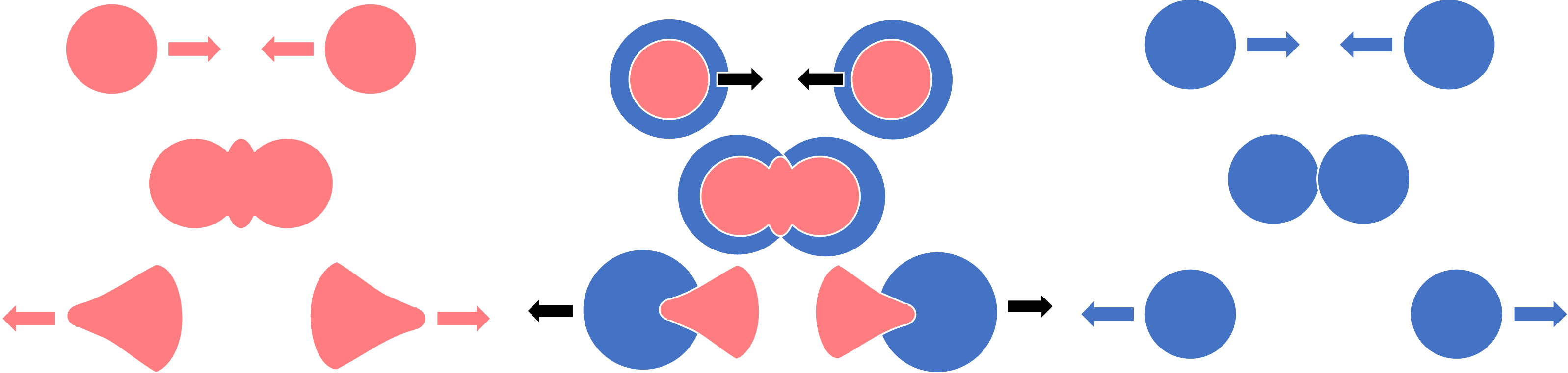}
\caption[Illustration of the offset between interacting and non-interacting matter after a cluster collision]{Illustration of the offset between interacting (red) and non-interacting (blue) matter after a cluster collision.}\label{fig:cluster_draw}
\end{figure}

The introduction of DM from the rotation curves argument may seems a little bit artificial or even unconvincing, albeit chronologically the first one, but this argument is far to be the only one. A little bit less than twenty years ago, the space telescope Chandra has observed vestiges of galaxy collisions in the Bullet cluster. Again, observations pointed out existence of additional invisible matter interacting only through gravity. In order to understand how such observations led to this conclusion, let us first resume how objects made of visible matter collide. Imagine two spheres of matter which are attracted to each other because of gravity, if these spheres are made of interacting matter, they can interact and there will be many impacts during collision between constituents. After the collision, shapes will be highly modified due to these impacts, this is represented in red in Figure \ref{fig:cluster_draw}. If these spheres interact only through gravity (as it is supposedly the case for DM or stars), there will be no contact interaction and spheres will only go through each other without actually "seeing" the other one. Shapes will remain spherical after the collision since the gravitational interaction is much feebler than all other interactions\footnote{The strength of the electromagnetic force between two protons is $\sim\mathcal{O}(10^{36})$ stronger than the strength of the gravitational force at the same distance.} (see blue in Figure \ref{fig:cluster_draw}).
\\

\begin{figure}[t]
\centering
\includegraphics[scale=0.2]{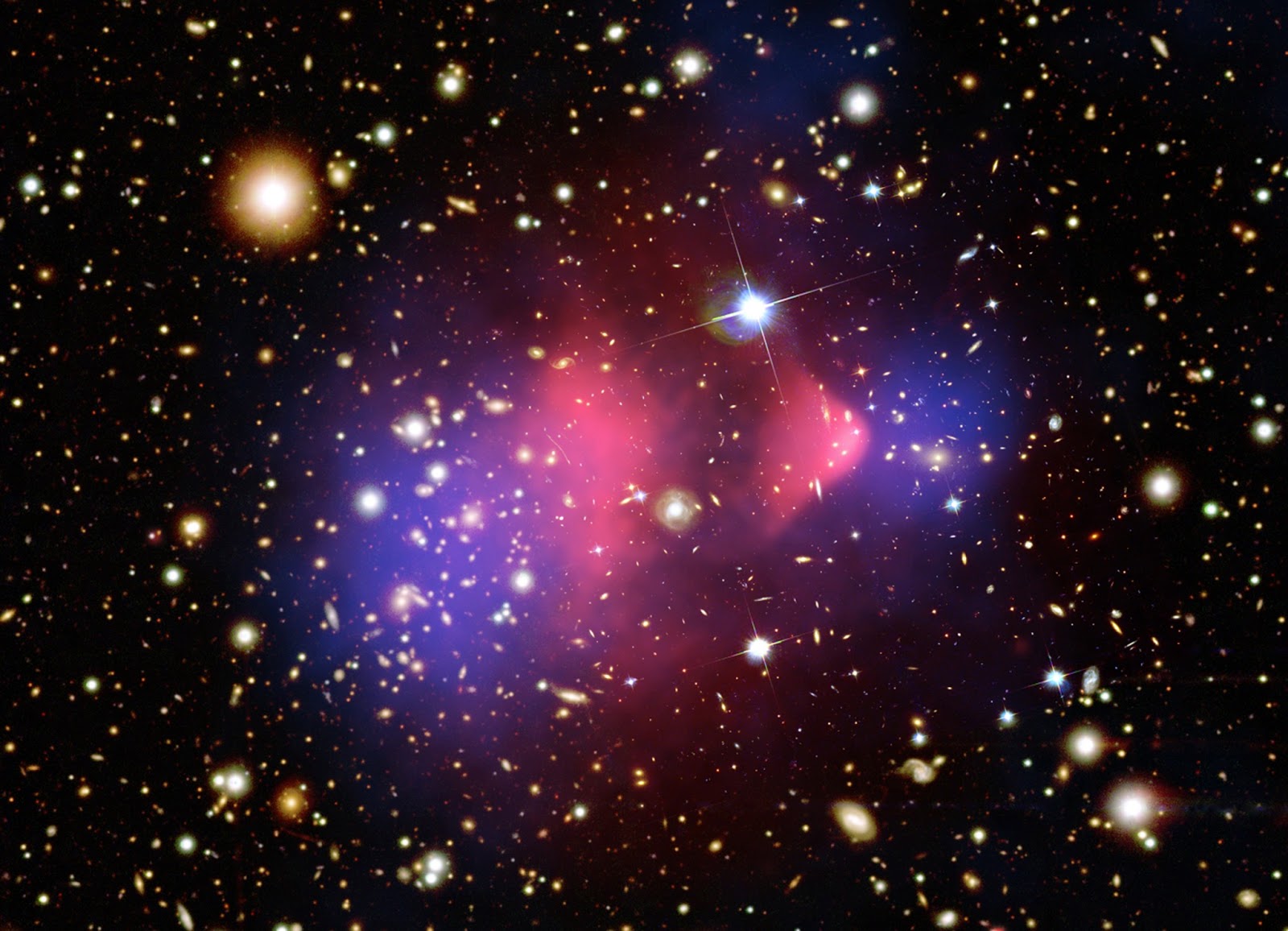}
\caption[Representation of the Bullet cluster]{Representation of the Bullet cluster. The pink area shows X-ray emission from visible matter while the blue area shows a reconstruction of the total mass from measurements of gravitational lensing. Credit: X-ray: NASA/CXC/CfA/ M.Markevitch et al.}\label{fig:cluster}
\end{figure}

If both visible and invisible matter are present, even if the centre of mass of both kind of matters remains unaffected after the collision, one should expect a shift between distributions of visible and invisible matter after the collision. In other word, if one is able to determine the centre of mass of the entire system (= visible and invisible matter), one should see that it is aligned with the centre of mass of stars and not with the centre of mass of the supposedly dominant component of the cluster, the gases. This is exactly what the space telescope Chandra observed in 2005 \cite{Vikhlinin:2005mp,Clowe:2006eq}, see Figure \ref{fig:cluster}.
\\

Figure \ref{fig:cluster} shows reconstructed data measured by the Chandra telescope. It represents a collision of galaxies in the Bullet cluster. Galaxies are made of stars and gas, but as distances between stars are huge, they never interact with each other and the stars distributions remain the same after the collision. However, the gases were playing the role of the interacting matter in Figure \ref{fig:cluster_draw} and the gas distribution was modified due to the collision tending to remain closer to the centre of the all system, according to Chandra measurements. Thanks to the gravitational lensing, the collaboration was able to measure the offset between the centre of mass of the luminous matter and the centre of gravity of the whole matter distribution. Indeed, the gravitational lensing techniques use a light property which is known since Einstein's theory of general relativity. In this framework, it is not only the mass energy which interacts with gravity but the whole energy content of an objects. Thus, even a light beam (made of massless particles, the photons) would be bended when passing close to massive objects. This bending intensity will be directly related to the mass of the gravitational source around which the light beam is travelling. One very powerful application of this effect is that if there is a luminous source behind (with respect to us) a massive (visible or invisible) object, one can determine the mass of this object by measuring the bending intensity of the light coming from behind the massive object. It is thanks to this technique that the distribution of invisible matter has been determined and that the offset between luminous matter and non-luminous matter in the Bullet cluster has been resolved.
\\

Moreover, it has also been shown that this very same offset cannot be explained by a modification of the theory of gravity \cite{Clowe:2006eq} such that the favoured conclusion of the Chandra collaboration results were the existence of DM.

\subsection{Cosmic Microwave Background}
We will now review the whole physics behind the Cosmic Microwave Background or CMB and we will see how its measurement today gives an additional strong hint for the existence of DM.

\subsubsection{Hydrogen recombination}
During the early Universe epoch, the temperature was so high that atoms could not stay bounded for long. Every time an electron ($e^{-}$) and a proton ($p^{+}$) interact together to form a bound state, a hydrogen atom ($H$), the very high kinetic energy of the bound state was high enough to make it unstable and break it. The Universe was then mostly composed of charged particles (electrons and protons\footnote{There were also light nuclei like we will see in the section dedicated to BBN, see Section \ref{sec:CST-BBN}. However, this has a negligible impact in this context.}) and photons ($\gamma$) always interacting strongly with the medium. But, as we learned from the Universe expansion, the temperature eventually dropped below the hydrogen ionisation temperature. The corresponding process (see Eq. \ref{eq:CST-ion}) frozen out and photons decoupled from the thermal bath (finally composed of neutral particles), evolving freely. The Universe became transparent. This is known as the recombination. This last emission happened when Universe's temperature was about the hydrogen ionisation energy, $E_{I}^{H}=13.6$ eV, or actually $\simeq 0.3$ eV because at the time there were already many more photons than electrons and protons so that the dissociation process stopped later than 13.6 eV.

\myeq{
e^{-} + p ^{+} \leftrightarrow H + \gamma\label{eq:CST-ion}
}

\begin{figure}
\centering
\includegraphics[scale=0.20]{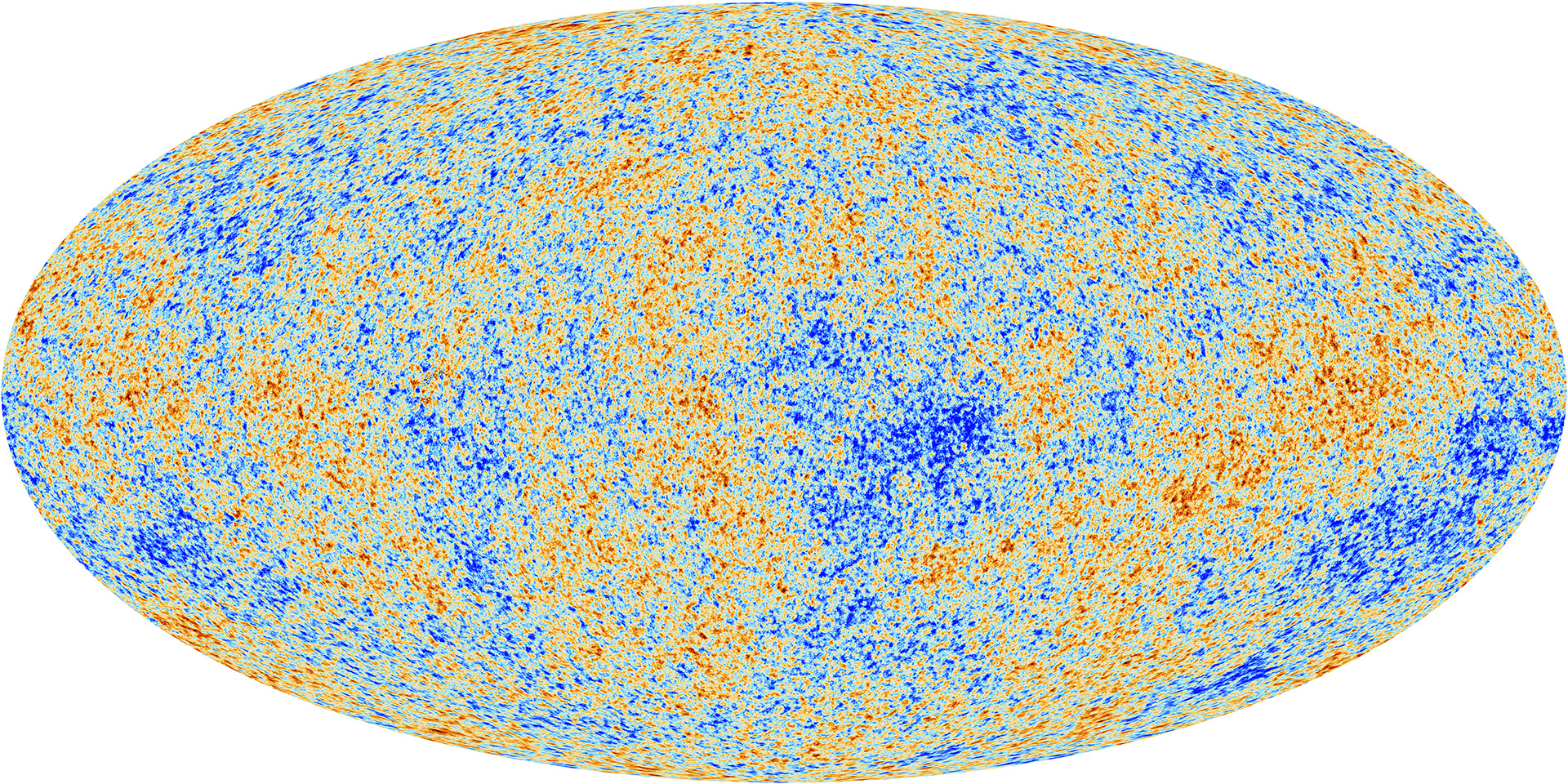}
\caption[Last scattering surface]{Last scattering surface seen by the Planck collaboration. Taken from \cite{PlanckPicture}} \label{fig:CST-planck}
\end{figure}

\noindent Photons which decoupled from the early Universe were then able to pursue their travel following the direction and the distribution they had just after their last scattering process. They compose the so called CMB and can be detected today. The last scattering surface is the "picture" of the actual measurement of the temperature of those photons. Figure \ref{fig:CST-planck} shows the CMB temperature distribution measured today by the Planck collaboration (picture taken from \cite{PlanckPicture}).
\\

In order to understand this picture, let us review the physics behind it. Making the hypothesis that the recombination was dominated by the hydrogen, Eq. \ref{eq:CST-ion}. At very high temperature, when $T\gg E_{I}^{H}$, all particles involved were in chemical equilibrium, and we had then the following relation between chemical potentials: $\mu_{e} + \mu_{p} = \mu_{H} + \mu_{\gamma}$, with $\mu_{x}$ the chemical potential of the $x$ species. If these quantities can take many different values in general, the photons chemical potential is always zero due to the non-conservation of the photon number\footnote{Looking at the following process: $e^{-}+\gamma \leftrightarrow e^{-}+\gamma+\gamma$. We have then $\mu_{e}+\mu_{\gamma}=\mu_{e}+2\mu_{\gamma}$ which implies that $\mu_{\gamma} = 0$.}. When the temperature reached the hydrogen ionisation energy, $T\simeq E_{I}^{H}$, the freeze-out of this process happened. At this moment, since all massive particles involved in the process ($e,p$ and $H$) had their mass being larger than the temperature ($m_{H}\simeq m_{p}\gg m_{e}\gg T\simeq 0.3$ eV), they all were non-relativistic and were following a Maxwell-Boltzmann distribution. The photons, as massless particles, are of course relativistic and follow the Bose-Einstein distribution, we write

\myeq{
n_{x} &= g_{x}\left(\frac{m_{x}T}{2\pi}\right)^{3/2}e^{-\frac{m_{x}-\mu_{x}}{T}},\hspace{1cm}\text{for }x=e^{-},p^{+},H,\\
n_{\gamma} &= g_{\gamma}\frac{\xi (3)}{\pi^{2}}T^{3},
}

\noindent where $\xi (3)\simeq 1.202$ and with $m_{x}$ and $g_{x}$ the mass and the degrees of freedom of the $x$ species respectively. Moreover, since we have $E_{I}^{H}\equiv m_{H}-m_{p}-m_{e}$ and $m_{p}\simeq m_{H}$, one can write Saha's equation,

\myeq{
\left(\frac{n_{e}n_{p}}{n_{H}}\right)_{eq}\simeq\eta\left(\frac{m_{e}T}{2\pi}\right)^{3/2}e^{-E_{I}^{H}/T},
}

\noindent where the "$eq$" subscript refers to quantities taken at equilibrium and where the baryon-to-photon ratio is given by $\eta\equiv\frac{n_{B}}{n_{\gamma}}$ in terms of the baryonic number which is approximately given by $n_{B}\simeq n_{p}+n_{H}$ since there were almost no other baryon at this time.
\\

We can look at the degree of ionisation of the bath by comparing the amount of (free) protons and hydrogen atoms (bounded protons) evolving in the bath: $X\equiv\frac{n_{p}}{n_{p}+n_{H}}$. This number can be related to the baryon-to-photon ratio, we have: $X\simeq\frac{1}{\eta}\frac{n_{p}}{n_{\gamma}}$. Assuming the photons kinetic equilibrium and adding the neutrality requirement\footnote{The Universe appears to be globally neutral today and thus, by charge conservation, was always globally neutral.} (i.e. $n_{e}=n_{p}$), we finally get,

\myeq{
\frac{X_{eq}^{2}}{1-X_{eq}}=\frac{\sqrt{\pi}}{4\sqrt{2}\eta\xi (3)}\left(\frac{m_{e}}{T}\right)^{3/2}e^{-E_{I}^{H}/T}.
}

\noindent In order to describe the photons decoupling, one can use a tool that we will develop in more details in Chapter \ref{ch:prod}, known as Boltzmann equation (Eq. \ref{eq:BE-n}) for electrons, protons and hydrogen atoms. In substance, the Boltzmann equation expresses the time evolution of the number density of a given species. It takes into account interactions the species could have and the dilution factor due to the expansion of the Universe. One has,

\myeq{
\frac{\diff n_{e}}{\diff t} +3H n_{e} &= n_{e}^{eq}n_{p}^{eq}\left\langle\sigma v\right\rangle_{ep\rightarrow H\gamma}\left(\frac{n_{H}n_{\gamma}}{n_{H}^{eq}n_{\gamma}^{eq}}-\frac{n_{e}n_{p}}{n_{e}^{eq}n_{p}^{eq}}\right),\label{eq:BE-e}\\
\frac{\diff n_{p}}{\diff t} +3H n_{p} &= n_{e}^{eq}n_{p}^{eq}\left\langle\sigma v\right\rangle_{ep\rightarrow H\gamma}\left(\frac{n_{H}n_{\gamma}}{n_{H}^{eq}n_{\gamma}^{eq}}-\frac{n_{e}n_{p}}{n_{e}^{eq}n_{p}^{eq}}\right),\label{eq:BE-p}\\
\frac{\diff n_{H}}{\diff t} +3H n_{H} &= n_{H}^{eq}n_{\gamma}^{eq}\left\langle\sigma v\right\rangle_{ep\rightarrow H\gamma}\left(\frac{n_{e}n_{p}}{n_{e}^{eq}n_{p}^{eq}}-\frac{n_{H}n_{\gamma}}{n_{H}^{eq}n_{\gamma}^{eq}}\right)\label{eq:BE-H},
}

\noindent where $\frac{\diff}{\diff t}$ refers to the time total derivative while $H\equiv\frac{1}{a}\frac{\diff a}{\diff t}$, the Hubble parameter, is the total expansion rate of the Universe with $a$ the scale factor. From Eqs. \ref{eq:BE-e}-\ref{eq:BE-H}, neutrality, photons kinetic equilibrium and matter conservation\footnote{The baryonic density Liouville operator has to be null: $\diff n_{b}/\diff t+3Hn_{b}=0$ since there is no source of baryonic matter at this stage.}, we have

\myeq{
\frac{\diff X}{\diff t} = \left\langle\sigma v\right\rangle_{ep\rightarrow H\gamma}\left(n_{b}\frac{X_{eq}^{2}}{1-X_{eq}}\left(1-X\right)-X^{2}n_{b}\right).
}

\noindent The cosmological redshift $z$, which is defined as the ratio of the observed wavelength and the emitted wavelength $1+z\equiv\lambda_{0}/\lambda$, is often used as a measure of the time. Indeed, because of the expansion of the Universe, objects are moving away to each other. Moreover, thanks to the Hubble law, we also know that more objects are far away from us, more they are moving away quickly. Thus, if one can measure the velocity at which an object is moving away from us, one is able to determine how far it is and thus one is able to determine the age of the image we are seeing. This is again due to the expansion of the Universe. More an object is far more it is old. At the end of the day, the redshift can be related to the age of the Universe (i.e. the time) and one can finally write the time evolution equation (in terms of the redshift) of the degree of ionisation $X$ of the thermal bath in the early Universe 
%\cite{1964ApJS....9..185B,1991A&A...251..680P}

\myeq{
\frac{\diff X}{\diff z} = C_{r}\frac{a}{H}\left[n_{b}A(z)X^{2}-B(z)\left(1-X\right)\right],\label{eq:BE-Xe}
}

\noindent where we have used the fact that the recombination is the most efficient for a de-excitement from the $2s$ to the $1s$ atomic level at a rate of $\Gamma_{2s}\simeq 8$ Hz \cite{PhysRevA.40.1185},

\myeq{
C_{r}(z) &\equiv \frac{\Gamma _{\alpha}+\Gamma _{2s}}{\Gamma _{\alpha}+\Gamma _{2s}+B(z)e^{E_{I}^{H}}/T(z)},\\
A(z) &\equiv \left\langle\sigma v\right\rangle _{2s}\simeq 9.78\frac{\alpha ^{2}}{m_{e}^{2}}\sqrt{\frac{E_{I}^{H}}{T(z)}}log\left(\frac{E_{I}^{H}}{T(z)}\right),\\
B(z) &\equiv \frac{X_{eq}^{2}}{1-X_{eq}}A(z)n_{b}(z),
}

\noindent with $\alpha \simeq 1/137$ is the fine structure constant and $m_{e} = 511$ keV. The correcting factor $C_{r}$ takes into account the possibility of de-excitements from the $2p$ to the $1s$ atomic level and of Lyman $\alpha$ emissions ($\Gamma_{\alpha}=\frac{9(E_{I}^{H})^{2}H}{2(1-X_{eq})n_{b}}$).
\\

\begin{figure}
\centering
\includegraphics[scale=0.6]{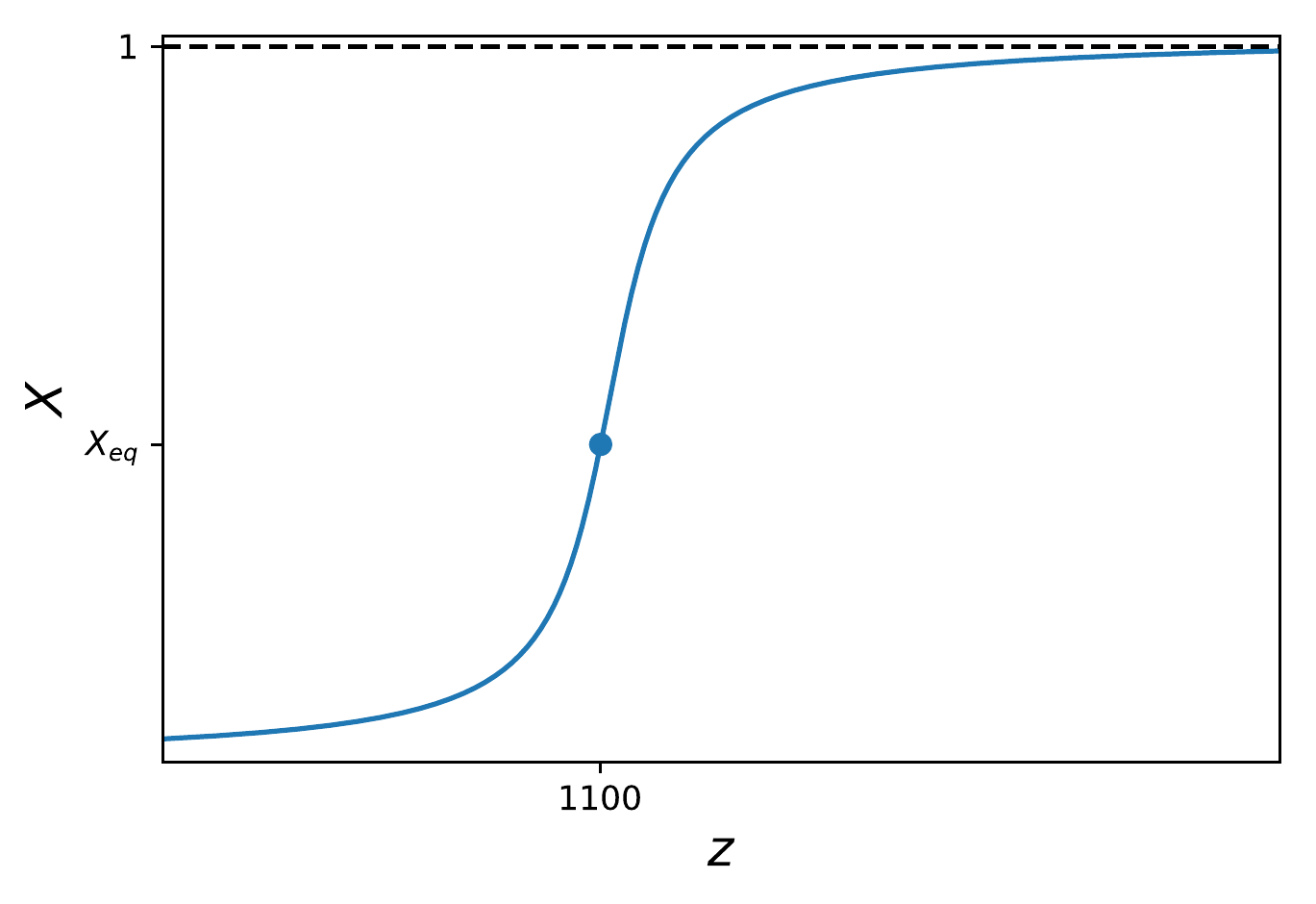}
\caption[Fraction of free electrons in the Universe as a function of the redshift]{Fraction of free electrons in the Universe as a function of the redshift.} \label{fig:CST-ionisation}
\end{figure}

The results of a numerical integration of Eq. \ref{eq:BE-Xe} are shown in Figure \ref{fig:CST-ionisation} where we highlighted when $X=X_{eq}$ which is around $z\simeq 1100$. This instant will be taken as the time of recombination: $z_{rec}\simeq 1100$ since it is the time when the fraction of ionised protons drops below its equilibrium value.

\subsubsection{Optical depth and photons decoupling}
Photons propagation in the early Universe thermal bath is highly related to the number of free electrons, due to the Thomson diffusion: $e^{-}\, +\,\gamma\rightarrow e^{-}\, +\,\gamma$. Above, we have seen that the number of free electrons as a function of the redshift is given by the number of free protons which is related to the fraction number $X$ through the baryon-to-photon ratio and the photon number: $n_{e}(z)=X(z)\eta (z) n_{\gamma}(z)$.
\\

\begin{figure}
\centering
\includegraphics[scale=1]{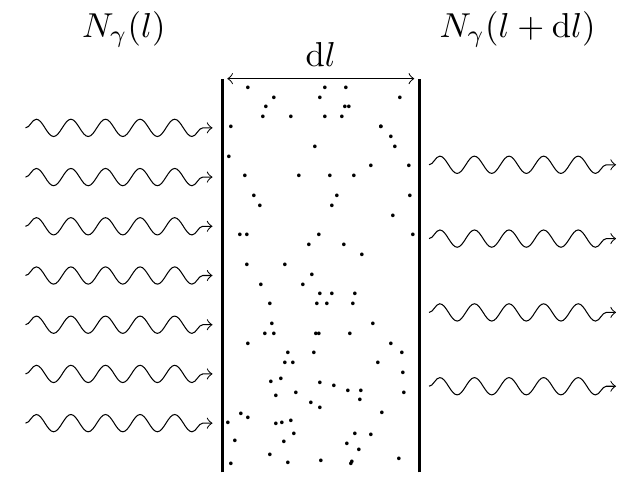}
\caption[Flux of photons going through an electron gas]{Flux of photons going through an electron gas of width $\diff l$.} \label{fig:CST-optical}
\end{figure}

Let us consider the following picture: a photon flux going through an electron gas, see Figure \ref{fig:CST-optical}. With a good approximation, the number of photons which were unaffected by the electron gas $N_{\gamma}(l+\diff l)$ is given by the number of photons before $N_{\gamma}(l)$ minus the number of photons which have been scattered by the electrons in the gas,

\myeq{
N_{\gamma}(l+\diff l) &= N_{\gamma}(l) - \sigma _{Thom.}N_{\gamma}(l)n_{e}(z)\diff l,\label{eq:CST-diff_photons}
}

\noindent where the Thomson diffusion cross section is given by $\sigma_{Thom.}=8\pi\alpha^{2}/3m_{e}^{2}\simeq6.65\times 10^{-25}$ cm$^{2}$ and where $l$ refers to the depth of the gas. From Eq. \ref{eq:CST-diff_photons}, one gets the evolution equation for the photon number:

\myeq{
\frac{\diff N_{\gamma}}{\diff l} &= -n_{e}(z)N_{\gamma}(l)\sigma_{Thom.}.
}

\noindent which can be converted into a differential equation on $N_{\gamma}$ as a function of the redshift $z$ thanks to the relationship between the depth and the redshift ($\diff z = -H\diff l/a$),

\myeq{
\frac{\diff N_{\gamma}}{\diff z} &= \frac{n_{e}(z)\sigma_{Thom.}}{(1+z)H(z)}N_{\gamma}(z).
}

\noindent This equation can be easily solved and gives,

\myeq{
N_{\gamma}(z) &= N_{\gamma}^{0}e^{-\tau (z)},\label{eq:Ng_sol}\\
\tau (z) &= \int_{0}^{z}\frac{n_{e}(x)\sigma_{Thom.}}{(1+x)H(x)}\diff x.
}

\noindent The decreasing exponential in Eq. \ref{eq:Ng_sol} represents the probability for a photon to not scatter between $z=0$ and $z$. Now, we can define the visibility function $g(z)$ as the probability density for a photon from the CMB to scatter one last time at $z\pm\diff z$,

\myeq{
g(z) &\equiv e^{-\tau (z)}\frac{\diff \tau}{\diff z}.\label{eq:visibility}
}

\noindent The thickness of the visibility function gives the thickness of the last scattering surface while the position of its maximum gives the moment where it is most probable for a photon to decouple. Both of those characteristics, the last scattering surface thickness and the redshift at recombination, will be observables and will provide strong constraints as we will see in the following.

\begin{figure}
\centering
\includegraphics[scale=0.6]{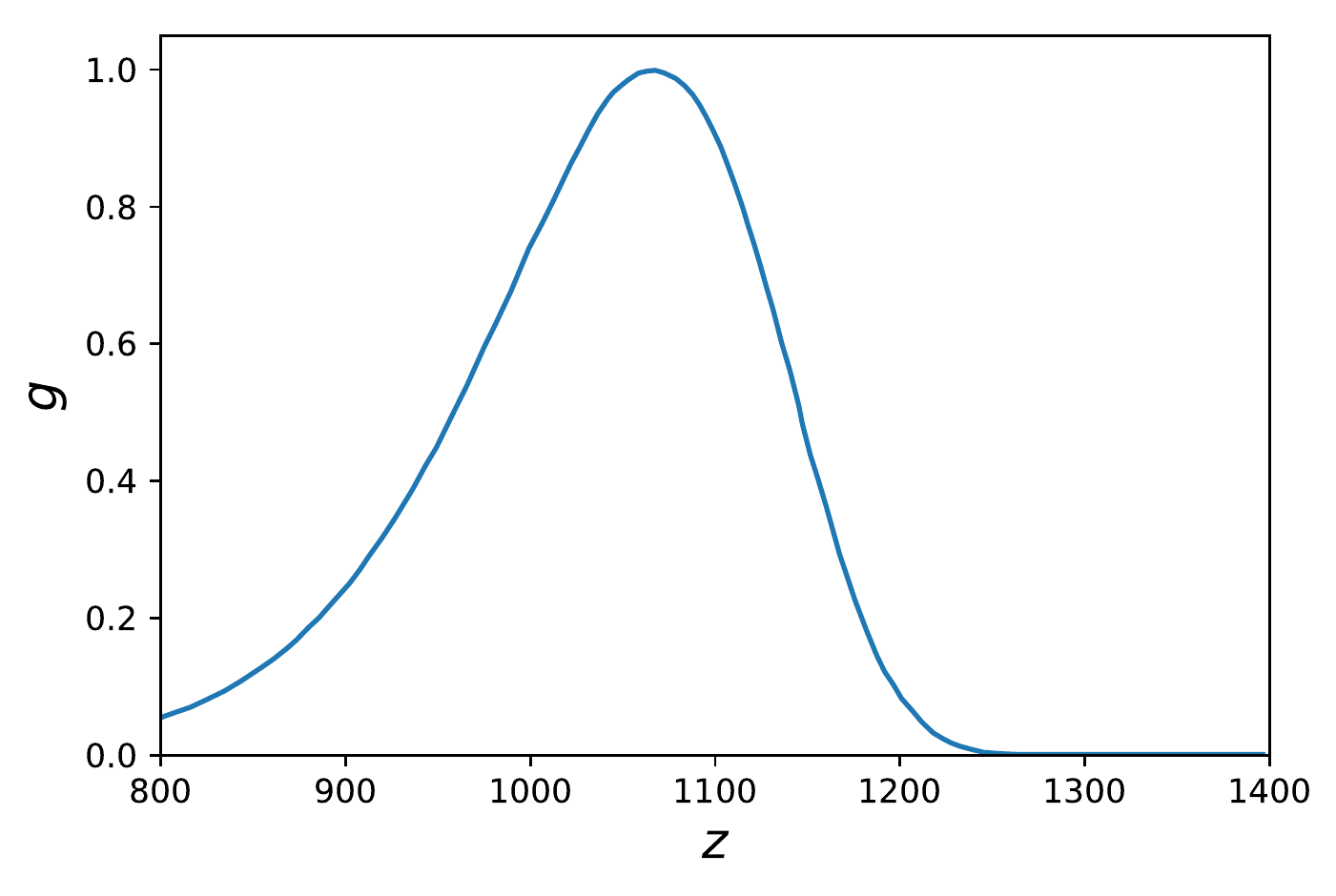}
\caption[Visibility function as a function of the redshift]{Visibility function as a function of the redshift.} \label{fig:CST-visibility}
\end{figure}

\subsubsection{CMB anisotropies}
Eqs. \ref{eq:BE-Xe} and \ref{eq:visibility} give a description of the decoupling and, thus, on how the last scattering surface was formed. Photons emitted from this last scattering surface, known as the CMB as already said above, have a specific spectrum which is measured today by the Planck collaboration. This spectrum gives many crucial information about the Universe from recombination until today. As a simple example, it gives the Earth motion with respect to the CMB. Indeed, there is a Doppler effect affecting the perception of the CMB on Earth because of its motion such that the speed of Earth in CMB's frame can be extracted from the dipolar term. Taking the Earth motion to be small relatively to the speed of light, this effect is simply given in terms of the Earth velocity $v_{\oplus}$, the average temperature $T_{0}$ and the polar angle $\theta$, by:

\myeq{
T &\simeq T_{0}\left(1+\frac{v_{\oplus}}{c}\cos\theta+\mathcal{O}\left(\frac{v_{\oplus}^{2}}{c^{2}}\right)\right).\label{eq:T_cmb_dipolar}
}

\noindent The measurement of the dipolar term gives $\delta T_{dip}=\left(3.346\pm0.017\right)\times 10^{-3}\cos\theta$ K, in Kelvin. We have then, $v_{\oplus}\simeq 368$ km/s\footnote{Note that in Eq. \ref{eq:T_cmb_dipolar} we explicitly used the light velocity $c\equiv299792.458$ km/s such that we express the Earth velocity in km/s. However, if not explicitly said, for the rest of this thesis we will fix $c=1$ such that the velocities given in the following will be expressed in units of $c$.} into the direction $\left(l,b\right)=\left(263.85^{\circ},48.25^{\circ}\right)$ where $l$, the galactic longitude, is the angle between the line of sight and our galaxy plane and $b$, the galactic latitude, is the azimuthal angle around our galaxy rotation axis, see \cite{Bennett:2003bz} for more details.
\\

After subtracting the dipolar term, tiny anisotropies were found and are the part of the CMB spectrum we are interested in. These anisotropies are really small as they are of the order of $\sqrt{\left\langle\left(\delta T/T_{0}\right)^{2}\right\rangle}\simeq 1.1\times 10 ^{-5}$ where we defined,

\myeq{
&\delta T\left(\theta,\phi\right) = T\left(\theta,\phi\right) - T_{0},\\
&T_{0} \equiv \frac{1}{4\pi}\int\int T\left(\theta,\phi\right)\sin\theta\diff\theta\diff\phi,
}

\noindent with $\theta\in\left[0;2\pi\right]$ and $\phi\in\left[0;\pi\right]$ which are the polar and the azimuthal angles respectively (in spherical coordinates). The actual average CMB temperature is known, $T_{0}\simeq 2.725\pm 0.001\text{ K}$.
\\

To understand these very small fluctuations, let us develop the CMB angular power spectrum. Taking $\vec{n}\in S^{2}$ such that $\vert\vec{n}\vert^{2}=1$, we can decompose the temperature contrast on the spherical harmonics basis, $Y^{l}_{n}$,

\myeq{
\frac{\delta T}{T_{0}}\left(\vec{n}\right) &= \sum_{l=0}^{\infty}\sum_{m=-l}^{l}A_{lm}Y^{l}_{n}\left(\vec{n}\right),
}

\noindent where the $A_{lm}\equiv\int\diff \Omega_{\vec{n}}\frac{\delta T}{T_{0}}\left(\vec{n}\right)Y^{\star}_{ln}\left(\vec{n}\right)$ are the coefficient associated to the Fourier transform. A very useful statistical tool is given by the two-points correlation function associated to the measured temperature of the CMB. This function encodes how the measured value of the temperature in two different directions $\vec{n}_{1}$ and $\vec{n}_{2}$ are correlated. This indicates if the temperature distribution is purely random or if there is some correlations even if small. Under the statistical isotropic hypothesis, the two-points correlation function, $C\left(\vec{n}_{1},\vec{n}_{2}\right)$, depends only on one parameter, the relative direction: $\mu\equiv\vec{n}_{1}\cdot\vec{n}_{2}$. It can thus be developed on the Legendre polynomial basis, $P_{l}$,

\myeq{
C(\mu) &= \left\langle\frac{\delta T}{T_{0}}\left(\vec{n}_{1}\right)\frac{\delta T}{T_{0}}\left(\vec{n}_{2}\right)\right\rangle\\
&= \sum_{l=0}^{\infty}\frac{2l+1}{4\pi}C_{l}P_{l}\left(\mu\right).
}

\noindent Then, one can relate the $C_{l}$ coefficients to the coefficient of the Fourier transform which form the so-called angular spectrum of temperature anisotropies,

\myeq{
\left\langle A_{l'm'}^{\star}A_{lm}\right\rangle &= C_{l}\delta_{ll'}\delta_{mm'}.
}

\begin{figure}
\centering
\includegraphics[scale=0.75]{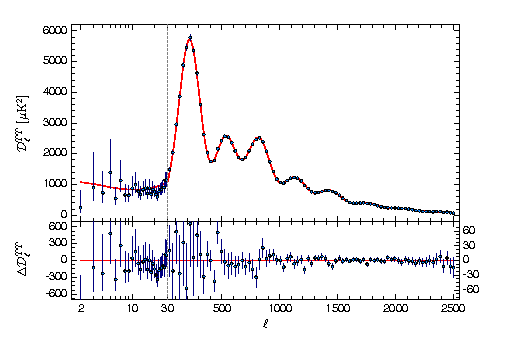}
\caption[CMB temperature anisotropies spectrum]{CMB temperature anisotropies spectrum, with $D_{l}\equiv l(l+1)C_{l}/2\pi$.} \label{fig:CST-CMB-an}
\end{figure}

\noindent Figure \ref{fig:CST-CMB-an} shows the CMB anisotropies angular power spectrum and its very characteristic shape. One can distinguish clearly a first pic at $l\simeq 200$ and that the curve starts to flatten at $l\gtrsim 1000$.

%\begin{figure}
%\centering
%\begin{minipage}{0.45\textwidth}
%  \centering
%\includegraphics[scale=0.1]{Planck_CMB.jpg}
%\end{minipage}
%\begin{minipage}{0.5\textwidth}
%  \centering
%\includegraphics[scale=0.55]{planck.png}
%\end{minipage}
%\caption{CMB.} \label{fig:CMB}
%\end{figure}

%In the previous chapter, we have explained how DM can be constrained with the help of CMB anisotropies measurements. If there is additional light particles, on top of the DM particle, the CMB can constrain further the model. Indeed, every physical phenomenon which would take place in the early Universe and which could modify the CMB could be theoretically probed. Adding a new light degree of freedom which couples to SM electromagnetically charged particles will automatically have an impact on the CMB. In this section, we will review the main constraints which are both on the DM annihilation cross section and on the decay rate (or equivalently the lifetime) of the new light particle \cite{Bernal:2015ova,Bringmann:2016din}.

\subsubsection{Relic density}
\begin{figure}
\centering
\includegraphics[scale=0.70]{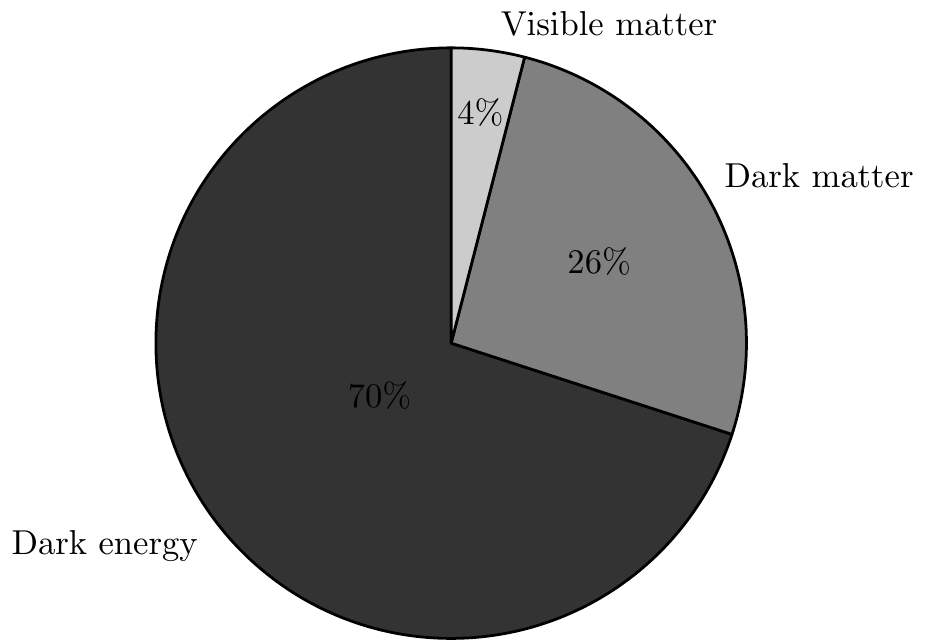}
\caption[Energy density content of the Universe]{Pie chart of the energy density content of the Universe that measurements of the CMB anisotropies pointed out.} \label{fig:Omega_pie}
\end{figure}

\begin{figure}
\centering
\includegraphics[scale=0.35]{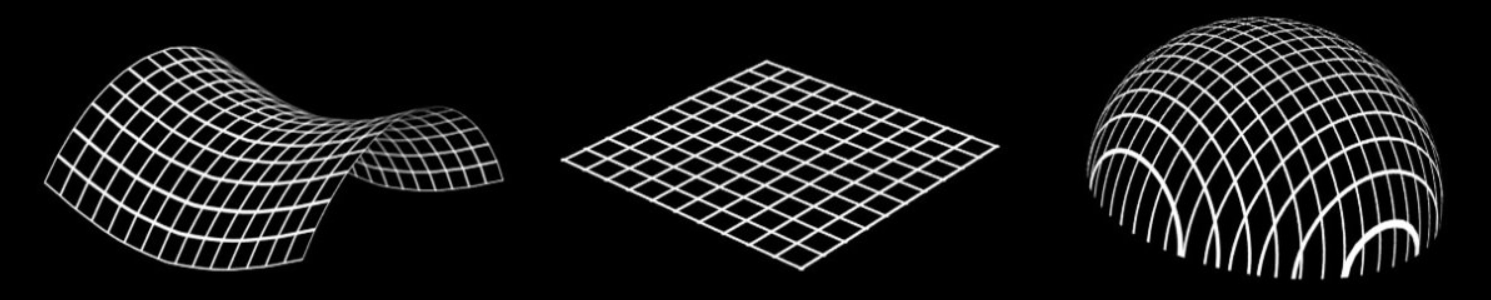}
\caption[Representation of an open/flat/closed Universe]{Two dimensional representation of an open/flat/closed Universe from left to right. CMB anisotropies strongly suggest a flat Universe.} \label{fig:flat}
\end{figure}

The position and the height of the pics visible in Figure \ref{fig:CST-CMB-an} depend on the proportion of each ingredient present in the early Universe (baryon, dark matter, dark energy, radiation). Observations of CMB anisotropies indicate that 26\% (see Figure \ref{fig:Omega_pie}) of the energy content of the Universe has to be in the form of a sector which interacts mostly through gravitational interactions with the rest of the Universe, but which is not ordinary matter as baryons for example, see \cite{Aghanim:2018eyx}. This requirement represents the strongest evidence for the introduction of DM. It also is the first theoretical constraint every DM model will inevitably have to satisfy and it gives in practice two constraints. First, the DM energy density $\rho _{\rm DM}$ constitutes 26\% of the total energy density today. This means that the sum of energy densities of all relic particles should represent 26\% of the whole energy content of the Universe. Thus, all those relic particles should constitute what we call DM. Second, if a model is composed of a more complex structure (i.e. other particles on top of the DM one), the energy density of any of those new particles $\sum_{i}\rho_{i}$ has to be negligible today compared to the DM one (see Figure \ref{fig:flat}), otherwise, it would be part of the DM.
\\

For the rest of this thesis, we will consider cases where there is only one species which constitutes the DM. Other additional particles $i$ will also be considered, but not as part of the DM. We have,

\myeq{
\Omega _{DM} &= 0.2645\pm 0.0050,\label{eq:CST-relic_density}\\
\sum_{i}\Omega _{i} &\ll \Omega _{DM}. 
}

\noindent If the first constraint is generically dubbed as the relic density constraint (as it refers to the energy density of DM particles today), the second one is referred as the overclosure constraint. This comes from the fact that if this last constraint is not satisfied, i.e. if $\sum_{i}\Omega _{i} \gtrsim \Omega _{DM}$, the universe would content too much energy coming from the additional particles $i$ and would overclose ($\Omega _{tot} > 1$). The total energy density today would be bigger than the one expected from standard cosmology which would indicate a closed Universe while it is known to be flat. Indeed, the CMB anisotropies measurements also indicate a flat Universe, see \cite{Aghanim:2018eyx}.

\section{Dark Matter relic density}\label{subsec:HS-rel_ab}
More concretely, in order to get more intuition on how the first constraint plays a role in a given model, let us consider two simple different production mechanisms (they will be detailed later on) as examples. But before proceeding to that we need to develop a coherent and well-established mathematical framework in which one can compute the evolution of a particle number density as a function of time.

\subsection{The Boltzmann equation}\label{sec:be}
The $\Lambda$CDM cosmological model describes an evolving Universe with a "beginning" (the Big Bang) followed by an inflation era. After this period of inflation, the Universe was constituted of a considerably hot and dense plasma. There was no bound state of particles at this stage since the Universe was too hot to allow such state to be stable as bound states would be right away disintegrated by plasma particles with kinetic energy larger than the bound state binding energy. Thus, there were only free elementary particles. One can show that, in the simplest picture one can think of, if 26\% of the energy content of the Universe today is indeed made of DM, the DM should have been already there in the early Universe.
\\

It would be very useful to be able to describe the DM production/depletion mechanism during the whole thermal history of the Universe. Then we would be able to quantify and to track the amount of DM from the early Universe epoch until now. In order to do so, we are going to consider the early Universe as a succession of thermal baths in which most of the particles are in thermal equilibrium. This hypothesis is well justified as the very high temperature favours thermalisation. Thus, it makes sense to use an equilibrium description for the early Universe. To describe accurately the evolution of a species in the thermal bath, one has to solve the Boltzmann equation which describes the microscopic evolution of the particle phase space distribution function. The Boltzmann equation, $\hat{\mathit{L}}\left[f\right]=\hat{\mathit{C}}\left[f\right]$, is built from two operators. On the left-hand side, we have the relativistic Liouville operator, $\hat{\mathcal{L}}$, which describes the global variation of the particle phase space distribution function:

\myeq{
\hat{\mathit{L}} = p^{\alpha}\frac{\partial}{\partial x^{\alpha}} - \Gamma^{\alpha}_{\beta\gamma}p^{\beta}p^{\gamma}\frac{\partial}{\partial p^{\alpha}},
}

\noindent where $x$ and $p$ are the position and momentum four vectors respectively and where $\Gamma^{\alpha}_{\beta\gamma}$ are the entries of the Christoffel symbols:

\myeq{
\Gamma^{\alpha}_{\beta\gamma} = \frac{1}{2}g^{\alpha\delta}\left(\frac{\partial}{\partial x^{\gamma}}g_{\delta\beta}+\frac{\partial}{\partial x^{\beta}}g_{\gamma\delta}-\frac{\partial}{\partial x^{\delta}}g_{\beta\gamma}\right),
}

\noindent with $g$ the metric tensor.
\\

On the right-hand side of the Boltzmann equation, we have the relativistic collision term , $\hat{\mathcal{C}}$, which in general is non-zero thanks to the interactions between particles in the bath. This operator can be very complicated, but under some well motivated assumptions, there is a way to rewrite the Boltzmann equation in a simpler form. Indeed, in the Friedmann-Lemaître-Robertson-Walker (FLRW) framework, the phase space distribution function is isotropic and homogeneous in the spatial directions. One can then integrate both sides of the Boltzmann equation $\hat{\mathit{L}}\left[f\right]=\hat{\mathit{C}}\left[f\right]$ over all possible momenta (i.e. over the whole phase space). The new Boltzmann equation obtained in this way will express the evolution of the number density instead of the phase space distribution function. Moreover, assuming CP invariance and considering Maxwell-Boltzmann distributions for all species in kinetic equilibrium, one can write this evolution equation for a given species $\chi$ under the following form:

\myeq{
\frac{\diff n_{\chi}}{\diff t} + 3Hn_{\chi} = -\left\langle\sigma v\right\rangle_{\bar{\chi}\chi\rightarrow XX}\left[n_{\chi}^{2}-\left(n^{eq}_{\chi}\right)^{2}\right],\label{eq:BE-n}
}

\noindent where $\left\langle\sigma v\right\rangle_{\bar{\chi}\chi\rightarrow XX}$ is the $\chi$'s thermally average annihilation cross section into some finale particles $X$,

\myeq{
\left\langle\sigma v\right\rangle_{\bar{\chi}\chi\rightarrow XX}\equiv\frac{\int\sigma v\,\diff n_{\chi}^{eq}\diff n_{\bar{\chi}}^{eq}}{\int\diff n_{\chi}^{eq}\diff n_{\bar{\chi}}^{eq}},\label{eq:th_av_sv}
}

\noindent with $n_{\chi}^{eq}$ the number density of $\chi$ at equilibrium and where the Hubble parameter can be expressed in terms of the total energy density of the Universe $\rho$ and the Planck mass $M_{pl}$: 

\myeq{
H(t)=\sqrt{\frac{8\pi}{3M_{pl}^{2}}\rho(t)}.
}

\noindent The interpretation of this equation is in fact simple. The $\left\langle\sigma v\right\rangle_{\bar{\chi}\chi\rightarrow XX}n_{\chi}^{2}$ term counts the number of annihilation of $\chi 's$ per unit of time per unit of volume, whereas the $\left\langle\sigma v\right\rangle_{\bar{\chi}\chi\rightarrow XX}\left(n^{eq}_{\chi}\right)^{2}$ term counts the number of production process of $\chi 's$ per unit of time per unit of volume. Equivalently, as the entropy $s$ evolves in the same way as the number density when there is no interaction changing its number, one can introduce the \textit{Yield}, a comoving quantity $Y\equiv n/s$ and $x\equiv m_{\chi}/T$ as well as the relationship between temperature $T$ and time $t$ during the radiation dominated era. This leads to,

\myeq{
\frac{\diff Y}{\diff x} = - \frac{\left\langle\sigma v\right\rangle_{\bar{\chi}\chi\rightarrow XX}s}{xH}\left[Y_{\chi}^{2}-\left(Y^{eq}_{\chi}\right)^{2}\right],\label{eq:BE_yield}
}

\noindent where $Y_{\chi}^{eq}$ is the yield of $\chi$ taken at equilibrium. If there is more than one annihilation or production channel, the equation on the yield gets one additional contribution for each new channel. Eq. \ref{eq:BE-n} will be our Boltzmann equation prototype that we will use and adapt to portal models in the following, see Chapter \ref{ch:som}.
\\

Note that the total energy density is dominated by the contribution of relativistic particles such that one can write,

\myeq{
\rho &= \sum_{i = rel}\frac{\pi ^{2}}{30}g_{i}T_{i}^{4}=\frac{\pi ^{2}}{30}g_{\ast}T^{4},
}

\noindent where we defined the effective number of degrees of freedom taking the difference between Bose-Einstein and Fermi-Dirac statistics into account

\myeq{
&g_{\ast}\equiv\sum_{i = rel}g^{B}_{i}\left(\frac{T_{i}}{T}\right)^{4}+\frac{7}{8}\sum_{i = rel}g^{F}_{i}\left(\frac{T_{i}}{T}\right)^{4}.\label{eq:g_star}
}

\noindent These useful way to condensate the expression of the energy density can also be applied to the entropy density which is also dominated by its relativistic components. We have,

\myeq{
s &= \sum_{i = rel}\frac{2\pi ^{2}}{45}g_{i}T_{i}^{3}=\frac{2\pi ^{2}}{45}g_{\ast}^{S}T^{3},
}

\noindent where this time we have,

\myeq{
&g_{\ast}^{S}\equiv\sum_{i = rel}g^{B}_{i}\left(\frac{T_{i}}{T}\right)^{3}+\frac{7}{8}\sum_{i = rel}g^{F}_{i}\left(\frac{T_{i}}{T}\right)^{3}.\label{eq:g_star_s}
}

\subsection{Freeze-out case}
If one assumes DM to be in thermal equilibrium with the SM in the early Universe, one should have a mechanism which could deplete the large amount of DM it implies. Since DM is actually matter, it will eventually dominate over radiation during the Universe's expansion. One would then require a very small amount of DM just after having decoupled from SM compared to thermal particles, otherwise one would get $\Omega_{\rm DM}\gg 0.26$. In the light of the mathematical framework we have just developed, one possibility would be to use a Boltzmann suppression which arises naturally for thermal particles once the temperature drops below their mass. Indeed, on the one hand, as long as the DM is relativistic, its number density goes like $\sim T^{3}$ such that its yield is constant with time. On the other hand, once the particle becomes non-relativistic, its number density fall exponentially $\sim\left(mT\right)^{3/2}e^{-m/T}$ and its yield with it. This mechanism is the so-called freeze-out (FO) mechanism where a particle (here the DM) is in thermal equilibrium with at least one other lighter species (say $X$). Because of the expansion of the Universe, the Universe's temperature decreases with time and will eventually drops below the DM mass. Then, as long as the DM is in thermal equilibrium, its number density will fall very quickly and, at some point, will become so suppressed that it will not be high enough anymore to satisfy the thermalisation condition. That is to say that the DM changing number reaction ($X\leftrightarrow$ DM) will stop to occur and the DM number density will freeze-out. 
\\

We will see in the third Chapter (see Section \ref{sec:phases}) more specifically that the DM yield can be written, in the instantaneous freeze-out approximation (see also Appendix \ref{app:inst_FO}), in terms of the annihilation cross section which is responsible for the FO mechanism. But this yield can also be easily related to the energy density and then to the DM relic abundance. We have for a s-wave annihilation process,

\myeq{
\Omega_{\text{DM}}h^{2} \simeq 8.77\times 10^{-11}\times\left(\frac{m_{DM}}{T_{\rm dec}}\right)\times\left(\frac{\text{GeV}^{-2}}{\left\langle\sigma v\right\rangle_{T=T_{\rm dec}}}\right)\times\left(\frac{\sqrt{g_{\star}^{eff}(T_{\rm dec})}}{g_{\star}^{S}(T_{\rm dec})}\right),\label{eq:Oh2_FO}
}

\noindent with $T_{\rm dec}$ the temperature of DM decoupling. This constraint fixes the value of the annihilation cross section at decoupling in terms of the DM mass and the decoupling temperature, $T_{\rm dec}$,

\myeq{
\left\langle\sigma v\right\rangle _{T=T_{\rm dec}} \simeq 7.38\times 10^{-10}\text{GeV}^{-2}\times\left(\frac{m_{\rm DM}}{T_{\rm dec}}\right)\times\left(\frac{0.1188}{\Omega_{\text{DM}}h^{2}}\right)\times\left(\frac{\sqrt{g_{\star}^{eff}(T_{\rm dec})}}{g_{\star}^{S}(T_{\rm dec})}\right).\label{eq:sv_FO}
}

\begin{figure}
\centering
\includegraphics[scale=0.70]{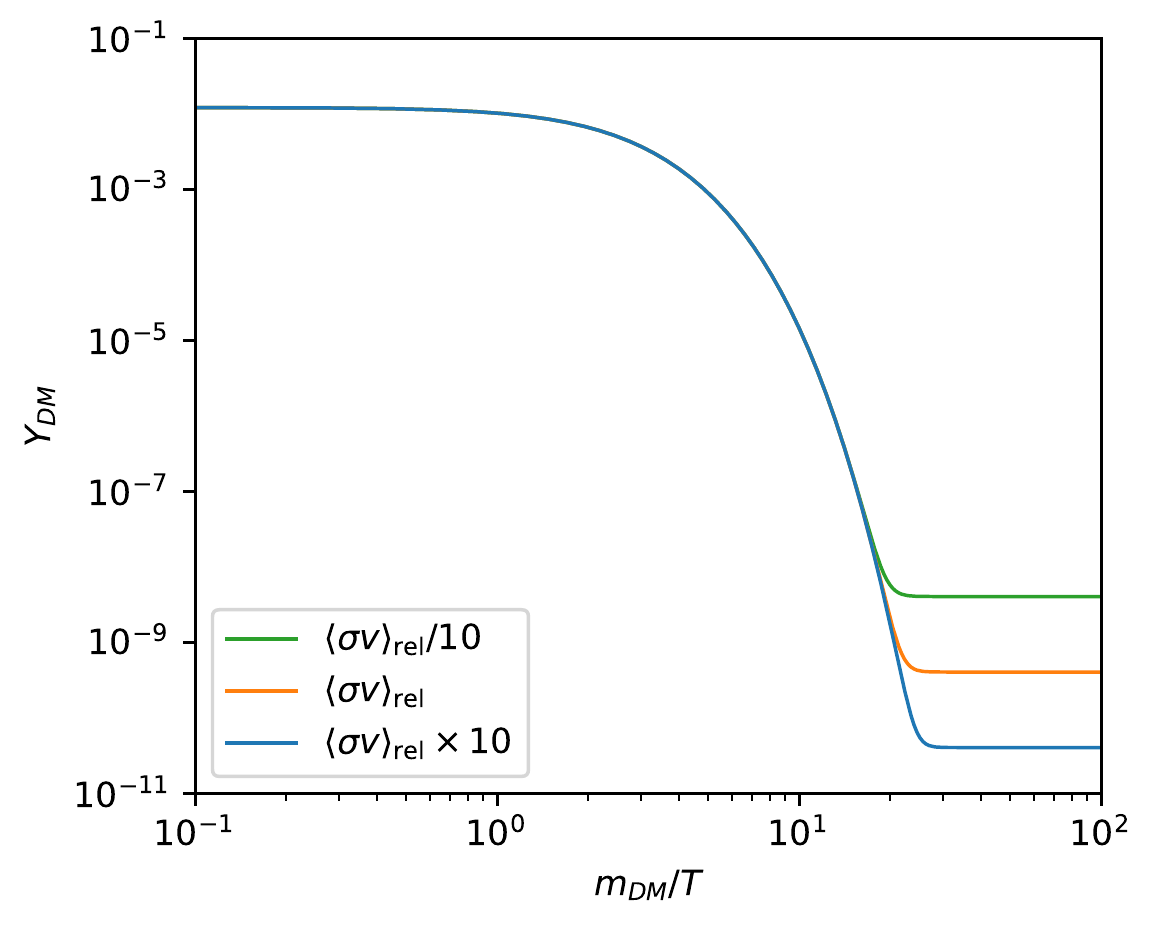}
\caption[DM yield time evolution in a freeze-out scenario]{Example of DM yield evolution as a function of the inverse temperature for three cross sections in a freeze-out scenario. The value of $\langle\sigma v\rangle_{\rm rel}$ has been set by Eq. \ref{eq:sv_FO}.} \label{fig:FO}
\end{figure}

\noindent The $m_{\rm DM}/T_{\rm dec}$ ratio is roughly constant\footnote{It depends on the inputs of the problem only through a logarithm, see Appendix \ref{app:inst_FO}.} and lies around $m_{\rm DM}/T_{\rm dec}\simeq 23$. This implies that the DM annihilation cross section responsible for the FO process should be, at the time of decoupling, of the order of $\left\langle\sigma v\right\rangle \sim 10^{-9}$ Gev$^{-2}$ and this whatever the DM mass and the model specificities. Figure \ref{fig:FO} shows the evolution of the DM yield as a function of $x\equiv m_{\rm DM}/T$ for three different values of the annihilation cross sections. One sees that the more the annihilation cross section is strong the longer the DM stays at equilibrium the more its number density is depleted as explained above.

\subsection{Freeze-in case}
\begin{figure}
\centering
\includegraphics[scale=0.70]{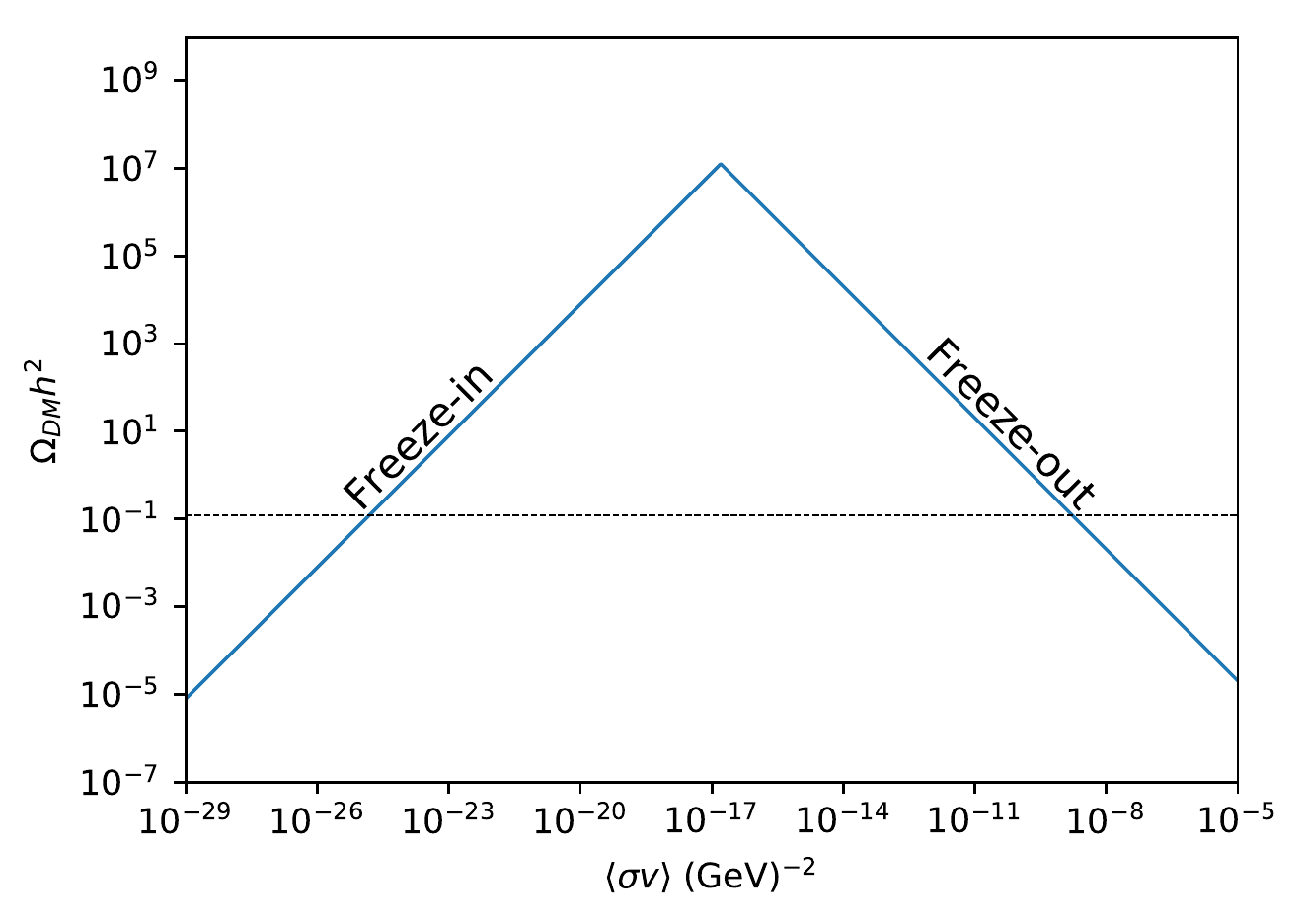}
\caption[DM relic abundance as a function of the cross section]{The freeze-out and freeze-in are two extremes of a same process. One with strong interactions, the freeze-out and the other with feeble interactions, the freeze-in.} \label{fig:volcano}
\end{figure}

Another simple way to have a small amount of DM at decoupling compared to thermal particles would be to have DM to be out of equilibrium. In this way, if the other particle $X$ which couples to DM is in kinetic equilibrium, it will slowly produce DM particles. However, when the temperature reaches the DM mass scale, as the $X$'s are lighter, they will not have enough kinetic energy to produce DM particles and the process stops. The DM number density freezes. This is the so-called freeze-in (FI) mechanism and is the other extreme (and simplest) way to account for the DM relic abundance. Here, the suppression does not come from a Boltzmann suppression as in the FO mechanism, but from the fact that in this case, the relic density is proportional to the production cross section which we assume to be driven by tiny couplings. Figure \ref{fig:volcano} shows the DM relic abundance as a function of the annihilation/production cross section in the instantaneous freeze-out and freeze-in cases. It shows that the FO and the FI mechanisms are two faces of a same phenomenon in which the DM abundance is set by the value of the annihilation/production cross section. The two production mechanisms merge when the cross section starts to be large enough for the DM to thermalise with its partner, it is the top of the "volcano" diagram shown in Figure \ref{fig:volcano}.
\\

We will also see in the following (see Section \ref{subsec:FI}) that the DM yield can be written, in the instantaneous freeze-in approximation, in terms of the production cross section which is responsible for the FI mechanism. Again, this yield can be related to the energy density and then to the DM relic abundance. We have for a s-wave annihilation process,

\myeq{
&\hspace{-0.25cm}\Omega_{\text{DM}}h^{2} \simeq 7.85\times 10^{26}\times\left(\frac{m_{\rm DM}}{\text{GeV}}\right)^{2}\times\left(\frac{\left\langle\sigma v\right\rangle_{T=m_{\rm DM}}}{\text{GeV}^{-2}}\right)\times\left(\frac{1}{g_{\star}^{S}(m_{\rm DM})\sqrt{g_{\star}^{eff}(m_{\rm DM})}}\right).\label{eq:Oh2_FI}
}

\noindent This constraint fixes the value of the production cross section at decoupling in terms of the DM mass,

\myeq{
\left\langle\sigma v\right\rangle _{T=m_{\rm DM}} &\simeq 1.51\times 10^{-28}\text{GeV}^{-2}\times\left(\frac{\text{GeV}}{m_{\rm DM}}\right)^{2}\times\left(\frac{\Omega_{\text{DM}}h^{2}}{0.1188}\right)\nn\\
&\hspace{2cm}\times\left(g_{\star}^{S}(m_{\rm DM})\sqrt{g_{\star}^{eff}(m_{\rm DM})}\right).\label{eq:sv_FI}
}

\begin{figure}
\centering
\includegraphics[scale=0.70]{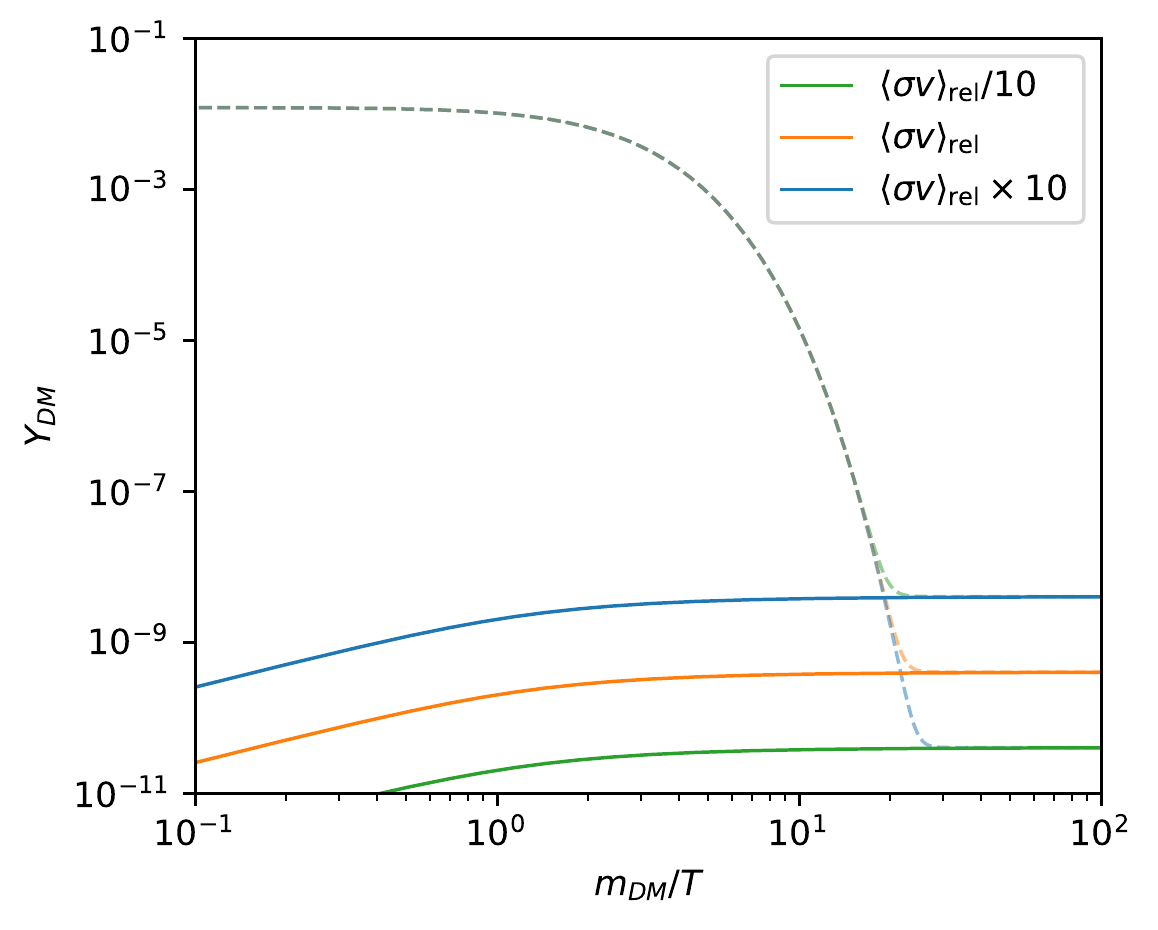}
\caption[DM yield time evolution in a freeze-in scenario]{Example of DM yield evolution as a function of the inverse temperature for three cross sections in a freeze-in scenario. We used $m_{\rm DM} = 1$ GeV.} \label{fig:FI}
\end{figure}

\noindent From this last expression, we can see that, in contrary with the FO case, in the FI case, the required value for the DM production cross section depends on the DM mass. Figure \ref{fig:FI} shows (for $m_{\rm DM} = 1$ GeV) the same as Figure \ref{fig:FO}, but for a freeze-in scenario. One sees from this figure that the stronger the production cross section, the more DM can be created before the process stops, as expected.
\\

These two simplified examples which are the FO and the FI mechanisms show how, given a production mechanism, the relic abundance requirement constrains explicitly a model (i.e. the cross section). When the simplified version is not precise enough and one has to solve the Boltzmann equation, it is solved numerically.

\section{Small scale structure}\label{sec:CST-small_scale_structure}
In section \ref{sec:CST-relic_censity}, we have seen that there are many different hints for DM at large scale in the Universe. In the current section, we will see that there are also hints for DM at small scale, cosmologically speaking. Let us review the most important ones.
 
\subsection{Too-big-to-fail}\label{subsec:too_big}
The \textit{Too-big-to-fail} problem comes from the solution of an older problem of $\Lambda$CDM: the \textit{missing satellites} problem. This last refers to the fact that $\Lambda$CDM simulations predict a significant bigger number of satellite galaxies for Milky-Way like galaxies \cite{Kauffmann:1993gv,Zavala:2009ms,Zwaan:2009dz} than what is actually observed \cite{Moore:1999nt,Klypin:1999uc}. A favoured solution to this problem argues that those satellite galaxies had their stars stripped from them during tidal interactions. Thus, if most stars have disappeared, this could explain why we are not seeing more dwarf galaxies. They are there but simply empty of visible matter.
\\

On the other hand, $\Lambda$CDM simulations predict relatively massive satellite galaxies and it appears that it is very unlikely that such galaxies have no visible stars \cite{BoylanKolchin:2011de,BoylanKolchin:2011dk,Tollerud:2014zha,Garrison-Kimmel:2014vqa}. This failure in the solution of the \textit{missing satellites} problem is what is called the \textit{too-big-to-fail} problem.
\\

\begin{figure}
\centering
\includegraphics[scale=0.5]{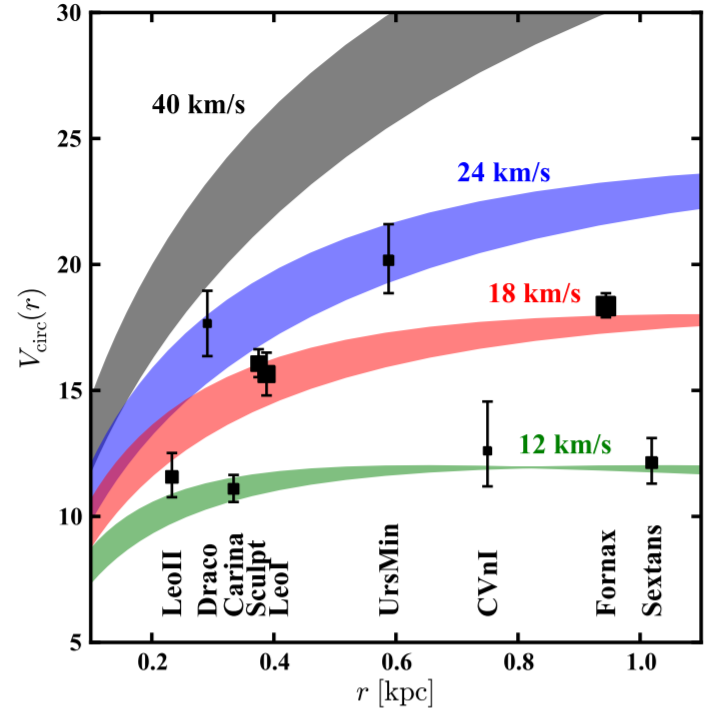}
\caption[Circular velocity of satellite galaxies as a function of their distance to the galactic centre]{Circular velocity of satellite galaxies as a function of their distance to the galactic centre, taken from \cite{BoylanKolchin:2011dk}. One can also distinguish rotation curves corresponding to the specified maximal velocity.}\label{fig:tbtf}
\end{figure}

Figure \ref{fig:tbtf} shows the measured rotation velocity of actual satellite galaxies as a function of their distance to the galactic centre \cite{BoylanKolchin:2011dk}. Even if we agree on the fact that most of the satellite galaxies have lost their stars, we should still be able to observe the heaviest ones. Since high rotation velocities indicate massive galaxies, we can see on this figure that we do not observe such massive dwarf galaxies as we should.

\subsection{Core-vs-cusp}\label{subsec:core}
$\Lambda$CDM simulations show that the DM energy density profile should slope as $1/r^{3}$ in the outer region of the DM halos while it should slope as $1/r$ closer to the centre \cite{Dubinski:1991bm,Navarro:1995iw,Navarro:1996gj}, in the inner part. This cuspy profile has been parametrised in several ways. As a concrete example, we give the Navarro-Frenk-White (NFW) profile,

\myeq{
\rho_{NFW}(r) &= \rho_{0}\times\frac{1}{\frac{r}{r_{s}}\left(1+\frac{r}{r_{s}}\right)^{2}},
}

\noindent with $\rho_{0}$ some energy density normalisation and $r_{s}$ the value of the critical radius for which the behaviour of the profile changes.
\\

\begin{figure}
\centering
\includegraphics[scale=0.45]{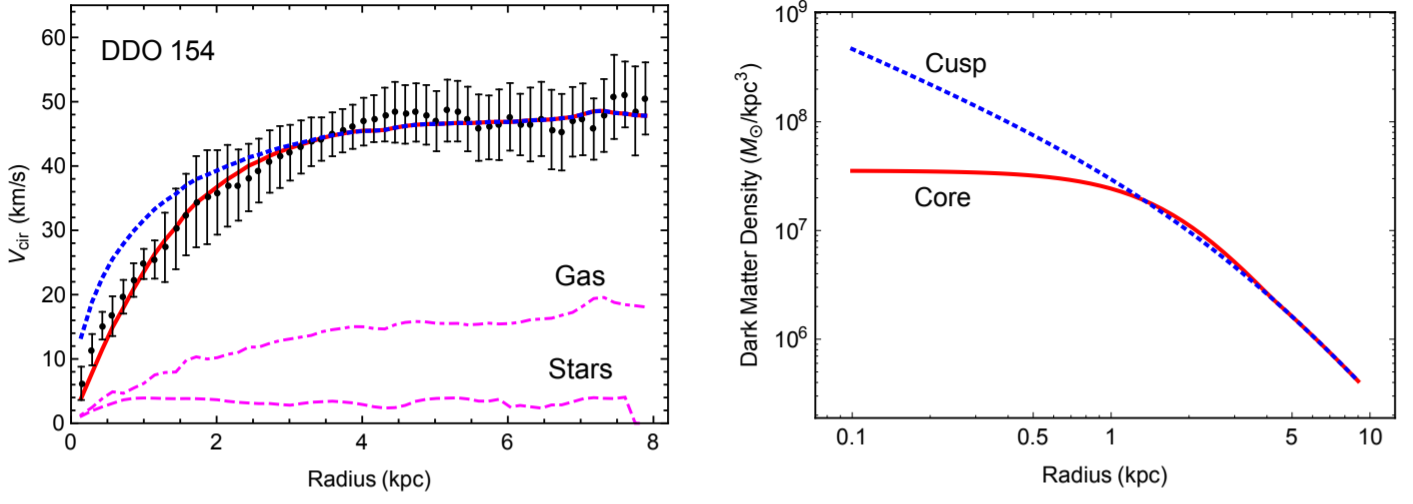}
\caption[Rotation curve of DDO 154 and core versus cusp galaxy profile]{Left: Data points show the measured rotation curve of DDO 154 \cite{Oh:2015xoa}. On the same figure, one can see expectation from a cuspy (blue) or a core (red) profile for DM. Right: The cuspy and core profiles associated to both fits. \cite{Tulin:2017ara}}\label{fig:cc}
\end{figure}

The problem arises once we compare this expected cuspy behaviour to observations. The outer regime behaviour agrees pretty well with it, but the inner part seems, according to observations of many halos, to flatten. Indeed, experiments show a core profile in the inner region of halos \cite{Flores:1994gz,Moore:1994yx,Moore:1999gc}. Figure \ref{fig:cc} shows discrepancies between simulations and observations (taken from \cite{Tulin:2017ara}, see also \cite{Kaplinghat:2015aga,Kamada:2016euw}). This \textit{core-vs-cusp} problem is a major issued of modern cosmology and has been studied extensively in the literature \cite{Burkert:1995yz,McGaugh:1998tq,vandenBosch:2000rza,Borriello:2000rv,deBlok:2001hbg,deBlok:2001rgg,Marchesini:2002vm,Gentile:2005de,Gentile:2006hv,KuziodeNaray:2006wh,KuziodeNaray:2007qi,Salucci:2007tm}.

\subsection{Diversity}\label{subsec:diver}
There is a last discrepancy between $\Lambda$CDM simulations and observations we would like to highlight. Looking again at the DM energy density profile in the inner region of halos of similar size and mass, observations pointed out many different behaviours \cite{Oman:2015xda,deNaray:2009xj}. While simulations give a small scatter in density profile for halos of similar size and mass \cite{Navarro:1996gj,Bullock:1999he}. This variety in the slope of the DM energy density profile is known as the diversity problem.

\subsection{DM hidden sector and portals}\label{subsec:portal}
The relic density issue above as well as the possible constraints from small scale structure have important implications on the DM particle model. Depending on the type and the strength of the interaction one assumes between the DM and the SM, one could constrain the DM mass, its coupling to SM particles as well as coupling to new particles that could exist. On top of these constraints, as we will see, there exist many additional constraints on the DM particle model coming from cosmology, astrophysics and particle physics experiments which suggest a feeble connection between the DM and SM particles. For this reason, in the following, we will consider a particular structure of DM models in which the DM particle is a singlet of all SM gauge groups. Additionally, there is no reason for the DM to have no interaction at all, actually we will see later in this chapter that small scale structure tensions could be alleviated if the DM particle self-interacts. Thus, one necessarily needs a new interaction which would allow the DM to self-interact. We will then consider DM models in which the DM self-interactions would be due to the introduction of a new light particle which would also be a singlet of $SU(3)_{c}\times SU(2)_{L}\times U(1)_{Y}$ and a singlet of all gauge groups which could arise in the HS. Thus, the DM particles together with the new light particles form a new self-interacting sector, independent of the one formed by SM particles (called visible sector or VS) which we will refer to as the hidden sector (HS). Moreover, the new particle will directly couple to one or more SM particles in addition to its coupling to the DM such that one has an indirect connection between the DM and the SM. These type of model in which the DM evolves on its own in a HS and couples to the VS through an additional degree of freedom is usually referred as portal models.
\\

It exists different types of portal models. Indeed, the only requirement is that a particle which is a singlet of all gauge groups of the theory (VS and HS) directly couples to a combination of DM fields and independently to a combination of SM fields which are both singlet of all possible gauge groups. There are three combination of SM fields which are singlet of all possible gauge groups such that $d<4$\footnote{The dimension $d$ of the operator made of SM fields has to be smaller than 4 if one wants a renormalisable interaction.}: $\bar{L}H$, $H^{\dagger}H$ and $F_{Y}^{\mu\nu}$ where $L$ is the lepton doublet, $H$ the Brout-Englert-Higgs scalar doublet and $F_{Y}^{\mu\nu}$ the strength field tensor associated to the hypercharge gauge boson. The first operator, $\bar{L}H$, can only couple to a fermion $\psi$ if one imposes the final operator to have a dimension $d\leq 4$. This portal, that one could write $\bar{L}H\psi$, is called the "neutrino portal" and will not be further studied in this thesis. The second operator, $H^{\dagger}H$, can couple to either a single real scalar or to a neutral combination of a complex scalar $\Phi$. The portal, $H^{\dagger}H\Phi$, $H^{\dagger}H\Phi^{2}$ or $H^{\dagger}H\Phi^{\dagger}\Phi$, is called the "Higgs scalar portal". Finally, the third operator, $F_{Y}^{\mu\nu}$, can only couple to another strength tensor field $F'_{\mu\nu}$ of a new gauge boson for example. This last portal, $F_{Y}^{\mu\nu}F'_{\mu\nu}$, is called the "kinetic mixing portal".
\\

Assuming a Dirac fermion DM $\chi$, the singlet combination of the DM fields is simply given by $\bar{\chi}\chi$ such that one ends up with only two possibilities for the portal (excluding the neutrino portal):

\myeq{
&\bullet\hspace{0.3cm} H^{\dagger}H\Phi\hspace{0.3cm}\&\hspace{0.3cm}H^{\dagger}H\Phi^{2}\hspace{0.3cm}\&\hspace{0.3cm}\bar{\chi}\chi\Phi,\label{eq:portal_2}\\
&\bullet\hspace{0.3cm} F_{Y}^{\mu\nu}F'_{\mu\nu}\hspace{0.3cm}\&\hspace{0.3cm}i\bar{\chi}\gamma^{\mu}\left(\partial_{\mu}-ie'A'_{\mu}\right)\chi,\label{eq:portal_3}
}

\noindent where in the last line, $e'$ and $A'$ are the gauge charge and the gauge field associated to a new $U(1)'$ symmetry and with $F'_{\mu\nu}\equiv\partial_{\mu}A'_{\nu}-\partial_{\nu}A'_{\mu}$. Note that we have joined the trilinear and the quadrilinear terms for the scalar option as if one of the two terms is theoretically allowed, the other is as well. We thus apply the principle saying that "everything which is not forbidden exists".
\\

In order to illustrate the constraints mentioned above, in the following, we will then consider both possibilities of DM portal models given in Eqs. \ref{eq:portal_2} and \ref{eq:portal_3} as they are generic and very well motivated models. Indeed, this type of model is well motivated in the context of scenarios where one assumes that DM undergoes self-interactions usually named Self-Interacting DM (SIDM), see Chapter \ref{ch:som}.

\subsubsection{Benchmark model A: Higgs portal DM model}\label{subsec:HP}
\begin{center}
\begin{figure}
\centering
\includegraphics[scale=0.8]{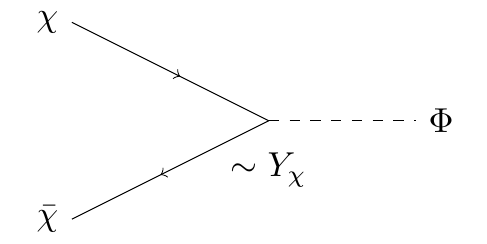}
\includegraphics[scale=0.8]{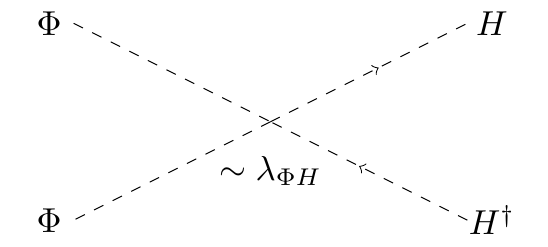}
\caption[Couplings between the DM and the mediator (left), and between the mediator and the visible sector (right) in the symmetric phase]{Couplings between the DM and the mediator, and between the mediator and the visible sector in the symmetric phase (i.e. no vacuum expectation value).}
\label{fig:HP-SYM}
\end{figure}
\end{center}

As a representative of a DM model with a Higgs portal, we take the Higgs scalar Portal (HP) where the DM is a singlet fermion $\chi$ and where the mediator role is played by a new real scalar field $\Phi$. The Lagrangian of this model is given by,

\myeq{
\mathcal{L}&= \mathcal{L}_{SM}+ i\bar{\chi}\slashed{D}\chi - m_{\rm DM}\bar{\chi}\chi + Y_{\chi}\Phi\bar{\chi}\chi-\mu_{\Phi}^{2}\Phi^{2}+\lambda_{\Phi}\Phi^{4}-\mu_{H}^{2}H^{\dagger}H\nn\\
&\hspace{1cm}+\lambda_{H}\left(H^{\dagger}H\right)^{2}+\lambda_{3}\Phi H^{\dagger}H+\lambda_{\Phi H}\Phi^{2}H^{\dagger}H,
\label{eq:lag_hp}
}

\noindent where the key interactions are the Yukawa interaction ruled by the $Y_{\chi}$, the Higgs portal interaction ruled by $\lambda_{\Phi H}$ and the scalar potential interactions (involving the purely non-derivative scalar interactions). Note that in Eq. \ref{eq:lag_hp}, we have explicitly written the scalar potential already included in the SM as we will need it in the following. From the expression of this Lagrangian, one can see that the connection between the hidden and the visible sectors goes through the last term and the Higgs portal parameter, $\lambda_{\Phi H}$. The two new couplings are represented in Figure \ref{fig:HP-SYM} and connect the DM to the mediator and the mediator to the visible sector.
\\

Moreover, if the new field $\Phi$ acquires a non-zero vacuum expectation value (VEV), a mixing occurs between the two real scalars of the theory, $\tilde{h}$ and $\tilde{\phi}$. The original fields can be expanded around their respective VEV ($v_{H}$ and $v_{\Phi}$):

\myeq{
H &=
\begin{pmatrix}
h^{+} \\ 
\frac{v_{H}+\tilde{h}+ig}{\sqrt{2}}
\end{pmatrix}
\hspace{1cm} \text{and} \hspace{1cm}
\Phi =\frac{v_{\Phi}+\tilde{\phi}}{\sqrt{2}},
}

\noindent where $\lambda_{3}$ has been absorbed in the definition of $v_{\Phi}$. Due to this mixing, the real scalars $\tilde{h}$ and $\tilde{\phi}$ are not mass eigenstates and a transformation is needed to diagonalise the mass matrix given by

\myeq{
M_{\Phi H}^{2} &=
\begin{pmatrix}
 2\lambda_{H}v_{H}^{2} & \lambda_{\Phi H}v_{H}v_{\Phi} \\ 
 \lambda_{\Phi H}v_{H}v_{\Phi} & 2\lambda_{\Phi}v_{\Phi}^{2}
\end{pmatrix}.
}

\noindent The diagonalisation of this matrix can be done with an usual rotation,

\myeq{
\begin{pmatrix}
h \\ 
\phi
\end{pmatrix}
&=
\begin{pmatrix}
 \cos\theta & -\sin\theta \\ 
 \sin\theta & \cos\theta
 \end{pmatrix}  
 \begin{pmatrix}
\tilde{h} \\ 
\tilde{\phi}
\end{pmatrix},
}

\noindent where the mixing angle $\theta$ has to be given by

\myeq{
\tan\left(2\theta\right) = \frac{\lambda_{\Phi H}v_{H}v_{\Phi}}{\lambda _{\Phi}v_{\Phi}^{2}-\lambda _{H}v_{H}^{2}}.\label{eq:mixing_angle}
}

\begin{center}
\begin{figure}
\centering
\includegraphics[scale=0.8]{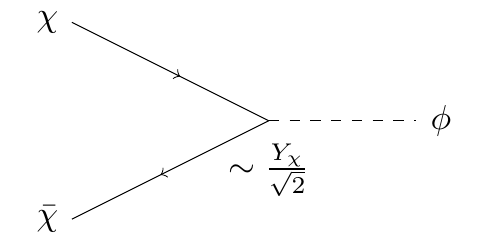}
\includegraphics[scale=0.8]{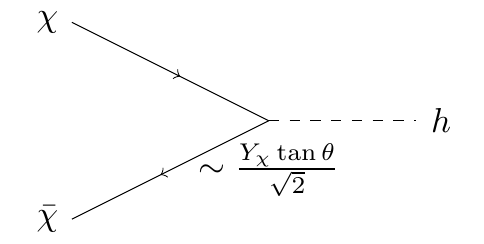}\\
\includegraphics[scale=0.8]{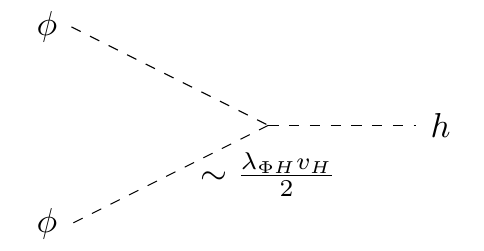}
\includegraphics[scale=0.8]{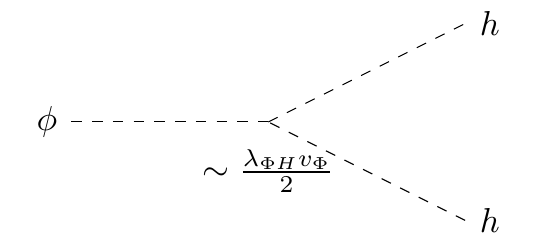}\\
\includegraphics[scale=0.8]{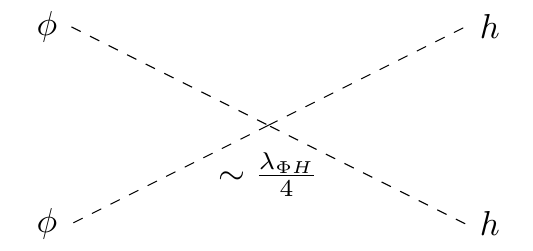}
\includegraphics[scale=0.8]{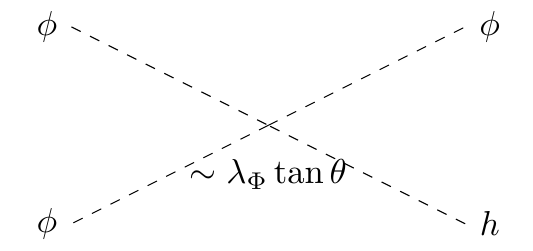}
\includegraphics[scale=0.8]{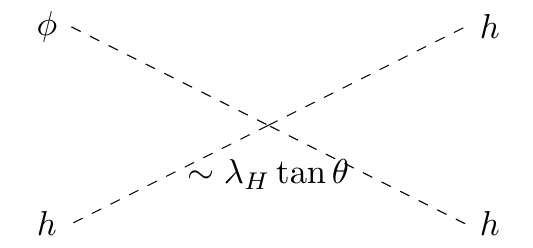}
\caption[Couplings between the DM, the mediator and the visible sector in the broken phase]{Couplings (at leading order for a small mixing angle) between the DM, the mediator and the visible sector in the broken phase.}
\label{fig:HP-BRO}
\end{figure}
\end{center}

\noindent We show in Figure \ref{fig:HP-BRO} the various couplings of this theory in terms of mass eigenstate fields in the case of a small mixing angle. We will see later in this chapter (see Sections \ref{sec:CST-CMB} to \ref{sec:CST-COL}) that small mixing angles are strongly required for this kind of model. Depending on the mass and VEV hierarchy, one can distinguish different cases. In the following, we will generally consider the fully broken cases: $m_{\rm med}<m_{\rm DM}<v_{H}<v_{\Phi}$. We will see in the following that the DM abundance is generally set at a temperature close to its mass $T\lesssim m_{\rm DM}$. Then, in the case where the DM mass is smaller than the two VEV's of the theory ($m_{\rm med}<m_{\rm DM}<v_{H}<v_{\Phi}$), the DM abundance will be set while the symmetries of the theory are broken ($T<v_{H}<v_{\Phi}$). Therefore, the hierarchy we are focusing on is such that the VS and the HS symmetries are broken when the DM is produced/depleted. We will discuss in Section \ref{sec:spec_scalar} other possible hierarchies and their implications.

\subsubsection{Benchmark model B: Kinetic Mixing portal DM model}\label{subsec:KM}
As a representative of a model based on a vector portal, we use the Kinetic Mixing portal (KM) where the DM is again a fermion denoted as $\chi$ but the mediator is the gauge field $B'_{\mu}$ of a new $U(1)'$ gauge symmetry. Such a new gauge symmetry is well motivated both experimentally and theoretically \cite{Jaeckel:2010ni,Essig:2013lka,Alexander:2016aln}. In this framework, the DM field is charged under this new symmetry which makes it naturally stable \cite{Foot:2014uba,Ackerman:mha,Feng:2008mu,Feng:2009mn,Hambye:2010zb}.
\\

The Lagrangian of this model is given by,

\myeq{
\mathcal{L}&= \mathcal{L}_{SM}+ i\bar{\chi}\slashed{D}\chi - m_{\rm DM}\bar{\chi}\chi-\frac{1}{4}\tilde{B}'^{\mu\nu}\tilde{B}'_{\mu\nu}  -\frac{\hat\epsilon}{2}\tilde{B}^{\mu\nu}\tilde{B}'_{\mu\nu} +\frac{1}{2}m_{\gamma'}^{2}\tilde{B}'^{\mu}\tilde{B}'_{\mu} + \ldots,
\label{eq:lag_km}
}

\noindent where the dots refer to what would be needed for giving a mass to the new mediator and the DM covariant derivative is $D_\mu = \partial_\mu+ie' \tilde{B}'_\mu$, where $\tilde{B}'_\mu$ and $e'$ are the U(1)' gauge field and coupling respectively. Such a mass for the new gauge boson could arise through the St\"uckelberg \cite{Stueckelberg:1900zz} or through the Brout-Englert-Higgs mechanism \cite{Englert:1964et,Higgs:1964pj}. In this model, the connection between the mediator and the visible sector goes through a kinetic mixing term (ruled by the kinetic mixing parameter $\hat{\epsilon}$) such that fields appearing in the Lagrangian are not propagation eigenstate fields. One must perform first a non-orthogonal transformation in order to have canonical kinetic terms. For a small mixing parameter (as it will be required later, see Sections \ref{sec:CST-CMB} to \ref{sec:CST-COL}), this transformations at leading order in $\hat{\epsilon}$ is simply,

\myeq{
\begin{pmatrix}
B'^{\mu} \\ 
B^{\mu}
\end{pmatrix}
&=
\begin{pmatrix}
 1 & \hat{\epsilon} \\ 
 0 & 1
 \end{pmatrix}  
 \begin{pmatrix}
\tilde{B}'^{\mu} \\ 
\tilde{B}^{\mu}
\end{pmatrix}.
}

\noindent After this transformation on the fields, we get for the relevant terms in the Lagrangian (still at the leading order in the mixing parameter),

\myeq{
\mathcal{L}&\supset -\frac{1}{4} B'^{\mu\nu} B'_{\mu\nu} +\frac{1}{2}m_{\gamma'}^{2} B'^{\mu} B'_{\mu} - \hat \epsilon\,  m_{\gamma'}^{2} B^{\mu} B'_{\mu} - e' \bar{\chi}\gamma^\mu\chi  ( B'_\mu  - \hat \epsilon B_\mu).\label{eq:lag_km_mass_mix}
}

\begin{center}
\begin{figure}
\centering
\includegraphics[scale=0.8]{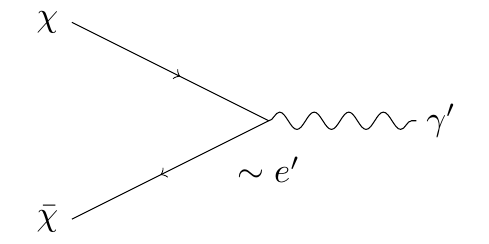}
\includegraphics[scale=0.8]{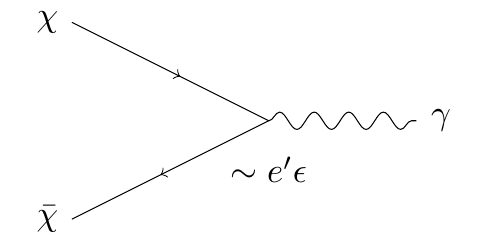}
\includegraphics[scale=0.8]{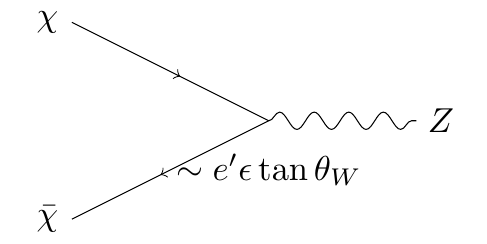}
\caption[Couplings between the DM, the mediator and the visible sector for a massless dark photon]{Couplings (at leading order for a small mixing parameter) between the DM, the mediator and the visible sector for a massless dark photon.}
\label{fig:KM-SYM}
\end{figure}
\end{center}

\noindent From here, one has to consider the massless and massive mediator cases separately as the continuous limit $m_{\gamma '}\rightarrow 0$ is not trivial and will be discussed later (see Section \ref{sec:spec_vector}). In the massless case, since fields are directly mass eigenstates, after having perform the Weinberg rotation,

\myeq{
\begin{pmatrix}
B^{\mu} \\ 
W^{3\mu}
\end{pmatrix}
&=
\begin{pmatrix}
 \cos\theta_{W} & -\sin\theta_{W} \\ 
 \sin\theta_{W} & \cos\theta_{W}
 \end{pmatrix}  
 \begin{pmatrix}
A^{\mu} \\ 
Z^{\mu}
\end{pmatrix},
}

\noindent one can identify $B'^{\mu}$, $A^{\mu}$ and $Z^{\mu}$ as the dark photon, the photon and the $Z$ boson respectively. Couplings between relevant particles are then shown in Figure \ref{fig:KM-SYM} where we have defined $\epsilon\equiv\hat{\epsilon}\cos\theta_{W}$. The final Lagrangian in the massless dark photon case is given by, 

\myeq{
\mathcal{L}&\supset -\frac{1}{4} B'^{\mu\nu} B'_{\mu\nu} - e' \bar{\chi}\gamma^\mu\chi B'_\mu + e'\epsilon \bar{\chi}\gamma^\mu\chi  (A_\mu  - \tan\theta_{W} Z_\mu).\label{eq:lag_km_massless_final}
}

\noindent In the massive case, fields are not mass eigenstates after having performed the Weinberg rotation,

\myeq{
\begin{pmatrix}
B^{\mu} \\ 
W^{3\mu}
\end{pmatrix}
&=
\begin{pmatrix}
 \cos\theta_{W} & -\sin\theta_{W} \\ 
 \sin\theta_{W} & \cos\theta_{W}
 \end{pmatrix}  
 \begin{pmatrix}
A^{\mu}_{0} \\ 
Z^{\mu}_{0}
\end{pmatrix}.
}

\noindent One needs a further orthogonal diagonalisation of the mass term which, at leading order in $\epsilon$, is

\myeq{
\begin{pmatrix}
B'^{\mu} \\ 
A^{\mu}_{0}
\end{pmatrix}
&=
\begin{pmatrix}
 1 & \epsilon \\ 
 -\epsilon & 1
 \end{pmatrix}  
 \begin{pmatrix}
A'^{\mu} \\ 
A^{\mu}
\end{pmatrix}.
}

\begin{center}
\begin{figure}
\centering
\includegraphics[scale=0.8]{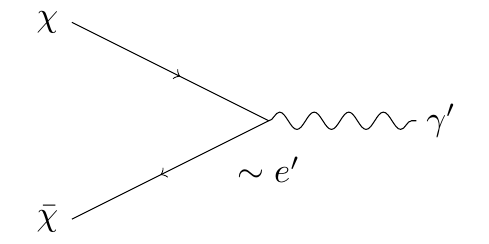}
\includegraphics[scale=0.8]{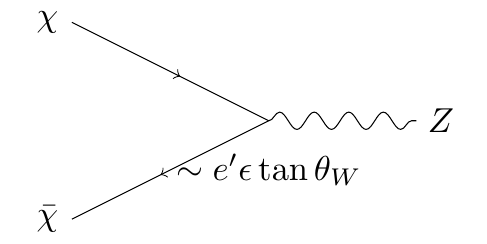}
\caption[Couplings between the DM, the mediator and the visible sector for a massive dark photon]{Couplings (at leading order for a small mixing parameter) between the DM, the mediator and the visible sector for a massive dark photon.}
\label{fig:KM-BRO}
\end{figure}
\end{center}

\noindent Such that the final Lagrangian in the massive dark photon case is given by,

\myeq{
&\mathcal{L}\supset -\frac{1}{4} F'^{\mu\nu} F'_{\mu\nu} - e' \bar{\chi}\gamma^\mu\chi A'_{\mu} - e'\epsilon\tan\theta_{W}\bar{\chi}\gamma^\mu\chi Z_{\mu} + \frac{1}{2}m_{\gamma'}^{2} A'^{\mu} A'_{\mu}\nonumber\\
&\hspace{4cm} + \epsilon\tan\theta_{W} m_{\gamma'}^{2} A'^{\mu} Z_{\mu},\label{eq:lag_km_massive_final}
}

\noindent with $F'^{\mu\nu}$ is the strength field tensor associated to the dark photon $A'^{\mu}$ and where the last term is not diagonal, but it will introduce a mass splitting between $A'^{\mu}$ and $Z_{0}^{\mu}$ of order $\mathcal{O}\left(\epsilon^{2}\right)$ and will then be neglected as our whole computation was already done at leading order in $\epsilon$. Then one can identify $A'^{\mu}$, $A^{\mu}$ and $Z_{0}^{\mu}$ to the dark photon\footnote{In the massive case, the new vector boson is sometimes called $Z'$ instead of $\gamma '$ (dark photon) as it is closer to "be" a $Z$ than a $\gamma$ as it is massive. However, in this thesis, we will keep the name "dark photon" whatever its mass is.}, the photon and the $Z$ boson, respectively. Two major consequences of this diagonalisation are that the DM field does not couple to the photon but only to the $Z$ boson and that the dark photon is distinguishable from the SM photon as one can read from the third term of Eq. \ref{eq:lag_km_massive_final}. The latter consequence will have a strong impact on the DM production mechanisms as we will see in Chapter \ref{ch:prod}. Couplings between different relevant particles can be seen in Figure \ref{fig:KM-BRO}.
\\

The treatment of the limit $m_{\gamma '}\rightarrow 0$ may seem discontinuous and leading to incoherences. Indeed, in the massive case, one distinguishes the dark photon propagation and mass eigenstate basis, while for $m_{\gamma '}\rightarrow 0$, the two fields are degenerate, and the distinction between the basis disappears. The proper way to treat this limit starting from a massive dark photon goes through the incorporation of thermal effects which can be important in some cases. This has been studied at length in the context of dark photon production, in particular in stars \cite{Redondo:2008aa,Jaeckel:2008fi,Redondo:2008ec,An:2013yfc,Redondo:2013lna,Fradette:2014sza}. As this problem is relevant for our work but not central, we only summarise the salient points in a dedicated section (see Section \ref{sec:spec_vector} and Appendix \ref{app:th} for more details on some technical issues) and we refer to \cite{Redondo:2008aa,Jaeckel:2008fi,Redondo:2008ec,An:2013yfc,Redondo:2013lna,Fradette:2014sza} for a more detailed explanation of the effects.

\subsection{Dark matter self-interaction cross section}\label{subsec:DM-SI_cross_section}
In Subsections \ref{subsec:too_big} and \ref{subsec:core}, we discussed the \textit{Too-big-to-fail} and the \textit{Core-vs-cusp} problems. Both problems arise when comparing results of small-scale structure simulations of galaxy formations which were performed with collisionless DM to observations of these structures. However, simulations also show that if the new form of matter, the DM, self-interacts relatively strongly, these problems could be alleviated. In practice, simulations show that this requires a self-interaction cross section divided by the DM mass lying within the range $0.1$ $\hbox{cm}^2/g \lesssim \sigma_T/m_{\rm DM} \lesssim 10$ $\hbox{cm}^2/g$ \cite{Vogelsberger:2012ku,Rocha:2012jg,Zavala:2012us,Peter:2012jh}. Moreover, observations of galaxy or cluster mergers can also provide an upper bound on the self-interaction cross section. Indeed, the lack of a visible offset between stars and DM tells us that the number of self-interactions cannot be as large as we want such that this quantity is bounded above. Concretely, the Bullet cluster gives $\sigma_T/m_{\rm DM} < 1.25$ $\hbox{cm}^2/\hbox{g}$ at 68$\%$ CL \cite{Clowe:2003tk,Markevitch:2003at,Randall:2007ph} while the most recent observations of cluster mergers lead to $\sigma_T/m_{\rm DM} < 0.47$ $\hbox{cm}^2/\hbox{g}$ at 95\% CL \cite{Harvey:2015hha} (or even smaller with $\sigma_T/m_{\rm DM} < 0.3$ $\hbox{cm}^2/\hbox{g}$ in \cite{Bondarenko_2018}). We will finally keep in mind that, if a model allows self-interactions, we must have that

\myeq{
0.10 \text{ cm}^2/\text{g} < \sigma_T/m_{\rm DM} < 0.47 \text{ cm}^2/\text{g}.\label{eq:CST-small_scale_structure}
}

\noindent Rewriting this constraint in particle physics units for a DM mass of the same order of the electroweak scale, one gets $\sigma_{T}\sim 10^{5}\left(\frac{m_{\rm DM}}{100\text{ GeV}}\right)$ GeV$^{-2}$. This order of magnitude is considerably larger than the value of the annihilation (or production) cross section considered in standard freeze-out (or respectively freeze-in) scenarios. Indeed, one can verify that typically, the cross section is 14 and 29 order of magnitudes smaller in a FO and FI scenario respectively, see Figure \ref{fig:volcano}. To account for this large difference between both type of cross sections (annihilation/production and self-interaction), one possibility turns out to assume the light mediator to be much lighter than the DM particle. This hierarchy between the two masses would boost the t-channel mediator exchange (i.e. the self-interaction) with respect to the annihilation cross section. Indeed, in this case, the so-called Sommerfeld effect (see Chapter \ref{ch:som}) will play a major role by increasing the self-interaction cross section at tree level by several order of magnitudes.
\\

\begin{figure}
\centering
\begin{minipage}{0.50\textwidth}
\centering
\includegraphics[scale=0.65]{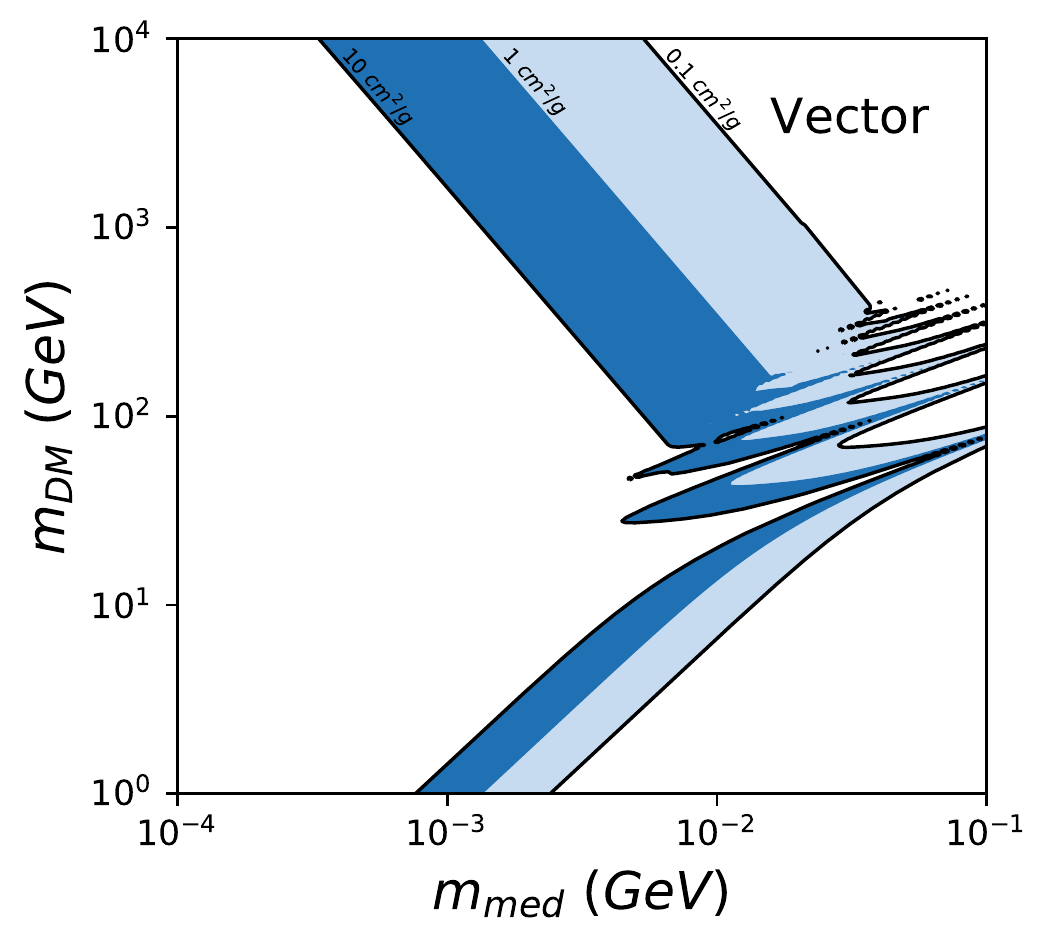}
\end{minipage}
\begin{minipage}{0.46\textwidth}
\centering
\includegraphics[scale=0.65]{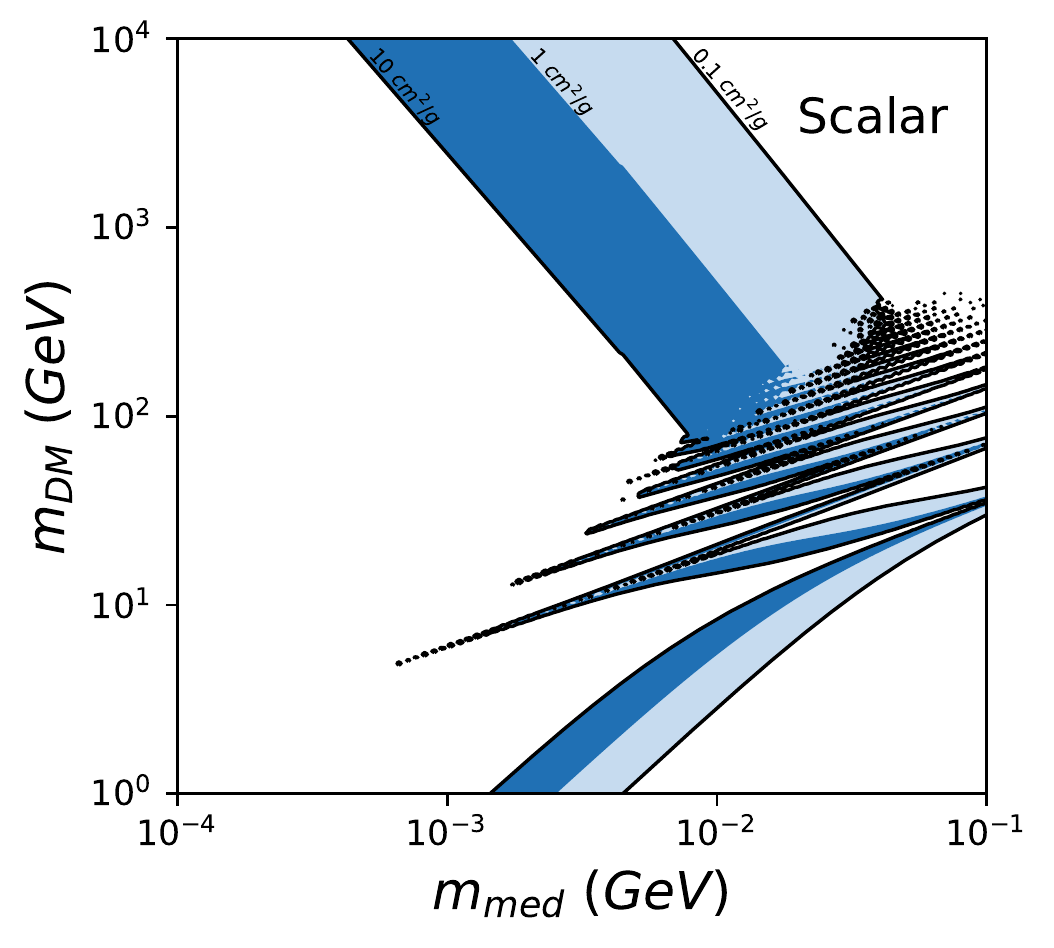}
\end{minipage}
\caption[Self interaction constraints for a Dirac DM and a vector or a scalar mediator]{Self interaction constraints for a Dirac DM and a vector (Left) or a scalar (Right) mediator. See text for more details.} \label{fig:SI_gammap_phi}
\end{figure}

This potential necessity of a large self-interaction consequently motivates models with light mediators, in particular the two portal benchmark models considered above (see Subsection \ref{subsec:portal}) and taking $m_{\rm med}\ll m_{\rm DM}$. More concretely and as we will see in Chapter \ref{ch:som}, one needs $v_{\rm DM}/2\ll \alpha_{\rm med}$\footnote{$\alpha_{\rm med} = \alpha' \equiv e'^{2}/4\pi$ in the vector portal and $\alpha_{\rm med} = \alpha_{\phi} \equiv y_{\phi}^{2}/4\pi$ in the scalar portal model.} and $m_{med}\ll\alpha_{\rm med}m_{\rm DM}$ \cite{Tulin:2013teo}. Under those conditions, the model enters the highly resonant regime and the Sommerfeld effect enhances drastically the self-interaction cross section. This is shown in Figure \ref{fig:SI_gammap_phi} where we have highlighted areas, in the mediator mass - DM mass plane, where $0.10 \text{ cm}^2/\text{g} < \sigma_T/m_{\rm DM} < 10 \text{ cm}^2/\text{g}$. We fixed the coupling ($e'$ or $y_{\phi}$) between the mediator and the DM imposing that it leads to the observed relic density (see Eq. \ref{eq:CST-relic_density}), assuming a standard freeze-out. From this figure, one can conclude that the allowed parameter space is globally the same in both cases even if the two models have two important differences which will be further detailed in Chapter \ref{ch:som}. First, the DM annihilation cross section is either s- or p-wave (in the vector or scalar portal model respectively). Second, in the vector mediator case, the self-interaction has an attractive and a repulsive contribution while there is only an attractive contribution in the scalar mediator case. This explain why the Sommerfeld enhancement resonances appears to be smoothed. Despite these relevant differences, the final allowed part of the parameter space is barely the same for both models.
\\

In the following, we will consider constraints which apply on these two self-interacting benchmark models. This will allow us to show that these constraints are very strong in most cases and thus will be the basis of Chapters \ref{ch:TpT} and \ref{ch:other_minimal} which are precisely devoted to the study of the various ways out. Most of these constraints, as we will see, are coming from the fact that the light mediator will be typically present in large numbers at relatively late time. That is to say that there are extra degrees of freedom in the thermal bath at late time and then potential additional production of SM particles\footnote{Through the decay or annihilation of the new degrees of freedom.}. This bounds, that we will see, typically apply when the number of light mediators is large prior to the decay, which is mostly the case when it decouples relativistically because in this case, it decouples when not Boltzmann suppressed. But a relativistic decoupling for the light mediator is generic of light mediator models because it will typically decouple at the same time than DM: $T\sim m_{\rm DM}>m_{\rm med}$.

\section{Cosmic microwave background}\label{sec:CST-CMB}
We have already discussed on how the measurements of the CMB anisotropies brought a tremendous evidence for DM. But the CMB also brings his own set of constraints on SIDM with light mediator. In this section, we will review how portal models can be constrained by the CMB.

\subsection{CMB constraint on DM annihilation rate}
If DM annihilates into electromagnetically charged particles (or directly into photons) during the recombination (around redshift $z_{rec}\simeq 1100$), it would inject energy into the thermal bath. The plasma composition (made of a mix of photons, electrons and baryons) could be perturbed. Indeed, because of this new energy source, a consistent proportion of neutral hydrogen could be ionised. This would increase the proportion of unbounded electrons in the bath. CMB photons would thus be more scattered which would thicken the last scattering surface. Therefore, correlations in the temperature perturbation spectrum would be attenuated. Depending on the quantitative impact on the spectrum, this attenuation could be seen in the temperature angular power spectrum (Figure \ref{fig:CST-CMB-an}).
\\

In practice, for a given model, this has strong consequences. If DM particles annihilate into electromagnetically charged particles (or directly into photons), the DM annihilation cross section is bounded from above by the CMB. If DM particles annihilate into neutral particles which can produce photons or charged particles, this bound obviously applies as well. In models such as benchmark portal models of Eqs. \ref{eq:lag_hp} and \ref{eq:lag_km}, the DM annihilates into a pair of light mediators and, thanks to the portal interaction, the light mediator particles decay into SM particles. Except if the annihilation into a pair of neutrinos (i.e. to not producing electromagnetic materials) is the only available channel, the constraint on the annihilation cross section at the time of recombination (i.e. at redshift $z\sim1100$) is then

\myeq{
\langle \sigma v\rangle_{rec}\lesssim N_\chi \cdot 4 \times 10^{-25} \hbox{ cm}^3 \hbox{s}^{-1} \Big(\frac{f_{eff}}{0.1}\Big)^{-1}\Big(\frac{m_\chi}{100\hbox{ GeV}}\Big)\,,
\label{cmbconstraint}
}

\noindent where $f_{eff}$ is related to the fraction of the released energy ending up in photons or electrons, with $f_{eff}\gtrsim 0.1$ for any SM final states except neutrinos (see e.g.~\cite{Slatyer:2015jla}), and where $N_\chi=1,2$ for Majorana and Dirac dark matter respectively.
\\

\begin{figure}
\centering
\includegraphics[scale=0.75]{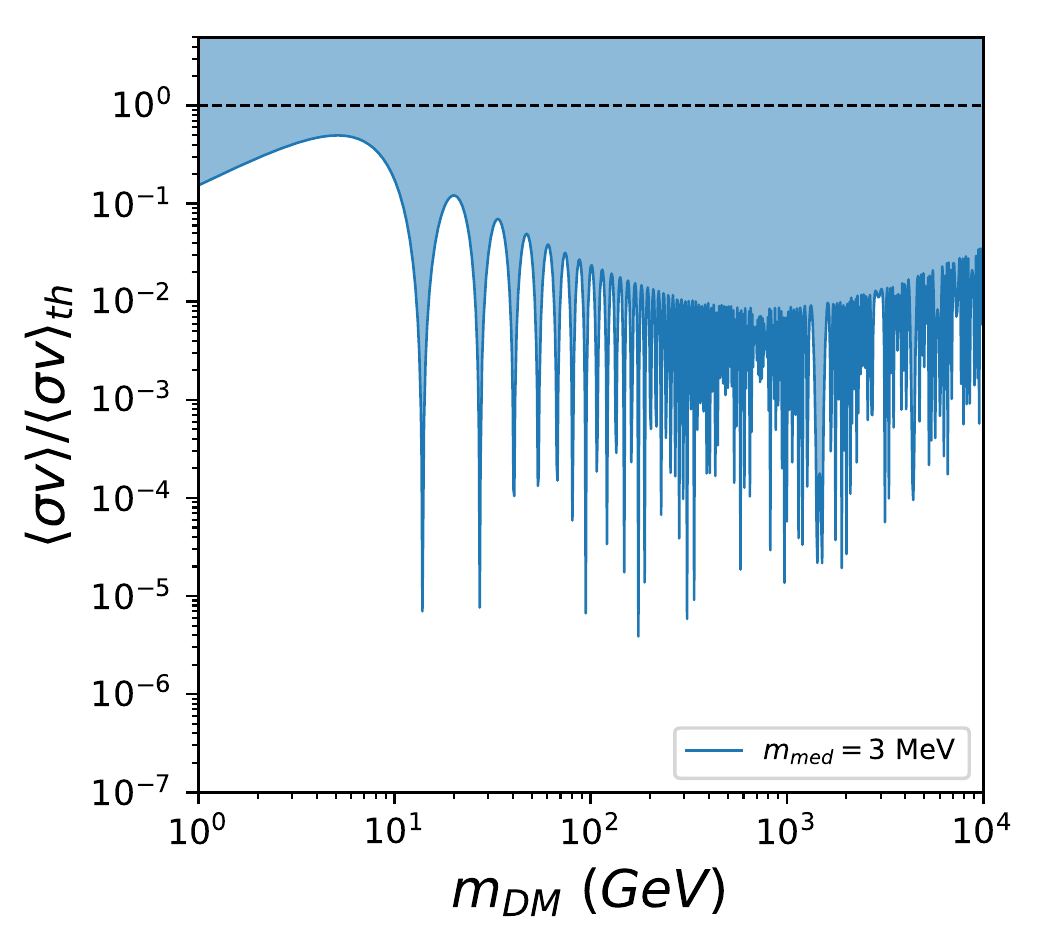}
\caption[CMB upper bound on the s-wave annihilation rate into light mediators cross section]{CMB upper bound on the s-wave annihilation rate into light mediators cross section, normalised to the thermal value. The dashed horizontal line corresponds to a cross section with the thermal value.} \label{fig:CST-CMB-1}
\end{figure}

The bound given in Eq. \ref{cmbconstraint} has many implications which have been analysed at length in \cite{Bringmann:2016din,Cirelli:2016rnw}. In particular, Eq. \ref{cmbconstraint} can be fully relevant when considering a specific model and fixing the coupling between the light mediator and the DM from to the constraint on the DM abundance today and assuming a secluded freeze-out (i.e. $\chi\bar{\chi}\rightarrow XX$, with $X$ the light mediator). On the one hand, in the case of the scalar portal, even if the light mediator decays into SM charged particles, since the DM annihilation process is p-wave (i.e. $\sim v^{2}$, with $v$ the DM relative velocity), the annihilation at recombination will be suppressed ($v_{rec}\ll 1$). This constraint will then be automatically satisfied in the framework of the scalar portal model. On the other hand, in the case of the vector portal, the DM annihilation process is s-wave (i.e. $\sim v^{0}$). Thus, the upper bound given by Eq. \ref{cmbconstraint} is strong enough to exclude most of the parameter space. Indeed, as we will see in Chapter \ref{ch:som}, the Sommerfeld effect largely enhance the annihilation process at this time (see Figure 1 of \cite{Bringmann:2016din}). This is due to the fact that DM particle at recombination time are highly non-relativistic ($v\lesssim 10^{-7}$) with respect to the annihilation time where $v\lesssim 1/4$. In Figure \ref{fig:CST-CMB-1}, we show constraints obtained in the vector portal case for a 3 MeV mediator mass. We have plotted the annihilation cross section at recombination divided by the one at freeze-out. The large number of resonances are characteristics of the Sommerfeld effect as we will see in Chapter \ref{ch:som}. Every point localised in the blue shaded area is excluded. One then sees that, in order to not disturb too much the CMB spectrum, one requires a much smaller value of the annihilation cross section at recombination than at freeze-out. For the whole parameter space to be totally excluded (or not) relies on the possibility for the mediator to by lighter than two electrons ($m_{\rm med}<2 m_e$). In this case, the light particle will essentially decay into neutrinos and will no longer affect the CMB, see Ref.~\cite{Bringmann:2016din} for a detailed discussion.

\subsection{CMB constraints on mediator decay}
\begin{figure}
\centering
\includegraphics[scale=0.75]{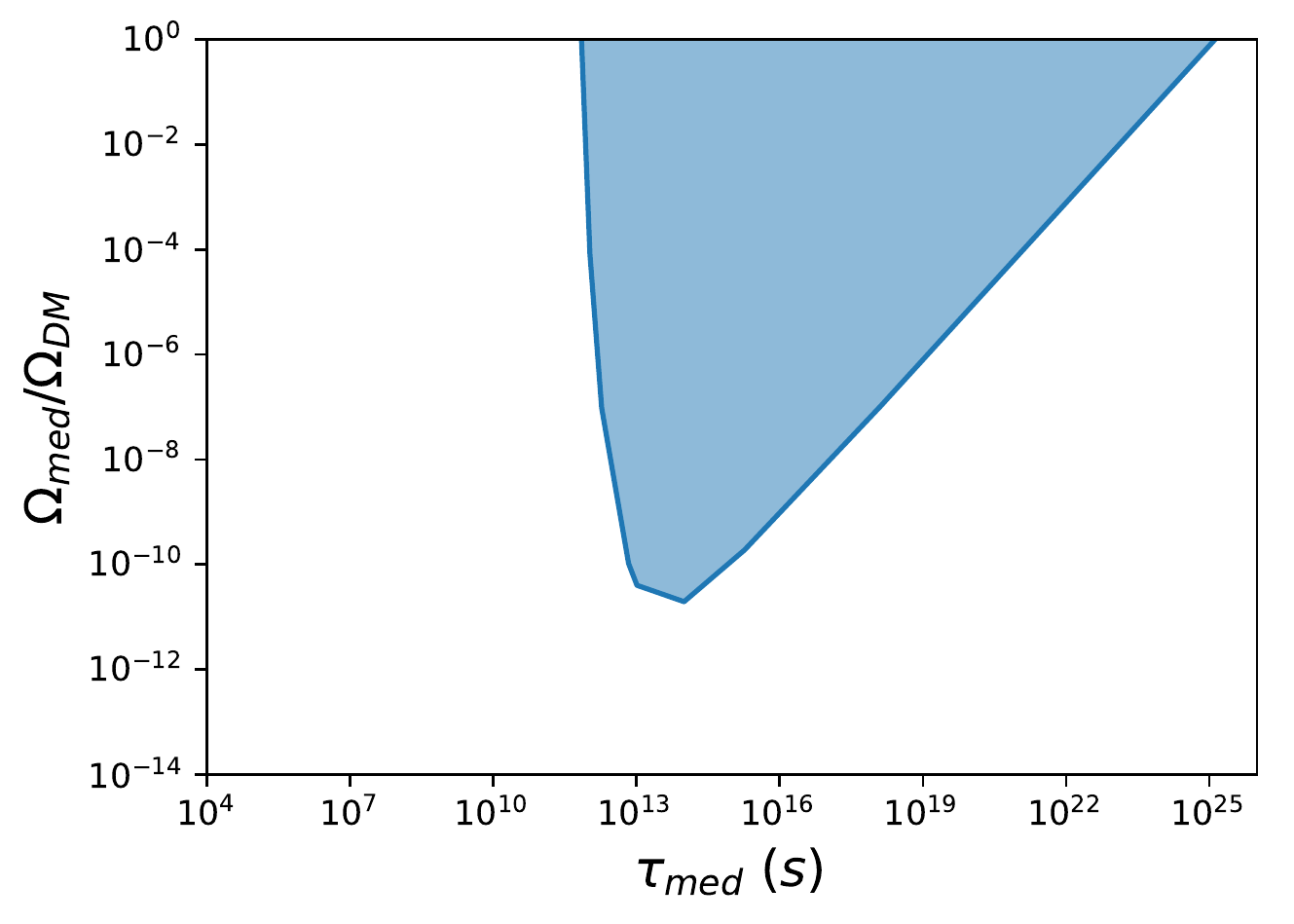}
\caption[Upper bound on the amount of mediator particles there would be today if there were no decay]{Upper bound on the amount of mediator particle there would be today if there were no decay \cite{Poulin:2016anj} (see also \cite{Slatyer:2012yq}).} \label{fig:CST-omega_med}
\end{figure}

We have seen in the previous subsection that CMB can provide an upper bound on the DM annihilation cross section as soon as the decay products can inject photons in the thermal bath. This means that if the light mediator is also present in a large proportion in the bath, its decay rate into SM particles (except neutrinos) could also be constrained in the same way (the decay rate and thus the lifetime will play an important role in this bound). If the light mediator is not much heavier than two electrons, it can decay only to $e^\pm$, neutrinos and photons. If, among these particles, the mediator decays mainly to electromagnetic channels ($e^+e^-$ or $\gamma \gamma$) and if it is still present in the thermal bath at the time of recombination (i.e. $\tau_{\rm med}\gtrsim 10^{12}$~sec), the energy the mediator will inject would change significantly the CMB anisotropy spectrum. This would have been already observed unless the light mediator has a short enough or a very long lifetime or if its abundance is low enough that the changes on the CMB anisotropy structure are smaller than experimental uncertainties. It is standard to illustrate this type of constraint in the $\tau_{\rm med}-\frac{\Omega_{\rm med}}{\Omega_{\rm DM}}$ plane where $\frac{\Omega_{\rm med}}{\Omega_{\rm DM}}$ expresses the light mediator abundance (it would have today if it was not decaying) divided by the DM abundance today. Figure \ref{fig:CST-omega_med} gives then the upper bound on this ratio as a function of the mediator lifetime.
\\

For a lifetime larger  than the age of the Universe today and $m_{\rm med}<2 m_e$ (i.e. not producing electron-positron pairs), there are additional constraints coming from X-ray observation, which in a conservative way require \cite{Essig:2013goa,Boddy:2015efa,Riemer-Sorensen:2015kqa}, 

\myeq{
\tau_{\rm med}\gtrsim 10^{28}\,\hbox{sec}\times \Omega_{\rm med}h^2/0.12\label{eq:CST-xray_med},
}

\noindent and which apply basically to any scenario as soon as the mediator decay produces SM particles (apart from neutrinos).

\subsection{CMB constraint on $N_{eff}$}
The precise measurement of the CMB anisotropies does not provide constraints only on the DM abundance today or on the number of electromagnetic materials a model can inject, but also on the Hubble rate at recombination. Since any modification of the Hubble rate is highly related to the number of relativistic degrees of freedom contained in the thermal bath, CMB also provides a stringent constraint on this quantity at the time of recombination. The number of relativistic degrees of freedom is generally expressed in terms of the effective number of neutrinos, $N_{eff}$. The most recent constraint on the minimum of effective neutrinos, given by the Planck satellite, is $2.66<N_{eff}< 3.33$ at 2$\sigma$ level \cite{Aghanim:2018eyx}. It is straightforward to translate these bounds into constraints on the light mediator lifetime and abundance at the time of neutrino decoupling. Indeed, light mediator particles which decay after neutrino decoupling reheat the photons but not the neutrinos. Thus, a particle decaying after neutrino decoupling, i.e. when $T<T^{\nu}_{\rm dec} \simeq 1$~MeV (equivalently $t>t_{\rm dec}^\nu\simeq 7$ sec), will change the amount of relativistic particles and change the value of $N_{eff}$ at this epoch. This constraint is thus only relevant if the decoupling time of the new light particle ($t_{\rm dec}$) is bigger than the neutrino decoupling time ($t_{\rm dec}^{\nu}$). $t_{\rm dec}$ can be written in terms of the light mediator lifetime ($\tau_{\rm med}$) and the Lorentz boost factor ($B_{\mathcal{L}}(T_{\rm dec})$): $t_{\rm dec}=B_{\mathcal{L}}(T_{\rm dec})\tau_{\rm med}$. With $B_{\mathcal{L}}(T_{\rm dec}<m_{\rm med})\simeq 1$ and $B_{\mathcal{L}}(T_{\rm dec}>m_{\rm med})\simeq T_{\rm dec}/m_{\rm med}$ in a good approximation. We will not review more this effect since a detailed study of it has already been performed in Ref. \cite{Hufnagel:2018bjp}. In this reference, the authors have chosen to constrain the mediator number density divided by the photon number density at the time of decoupling $\left(n_{\rm med}/n_{\gamma}\right)\vert_{T_{\rm dec}}$, assuming a chemical decoupling of the light mediator from the thermal bath at $T_{\rm dec}=10$ GeV. Thanks to the following conversion formula, it is easy to translate this bound into a bound on $\Omega_{\rm med}/\Omega_{\rm DM}$ (same ratio than previously).

\myeq{
\Omega_{\rm med}h^{2}&\leq 8\times 10^{4}\left(\frac{g_{\ast}^{eff}(T_{\rm dec})}{g_{\ast}^{S}(T_{\rm dec})}\right)\left(\frac{m_{\rm med}}{\text{MeV}}\right)\left(\frac{n_{\rm med}(T_{\rm dec}')}{n_{\gamma}(T_{\rm dec})}\right)\left(\frac{T_{\rm dec}'}{T_{\rm dec}}\right)^{3},\label{eq:conv_Omed_nmed}
}

\begin{figure}
\centering
\includegraphics[scale=0.75]{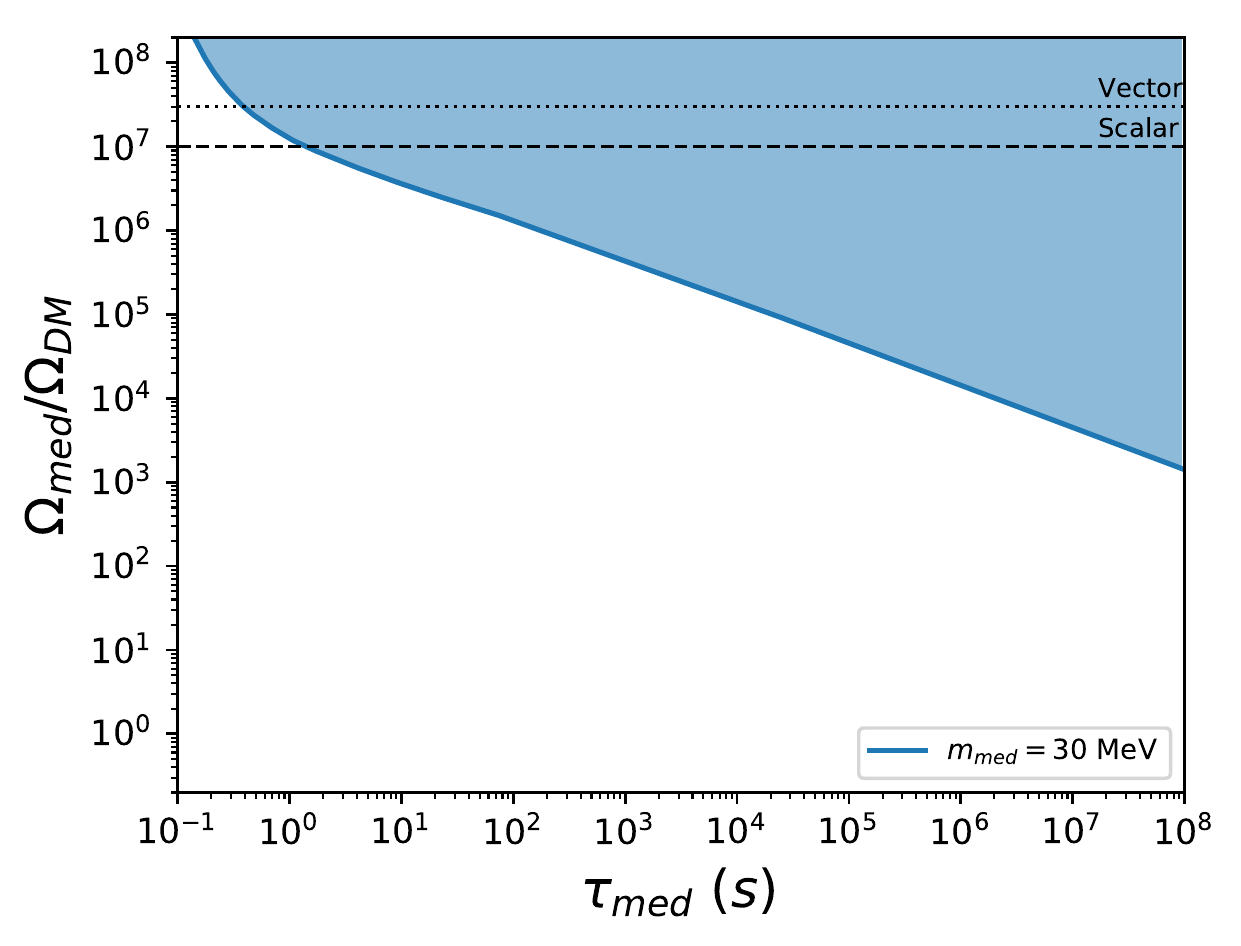}
\caption[Upper bound on the amount of mediator particles there would be today if there was no decay from the $N_{eff}$ constraint]{Upper bound on the amount of mediator particle there would be today if there was no decay \cite{Hufnagel:2018bjp} from the $N_{eff}$ constraint. The dotted (dashed) line shows the value of $\Omega_{\rm med}/\Omega_{\rm DM}$ one would expect from a relativistic decoupling of a non-decaying vector (resp. scalar) particle.} \label{fig:CST-omega_neff}
\end{figure}

\noindent We show in Figure \ref{fig:CST-omega_neff} the upper bound on $\Omega_{\rm med}/\Omega_{\rm DM}$ as a function of the mediator lifetime for $m_{\rm med}=60$ MeV (taken from \cite{Hufnagel:2018bjp}). This constraint is fully relevant for a particle decaying while still relativistic. Indeed, using Eq. \ref{eq:conv_Omed_nmed} one can find from Ref. \cite{Hufnagel:2018bjp} that the light mediator number density given by its relativistic decoupling is perfectly compatible with these constraints if its lifetime is small enough. We give in Table \ref{table:Neff_Az}, for both scalar and vector portal models and for various values of $m_{\rm med}$, the upper bound on the light mediator lifetime.
\\

All these lifetimes have been obtained from Ref. \cite{Hufnagel:2018bjp} assuming here too that  $T_{\rm dec}=10$ GeV. For another decoupling temperature, the bound will be moderately affected by a factor of the relativistic degrees of freedom contributing to the entropy, $g_{\ast}^{S}(10\hbox{ GeV})/g_{\ast}^{S}(T_{\rm dec})$ due to decoupling of relativistic species between both temperatures. 

\begin{table*}[t]
\centering
\resizebox{\textwidth}{!}{%
\begin{tabular}{|c|c|c|c|c|c|c|c|c|c|c|}
\hline 
$m_{\gamma '}$ (MeV)            & 0.01    & 0.03    & 0.1     & 0.3   & 1    & 3    & 10    & 30    & 100   & 300   \\ \hline
$N_{eff}$            & $10^{6.3}$       & $10^{5.3}$      & $10^{4.3}$     & $10^{3.3}$     & $10^{2.2}$      & $10^{1.1}$      & $10^{0.1}$     & $10^{-0.4}$     & $10^{-0.7}$    & $10^{-0.9}$     \\
Photodis./Entropy inj.         & $10^{5.9}$       & $10^{4.9}$      & $10^{3.9}$     & $10^{3.1}$     & $10^{2.4}$      & $10^{2.0}$      & $10^{1.0}$     & $10^{0.0}$     & $10^{-0.4}$    & $10^{-0.5}$     \\ \hline
\hline 
$m_{\phi}$~(MeV)            & 0.01    & 0.03    & 0.1     & 0.3    & 1    & 3    & 10    & 30    & 100   & 300   \\ \hline
$N_{eff}$            & $10^{7.3}$       & $10^{6.3}$      & $10^{5.3}$     & $10^{4.3}$     & $10^{3.2}$      & $10^{2.1}$      & $10^{1.1}$     & $10^{0.1}$     & $10^{-0.5}$    & $10^{-0.7}$     \\
Photodis./Entropy inj.         & $10^{6.8}$       & $10^{5.9}$      & $10^{4.8}$     & $10^{3.9}$     & $10^{3.0}$      & $10^{2.4}$      & $10^{2.0}$     & $10^{1.0}$     & $10^{-0.1}$    & $10^{-0.4}$     \\ \hline
\end{tabular}
}
\caption[Upper bound on the light mediator lifetime from CMB and BBN constraints]{Upper bound on the light mediator lifetime (in seconds) from $N_{eff}$, photodisintegration and Hubble constant/entropy injection constraints, assuming a relativistic decoupling of the light mediator for the vector portal model (top) and the scalar portal model (bottom). A value of $T'/T=1$ has been assumed at DM freeze-out time.}\label{table:Neff_Az}
\end{table*}

\section{Big bang nucleosynthesis}\label{sec:CST-BBN}
New light degrees of freedom does not affect only the thermal history of the Universe at recombination, but may also easily affect the \textit{Big Bang Nucleosynthesis} (BBN) process. In this section, after a brief explanation of what occurs during the BBN, we will review the two main ways for additional relativistic degrees of freedom to affect the BBN process. First, we will discuss the photodisintegration of light nuclei during and after the BBN process from decay of the light particle. Then we will see how the modification of light nuclei relative abundances from modification of the Hubble constant and entropy injection by the light mediator can constrain portal models.

\subsection{Production of light nuclei}
The BBN refers to the epoch, during the early Universe, when the light nuclei were formed. Let us recap the few steps which led to the formation of the lightest nuclei $^{4}He^{++}$, $^{3}He^{++}$, $^{7}Li^{+++}$ and $^{2}H^{+}$.
\\

At high temperature, there was no free proton nor free neutron in the thermal bath. The Universe was exclusively made of elementary particles like quarks, electrons, .... But, due to the expansion of the Universe, its temperature continued to cool down such that, at some point, it became cold enough for the QCD\footnote{The Quantum Chromodynamics (QCD) is the theory of strong interactions. It describes how quarks and gluons interact and how they may form hadrons such as protons, neutrons, pions, kaons, ....} phase transition to occur, when $T\sim 1$ GeV, at which point protons and neutrons were formed by the fusion of quarks. Even if the neutron particle is unstable and decays into proton, electron and anti-neutrino, the bath was sufficiently hot to ensure the reverse process to occur as fast as the direct process. That is to say that the neutron decay was is thermal equilibrium with the bath: $\Gamma_{n\rightarrow p+e+\bar{\nu}_{e}}\gg H$.
\\

Thus, as long as the neutron decay and the conversion process (i.e. $n + e^{+}\rightarrow p^{+}+\bar{\nu}_{e}$) were in thermal equilibrium with the rest of the bath, the number of neutrons was equal to the number of protons in the bath, $n_{n}=n_{p}$. However, this equilibrium did not last for ever and protons started to be dominant. This happened when $T\lesssim\Delta m\equiv m_{n}-m_{p}$ where the number of neutrons started to decrease: $n_{n}\propto e^{-\Delta m/T}$. Later on, when the inverse decay stopped to be as efficient as the decay, the number of neutron followed an usual decay law: $n_{n}\propto e^{-(t-t_{\rm dec})/\tau_{n}}$ with $t_{\rm dec}$ the decay time and $\tau_{n}$ the neutron lifetime. The most favourable (energetically speaking) configuration for neutron to survive was to bind with protons and form deuterium. It was the beginning of a small chain of reactions which produced the lightest stable nucleus.

\myeq{
& ^{1}H^{+} + \, ^{1}H^{+} \rightarrow \, ^{2}He^{2+} \rightarrow \, ^{2}H^{+} + e^{+} + \nu_{e}\\
& ^{2}H^{+} + \, ^{1}H^{+} \rightarrow \, ^{3}He^{2+} + \gamma\\
& ^{3}He^{2+} + \, ^{3}He^{2+} \rightarrow \, ^{6}Be^{4+} \rightarrow \, ^{4}He^{2+} + \, ^{1}H^{+} + \, ^{1}H^{+}
}

\noindent Theoretically, this chain could have continued and form heavier nuclei since the temperature was still high enough to allow more fusions but, in practice it stopped. Indeed, the formation of $^{4}He^{2+}$ occurs very late because it has to wait for the deuterium to be first produced (this is called the "deuterium bottleneck"). The deuterium formation occurs also very late due to the fact that its binding energy is very small and that during a long time there were still too many photons with enough energy to dissociate the deuterium.\footnote{The photons average kinetic energy was smaller than the deuterium binding energy, but there were so many photons with respect to deuterium than even the small proportion of photons with enough energy to dissociate the deuterium was large: $n_{\gamma}/n_{B}\sim 10^{10}$.} Moreover, the fusion of two $^{4}He^{2+}$ or of a $^{4}He^{2+}$ and a proton would produce a nucleus with an atomic mass of 8 or 5 respectively and there is no such such nucleus. However, there is a small fraction of the $^{3}He^{2+}$ population which has been able to reacts with some $^{4}He^{2+}$ to produce a small proportion of $^{7}Li^{3+}$.
\\

\begin{figure}
\centering
\includegraphics[scale=0.5]{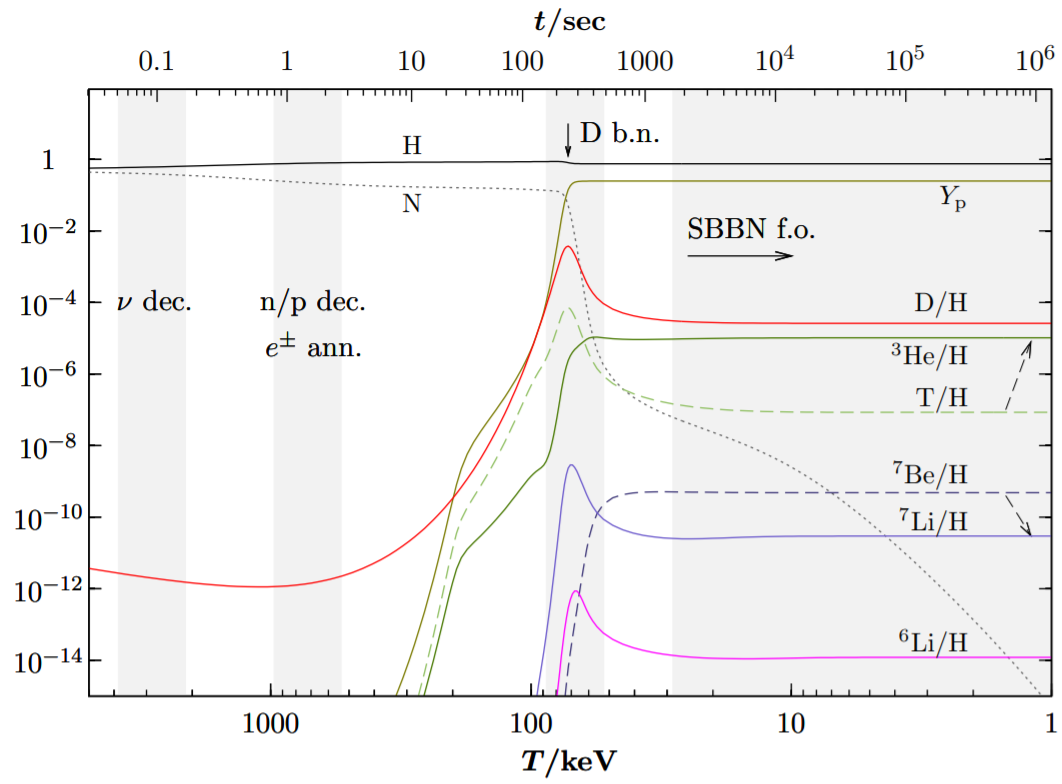}
\caption[Time evolution of abundances of light nuclei produced during BBN]{Evolution of abundance of light nuclei produced during BBN with respect to time and temperature, taken from \cite{Pospelov:2010hj}.} \label{fig:CST-bbn}
\end{figure}

Figure \ref{fig:CST-bbn} gives the light nuclei relative abundances with respect to time (as the temperature of the Universe decreases with time) \cite{Pospelov:2010hj}. One can see that the BBN started while the Universe had $T\sim 0.2$ MeV and stopped when the temperature reached a few keV. Afterwards, the relative abundance of light nuclei was fixed and these relative abundances provide strong constraints on any new physics modifying the thermal bath at these energies.

\subsection{Photodisintegration BBN constraints}\label{subsec:BBN-CST}
If, by decaying, the light mediator produces eventually photons (either directly or indirectly from decay products), these photons could potentially dissociate the light nuclei formed during BBN. This would then change the relative abundances of light nuclei today which are known today with a good accuracy for some of them (see e.g. \cite{Sarkar:1998gx,Kirkman:2003uv}) such that constraints can be stringent as we will see.
\\

This effect of photodisintegration will be particularly relevant if the light mediator decays into $e^{+}e^{-}$ or $\gamma\gamma$ directly and also if these particles are energetic enough to dissociate light elements $^{2}H$, $^{3}H$ and $^{4}He$. The minimum energy to dissociate these elements are $E_\gamma^{^2H}=2.22$~MeV, $E_\gamma^{^3H}=6.92$~MeV and $E_\gamma^{^4He}=28.3$ MeV respectively. These thresholds imply that if the mass of the decaying particle lies below the lowest of the three (i.e. if $m_{\rm med}<4.4$ MeV) there is no constraint coming from photodisintegration. Moreover, due to the deuterium bottleneck that we have seen in the previous subsection, the light nuclei predominantly form only when $t\geq180$ sec. Constraints are then much weaker for shorter lifetime or actually even up to $\sim 10^{3.5}$ sec.

\subsection{BBN constraints from modification of the Hubble constant and entropy injection}
Like in the case of CMB constraints, as long as the mediator has not decayed it modifies the Hubble constant. While when it eventually decays, it will inject entropy into the thermal bath. Both the Hubble constant modification and the entropy injection modify the relation between time and temperature, which can lead to significant modification of the light nuclei abundances when integrating the Boltzmann equations for the abundances of the nuclei. Since observations are able to precisely determine relative abundances of some of the light nuclei, it is then possible to constrain the light mediator number density prior to decay or, equivalently, the abundance it would have today if it was not decaying, as a function of its lifetime.
\\

As for the previous constraint, this one is also relevant only if a sizeable fraction of the light mediator decays while the light elements are already dominantly formed. That is to say when $t>t_{\rm BBN}\simeq 180$ sec, or equivalently when $T<T_{\rm BBN}\simeq 0.07$~MeV. Which means that this constraint applies only if $t_{\rm dec}=B_{\mathcal{L}}(T_{\rm dec})\tau_{\rm med}\gtrsim t_{\rm BBN}$ where $B_{\mathcal{L}}(T_{\rm dec})\simeq T(t_{\rm dec})/m_{\rm med}$ is the approximate relativistic Lorentz boost factor which applies when $m_{\rm med}\lesssim T^{BBN}$.
\\

\begin{figure}
\centering
\includegraphics[scale=0.75]{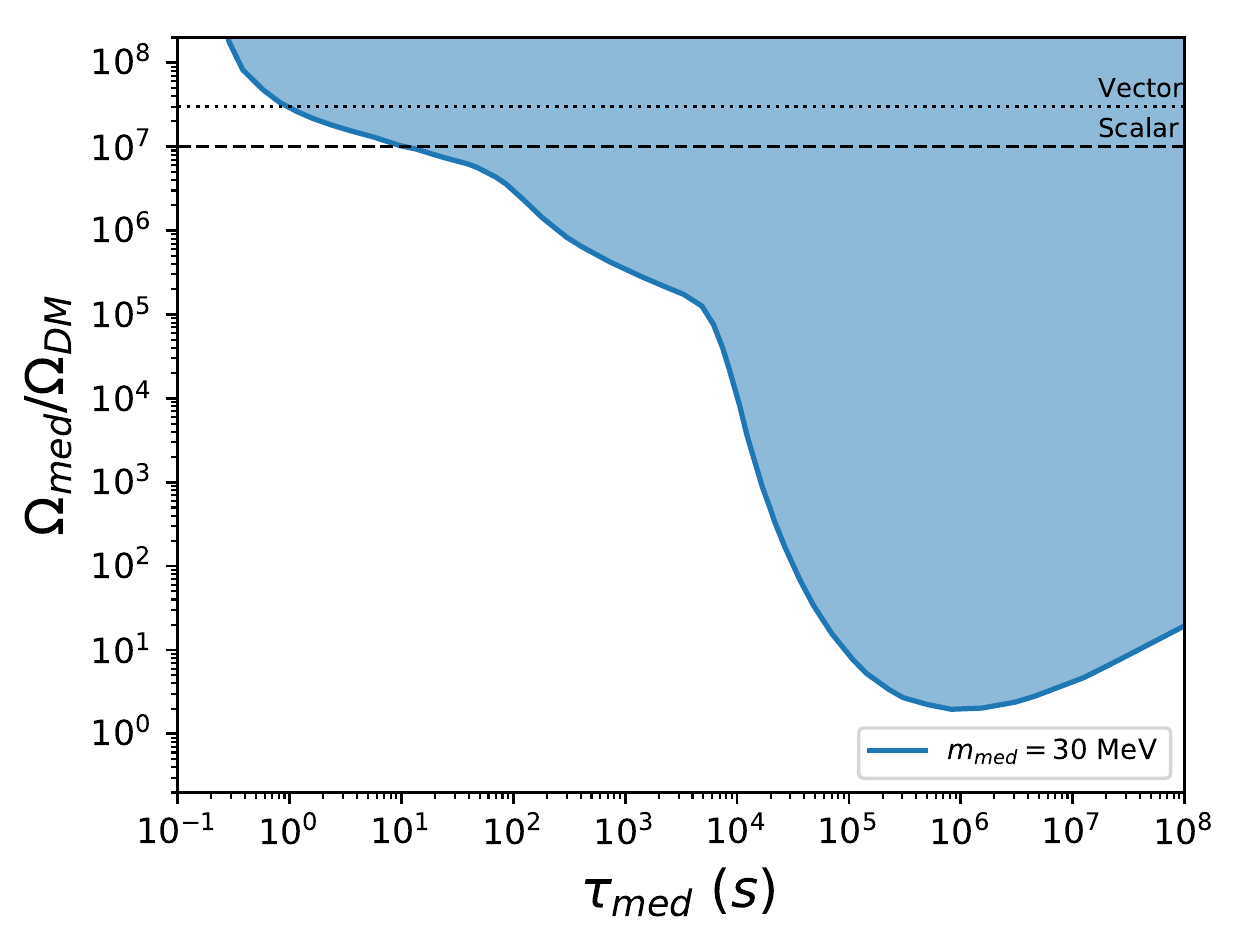}
\caption[Upper bound on the amount of mediator particle there would be today if there was no decay from the BBN constraint]{Upper bound on the amount of mediator particle there would be today if there was no decay \cite{Hufnagel:2018bjp} from the BBN constraint. The dotted (dashed) line shows the value of $\Omega_{\rm med}/\Omega_{\rm DM}$ one would expect from a relativistic decoupling of a non-decaying vector (resp. scalar) particle.} \label{fig:CST-omega_bbn}
\end{figure}

Similarly to the $N_{eff}$ bound, we used in Table \ref{table:Neff_Az} the results taken from \cite{Hufnagel:2018bjp}, to give the upper bounds on the light mediator lifetime that the Hubble constant/entropy injection constraints, together with the previous constraint, require. Results in this table are given assuming that the light mediator decouples relativistically at $T_{\rm dec}=10$ GeV. Looking at all constraints given in Table \ref{table:Neff_Az}, one can conclude that, when the decaying particle decouples while still relativistic, the Hubble constraint and the entropy injection constraints turn out to be always more stringent than the constraint coming from the photodisintegration. We present in Figure \ref{fig:CST-omega_bbn} all constraints given by the BBN \cite{Hufnagel:2018bjp} in the plane $\tau_{\rm med}-\frac{\Omega_{\rm med}}{\Omega_{\rm DM}}$ for $m_{\rm med}=30$ MeV.

\section{Supernovae 1987a}\label{sec:CST-SN1987a}
Astrophysical observations are also considerably helpful in the quest to constrain DM portal models. Indeed, DM models with MeV-scale mediator can be probed by observations of supernovae collapse, but these bounds are model-dependant. 
\\

\begin{figure}
\centering
\includegraphics[scale=0.565]{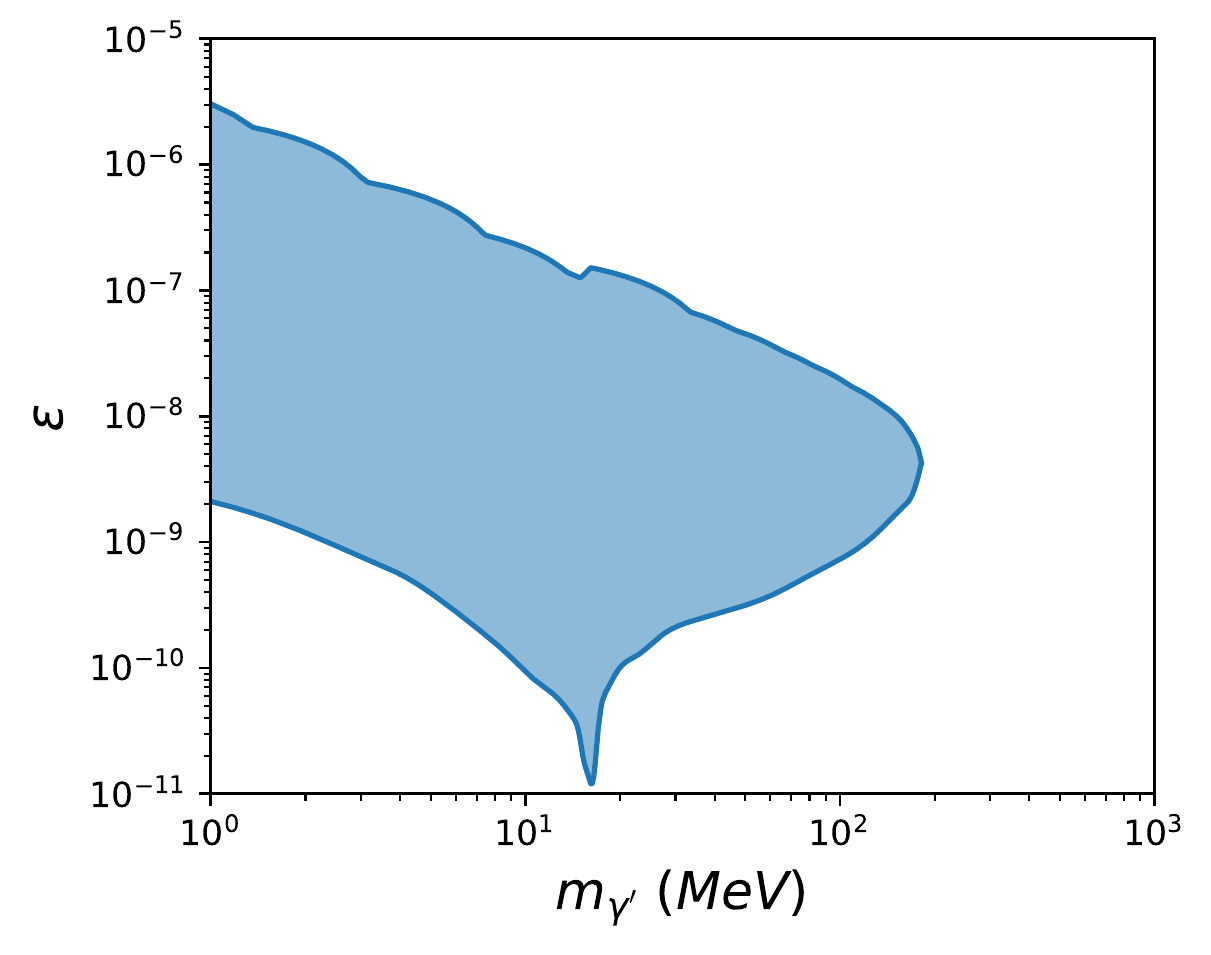}
\includegraphics[scale=0.565]{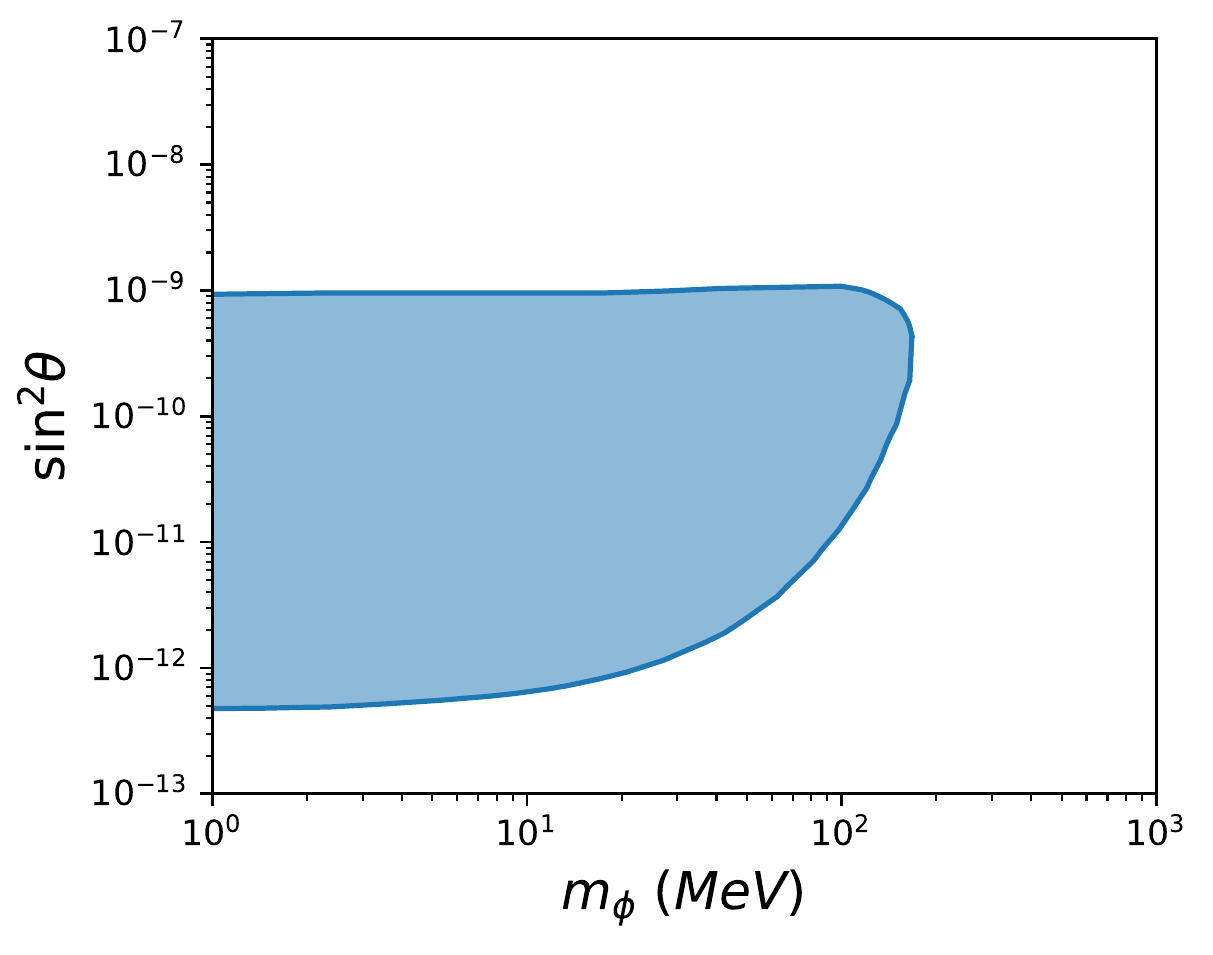}
\caption[Constraints from observation of SN1987A for the vector portal model and the scalar portal model]{Constraints from observation of SN1987A for the vector portal model (left) (see e.g. \cite{Kazanas:2014mca,Chang:2016ntp,Mahoney:2017jqk,Chang:2018rso}) and the scalar portal model (right) (see e.g. \cite{PhysRevD.36.2201,PhysRevLett.60.1797,PhysRevD.39.1020,Essig:2010gu,Cline:2013gha,Krnjaic:2015mbs,Kahlhoefer:2017ddj,Chang:2018rso})} \label{fig:CST-SN}
\end{figure}

Inside supernovae, in particular for the SN1987A supernovae observed in 1987, proton-proton and proton-neutron scattering occur very often and can produce photons in the final state. In the kinetic-mixing portal model, those photons have a non-zero probability to oscillate into dark photons (i.e. the light mediator in this model). This effect can have a strong impact on the supernovae luminosity, on its internal dynamics or on the observable gamma ray flux. The one of these effects which will be dominant depends on where the oscillation mostly happen and where the dark photons mostly decay: in the core of the supernovae, between the core and the external surface or outside the object already travelling toward us. Observations of SN1987A can constrain the strength of the mixing parameter and the mediator mass since they are parameters ruling the conversion and the decay rates. It is then convenient to present bounds from supernovae in the $m_{\gamma'}-\epsilon$ plane for this portal model, see left panel of Figure \ref{fig:CST-SN}. This figure can be summarised by requiring that $\epsilon\notin [10^{-10},10^{-6}]$ \cite{Kazanas:2014mca,Chang:2016ntp,Mahoney:2017jqk,Chang:2018rso}.
\\

For the case of the scalar portal, the new light mediator could also be produced during the supernovae explosion and could contribute to its energy loss. The observable neutrino pulse which is emitted during the collapse of the core would thus be shortened. This kind of constraint has been used to reduce the viable parameter space of axions and axion-like particles for which the production in supernovae is dominated by radiative production off nucleons.  Right panel of Figure \ref{fig:CST-SN} shows this constraint in the $m_{\phi}-\sin^{2}\theta$ plane and can be summarised by imposing $\lambda_{\phi H}\times \left(v_{\phi}/\text{GeV}\right)\notin\left[3\times 10^{-5},2\times 10^{-3}\right]$ \cite{PhysRevD.36.2201,PhysRevLett.60.1797,PhysRevD.39.1020,Essig:2010gu,Cline:2013gha,Krnjaic:2015mbs,Kahlhoefer:2017ddj,Chang:2018rso}.
\\

However, note that physics of supernovae is subject to systematic uncertainties and such constraints (for both the vector and the scalar portal models) are generally considered as less robust that those based on cosmological production and decay of light mediator that we have seen in previous sections.

\section{Direct detection}\label{sec:CST-DD}
Aside from Cosmology and Astrophysics, if the DM is made of new particles, it should be possible to constrain a DM model from the particle physics perspective. There are many different particle physics experiments trying to constrain DM models, but most of them can be organised in three categories: direct detection, indirect detection and collider experiments. Let us start by developing the first category. Direct detection experiments try to constrain DM models by looking at heavy nucleus in ordinary matter which could be hit by some extra-terrestrial particles. In this case, the nucleus will move backward and emit an electromagnetic signal which will propagate in the material and is proportional to the strength of the collision. The energy recoil is then measured and, comparing it to what is expected from the known background, it is possible to provide an upper bound on the scattering cross section of DM on the chosen nucleus (or on nucleon).
\\

It is usual for direct detection experiments to report their results in the plane: elastic collision cross section on nucleon ($\sigma_{\text{DM,n}}$) and DM mass ($m_{\rm DM}$). The considered cross section can be nucleus spin-dependent or spin-independent, but we choose to focus on the spin-independent cross section case as it provides the strong-est constraint. Experimental constraints on the DM-nucleon spin-independent cross section are generally given assuming a contact interaction. It means that collaborations usually assume a short-range interaction between DM and the nucleon mediated by a heavy mediator. This assumption has as the consequence that the cross section can be taken as independent of the nucleus recoil energy ($E_{R}$):

\myeq{
{\sigma_{{\rm DM},n}\sim \frac{1}{\left(t- m_{\rm med}^2\right)^{2}} }= \frac{1}{\left(2m_N E_R+ m^2_{\rm med}\right)^{2}}\simeq \frac{1}{m_{\rm med}^{4}}.\label{eq:DD-cross-section}
}

\begin{figure}
\centering
\includegraphics[scale=0.75]{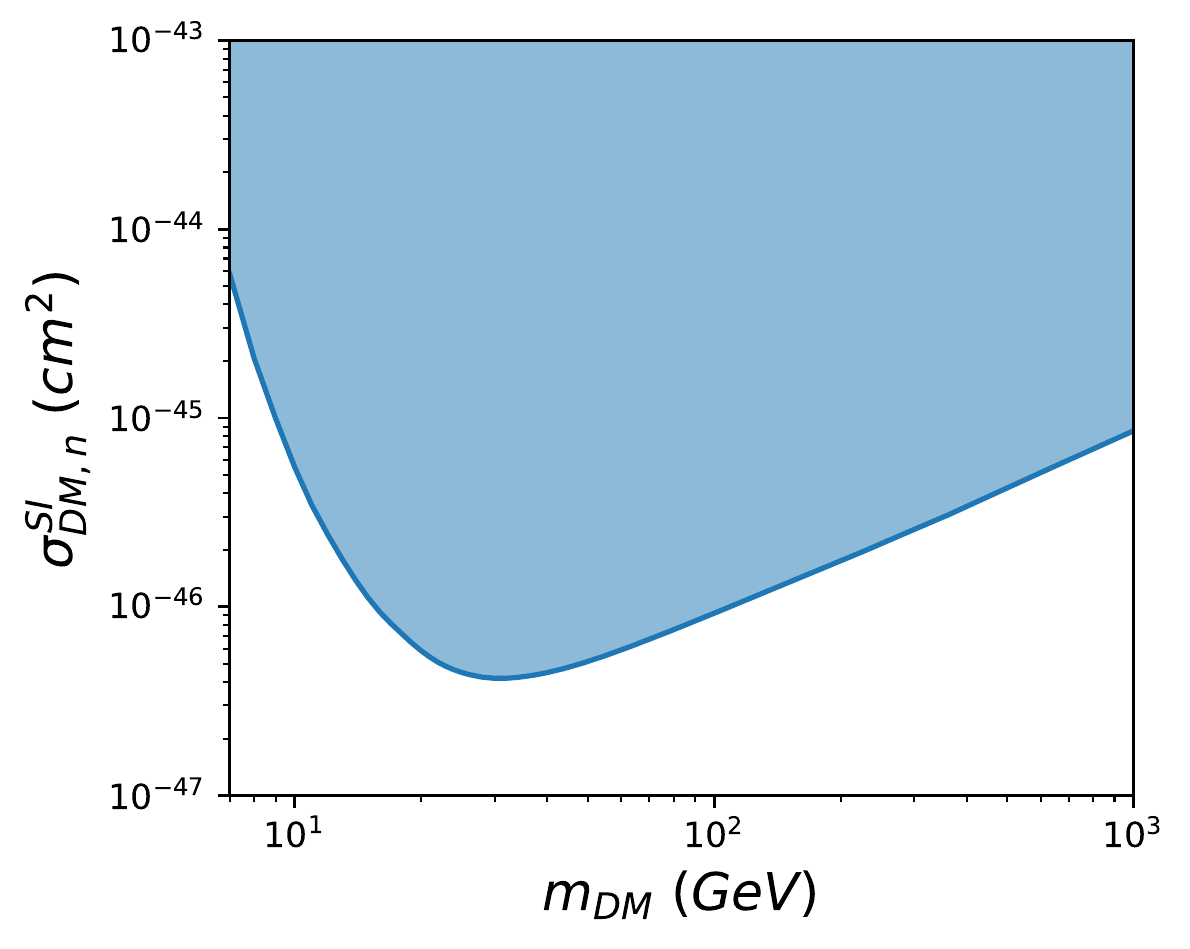}
\caption[Upper bound on the DM-nucleon collision spin-independent cross section for a contact interaction from Xenon1T]{Upper bound on the DM-nucleon collision spin-independent cross section for a contact interaction from Xenon1T \cite{Aprile:2018dbl}.} \label{fig:CST-XENON1T}
\end{figure}

\noindent The contact interaction approximation holds when $m_{\rm med}>\sqrt{2m_N E_R}\sim \mathcal{O}\left(40\text{ MeV}\right)$. Where we have used the typical value of the measured recoil energy which applies for the most sensitive current experiment, Xenon1T, which uses xenon material with $E_{R}\gtrsim 5$ keV and $m_{^{131}_{54}Xe} \simeq 123$ GeV. We show in Figure \ref{fig:CST-XENON1T} the latest results of the Xenon1T collaboration which follow a 1 tonne$\times$year exposure, considering this contact interaction assumption \cite{Aprile:2018dbl}. We can see on this figure that, for a 30 GeV DM mass, it goes down to $7\times 10^{-47}$ cm$^{2}$.
\\

In this work, we are more interested in the long range interaction scenario (i.e. light mediator) where the cross section goes like $\sigma_{{\rm DM},n}\propto 1/\left(m_{N}E_{R}\right)^{2}$ for $m_{\rm med}< \mathcal{O}\left(40\text{ MeV}\right)$ as we can see from Eq. \ref{eq:DD-cross-section}. Then, the upper bound from Figure \ref{fig:CST-XENON1T} cannot be directly applied to the light mediator case and need to be recast since, surprisingly, direct detection experiments like Xenon1T did not do an analysis devoted to the light mediator assumption. This is what we did and, even if this chapter is mostly introductory, we will here already present this original part. In fact, since the experimental recoil energy is typically of the order of a few keV, one understands that the scattering cross section is highly boosted in the light mediator case with respect to the heavy mediator case. To this end, we start from the differential rate of collisions (the number of events per second per unit of recoil energy)

\myeq{
\frac{\mathrm{d}R}{\mathrm{d}E_R}={N_T \,n_{\rm DM} \int \frac{\mathrm{d}\sigma}{\mathrm{d}E_R} {v} f_{\mathrm{\oplus}}\left(\vec{v}\right)\mathrm{d}^3v},
\label{eq:Diff-Rate}
}

\noindent with $N_T$ the number of target nucleus, $n_{\rm DM}$ the local number density of DM and $f_{\oplus}\left(\vec{v}\right)$ its velocity distribution in the Earth frame, which we take to be Maxwellian with r.m.s\footnote{Root mean velocity: $\sigma_{v}\equiv \sqrt{\langle v^{2}\rangle}$.} velocity $\sigma_v=270\text{ km}/\text{s}$ in the Galactic reference frame. The integration is made on $v\in\left[v_{min}, c\right]$ where $v_{\text{min}}=\sqrt{{m_N E_R}/{2\mu_{\chi N}^2}}$ with $\mu_{\chi N}$ the DM-N reduced mass and $c$ the speed of light. The DM-nucleus differential cross section given as a function of the recoil energy can be written as \cite{Fornengo:2011sz}

\myeq{
\frac{\mathrm{d}\sigma}{\mathrm{d}E_R}=\frac{m_N}{2\mu_{\chi p}^2}\frac{1}{v^2}\sigma_{\chi p}(E_R) Z^2 F^2\left(q\, r_A\right),\label{eq:dsigmadE-KM}\\
\frac{\mathrm{d}\sigma}{\mathrm{d}E_R}=\frac{m_N}{2\mu_{\chi n}^2}\frac{1}{v^2}\sigma_{\chi n}(E_R) A^2 F^2\left(q\, r_A\right).\label{eq:dsigmadE-HP}
}

\begin{figure}
\centering
\includegraphics[scale=1]{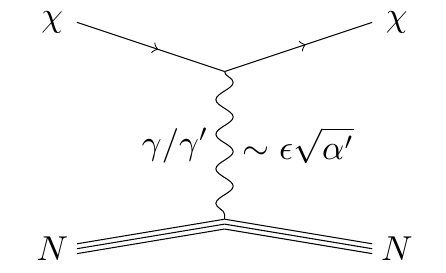}
\includegraphics[scale=1]{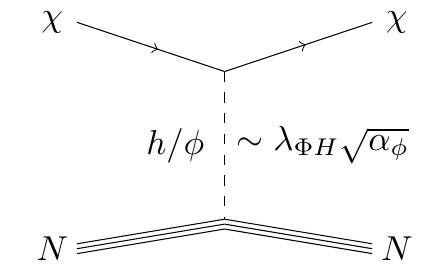}
\caption[Feynman diagrams of DM elastic scattering on nucleus.]{Feynman diagrams of DM elastic scattering on nucleus for the vector (left) and scalar (right) portal model.} \label{fig:CST_DD}
\end{figure}

\noindent Where $\mu_{\chi p}$ (resp. $\mu_{\chi n}$) is the DM-proton (resp. DM-nucleon) reduced mass and $Z$ (resp. $A$) the nucleus atomic number (resp. the nucleus mass number). The $F\left(q r_A\right)$ is the nucleus form factor, which for concreteness we take from \cite{Fornengo:2011sz,PhysRev.104.1466}. The first of the two cross sections above is used if the mediator couples only to the proton, like in the kinetic mixing portal model where the interaction goes through the electromagnetic charge. The second is used when the mediator also couple to the neutrons. Like in the scalar portal model where the interaction goes through the Higgs coupling, the DM couples then in the same way to the proton and to the neutron (modulo small isospin breaking effects). For concreteness, these elastic collision cross sections of Figure \ref{fig:CST_DD} are given by,

\myeq{
&\sigma_{\chi p}(E_R)=\frac{16\pi \mu^2_{\chi p}\alpha^2\kappa'^2}{\left(2m_NE_R+m_{\gamma'}^2\right)^2},\label{eq:cross-section-KM}\\
&\sigma_{\chi n}(E_R)=\frac{\mu^2_{\chi n}y_{hnn}^2\kappa_{\phi}^2}{\pi\left(2m_NE_R+m_{\phi}^2\right)^2},\label{eq:cross-section-HP}
}

\noindent with $\alpha$ the QED fine structure constant and $y_{hnn}\simeq 1.2\times 10^{-3}$ the $h$-nucleon-nucleon coupling constant \cite{Cline:2013gha}. Note that in Eqs. \ref{eq:cross-section-KM} and \ref{eq:cross-section-HP}, we have introduced the DM-to-SM coupling which is, as we will see in Chapter \ref{ch:prod}, a combination of the SM-to-med and DM-to-med couplings: $\kappa' \equiv \epsilon\sqrt{\alpha'/\alpha}$ in the vector portal model and $\kappa_{\phi} \equiv Y_{\chi}\sin\left(2\theta\right)/2$ in the scalar portal model. When the mediator is lighter that the threshold given above (i.e. $m_{\rm med}<40$ MeV), direct detection constraints depend only on two parameters: the DM mass ($m_{\rm DM}$) and the coupling between the DM and the SM ($\kappa'$ and $\kappa_{\phi}$). It is then natural to represent these constraints in the plane $m_{\rm DM}-\kappa$. 
\\

In order to convert constraints on $m_{\rm DM}-\sigma_{\text{DM,n}}$ for a contact interaction into constraints on $m_{\rm DM}-\kappa$ for a long range interaction, we will have to look a little bit closer to the shape of the differential rate. Indeed, for a same DM mass, the shape of the differential rate in the light mediator case can behave differently than in the heavy mediator case. Thus, the limit on the coupling strength for one value of the DM mass in the light mediator case, $m_{\rm DM}$, cannot be obtained directly from the limit in the heavy case. Instead, one must look for the value of the DM mass in the heavy case, $m_{\rm DM}^{heavy}$, for which the differential rate matches the best the differential rate obtained in the light mediator case for the DM mass considered, i.e. $m_{\rm DM}$. This must be done within the range of the energy recoil where the differential rate is the highest as given by the efficiency form factor. The measured differential rates fall rapidly at low and high recoil energies and this regardless of the type of interactions. These experimental features allow to convert every couple $(m_{\rm DM}^{heavy},\sigma_{\text{DM,n}})$ from Figure \ref{fig:CST-XENON1T} onto a new couple $(m_{\rm DM},\kappa)$. In practice, to get a proxy for the observable differential rate, 

\myeq{
\Big(\frac{\mathrm{d}R}{\mathrm{d}E_R}\Big)_{\text{exp}}=\epsilon\left(E_R\right) \frac{\mathrm{d}R}{\mathrm{d}E_R},
\label{eq:dRdE}
}

\noindent where $\epsilon(E_R)$ is the detector efficiency from Figure 1 in \cite{Aprile:2018dbl} and with ${\mathrm{d}R}/{\mathrm{d}E_R}$ given by Eq. \ref{eq:DD-cross-section}. The differential cross section which appears in the expression of the differential rate for a short range interaction is the same as given in Eq. \ref{eq:cross-section-HP} but with a constant cross section. For every couple $(m_{\rm DM}^{heavy}, \sigma_{\text{DM,n}})$ along the upper bound line in Figure \ref{fig:CST-XENON1T}, we then determine the couple $(m_{\rm DM}, \kappa)$ that would have a similar observable differential rate $(dR_{\rm DM}/dE_R)_{\text{exp}}$. Concretely, we consider the couple $(m_{\rm DM}, \kappa)$ that minimizes the quadratic distance between the two rates
\myeq{
\Delta^2_{\rm DR} = {1\over R_{\text{exp}}^2} \int \mathrm{d}E \,\epsilon\!\left(E\right)^2 \left(\Big(\frac{\mathrm{d}R}{\mathrm{d}E}\Big)-\Big(\frac{\mathrm{d}R_{\rm DM}}{\mathrm{d}E}\Big)\right)^2,
\label{eq:Error}
} 

\noindent where $R_{\text{exp}}$ is the total measurable rate from Eq. \ref{eq:dRdE} and $dR_{\rm DM}/d E_R$ is the differential rate corresponding to a candidate with a light mediator.
\\

\begin{figure}[h]
\centering
\includegraphics[width=8cm]{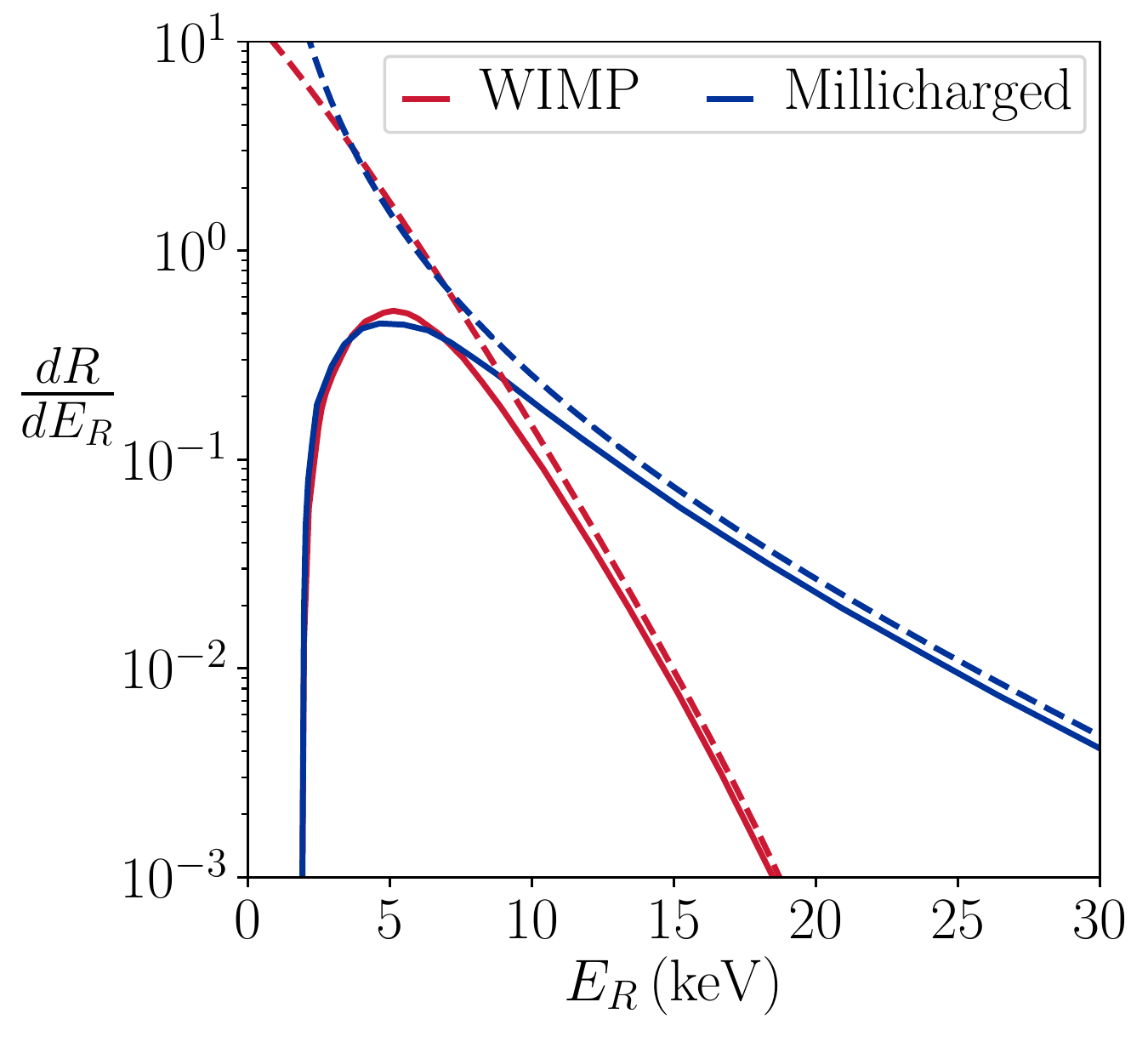}
\caption[Comparison of DM scattering on nucleon differential rate between short and long range interactions]{Red solid: differential rate for a $\left(m_{\text{DM}},\sigma_{{\rm DM},n}\right)=\left(15 \text{ GeV}, 10^{-46}\text{ cm}^2\right)$. Blue solid: best fit for a DM candidate with long range interactions with $\left(m_\chi,\kappa\right)=\left(70\text{ GeV},3.1\times 10^{-11}\right)$. The error is $\Delta_{\text{DR}} \approx 25\%$. The dashed curves are for the same candidates, but not taking into account Xenon1T efficiency. Taken from \cite{Hambye:2018dpi}.}
\label{fig:DiffRateFit}
\end{figure}

In Figure \ref{fig:DiffRateFit} is shown an example of this recasting procedure according to our criteria (Eq. \ref{eq:Error}). Starting from the couple $\left(m_{\text{DM}}^{heavy},\sigma_{{\rm DM},n}\right)=\left(15 \text{ GeV}, 10^{-46}\text{ cm}^2\right)$ from the Xenon1T upper bound (see Figure \ref{fig:CST-XENON1T}), we have computed the theoretical (dashed red) and observable (solid red) differential rates. The latter is obtained from the former by multiplying by the experimental efficiency curve. Taking the kinetic mixing model (i.e. Eqs \ref{eq:dsigmadE-KM} and \ref{eq:cross-section-KM}) for example, our procedure of Eq. \ref{eq:Error} tells us that the candidate $\left(m_\chi,\kappa\right)=\left(70\text{ GeV},3.1\times 10^{-11}\right)$ produces a differential rate which minimizes $\Delta_{\text{DR}}$. Thus, we show also on Figure \ref{fig:DiffRateFit} the theoretical (dashed blue) and observable (solid blue) differential rates. At this point, we must emphasise that both differential rates (from heavy and light mediator cases) match mostly at low recoil energies $E_R<10$ keV which correspond to higher expected number of events. In the example we are showing in Figure \ref{fig:DiffRateFit} \cite{Hambye:2018dpi}, the error is $\Delta_{\rm DR} \simeq 25 \%$ while the error on the total number of event (see Eq. \ref{eq:Error-total} below) is $\Delta_{\rm TR} \simeq 20 \%$.
\\

Defining the error on the total number of event $N_{\rm DM}$ as,

\myeq{
\Delta_{\rm TR} = \frac{N_{\rm DM} - N_{\rm DM}^{heavy}}{N_{\rm DM}^{heavy}},\label{eq:Error-total}
}

\noindent where $N_{\rm DM}^{heavy}$ is the total number of event obtained in the heavy mediator case. One can then compare the two errors $\Delta_{\rm DR}$ and $\Delta_{\rm TR}$. As an illustration, we give in Table \ref{table1} several DM candidates from the heavy mediator case and the one it corresponds to in the light mediator case. With the error on the total rate $\Delta_{\rm TR}$ and the error on the differential rate $\Delta_{\rm DR}$. We emphasise that the errors on the total rate are positive, meaning that the total number of events is always larger for the $\chi$ particle than for the corresponding DM particle with a massive mediator. Reducing the error on the total number of events would thus require decreasing the parameter $\kappa$. In that sense, we deem our constraints on $\kappa$ to be conservative.
\\

\begin{table}
\centering
\begin{tabular}{|c|c|c||c|c||c|c|}
\hline
  $m_{\rm DM}$ (GeV)& $\kappa'\, (10^{-11})$ & $\kappa_{\phi}\, (10^{-9})$ & $m_{\rm DM}^{heavy}$ (GeV) &$\sigma_{{\rm DM},n}$ (cm$^2$) & $\Delta_{\rm DR}$ & $\Delta_{\rm TR}$   \\
 \hline
  $15$  & $3.0$ & $1.0$ & $10$ & $5.6\times 10^{-46}$ &   $16 \%$ &  $15 \%$  \\
$70$   & $3.1$ & $1.0$ & $15$ & $1.1\times 10^{-46}$ & $23 \%$ & $22 \%$\\
 $200$  & $5.2$ & $1.8$ &  $20$& $5.9\times 10^{-47}$ &  $22 \%$ &  $13 \%$  \\
 $ 500$ & $8.2$ & $2.8$ & $22 $& $5.3 \times 10^{-47} $& $26\%$& $3\%$\\
 \hline
\end{tabular}
\caption[Upper bounds on the mixing parameter for four DM masses, based on the correspondence with WIMP exclusion limits]{Upper bounds on the mixing parameter $\kappa$ (2nd and 3rd columns) for four DM masses (1st column), based on the correspondence with WIMP exclusion limits (4th and 5th columns). The last two column give respectively the error on the differential rate $\Delta_{\rm DR}$ and total rate $\Delta_{\rm TR}$. 
\label{table1}
}
\end{table}

\begin{figure}
\centering
\includegraphics[scale=0.70]{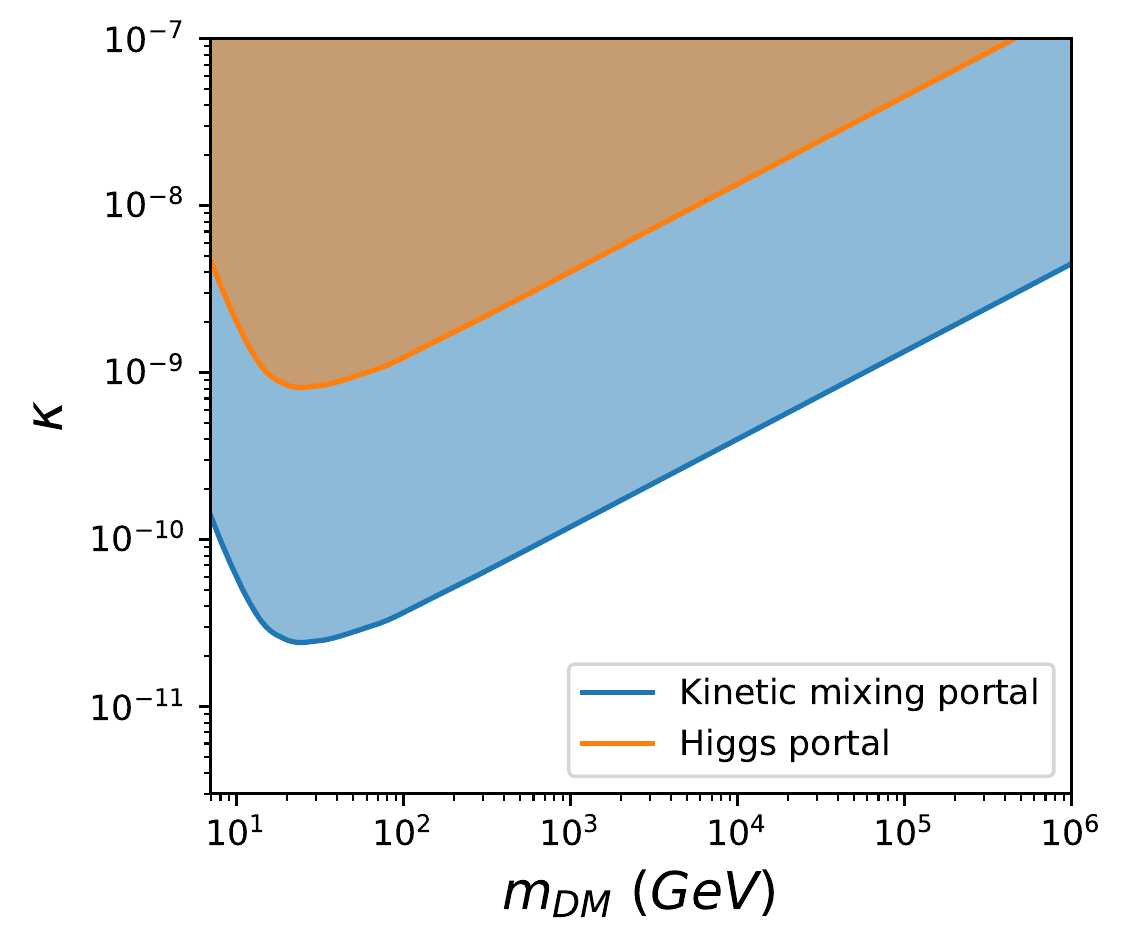}
\caption[Direct detection constraints for an interaction mediated by a light mediator]{Direct detection constraints for an interaction mediated by a light mediator (i.e. $m_{\rm med}\lesssim 30$~MeV). The blue (orange) solid line is the current constraints from Xenon 1T experiment \cite{Aprile:2018dbl} from \cite{Hambye:2018dpi}.}\label{fig:DD_Constraints}
\end{figure}

We show in Figure \ref{fig:DD_Constraints}, the bound from Figure \ref{fig:CST-XENON1T} adapted for a light mediator scenario following our procedure we have just explained. From this figure, we can conclude that the direct detection constraint requires $\kappa_{\phi} \lesssim  10^{-9}$ and $\kappa'\lesssim 3\times 10^{-11}$ for a Higgs portal and a kinetic mixing portal respectively. This applies as soon as the light mediator below is lighter than $m_{\rm med}\ll 40$ MeV and the DM is heavier than $m_{\rm DM}\gtrsim$ GeV, which is usually the case while considering self-interactions constraints. To consider a value of $m_{\rm med}\sim100$ MeV hardly relaxes this bound. The very tiny values of the coupling, that direct detection can probe, show that this type of experiments is an extremely powerful tool to constrain light mediator scenarios. For comparison, in the heavy mediator scenario, with the mediator mass of order of the electroweak scale, direct detection experiments are able to probe coupling down to typically $10^{-1}$ or $10^{-2}$. Let us finally mention that in \cite{Hambye:2018dpi}, we have shown that these bounds on the DM-to-SM coupling in the light mediator scenario allow to constrain FI scenario as the one described in Subsection \ref{subsec:FI}. This will not be more detailed in this thesis, see \cite{Hambye:2018dpi} for more details.

\section{Indirect detection}\label{sec:CST-ID}
We have already insisted on the key role of the mediator in order to constrain DM portal models (for CMB, BBN and particularly for direct detection as we just discussed), especially if the mediator is much lighter than the DM. This mass hierarchy turns out to play an important role for indirect detection experiments too. That is to say for intensity of the flux of particles that could result from DM annihilation or decay happening today in the Milky Way or beyond, see Section \ref{sec:somm_id}. A mediator much lighter than the DM particle will boost the DM annihilation today into lighter particles (SM or not) via the Sommerfeld effect. This effect is known for being crucial in indirect detection experiments for s-wave annihilation. It will considerably enhance indirect detection rates, as a result of the small dark matter particle velocity today. On the other hand, for a p-wave annihilation, indirect detection signals are usually not studied because considered as hopeless as they are suppressed by 2 powers of the velocity $v$ (instead of 0 power for the s-wave annihilation). Nevertheless, in presence of an extremely strong mass hierarchy between the DM and the light mediator, the Sommerfeld effect can compensate for this suppression, a property which has been hardly considered (see \cite{Das:2016ced} for an example of non self-interacting model). The Sommerfeld enhancement factor, which multiplies the annihilation cross section at tree level, goes like $1/v^3$ for p-wave annihilation, giving an overall $1/v$. This gives, in both s-wave and p-wave cases, an overall $1/v$ dependence. This arises in the same way as for the Sommerfeld boost enhancing the cross section at the CMB time. The difference comes from the fact that at recombination the velocity is so small that the scaling in $v$ does not go anymore in $1/v$ as it is the case today but scales as $v^0$ and $v$ for s-wave and p-wave respectively \cite{Bringmann:2016din}. Thus, for p-wave annihilations, the CMB constraint is clearly worse than for s-wave annihilations, but for indirect detection, this is not the case. This relevant property will be further studied below in Chapter \ref{ch:som}.
\\

\begin{figure}
\centering
\includegraphics[scale=0.70]{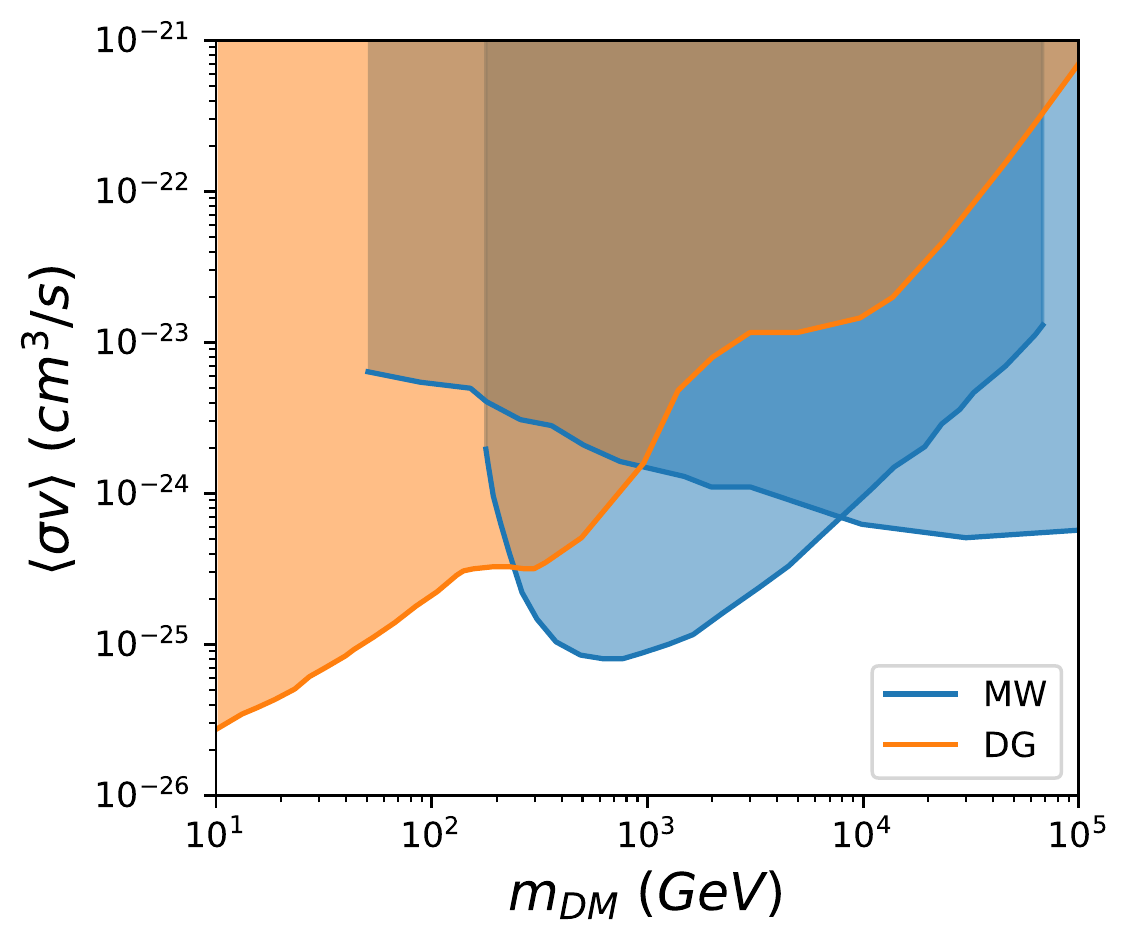}
\caption[Indirect detection constraints on the DM annihilation cross section]{Indirect detection constraints on the DM annihilation cross section into a $e^{+}/e^{-}$ pair. Blue solid lines show current constraints for annihilation in the Milky Way and have been taken from Fermi-LAT, H.E.S.S and Antares experiments \cite{Ackermann_2012,Abramowski_2013,Ackermann_2015,Abdallah_2016,Albert:2016emp}. Orange solid line shows current constraints for annihilation in dwarf galaxies, taken from Fermi-LAT and MAGIC experiments \cite{2016}. All these constraints assume a NFW profile.}\label{fig:ID_Constraints}
\end{figure}

We present in Figure \ref{fig:ID_Constraints} the current experimental upper bounds on the annihilation cross section today. The distinction between annihilation of DM in the Milky Way (MW) galactic centre and in dwarf galaxies (DG) comes from the fact that indirect detection observations both look at a flux of particles coming from the MW or DG, but that the DM velocity is not expected to be the same in both environments. In the Milky Way, $v_{\rm DM}^{MW}\simeq 2\times 10^{-3}$ while in dwarf galaxies, $v_{\rm DM}^{DG}\simeq 1\times 10^{-5}$.

\section{Production at colliders}\label{sec:CST-COL}
Last but not least, instead of detecting directly or indirectly DM particles, one can think to produce DM particles directly in collider experiments. In practice, since it is unlikely to detect a particle which supposedly interacts feebly with the SM, collider experiments are looking for signals with missing energy or with a lack of SM particles where more were expected if there was no interaction with DM. For example, in the light scalar mediator case, the Higgs portal interaction, Eq. \ref{eq:lag_hp}, induces an invisible decay channel for the Higgs boson, $H\rightarrow \phi\phi$. The Higgs decay rates into SM particles are well known theoretically and the corresponding lifetime and branching ratios are now measured with a good accuracy. It is then possible to constrain the Higgs invisible decay width with collider experiments. The current LHC bound is obtained combining two constraints given by CMS and ATLAS collaboration at LHC\footnote{Large Hadron Collider.}. The first comes from the fact that expected SM decay products are missing while looking at Higgs decay events. This gives an upper bound on the branching ratio of the Higgs invisible decay channel: $\text{BR}(h\rightarrow\text{inv.})<0.19$ (95\% C.L.) \cite{Sirunyan:2018owy}. This upper bound can be easily translated into an upper bound on the invisible decay width: $\Gamma_{\text{inv}}<0.96$ MeV. The second constraint comes from the observed Higgs production rate and impose an upper bound on the ratio of experimental and theoretical Higgs production rates: $\mu \equiv \left[\sigma_{h}\text{BR}(h\rightarrow\text{SM})\right]_{\text{exp}}/\left[\sigma_{h}\text{BR}(h\rightarrow\text{SM})\right]_{\text{SM}}<0.89$ (95\% C.L.) \cite{Khachatryan:2016vau} where $\sigma_{h}$ gives the Higgs production cross section. Again, this last bound is translated into $\Gamma_{\text{inv}}<0.50$ MeV. This value has to be compared to the decay width induced by the scalar portal interaction:

\myeq{
\Gamma _{H\rightarrow \phi\phi}&\simeq 0.50 \left(\frac{\lambda _{\phi H}}{0.01}\right)^{2}\text{ MeV}.
}

\noindent If the decay into two light mediator particles is the only invisible decay channel, this bound can be converted into an upper bound on the portal strength parameter: $\lambda _{\phi H}<0.01$.
\\

It is also possible to constrain the scalar portal scenario by looking for meson decays involving the light mediator in the final state (see e.g. Figure 3 of \cite{Bondarenko:2019vrb}). As a summary, this constraint requires that $\sin\theta<3\times 10^{-4}$ for a mediator mass smaller than $m_{\phi}\leq 100$ MeV, but does not constrain the model for heavier mediator masses.
\\

For the kinetic mixing portal, there are many constraints applying on the mixing parameter $\epsilon$, but all of them are relevant for rather large kinetic mixing value. As an example, the LHCb detector is looking for dark photon decays ($\gamma'\rightarrow\mu^{+}\mu^{-}$) which could have been produced in proton-proton collisions with a centre-of-mass energy of 13 TeV. This experiment is able to exclude kinetic mixing of order $\epsilon\sim 10^{-5},10^{-4}$ for dark photon masses of about $m_{\gamma'}\sim 200,300$ MeV \cite{Aaij:2019bvg}. Other constraints can be found in e.g. \cite{Berezhiani:2000gw,Goodsell:2009xc,Jaeckel:2010ni,Berezhiani:2008gi,McDermott:2010pa}.

\section{A global picture}\label{sec:global_picture}
\begin{figure}
\centering
\includegraphics[scale=0.75]{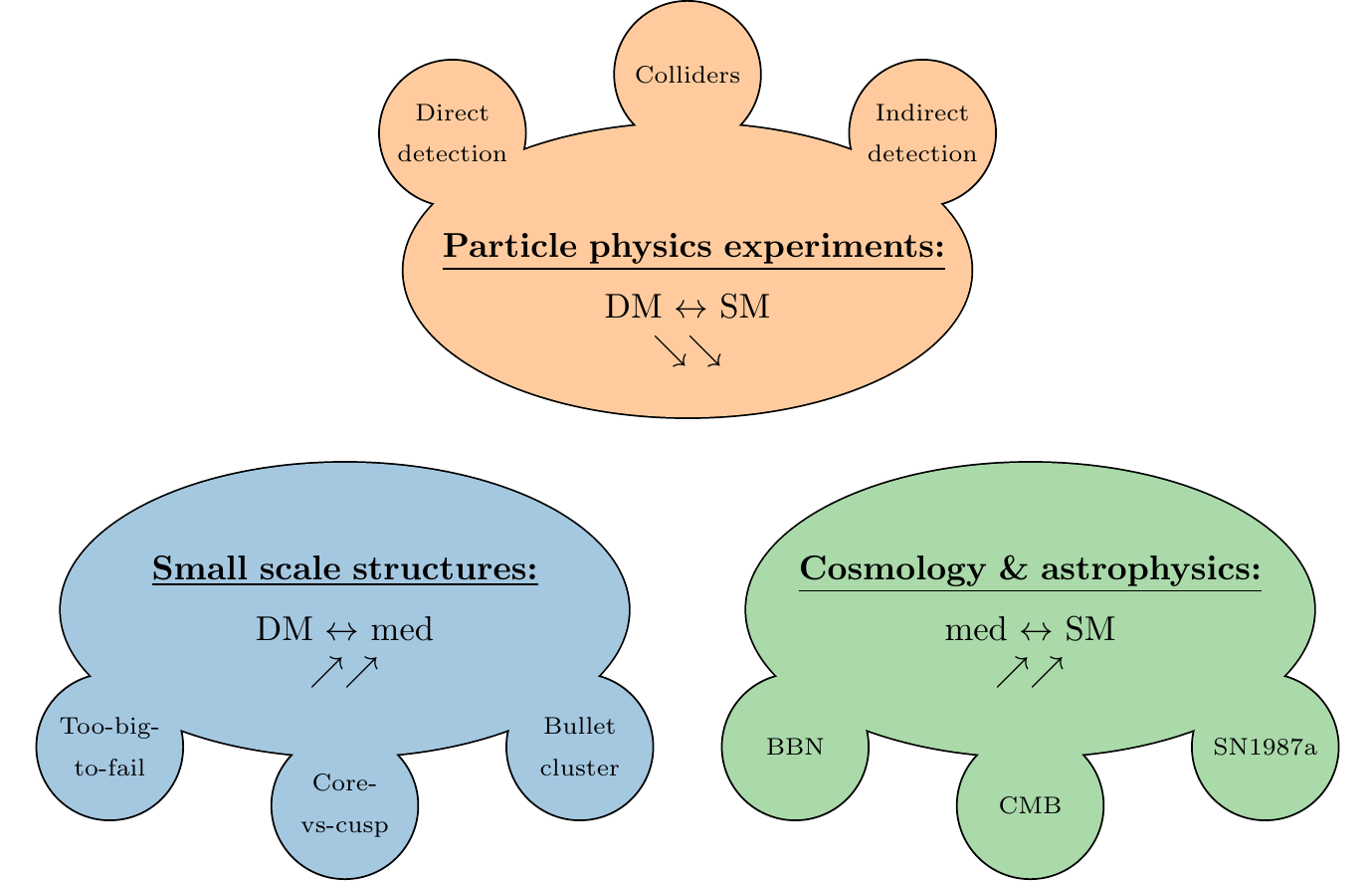}
\caption[Illustration of tensions between all observations on SIDM with light mediator]{Illustration of tensions between all observations on SIDM with light mediator.}\label{fig:CST-SIDM_pic}
\end{figure}

Finally, we conclude this long introducing chapter by collecting all the above information to emphasise how self-interacting DM models are constrained in many ways. Indeed, throughout this chapter, we have seen that there are three kinds of constraints applying on DM models (illustrated in Figure \ref{fig:CST-SIDM_pic}):

\begin{itemize}
\item Particle physics experiments dominantly constrain the strength of the interaction between DM and SM particles. They require a small interaction strength since no DM candidate has been discovered for now;
\item Small scale structures probe the strength of the interaction between DM and the mediator, i.e. the HS interactions. Problems at small scale can be alleviate if the DM self-interaction cross section is strong enough;
\item Cosmological and astrophysical observations test, above all, the interaction between the mediator and SM particles. They exclude a range of mediator lifetime or, for example, small couplings unless if very tiny.
\end{itemize}

\noindent All in all, small scale structures constraints together with cosmological and astrophysical observations suggest that DM should couple significantly to some SM particles while particle physics experiments forbid it. This tension seems inevitable and seems very difficult to alleviate in minimal DM models. This was one of the starting points of this thesis. In the following chapters, we will see how these many constraints can be partially or totally evaded and what is the price to pay for that in terms of assumptions to be made, extra particles to assume, etc ....

\chapter{The Sommerfeld effect}\label{ch:som}
\begin{figure}[h]
\centering
\includegraphics[scale=1]{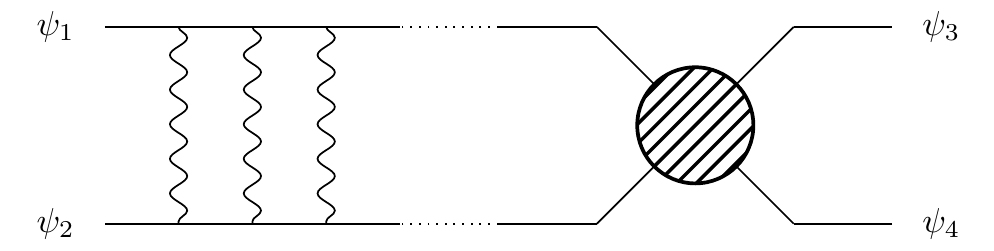}
\caption[Illustration of the Sommerfeld effect for a two-to-two process]{Illustration of the Sommerfeld effect for a two-to-two process.}\label{fig:app_som}
\end{figure}

\yinipar{D}uring an interaction between two non-relativistic particles, if they are close enough to each other\footnote{To be define quantitatively below.}, some non-perturbative effects can be significant. The Sommerfeld effect describes how two particles, if they couple to an additional lighter particle, can exchange this particle many times before that the proper interaction takes place \cite{ANDP:ANDP19314030302} as depicted in Figure \ref{fig:app_som}. The exchange of those particles can be determined from the non-relativistic potential holding between the incident particles. This potential can be either attractive or repulsive depending on the types and natures of the interactions and the involved particles.
\\

This Sommerfeld effect has an important impact for DM in many different contexts as DM freeze-out, DM self-interaction and DM indirect detection, where all of those situations involve non-relativistic scattering of two DM particles. This will have a major impact on all subsequent chapters. Results discussed in this second chapter are partially new and unpublished while others have been detailed in \cite{Hambye_2020}.

\section{Self-scattering case}
\subsection{The dimensionless Schrödinger equation}
Non-relativistic scattering of two DM particles can be determined from the Schrödin-ger equation which describes how a wave function (associated to the motion of particle) is modified by the action of an external potential well. Thus, this equation naturally encodes the Sommerfeld enhancement such that one must solve the Schrödinger equation for the reduced DM two-particle system. Moreover, we have learned from scattering theory that the differential cross section can be related to the out-coming wave function. Indeed, if the out-coming particles are of the same nature of the incoming particles (i.e. a simple self-scattering), one can write the out-coming wave function $\psi(\vec{r})$ as the incoming wave (which is simply a spherical wave) plus a perturbation:

\myeq{
\lim_{r\rightarrow \infty}\psi(\vec{r})\simeq e^{i\vec{k}\cdot\vec{r}}+f(\theta)\frac{e^{ikr}}{r},\label{eq:schro_BC}
}

\noindent where $\vec{r}$ is the position vector of both particles relatively, $\vec{k}$ the relative impulsion vector and $f(\theta)$ is a function of the azimuthal angle of the scattering in spherical coordinates. The $f$ function could be very complicated in general, but assuming an isotropic, short-ranged and elastic scattering, it can be expressed using the Legendre polynomial basis:

\myeq{
f(\theta) = \sum_{l=0}^{\infty}(2l+1)f_{l}(k)P_{l}(\cos\theta),\label{eq:f_dev}
}

\noindent with $f_{l}$ the partial wave scattering amplitudes for $l$ the angular momentum quantum number of the incoming wave and $P_{l}$ the Legendre polynomials. It is usual to use the phase shift parametrisation of the partial wave scattering amplitudes for each wave: $f_{l}(k)=e^{i\delta_{l}(k)}/k\sin\delta_{l}(k)$ such that the differential cross section can be written in terms of the phase shifts. Indeed, one can show that the differential cross section can be related to this $f(\theta)$ function by 

\myeq{
\frac{\diff \sigma}{\diff \Omega}=\vert f(\theta)\vert^{2}.
}

\noindent We then finally have for the differential cross section,

\myeq{
\frac{\diff \sigma}{\diff \Omega}=\frac{4\pi}{k^{2}}\sum_{l=0}^{\infty}(2l+1)\sin^{2}\delta_{l}(k).\label{eq:diff_s_diff_omega}
}

\noindent From this last expression, we understand that one must solve the Schrödinger equation for the reduced (incoming) two-particle system if one wants the full self-scatter-ing cross section with the Sommerfeld enhancement taken into account. Indeed, since the Schrödinger equation describes the evolution of the reduced two-particle wave function, solving it gives the wave function which, as we just saw, can be related to the differential cross section through the phase shift $\delta_{l}$. Then, one has, in principle, to integrate the differential cross section to get the total self-interacting cross section: 

\myeq{
\sigma = \int \frac{\diff\sigma}{\diff\Omega}\diff\Omega.
}

\noindent However, this cross section develops a divergence for the light mediator case known as the forward-scattering divergence. This divergence comes from the fact that at very small relative velocities, that is to say when angle between the trajectories of both incoming particles tends to zero, $\cos\theta\rightarrow 0$, the trajectories remain unchanged after the interaction. To regulate this divergence, assuming classical distinguishibility between incoming particles\footnote{This assumption will be always valid for what is our concern since we consider only DM Dirac fermion, see \cite{PhysRevA.60.2118,Tulin:2013teo} for more details.}, one can instead consider the transfer cross section which is obtained from the differential cross section, but weighted by the fractional longitudinal momentum transfer:

\myeq{
\sigma_{T} = \int \left(1-\cos\theta\right)\frac{\diff\sigma}{\diff\Omega}\diff\Omega.\label{eq:sT_def}
}

\noindent The Schrödinger equation describing the evolution of the reduced two-particle wave function for particles of mass $m_{\psi}$ and impulsion $\vec{k}$ in a potential $V$ is given by,

\myeq{
\frac{1}{m_{\psi}}\nabla^{2}\psi_{k}-V(r)\psi_{k}&=-\frac{k^{2}}{m_{\psi}}\psi_{k},\label{eq:app_Seq}
}

\noindent where $r$ is the radial coordinate in spherical coordinates and $\nabla$ is the gradient operator. Note the missing factor of two wherever the mass appears compared to the Schrödinger equation of a lonely free particle. This is due to the fact that we are considering the two-particle system wave function and thus its mass is the reduced mass which corresponds to $\frac{1}{2}m_{\psi}$. To this equation we impose as boundary conditions that the waves produced by the perturbation at $r=0$ are out-going: $\lim_{r\rightarrow \infty}\psi_{k}(r)\sim e^{ikx}+f(\theta)\frac{e^{ikr}}{r}$. Where we assumed that the incoming wave is going along the x-axis and where $f=f(\theta)$ can be written as in Eq. \ref{eq:f_dev}. In other words, before entering inside the zone of influence of the potential, the incoming wave is a plane wave (as it should) which is modified only when close to the perturbation at $r=0$. Later, far from the perturbation, the out-going wave is the original plane wave plus a spherical wave due to the central potential as required by Eq. \ref{eq:schro_BC}. Assuming a spherical symmetry for the potential, the wave function can be developed on the spherical harmonic basis \cite{cohen1973mecanique2},

\myeq{
\psi_{k}(r,\theta,\phi) &=\sum_{l=0}^{\infty}\sum_{m=-l}^{l}R_{k,l}(r)Y_{l}^{m}(\theta,\phi).\label{eq:psi_k}
}

\noindent Imposing invariance under rotation for the wave function, \ref{eq:app_Seq} becomes,

\myeq{
&\frac{1}{m_{\psi}r^{2}}\frac{\partial}{\partial r}\left(r^{2} \frac{\partial R_{k,l}}{\partial r}\right)-\frac{l(l+1)}{m_{\psi}r^{2}}-V(r)R_{k,l} = -\frac{k^{2}}{m_{\psi}}R_{k,l},\label{eq:shro_Rkl}\\
&R_{k,l}\left(r\rightarrow\infty\right)\rightarrow\frac{1}{r}\sin\left(kr-\frac{l\pi}{2}+\delta_{l}(r)\right),
}

\noindent where the second equation which normalises the function $R_{k,l}$ stems from the fact that the wave function oscillates and goes to zero very far from the potential. The phase shift $\delta_{l}$ is there to regularise the wave function such that it does not diverge when $r\rightarrow 0$. Equation \ref{eq:shro_Rkl} can be written as,

\myeq{
&\left(\frac{1}{r^{2}}\frac{\partial^{2}}{\partial r^{2}}+\frac{2}{r}\frac{\partial}{\partial r}-\frac{l(l+1)}{r^{2}}-m_{\psi}V(r) +k^{2}\right)R_{k,l}=0,\label{eq:shro_Rkl_2}
}

\noindent In the simplest scenario where the mediator particle responsible for the self-scatter-ing is a boson and considering a Dirac fermion as interacting particles, the potential interaction can be represented by a Yukawa potential at leading order. It is the case for the scalar portal model and the vector portal model developed in Eqs. \ref{eq:lag_hp} and \ref{eq:lag_km} respectively. As just said, in those scenarios the potential is the ordinary Yukawa potential:

\myeq{
V_{\phi}(r) = -\frac{\alpha_{\phi}}{r}e^{-m_{\phi}r},\label{eq:yukawa_s}\\
V_{\gamma '}(r) = \pm\frac{\alpha'}{r}e^{-m_{\gamma '}r}.\label{eq:yukawa_v}
}

\noindent The minus signs indicate an attractive potential while the plus sign indicates a repulsive one. The scalar potential can be only attractive as it allows only particle/antiparticle scatterings while the vector potential also allows particle/particle and antiparticle/antiparticle scatterings such that the potential in this case can also be repulsive. One can now use the explicit form of the potential into Eq. \ref{eq:shro_Rkl_2} and define a dimensionless function and a dimensionless variable as suggested in \cite{Tulin:2013teo}:

\myeq{
&\chi_{l}\equiv r R_{k,l},\\
&x\equiv \alpha_{\rm med} m_{\psi}r,
}

\noindent where $\alpha_{\rm med}$ can be either $\alpha_{\phi}$ in the scalar portal model or $\alpha'$ in the vector portal model. Eq. \ref{eq:shro_Rkl_2} becomes,

\myeq{
\left(\frac{\diff^{2}}{\diff x^{2}}+\frac{v^{2}}{4\alpha_{\rm med}^{2}}-\frac{l(l+1)}{x^{2}}\pm\frac{1}{x}e^{-xm_{\rm med}/\alpha_{\rm med}m_{\psi}}\right)\chi_{l}(x) = 0.\label{eq:shro_Xl}
}

\noindent Here, we would like to emphasise the fact that Eq. \ref{eq:shro_Xl} depends only on three dimensionless parameters: $l$, $v/2\alpha_{\rm med}$ and $\alpha_{\rm med}m_{\psi}/m_{\rm med}$. As already mentioned above, $l$ is the angular momentum quantum number and indicates in which configuration the incoming wave is. The two other parameters can also be studied, but first let us define for clarity,

\myeq{
&a\equiv\frac{v}{2\alpha_{\rm med}},\label{eq:def_a}\\
&b\equiv\frac{\alpha_{\rm med}m_{\psi}}{m_{\rm med}}.\label{eq:def_b}
}

\noindent The first parameter, $a$, regulates the strength of the bound of the incoming particles. Indeed, the larger the relative velocity is, the more it is difficult for the incoming particles to interact with each other as they have "less time" to interact. Moreover, the smaller the coupling is, the less sensitive to the potential well the particles are. In other words, on the one hand, a large value of the $a$ parameter indicates for each of the two incoming particles a small sensitivity to the other particle presence such that the Sommerfeld enhancement will be small. On the other hand, a small value means a strong Sommerfeld effect.
\\

The second parameter, $b$, regulates the mass hierarchy between the incoming particles and the exchanged one and more precisely the sensitivity to this hierarchy. Larger the mass hierarchy is, easier it is for the incoming particles to produce the mediator a large number of time as represented in Figure \ref{fig:app_som}. Note that $b$ is not exactly the mass ratio, but contains the DM-to-med coupling. This weights the DM mass by $\alpha_{\rm med}$ such that it is in practice the product $\alpha_{\rm med}m_{\rm DM}$ which has to be compared with the mediator mass $m_{\rm med}$. This can be understood as following: the larger the DM-to-med coupling is, the less heavy the DM need to be compared to the mediator in order to be sensitive to the potential well. Indeed, if the mediator is much lighter than the DM, but the DM-to-med coupling is approximately zero, no Sommerfeld effect can enhance the cross section. Thus, this parameter indicates how strongly coupled to the potential well the incoming particles are.
\\

In terms of these dimensionless parameters, the Schrödinger equation for a two-particle system which interacts through a Yukawa potential can be written,

\myeq{
\left(\frac{\diff^{2}}{\diff x^{2}}+a^{2}-\frac{l(l+1)}{x^{2}}\pm\frac{1}{x}e^{-x/b}\right)\chi_{l}(x) = 0.\label{eq:shro_Xl_2}
}

\subsection{Relic density requirement for a freeze-out}\label{subsec:SI_relic_fo}
Before applying what we just have seen to the portal models we consider, let us use the relic density constraint in a freeze-out scenario (see Eq. \ref{eq:Oh2_FO}) to fix the DM-to-med coupling $\alpha'/\alpha_{\phi}$. To do so, one needs to compute the thermally averaged cross section as given in Eq. \ref{eq:th_av_sv}. Albeit this expression could be in general very complicated, the square of the M\o ller velocity\footnote{The M\o ller velocity is defined in terms of the velocity of the two incoming particles $\vec{v}_{1}$ and $\vec{v}_{2}$ by $v\equiv\sqrt{\vert\vec{v}_{1}-\vec{v}_{2}\vert^{2}-\vert\vec{v}_{1}\times\vec{v}_{2}\vert^{2}}$. It can be simplified to $v= 2\sqrt{1-4m_{\rm DM}^{2}/s}$ with $s$ the Mandelstam variable for pair annihilation.} $v$ of the incoming particles is small at DM decoupling $v_{\rm dec}^{2}/c^{2}\simeq 0.24^{2}\ll 1$ such that one can expand the annihilation cross section in power of $v^{2}$, see \cite{Gondolo:1990dk,Kolb:1990vq} and references therein. One has,

\myeq{
\langle\sigma v\rangle \simeq \langle\sigma v\rangle_{s} + \langle\sigma v\rangle_{p}v^{2}+\langle\sigma v\rangle_{d}v^{4}+\cdots ,\label{eq:sv_expansion}
}

\noindent where $s,p,d$ stand for s-wave, p-wave and d-wave respectively which corresponds to an angular momentum quantum number of $l=0,1,2$ respectively\footnote{The wave naming follows the nomenclature of atomic physics.}.
\\

\begin{center}
\begin{figure}
\centering
\includegraphics[scale=1]{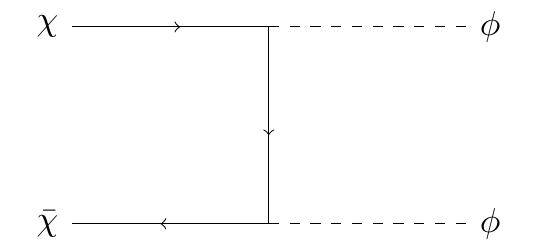}
\includegraphics[scale=1]{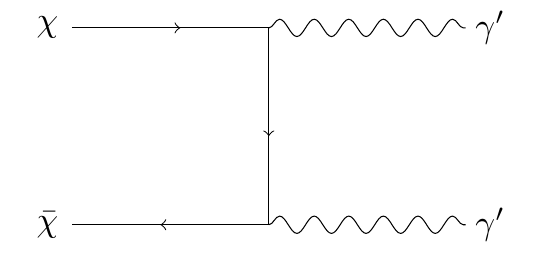}
\caption[Feynman diagram of DM annihilation]{Feynman diagram corresponding to the DM annihilation process responsible for the freeze-out mechanism for the Higgs portal model (left) and the kinetic mixing portal model (right).}
\label{fig:diag_DM_annihilation_s}
\end{figure}
\end{center}

The thermally averaged DM annihilation in the scalar and vector portal models as depicted in Figure \ref{fig:diag_DM_annihilation_s} are given by

\begin{figure}
\centering
\includegraphics[scale=0.64]{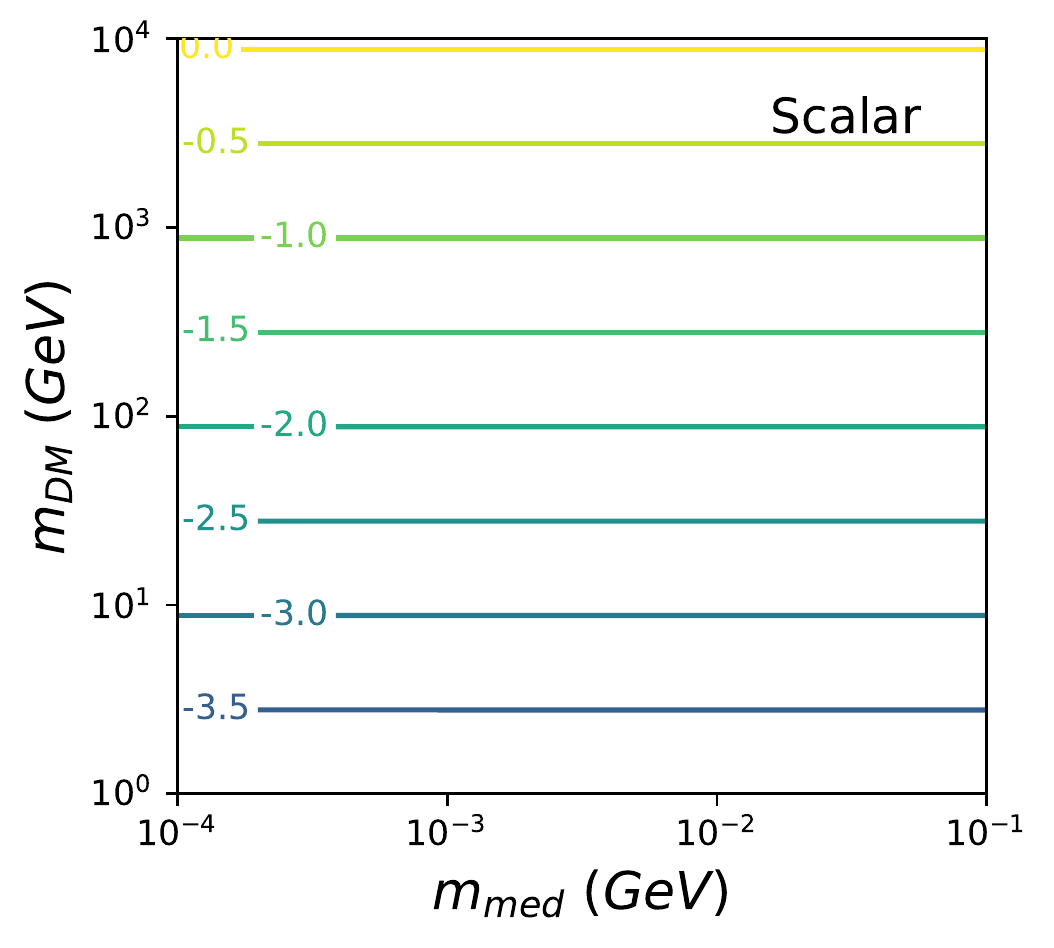}
\includegraphics[scale=0.64]{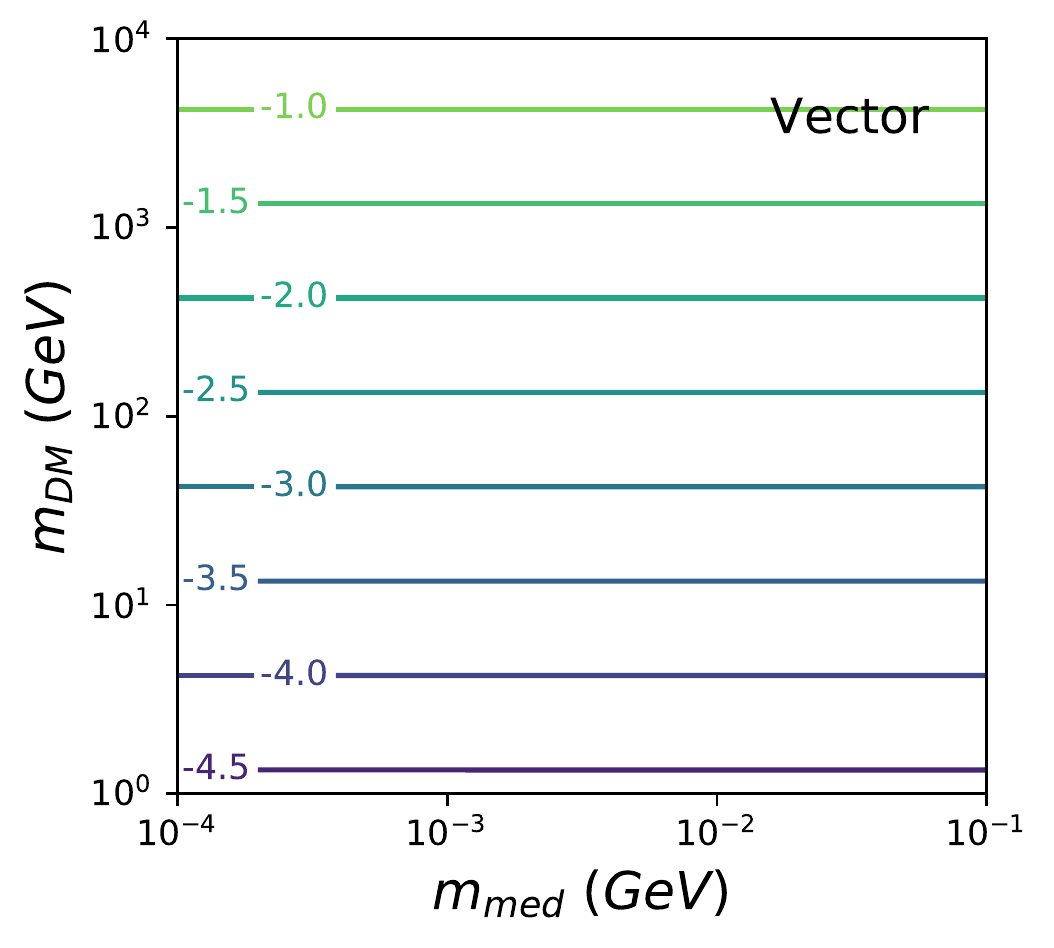}
\caption[Contour of the DM-to-med coupling in the DM versus mediator mass plane]{Contour of $\log\left(\alpha_{\rm med}\right)$ fixed by the relic density constraint in the DM versus mediator mass plane for the scalar (left) and vector (right) portal models.}\label{fig:SI_relic_density}
\end{figure}

\myeq{
&\left\langle\sigma v\right\rangle_{\bar{\chi}\chi\rightarrow \phi\phi} = \frac{3v^{2}}{4}\frac{\pi \alpha_{\phi}^{2}}{m_{\rm DM}^{2}}\sqrt{1-\frac{m_{\phi^{2}}}{m_{\rm DM}^{2}}},\label{eq:sv_HP_SI}\\
&\left\langle\sigma v\right\rangle_{\bar{\chi}\chi\rightarrow \gamma '\gamma '} = \frac{\pi \alpha '^{2}}{m_{\rm DM}^{2}}\sqrt{1-\frac{m_{\gamma'^{2}}}{m_{\rm DM}^{2}}},\label{eq:sv_KM_SI}
}

\noindent at leading order in $v^{2}$ and at tree level. Note that, as we will see in Section \ref{sec:somm_id}, Sommerfeld enhancement is totally negligible for DM annihilation at decoupling such that we neglected it while computing the DM relic density. Plugging Eqs \ref{eq:sv_HP_SI} and \ref{eq:sv_KM_SI} into the expression of the DM relic density in terms of the DM annihilation cross section (Eq. \ref{eq:Oh2_FO}), one is then able to fix the DM-to-med coupling as a function of the two left couplings: the DM and mediator masses (the velocity being fixed at DM decoupling as mentioned above). Figure \ref{fig:SI_relic_density} gives contours of the DM-to-med coupling in the DM versus mediator mass plane for both benchmark models we consider.

\subsection{Dark Matter self-interactions}
Now that we have seen how to compute the DM self-interactions and that we have fixed the DM-to-med coupling in Figure \ref{fig:SI_relic_density}, we are able to actually constrain the DM versus mediator mass plane by the small scale structure constraint of Section \ref{sec:CST-small_scale_structure} for both the scalar and the vector portal models presented in Chapter \ref{ch:constr} and studied in the rest of this thesis. 
\\

\begin{figure}
\centering
\includegraphics[scale=1]{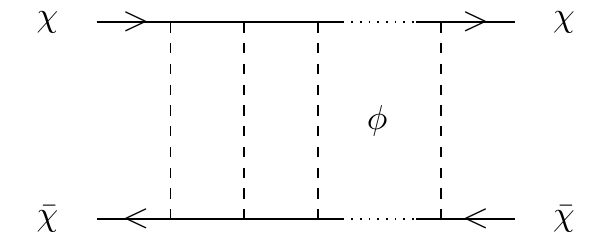}
\caption[Ladder diagram of a DM self-scattering process in the scalar portal model]{Ladder diagram of a DM self-scattering process in the scalar portal model.}\label{fig:SI_s}
\end{figure}

We saw above in Eq. \ref{eq:yukawa_s} that in the scalar portal model, there is only an attractive contribution to self-interactions from the Yukawa potential such that only one of the two Schrödinger equations of Eq. \ref{eq:shro_Xl_2} has to be solved (process depicted in Figure \ref{fig:SI_s})\footnote{The equation corresponding to the attractive case is the one with a plus sign in Eq. \ref{eq:shro_Xl_2} as the potential comes with a minus sign in the Schrödinger equation.}. 
\\

\begin{figure}
\centering
\includegraphics[scale=1]{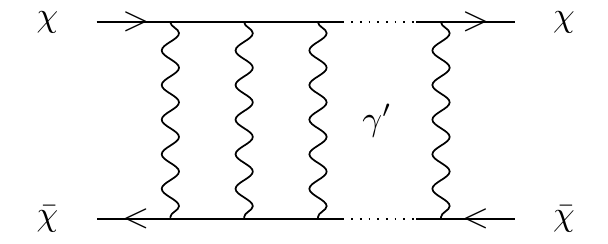}
\includegraphics[scale=1]{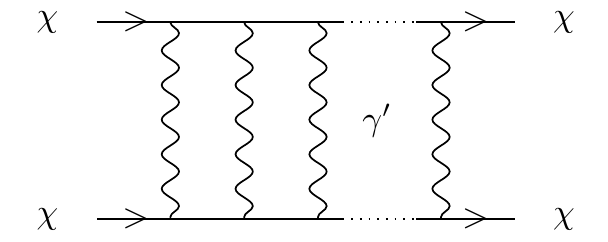}
\caption[Ladder diagrams of a DM self-scattering process in the vector portal model for the attractive and the repulsive contributions]{Ladder diagrams of a DM self-scattering process in the vector portal model for the attractive (left) and the repulsive (right) contributions.}\label{fig:SI_v}
\end{figure}

However, in the vector portal model, DM self-interactions get both attractive and repulsive contributions from the Yukawa potential such that both Schrödinger equations of Eq. \ref{eq:shro_Xl_2} have to be solved (processes depicted in Figure \ref{fig:SI_v}).
\\

Following the numerical resolution method of \cite{Tulin:2013teo}\footnote{One can also use the analytical formulas given in \cite{Tulin:2013teo} as very good proxy of the solutions of the Schrödinger equation.} one can thus solve Eq. \ref{eq:shro_Xl_2} in the DM versus mediator mass plane as those are the only free parameters. Indeed, the DM-to-med coupling is fixed by the relic density and the DM relative velocity which is pointed out by simulations to alleviate tensions at small scales is approximately fixed to $v\simeq 3\times 10^{-5}$ \cite{Feng:2009mn,Ackerman:mha,Feng:2009hw,Ibe:2009mk,Loeb:2010gj,Tulin:2012wi}. 
\\

\begin{figure}
\centering
\includegraphics[scale=0.65]{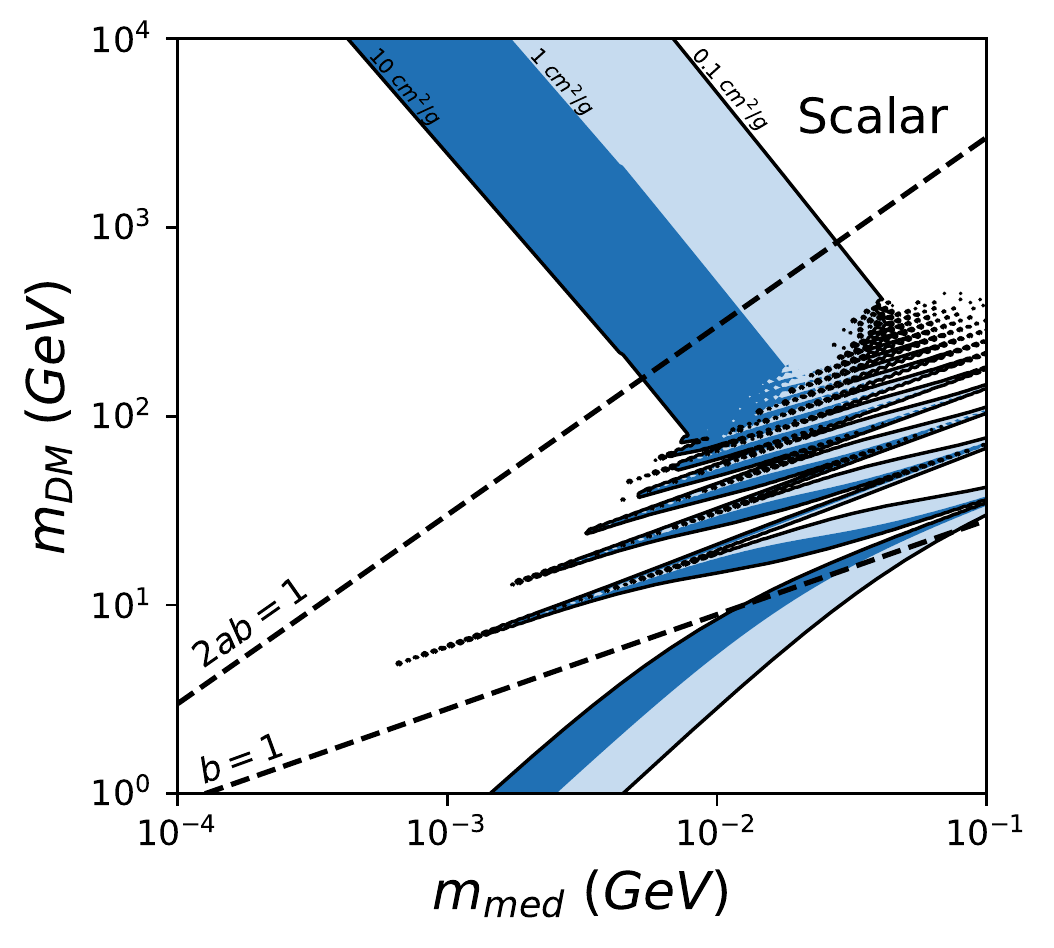}
\includegraphics[scale=0.65]{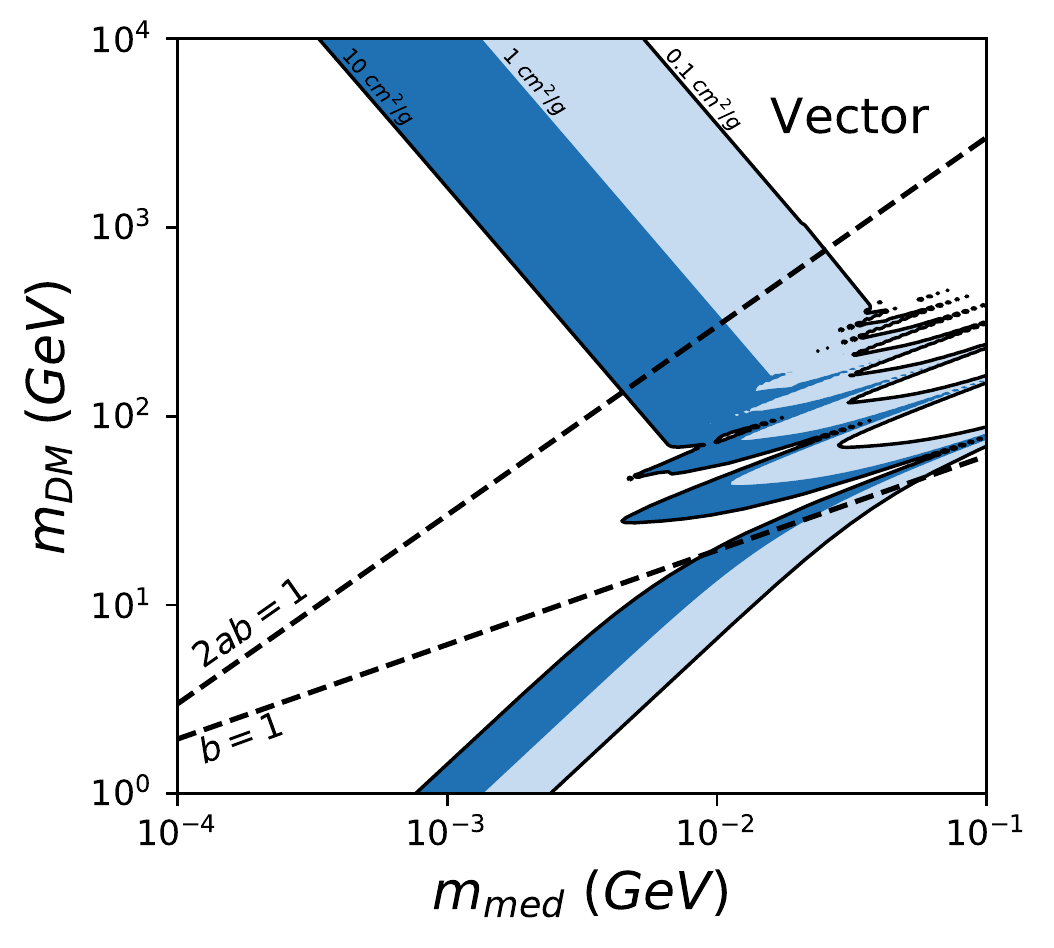}
\caption[Contour of the self-scattering cross section divided by the DM mass for the scalar and vector portal models]{Contour of the self-scattering cross section divided by the DM mass for the scalar (left) and vector (right) portal models. The DM-to-med coupling as been fixed requiring that the self-interacting particle constitute 100\% of the DM today.}\label{fig:SI_Oh2}
\end{figure}

Figure \ref{fig:SI_Oh2} shows then in this plane where the self-scattering transfer cross section satisfy the small scale structure constraints, i.e. where $0.1$ cm$^{2}$/g $\leq \sigma_{T}/m_{\rm DM}\leq 10$ cm$^{2}$/g (see Section \ref{sec:CST-small_scale_structure} on small scale structure constraints). These results have been obtained using the tools developed in \cite{Tulin:2013teo} for DM candidate, that is to say fixing the DM-to-med coupling by the relic density constraint as explained in the previous subsection. Left panel of Figure \ref{fig:SI_Oh2} presents the self-interacting cross section corresponding to the diagram shown in Figure \ref{fig:SI_s} taking the Sommerfeld enhancement into account for the scalar portal model. While the right panel of the very same Figure shows results for the vector portal model that is to say from both attractive and repulsive contributions in this case, see Figure \ref{fig:SI_v}.
\\

From left and right panels of Figure \ref{fig:SI_Oh2}, one can distinguish three regimes delimited by the values of $a$ and $b$ defined above in Eqs. \ref{eq:def_a} and \ref{eq:def_b}. This is why the two lines defined by $2ab=1$ and $b=1$ are shown (dashed lines on Figure \ref{fig:SI_Oh2}). These lines arises for similar values of the DM and mediator particles masses in both plots of Figure \ref{fig:SI_Oh2}. Because the annihilation cross sections responsible for the DM freeze-out are similar (see Eqs. \ref{eq:sv_HP_SI} and \ref{eq:sv_KM_SI}). Indeed, the major difference comes in the factor of $3v^{2}/4$ in the scalar case which does not appear in the vector case (p-wave annihilation versus s-wave). This only brings a global moderate numerical factor once fixing the DM-to-med coupling from the DM relic abundance constraint. As a result, the required coupling is bigger in the scalar case than in the vector case for a chosen set of masses as this extra factor is smaller than unity at DM decoupling: $3v^{2}_{\rm dec}/4\simeq 0.043$. A larger DM-to-med coupling is then needed to compensate this additional suppression factor in the cross section. This explain why the $b=1$ line is a little bit closer to the top for the vector case compared to the scalar one. Let us now discuss these regimes one by one. Starting from the bottom of left and right panels of Figure \ref{fig:SI_Oh2}, both the DM-to-med coupling and the DM mass are small such that $b<1$\footnote{We recap that the DM-to-med coupling is fixed by the relic density constraint such that the DM-to-med coupling increases with the DM mass.}. In such a case, the exponential term of Eq. \ref{eq:shro_Xl_2} coming from the Yukawa potential is suppressed with respect to the other terms so its impact is small. Thus, for both attractive and repulsive cases, one enters into the so-called \textit{Born} regime in which the Born approximation holds. In this regime, one can compute the self-scattering without considering the full ladder diagram, that is to say without taking any Sommerfeld enhancement into account. Indeed, this regime corresponds to a feeble Yukawa potential and a weak mass hierarchy as $\alpha_{\rm med}m_{\rm DM}<m_{\rm med}$. Solving Eq. \ref{eq:shro_Xl_2} and plugging its solution for $\chi_{l}$ in Eq. \ref{eq:diff_s_diff_omega}, one gets a cross section which is the same for both the attractive and the repulsive potentials \cite{Feng:2009hw},

\myeq{
\sigma_{T}^{\text{Born}}=\frac{2\pi}{a^{2}m_{\rm med}^{2}}\frac{1}{(2ab)^{2}}\left(\log\left(1+(2ab)^{2}\right)-\frac{(2ab)^{2}}{1+(2ab)^{2}}\right).
}

\noindent In this regime, one does not expect any specific features as there is no enhancement of the cross section. The self-scattering cross section divided by the DM mass increases only with $b=\alpha_{\rm med}m_{\rm DM}/m_{\rm med}$. On the one hand, for a fixed mediator mass, one expects the self-scattering cross section divided by the DM mass to increase with the DM mass. On the other hand, for a fixed DM mass one expects this quantity to increase if the mediator mass decrease. These behaviours can be seen in both panels of Figure \ref{fig:SI_Oh2} in the region $b<1$, below the $b=1$ dashed line.
\\

The intermediate regime for which $b>1$ and $2ab\ll 1$ shows up when all terms in Eq. \ref{eq:shro_Xl_2} have comparable impact and no effect can be neglected. However, solving Eq. \ref{eq:shro_Xl_2} in this case is not easy. One can nevertheless solve a proxy of this equation using the Hulthén potential instead of the Yukawa potential:

\myeq{
V^{\text{Hulthèn}}(x) = \pm a\eta\frac{m_{\rm med}^{2}}{m_{\rm DM}}\frac{e^{-\eta x/b}}{1-e^{-\eta x/b}},
}

\noindent where $\eta$ is a numerical factor used to fit the Yukawa solution, $\eta\simeq 1.6$ \cite{Cassel_2010,Tulin:2013teo} and the minus or plus signs stand for the attractive and repulsive potential respectively. In that case, Eq. \ref{eq:shro_Xl_2} can be solved analytically for the $l=0$ contribution which dominates in this regime. We have \cite{Tulin:2013teo},

\myeq{
\sigma_{T}^{\text{Hulthèn}} = \frac{16\pi}{m_{\rm med}^{2}}\frac{1}{(2ab)^{2}}\sin^{2}\left(\text{arg}\left(\frac{i\Gamma(i2ab/\eta)}{\Gamma(\lambda_{+})\Gamma(\lambda_{-})}\right)\right),
}

\noindent where $\Gamma$ is the Euler function and the parameters $\lambda_{\pm}$ are given by,

\myeq{
\lambda_{\pm} = 
\begin{cases}
1+\frac{iab}{\eta}\pm\sqrt{\frac{a}{\eta}-\left(\frac{ab}{\eta}\right)^{2}}\hspace{1cm}\text{(attractive potential)},\\
1+\frac{iab}{\eta}\pm i\sqrt{\frac{a}{\eta}+\left(\frac{ab}{\eta}\right)^{2}}\hspace{1cm}\text{(repulsive potential)}.
\end{cases}
}

\noindent It can be shown that these analytical results give a good approximation of the exact numerical results, see \cite{An:2009vq,Tulin:2013teo}. In this regime, the self-interacting cross section depends periodically and strongly on the values of the parameters $a$ and $b$. This is due to the fact that this regime for the attractive case displays quantum resonances especially in the attractive scalar case, but also in the attractive vector case in a less significant way. The addition of the repulsive component for the vector case explains why the resonances are smoothed in the right panel of Figure \ref{fig:SI_Oh2}. Moreover, as we already mentioned above, the DM-to-med coupling is sizeably larger in the scalar case than in the vector one due to the extra numerical factor in the DM annihilation cross section at freeze-out. Thus, if for a given set of DM and mediator masses, the coupling is larger in the scalar case, one expects a stronger self-interaction cross section as well as a wider resonant regime (i.e. more resonances) in the scalar model. This specific difference in the two models can be seen comparing left and right panels of Figure \ref{fig:SI_Oh2} and more specifically the number of resonances which is bigger in the scalar case as expected.
\\

Finally, the last regime is defined for $2ab>1$, i.e. for $m_{\rm DM}v>m_{\rm med}$ and describes the so-called classical regime. The behaviour of the self-scattering cross section divided by the DM mass is here much simpler than in the resonant regime. Indeed, in the classical regime, the mediator is much much lighter than the DM such that the Yukawa potential can be associated to a Coulombian-like potential as a proxy. Thus, in the limit $2ab\gg 1$, the self-scattering rate reduces to the case we obtained in the classical regime of a particle propagating in a classical Coulombian potential. This results in an analytic solution for the transfer cross section (see \cite{Tulin:2013teo,Feng:2009hw,PhysRevLett.90.225002,PhysRevE.70.056405}) which for the attractive potential case is

\myeq{
\sigma_{T}^{\text{Classical}} =
\begin{cases}
\frac{4\pi}{a^{2}m_{\rm med}^{2}}\frac{1}{(2ab)^{2}}\log\left(1+2a^{2}b\right)\hspace{2.35cm}10\leq 2a^{2}b,\\
\frac{8\pi}{a^{2}m_{\rm med}^{2}}\frac{1}{(2ab)^{2}}\left(1+1.5(2a^{2}b)^{-1.65}\right)^{-1}\hspace{1cm}10^{-3}\leq 2a^{2}b\leq 10,\\
\frac{\pi}{m_{\rm med}^{2}}\left(1-\log(2a^{2}b)+\frac{1}{2\log(2a^{2}b)}\right)^{2}\hspace{1.1cm}2a^{2}b\leq 10^{-3},
\end{cases}
}

\noindent while for the repulsive potential case it is instead

\myeq{
\sigma_{T}^{\text{Classical}} =
\begin{cases}
\frac{2\pi}{a^{2}m_{\rm med}^{2}}\frac{1}{(2ab)^{2}}\log\left(1+a^{2}(2ab)^{2}\right)\hspace{2.15cm}1\leq 2a^{2}b,\\
\frac{\pi}{m_{\rm med}^{2}}\left(\log(1/a^{2}b)-\log\log(1/a^{2}b)\right)^{2}\hspace{1.1cm}2a^{2}b\leq 1.
\end{cases}
}

\noindent For a fixed DM mass we recover a behaviour similar to the one obtained in the Born regime: the self-scattering cross section divided by the DM mass increases if the mediator mass decreases but slower than in the Born regime (i.e. the interaction is less sensitive to the mediator mass). Moreover, for a fixed mediator mass, the self-scattering cross section divided by the DM mass decreases if the DM mass increases. This is because at high DM mass there is only one relevant scale, the DM mass. Thus the self-scattering cross section has no other choice to go like $\sigma_{T}\sim 1/m_{\rm DM}^{2}$ such that it decreases with the DM mass. This can be seen also from the fact that the self-interacting cross section divided by the DM mass does not depend on the mass ratio $m_{\rm DM}/m_{\rm med}$ in this regime. In other words, the mediator is so light that in practice it is massless and the solution of the Schrödinger equation cannot depend on this ratio like in the case of a coulombian potential. Once again, this behaviour can be seen in both panels of Figure \ref{fig:SI_Oh2} for the scalar and the vector portal models.

\subsection{Smaller annihilation cross section}\label{subsec:SI_over_relic_fo}
\begin{figure}
\centering
\includegraphics[scale=0.6]{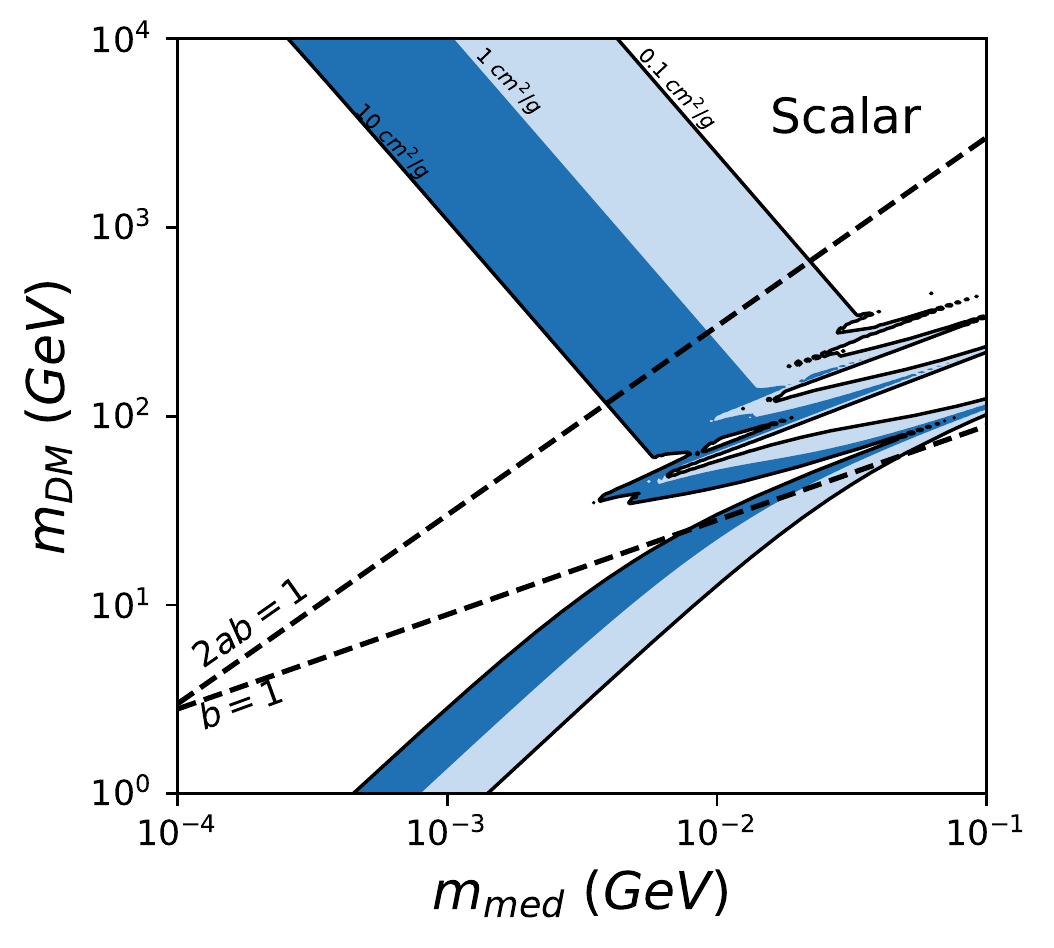}
\includegraphics[scale=0.6]{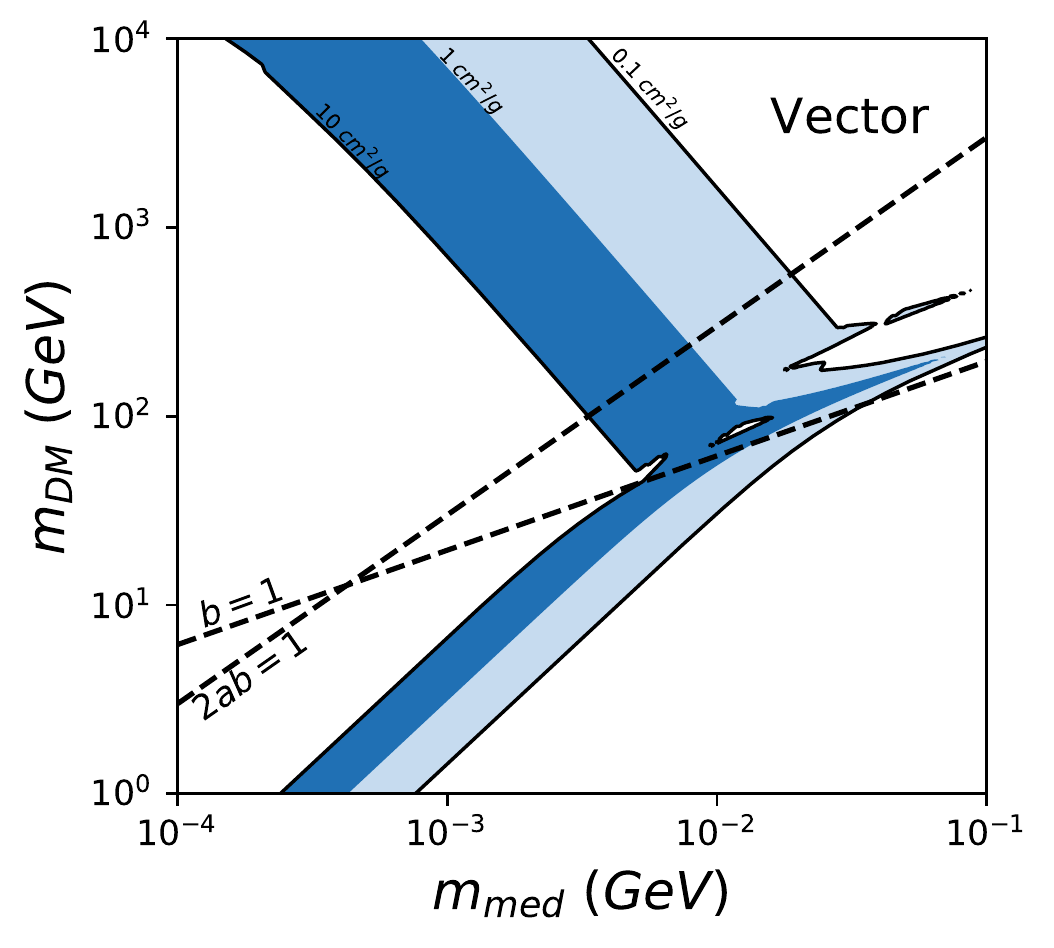}
\includegraphics[scale=0.6]{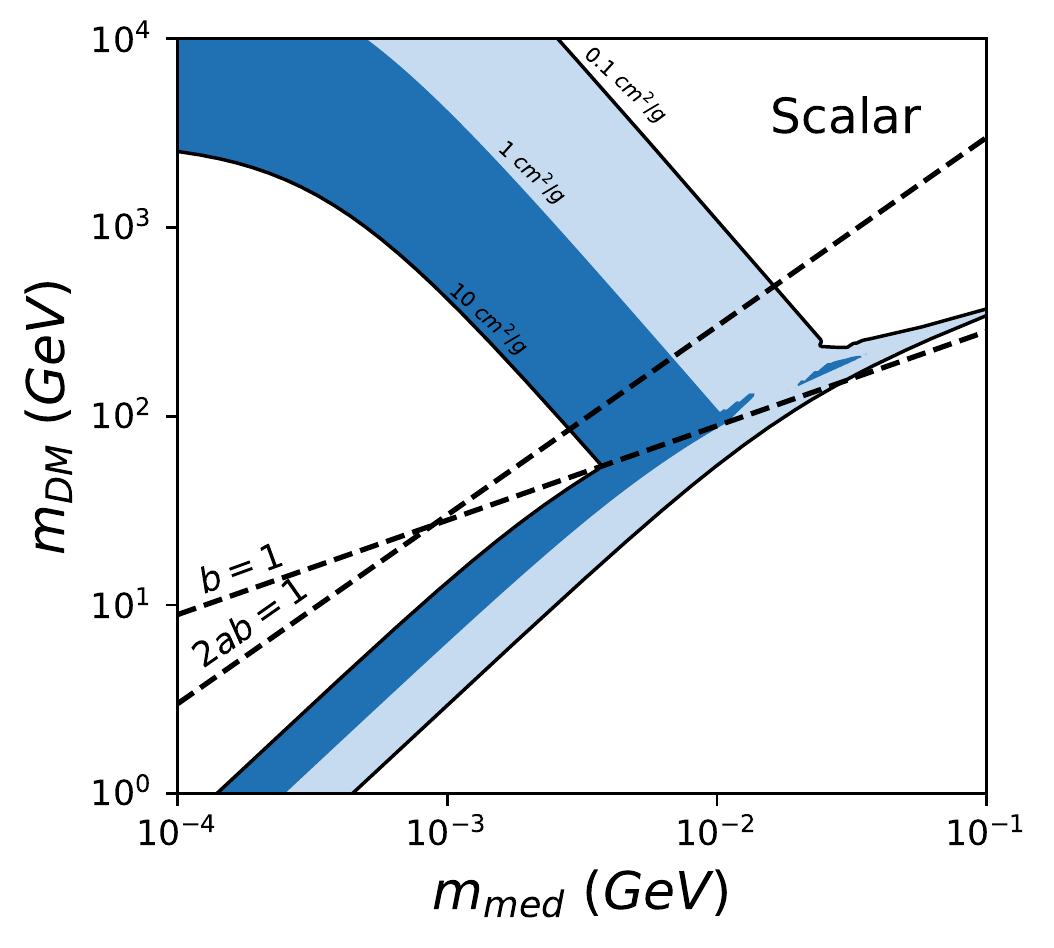}
\includegraphics[scale=0.6]{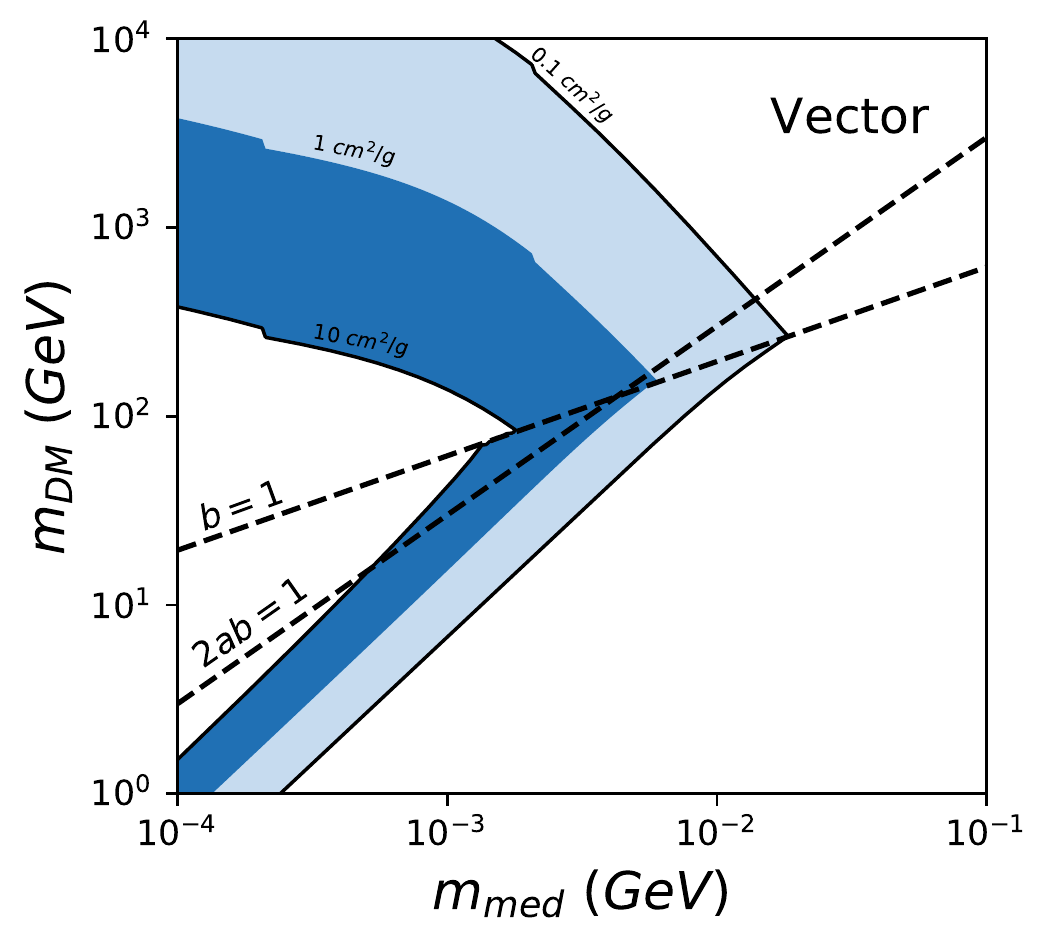}
\includegraphics[scale=0.6]{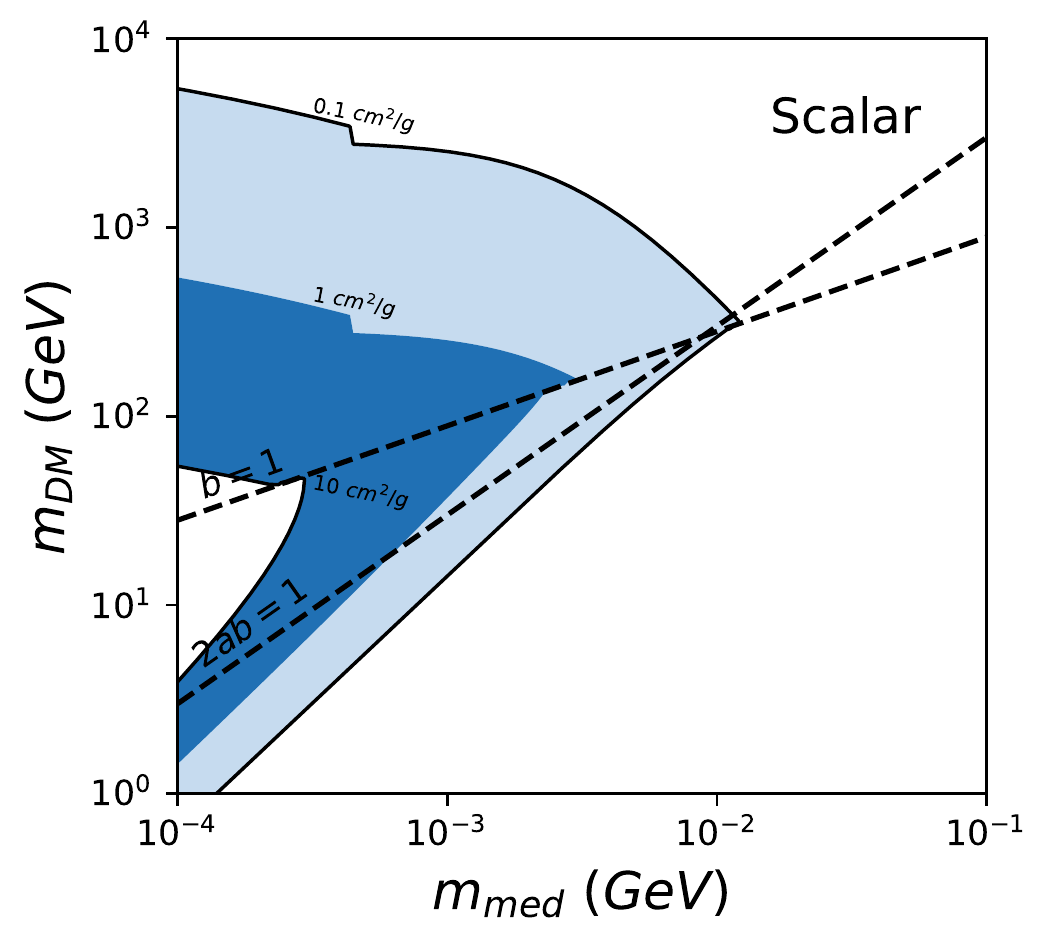}
\includegraphics[scale=0.6]{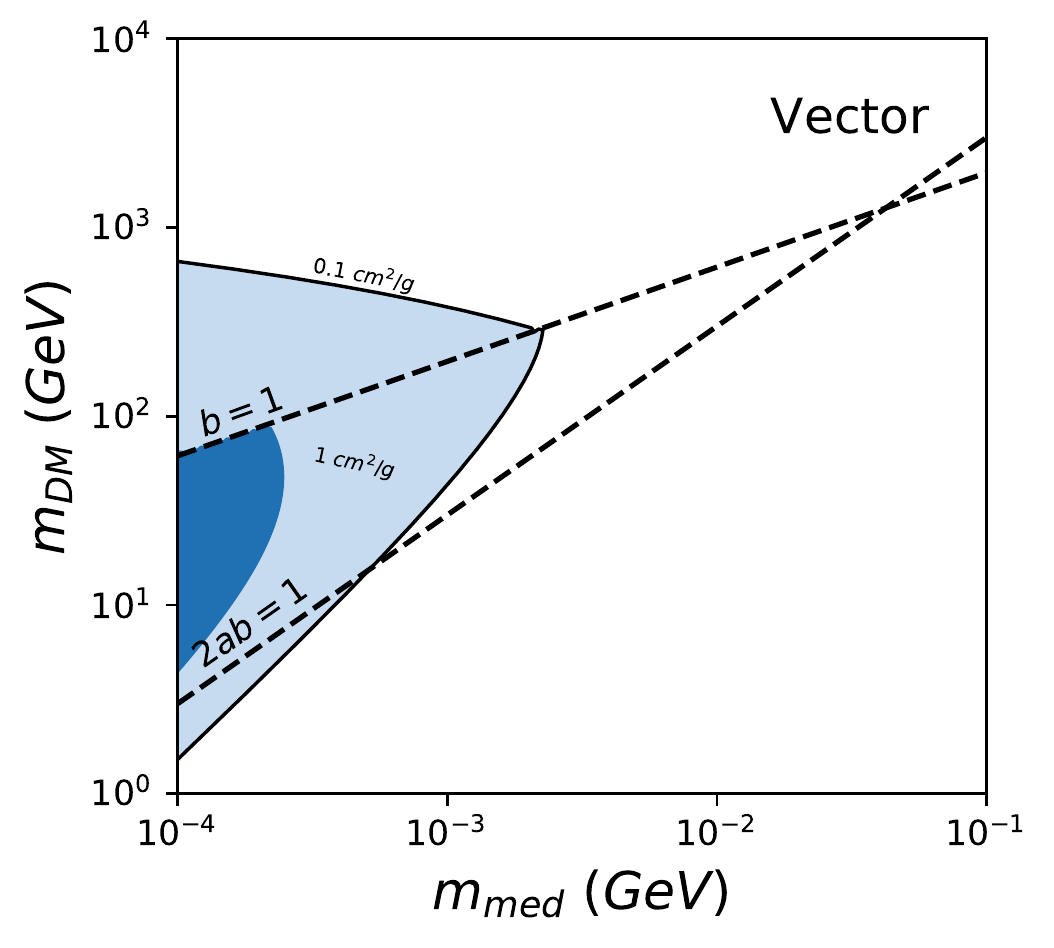}
\caption[Contour of the self-scattering cross section divided by the DM mass for the scalar and vector portal models for a subdominant DM component]{Contour of the self-scattering cross section divided by the DM mass for the scalar (left) and vector (right) portal models. The DM-to-med coupling as been fixed requiring that the self-interacting particle abundance today is $10^{2}\times\Omega_{\rm DM}$ (top), $10^{4}\times\Omega_{\rm DM}$ (middle) and $10^{6}\times\Omega_{\rm DM}$ (bottom).}\label{fig:SI_Oh2_sub}
\end{figure}

Until now, one has assumed that the interaction at the origin of the self-interactions is also the one leading to the observed relic density through the FO mechanism. However, this does not have to be necessarily the case. One could instead consider smaller DM-to-med couplings leading to a smaller DM annihilation cross section at FO. This interaction alone would then lead to a larger DM relic density than the one observed today and needs an additional annihilation process to take care of the FO mechanism. In this case, one expects the self-interactions to decrease too since the DM-to-med coupling decreases. Then, the dimensionless parameter $b$ (resp. $a$) will decrease (resp. increase) while the combination $2ab$ will remain unchanged. Thus looking at the DM versus mediator mass plane, even if the $2ab=1$ line does not move, the $b=1$ line will go up and this will shrink the resonant regime. However, we found that, in such a case, small scale structure constraints can still be satisfied for a wide range of the parameter space even for an annihilation cross section suppressed by a factor of one million. Figure \ref{fig:SI_Oh2_sub} shows what the DM versus mediator mass plane looks like if one considers a smaller annihilation cross section at freeze-out than the one required by the DM relic abundance constraint in a standard scenario. By standard we mean a scenario where the HS is in thermal equilibrium with the VS and in which the DM undergoes an usual freeze-out. In Figure \ref{fig:SI_Oh2_sub} for both the scalar (left) and vector (right) portal models, we fixed the DM-to-med coupling requiring that the annihilation process alone would give a DM relic density $\Omega_{\rm DM}$ larger by a factor of $10^{2}$, $10^{4}$ and $10^{6}$ for top, middle and bottom respectively.
\\

This shows how easy it is to satisfy the small scale structure constraints (see Section \ref{sec:CST-small_scale_structure}) even if the DM annihilation cross section is much smaller than the one usually expected, that is to say if $\left\langle\sigma v\right\rangle_{\rm dec}\ll 2.2\times 10^{-9}$ GeV$^{-2}$. We will make a great use of this results in the following when we will study minimal ways out for self-interacting DM models with light mediators, see Chapters \ref{ch:TpT} and \ref{ch:other_minimal}.

\section{Effect of the Sommerfeld effect on DM annihilations}\label{sec:somm_id}
In the previous section we were focused on how to take into account the Sommerfeld effect for a self-scattering process in which the particle responsible for the self-scattering is the same as the one responsible for the Sommerfeld effect. However, the Sommerfeld effect can also be very relevant for all scattering processes, particularly for DM annihilations. Indeed, before the proper interaction (here the annihilation), the DM incoming particles may interact in the exact same way they do in self-interactions. The computation of the impact of the Sommerfeld enhancement in the case of an annihilation is a little bit different than for purely self-interacting processes. One has to distinguish two processes: the annihilation itself and the Sommerfeld effect which takes into account the self-interactions which may occur before the annihilation. Ideally, one should then solve the full Schrödinger equation which would take the Yukawa potential (responsible for the self-interaction) and the final annihilation process into account all together, but this is rather complicated as the former is a non-perturbative effect and the latter can only be computed in a perturbative theory. However one may assume that these two processes are independent such that one can factorise the Sommerfeld effect from the thermally average annihilation cross section: $\langle\sigma v\rangle\simeq S_{F}\times\langle\sigma v\rangle_{TL}$ where $TL$ stands for the cross section at tree level, i.e. without taking the Sommerfeld effect into account and $S_{F}$ is the Sommerfeld factor which encodes the contribution of the Sommerfeld effect in the process, see below.
\\

As we are now interested in DM annihilation processes, one must pay attention to the typical values of the DM relative velocity in such processes. As we have seen in Chapter \ref{ch:constr}, DM annihilations are relevant at the DM decoupling in FO scenario, but also later on at CMB recombination (see Section \ref{sec:CST-CMB}) or today for indirect detection when annihilating in the centre of the Milky Way or in dwarf galaxies (see Section \ref{sec:CST-ID}). In all of these contexts, DM relative velocity is much smaller than at DM decoupling:

\myeq{
v\sim
\begin{cases}
2\times 10^{-3}\hspace{1cm}\text{in the Milky Way,}\\
3\times 10^{-5}\hspace{1cm}\text{in dwarf galaxies,}\\
10^{-7}\hspace{1.68cm}\text{at CMB recombination.}
\end{cases}\label{eq:vel}
}

\noindent Then, as the DM relative velocity is small, one can always consider the leading order in the expansion of the annihilation cross section given in Eq. \ref{eq:sv_expansion}. In this case, as the final process is largely dominated by the leading wave, 
it is not necessary to solve the Schrödinger equation for all waves as we did for the self-interaction cross section, but simply solve the one corresponding to the dominant wave in the process.

\subsection{The Sommerfeld factor}
The result of the above assumption that one can factorise the Sommerfeld effect from the tree-level annihilation process is that one can define the Sommerfeld factor as the ratio between the full cross-section including the Sommerfeld effect ($\sigma$) and the tree-level cross-section without the Sommerfeld enhancement ($\sigma_{TL}$),

\myeq{
S_{k}\equiv\frac{\sigma}{\sigma_{TL}},\label{eq:app_def_SF}
}

\noindent with $\vec{k}$ the relative impulsion of the incoming particles.
\\

The ratio given in Eq. \ref{eq:app_def_SF} can be related to the wave function found with and without the Sommerfeld effect. Indeed, as we already discussed above, the Sommerfeld effect does nothing but to modify the Schrödinger equation of the incoming wave function by adding a potential term which represents the exchange of particles in the initial state. Moreover, since the cross section represents the probability for a process to occur, it is proportional to the square of the module of th wave function evaluated at the interaction point: $\sigma\propto\vert\psi(0)\vert ^{2}$. Thus, the Sommerfeld factor is also given by,

\myeq{
S_{k}=\frac{\vert\psi_{k}(0)\vert ^{2}}{\vert\psi_{k,TL}(0)\vert ^{2}},
}

\noindent where $\psi_{k,TL}$ stands for the wave function without any perturbation (i.e. at tree level) and $\psi_{k}$ takes the Sommerfeld effect into account. Performing the same expansion of the wave function in terms of the partial wave as done in Eq. \ref{eq:psi_k}, one can compute the Sommerfeld factor corresponding to each wave in terms of the solution of Eq. \ref{eq:shro_Xl_2} taken at $x\rightarrow 0$, see \cite{Slatyer_2010} for an in depth discussion,

\myeq{
S_{l} = \left\vert \frac{(2l+1)!!}{2^{l+1}(l+1)!}\frac{\lim_{x\rightarrow 0}\chi^{(l+1)}(x)}{a^{l+1}} \right\vert, \label{eq:S_l}
}

\noindent where $\chi^{n}$ stands for the nth derivative of $\chi$ with respect to the dimensionless variable $x$. At the end of the day, the final annihilation cross section which takes the Sommerfeld enhancement into account can be written as,

\myeq{
\langle\sigma v\rangle \simeq \sum_{l=0}^{\infty}S_{l}\langle\sigma v\rangle_{l}v^{2l}.
}

\subsection{s-wave annihilation}
The Sommerfeld factor expressed in Eq. \ref{eq:S_l} can be close to unity or be very large depending on the wave and on the values of the dimensionless parameters $a$ and $b$\footnote{Let us recall that $S_{l}$ does depend on $a$ and $b$ through $\lim_{x\rightarrow 0}\chi^{(l+1)}(x)$ which is the solution of Eq. \ref{eq:shro_Rkl_2}.}. Thus, let us first determine in more details what are the values that can take this factor for the simplest case, the s-wave.
\\

\begin{figure}
\centering
\includegraphics[scale=0.62]{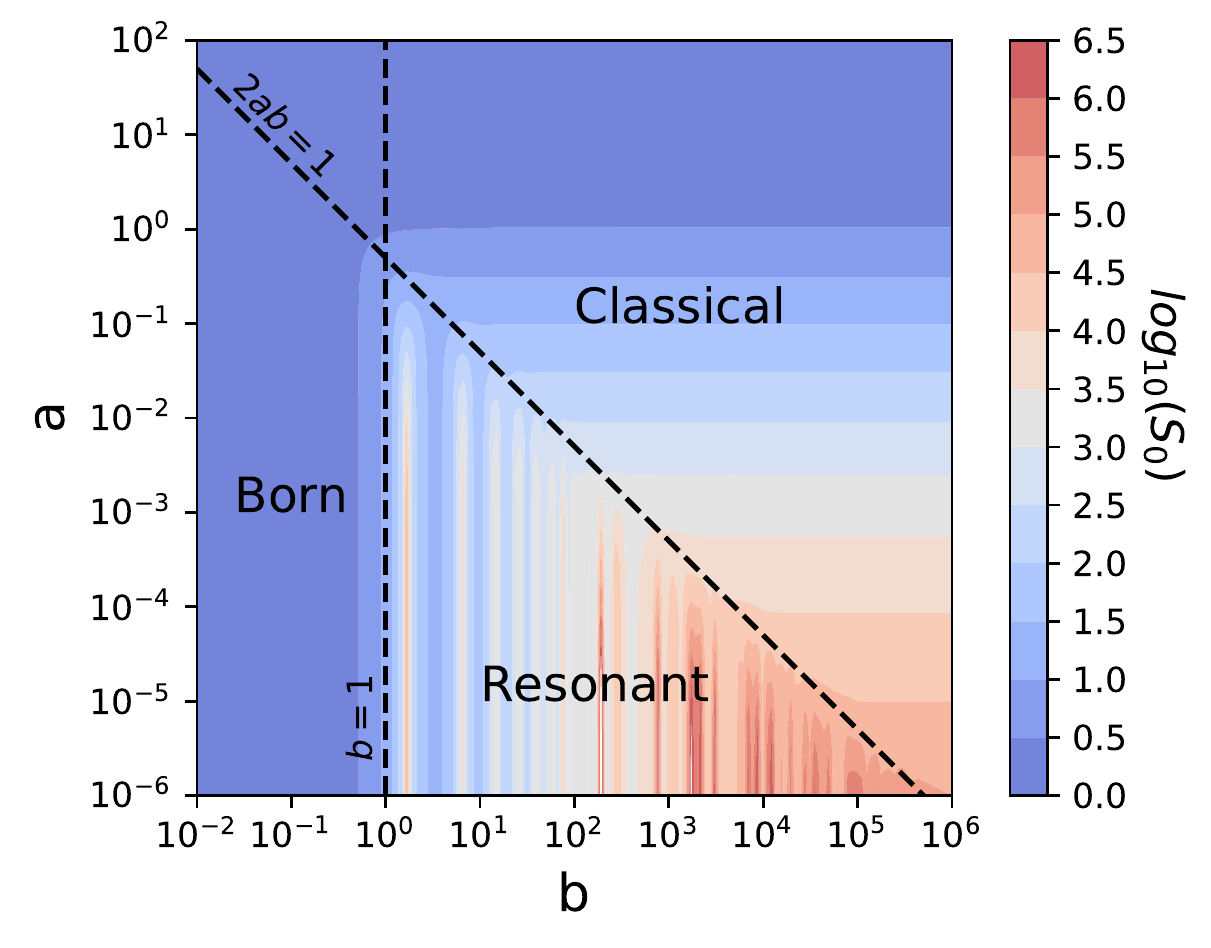}
\includegraphics[scale=0.62]{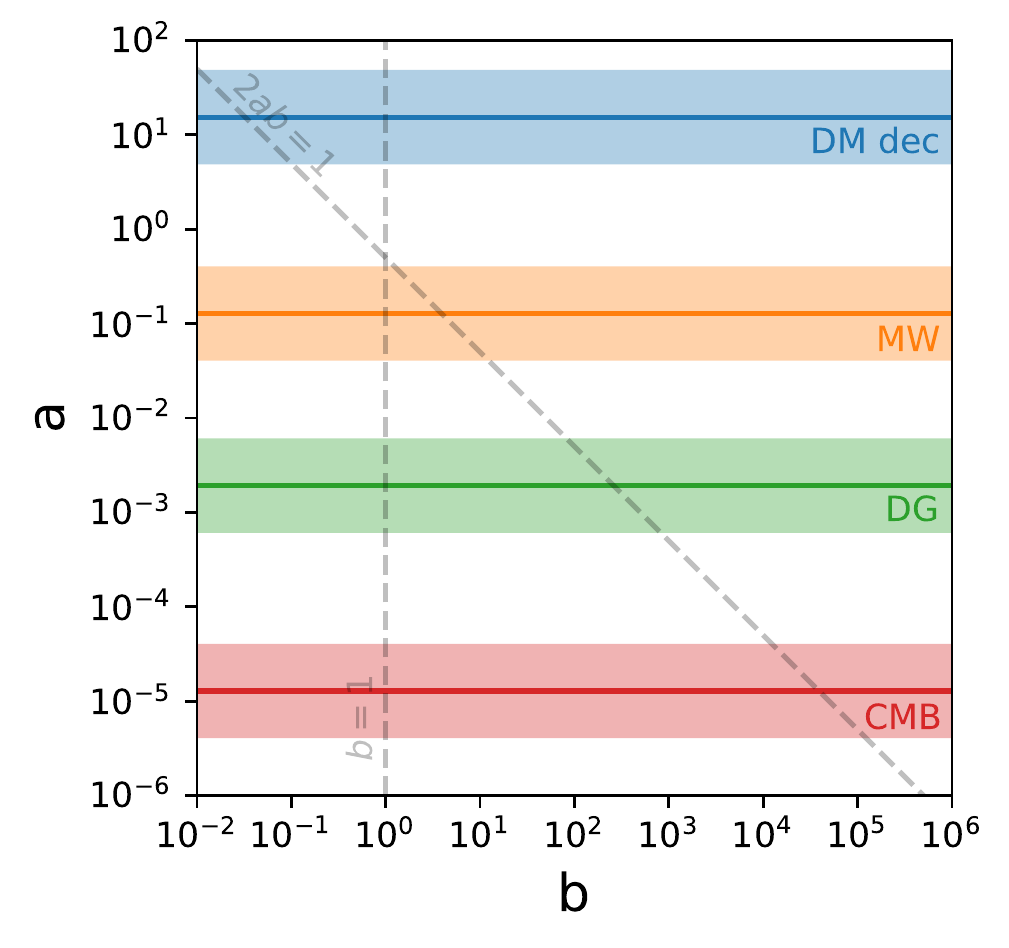}
\caption[Sommerfeld factor for s-wave annihilation]{Left: contour of the Sommerfeld factor for s-wave annihilation in the Born, resonant and classical regimes in the $a-b$ plane. Right: relevant regions of the $a-b$ plane when considering DM freeze-out (blue), DM indirect detection in the Milky Way (orange) and in dwarf galaxies (green) and CMB constraints on DM annihilation (red) with $\alpha_{\rm med}\simeq \alpha_{EW}$.}\label{fig:ab_s}
\end{figure}

The left panel of Figure \ref{fig:ab_s} shows contours of the Sommerfeld factor for the s-wave annihilation in the $a-b$ plane. One can also distinguish in Figure \ref{fig:ab_s} the two lines already met above defined by $b=1$ and $2ab=1$ which delimit the Born, the resonant and the classical regimes of the Sommerfeld effect. The values of the Sommerfeld factor shown in this Figure are obtained by solving numerically Eq. \ref{eq:shro_Xl_2} for the s-wave contribution ($l=0$) and plugging the solution into Eq. \ref{eq:S_l}, that is to say that we did not use any analytic approximation. From this panel, we see that we recover behaviours similar to the ones observed for the self-interacting cross section. Indeed, in the Born regime the Sommerfeld factor for the s-wave contribution is constant $S_{0}^{\rm Born}\simeq 1$. In the resonant regime the factor highly depends on the values of $a$ and $b$ and lies between $S_{0}^{\rm Resonant}\gtrsim 10$ and $S_{0}^{\rm Resonant}\lesssim 10^{7}$. This factor could of course be even larger if one does consider larger values of $b$ and smaller values of $a$, but this would require a very large coupling which may become non-perturbative at some point. Finally, in the classical regime the Sommerfeld factor does not depend on $b=\alpha_{\rm med}m_{\rm DM}/m_{\rm med}$ since the mediator is basically massless when compared to the DM and the mass ratio does not play any role. We recover the solution one can obtain with a Coulombian potential.
\\

We have briefly mentioned that the DM relative velocity is not the same in the various contexts where the Sommerfeld effect can be relevant for DM annihilations, see Eq. \ref{eq:vel}. Thus, in general, different regions of the $a-b$ plane are relevant. On the right panel of Figure \ref{fig:ab_s}, we show which region of the parameter space (i.e. the $a-b$ plane) is relevant for which type of situation. We considered DM annihilation at freeze-out (blue), in the Milky way or in dwarf galaxies today (orange and green respectively) and at CMB recombination (red). In order to delimit an area for each of those situations such as we did in this plot, we fixed the DM-to-med coupling to the electroweak fine structure constant $\alpha_{\rm med}=\alpha_{EW}\simeq 1/128$. Indeed, as the velocity is fixed for each situation, fixing the DM-to-med coupling allow us to fix the dimensionless parameter $a$. The wide bands correspond to a variation of one order of magnitude in the DM-to-med coupling. Hence, in each coloured region of right panel of Figure \ref{fig:ab_s}, the DM-to-med coupling lies in the following range: $\alpha_{\rm med}\in\left[\alpha_{EW}/\sqrt{10},\alpha_{EW}\sqrt{10}\right]$.
\\

Thanks to the right panel of Figure \ref{fig:ab_s}, we see that in the s-wave case, for a DM-to-med coupling of order of the electroweak fine structure constant, the Sommerfeld effect is negligible when considering the DM freeze-out (as already mentioned in Section \ref{subsec:SI_relic_fo}) and starts to be relevant when considering indirect detection experiments looking for DM annihilation in the Milky Way. Moreover, it becomes really important and could change the whole picture for DM annihilation in dwarf galaxies and when considering CMB constraints on the DM annihilation cross section\footnote{Note that it is due to this effect that s-wave annihilation processes during the recombination epoch can be constrained by CMB, see Section \ref{sec:CST-CMB} and Eq. \ref{cmbconstraint}}. Considering a smaller or a larger DM-to-med coupling would move up or move down respectively the four coloured regions of right panel of Figure \ref{fig:ab_s} simultaneously.

\subsection{p-wave annihilation}\label{subsec:p_wave_ID}
\begin{figure}
\centering
\includegraphics[scale=0.62]{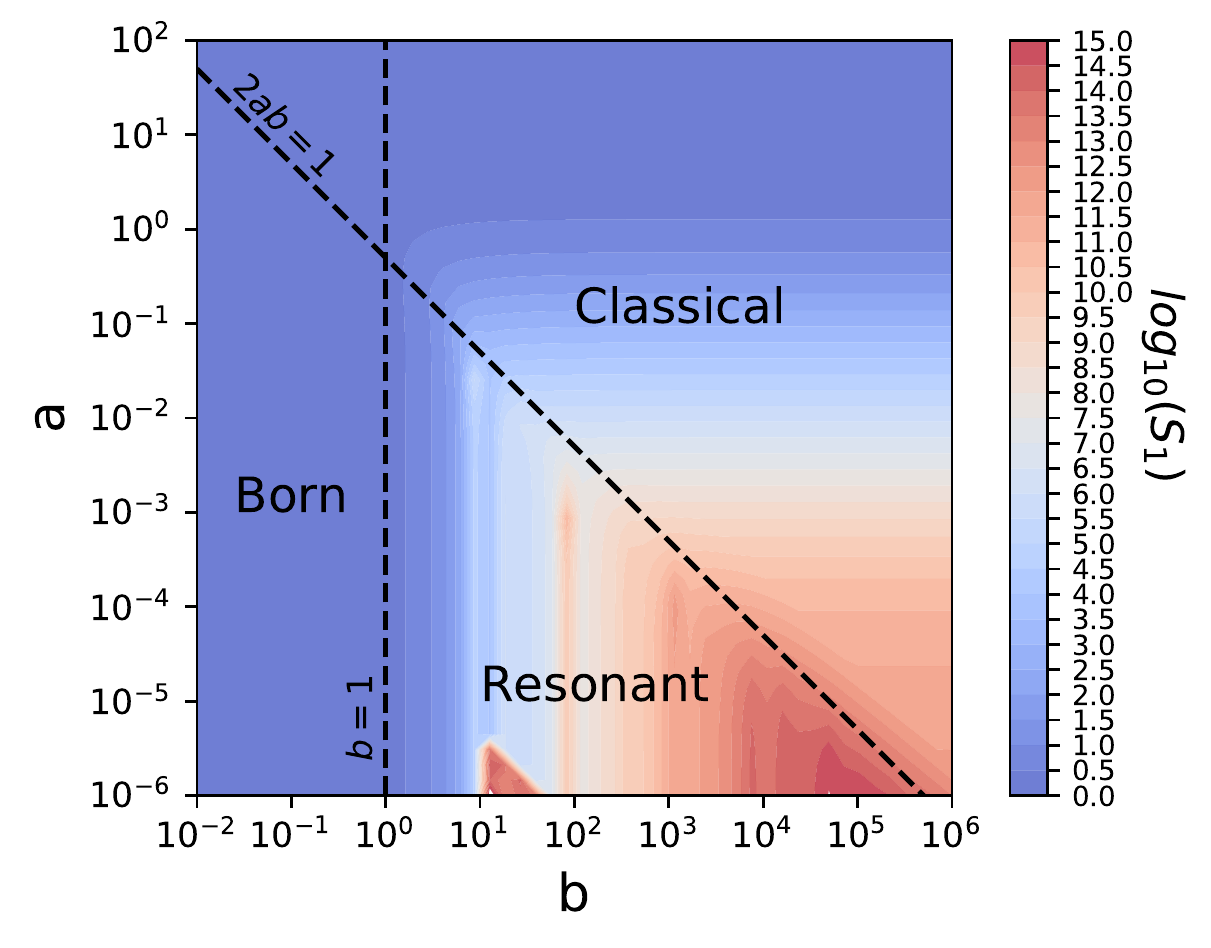}
\includegraphics[scale=0.62]{ab_regime.pdf}
\caption[Sommerfeld factor for p-wave annihilation]{Left: contour of the Sommerfeld factor for p-wave annihilation in the Bron, resonant and classical regimes in the $a-b$ plane. Right: relevant regions of the $a-b$ plane when considering DM freeze-out (blue), DM indirect detection in the Milky Way (orange) and in dwarf galaxies (green) and CMB constraints on DM annihilation (red) with $\alpha_{\rm med}\simeq \alpha_{EW}$.}\label{fig:ab_p}
\end{figure}

Now that we have learned more about the simplest case, the s-wave, we can take a look to the next step: the p-wave case. Once one consider non-zero angular momentum quantum number ($l>0$), since the angular momentum starts to play a role, this make the Schrödinger equation of Eq. \ref{eq:shro_Xl_2} more unstable (numerically speaking). However, it still can be solved numerically relatively easily for the p-wave contribution. We give in the left panel of Figure \ref{fig:ab_p} contours of the Sommerfeld factor in the p-wave case ($l=1$) in the $a-b$ plane. Note that the colour scale is different than in the s-wave case of Figure \ref{fig:ab_s}. Here the Sommerfeld factor goes up to $10^{15}$.
\\

We give in the right panel of Figure \ref{fig:ab_p} the regions of the $a-b$ plane that are relevant for the various contexts similarly to the right panel of Figure \ref{fig:ab_s}. Then one can compare both panels of Figure \ref{fig:ab_p} and see that, as in the s-wave case, the Sommerfeld factor is not much larger than one at DM decoupling if we fix the DM-to-med coupling to the electroweak fine structure constant. However, as the Sommerfeld factor is globally much much larger in the p-wave case than in the s-wave case, one could think that it becomes highly non-negligible even for DM annihilation in the Milky Way today. Indeed, the Sommerfeld factor can already multiply the p-wave annihilation cross section by 3 or even 4 orders of magnitude. It goes even to 10 orders of magnitude for DM annihilations in dwarf galaxies and up to 15 for annihilations at CMB recombination.
\\

\begin{figure}[t]
\centering
\includegraphics[scale=0.56]{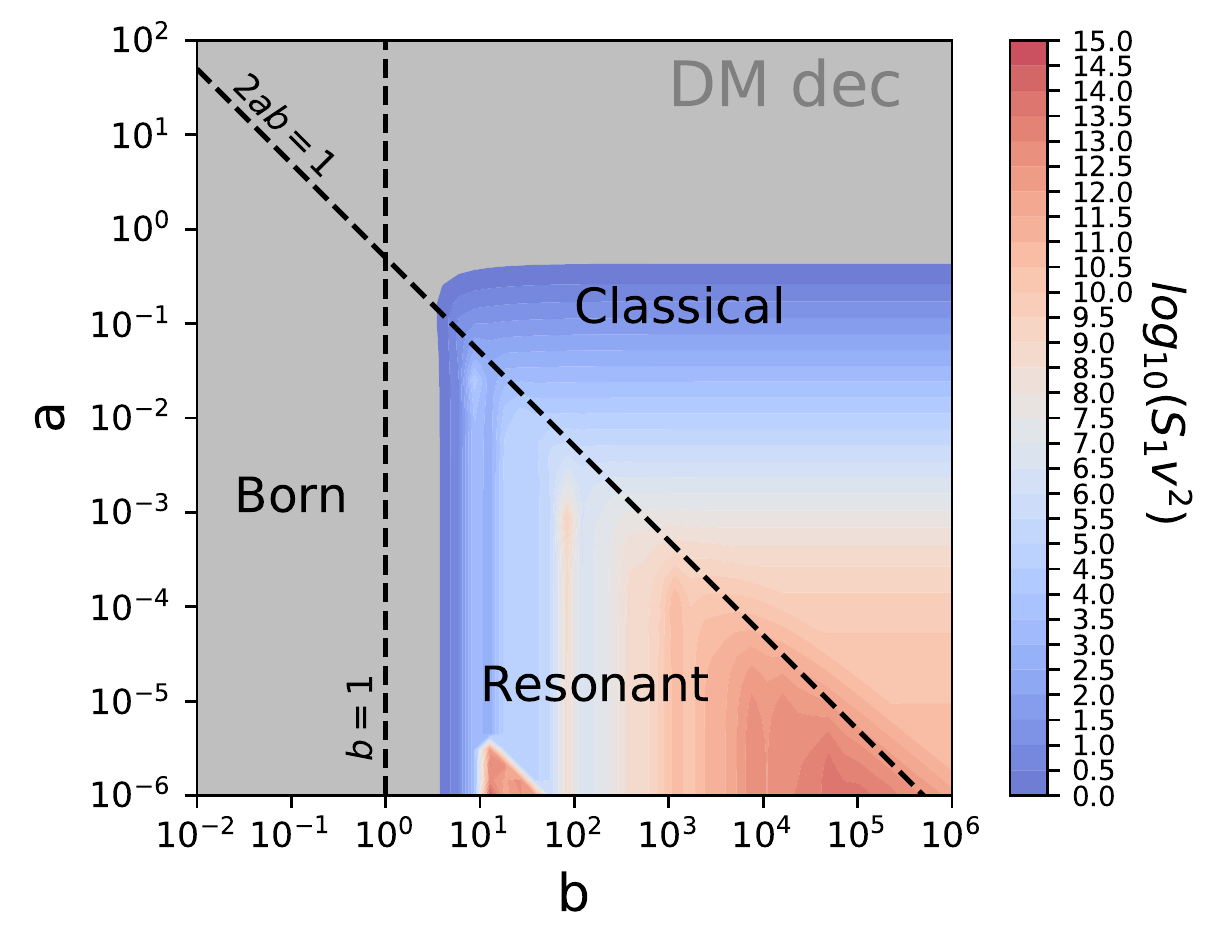}
\includegraphics[scale=0.56]{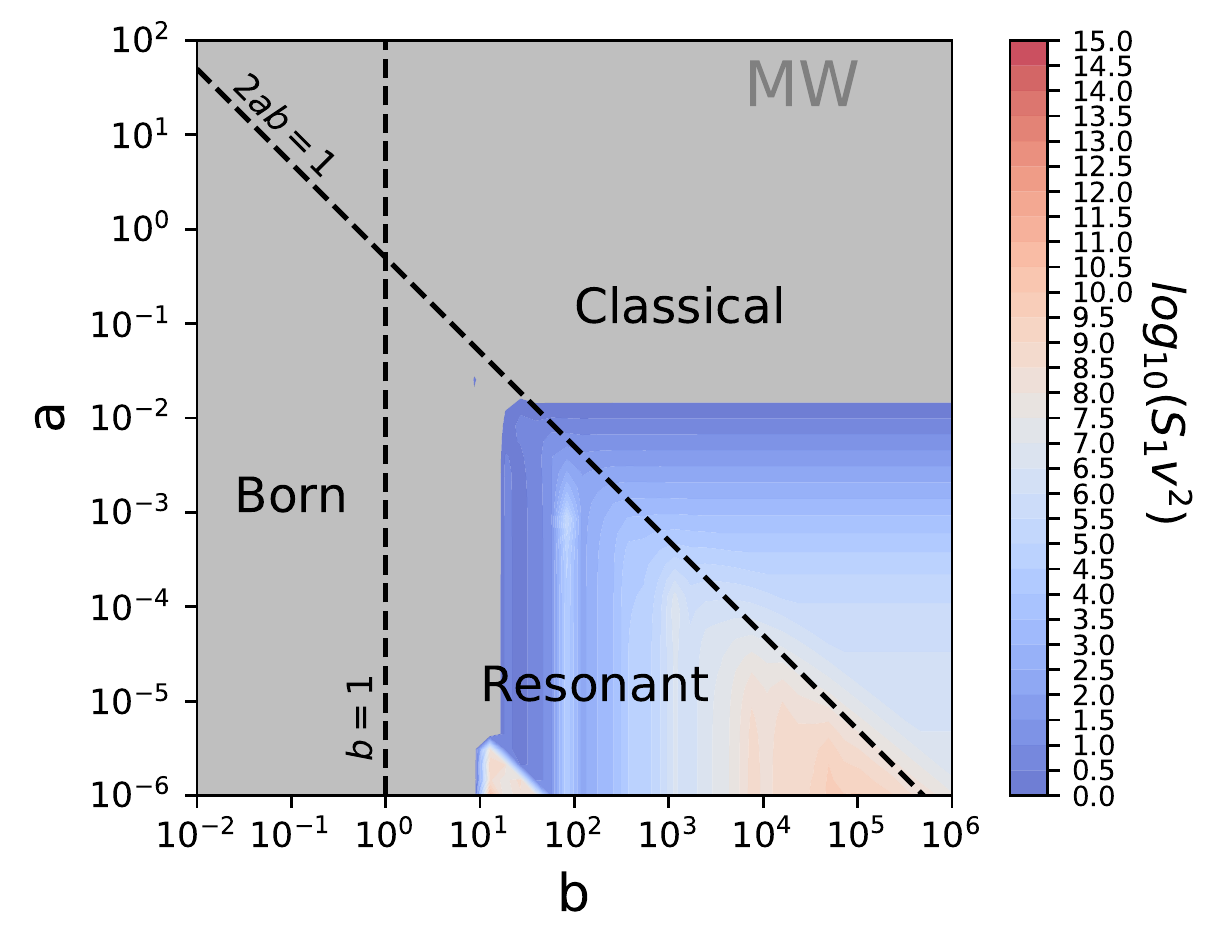}
\includegraphics[scale=0.56]{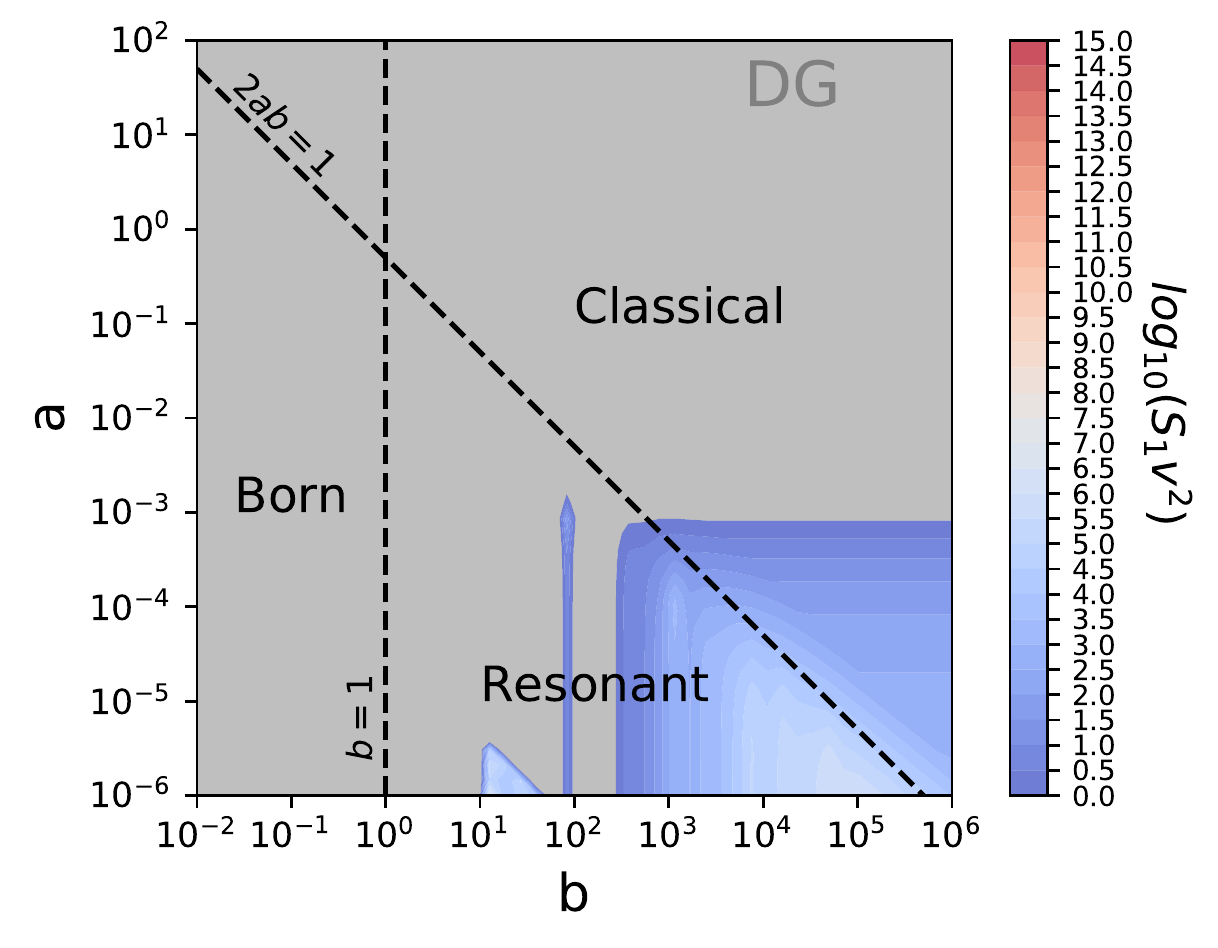}
\includegraphics[scale=0.56]{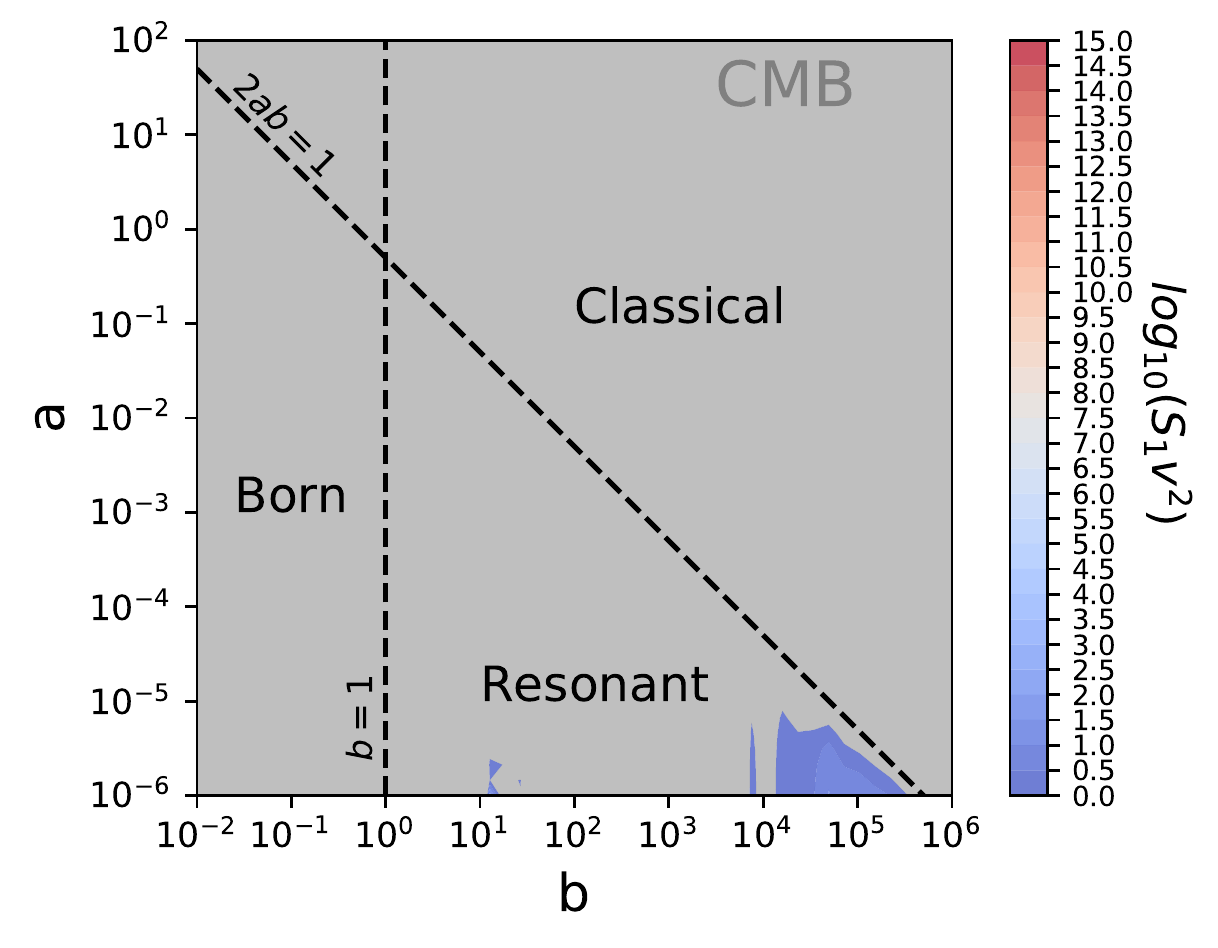}
\caption[Sommerfeld factor times the velocity squared for p-wave annihilation]{Contour of the Sommerfeld factor times $v^{2}$ for p-wave annihilation in the Born, resonant and classical regimes in the $a-b$ plane.}\label{fig:ab_pv2}
\end{figure}

However, one has to be more careful as in the p-wave case the annihilation cross section contains an additional factor of $v^{2}$ which compensates the Sommerfeld factor as it is well known. This is why indirect detection constraints are usually neglected in DM models in which the DM annihilates through a p-wave. Thus, in order to know if the Sommerfeld enhancement is in fact relevant or not for a given wave, one should study $S_{l}v^{2l}$ instead of $S_{l}$ alone. This is what we do in Figure \ref{fig:ab_pv2} which shows, for the four situations we considered individually, contours of $S_{1}v^{2}$ in the $a-b$ plane. In the four panels of Figure \ref{fig:ab_pv2}, the grey regions are where $S_{1}v^{2}<1$ such that the p-wave contribution is negligible. 
\\

Now, we can compare the results for p-wave annihilation presented in Figure \ref{fig:ab_pv2} to the one obtained for s-wave annihilation shown in Figure \ref{fig:ab_s}. For a given choice of the dimensionless parameters $a$ and $b$, it is clear from this comparison that if both s-wave and p-wave are present, the s-wave annihilation always easily dominates over the p-wave annihilation. Thus, one can always neglect the p-wave contribution even taking the Sommerfeld effect into account. However, if the s-wave annihilation is forbidden\footnote{This can be the case due to angular momentum conservation considerations for specific initial and final states. For example this is the case for the annihilation of a pair of Majorana fermions into a pair of Dirac fermions or in annihilation of a pair of Dirac fermions into a pair of real scalars.}, the DM p-wave annihilation cross section could be largely enhanced by the Sommerfeld effect if the DM-to-med coupling is large enough. This can be seen on the bottom left panel of Figure \ref{fig:ab_pv2} in which the product $S_{1}v^{2}$ can be of order of $\sim 10$ in dwarf galaxies if $\alpha_{\rm med}\sim\alpha_{EW}\times \sqrt{10}$. This effect can thus be relevant for a DM candidate as we can see in the scalar portal model for example. Indeed, this can be seen for example if we fix the DM-to-med coupling from the DM relic density constraint of Section \ref{sec:CST-relic_censity}, if we fix the mediator (a real scalar) mass to $m_{\phi}=1$ GeV for example\footnote{The only requirement is that the light mediator must be much lighter than the DM.} and then if we compute the DM annihilation cross section as a function of the DM mass, taking the Sommerfeld enhancement into account.
\\

\begin{figure}
\centering
\includegraphics[scale=0.6]{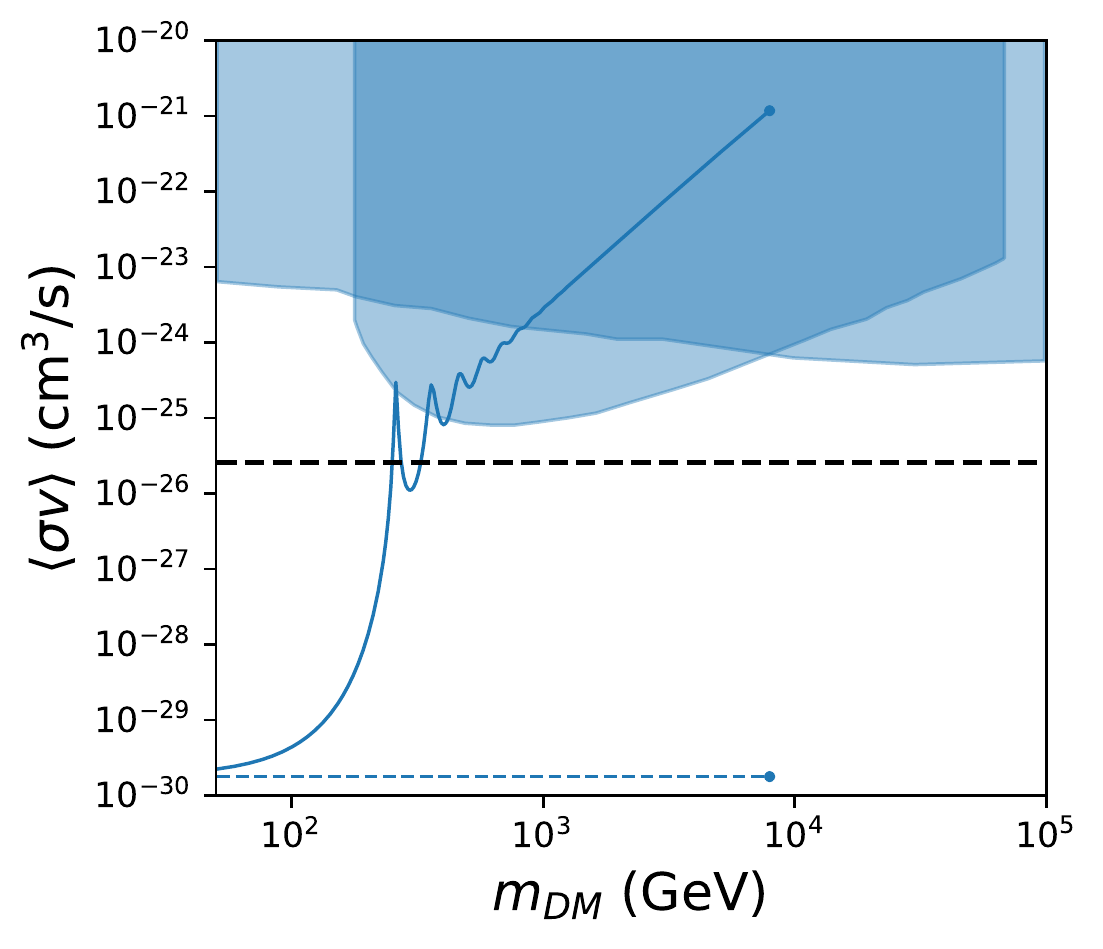}
\includegraphics[scale=0.6]{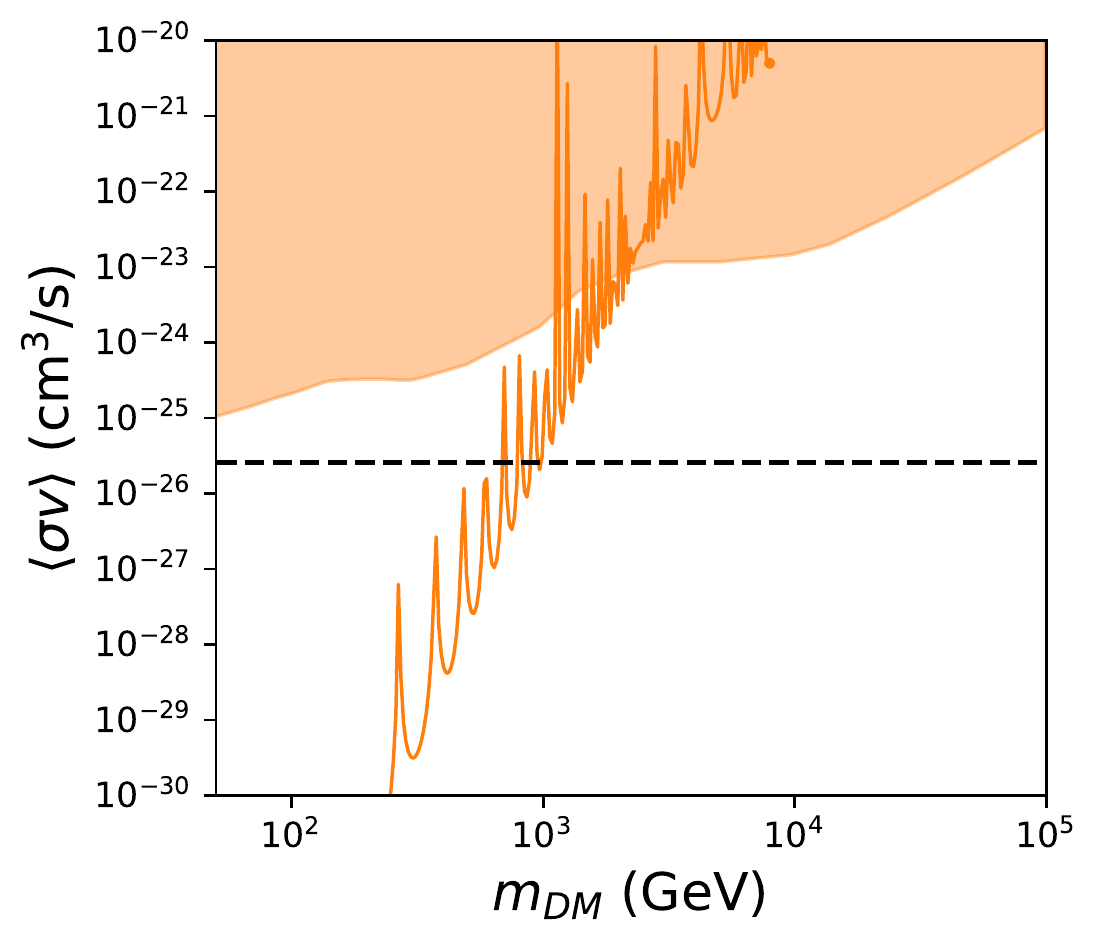}
\caption[DM annihilation cross section with and without the Sommerfeld factor as a function of the DM mass for annihilation in the Milky way and in dwarf galaxies]{DM annihilation cross section with (solid line) and without (dashed coloured line) the Sommerfeld factor as a function of the DM mass for annihilation in the Milky way (left) and in dwarf galaxies (right). Are also shown the constraints coming from indirect detection experiments as explained in Section \ref{sec:CST-ID} and the value of the annihilation cross section at freeze-out (black dashed).}\label{fig:svp}
\end{figure}

The result is shown in Figure \ref{fig:svp} which displays the values of the DM annihilation cross section with (solid coloured) and without (dashed coloured) the Sommerfeld factor for DM annihilations in the Milky Way (left panel) and in dwarf galaxies (right panel) together with the corresponding constraints coming from indirect detection experiments (see Section \ref{sec:CST-ID} for more details on these constraints). The colour scheme of Figure \ref{fig:svp} follows the one of Figure \ref{fig:ID_Constraints} for simplicity. The solid dots ending the curves of both panels of Figure \ref{fig:svp} indicate where the DM relic density constraint requires a DM-to-med coupling larger than unity. Continuing the line beyond this dot would then be untrustworthy as a perturbative calculus. On the left panel (and less easily on the right panel) of this Figure, one can recognize the previously discussed Born, resonant and classical regimes. Indeed, the annihilation cross section starts by being flat because the mediator mass is not much lighter than the DM one and the DM-to-med coupling is still quite low\footnote{We recall that fixing this coupling from the DM relic density constraint makes the coupling to increase linearly with the DM mass.}. The Sommerfeld factor is thus close to one as the solid and dashed coloured line merge. At higher DM mass, the mediator versus DM mass ratio decrease and the DM-to-med coupling increase such that one enters the classical regime. The Sommerfeld factor now strongly depends on the precise values of the parameters. At even higher DM mass, the mediator is effectively massless, this is the classical regime where the curve flatten.
\\

These results show a concrete and very simple example along which, contrary to what is usually admitted in the literature, a p-wave annihilation can be constrained by indirect detection. As we have seen in the first chapter of this thesis, this model is nevertheless already excluded in many ways before being constrained by indirect detection, but this simply illustrates the importance of the Sommerfeld enhancement even for p-wave annihilations. We will see in the following that this effect can be fully relevant for other concrete p-wave models in the context of self-interacting DM. For which some part of the parameter space will be constrained precisely by indirect detection, see Chapter \ref{ch:other_minimal}.
\\

In summary, in this chapter we have introduced the Sommerfeld effect and emphasised the fact that it is important in several different contexts, for self-interactions in particular as well as for DM annihilation processes that could take place at CMB recombination or in the MW and DG. We finished this chapter in showing that contrary to common belief in general in the literature, p-wave annihilation scenarios can be tested by indirect detection experiments thanks to the Sommerfeld enhancement.
\part{DM production and Hidden Sector}
\chapter{Dark Matter production mechanisms}\label{ch:prod}
\yinipar{I}n this third Chapter, we will present how to account for the relic abundance of DM when it interacts with the SM through portals as we did in \cite{Hambye:2019dwd,Vanderheyden:2021tih}\footnote{These works threaten the vector portal case, the scalar portal case is then an unpublished original work.}. In particular, we consider the possibility for the mediator to be massive and its implications on the DM relic abundance production mechanism in a generic way for both the scalar and the vector portal models of Subsection \ref{subsec:portal}. In this context and using the formalism of the Boltzmann equation developed above (see Subsection \ref{sec:be}), we will detail the phase diagram of DM production. Next, we will discuss some specificities of the two portal models we consider.

\section{The three sectors and their connections}\label{subsec:Triangle}
The underlying structure is the same for the Higgs portal model in the fully broken case (i.e. $0<m_{\phi}<m_{\rm DM}<v_{H}<v_{\Phi}$) and for the kinetic mixing portal model in the massive mediator case (i.e. $0<m_{\gamma'}<m_{\rm DM}$). This structure is composed of three sectors connected to each other by three different couplings resulting from two interactions. This can be seen in Figure \ref{fig:Triangle}.
\\

\begin{center}
\begin{figure}[h!]
\centering
\includegraphics[scale=1.0]{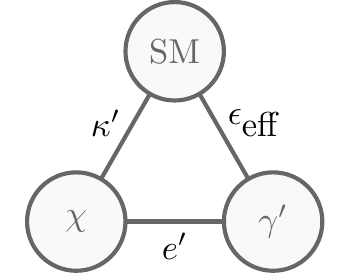}
\includegraphics[scale=1.0]{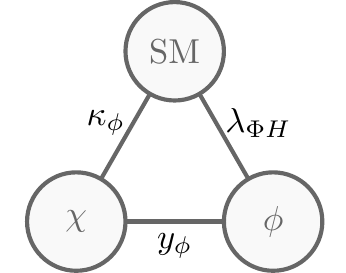}
\caption[The three sectors and their three connections for the vector and scalar portal models]{The three sectors (blobs) and their three connections (lines) for the vector (left) and scalar (right) portal models.}
\label{fig:Triangle}
\end{figure}
\end{center}

We would like to draw the reader's attention on the fact that this three sector and three connections structure (depicted in Figure \ref{fig:Triangle}) for the kinetic mixing portal model is specific to the massive mediator case. Indeed, if the dark photon is massless, one can see from Eq. \ref{eq:lag_km_mass_mix} that the dark photon does not couple to any SM particle such that the only way to produce dark photon is to go through the production of DM particles. As already said above, the smooth transition between massive and massless mediator cases goes through the proper treatment of thermal effects. This will be done in Section \ref{sec:spec_vector} where we will see that the effective coupling $\epsilon_{\rm eff}$ in Figure \ref{fig:Triangle} goes to zero with the dark photon mass. That is to say that in the massless mediator case, the line (i.e. the connection) between the SM and dark photon baths no longer exists.
\\

\begin{center}
\begin{figure}[h!]
\centering
\includegraphics[scale=1.0]{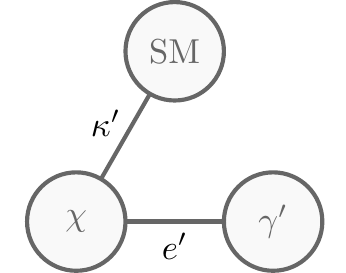}
\caption[The three sectors and their two connections for the kinetic mixing portal model]{The three sectors (blobs) and their two connections (lines) for the kinetic mixing portal model in the massless dark photon case.}
\label{fig:Triangle}
\end{figure}
\end{center}

\section{Phase diagram of dark matter production}\label{sec:phases}
We will now discuss how the DM relic abundance depends on this "three sectors-three connectors" structure. We will start with an empty hidden sector (i.e. no DM and no mediator in the thermal bath at the end of inflation or a negligible amount of them), and from this initial condition we will go through all production regimes (or "phases") all the way from phases where the DM never thermalises with any other particles to phases where all particles thermalise with each other\footnote{If the visible and hidden sectors are feebly coupled with each others, it would not be surprising that the reheating at the end of inflation took place mostly into one of the sectors rather than both of them. We assume here that reheating happened in the VS.}.
\\

In this chapter, we aim to illustrate the fact that the final phase diagram of DM production is quite general for portal models which present a "three sectors-three connectors" behaviour. We will make this discussion for both vector and scalar portal models. Note that for the sake of clarity and generality, thermal effects will not be included in the analysis as they are specific to the dark photon portal model. However we will discuss results including thermal effects in Section \ref{sec:spec_vector} and refer to \cite{Hambye:2019dwd} for a more in-depth analysis of consequences of thermal effects on DM production mechanisms.
\\

\begin{center}
\begin{figure}[h!]
\centering
\includegraphics[scale=0.8]{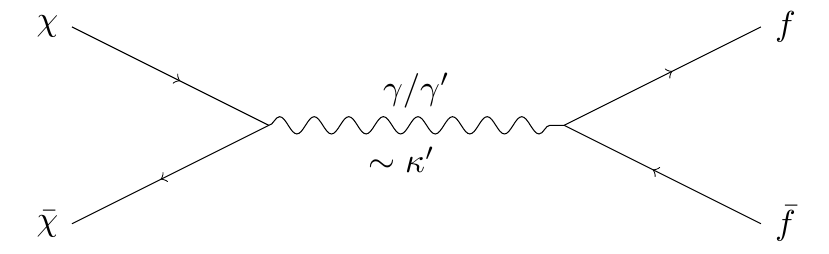}
\includegraphics[scale=0.8]{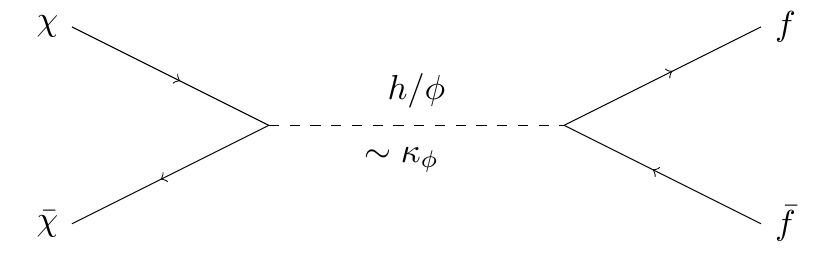}
\caption[Feynman diagram of DM annihilation into SM fermions]{Connection between the DM and SM baths goes through the production of the mediator in the s-channel for both the kinetic mixing portal (left) and the Higgs portal (right) models. $f$ indicates a SM fermion.}
\label{fig:diag_DM-to-SM}
\end{figure}
\end{center}

There are only four free parameters on which the DM relic abundance can depend on: the DM and the mediator masses ($m_{\gamma'}$ or $m_{\phi}$) and the two couplings between the DM and the mediator ($\alpha'$ or $\alpha_{\phi}$) and between the mediator and the SM ($\epsilon$ or $\lambda_{\phi H}$). These parameters rule all possible interactions of the theory. In particular, the connection between the DM and SM baths is going through the production of the mediator in the s-channel, see Figure \ref{fig:diag_DM-to-SM}.
\\

The strength of these interactions is set by the DM-to-SM connector:

\myeq{
\kappa ' &\equiv \epsilon\sqrt{\alpha '/\alpha},\label{eq:kappa_KM}\\
\kappa_{\phi} &\equiv \sin\left(2\theta\right)\sqrt{\pi\alpha_{\phi}}.\label{eq:kappa_HP}
}

\begin{center}
\begin{figure}[h!]
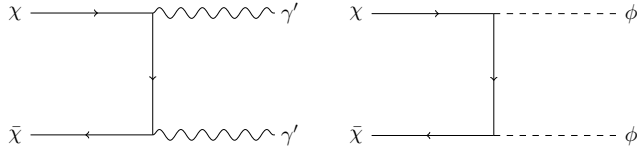

\centering
\includegraphics[scale=0.8]{XXApAp.pdf}
\includegraphics[scale=0.8]{hp_ap_BRO_XXPP.pdf}
\caption[Feynman diagram of DM annihilation into light bosons]{Connection between the DM and mediator baths goes through the annihilation of two DM particles into two mediator particles in the t-channel for both the kinetic mixing portal (left) and the Higgs portal (right) models.}
\label{fig:diag_DM-to-med}
\end{figure}
\end{center}

\noindent In the hidden sector, the connection between the DM and the mediator baths is set by a DM annihilation in the t-channel which is driven by $\alpha'$ and $\alpha_{\phi}$ for the vector and scalar portal respectively, see Figure \ref{fig:diag_DM-to-med}. 
\\

\begin{center}
\begin{figure}[h!]
\centering
\includegraphics[scale=0.8]{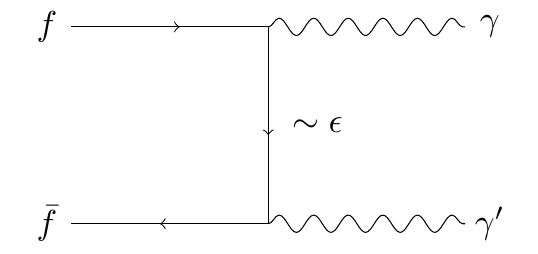}
\includegraphics[scale=0.8]{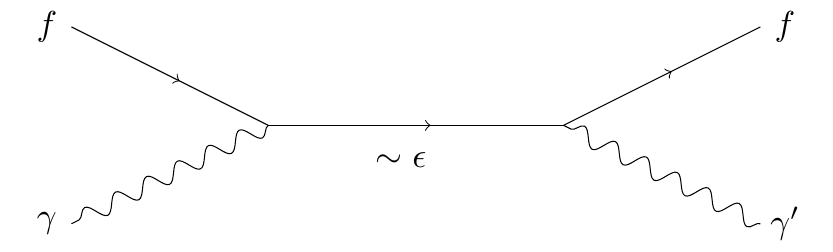}\\
\includegraphics[scale=0.8]{PPHH_b.pdf}
\caption[Feynman diagram of dominant processes connecting the SM and mediator baths]{Dominant processes connecting the SM and mediator baths for the kinetic mixing portal (top) and the Higgs portal (bottom) models. $f$ indicates a SM fermion.}
\label{fig:diag_SM-to-med}
\end{figure}
\end{center}

Finally, the connection between the mediator and SM baths goes through several processes for which the dominant ones are shown in Figure \ref{fig:diag_SM-to-med}. We recall here that for the vector portal model, the processes given in Figure \ref{fig:diag_SM-to-med} are valid only in the massive mediator case as the connection between SM fermions and the dark photon no longer exists in the massless dark photon case. However, these processes are driven by the mixing parameter $\epsilon$ and $\lambda_{\Phi H}$ for the kinetic mixing portal and the Higgs portal models respectively such that they are not independent of the two other connectors. Indeed, as we have introduced the SM-to-DM connectors $\kappa'$ and $\kappa_{\phi}$ in Eqs. \ref{eq:kappa_KM} and \ref{eq:kappa_HP}, one can see the SM-to-med couplings as function of the DM-to-SM and the DM-to-med connectors:

\myeq{
\epsilon ' &= \kappa '\sqrt{\alpha/\alpha '},\label{eq:epsilon_KM}\\
\lambda_{\Phi H} &= \frac{\kappa_{\phi}}{\sqrt{\pi\alpha_{\phi}}}\left(\frac{m_{H}^{2}-m_{\Phi}^{2}}{v_{\Phi}v_{H}}\right),\label{eq:lambda_HP}
}

\noindent where in the second equation we have used the definition of the mixing angle given in Eq. \ref{eq:mixing_angle}.
\\

As a consequence of this structure, for a given set of masses, one can represent the DM relic abundance as contour line in the $\kappa'-\alpha'$ (or $\kappa_{\phi}-\alpha_{\phi}$) plane, that is to say as a function of the strength of the SM-to-DM connector and purely hidden sector interactions. This will lead to a "phase diagram" with various interesting regimes. To get this phase diagram we need to integrate a set of Boltzmann equations for the DM and the mediator yields. Once it is done, one can distinguish nine regimes along five distinct dynamical mechanisms\footnote{In the massless mediator case, five regimes and four dynamical mechanisms were already found and discussed in \cite{Chu:2011be} for the Kinetic Mixing model and another type of scalar portal.}.
\\

\begin{center}
\begin{figure}[h!]
\centering
\includegraphics[scale=0.65]{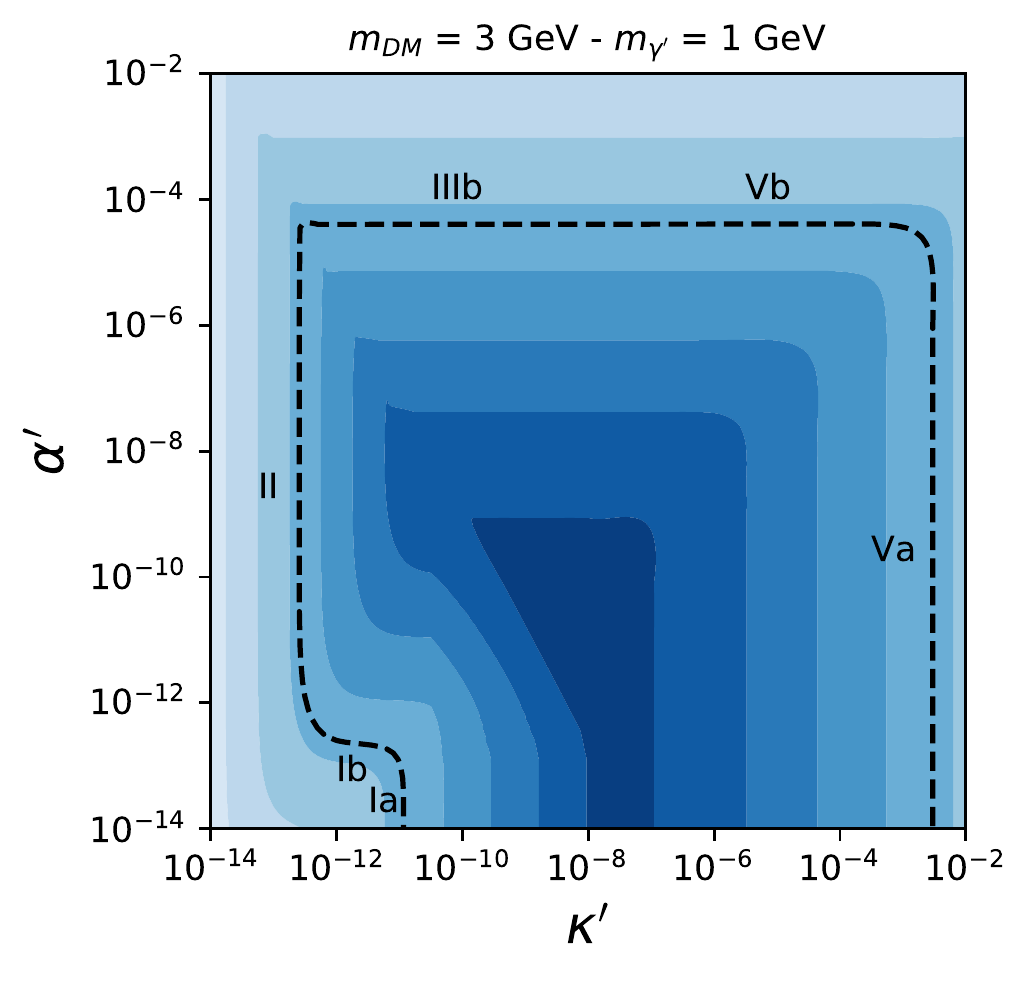}
\includegraphics[scale=0.65]{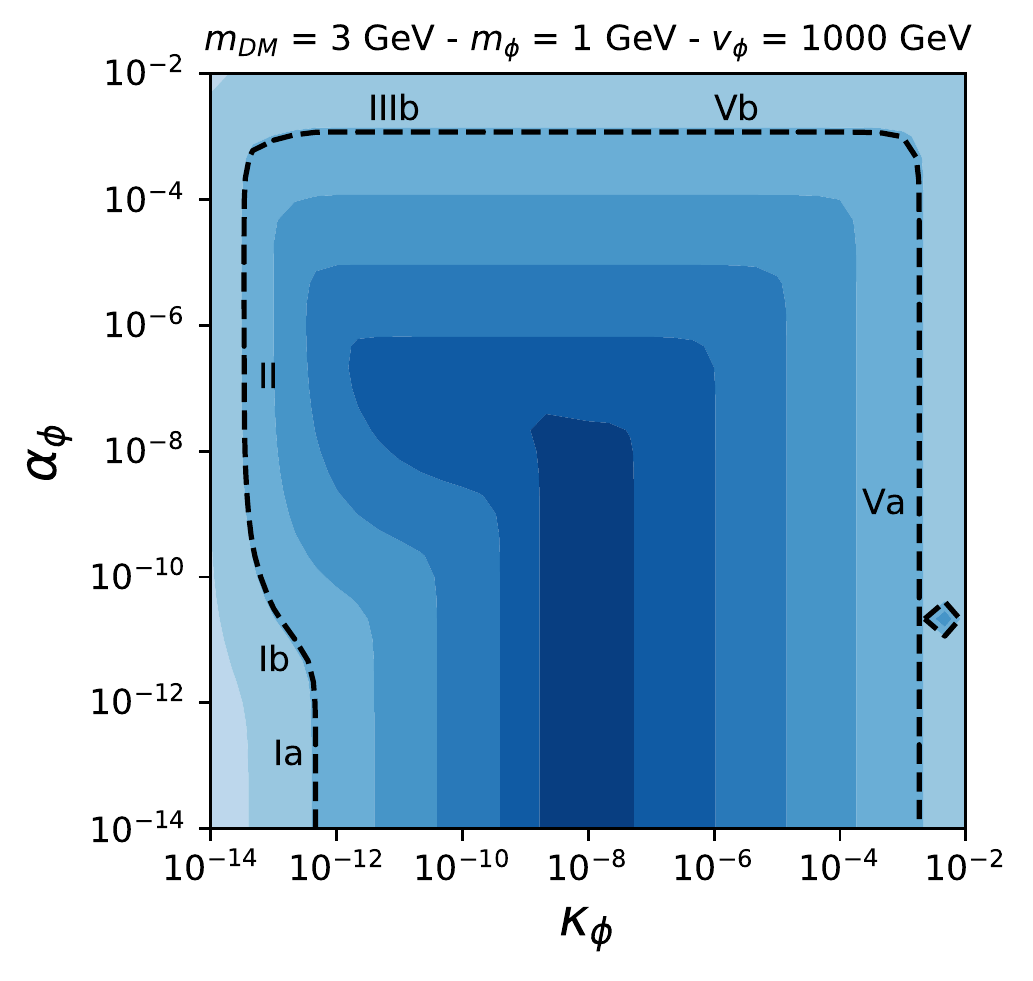}
\caption[Vector and scalar portal models global phase space]{DM relic density obtained as a function of $\kappa'/\kappa_{\phi}$ and $\alpha'/\alpha_{\phi}$ for $m_{\rm DM}=3$ GeV and $m_{\rm med}=1$ GeV. This diagram displays the various different production regimes which can lead to the observed relic density. For this particular choice of masses one obtains 6 different production regimes along 4 dynamical ways: freeze-in (Ia $\&$ Ib), sequential freeze-in (II), reannihilation (IIIb) and freeze-out (Va $\&$ Vb).}\label{fig::DogDiagram1}
\end{figure}
\end{center}

These five mechanisms are freeze-in (I), sequential freeze-in (II), reannihilation (III), secluded freeze-out (IV) and freeze-out (V). These regimes differ depending on which of the three connections thermalises the two reservoirs it connects or instead remains out of equilibrium at all times. In the latter case, a connection can be either totally irrelevant or, on the contrary, induce a relevant out-of-equilibrium production of a reservoir from another one. The most generic sequence of regimes, in the massive mediator case, appearing along the phase diagram is the one shown in Figures \ref{fig::DogDiagram1}. Left (right) panel of this figure shows contour lines of the DM relic abundance in the $\kappa'-\alpha'$ ($\kappa_{\phi}-\alpha_{\phi}$) plane for the vector (scalar) portal model. We have always assumed DM to be heavier than the mediator and we present results obtained with a weak mass hierarchy.
\\

\begin{figure}[h!]
\centering
\stackunder[5pt]{\includegraphics[scale = 0.9]{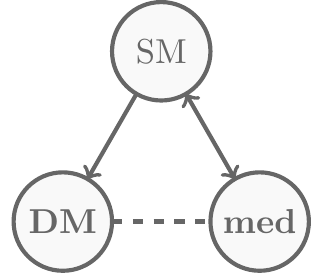}}{Ia}
\stackunder[5pt]{\includegraphics[scale = 0.9]{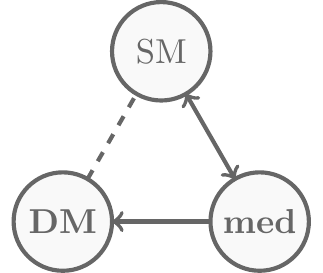}}{Ib}
\stackunder[5pt]{\includegraphics[scale = 0.9]{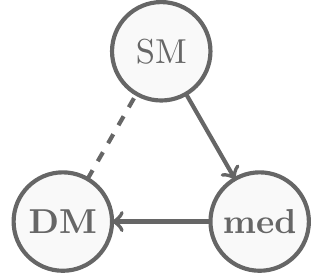}}{II}\\
\vspace{0.5cm}
\stackunder[5pt]{\includegraphics[scale = 0.9]{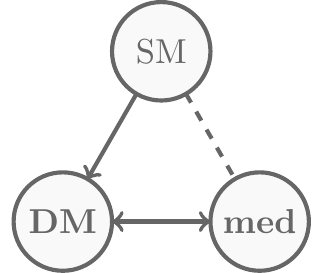}}{IIIa \& IVa}
\stackunder[5pt]{\includegraphics[scale = 0.9]{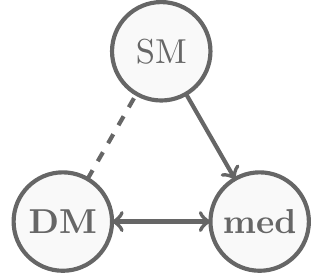}}{IIIb \& IVa}
\stackunder[5pt]{\includegraphics[scale = 0.9]{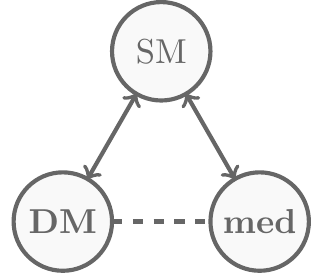}}{Va}
\stackunder[5pt]{\includegraphics[scale = 0.9]{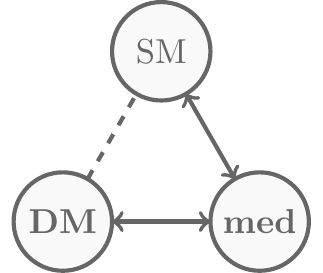}}{Vb}\\
\caption[Schematic representation of all 9 possible DM production regimes in the three sectors three connectors scenario]{The 9 possible DM production regimes in the three sectors three connectors scenario. A double sided arrow means that the two corresponding sectors have reached chemical equilibrium; a single sided arrow indicates slow out-of-equilibrium production of one sector by the other one; a dashed line corresponds to a subdominant (i.e. irrelevant) interaction between the sectors. Regimes Ia and Ib, II, IIIa and IIIb, IVa and IVb, Va and Vb are associated to 5 distinct mechanisms to produce the DM abundance:  the freeze-in (I), sequential freeze-in (II), reannihilation (III), secluded freeze-out (IV) and freeze-out (V) mechanisms respectively (see Sections \ref{subsec:FI} to \ref{subsec:FO}). Note that the diagrams are identical for the reannihilation and secluded freeze-out mechanisms.}
 \label{fig::7Triangle}
\end{figure}

This sequence of regimes can be illustrated by the following chain,

\myeq{
{\rm Ia \rightarrow Ib \rightarrow II \rightarrow IIIb \rightarrow Vb \rightarrow Va},
}

\noindent where the "a" and "b" labels refer to a process going through the visible sector or the hidden sector respectively and will be defined properly below\footnote{Note however that the "a" and "b" notations here differ from the one defined in \cite{Hambye:2019dwd}}. Using the "three sectors-three connectors" representation as in Figure \ref{fig:Triangle}, the Figure \ref{fig::7Triangle} gives all mechanisms we found, to be explained below.
\\

This specific structure involves three new regimes (Ib, II and IIIb), on top of those already existing in the massless case (Ia, Va and Vb). The fourth new regime (IVb) and the two other regimes existing in the massless case (IIIa and IVa) will appear for other models. In all cases, the shape of the phase diagram is generic.

%This structure, on top of the regimes already existing in the massless case (Ia, Va and Vb), involve 3 new regimes: Ib, II and IIIa. The fourth new regime, IVa, as well as the 2 other regimes already existing in the massless case (IIIb and IVb), 
%can appear for other choices of the parameter and more generally for other models. The four new regimes arise from the extra SM production of dark photons.
%In all cases the mooring bollard shape of the phase diagram is a generic signature.

\subsection{Set of Boltzmann equation}
Let us now discuss how the parameter space (i.e. the $\kappa'/\kappa_{\phi}-\alpha'/\alpha_{\phi}$ plane) is naturally divided in different phases. In order to understand which process will dominate and when it will, we present in this subsection the set of Boltzmann equations in a three sectors three connectors scenario. Since the global behaviour of the parameter space is the same for both portal models, we will use a generic notation for the analysis and we will present results for both the vector and the scalar portal models.
\\

The set of Boltzmann equations that determine the time evolution of the DM and the light mediator yields is given by\footnote{Here and in subsequent Boltzmann equations, we have included  factors of $1/2$, typical of Dirac DM particles, into the definitions of the cross-sections \cite{Gondolo:1990dk}.}

\myeq{
xHs\frac{\diff Y_{\rm DM}}{\diff x}&=\langle \sigma_{\rm med\rightarrow DM} v\rangle n^2_{\rm med}-\langle \sigma_{\rm DM\rightarrow med} v\rangle n^2_{\rm DM}\nonumber\\
&\hspace{-2cm}+\langle \sigma_{\rm DM \rightarrow SM} v\rangle \left[({n_{\rm DM}^{\rm eq}})^2-n^2_{\rm DM} \right]+\langle \Gamma^D_{\rm SM \rightarrow DM} \rangle\frac{n_{\rm SM}^{\rm eq}}{(n_{\rm DM}^{\rm eq})^{2}} \left[{(n_{\rm DM}^{\rm eq}})^2-n^2_{\rm DM} \right] + \dots, \label{eq:YDM}\\
xHs\frac{\diff Y_{\rm med}}{\diff x}&=\langle \sigma_{\rm DM\rightarrow med} v\rangle n^2_{\rm DM}-\langle \sigma_{\rm med\rightarrow DM} v\rangle n^2_{\rm med}\nonumber\\
&\hspace{-2cm}+\langle \sigma_{\rm med \rightarrow SM} v\rangle \left[({n_{\rm med}^{\rm eq}})^2-n^2_{\rm med} \right]+\langle \Gamma^D_{\rm SM \rightarrow med} \rangle\frac{n_{\rm SM}^{\rm eq}}{(n_{\rm med}^{\rm eq})^{2}} \left[{(n_{\rm med}^{\rm eq}})^2-n^2_{\rm med} \right] + \dots. \label{eq:Ymed}
}

%\myeq{
%xHs\frac{\diff Y_{\rm DM}}{\diff x}&=\langle \sigma_{\rm DM \rightarrow SM} v\rangle \left[({n_{\rm DM}^{\rm eq}})^2-n^2_{\rm DM} \right]+\langle \Gamma^D_{\rm SM \rightarrow DM} \rangle\frac{n_{\rm SM}^{\rm eq}}{(n_{\rm DM}^{\rm eq})^{2}} \left[{(n_{\rm DM}^{\rm eq}})^2-n^2_{\rm DM} \right]\nonumber\\
%&\hspace{1cm}+\langle \sigma_{\rm med\rightarrow DM} v\rangle n^2_{\rm med}-\langle \sigma_{\rm DM\rightarrow med} v\rangle n^2_{\rm DM}, \label{eq:YDM}\\
%xHs\frac{\diff Y_{\rm med}}{\diff x}&=\langle \sigma_{\rm med \rightarrow SM} v\rangle \left[({n_{\rm med}^{\rm eq}})^2-n^2_{\rm med} \right]+\langle \Gamma^D_{\rm SM \rightarrow med} \rangle\frac{n_{\rm SM}^{\rm eq}}{(n_{\rm med}^{\rm eq})^{2}} \left[{(n_{\rm med}^{\rm eq}})^2-n^2_{\rm med} \right]\nonumber\\
%&\hspace{1cm}+\langle \sigma_{\rm DM\rightarrow med} v\rangle n^2_{\rm DM}-\langle \sigma_{\rm med\rightarrow DM} v\rangle n^2_{\rm med}, \label{eq:med}
%}

\noindent where $\Gamma^D$ refers to the decay rate of a SM particle into two DM particles or two mediator particles. We kept a sum over all $\rm SM\leftrightarrow DM$ and $\rm SM\leftrightarrow med$ channels implicit for clarity. The dots in equations \ref{eq:YDM} and \ref{eq:Ymed} stand for possible additional terms which could arise in one model or another\footnote{For example, in the vector portal model we have $f+\gamma'\rightarrow f+\gamma$ for $f$ a SM fermion. In the scalar portal, we have $h+h\rightarrow h+\phi$ for example.}. However, in this thesis, we will limit ourself to leading order processes on the relevant couplings\footnote{For example, in the vector portal model, we have $f+f\rightarrow \gamma'+\gamma$ which dominates over $f+f\rightarrow \gamma'+\gamma'$ as only one mixing parameter is required in the first and two are required in the second process.}. Moreover, in order to avoid cluttering of the equations let us define

\myeq{
\gamma_{\rm SM \leftrightarrow DM}^{\rm eq} \equiv  \langle \sigma_{\rm DM \rightarrow SM}\, v\rangle ({n_{\rm DM}^{\rm eq}})^2+ {\langle \Gamma^D_{\rm SM \rightarrow DM} \rangle} {n_{\rm SM}^{\rm eq}} +\dots,
}

\noindent and

\myeq{
\gamma_{\rm SM\leftrightarrow med}^{\rm eq} \equiv \langle \sigma_{\rm med \rightarrow SM}\, v\rangle ({n_{\rm med}^{\rm eq}})^2+ {\langle \Gamma^D_{\rm SM \rightarrow med} \rangle} {n_{\rm SM}^{\rm eq}} +\dots.
}

\noindent Although the Boltzmann equations contain in general many terms, only one or two terms are relevant for production regimes we are interested in. At the opposite, these equations are not sufficient to determine the amount of DM produced in the reannihilation regimes. Indeed, these regimes are characterised by a different temperature for the hidden and the visible sectors. Thus, we also need to evaluate the amount of energy which has been transferred from one sector to the other. We will discuss the additional needed equation when we will discuss the reannihilation regimes (see subsection \ref{subsec:rean}).
\\

The parameter space can be divided in several regions depending on whether the various connecting processes lead, or not, to thermalisation. Delimiting those regions will help us to understand the distinction between the various production regimes. Notice also that kinetic and chemical equilibrium are reached almost at the same time for relativistic species \cite{Chu:2011be} such that one can use the kinetic equilibrium condition to check if two baths thermalised prior to DM decoupling. The "three sector - three connectors" structure gives us three thermalisation conditions:

\myeq{
\Gamma_{\rm SM \leftrightarrow DM} \gtrsim H\hspace{0.5cm}-\hspace{0.5cm}\Gamma_{\rm SM \leftrightarrow med} \gtrsim H\hspace{0.5cm}-\hspace{0.5cm}\Gamma_{\rm DM \leftrightarrow med} \gtrsim H.\label{eq:thermalisations}
}

\noindent Note that, for the case of SM bath thermalising with one of the two "dark" particles (DM or mediator), the thermalisation conditions of Eq. \ref{eq:thermalisations} have to be evaluated at $T\sim m_{DM}$ since we aim to look for thermalisation when the DM abundance will freeze which typically happens when the temperature cools down to the DM mass. On the other hand, for the last case, where we focus on the possible thermalisation within the hidden sector, we could face two distinct temperatures $T$ and $T'$ for the visible and hidden sector respectively. This would happen if the hidden sector thermalises but is not in thermal equilibrium with the SM, i.e. the two first conditions of Eq. \ref{eq:thermalisations} are not satisfied while the last one is. In this case, the last thermalisation conditions of Eq. \ref{eq:thermalisations} has to be evaluated at $T'\sim m_{DM}$. The interaction rates in Eq. \ref{eq:thermalisations} are given by,

\myeq{
\Gamma_{\rm SM \leftrightarrow DM} &= \frac{\gamma^{eq}_{\rm SM \leftrightarrow DM}}{n_{\rm DM}^{eq}},\\
\Gamma_{\rm SM \leftrightarrow med} &=\frac{\gamma^{eq}_{\rm SM \leftrightarrow med}}{n_{\rm med}^{eq}},\\
\Gamma_{\rm DM \leftrightarrow med} &=\left\langle\sigma_{\rm med \rightarrow \rm DM}v\right\rangle n_{\rm med}.\label{eq:G_DM_gammap}
}

\begin{center}
\begin{figure}[h!]
\centering
\includegraphics[scale=0.75]{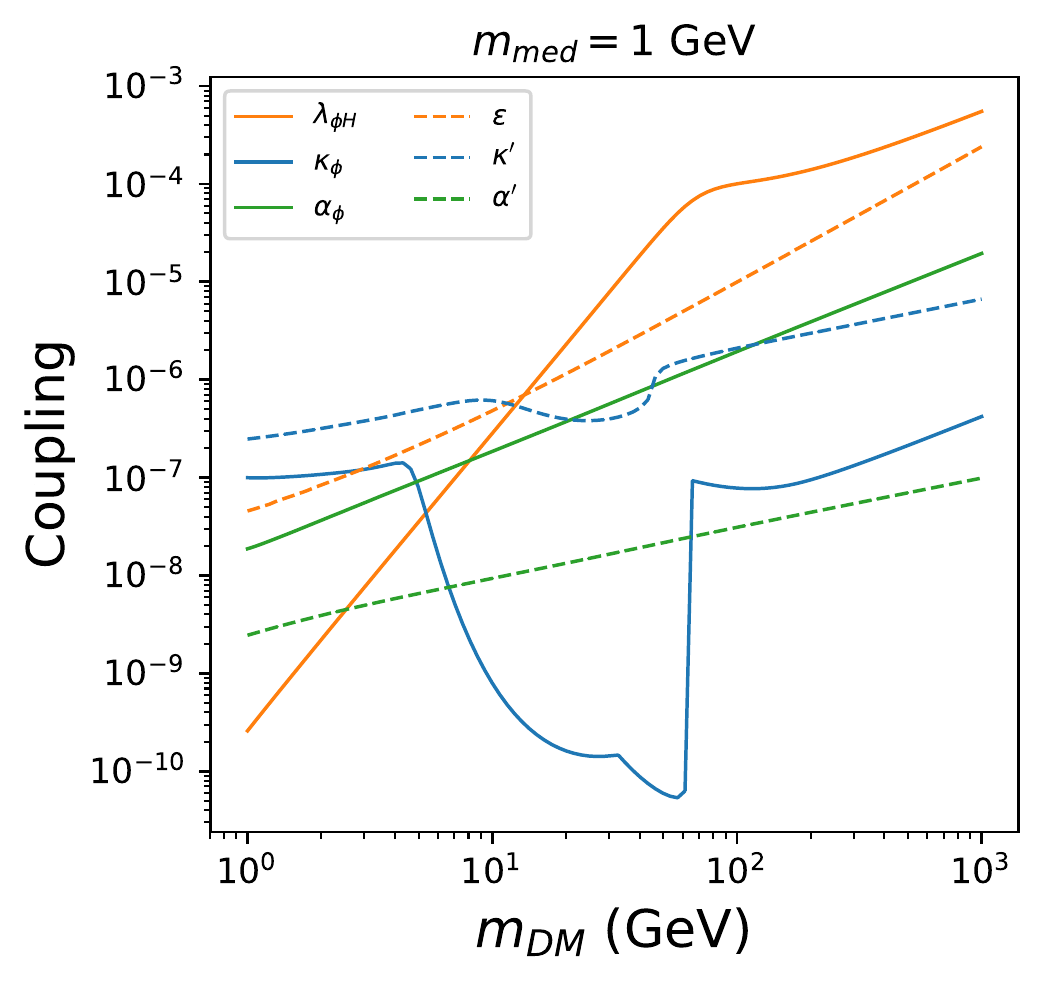}
\caption[Couplings required for thermalisation as a function of the DM mass]{Value of the couplings needed to have thermalisation between DM and SM baths (blue), DM and mediator baths (green) and SM and mediator baths (orange). The solid and dashed lines show results for the vector and scalar portal models respectively.}
\label{fig:th_lines}
\end{figure}
\end{center}

\noindent Those thermalisation conditions can all be converted into a condition on the involved coupling. Indeed, we have seen from the relevant Feynman diagrams of Figures \ref{fig:diag_DM-to-SM}, \ref{fig:diag_DM-to-med} and \ref{fig:diag_SM-to-med} that these three thermalisation conditions are driven by $\kappa'/\kappa_{\phi}$, $\epsilon/\lambda_{\Phi H}$ and $\alpha'/\alpha_{\phi}$ respectively such that one can extract a lower bound on these couplings from thermalisation conditions. Figure \ref{fig:th_lines} gives the minimal value for each coupling required for thermalisation in both the vector and scalar portal models, shown in dashed and solid respectively.
\\

The reason that in Eq. \ref{eq:G_DM_gammap} we do not evaluate quantities at equilibrium is due to fact that mediator particles are still being produced by slow $\rm SM\rightarrow med$ processes. So, we could have situation where neither the mediator nor the DM is at equilibrium and we need to consider out-of-equilibrium quantities. We then have two situations depending on if the mediator has reached equilibrium or not. If the SM and the mediator bath are in thermal equilibrium, quantities appearing in \ref{eq:G_DM_gammap} are at equilibrium and the value of $\alpha'/\alpha_{\phi}$ required for thermalisation does not depend on the connector between the SM and the mediator, $\epsilon/\lambda_{\Phi H}$. On the other hand, if the SM and mediator baths are not in thermal equilibrium, there is a slow production of mediator particles from the SM which is still occurring and the value of $\alpha'/\alpha_{\phi}$ required for thermalisation depends on $\epsilon/\lambda_{\Phi H}$.
\\

\begin{center}
\begin{figure}[h!]
\centering
\includegraphics[scale=0.75]{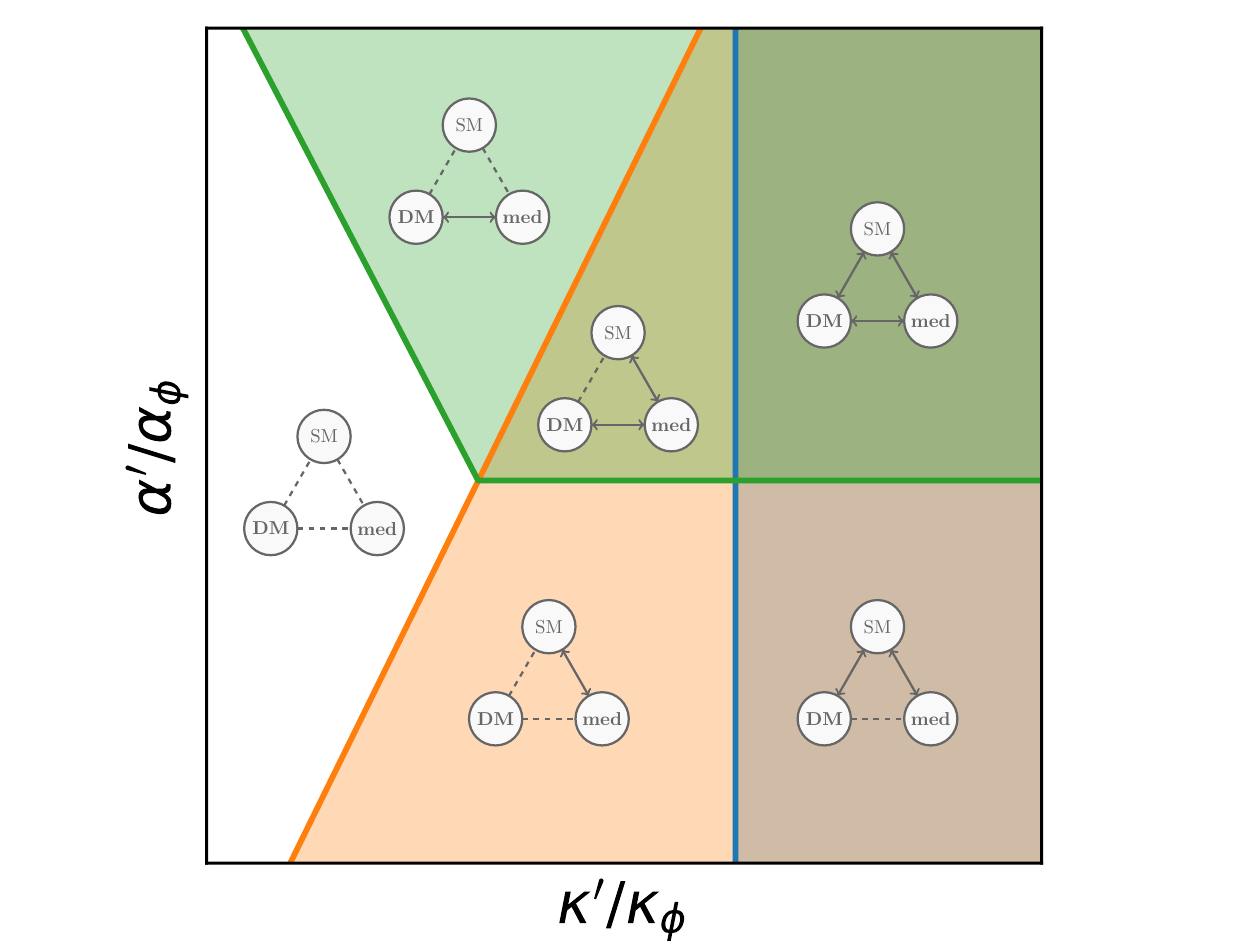}
\caption[Phase space of portal models subdivided by the three thermalisation lines]{DM-to-med coupling versus DM-to-SM coupling and subdivision made by the three thermalisation conditions. In blue: the SM and DM baths thermalise, in orange: the SM and mediator baths thermalise and in green: the DM and mediator baths thermalise.}
\label{fig:mesa_regimes}
\end{figure}
\end{center}

If we consider thermalisation or not, there are only eight possibilities with three sectors. Figure \ref{fig:mesa_regimes} shows how the three thermalisation lines generically divide the parameter space for a chosen set of DM and mediator masses. We do not show scale on the x- and y-axis as we aim to provide a qualitative understanding for now. Six of the eight possibilities are visible on the Figure \ref{fig:mesa_regimes}, but two are missing. The one which connects all sector but the mediator to the SM lies above the part of the parameter space we are presenting. Indeed, it lies in the non-perturbative region ($\alpha'/\alpha_{\phi}>1$) and will not be considered. The other possibility we are missing requires a thermalisation only between the DM and the SM baths. This is not possible as the value of $\kappa'/\kappa_{\phi}$ it requires to thermalise implies either a thermalisation between the SM and the mediator baths either a thermalisation between the DM and the mediator baths. Thus, we argue here that Figure \ref{fig:mesa_regimes} shows, in a qualitative way, all possible combination of thermalisations which could arise in a "three sectors-three connectors" structure. Note that since $\epsilon\propto 1/\sqrt{\alpha'}$ (resp. $\lambda_{\Phi H}\propto 1/\sqrt{\alpha_{\phi}}$), the smaller $\alpha'$ (resp. $\alpha_{\phi}$), the larger $\epsilon$ (resp. $\lambda_{\Phi H}$) for fixed values of $\kappa'$ (resp. $\kappa_{\phi}$). This explains why, in Figure \ref{fig:mesa_regimes}, we have that SM-to-med interactions thermalise more easily for small values of $\alpha'$ (resp. $\alpha_{\phi}$).
\\

\begin{center}
\begin{figure}[h!]
\centering
\includegraphics[scale=0.75]{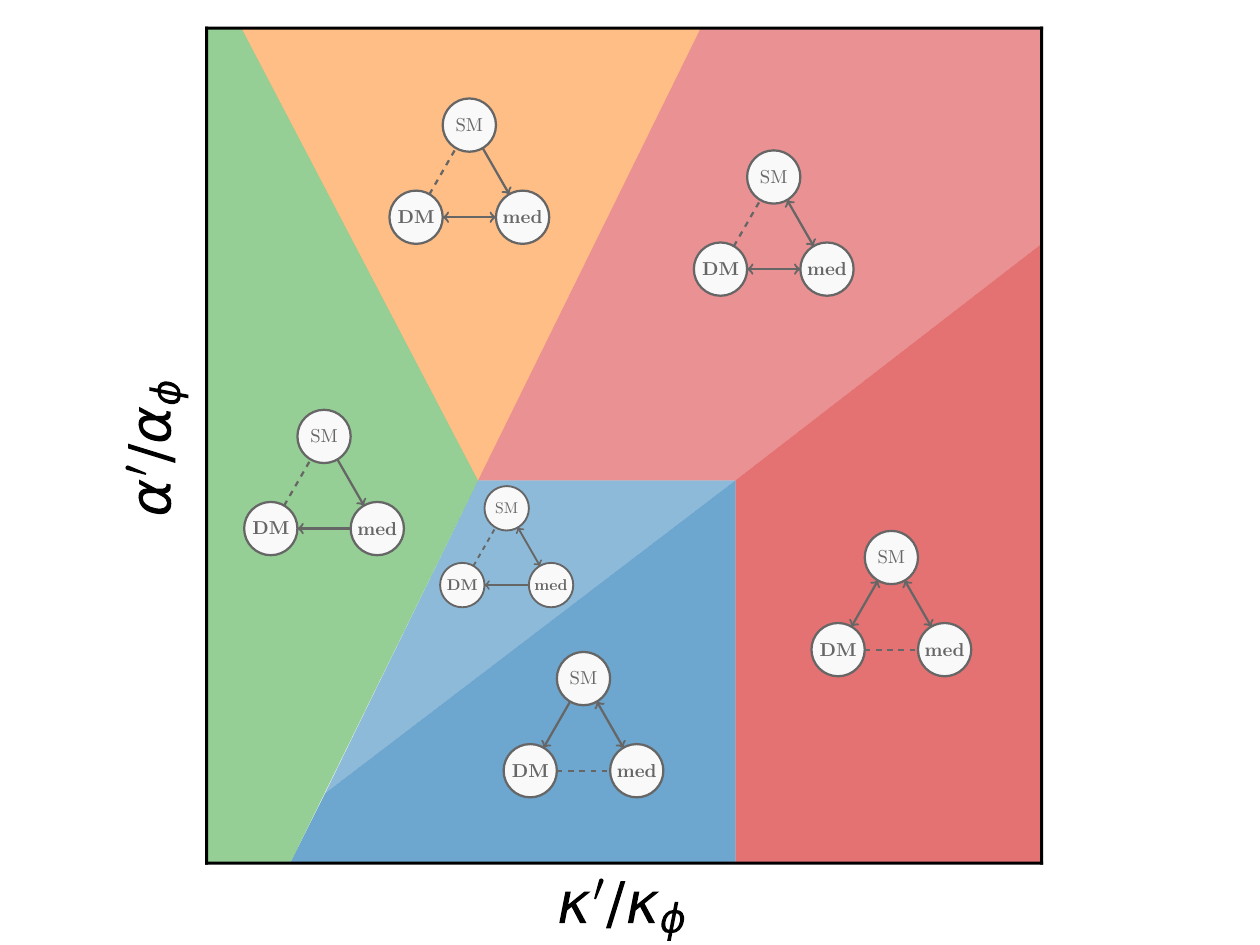}
\caption[Dominant processes in the phase space of portal models subdivided by the three thermalisation lines]{DM-to-med coupling versus DM-to-SM coupling and subdivision made by the three thermalisation conditions and considering the dominant connection to the DM bath only.}
\label{fig:mesa_dynamical}
\end{figure}
\end{center}

In order to know how the relic abundance is generated, it is not enough to know which interaction does thermalise. As mentioned already above, out-of-equilibrium processes can also play a crucial role. The later can slowly produce particles from a filled reservoir to a more empty reservoir. Thus in Figure \ref{fig:mesa_regimes}, one must also indicate these processes whenever they are relevant, pointing in which direction they are relevant. One would also leave the irrelevant out-of-equilibrium processes in dashed line. This is what we do in Figure \ref{fig:mesa_dynamical}. This implies that some regions of Figure \ref{fig:mesa_regimes} will be now divided in two. For instance, the orange only region in Figure \ref{fig:mesa_regimes} will be split into two regions depending on which of the two out-of-equilibrium channels (DM-to-SM or DM-to-med) will dominate. Indeed, when the DM is connected in the same way to both the SM and the mediator baths\footnote{i.e. if the lines connecting DM to the SM and DM to the mediator in Figure \ref{fig:mesa_regimes} are both dashed (no thermalisation) or both solid (thermalisation).}, we have to know the hierarchy of those interactions. Is the DM more connected to the SM or to the mediator? This can be answered by looking at the following condition,

\myeq{
&\gamma_{\rm DM\leftrightarrow med}=\gamma_{\rm DM\leftrightarrow SM}.\label{eq:g_SM-g_med}
}

\noindent We face the same issue in regimes where several processes are in thermal equilibrium. These regimes have also to be divided into two regions depending on which channel is dominant. Then, in these cases, these divisions in sub-regions are driven by,

\myeq{
&\gamma_{\rm DM\leftrightarrow med}^{{\rm eq}}=\gamma_{\rm DM\leftrightarrow SM}^{{\rm eq}},\label{eq:g_SM-g_med_EQ}
}

\noindent with $\gamma_{\rm DM\leftrightarrow med}^{{\rm eq}}\equiv \left\langle\sigma_{\rm DM\leftrightarrow med}v\right\rangle\left(n_{\rm med}^{eq}\right)^{2}$.
\\

Let us detail a little bit more the transition from Figure \ref{fig:mesa_regimes} to Figure \ref{fig:mesa_dynamical}. Indeed, we still have six different regions in the latter one but they are not exactly the same than in the first one. Indeed, let us look at the orange subdivision in Figure \ref{fig:mesa_regimes} where the SM and mediator baths are in thermal equilibrium but not with the DM bath. The condition given by Eq. \ref{eq:g_SM-g_med} divide this subsection in two in Figure \ref{fig:mesa_dynamical}: the dark blue regime where the DM-to-SM connection is stronger than the DM-to-med connection and the light blue regime where the DM-to-med connection is stronger than the DM-to-SM connection.
\\

The white region in Figure \ref{fig:mesa_regimes}, where there is no thermal contact between any of the three baths, is unchanged as well as the green only region where only the DM and mediator baths are in thermal contact (green and orange respectively in Figure \ref{fig:mesa_dynamical}). In the orange and green region in Figure \ref{fig:mesa_regimes} the mediator bath thermalise with the DM and SM ones. Now, the fully thermalised region (orange, green and blue in Figure \ref{fig:mesa_regimes}), because of the condition given in Eq. \ref{eq:g_SM-g_med}, is split in two and in one of these two subdivision, the DM stronger connects to the mediator bath than to the SM one. Thus, in this region, one can neglect the effect of the DM connection to the SM bath and one gets back to the same physical situation than in the orange and green region. Moreover, in the other part of the fully thermalised region, the DM is stronger connected to the mediator bath and one gets back to the same physical situation than in the orange and blue region where the DM and mediator baths are not thermally connected. This behaviour explains the rearrangement of these three subdivision in Figure \ref{fig:mesa_regimes} into two in Figure \ref{fig:mesa_dynamical}.
\\

Now that we have an overall idea of the picture, let us analyse all of the six regimes shown in Figure \ref{fig:mesa_dynamical}.

\subsection{Freeze-in: regimes Ia and Ib}\label{subsec:FI}
Let us first consider very small values for the DM-to-SM ($\kappa'/\kappa_{\phi}$) and DM-to-med ($\alpha'/\alpha_{\phi}$) couplings. In this case, the DM-to-SM and DM-to-med processes do not thermalise, but nevertheless the SM-to-med processes do thermalise because $\epsilon\sim\kappa'/\sqrt{\alpha'}$ (or $\lambda_{\Phi H}\sim\kappa_{\phi}/\sqrt{\alpha_{\phi}}$). Thus, the mediator is a part of the SM thermal bath. Then, either the DM bath is already there just after the inflation and the DM relic abundance has to be set by the initial condition. Or the DM reservoir was empty or negligible at the end of inflation. The DM could then only be slowly produced through out-of-equilibrium processes SM$\rightarrow$DM and med$\rightarrow$DM parametrised by $\kappa'/\kappa_{\phi}$ and $\alpha'/\alpha_{\phi}$ respectively. This is the so-called Freeze-In regime that we have already met in Subsection \ref{subsec:FI}. The Boltzmann equation for the DM yield (Eq. \ref{eq:YDM}) becomes:

\myeq{
\hbox{\underline{Regime I}}\,:\qquad xHs\frac{\diff Y_{\rm DM}}{\diff x} \simeq \gamma_{\rm SM \leftrightarrow DM}^{\rm eq}(x) + \gamma_{\rm med \leftrightarrow DM}^{\rm eq}(x),
}

\noindent where we simplified the right hand side as the DM yield will always be much smaller than equilibrium yields. Now, depending on which process dominates (i.e. on which side of the line defined by Eq. \ref{eq:g_SM-g_med} we sit), we get two different dynamical production mechanisms for DM. The first one is:

\myeq{
\hbox{\underline{Regime Ia}}\,:\qquad xHs\frac{\diff Y_{\rm DM}}{\diff x} \simeq \gamma_{\rm SM \leftrightarrow DM}^{\rm eq}(x).
}

\begin{center}
\begin{figure}[h!]
\centering
\includegraphics[scale=0.75]{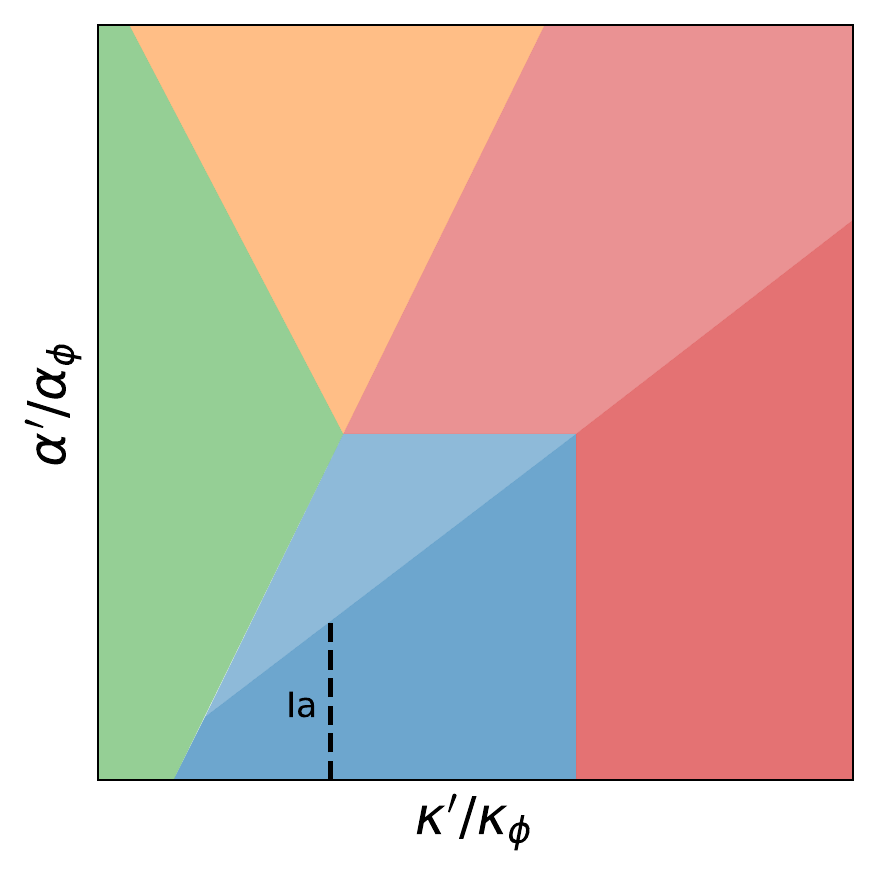}
\includegraphics[scale=0.75]{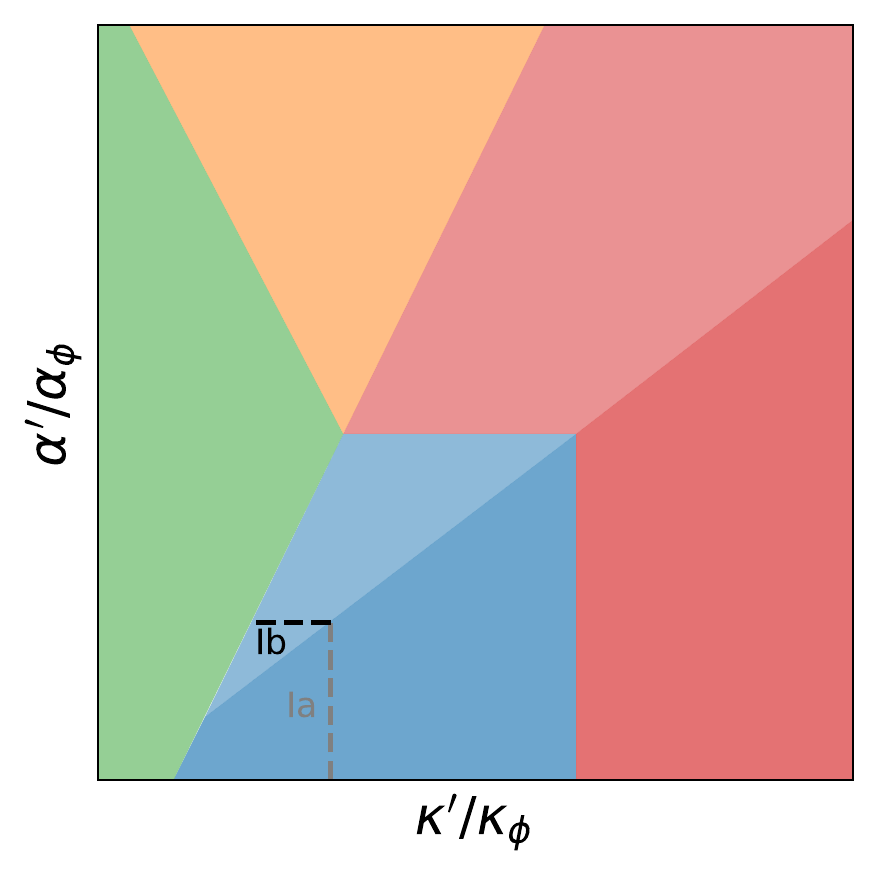}
\caption[Qualitative representation of the parameter space of a three sectors three connectors scenario for the regime Ia and Ib]{Qualitative representation of the parameter space of a three sectors three connectors scenario for the regime Ia (left) and Ib (right).}
\label{fig:mesa_I}
\end{figure}
\end{center}

\noindent In the first case (Ia), DM production is dominated by slow SM$\rightarrow$DM out-of-equilibrium processes (in dark blue in Figure \ref{fig:mesa_dynamical}). These processes depend only on the connector $\kappa'/\kappa_{\phi}$. As a consequence, the DM relic abundance depends only on $\kappa'/\kappa_{\phi}$ and it gives a vertical line in the phase diagram, see left panel of Figure \ref{fig:mesa_I}. 
\\

\begin{figure}
\centering
\includegraphics[scale=0.63]{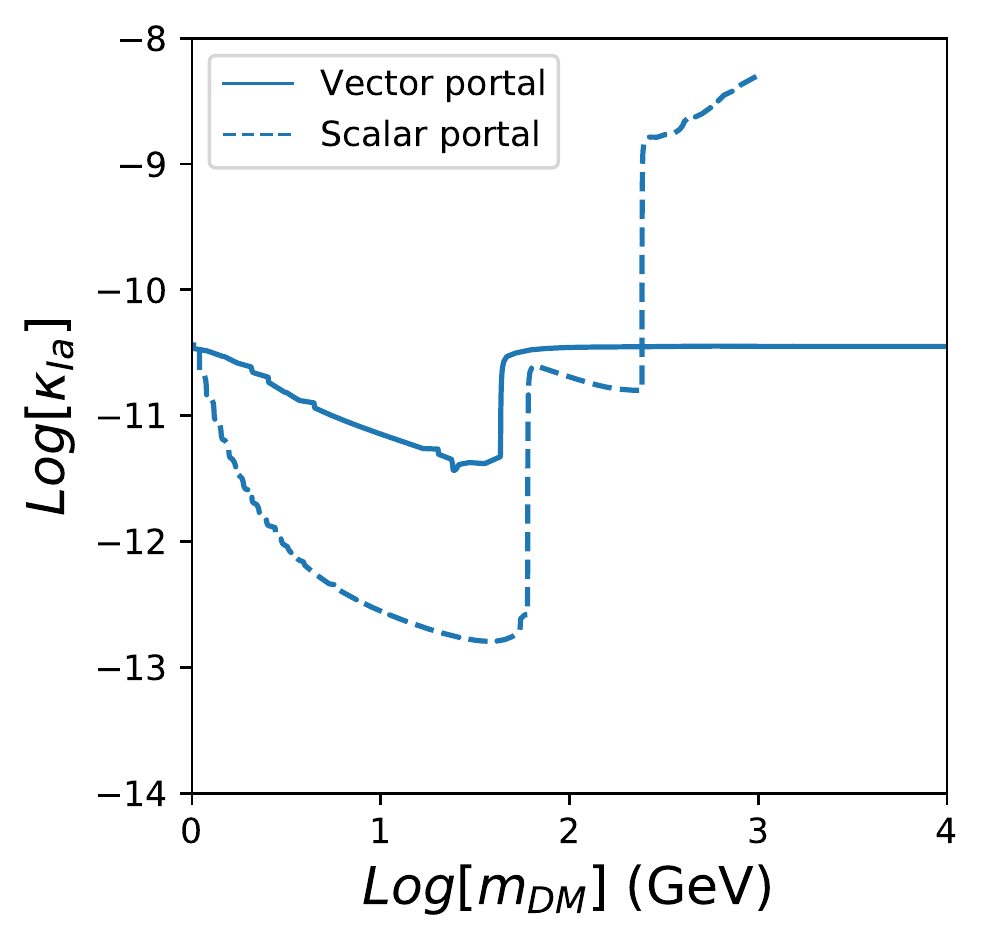}
\includegraphics[scale=0.63]{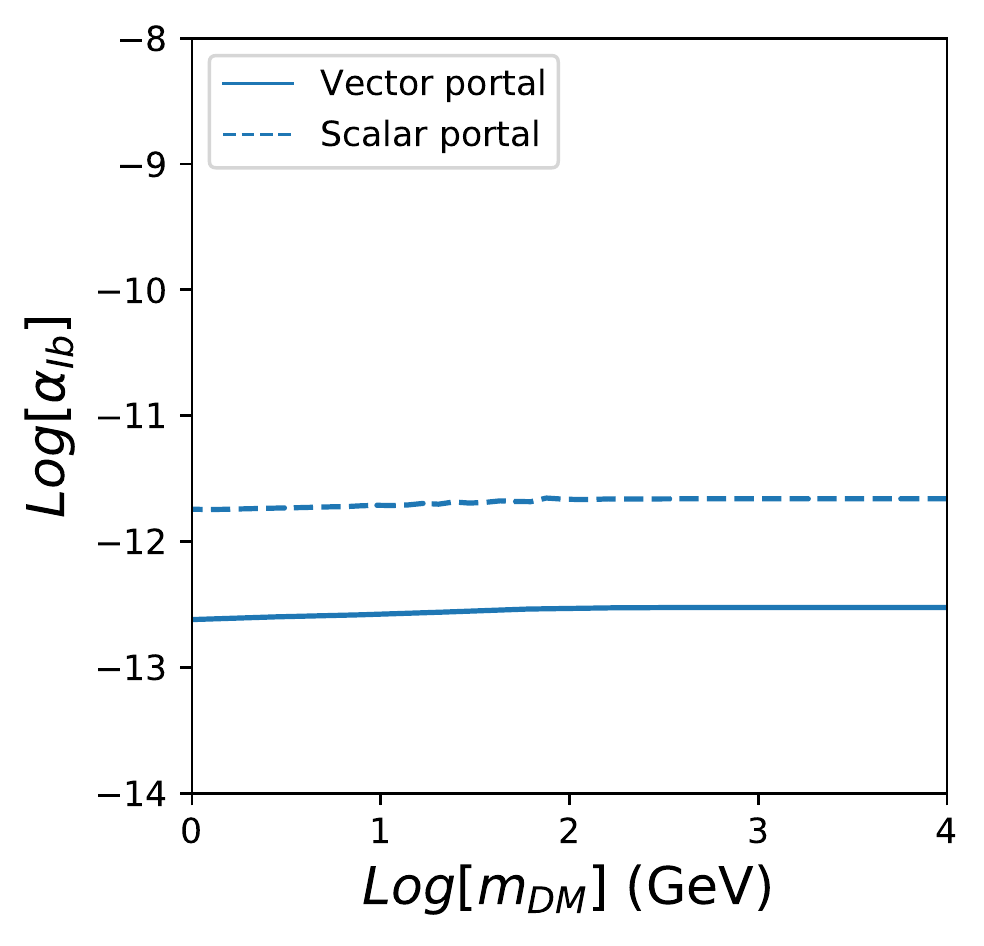}
\caption[Couplings as a function of the DM mass fixed by the relic density in regime Ia and Ib]{Left: Values of $\kappa'/\kappa_{\phi}$ needed to account for the observed relic density, as a function of the DM mass $m_{DM}$, for the standard freeze-in (regime Ia) for the vector portal (solid) and the scalar portal (dashed) models. Right: Values of $\alpha'/\alpha_{\phi}$ needed to account for the observed relic density, as a function of the DM mass $m_{DM}$, for the freeze-in from mediators (regime Ib) for the vector portal (solid) and the scalar portal (dashed) models.}
\label{fig:Freeze_In}
\end{figure}

Note that the fact that there is here thermalisation between the SM and the mediator baths does not change anything to the dynamics of this freeze-in regime. Indeed, since the DM particles are dominantly produced by the SM particles it does not matter if the mediator bath thermalises with the SM. The only effect this thermalisation can have is that the thermalised mediator particles will modify the number of relativistic degrees of freedom and thus the Hubble expansion rate. The modification of the Hubble rate is of order\footnote{We remind here that $g_{\phi} = 0$ and $g_{\gamma '} = 3$ (in the massive vector case).} $g_{\rm med}/g^{\rm eff}_\ast \sim 10^{-2}$. As this effect is of percent level, we neglect it. In the left panel of Fig.~\ref{fig:Freeze_In} we show as a function of $m_{\rm DM}$ the value of the SM-to-DM connector required to reach the observed relic abundance in the Ia regime, which we note $\kappa_{\rm Ia}$\footnote{Note that we have shown in Section \ref{sec:CST-DD} and Figure \ref{fig:DD_Constraints} that direct detection can probe values of $\kappa$ down to $3\times 10^{-11}$ and $10^{-9}$ for the vector and scalar portal respectively. We see now that it implies that direct detection experiments are able to probe FI regime, see \cite{Hambye:2018dpi} for more details.}. For the vector portal model, as the thermalised dark photons role is negligible in the production of the DM particles, this solid blue line is essentially the same as in the massless dark photon case \cite{Chu:2011be}. The step in this figure is due to the closing of the production of DM through the $Z$ boson channel. Indeed, if $m_{\rm DM}>m_{Z}$, this production channel is no longer available. The production of DM from SM is then less efficient and a bigger value for the coupling is required. For the scalar portal model, the first step is due to the closing of the production of DM through the $H$ boson channel. The second step is due to the closing of channels through the SM scalar VEV ($v_{H}$). The fact that the line stops at $m_{\rm DM}=1000$ GeV in the scalar case is due to the choice of the new scalar VEV ($v_{\Phi}=1000$ GeV) in this example. Indeed, when $m_{\rm DM}>v_{\Phi}$, the freeze-in process stops to be efficient before the symmetry breaking because there is no scalar mixing in this case. It is then no longer possible to produce DM from SM and the freeze-in regime does not exist for $m_{\rm DM}>v_{\Phi}$. See section \ref{sec:spec_scalar} for a more in depth analysis.
\\

In both models, the dependence of $\kappa_{\rm Ia}$ on the DM mass can be complicated because the dominant production channels depend also on $m_{\rm DM}$. But the number of DM particles produced through $\rm SM_i \,SM_i\rightarrow DM\,DM$ is related in a simple way to equilibrium quantities evaluated at a temperature $T\ll {\rm max}\{m_i,m_{\rm DM}\}$, 

\myeq{
&Y_{\rm DM}^{Ia}(x)=\sum_i c_i\,\frac{({n_i^{\rm eq}})^2\,\langle \sigma_{\rm SM_i\rightarrow DM} v\rangle}{Hs}\Big|_{T={\rm max}\{T,m_i,m_{\rm DM}\}},
\label{nDMFIa}
}

\noindent where the $c_i$ are order unity coefficients \cite{Chu:2011be}. Eq. \ref{nDMFIa} allows us to understand easily why $\Omega_{\rm DM}=26\%$ requires a small value of the DM-to-SM coupling. Indeed, as already mentioned, one needs a suppressed amount of DM with respect to the amount it has if it thermalises and is relativistic. Thus one requires a small value for the DM yield and then a small production cross section $\langle \sigma_{\rm SM_i\rightarrow DM} v\rangle\sim\kappa'^{2}$ (or $\kappa_{\phi}^{2}$).
\\

Going up along the vertical line of the Ia regime depicted in Fig.~\ref{fig:mesa_I}, the value of $\alpha'/\alpha_{\phi}$ increases while that of $\epsilon/\lambda_{\Phi H}$ decreases. Thus, at some point, the SM and the mediator baths stop to thermalise, $\epsilon/\lambda_{\Phi H}<\epsilon_{\rm th}/\lambda_{\Phi H_{\rm th}}$. However, before this could happen and while the mediator particles are still in thermal equilibrium with the SM bath, $\alpha'/\alpha_{\phi}$ becomes large enough for the med$\rightarrow$DM processes to overcome the SM$\rightarrow$DM production and to dominate the DM production. Clearly, in this case a smaller value of the DM-to-SM connector $\kappa'/\kappa_{\phi}$ is required to avoid over-production of DM particles and moreover, the DM relic abundance depends only on $\alpha'/\alpha_{\phi}$. This gives rise to the horizontal line depicted in Fig.~\ref{fig:mesa_I}\footnote{This regime has been briefly discussed in \cite{Kane:2015qea}, where it is dubbed "inverse annihilation", and in \cite{Klasen:2013ypa}, in a model with a scalar singlet that mixes with the Higgs. It is also considered in a scenario with a $Z'$ based on a $B-L$ gauge symmetry, which appeared simultaneously with our work \cite{Heeba:2019jho}.}.

\myeq{
\hbox{\underline{Regime Ib}}\,:\qquad xHs\frac{\diff Y_{\rm DM}}{\diff x} \simeq \gamma_{\rm med \leftrightarrow DM}^{\rm eq}(x).
}

\noindent In other words for this second case (Ib), the DM production is dominated by the med$\rightarrow$DM process (in light blue in Figure \ref{fig:mesa_dynamical}), and the process depends only on the connector $\alpha'/\alpha_{\phi}$. As said above, the DM relic abundance depends only on $\alpha'/\alpha_{\phi}$ and it gives a horizontal line in the phase diagram, see right panel of Figure \ref{fig:mesa_I}. In the right panel of Fig.~\ref{fig:Freeze_In} we show as a function of $m_{\rm DM}$ the value of the med-to-DM connector required to reach the observed relic abundance in the Ib regime, which we note $\alpha_{\rm Ib}$. This time, there is no particular feature since there is only one production channel which is an annihilation and that it occurs in the t-channel (i.e. no \textit{on-shell} resonance).
\\

As in the Ia regime, the DM yield can be simply written in the Ib regime,

\myeq{
&Y_{\rm DM}^{Ib}(x) =c_{\rm med}\,\frac{({n_{\rm med}^{\rm eq}})^2(x)\,\langle \sigma_{\rm med \rightarrow DM} v\rangle}{Hs}\Big|_{T = Max[T,m_{\rm DM}]},
\label{nDMFIb}
}

\noindent with $c_{\rm med} = {\cal O}(1)$.
\\

Note that this production regime is not new as a dynamical regime (it is a simple freeze-in) but is new in the sense that source particles here are not SM particles but hidden sector particles.

\subsection{Sequential freeze-in: regime II}\label{subsec:SFI}
As we move toward smaller values of the DM-to-SM coupling $\kappa'/\kappa_{\phi}$, the value of the med-to-SM coupling $\epsilon/\lambda_{\Phi H}$ decreases. This will imply that, at some point, the value of this last connector will not be high enough to allow thermalisation between the SM and the mediator baths anymore: $\epsilon/\lambda_{\Phi H}<\epsilon_{\rm th}/\lambda_{\Phi H_{\rm th}}$. At this moment, the mediator yield is no longer given by the equilibrium quantity and one has to compute the actual mediator yield as a function of time. In this new regime, shown in green in Figure \ref{fig:mesa_dynamical}, none of the three sector thermalises with an other one.
\\

One could think that, like in the previous regime, this regime could be split in two distinct phases depending on to which bath the DM one is more coupled to (see Eq. \ref{eq:g_SM-g_med}). But, if the DM bath is more connected to the SM bath, as the DM-to-SM coupling is already much smaller than the one required in the standard freeze-in phase, $\kappa'/\kappa_{\phi}\ll\kappa_{Ia}$, the DM relic abundance could not be produced directly from SM. Thus, there is, in practice, only one possibility for the phase "II". It turns out that this slow out-of-equilibrium production of mediator particles from SM directly followed by the slow out-of-equilibrium production of DM particles from these unthermalised mediator particles can produce the observed DM relic abundance.
\\

This chain of successive freeze-in processes has been dubbed "sequential freeze-in" and constitue a new dynamical way of accounting for the DM relic density. As in standard freeze-in, the reverse processes can be neglected as they have a minor impact on the final DM yield. The Boltzman equations take then the following form,

\myeq{
&\hbox{\underline{Regime II}}\,:\qquad 
\begin{cases}
xHs\frac{\diff Y_{\rm med}}{\diff x}\simeq \gamma ^{\rm eq}_{\rm SM\leftrightarrow med},\\
xHs\frac{\diff Y_{\rm DM}}{\diff x}\simeq \gamma_{\rm med\leftrightarrow DM}.
\end{cases}
\label{BoltzEqII}
}

\begin{center}
\begin{figure}[h!]
\centering
\includegraphics[scale=0.75]{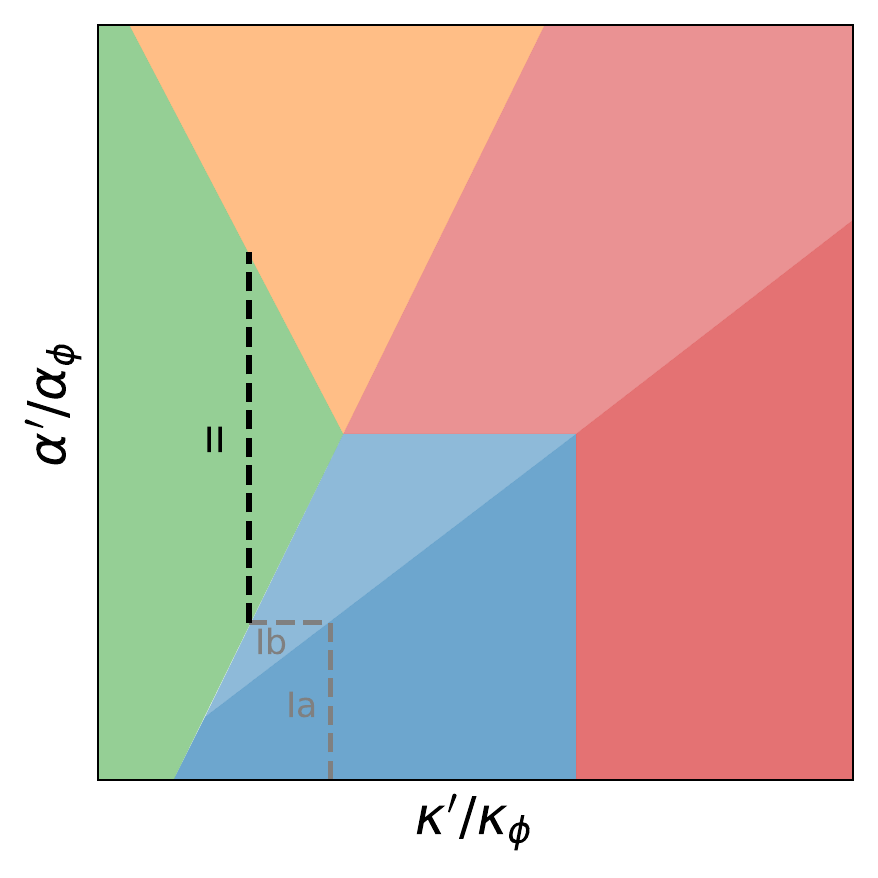}
\caption[Qualitative representation of the parameter space of a three sectors three connectors scenario for the regime II]{Qualitative representation of the parameter space of a three sectors three connectors scenario for the regime II.}
\label{fig:mesa_II}
\end{figure}
\end{center}

\noindent There is a suppression factor on this reaction density with respect to the equilibrium density, as the DM production rate will be proportional to $\propto(n_{\rm med}/n_{\rm med}^{\rm eq})^2$ with $n_{\rm med}$ determined by the first Boltzmann equation. The number of mediator produced by freeze-in from SM particles is proportional to $\epsilon^{2}$ and $\lambda_{\Phi H}^{2}$ in the vector and scalar portal models respectively. Putting it in the DM particles production rate which is proportional to $n_{\rm med}^2 \,\langle \sigma_{\rm med\rightarrow DM}\,v\rangle$, in the vector portal model we get $n_{\rm DM}\sim \epsilon^{4}\alpha'^{2}\sim\kappa'^{4}$ and, in the scalar portal model, $n_{\rm DM}\sim\lambda_{\Phi H}^{4}\alpha_{\phi}^{2}\sim\kappa_{\phi}^{4}$. We get then that, in the sequential freeze-in regime, the DM relic density depends on the magnitude of the DM-to-SM connector only. This feature is again translated into a vertical line in the phase diagram, see Figure \ref{fig:mesa_II}.
\\

\begin{figure}
\centering
\includegraphics[scale=0.63]{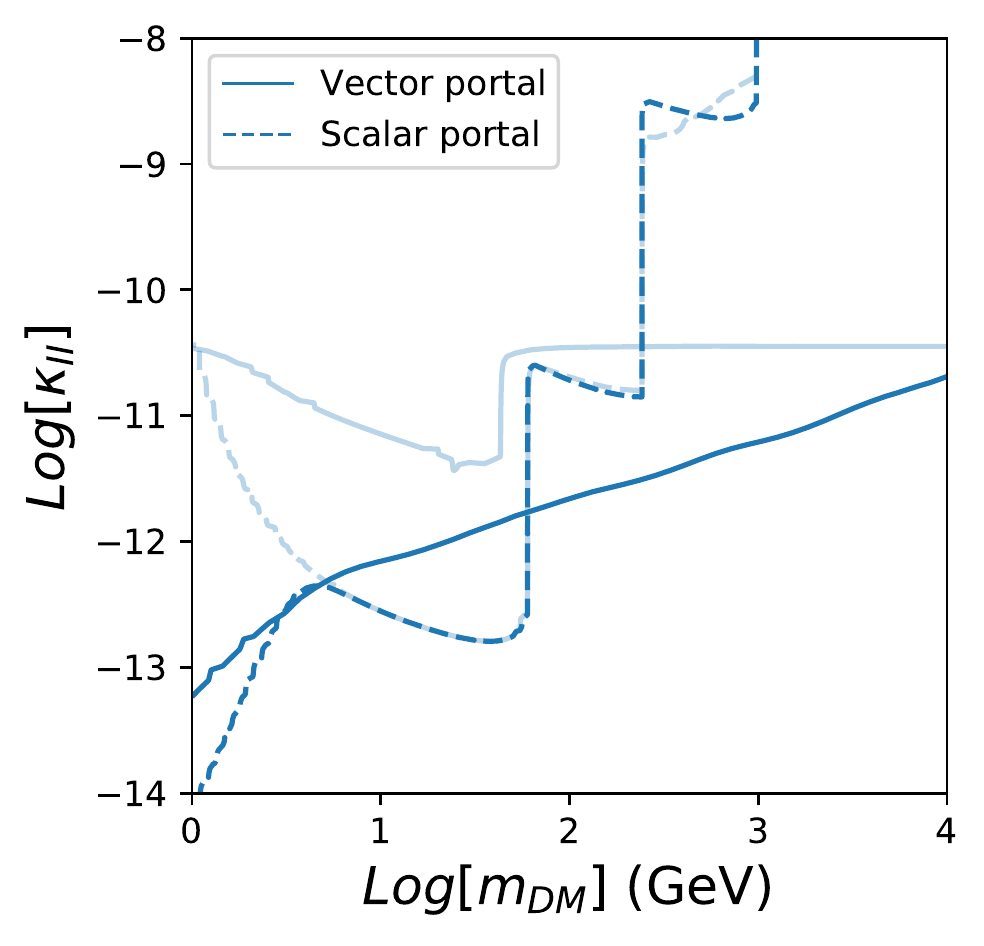}
\caption[Couplings as a function of the DM mass fixed by the relic density in regime II]{Values of $\kappa'/\kappa_{\phi}$ needed to account for the observed relic density, as a function of the DM mass $m_{DM}$, for the sequential freeze-in (regime II) for the vector portal (solid) and the scalar portal (dashed) models. Solutions obtained in the regime Ia are also shown in light blue.}
\label{fig:Sequential_Freeze_In}
\end{figure}

These results show that along the sequential freeze-in it is possible to have a DM-to-SM connector orders of magnitude below the one required in the standard freeze-in regime. Figure \ref{fig:Sequential_Freeze_In} gives the required value of the DM-to-SM connector $\kappa'/\kappa_{\phi}$, dept $\kappa_{II}$, to produce the observed DM relic abundance in the sequential freeze-in regime (II). We also show results in the standard freeze-in regime (Ia) for comparison. One can see that for large values of $m_{\rm DM}/m_{\rm med}$, $\kappa_{II}$ tends to $\kappa_{Ia}$. The merging of the sequential freeze-in regime (II) with the very well known standard freeze-in regime (Ia) is easy to understand. Heavier DM particles stop to be produced at higher temperature and thus earlier in the Universe history. The SM had then less time to produce mediator particles prior the temperature $T$ reaches the DM mass scale. With less mediator particles in the bath, one need to increase the sequential freeze-in efficiency. As we have seen, this efficiency is ruled only by the DM-to-SM connector $\kappa'/\kappa_{\phi}$. Then, increasing the mass ratio $m_{\rm DM}/m_{\rm med}$ requires to increase $\kappa_{II}$. Increasing $\kappa_{II}$ will imply that, at some point, the production of DM directly from SM particles will be comparable to the sequential freeze-in production such that the two regimes will merge. This feature can also be understood looking at the $m_{\rm DM}/m_{\rm med}$ mass ratio. As $m_{\rm DM}/m_{\rm med}\rightarrow\infty$, one gets back to the massless mediator case where the mediator cannot be directly produced from the SM. The sequential FI regime does not exist or is much less efficient in the case of the scalar portal model. Indeed, one sees that the sequential freeze-in regime exists for $m_{\rm DM}>v_{\Phi}$ while it was not the case for the standard freeze-in regime. This is due to the fact that mediator particles can be produced without scalar mixing, directly from the quartic coupling. Thus, enough mediator particles could have been produce at high temperature for the DM to freeze-in from the mediator bath, even before the hidden and visible sector symmetry breaking temperature. The difference at high DM masses between the curves for the scalar model in the regime Ia (light dashed blue) and the regime II (dark dashed blue) stems from the fact that sequential freeze-in starts to be relevant for DM particles lighter than the hidden sector symmetry breaking and this was not taken into account in the pure freeze-in regime (regime Ia). That is to say that the pure freeze-in curve in Figure \ref{fig:Sequential_Freeze_In} (dashed light blue) is pure freeze-in and does not take sequential freeze-in contribution into account.

%%%%%%%%%%%%%%%%%%%%%%%%%%%%%%%%%%%%%%%% 
\subsection{Reannihilation: regimes IIIa and IIIb}\label{subsec:rean}
In the regime II, if we increase the DM-to-med connector $\alpha'/\alpha_{\phi}$, the DM and mediator baths will start to thermalise at some point. Here, one has two different thermal baths because the HS does not thermalise with the SM. That is to say that the HS bath has no reason to have the same temperature as the SM one, we have $T_{\rm HS}\equiv T'\neq T_{\rm SM}\equiv T$. In this regime, clearly since the DM thermalises with the mediator, its abundance will be Boltzmann suppressed when the temperature goes below its mass ($T'<m_{\rm DM}$). But then it turns out that it will not undergo a standard freeze-out mechanism and this for two reasons: first this happens in a sector with $T'\neq T$. Second, at the time the DM abundance gets Boltzmann suppressed, HS particles are still slowly produced out-of-equilibrium from the SM. One faces thus a DM production regime called "reannihilation" in which this small source of dark sector particles forced the DM to freeze-out somewhat later than during standard freeze-out (actually when $T\sim m_{\rm DM}$ rather than when $T'\sim m_{\rm DM}/(\text{a few}$). This regime takes place already in the vector portal model for a massless mediator (see \cite{Chu:2011be}).
\\

As in the freeze-in regime (I), the reannihilation regime (III) is also made of two sub-regimes. Indeed, the source of dark sector particles could produce either DM particles (IIIa) either mediator particles (IIIb), depending on which connection is the strongest one. In practice, in the two models we are interested in, moving towards larger values of $\alpha'/\alpha_{\phi}$, the system enters into the regime IIIb and not the regime IIIa. This is due to the value of the DM-to-SM connector which is very small in this part of the parameter space. Indeed, the value of $\kappa'/\kappa_{\phi}$ is smaller than the one required in the regime Ia. That is to say that DM cannot be efficiently produced from SM in this region of the parameter space. Thus, we first discuss the regime IIIb and discuss next the regime IIIa.
\\

As said above, in the regime IIIb (shown in orange in Figure \ref{fig:mesa_dynamical}) one has to consider quantities in the hidden sector at a different temperature $x'\neq x$ and a source term from SM for the mediator. The Boltzmann equations can thus be written as,

\myeq{
&\hbox{\underline{Regime IIIb}}\,:\quad 
\begin{cases}
xHs\frac{\diff Y_{\rm med}}{\diff x} \simeq \gamma^{\rm eq}_{\rm SM \leftrightarrow med} (x)-\gamma^{\rm eq}_{\rm med\leftrightarrow DM}  (x')   \left[1-\left(\frac{Y_{\rm DM}}{Y_{\rm DM}^{\rm eq}\left(x'\right)}\right)^2\right],\\
xHs\frac{\diff Y_{\rm DM}}{\diff x} \simeq \gamma^{\rm eq}_{\rm med\leftrightarrow DM}  (x') \left[1-\left(\frac{Y_{\rm DM}}{Y_{\rm DM}^{\rm eq}\left(x'\right)}\right)^2\right].
\end{cases}
\label{BoltzEqIIIb}
}

\noindent Clearly, these equations can be solved only if one knows $T'$ as a function of $T$. Thus, one also needs to provide the dark sector temperature as a function of the temperature of the visible sector. This is done by integrating the Boltzmann equation for the SM-to-hidden sector energy transfer (see \cite{Chu:2011be}),

\myeq{
xH\frac{d\rho '}{\diff x}+4H(\rho '+p') \simeq ({n_{\rm SM}^{\rm eq}}(x))^2 \langle \sigma_{\rm SM\rightarrow med}v\Delta E \rangle.
}

\noindent One can plug into this equation, the hidden sector equation of state $p'=p'(\rho')$ which is given by 

\myeq{
&p' = \frac{1}{3}\left(\rho'-m_{\rm DM}Y_{\rm DM}s\right),\\
&\rho_{\rm DM}(x') = \rho^{\rm eq}_{\rm DM}(x)\frac{Y_{\rm DM}}{Y^{\rm eq}_{\rm DM}(x)}.
}

\noindent The final DM relic density obtained in the regime IIIb approximately scale as

\begin{center}
\begin{figure}[h!]
\centering
\includegraphics[scale=0.75]{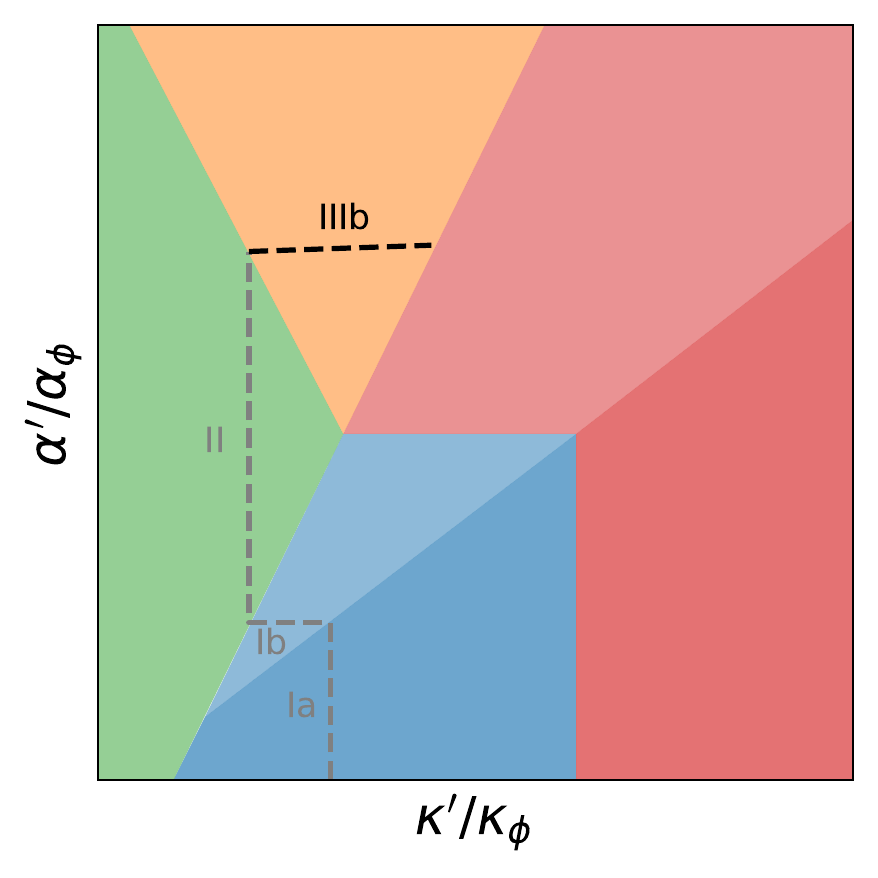}
\caption[Qualitative representation of the parameter space of a three sectors three connectors scenario for the regime IIIb]{Qualitative representation of the parameter space of a three sectors three connectors scenario for the regime IIIb.}
\label{fig:mesa_III}
\end{figure}
\end{center}

\myeq{
\Omega_{\rm DM} \propto \frac{\log\left(\sqrt{\left\langle\sigma_{\rm SM \rightarrow med}v\right\rangle\left\langle\sigma_{\rm DM \rightarrow med}v\right\rangle}\right)}{\left\langle\sigma_{\rm med \rightarrow DM}v\right\rangle},
}

\noindent which goes like $\propto\log\left(\alpha'\epsilon\right)/\alpha'^{2}$ in the vector portal model or like $\propto\log\left(\alpha_{\phi}\lambda_{\Phi H}\right)/\alpha_{\phi}^{2}$ in the scalar portal model. This explains why this regime leads to a line which is close to be horizontal in the phase diagram, see Figure \ref{fig:mesa_III}.
\\

%Figure \ref{fig:reannihilation} shows values of the DM-to-mediator connector needed to account the observed DM relic density in the reannihilation regime where the hidden sector is populated from SM-to-mediator process (regime IIIb).
%
%\begin{figure}
%\centering
%\includegraphics[scale=0.63]{Ia.pdf}
%\caption{Values of $\kappa'/\kappa_{\phi}$ needed to account for the observed relic density, as a function of the DM mass $m_{DM}$, in the reannihilation from mediator regime (regime III) for the vector portal (solid) and the scalar portal (dashed) models.}
%\label{fig:reannihilation}
%\end{figure}

The IIIb reannihilation regime is new\footnote{It has not been discussed in \cite{Chu:2011be} as it does not exist in the massless dark photon case and as authors did not considered the scalar portal we are considering. Indeed, in this work, authors considered a scalar DM directly mixing with the SM scalar.} as it does not occur in the massless mediator case. Alternatively, in the massless mediator case, the system enters in the regime IIIa instead of IIIb. This stems from the fact that the massless mediator does not directly couple to SM particles such that the source of dark sector particles goes through SM$\rightarrow$DM slow out-of-equilibrium processes. In this regime IIIa, the hidden sector is then populated from SM-to-DM processes (see \cite{Chu:2011be}) and the Boltzmann equations are given by,

\myeq{
&\hbox{\underline{Regime IIIa}}\,:\quad 
\begin{cases}
xHs\frac{\diff Y_{\rm med}}{\diff x} \simeq -\gamma^{\rm eq}_{\rm med\leftrightarrow DM}  (x')   \left[1-\left(\frac{Y_{\rm DM}}{Y_{\rm DM}^{\rm eq}\left(x'\right)}\right)^2\right],\\
xHs\frac{\diff Y_{\rm DM}}{\diff x} \simeq \gamma^{\rm eq}_{\rm SM \leftrightarrow DM} (x) +\gamma^{\rm eq}_{\rm med\leftrightarrow DM}  (x') \left[1-\left(\frac{Y_{\rm DM}}{Y_{\rm DM}^{\rm eq}\left(x'\right)}\right)^2\right].
\end{cases}
\label{BoltzEqIIIa}
}

\noindent As in the regime IIIb, one has also to compute the transferred energy from one sector to the other. This time, this transfer is dominated by the SM-to-DM process,

\myeq{
xH\frac{d\rho '}{\diff x}+4H(\rho '+p') \simeq ({n_{\rm SM}^{\rm eq}}(x))^2 \langle \sigma_{\rm SM\rightarrow DM}v\Delta E \rangle.
}

\noindent In the massive mediator case, the slow production of the mediator from SM is always more efficient than the slow production of DM from SM in this part of the parameter space. Thus, this regime can only show up in the massless mediator case where the production of mediator from SM is not possible. Thus, we will not discuss the regime IIIa further, see \cite{Chu:2011be} for a detailed analysis. We will just mention that in this regime, as said in \cite{Chu:2011be}, the final DM abundance, which is set at $T\sim m_{\rm DM}$ rather than at $T'\sim m_{\rm DM}/(\text{a few})$, scales as $\sim\log\left(\alpha'\kappa'\right)/\alpha'^{2}$ in the vector portal model and we found a similar behaviour in the scalar portal model $\sim\log\left(\alpha_{\phi}\kappa_{\phi}\right)/\alpha_{\phi}^{2}$.

\subsection{Secluded freeze-out: regimes IVa and IVb}\label{subsec:secluded}
As we have seen in the previous subsection, the two reannihilation regimes (IIIa and IIIb) occur when a slow production of hidden sector particles from SM particles is still active when the DM freezes-out, that is to say when $T'$ becomes smaller than $m_{\rm DM}$. If such a source term  becomes negligible before $T'\sim m_{\rm DM}$, the system does not enter in one of the two reannihilation regimes. Here too, once the source term from the SM has created enough DM or mediator particles, these particles thermalise with each other forming a hidden thermal bath with temperature $T'$. But, since the same term from SM has stopped to be active at $T'\gtrsim m_{\rm DM}$, there is no period of reannihilation. Instead, the DM particles undergo a simple ``secluded freeze-out''. Here we refer to standard text-books freeze-out, except that it takes place in a hidden sector, characterised by a temperature $T'$ which differs from that of the visible sector, $T$ (see Ref.~\cite{Chu:2011be} for details).
\\

In practice, in the mass ranges we are considering in this chapter, secluded freeze-out does not occur in the instances depicted in Figures \ref{fig::DogDiagram1}. This stems from the fact that, for the considered mass ranges, the SM-to-DM source term is always still active at $T'\sim m_{\rm DM}$. Indeed, since we are always considering DM candidates heavier than the electron, the slow out-of-equilibrium production of DM particles from electron annihilation is always occurring at $T'\sim m_{\rm DM}$\footnote{Other production channels are still active if the DM is heavier than other SM leptons.}. Nevertheless, there are of course values of the masses for which the secluded freeze-out regime clearly dominates. This is the case for example if $m_{\rm med}\ll m_{\rm DM}$ and if the hidden sector was produced by some processes at the end of inflation. In fact, a general way for the secluded freeze-out to dominate would be to have an IR\footnote{Infrared.} mass scale which would cut-off the hidden sector particles source term at a higher temperature, $T'> m_{DM}$. An other example, would be if the mass hierarchy is as following: $m_{\rm med}< m_{\rm DM}<m_{e}$, the SM-to-DM and SM-to-mediator potential source terms would be cut when the SM temperature reach the electron mass, $T\sim m_{e}$.
\\

As in the previous dynamical ways, one can in principle distinguish two regimes, depending on whether the dominant dark sector particles source term produces DM particles or mediator particles, corresponding to IVa and IVb regimes respectively. The Boltzmann equations for the DM yield in the secluded freeze-out regimes are similar than the one in the reannihilation regimes. Here too, once the source term has stopped to be active, the Boltzmann equations for the DM yield are the same in both IVa and IVb regimes:

\myeq{
&\hbox{\underline{Regimes IVa \& IVb}}\,:\quad 
\begin{cases}
xHs\frac{\diff Y_{\rm med}}{\diff x} \simeq -\gamma^{\rm eq}_{\rm med\leftrightarrow DM}  (x')   \left[1-\left(\frac{Y_{\rm DM}}{Y_{\rm DM}^{\rm eq}\left(x'\right)}\right)^2\right],\\
xHs\frac{\diff Y_{\rm DM}}{\diff x} \simeq \gamma^{\rm eq}_{\rm med\leftrightarrow DM}  (x') \left[1-\left(\frac{Y_{\rm DM}}{Y_{\rm DM}^{\rm eq}\left(x'\right)}\right)^2\right].
\end{cases}
\label{BoltzEqIVa}
}

\noindent In regimes regimes IVa and IVb, one can neglect the production terms at time DM freezes, see eqs \ref{BoltzEqIVa}. In all cases, the relic density is essentially determined by the value of the DM-to-med coupling ($\alpha'$ or $\alpha_{\phi}$), which would lead to a horizontal line in the phase diagram (if the choice of masses was allowing these regimes), see \cite{Chu:2011be}.

\subsection{Freeze-out: regimes Va and Vb}\label{subsec:FO}
Finally, from the reannihilation or secluded freeze-out regimes, if we keep increasing the DM-to-SM connector $\kappa'/\kappa_{\phi}$ (and consequently the SM-to-med connector $\epsilon/\lambda_{\Phi H}$), at some point all particles will thermalise and will form a unique thermal bath, characterised by a unique temperature $T$. Thus, DM particles will have no other choice to undergo a standard freeze-out. This freeze-out regimes will happen when, on top of the $\alpha'/\alpha_{\phi}$ driven processes which were already in thermal equilibrium in the previous regimes, the $\kappa'/\kappa_{\phi}$ driven processes and/or the $\epsilon/\lambda_{\Phi H}$ driven processes thermalise. Thus, such a transition into the freeze-out regime takes place when either $\kappa'/\kappa_{\phi}$ becomes larger than $\kappa_{\rm th}$ or $\epsilon/\lambda_{\Phi H}$ becomes larger than $\epsilon_{\rm th}/\lambda_{\Phi H_{\rm th}}$, see Figure \ref{fig:th_lines}.
\\

Once again, the freeze-out regime can be dominated either by the $\rm med \leftrightarrow DM$ annihilation process (regime Vb, in light red in Figure \ref{fig:mesa_dynamical}) with the following Boltzmann equation,

\myeq{
&\hbox{\underline{Regime Vb}}\,:\quad xHs\frac{\diff Y_{\rm DM}}{\diff x} \simeq \gamma^{\rm eq}_{\rm med\leftrightarrow DM}  (x) \left[1-\left(\frac{Y_{\rm DM}}{Y_{\rm DM}^{\rm eq}\left(x\right)}\right)^2\right],
\label{BoltzEqVb}
}

\noindent or by the $\rm SM\leftrightarrow DM$ processes for larger values of  $\kappa'/\kappa_{\phi}$ (regime Va, in dark red in Figure \ref{fig:mesa_dynamical}) with

\myeq{
&\hbox{\underline{Regime Va}}\,:\quad xHs\frac{\diff Y_{\rm DM}}{\diff x} \simeq \gamma^{\rm eq}_{\rm SM\leftrightarrow DM}  (x) \left[1-\left(\frac{Y_{\rm DM}}{Y_{\rm DM}^{\rm eq}\left(x\right)}\right)^2\right].
\label{BoltzEqVb}
}

\noindent These two Boltzmann equations are very similar to the one applying in the secluded regimes except that this time there is only an equation for the DM yield since the mediator thermalises with both the DM and the SM particles. The other difference comes in the unique variable $x$ instead of having both $x$ and $x'$. This stems again from the fact that all populations thermalise with each other such that there is a unique temperature $T$.
\\

\begin{center}
\begin{figure}[h!]
\centering
\includegraphics[scale=0.75]{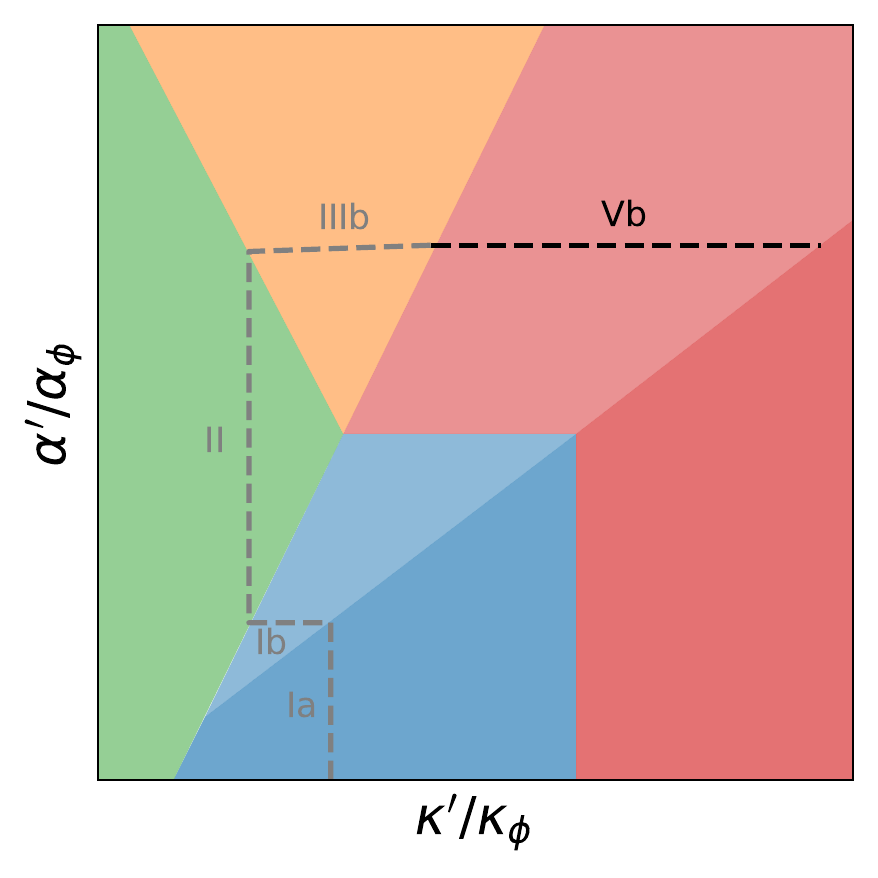}
\includegraphics[scale=0.75]{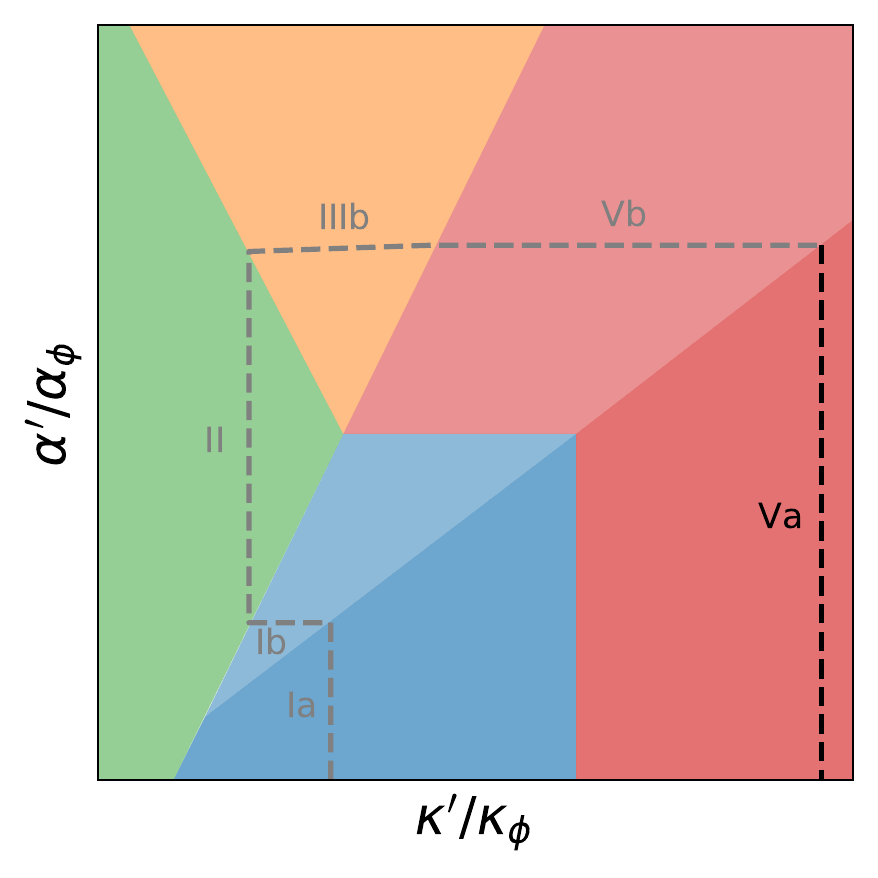}
\caption[Qualitative representation of the parameter space of a three sectors three connectors scenario for the regimes Va and Vb]{Qualitative representation of the parameter space of a three sectors three connectors scenario for the regime Vb (left) and Va (right).}
\label{fig:mesa_V}
\end{figure}
\end{center}

Depending on which process dominates, the DM relic abundance will be either driven by $\alpha'/\alpha_{\phi}$ leading to a horizontal line in the phase diagram for the regime Vb (see left panel of Figure \ref{fig:mesa_V}), either driven by $\kappa'/\kappa_{\Phi H}$ leading to a vertical line in the phase diagram for the regime Va (see right panel of Figure \ref{fig:mesa_V}). In the latest case, if we keep decreasing the DM-to-med connector $\alpha'/\alpha_{\phi}$, the hidden sector will eventually stop to thermalise, but this will have no impact on the DM final relic abundance since the connection of the DM to the mediator bath was already subdominant in this part of the parameter space.
\\

\begin{figure}[h!]
\centering
\includegraphics[scale=0.63]{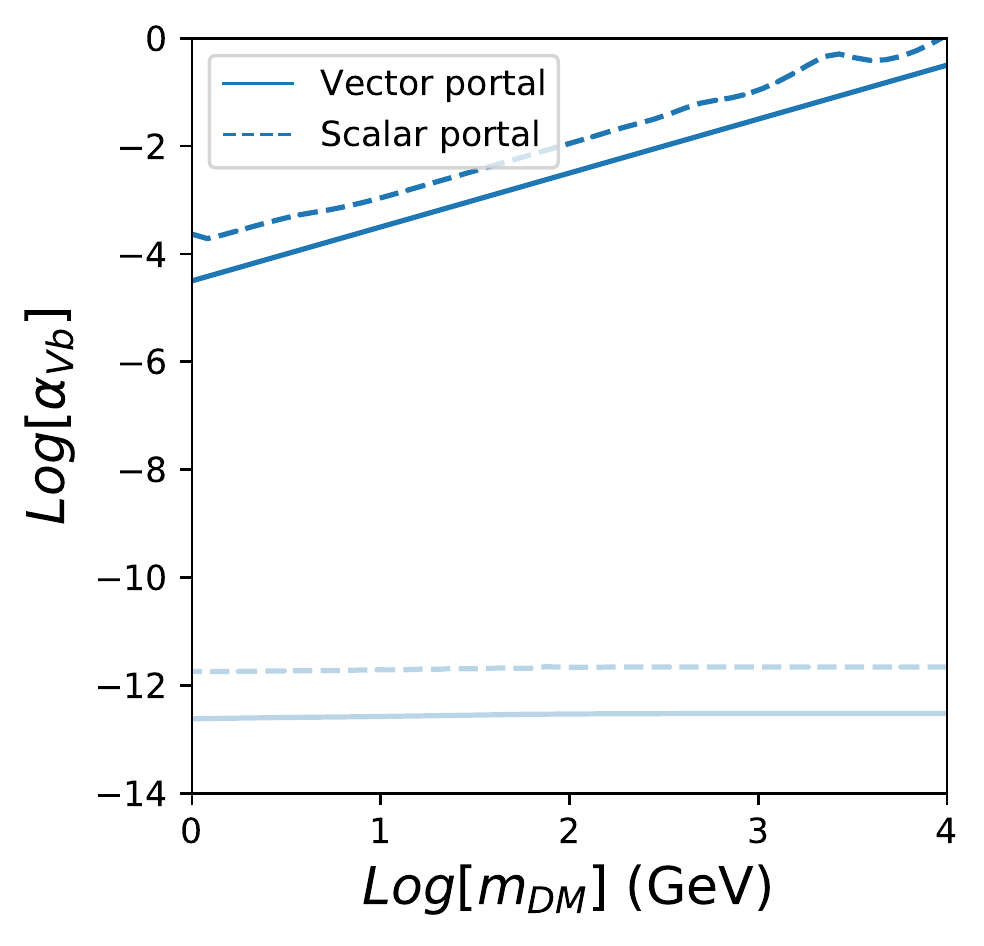}
\includegraphics[scale=0.63]{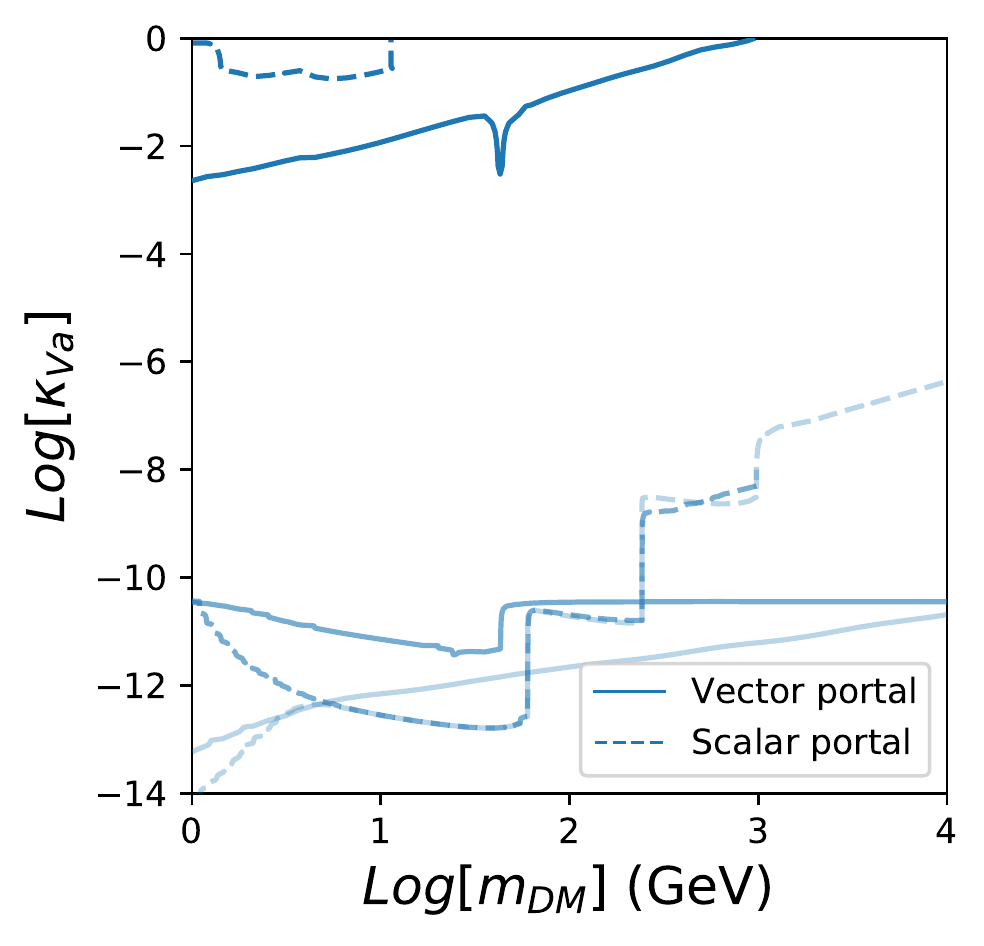}
\caption[Couplings as a function of the DM mass fixed by the relic density in regimes Va and Vb]{Values of $\kappa'/\kappa_{\phi}$ needed to account for the observed relic density, as a function of the DM mass $m_{DM}$, in the freeze-out regimes: regime Vb (left) and regime Va (right) for the vector portal (solid) and the scalar portal (dashed) models. Solutions obtained in the regime Ia and II/Ib are also shown for comparison in light blue and very light blue respectively.}
\label{fig:freeze_out}
\end{figure}

We show in Figure \ref{fig:freeze_out} the needed values for the relevant coupling in order to account for the observed DM relic density in both regimes Vb (left) and Va (right).

\section{Specificity of the vector portal model}\label{sec:spec_vector}
\subsection{Thermal effects}
The vector portal model, also known as the dark photon model, is known for displaying thermal effects. These effects can be important when considering the dark photon production rates (in particular in stars, see Refs. \cite{Redondo:2008aa,Jaeckel:2008fi,Redondo:2008ec,An:2013yfc,Redondo:2013lna,Fradette:2014sza}). As a consequence, taking these thermal effects into account could be very relevant in the determination of the DM final relic abundance. We will thus now review these thermal effects, in particular for dark photons production during the early Universe. In the next subsections, we will see what these effects imply on the DM relic abundance.
\\

As we briefly mention in Subsection \ref{subsec:KM}, the most important issue when considering thermal effects concerns how to treat the massless dark photon limit (i.e. $m_{\gamma'}\rightarrow~0$). Indeed, on the one hand, we have seen that, in the case of a massive dark photon, one can dissociate the dark photon propagation eigenstate basis from the mass eigenstate basis. On the other hand, we have also seen that in the massless dark photon limit, the propagation and mass eigenstates are degenerate and one cannot distinguish the two basis anymore. This major difference between the massless and massive dark photon cases is expressed, in practice, by the fact that the dark photon does not couple directly to any SM particles in the massless case. Naively, taking the massless limit of the massive case does not display this property. This apparently contradictory phenomenon has been extensively studied in the literature in presence of a medium\footnote{As it is the case in stars or in the early Universe.} (see Refs. \cite{Redondo:2008aa,Jaeckel:2008fi,Redondo:2008ec,An:2013yfc,Redondo:2013lna,Fradette:2014sza}). For our purpose, the important practical consequence of this absence of SM-to-med interaction in the massless case is that it imply that mediator particles cannot be produced directly from the SM bath. As a consequence, the sequential freeze-in regime (II) does not exist.
\\

\begin{center}
\begin{figure}[h!]
\centering
\includegraphics[scale=.2]{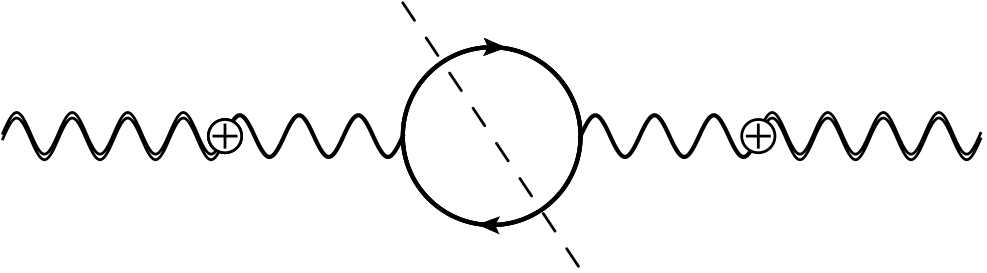}
\caption[The imaginary part of the dark photon propagator in a medium]{The imaginary part of the dark photon propagator (double wiggly lines) in a medium includes both its decay rate and creation rates.}
\label{fig:fg1}
\end{figure}
\end{center}

When a dark photon propagates through a thermal bath, it can oscillate into a SM photon and interact with charged particles from the plasma. This interaction of a propagating dark photon with the plasma is illustrated in Figure \ref{fig:fg1}. The single and double wiggly lines represent a propagating photon and dark photon respectively. The oscillation from one particle to the other is shown by crossed circles while the photon polarisation in the thermal bath is depicted by the blob (made of SM charged fermions for example). The cut in the blob is associated to the imaginary part of the photon polarisation tensor. Indeed, in vacuum this cut is related to the dark photon decay rate and in a medium it also takes the dark photon production rate (coalescence) into account \cite{Redondo:2008aa,Weldon:1983jn}. In the same way, the dark photon production rate from Compton scattering and pair annihilation would be considered by a cut in the two-loop diagram with photon exchange within the blob.
\\

When considering thermal effect, one has also to consider separately the transverse and the longitudinal components of the photon polarisation tensor. Indeed, in a medium, the longitudinal mode corresponds to the proper excitation of the medium which is known as plasmons (see \cite{Bellac:2011kqa} for example). In most of the dark photon mass range we are interested in, the dark photon production is mostly dominated by production through transverse photons \cite{An:2013yfc,Redondo:2013lna}. In a thermal bath, the transverse mode of propagating photons behave essentially like massive particles with a thermal mass (noted $\omega_T$) \cite{Braaten:1993jw}.

\myeq{
\operatorname{Re}\Pi_{\gamma,T} \equiv \omega^2_{T} \sim \sum_i e^2_i T^2,
}

\noindent is the real part of $\Pi_{\gamma,T}$ the self-energy of the transverse photons with $e_i$ the relativistic particles electric charge. Taking this into account, the creation of transverse dark photons proceeds through an effective mixing parameter $\epsilon \rightarrow \epsilon_{\rm eff}$, with 

\myeq{ 
\label{eq:substitution}
\epsilon \rightarrow \epsilon_{\rm eff} = \epsilon\times\frac{m_{\gamma'}^2}{m_{\gamma'}^2 - \Pi_{\gamma,T}},
}

\noindent where the denominator comes from the propagator in Figure \ref{fig:fg1} and the numerator comes from the $\gamma\rightarrow\gamma'$ transition in this figure too, see Eq. \ref{eq:lag_km_mass_mix} (see also Appendix \ref{app:th} for more details on $\epsilon_{\rm eff}$). This effective coupling displays a resonance, when $m_{\gamma'} \approx \omega_{\gamma,T}$, one should also consider the transverse photon modes finite width in a medium: $\propto \operatorname{Im}\Pi_{\gamma,T} \ll \operatorname{Re}\Pi_{\gamma,T}$. This effective coupling also displays a suppression at high temperature (or law dark photon mass, $m_{\gamma'} \ll \omega_{T}$) of the dark photon production from SM coming from the fact that in the numerator we have $m_{\gamma'}^{2}\sim T^{2}$. This latter behaviour, moreover also implies a smooth transition in the physical massless dark photon limit as the effective coupling goes to zero in this limit. Indeed, when $m_{\gamma'}\rightarrow 0$, the dark photon does not couple to the SM anymore since $\epsilon_{\rm eff}\rightarrow 0$, just as in the massless case. In the opposite limit, when $m_{\gamma'} \gg \omega_{T}$, the effective mixing parameter goes back to the usual mixing parameter we have in vacuum, $\epsilon_{\rm eff} \rightarrow \epsilon$. For off-shell dark photon, this effective coupling is simply given by $\epsilon$ since such process are essentially insensitive to the dark photon mass quicker (i.e. for smaller DM masses) than without taking thermal effects into account.
\\

Practically, the major consequence of the above is that the dark photon production rate is highly suppressed at high temperature ($T\gg m_{\gamma '}$) and is strongly enhanced when the temperature approximately reaches the dark photon mass (i.e. when $\omega_T\approx m_{\gamma '}$). Thus, if the DM is much heavier than the dark photon, the processes which will set the DM relic abundance will freeze before that the dark photons are significantly produced from the SM bath and thermal effects are irrelevant such that one recovers the massless case results.
\\

\begin{figure}[h!]
\centering
\includegraphics[scale=0.5]{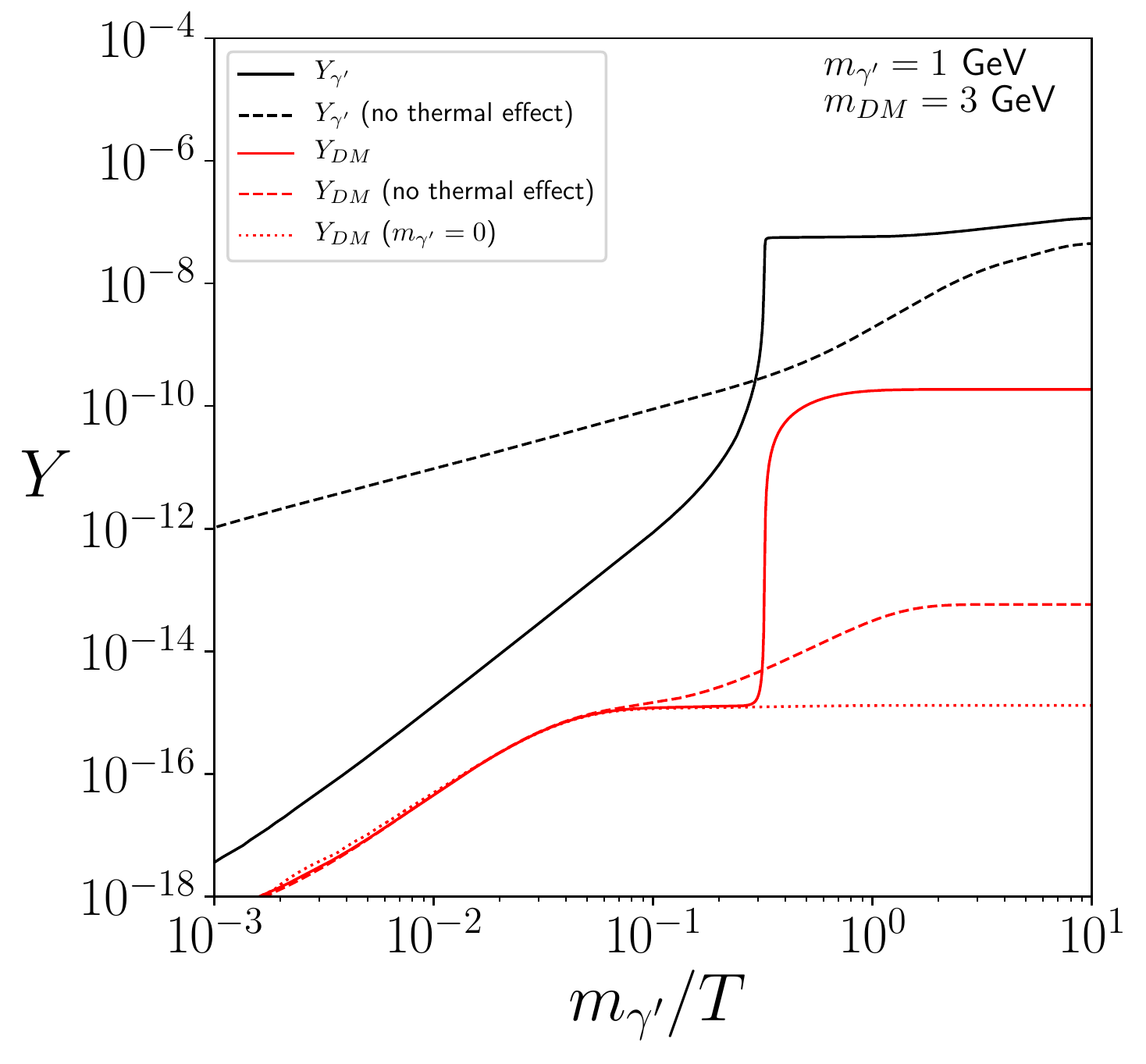}
\caption[Comparison of the evolution of the dark photon and DM abundances, with and without thermal effects on dark photon production]{Comparison of the evolution of the dark photon (black lines) and DM abundances (red lines), with (solid) and without (dashed) thermal effects on dark photon production. The dotted lines give the abundance of dark photon in the massless limit. We consider values of the couplings which lead to the observed relic density when the thermal corrections are taken into account (solid red lines): $\kappa_{\rm II}=3.6\times 10^{-14}$ and $\epsilon = 10^{-11}$.}
\label{fig::Yields}
\end{figure}

Figure \ref{fig::Yields} shows the difference in the dark photon and DM yields ($Y_{\gamma'}$ and $Y_{\rm DM}$ respectively) evolution with (solid) and without (dashed) taking thermal effects into account for dark photon production. On this figure we see how thermal effects can be relevant when the DM and dark photons masses are taken to be close, here we fixed: $m_{\gamma '}=1$ GeV, $m_{\rm DM}=3$~GeV. For this particular example, the DM-to-SM connector has been fixed such that this set of values lead to the observed relic density taking into account thermal corrections: $\kappa_{\rm II}=3.6\times 10^{-14}$ and $\epsilon = 10^{-11}$. One can see that most of the dark photons are produced at a temperature close to the DM mass scale. As the DM production rate from dark photons is not yet Boltzmann suppressed at this time, the DM yield is strongly enhanced and follows the dark photon yield curve. Thus, the DM production from dark photon is strongly enhanced compared to the massive case without thermal correction and to the massless case. See Appendix \ref{app:th} for more technical details of the effect of the resonance on the DM production. 
\\

Moreover, the dark photon production rate is not the only rate which is impacted by thermal effects. Indeed, it also affects the dark photon thermalisation with the SM bath. It will be easier for them to thermalise with SM particles as there production is boosted. Thus, a smaller value of the SM-to-med connector will be required to thermalise.
\\

\begin{center}
\begin{figure}[h!]
\centering
\includegraphics[scale=0.75]{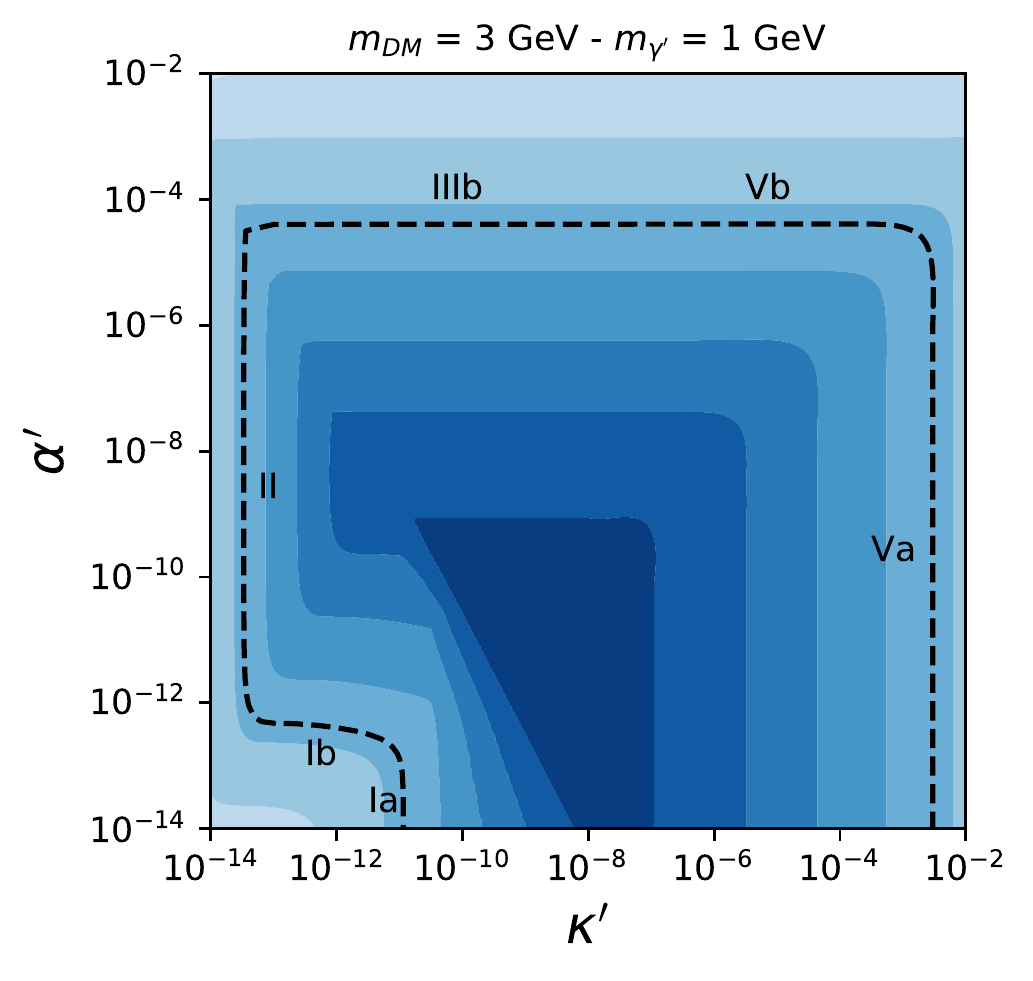}
\caption[Vector portal model global phase space including thermal effects]{Same as Figure \ref{fig::DogDiagram1},  $m_{\rm DM} = 3 $ GeV and $m_{\gamma'} = 1$ GeV, but taking into account thermal effects on dark photon production.}\label{fig::MESA_3_1} 
\end{figure}
\end{center}

Figure \ref{fig::MESA_3_1} shows the same phase diagram than the one depicted in the left panel of Figure \ref{fig::DogDiagram1}, but taking thermal effects into account. We will now review what these thermal effects change for every regime one by one.

\subsection{Freeze-in regimes}
In the first regime we considered, the regime Ia, the DM is produced by slow out-of-equilibrium processes from annihilation or decay of SM particles. On the one hand, in this regime, the dark photon bath plays a negligible role and taking thermal effects into account has no impact on the DM final relic abundance. On the other hand, in the second freeze-in regime, the regime Ib, the DM is produced by a slow out-of-equilibrium process from annihilation of dark photons. Since these dark photons are in thermal equilibrium with the SM, thermal effects will play a role and have an impact on the DM abundance. Indeed, if $m_{\rm DM}\gtrsim m_{\gamma '}$, the dark photon number density will be strongly enhanced by the resonance such that dark photons will be able to reach equilibrium with SM for even smaller values of the SM-to-med coupling $\epsilon$. Thus, if dark photons are in thermal equilibrium with the SM for smaller $\epsilon$, the regime Ib will be relevant for smaller $\epsilon$ or, equivalently, smaller DM-to-SM coupling $\kappa'$.

\subsection{Sequential freeze-in regime}
\begin{figure}[t]
\centering
\includegraphics[width=8cm]{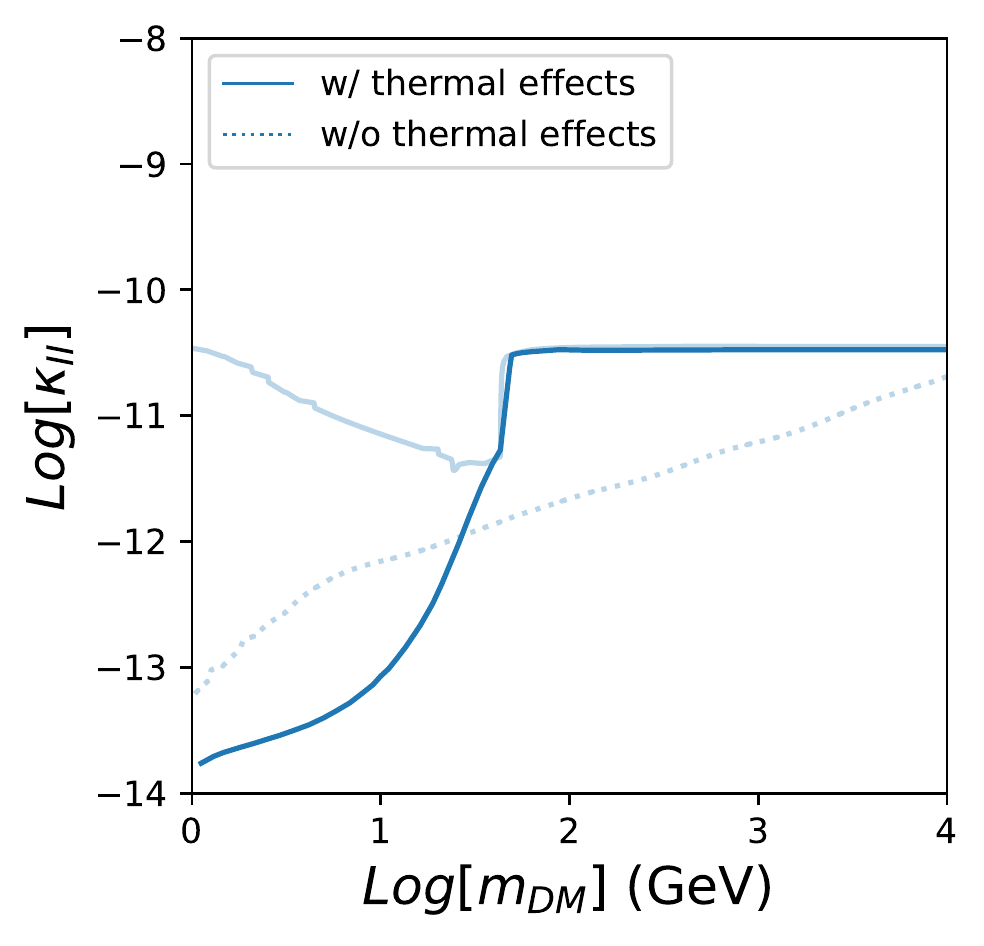}
\caption[Couplings as a function of the DM mass fixed by the relic density in regime II  including thermal effects]{Dark matter candidates as shown in Figure \ref{fig:Sequential_Freeze_In} but taking thermal effects in the dark photon production rates. The additional dashed lines are given for the sake of comparison. They correspond to the curves shown in Fig.~\ref{fig:Sequential_Freeze_In}, thus without thermal effects.}
\label{fig:Sequential_Freeze_In_TH}
\end{figure}

As in the regime Ib, thermal corrections will play an important role in the sequential freeze-in regime, regime II. For a DM particle not much heavier than the dark photon, the slow out-of-equilibrium production of DM particles from dark photon will coincide with the resonant enhancement of the slow out-of-equilibrium production of dark photons from the annihilation or the decay of SM particles. This will then make this chain of freeze-in more efficient and one will have to lower the DM-to-SM connector in order to not overproduce DM. This will lead to an even more pronounced extension towards the left of the mesa shaped phase diagram. This feature can also be seen on the Figure \ref{fig:Sequential_Freeze_In_TH} which shows the same as Figure \ref{fig:Sequential_Freeze_In} taking thermal effects into account. We see that thermal effects are mostly relevant when the dark photon is not much lighter than the DM. Otherwise, we get approximately back to the massless dark photon case.

\subsection{Reannihilation regimes}
As for the regime Ib, in the reannihilation from dark photon regime, regime IIIb, since the production of dark photon from SM is enhanced, this regime will dominates over the production of DM from SM. The regime IIIb will thus be relevant for a wider range of the DM-to-SM coupling. On the other hand, the regime IIIa in which the SM sources the dark sector producing directly DM particles is unchanged as the production rate of dark photons from SM is negligible.

\subsection{Secluded freeze-out regimes}
For $m_{\gamma'}< m_{\rm DM}<m_e$, the regime IVb is not relevant anymore with thermal corrections. This stems from the fact that at a temperature much bigger than the dark photon mass scale, the dark photon production is considerably suppressed once taking thermal effects into account and that at a temperature close to the dark photon mass scale, the production is Boltzmann suppressed as the dark photons are lighter than the electrons\footnote{The electrons are thus already Boltzmann suppressed when the resonant dark photon production should have started.}. Remember that the production stops when $T'\sim m_{\rm DM}$. Nevertheless, the regime IVa still apply for scenario where dark photon particles are much lighter than the DM particles, as in the massless case (see see \cite{Chu:2011be}).

\subsection{Freeze-out regimes}
Since  all particles are thermalised in a single thermal bath, the effects of the above thermal corrections on the dynamics of these regimes are negligible.

\section{Specificity of the scalar portal model}\label{sec:spec_scalar}
Introducing a new scalar and a new VEV bring some new dynamics to the system. As we have already mentioned earlier, depending on the value of this new VEV and the value of the DM mass with respect to the two VEV's, we will face different phases of the model. In practice, we have considered the new VEV to always be bigger than the electroweak symmetry breaking scale ($v_{H}<v_{\Phi}$). Indeed, a new symmetry breaking is not expected at the electroweak scale. Thus, there are essentially three possibilities,

\begin{itemize}
\item The broken phase: $m_{\rm DM} < v_{H} < v_{\Phi}$
\item The semi-broken phase: $v_{H} < m_{\rm DM} < v_{\Phi}$
\item The symmetric phase: $v_{H} < v_{\Phi} < m_{\rm DM}$
\end{itemize}

\noindent We will see in the following that for most of the dynamical regimes we studied in Section \ref{sec:phases}, only the broken phase is relevant and this is why this is the phase we focused on in Section \ref{sec:phases}. However, we will also see that for two freeze-in regimes (Ia and II), discussing in which phase a DM candidate sits is relevant. For this reason, we will go through the three phases (broken, semi-broken and symmetric) one by one and explain, for each of them, what does it imply for regimes Ia and II.

\subsection{The broken phase}
\begin{center}
\begin{figure}[h!]
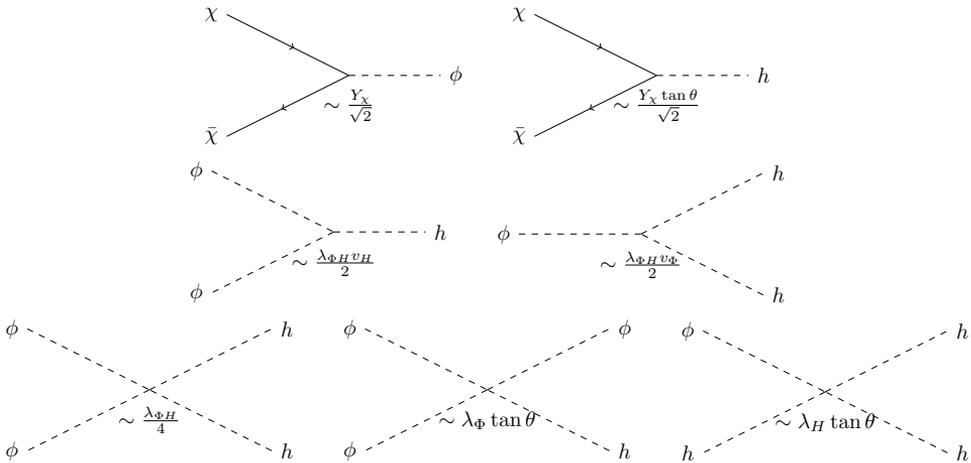

\centering
\includegraphics[scale=0.8]{XXP_b.pdf}
\includegraphics[scale=0.8]{XXH_b.pdf}\\
\includegraphics[scale=0.8]{PPH_b.pdf}
\includegraphics[scale=0.8]{PHH_b.pdf}\\
\includegraphics[scale=0.8]{PPHH_b.pdf}
\includegraphics[scale=0.8]{PPPH_b.pdf}
\includegraphics[scale=0.8]{PHHH_b.pdf}
\caption[Couplings between the DM, the mediator and the visible sector in the fully broken phase]{Couplings (at leading order for a small mixing angle) between the DM, the mediator and the visible sector in the fully broken phase.}
\label{fig:HP-BRO_2}
\end{figure}
\end{center}

If the DM mass is smaller than the new and SM scalar VEV's (i.e. $m_{\rm DM} < v_{H} < v_{\Phi}$), the DM production mechanism through which the DM abundance is set (see Section \ref{sec:phases}) will occur when both hidden and visible sector symmetries are broken: $T\lesssim m_{\rm DM}< v_{H}< v_{\Phi}$. Thus, the mixing between the two scalars of the theory is achieved as we have seen in Subsection \ref{subsec:HP} for which we remind all relevant couplings in Figure \ref{fig:HP-BRO_2}.
\\

\begin{center}
\begin{figure}[h!]
\centering
\includegraphics[scale=0.8]{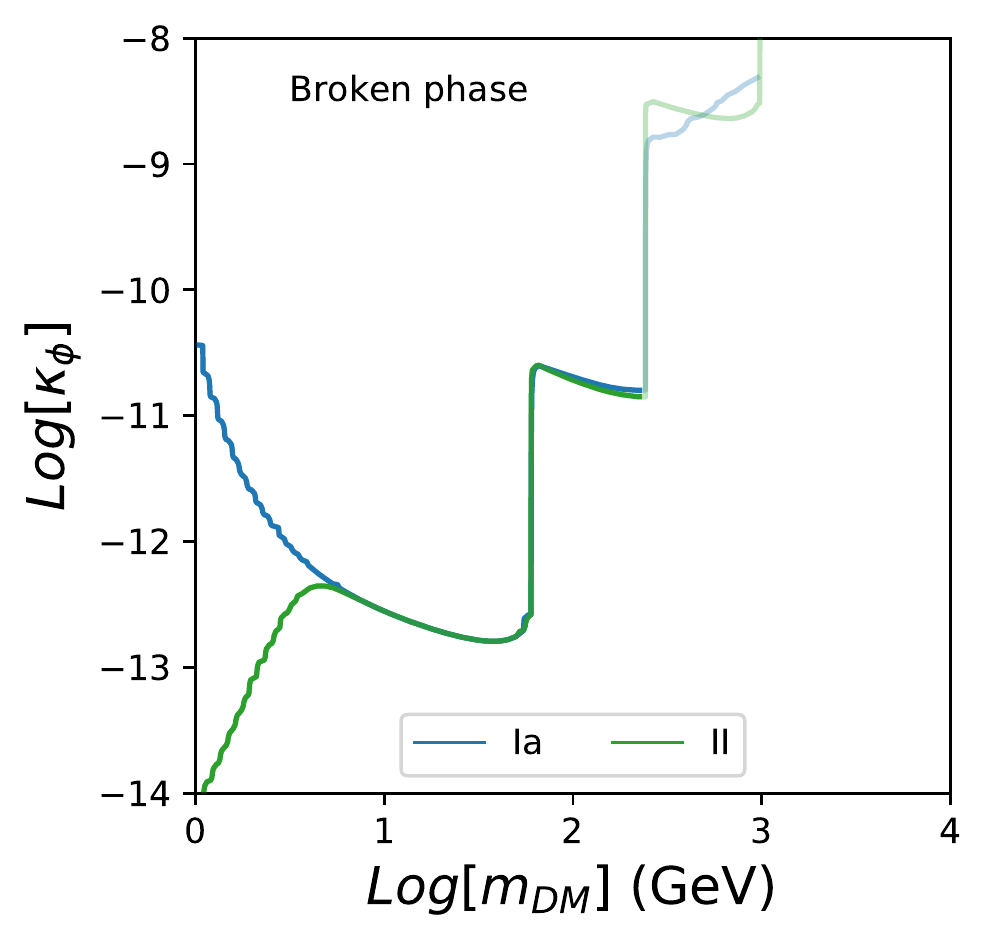}
\caption[Couplings as a function of the DM mass fixed by the relic density in the broken phase]{Values of the DM-to-SM connector to account for the observed DM relic abundance in the broken phase in the regimes Ia (blue) and II (green). The mediator mass has been taken to $m_{\rm med}= 1$ GeV.}
\label{fig:HP-DD-BRO_2}
\end{figure}
\end{center}

Results have already been discussed at length in Section \ref{sec:phases} where we have seen that this phase is allowing all regimes we discussed in in this section. Focusing on the low DM-to-SM connector region, we recap in Figure \ref{fig:HP-DD-BRO_2} results we had in Section \ref{sec:phases} highlighting the broken phase and to which we refer for more explanations on this figure.
 
\subsection{The semi-broken phase}
If the DM mass is bigger than the SM scalar VEV but smaller than the new scalar VEV (i.e. $v_{H} < m_{\rm DM} < v_{\Phi}$), the DM abundance will be set at a temperature higher than the SM broken symmetry scale such that the SM will still be in his symmetric phase while the hidden sector symmetry will be broken. As a consequence, only the new scalar $\Phi$ acquires a VEV,

\myeq{
\Phi =\frac{v_{\Phi}+\tilde{\phi}}{\sqrt{2}},
}

\noindent where the $\lambda_{3}$ parameter from Eq. \ref{eq:lag_hp} is again traded for the $v_{\Phi}$ one, as in Subsection \ref{subsec:HP}. There is then no mixing since the mass matrix, given by

\begin{center}
\begin{figure}[h!]
\centering
\includegraphics[scale=0.8]{XXP_b.pdf}
\includegraphics[scale=0.8]{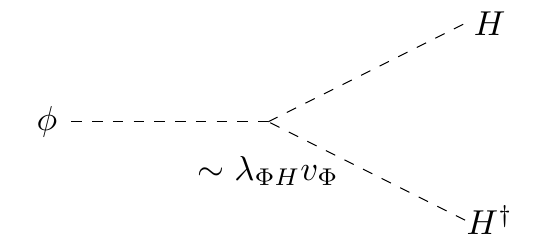}
\includegraphics[scale=0.8]{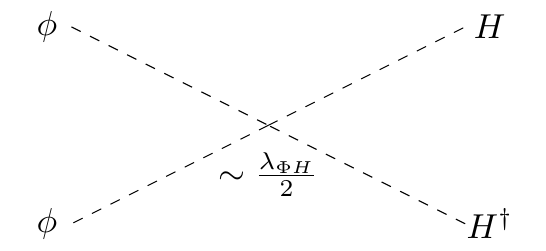}
\caption[Couplings between the DM, the mediator and the visible sector in the broken hidden sector phase]{Couplings between the DM, the mediator and the visible sector in the broken hidden sector phase.}
\label{fig:HP-HS_BRO_2}
\end{figure}
\end{center}

\myeq{
M_{\Phi H}^{2} &=
\begin{pmatrix}
 0 & 0 \\ 
 0 & 2\lambda_{\Phi}v_{\Phi}^{2}
\end{pmatrix},
}

\noindent is already diagonal. One can then get rid of the \textit{tilde} notation $\phi\equiv\tilde{\phi}$ and has now to consider the real scalar $\phi$ and the complex scalar $H$. The relevant couplings in this phase are shown in Figure \ref{fig:HP-HS_BRO_2}.
\\

\begin{center}
\begin{figure}[h!]
\centering
\includegraphics[scale=0.8]{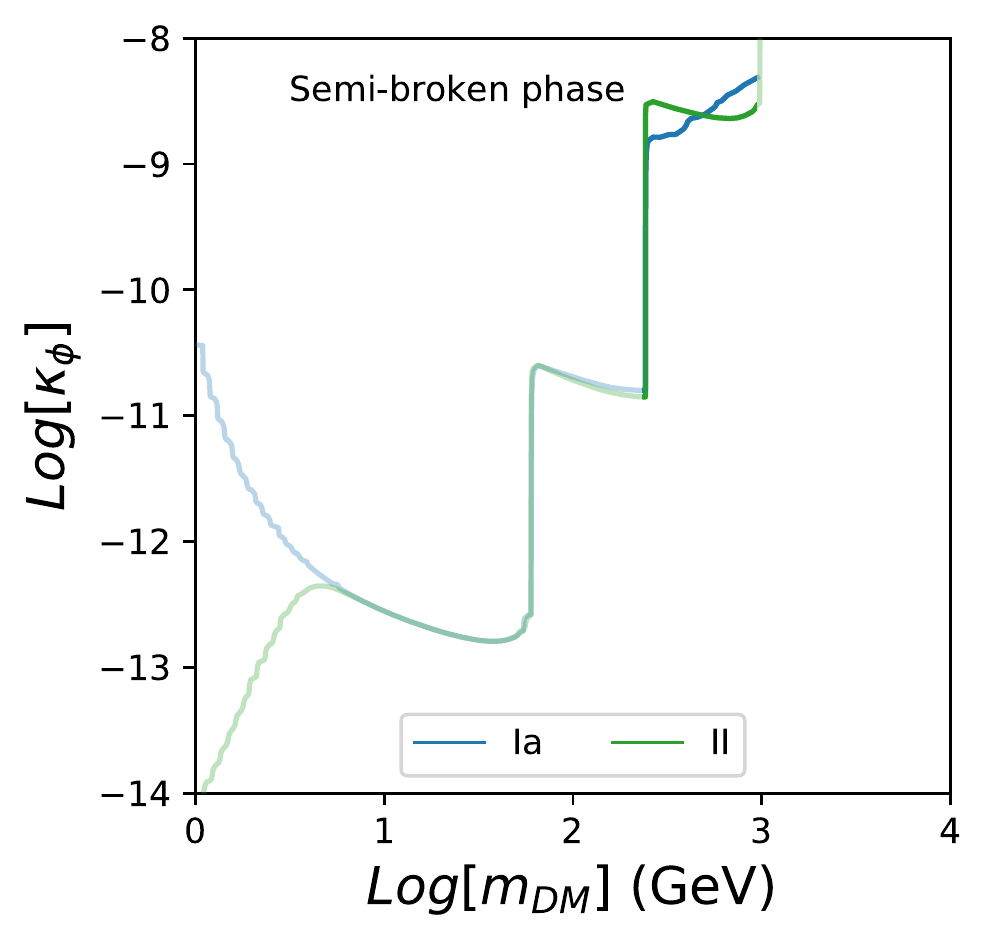}
\caption[Couplings as a function of the DM mass fixed by the relic density in the semi-broken phase]{Values of the DM-to-SM connector to account for the observed DM relic abundance in the semi-broken phase in the regimes Ia (blue) and II (green).}
\label{fig:HP-DD-BRO_HS_2}
\end{figure}
\end{center}

In this phase, it is still possible to produce DM directly from SM and the standard freeze-in regime still exists (regime Ia). Indeed, one can build from Figure \ref{fig:HP-HS_BRO_2} an annihilation process of two $H$'s into two DM particles through the production of a $\phi$ particle in the s-channel. Since this is the only way to produce DM directly from SM, the required value for the DM-to-SM coupling $\kappa_{\phi}$ has to be somewhat increased compared to the one required in the broken phase in which there are more DM production channels. The sequential freeze-in, regime II, is unchanged as it is still possible to produce the new scalar from slow out-of-equilibrium annihilation of two $H$'s through the quartic coupling and since the DM can still be frozen-in from slow out-of-equilibrium annihilation of two $\phi$'s through the Yukawa coupling. This can be seen in Figure \ref{fig:HP-DD-BRO_HS_2}.

\subsection{The symmetric phase}
\begin{center}
\begin{figure}[h!]
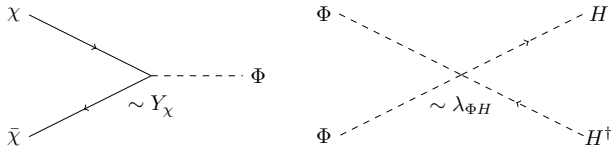

\centering
\includegraphics[scale=0.8]{XXP.pdf}
\includegraphics[scale=0.8]{PPHH.pdf}
\caption[Couplings between the DM to the mediator and the mediator to the visible sector in the symmetric phase]{Couplings between the DM to the mediator and the mediator to the visible sector in the symmetric phase.}
\label{fig:HP-SYM_2}
\end{figure}
\end{center}

Finally, if the DM mass lies above both VEV's of the theory, the DM abundance will be frozen before any symmetry breaking could occur. As a consequence, the only remnant couplings are the one directly readable from the Lagrangian given in Eq. \ref{eq:lag_hp} and shown in Figure \ref{fig:HP-SYM_2}.
\\

\begin{center}
\begin{figure}[h!]
\centering
\includegraphics[scale=0.8]{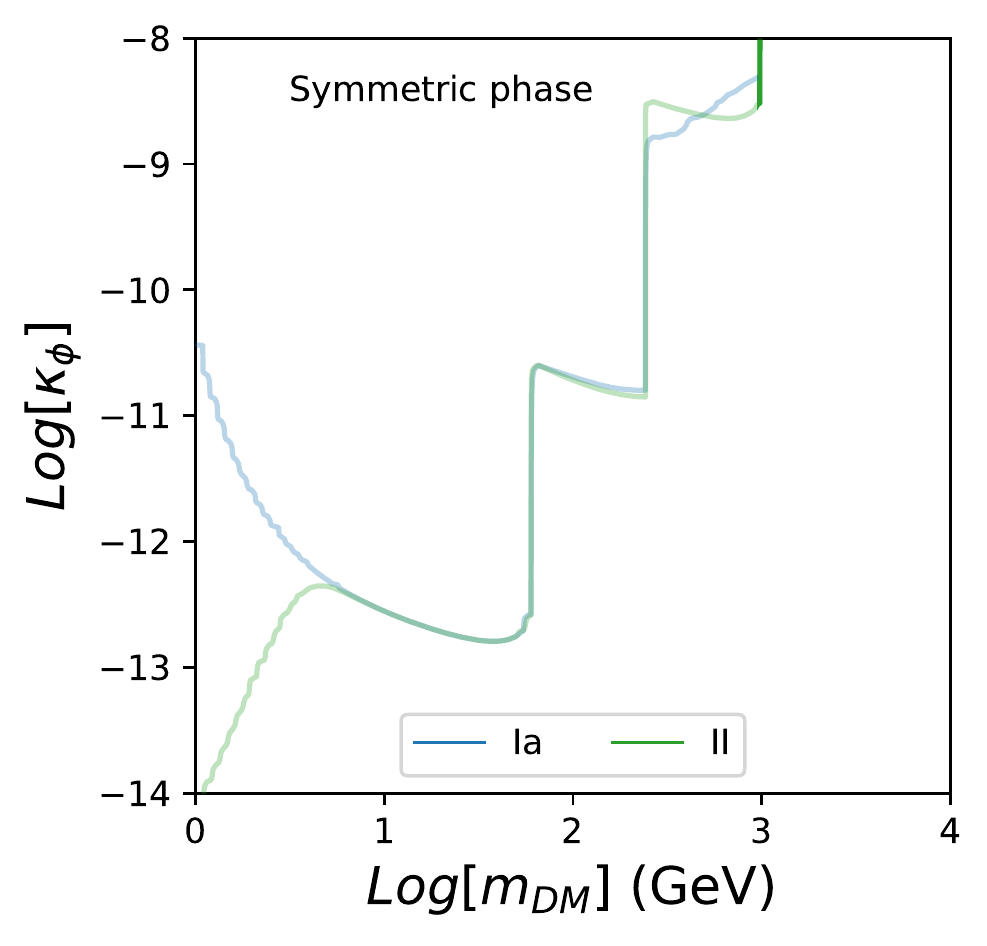}
\caption{Values of the DM-to-SM connector to account for the observed DM relic abundance in the symmetric phase in the regime II (green).}
\label{fig:HP-DD-SYM_2}
\end{figure}
\end{center}

From this figure, one can conclude that it is no longer possible to produce DM directly (i.e. in an unsuppressed way) from SM in the symmetric phase. Thus, the only way to account for the observed DM relic abundance starting from an empty hidden sector is through the sequential freeze-in, regime II. Thus, this regime we discovered plays a crucial role for heavy DM candidates. Again, as it is now the only way to produce DM, one needs a stronger DM-to-SM connector in order to produce enough DM particles with respect to couplings required in the semi-broken phase. Results one would obtain in this phase are shown in Figure \ref{fig:HP-DD-SYM_2} where there is no dark blue line as there is no solution in the regime Ia.

\chapter{Thermally disconnected HS}\label{ch:secl}
\yinipar{I}n Chapter \ref{ch:prod}, we have seen that, even if the DM particles have never been in thermal equilibrium with the SM thermal bath, one can account for the DM relic density in several ways. Indeed, the DM could be part of a HS made of particles which would not be in thermal equilibrium with the SM, the VS. In such scenarios, we have seen that there are various DM production mechanisms: FI, sequential FI, reannihilation and secluded FO, see Subsections \ref{subsec:FI}, \ref{subsec:SFI}, \ref{subsec:rean} and \ref{subsec:secluded} respectively. All this analysis we have made in Chapter \ref{ch:prod} was based on the assumption that at the end of inflation, the HS was negligible and that the HS has been created from the SM thermal bath afterwards. Thus, the portal was essential, thermalising or not both sectors with each other, depending on the regime considered. Here, in this chapter based on \cite{Hambye_2020,Hambye:2020lvy,Coy:2021ann}, we would like to consider a scenario which, like everywhere in this thesis, is based on a VS-HS portal structure, but which is different in the sense that, here, we will neglect the portal interactions. We assume instead that the HS was already there at the end of inflation or, at least, that the portal ceased to be relevant long before the DM relic density was frozen. Neglecting the portal in this way, we will also assume that DM particles thermalise with an additional particle within the HS. Thus, one looks at the situation where we have two distinct and thermally disconnected baths, one composed of SM particles (VS) and one composed of dark sector particles such as DM and possibly other new particles (HS). Each bath would thus have its own temperature, $T$ and $T'$ for the VS and HS respectively. 
\\

Under these assumptions, one is left with a scenario where the DM freezes-out within the HS which gives rise to two possibilities. Either, the DM decouples while non-relativistic as in the secluded FO of Subsection \ref{subsec:secluded} (since the portal ceases to have any effect long before DM decouples in the thermal HS bath, as we have seen), or it freezes-out while still relativistic. It turns out that these simple and very generic scenarios have never been studied in much detail. In this chapter, we determine the full DM mass and hidden-to-visible temperature ratio ranges which are allowed in such case. We call this available parameter space the "domain" of all DM thermal candidates.

\section{Theoretical constraints}\label{sec:2d_theoretical}
\subsection{Relativistic decoupling floor}\label{subsec:rel_floor}
Before discussing concrete models in which the DM thermalises within a hidden sector with its own temperature $T'$, let us study the DM relic density constraint in a model independent way in the context of thermally disconnected HS. In this section, we will see that this constraint gives, for the case of a relativistic decoupling in the HS, a lower bound on the hidden-to-visible temperature ratio. This is to be expected since a too low temperature ratio means a too small amount of DM particles with respect to SM particles and since the maximum of DM particles one can get when thermalised within the HS is when it decouples relativistically. Indeed, in the non-relativistic decoupling case, the number of DM particles left is Boltzmann suppressed. This lower bound on the hidden-to-visible temperature ratio can be translated in a lower bound on the DM mass for a given temperature ratio. This would thus give the generalisation of the Cowsik-McClelland \cite{Cowsik:1972gh} bound to cases where $T'/T \neq 1$.
\\

If the DM decouples relativistically in the HS, the DM number density before decoupling is simply given by its relativistic value

\myeq{
n_{\rm DM}= \frac{\zeta(3)}{\pi^2} g^{n}_{\rm DM} T'^3,\label{eq:rel_nDM}
}

\noindent where the effective degeneracy of the relativistic degrees of freedom are as usual: $g^{n}_{\rm DM}=g_{\rm DM}$ for a boson and $g^{n}_{\rm DM}=3g_{\rm DM}/4$ for a fermion. The Riemann zeta function of three is approximately given by $\zeta(3)\simeq 1.202$. This scenario has the nice feature of giving a relic density which does not depend on the value of the DM annihilation cross section, simply because Eq. \ref{eq:rel_nDM} is independent of it. Imposing that the DM number density given in Eq. \ref{eq:rel_nDM} leads to (at least) the observed DM relic density, we get the following lower bound

\myeq{
\frac{T'_{\rm dec}}{T_{\rm dec}}\geq 2.46\times 10^{-4}\times \left(\frac{100\text{ GeV}}{m_{\rm DM}}\right)^{1/3}\times  \left( \frac{G^{S}_{\ast}(T_{\rm dec})}{g^{n}_{\rm DM}(T'_{\rm dec})} \right)^{1/3},\label{eq:TpTlowerbound}
}

\noindent where $T_{\rm dec}$ and $T'_{\rm dec}$ indicate the value of the VS and the HS temperature at DM decoupling respectively while $G^{S}_{\ast}$ stems for the total effective relativistic degrees of freedom contained in both the visible and hidden sectors\footnote{Depending on which sector dominates the energy and entropy distributions of the Universe implies two types of solutions. See  Subsection \ref{subsec:uni_wall} for more details.}. It is defined in the same way than $g^{S}_{\ast}$ (see Eq. \ref{eq:g_star_s}), but taking HS relativistic degrees of freedom into account. This bound is saturated for a relativistic decoupling, but not for a non-relativistic one.
\\

Note that such relativistic decoupling can be realised in two different ways. In the first possibility, which is analogous to the neutrino decoupling in the SM, the DM interaction rate could have been $\propto T^{\prime 5}/\Lambda^4$ below some scale $T^\prime \sim \Lambda$, \footnote{$\Lambda$ would typically be the mass of some heavy mediator.}, with $\Lambda {> T'_{\rm dec} } \gg m_{\rm DM}$ in the same way as the interaction rate for SM neutrinos changed from $\Gamma \propto \alpha_W^2 T$ to ${\Gamma} \propto G_F^2 T^5$ for $T \lesssim M_W$ \cite{Kolb:1990vq}. In the second possibility, the mass of particles the DM is annihilating into could play the role of the heavy "cut-off" scale. If these particles are heavier than the DM, the annihilation rate became Boltzmann suppressed when $T'$ went below their mass. Schematically, this relativistic freeze-out possibility is analogous to non-relativistic DM decoupling in the sense that decoupling was due to a Boltzmann suppression. Then, in this case, the DM is reheated just before its decoupling since the heavier particle (final states) are becoming non-relativistic\footnote{This reheating is similar to the photon reheating which occurred during the $e^+e^-$ annihilation catastrophe in the VS}. Then, the initial value of the hidden-to-visible temperature ratio $T'_{\rm in}/T_{\rm in}$, i.e. before the heavier particle became non-relativistic, differs from the value of this ratio at DM decoupling, $T'_{\rm dec}/T_{\rm dec}$. They differ by

\myeq{
\frac{T'_{\rm dec}}{T_{\rm dec}}= \left({g'^{S}_{\ast}(T'_{\rm in}) \over g'^{S}_{\ast}(T'_{\rm dec})} \right)^{1/3} \times\frac{T'_{\rm in}}{T_{\rm in}}.
}\label{eq:TpT_in_to_dec}

\noindent In the first scenario instead there is no such reheating of the HS and $T'_{\rm in}/T_{\rm in}=T'_{\rm dec}/T_{\rm dec}$ as $g'^{S}_{\ast}(T'_{\rm in})=g'^{S}_{\ast}(T'_{\rm dec})$. 
\\

Let us emphasises the fact that the factor of $g'^{S}_{\ast}$ in Eq. \ref{eq:TpT_in_to_dec} takes into account all relativistic degrees of freedom contained in the HS and not only the DM. This is due to the fact that the HS may be composed of numerous particles on top of the DM one, such that theoretically, the HS could be made of several decoupled sector.
\\

\begin{center}
\begin{figure}[h!]
\centering
\includegraphics[scale=0.825]{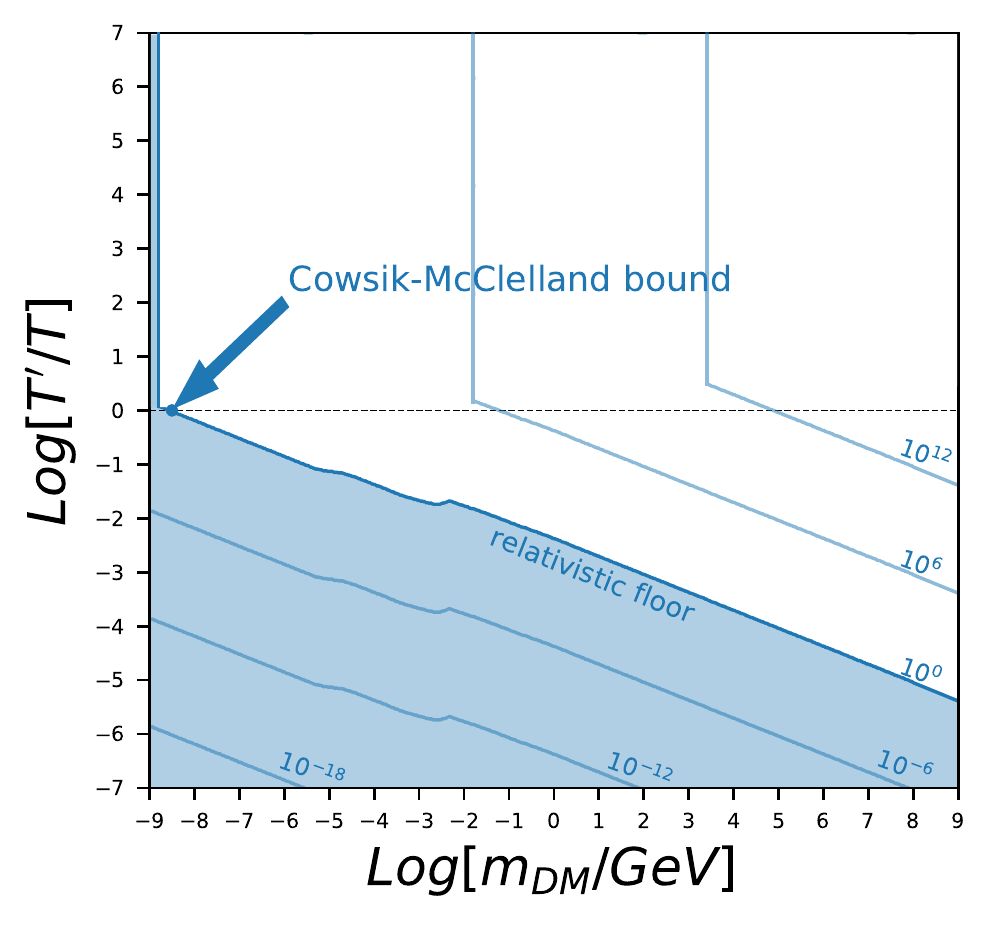}
\caption[Contour of the normalised DM relic abundance in the temperature ratio versus DM mass plane for a relativistic decoupling]{Contour of the normalised DM relic abundance $\Omega_{\rm DM}h^{2}/0.1188$ in the $T'/T$ vs DM mass plane for a relativistic decoupling.}
\label{fig:rel_floor}
\end{figure}
\end{center}

Figure \ref{fig:rel_floor} shows contours of the normalised DM relic abundance $\left(\Omega_{\rm DM}h^{2}/0.1188\right)$ in the $T'/T$ vs DM mass plane for a relativistic decoupling. The blue shaded area indicates where the initial amount of DM (i.e. before it decouples relativistically) lies already below the observed abundance today. In that sense and for a chosen DM mass, the contour $\Omega_{\rm DM}h^{2}=0.1188$ (depicted by the solid dark blue line in Figure \ref{fig:rel_floor}) gives the minimal temperature ratio one can consider for a DM thermal candidate as given in Eq. \ref{eq:TpTlowerbound}. Alternatively, for a given hidden-to-visible temperature ratio, the very same contour gives the minimal DM mass for a thermal candidate. One can also see in Figure \ref{fig:rel_floor}, the Cowsik-McClelland \cite{Cowsik:1972gh} bound which gives a lower bound on the DM mass for a thermal candidate which was in equilibrium with the visible sector (i.e. only one bath with $T'=T$, black dashed line in Figure \ref{fig:rel_floor}). Note that this very simple relativistic floor scenario has, to our knowledge, surprisingly not been previously presented in the literature in any published paper. After presenting it, we became aware of a preprint \cite{Sigurdson:2009uz} where this scenario has been partially presented.

\subsection{Unitarity wall}\label{subsec:uni_wall}
After having considered the extreme case of a relativistic decoupling, let us consider the opposite case in which the DM decouples while well non-relativistic (i.e. $m_{\rm DM}/T'_{\rm dec}\gg 1$). In this type of scenario, the final DM abundance depends on the annihilation cross section as we have seen in Subsection \ref{subsec:FO}. However, the expression given in Eq. \ref{eq:Oh2_FO} is only valid in a standard FO scenario in which the DM thermalises with the SM, i.e. $T'=T$. Thus, in Appendix \ref{app:inst_FO}, we generalise the instantaneous freeze-out approximation for the thermally disconnected HS case, i.e. fixing $x'_{\rm dec}=m_{\rm DM}/T'_{\rm dec}$ from the condition $\Gamma/H|_{T'=T'_{\rm dec}}=1$ and determining $Y_{\rm DM}$ by assuming that the yield after freeze-out is equal to $n_{\rm DM}^{eq}/s|_{T'=T'_{\rm dec}}$. Here, the outcome of this scenario is somewhat more complicated than for the standard FO case because the Universe expansion rate as well as its energy and entropy densities get contributions from both the visible and the hidden sectors,

\myeq{
& H(x') = \sqrt{\frac{8\pi}{3M_{\rm pl}^{2}}}\sqrt{\rho_{\rm VS}(x')+\rho_{\rm HS}(x')},\label{eq:H_universe}\\
& s(x') = s_{\rm VS}(x')+s_{\rm HS}(x')\label{eq:s_universe}.
}

\noindent In terms of the inverse hidden temperature $x'\equiv m_{\rm DM}/T'$ and the hidden-to-visible temperature ratio $\xi\equiv T'/T$, the energy and entropy densities of both sectors are given by,

\myeq{
&\rho_{\rm VS}(x') = \frac{\pi^{2}}{30}g_{\ast}^{\rm eff}\left(m_{\rm DM}/x'\xi\right)\left(\frac{m_{\rm DM}}{x'\xi}\right)^{4},\label{eq:rho_VS}\\
&\rho_{\rm HS}(x') = \frac{\pi^{2}}{30}g'^{\rm eff}_{\ast}\left(m_{\rm DM}/x'\right)\left(\frac{m_{\rm DM}}{x'}\right)^{4},\label{eq:rho_HS}\\
&s_{\rm VS}(x') = \frac{2\pi}{45}g_{\ast}^{S}\left(m_{\rm DM}/x'\xi\right)\left(\frac{m_{\rm DM}}{x'\xi}\right)^{3}\label{eq:s_VS},\\
&s_{\rm HS}(x') = \frac{2\pi}{45}g'^{S}_{\ast}\left(m_{\rm DM}/x'\right)\left(\frac{m_{\rm DM}}{x'}\right)^{3}\label{eq:s_HS}.
}

\noindent Using the instantaneous freeze-out approximation described above and presented in Appendix \ref{app:inst_FO}, one obtains

\myeq{
&\Omega _{\rm DM} h^{2} = 4.7\cdot10^8 \frac{g_{\rm DM}^{n}\sqrt{G^{\rm eff}_{\ast}(T_{\rm dec})}x'_{\rm dec}}{G^{S}_{\ast}(T_{\rm dec})M_{\rm pl}\langle\sigma v\rangle\,{\rm GeV}}\times\frac{T'_{\rm dec}}{T_{\rm dec}}
\label{eq:Oh2_FO_TpT}
}

\noindent with $x'_{\rm dec}$ given by (see Appendix \ref{app:inst_FO}),

\myeq{
&x'_{\rm dec} \simeq \ln\left[0.038\left(\frac{T'_{\rm dec}}{T_{\rm dec}}\right)^2\left\langle\sigma v\right\rangle m_{\rm pl}M_{\rm DM}\left(\frac{g_{\rm DM}}{\sqrt{G^{\rm eff}_{\ast}(T_{\rm dec})}}\right)\right]\nonumber\\
&\hspace{0.5cm}+\frac{1}{2}\ln\ln\left[0.038\left(\frac{T'_{\rm dec}}{T_{\rm dec}}\right)^2\left\langle\sigma v\right\rangle M_{\rm pl}m_{\rm DM}\left(\frac{g_{\rm DM}}{\sqrt{G^{\rm eff}_{\ast}(T_{\rm dec})}}\right)\right].\label{eq:xp_dec}
}

\noindent In the non-relativistic regime, the initial amount of DM may lie well above the abundance we will ultimately need such that one needs a large Boltzmann suppression, i.e. a large enough annihilation cross section. For large DM masses, the needed DM annihilation cross section could eventually violate the unitarity constraint \cite{Griest:1989wd}. The unitarity upper bound on the cross section is given by \cite{Griest:1989wd},

\myeq{
\sigma_{\rm ann.}v<\frac{\pi (2l+1)}{p^2_{\rm DM}}v=\frac{\pi(2l+1)}{m^2_{\rm DM}}\frac{(1-v^2/4)}{v/4},
\label{eq:unitaritygeneral}
}

\noindent where $v$ is the relative velocity between both annihilating DM particles and $l$ is the angular momentum quantum number, as in Chapter \ref{ch:som}, between the in-going particles. Since in Eq. \ref{eq:Oh2_FO_TpT}, it is the thermally average cross section which is involved, we have to take the thermally average of Eq. \ref{eq:unitaritygeneral}. This can be done by integrating Eq. \ref{eq:unitaritygeneral} over all possible velocities

\myeq{
\left\langle \sigma v\right\rangle \equiv \frac{\int\sigma_{\rm ann.} v f_{v}(E_{1})f_{v}(E_{2})\diff p_{1}^{3}\diff p_{2}^{3}}{\int f_{v}(E_{1})f_{v}(E_{2})\diff p_{1}^{3}\diff p_{2}^{3}},\label{eq:thermalaverage}
}

\noindent where 1 and 2 refer to the two in-going particles of energy $E_{1,2}$ and momentum $\vec{p}_{1,2}$. The velocity distribution, $f_{v}$, can be either the Fermi-Dirac or the Bose-Einstein distribution depending on the spin of the incident particles. This integration gives the following upper bound on the annihilation cross section,

\myeq{
\left\langle\sigma v\right\rangle<\frac{\pi (2l+1)}{4m_{\rm DM}^2}x'^2\mathcal{I}_{\epsilon}(x'),
\label{eq:unitaritygeneralaverage}
}

\noindent with $\epsilon = \pm 1$ for a fermion or a boson respectively and where the function $\mathcal{I}$ in Eq. \ref{eq:unitaritygeneralaverage} is defined as following,

\myeq{
\mathcal{I}_{\epsilon}(x')\equiv \frac{1}{N^{2}}\cdot \int_{4x'^{2}}^{\infty}\diff w\int_{\sqrt{w}}^{\infty}\diff k_{+}\int_{-k_{-,max}}^{k_{-,max}}\diff k_{-}\times\left\lbrace\frac{\sqrt{w/(w-4x'^{2})}}{\left(e^{\frac{k_{+}+k_{-}}{2}}+\epsilon\right)\left(e^{\frac{k_{+}-k_{-}}{2}}+\epsilon\right)}\right\rbrace.
}

\noindent with $k_{\pm}\equiv (E_{1}\pm E_{2})/T_{\rm dec}'$, $w\equiv s/T_{\rm dec}'^{2}$,  $k_{-,max}\equiv \sqrt{1-4x'^{2}/w}\sqrt{k_{+}^{2}-w}$, $s\equiv (p_1+p_2)^2$ and $N\equiv \int_{x'}^{\infty}\frac{\sqrt{k^{2}-x'^{2}}}{e^{k}+\epsilon}k \diff k$. Note that in the classical regime where one can use the Maxwell-Boltzmann distribution instead of the Fermi-Dirac or Bose-Einstein ones, one can take the relativistic and the non-relativistic limits. This would greatly simplify this expression for $\mathcal{I}_{\epsilon}$. In the relativistic regime we get,

\myeq{
\left\langle\sigma v\right\rangle \simeq {n_\epsilon\over T^{\prime 2}},
}

\noindent with $n_\epsilon = 5\pi/12$ for a fermion and $n_\epsilon = 15\pi/16$ for a boson in the relativistic limit $T^\prime \gg m_{\rm DM}$. In the non-relativistic limit $T^\prime \ll m_{\rm DM}$, we have instead

\myeq{
\left\langle\sigma v\right\rangle \simeq {4 \pi \over m_{\rm DM}^{2}}\frac{K_2(2 x')}{K_2^2(x')},
}

\noindent with $x' = m_{\rm DM}/T'$.
\\

Now, one can use the unitarity upper bound of Eq. \ref{eq:unitaritygeneralaverage} on the DM annihilation cross section to constraint the hidden-to-visible temperature ratio at DM decoupling as a function of the DM mass. Depending on which of the visible and the hidden sectors dominates at DM decoupling, one face two distinct physical situations. To understand which one dominates, let us compare their energy and entropy densities (see Eqs. \ref{eq:rho_VS} to \ref{eq:s_HS}):

\myeq{
&r_{\rho}(x')\equiv \frac{\rho_{\rm HS}(x')}{\rho_{\rm VS}(x')},\\
&r_{S}(x')\equiv \frac{s_{\rm HS}(x')}{s_{\rm VS}(x')}.
}

\noindent As a result, the HS dominates the expansion rate and entropy density of the Universe if,

\myeq{
&r_{\rho}>1\hspace{0.5cm}\Rightarrow\hspace{1cm}\left(\frac{T'}{T}\right)>\left(\frac{g^{\rm eff}_{\ast}(m_{\rm DM}/x'\xi)}{g'^{\rm eff}_{\ast}(m_{\rm DM}/x')}\right)^{1/4},\label{eq:r_rho}\\
&r_{S}>1\hspace{0.5cm}\Rightarrow\hspace{1cm}\left(\frac{T'}{T}\right)>\left(\frac{g^{S}_{\ast}(m_{\rm DM}/x'\xi)}{g'^{S}_{\ast}(m_{\rm DM}/x')}\right)^{1/3}.\label{eq:r_s}
}

\noindent The transition between a Universe dominated by the HS or by the VS is model dependent as it depends on the number of relativistic degrees of freedom in the HS. However, this number cannot be smaller than one and in a VS where there are only SM particles it is smaller than $\sim 106$,. In this case, the Eqs. \ref{eq:r_rho} and \ref{eq:r_s} hold as soon as $T'/T$ is above $\sim$ (a few)\footnote{Note that the two conditions $r_{\rho}>1$ and $r_{S}>1$ are very similar and differ from a factor of order unity.}. 
\\

Now that we know that, if $T'/T<$ (a few), the Universe expansion rate and its entropy density are dominated by the VS and by the HS otherwise, one can plug the maximum annihilation cross section allowed by unitarity in Eq. \ref{eq:Oh2_FO_TpT} for both cases. Saturating Eq. \ref{eq:unitaritygeneralaverage} gives,

\myeq{
\frac{T'_{\rm dec}}{T_{\rm dec}} = 1.18\times 10^{5}\times\left(\frac{100 \text{ GeV}}{m_{\rm DM}}\right)^{2}\times\left(\frac{g_{\ast}^{S}(T_{\rm dec})}{\sqrt{g_{\ast}^{\rm eff}(T_{\rm dec})}}\right)\times\mathcal{I}_{\epsilon}(x'_{\rm dec})\times x'_{\rm dec},\label{eq:UB_VS}
}

\noindent for a Universe dominated by the VS and

\myeq{
&m_{\rm DM} = 35\, \text{TeV}\times\left(\frac{g'^{S}_{\ast}(T'_{\rm dec})}{\sqrt{g'^{\rm eff}_{\ast}(T'_{\rm dec})}}\right)^{1/2}\times\left(\mathcal{I}_{\epsilon}(x'_{\rm dec})\times x'_{\rm dec} \right)^{1/2},\label{eq:UB_HS}
}

\noindent for a Universe dominated by the HS. Let us emphasise that in the latter case, the DM abundance does not depend on the hidden-to-visible temperature ratio. This is to be expected as in this case, the VS contribution can be totally neglected, i.e. neither the energy nor the entropy densities depend on this ratio (see Eq. \ref{eq:rho_HS} and \ref{eq:s_HS} respectively). As a consequence, the maximal DM mass has a unique value which does not depend on the temperature ratio.
\\

\begin{center}
\begin{figure}[h!]
\centering
\includegraphics[scale=0.825]{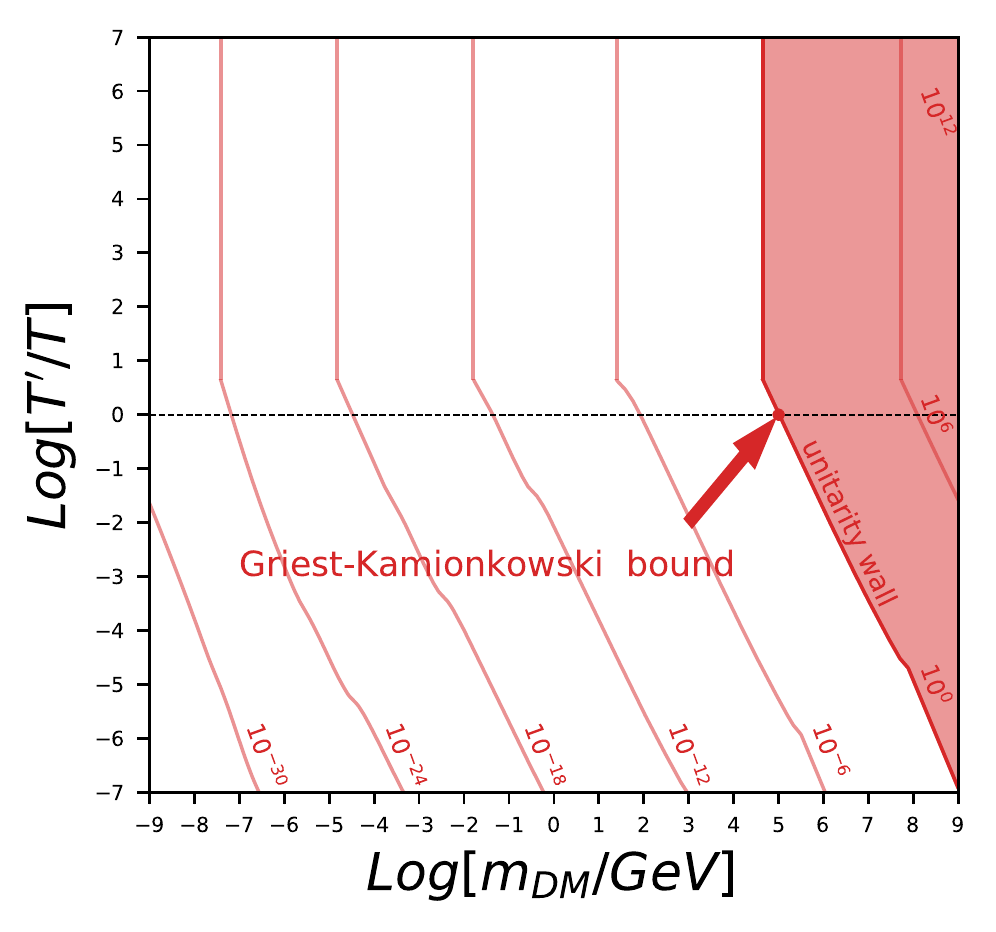}
\caption[Contour of the normalised DM relic abundance in the temperature ratio versus DM mass plane for a non-relativistic decoupling]{Contour of the normalised DM relic abundance $\Omega_{\rm DM}h^{2}/0.1188$ in the $T'/T$ vs DM mass plane for a non-relativistic decoupling in which the annihilation process saturates the unitarity bound.}
\label{fig:uni_ceiling}
\end{figure}
\end{center}

Figure \ref{fig:uni_ceiling} shows contours of the normalised DM relic abundance $\Omega_{\rm DM}h^{2}/0.1188$ in the $T'/T$ vs DM mass plane for a non-relativistic decoupling in which the annihilation process saturates the unitarity bound. This figure indicates how the temperature ratio versus DM mass plane is constrained by the unitarity bounds obtained in the case of a Universe dominated by the VS, $T'/T<$ (a few), and the case where it is dominated by the HS, $T'/T>$ (a few). On the same Figure, is also shown the Griest-Kamionkowski bound \cite{Griest:1989wd} which is the very well known upper bound on the mass of a DM thermal candidate, resulting from unitarity, in the case where there is only one bath, i.e. everything thermalised together such that $T'=T$.

\subsection{Thermalised Hidden Sector requirement}\label{sec:therm}
Now that we have seen how the parameter space is constrained in a model independent way by the \textit{relativistic floor} and the \textit{unitarity wall}, one should add one last theoretical constraint to this picture. Indeed, as we are interested in thermal DM candidate, we should impose that the HS thermalised at some point in the early Universe. This can be done using the maximal cross section allowed by unitarity (Eq. \ref{eq:unitaritygeneralaverage}) and imposing that the interaction rate was larger than the expansion rate at some point before DM decoupling (i.e. for $T'_{\rm th}> T'_{\rm dec}$ with $T'_{\rm th}$ the temperature of thermalisation). In practice we checked for every point of the parameter space if the following condition was fulfilled,

\myeq{
\left(n_{\rm DM}\langle\sigma v\rangle\right) _{T'>T'_{\rm dec}}>H\vert_{T>T_{\rm dec}},\label{eq:therm_condition}
}

\noindent where $\langle\sigma v\rangle$ is given by Eq. \ref{eq:unitaritygeneralaverage}, $H$ by Eq. \ref{eq:H_universe} and $n_{\rm DM}$ by its full non-relativistic form,

\myeq{
n_{\rm DM}(T') = \frac{g_{\rm DM}T'^{3}}{2\pi^{2}}\int_{x'}^{\infty}\frac{\sqrt{w^{2}-x'^{2}}}{e^{w}\pm 1}w\diff w,\label{eq:nrel_nDM}
}

\noindent where the $\pm$ sign stems for a fermion and a boson respectively.
\\

\begin{figure}
\centering
\includegraphics[height=7.5cm]{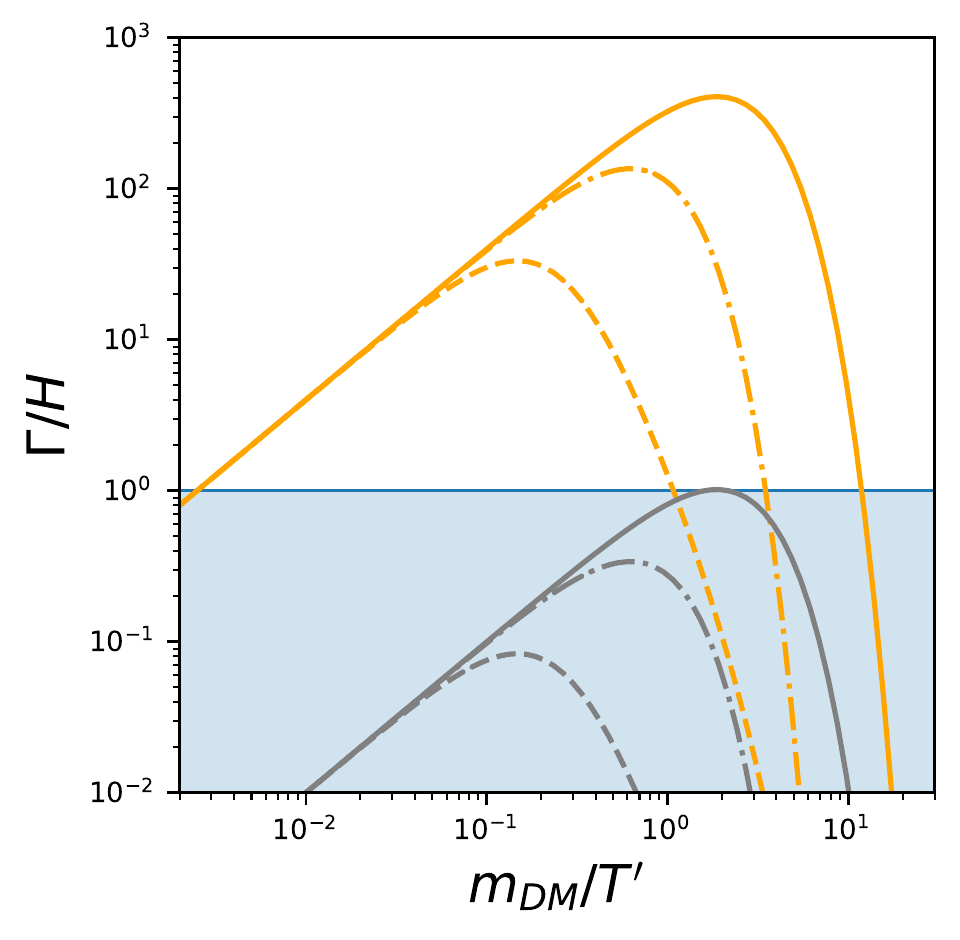}
\caption[Schematic behaviour of interaction rates]{Schematic behaviour of interaction rates, $\Gamma$, with respect to the expansion rate, as function of $m_{\rm DM}/T'$ for three scenarios for DM decoupling: the DM particles become non-relativistic (solid lines), the DM annihilates into heavier particles of mass $m'$ (here chosen to be $m'/m_{\rm DM}=3$) (dot-dashed lines) and the DM interacts through some heavy mediator and decouples while still relativistic (dashed lines). We show these three cases for the same DM mass but for two choices of $T'/T$ (orange and grey lines, respectively), see text for more details.
}
\label{fig:rates}
\end{figure}

Even though we aim to be as model independent as possible, the way that DM decouples is model dependent such that we will here consider three generic scenarios. The first possibility we consider is the usual non-relativistic DM decoupling, see solid lines in Figure \ref{fig:rates} which shows $\Gamma/H$ as a function of $m_{\rm DM}/T'$ for the three DM decoupling scenarios we consider here and for two choices of $T'/T$. That is to say that the DM particles were still in thermal equilibrium when $T'\sim m_{\rm DM}$ and that its number density changed from $n_{\rm DM} \propto T'^3$ to being Boltzmann suppressed $n_{\rm DM} \propto \exp(- m_{\rm DM}/T')$. The two other DM decoupling possibilities we consider are the two relativistic decoupling scenarios already mentioned in Subsection \ref{subsec:rel_floor}: the DM annihilation process is cut off either by a heavy mediator mass or by the mass of the final state particles, see the dashed and dot-dashed lines respectively in Figure \ref{fig:rates}.
\\

Figure \ref{fig:rates} shows the three DM decoupling scenarios just introduced assuming the same DM mass and for two choices of the hidden-to-visible temperature ratio $T'/T$ (orange and grey lines). The orange curves describe the behaviour of DM candidate which where in thermal equilibrium at a temperature $T'$ with $T'_{\rm eq}>T'>T'_{\rm dec}$. Decreasing (increasing) the temperature ratio would move down (up) the orange curves such that one can conclude from Figure \ref{fig:rates} that the temperature range within which the DM was in chemical equilibrium shrinks (resp. expands) as $T'/T$ decreases (resp. increases). If one keep decreasing $T'/T$, one will end up with the limit case where the equilibrium and decoupling temperatures are equal $T_{\rm eq}=T_{\rm dec}$, see the solid grey curve in Figure \ref{fig:rates}. This corresponds to the case of a mildly non-relativistic DM particle freeze-out such that the candidate lies close to the relativistic floor and such that the relativistic form of Eq. \ref{eq:nrel_nDM} given in Eq. \ref{eq:rel_nDM} can be used as a very good proxy. In such an approximation, one can solve analytically Eq. \ref{eq:therm_condition} and get the temperature ratio as a function of the DM mass for which the unitarity cross section is not strong enough to allow any thermalisation within the HS. We have,

\myeq{
\frac{T'_{\rm th}}{T_{\rm th}} > 1.2\times 10^{-8}\times\left(\frac{m_{\rm DM}}{100 \text{ GeV}}\right)^{1/2}\left(\frac{\sqrt{G^{\rm eff}_{\ast}(T_{\rm th})}}{g_{\rm DM}}\right)^{1/2}\left(\frac{1}{x'_{\rm th}\mathcal{I}_{\epsilon}(x'_{\rm th})}\right)^{1/2}.\label{eq:TH}
}

\begin{center}
\begin{figure}[h!]
\centering
\includegraphics[scale=0.825]{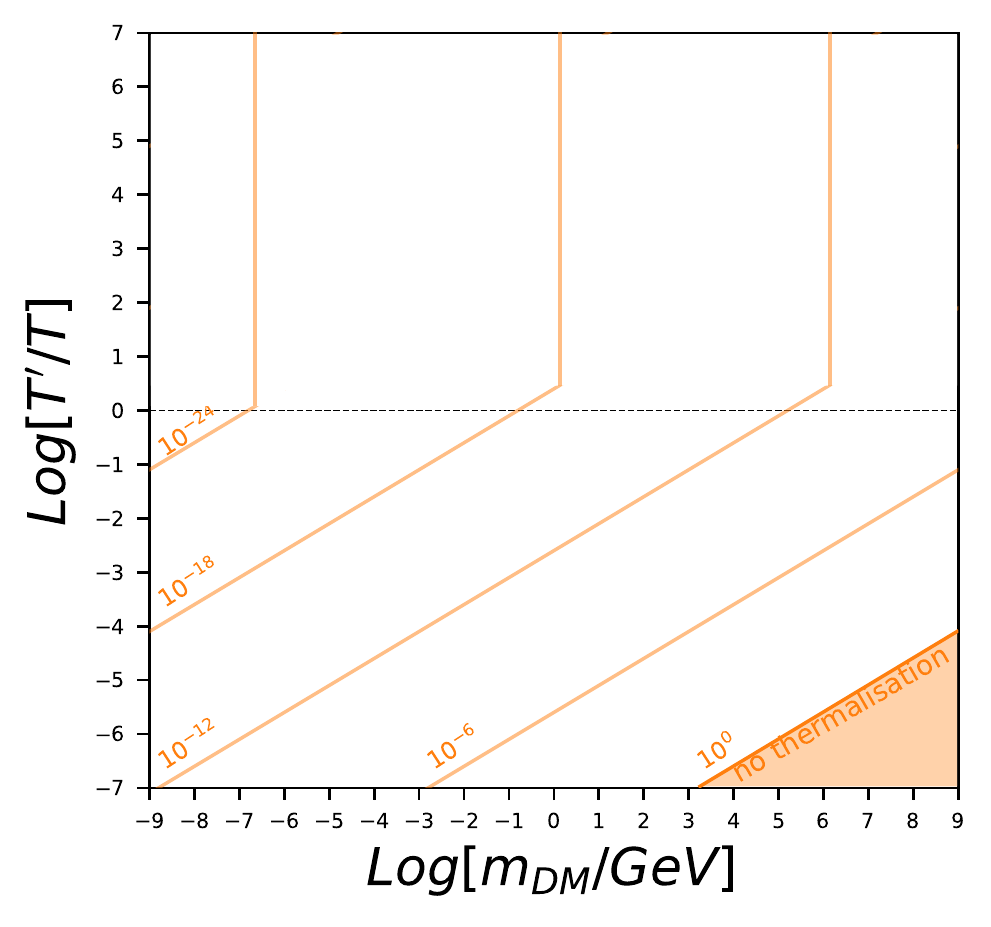}
\caption[Contour of the normalised DM annihilation cross section in the temperature ratio versus DM mass plane for a non-relativistic decoupling]{Contour of the normalised DM annihilation cross section $\langle\sigma v\rangle/\langle\sigma v\rangle_{ub}$ (with $\langle\sigma v\rangle_{ub}$ given by Eq. \ref{eq:unitaritygeneralaverage}) in the $T'/T$ vs DM mass plane for a non-relativistic decoupling.}
\label{fig:thermalisation}
\end{figure}
\end{center}

\noindent This gives the diagonal and vertical dark orange lines in Figure \ref{fig:thermalisation}. Note also that for smaller values of the DM annihilation cross section than the one allowed by unitarity of Eq. \ref{eq:unitaritygeneralaverage}, the thermalisation condition requires the DM mass to be smaller. This is also shown in Figure \ref{fig:thermalisation} for various values of the annihilation cross section with respect to the maximum value of it allowed by unitarity (Eq. \ref{eq:unitaritygeneralaverage}).

\subsection{Thermal DM mass range}
\begin{center}
\begin{figure}[h!]
\centering
\includegraphics[scale=0.825]{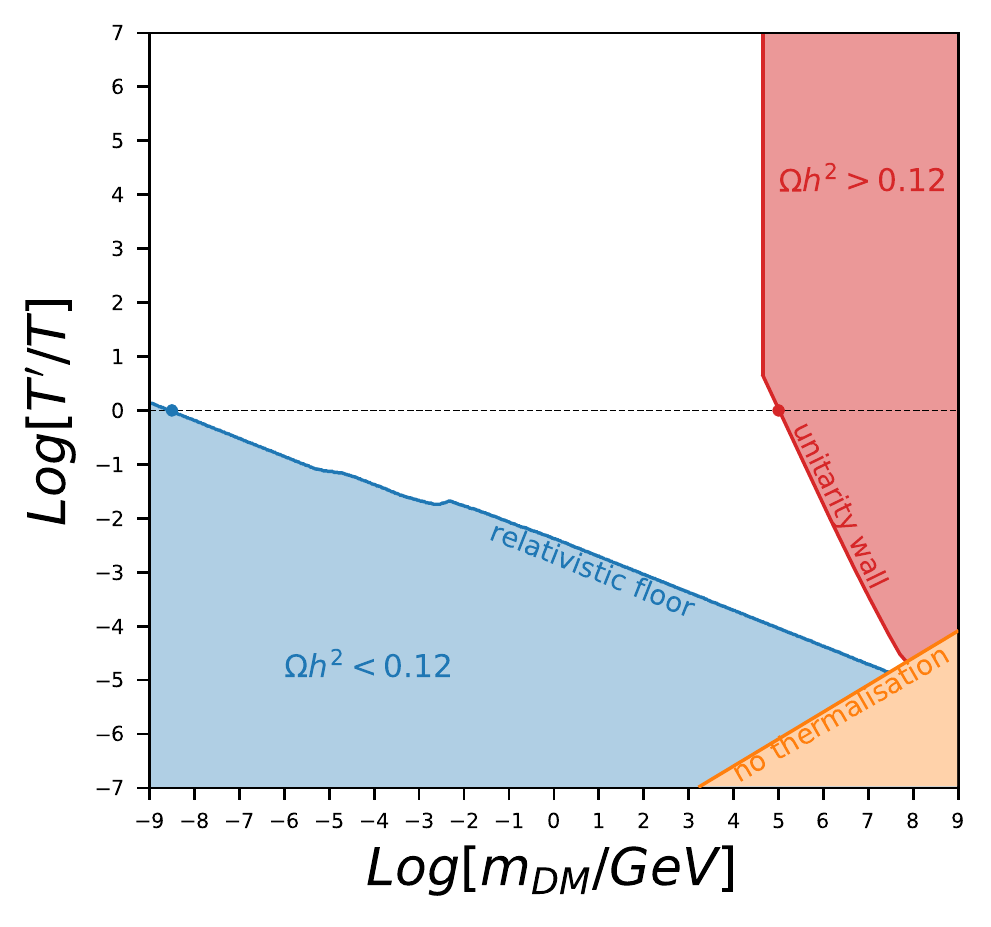}
\caption[Theoratical constraints from the relativistic floor, the unitarity wall and the thermalisation condition on the temperature ratio versus DM mass plane]{Theoratical constraints from the relativistic floor (blue), the unitarity wall (red) and the thermalisation condition (orange) on the temperature ratio versus DM mass plane.}
\label{fig:theroretical_bounds}
\end{figure}
\end{center}

One can now bring together the three model independent constraints on the temperature ratio versus DM mass plane. This has been done in Figure \ref{fig:theroretical_bounds} where one can see how the relativistic floor (blue), the unitarity wall (red) and the thermalisation condition (orange) constrain the parameter space. From Figure \ref{fig:theroretical_bounds}, one can see how the mass range of a thermal DM candidate evolves with the hidden-to-visible temperature ratio. As said above, for a unique bath in which the DM thermalises with the SM, the DM mass was constrained below by the Cowsik-McClelland bound \cite{Cowsik:1972gh} and above by the Griest-Kamionkowski bound \cite{Griest:1989wd},

\myeq{
\text{ eV}\lesssim m_{\rm DM}\lesssim 100\text{ TeV}\hspace{1cm}\text{for }T'=T.\label{eq:mass_range_1}
}

\noindent Thanks to Eqs. \ref{eq:TpTlowerbound} and \ref{eq:UB_VS} one can now constrain the mass range of a DM thermal candidate for a more generic scenario in which the DM did not necessarily thermalise with the VS, but did within the HS. We have, $m_{\rm CM}\left(\frac{T'_{\rm dec}}{T_{\rm dec}}\right)\lesssim m_{\rm DM}\lesssim m_{\rm GK}\left(\frac{T'_{\rm dec}}{T_{\rm dec}}\right)$, where the generalisation of the Cowsik-McClelland and Griest-Kamion-kowski masses are function of the hidden-to-visible temperature ratio at DM decoupling. These masses are defined for a Universe dominated by the VS as

\myeq{
&m_{\rm CM} \equiv 1.5\text{ eV }\times\left(\frac{T_{\rm dec}}{T'_{\rm dec}}\right)^{3}\times\frac{g^{S}_{\ast}(T_{\rm dec})g'^{S}_{\ast}(T'_{0})}{g^{n}_{\rm DM}(T'_{\rm dec})g'^{S}_{\ast}(T'_{\rm dec})},\label{eq:GK_TpT_VS}\\
&m_{\rm GK} \equiv 34\text{ TeV }\times\left(\frac{T_{\rm dec}}{T'_{\rm dec}}\right)^{1/2}\times\left(\frac{g^{S}_{\ast}(T_{\rm dec})}{\sqrt{g^{\rm eff}_{\ast}(T_{\rm dec})}}\right)^{1/2}\times\left(\mathcal{I}_{\epsilon}(x'_{\rm dec})\times x'_{\rm dec}\right)^{1/2},\label{eq:GK_TpT_VS}
}

\noindent while for a Universe dominated by the HS, we have instead

\myeq{
&m_{\rm CM} \equiv 1.5\text{ eV }\times\left(\frac{T_{\rm dec}}{T'_{\rm dec}}\right)^{3}\times\frac{g^{S}_{\ast}(T_{\rm dec})g'^{S}_{\ast}(T'_{0})}{g^{n}_{\rm DM}(T'_{\rm dec})g'^{S}_{\ast}(T'_{\rm dec})},\label{eq:CM_TpT_HS}\\
&m_{\rm GK} \equiv 35\, \text{TeV}\times\left(\frac{g'^{S}_{\ast}(T'_{\rm dec})}{\sqrt{g'^{\rm eff}_{\ast}(T'_{\rm dec})}}\right)^{1/2}\times\left(\mathcal{I}_{\epsilon}(x'_{\rm dec})\times x'_{\rm dec} \right)^{1/2}.\label{eq:GK_TpT_HS}
}

\noindent Note that the width of the mass range for a thermal DM candidate decreases with the hidden-to-visible temperature ratio (as it can be seen in Figure \ref{fig:theroretical_bounds}), but this mass range move to higher masses.
\\

Finally, let us emphasise that the maximum mass allowed for a DM thermal candidate does not exactly correspond to the minimum allowed value of the hidden-to-visible temperature ratio. Indeed, the maximum DM mass one can reach corresponds to the intersection of the unitarity wall and the thermalisation line of Figure \ref{fig:theroretical_bounds} and is given by Eq. \ref{eq:GK_TpT_VS}. One gets

\myeq{
& m_{\rm DM}^{max} \simeq 52\text{ PeV},
}

\noindent which corresponds to a temperature ratio at decoupling of

\myeq{
& \frac{T'_{\rm dec}}{T_{\rm dec}} \simeq 6.9\times 10^{-5}.
}

\noindent Instead, the minimum allowed value of the temperature ratio corresponds to the intersection of the relativistic floor and the thermalisation line. It can be obtained using Eqs. \ref{eq:TpTlowerbound} and \ref{eq:TH}, we have

\myeq{
& \left(\frac{T'_{\rm dec}}{T_{\rm dec}}\right)_{min} \simeq 1.4\times 10^{-5},
}

\noindent which corresponds to a slightly lower DM mass of

\myeq{
& m_{\rm DM} \simeq 30\text{ PeV}.\label{eq:m_max_rel}
}

\noindent Although this 30 PeV is not an absolute upper bound on the DM mass, it nevertheless constitutes an upper bound on the DM mass for all relativistic decoupling scenario discussed above.

\section{Observational constraints}\label{sec:2d_observational}
Let us now take a look at some observations which can further constrain the DM thermal candidate parameter space, still in a model independent way.

\subsection{\textbf{$N_{\rm eff}$}}\label{sec:neff}
We have already discussed the $N_{\rm eff}$ constraint from CMB data in Section \ref{sec:CST-CMB} showing that the effective number of relativistic degrees of freedom can play an important role in the Universe expansion. Indeed, it is well known that the Universe used to be in a radiation dominated era at the time of BBN. Thus, adding new relativistic degrees of freedom in the early Universe would increase the total energy density of the Universe and, consequently, the Hubble expansion rate. This would therefore modify the well measured relative abundances of light nuclei which were fixed by the BBN process. Thus, to avoid that those additional particles are still relativistic at the time of BBN, one could require that $T'_{\rm dec}>T'_{\rm BBN}$ such that one would be sure, in a model independent way, that DM was non-relativistic at BBN. However, we will see in Chapters \ref{ch:TpT} and \ref{ch:other_minimal} that it is not necessarily required. There are other ways to fulfil the $N_{\rm eff}$ constraint making the DM abundance negligible at BBN, but this is model dependent, see below.
\\

Focusing on a model independent $N_{\rm eff}$ constraint, we write the extra contribution to $N_\text{eff}$ due to DM at $T_{\rm BBN} \simeq 1$ MeV using $\Delta N_{\rm eff}\equiv \rho_{\rm DM}(T'_{\rm BBN})/\rho_{\nu}(T_{\rm BBN})$. The DM energy density is given by its integral form at equilibrium while the neutrino energy density is simply given in terms of the photon energy density as they already have decoupled from the SM thermal bath, $\rho_{\nu}=\frac{7}{8}\left(\frac{4}{11}\right)^{4/3}\rho_{\gamma}$. We have,

\myeq{
&\rho_{\rm DM}(T') = \frac{g_{\rm DM}T'^{4}}{2\pi^{2}}\int_{x'}^{\infty}\frac{\sqrt{w^{2}-x'^{2}}}{e^{w}\pm 1}w^{2}\diff w,\label{eq:rhorel_nDM}\\
&\rho_{\nu}(T) = \frac{7\pi^{2}}{240}\left(\frac{4}{11}\right)^{4/3}g_{\gamma}T^{4},
}

\noindent where $g_{\gamma}=2$ is the photon number of degrees of freedom. The additional contribution to $N_{\rm eff}$ from DM is therefore given by,

\myeq{
\Delta N_\text{eff} &\simeq  g_{\rm DM}\frac{60}{7\pi^4}\left(\frac{11}{4}\right)^{4/3}\left(\frac{T'_{\rm BBN}}{T_{\rm BBN}}\right)^{4} \int_{x'_{\rm BBN}}^\infty \frac{\sqrt{w^2 - x'^2_{\rm BBN}}}{e^{w} \pm 1}w^{2}\diff w,
\label{eq:neff}
}

\noindent where $x'_{\rm BBN}\equiv m_{\rm DM}/T'_{\rm BBN}$. The latest constraint given by the Planck collaboration \cite{Aghanim:2018eyx} is $N_\text{eff} = 2.99 \pm 0.17$ which is similar to the one obtained from BBN \cite{Fields:2019pfx}. SM predictions give $N_\text{eff} = 3.04$ (see \cite{Bennett:2019ewm} for example) such that one can constrain the quantity given in Eq. \ref{eq:neff}: $\Delta N_\text{eff} \leq 0.29$ at $2\sigma$.
\\

We emphasise the fact that we are, for the moment, considering only contributions to $\Delta N_{\rm eff}$ from the DM. Indeed, contributions from additional degrees of freedom within the HS are possible, but are also much more model dependent. For this reason, we will start discussing, in a model independent way, contributions from DM and later we will discuss how the picture would change if there are additional degrees of freedom within the HS.
\\

Applying the constraint $\Delta N_\text{eff} \leq 0.29$ on the DM contribution leads to the green region of Figure \ref{fig:polygon} which presents two distinct features. These features can be understood if one considers the relativistic and non-relativistic behaviour of Eq. \ref{eq:neff}. Indeed, lower the DM mass is (left hand side of Figure \ref{fig:polygon}) more relativistic the DM was at BBN. In the limit were the DM was relativistic at BBN, Eq. \ref{eq:neff} and $\Delta N_\text{eff} \leq 0.29$ give

\myeq{
\frac{T'_{\rm BBN}}{T_{\rm BBN}}\leq 0.44\times\left(\frac{3.5}{g_{\rm DM}^{\rm eff}}\right)^{1/4},\label{eq:neff_rel}
}

\noindent where $g_{\rm DM}^{\rm eff}=g_{\rm DM}$ for a boson and $g_{\rm DM}^{\rm eff}=7g_{\rm DM}/8$ for a fermion. This gives the constraint on the temperature ratio at DM decoupling, 

\myeq{
\frac{T'_{\rm dec}}{T_{\rm dec}}\leq 0.44\times\left(\frac{3.5}{g_{\rm DM}^{\rm eff}}\right)^{1/4}\times\left(\frac{g'^{S}_{\ast}(T'_{\rm BBN})}{g'^{S}_{\ast}(T'_{\rm dec})}\right)^{1/3}\left(\frac{g'^{S}_{\ast}(T'_{\rm dec})}{g^{S}_{\ast}(T_{\rm BBN})}\right)^{1/3},\label{eq:neff_rel_dec}
}

\begin{center}
\begin{figure}[h!]
\centering
\includegraphics[scale=0.825]{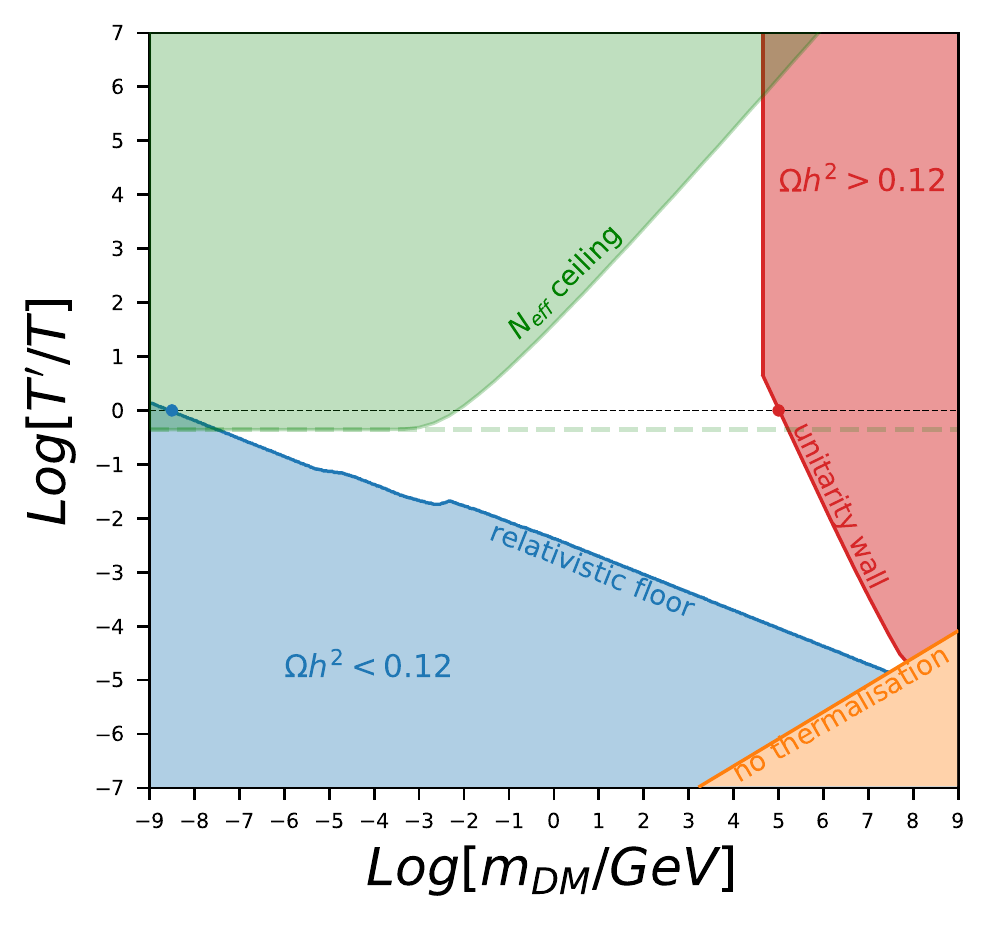}
\caption[$N_{\rm eff}$ exclusion area together with the relativistic floor, unitarity wall and thermalisation condition]{$N_{\rm eff}$ exclusion area (green) together with the relativistic floor, unitarity wall and thermalisation condition.}
\label{fig:Neff_HS}
\end{figure}
\end{center}

\noindent From Eq. \ref{eq:neff_rel_dec}, one can see that the bound on the hidden-to-visible temperature ratio at DM decoupling does not depend on the DM mass for a relativistic decoupling as expected. This regime is relevant for a DM which is relativistic at BBN, that is to say for $m_{\rm DM}\lesssim$ MeV and it can be seen in Figure \ref{fig:Neff_HS} as the horizontal plateau of the green region.
\\

On the other hand, if the DM mass lies above the MeV scale, the DM may have been already non-relativistic at BBN and its abundance would have been further suppressed. In this regime, using Eq. \ref{eq:neff} the constraint at BBN is 

\myeq{
\frac{T'_{\rm BBN}}{T_{\rm BBN}}\leq 0.51\times\left(\frac{4}{g_{\rm DM}}\right)^{1/4}(x'_{\rm BBN})^{-5/8}e^{x'_{\rm BBN}/4},\label{eq:neff_non_rel}
}

\noindent while at DM decoupling we have,

\myeq{
&\hspace{-0.1cm}\frac{T'_{\rm dec}}{T_{\rm dec}}\leq 0.51\times\left(\frac{4}{g_{\rm DM}}\right)^{1/4}(x'_{\rm BBN})^{-5/8}e^{x'_{\rm BBN}/4}\times\left(\frac{g'^{S}_{\ast}(T'_{\rm BBN})}{g'^{S}_{\ast}(T'_{\rm dec})}\right)^{1/3}\left(\frac{g'^{S}_{\ast}(T'_{\rm dec})}{g^{S}_{\ast}(T_{\rm BBN})}\right)^{1/3}.\label{eq:neff_non_rel_dec}
}

\noindent Eq. \ref{eq:neff_non_rel_dec} gives us a more complicated upper bound on the temperature ratio as it depends non trivially on the DM mass (through $x'_{\rm BBN}$). However, one can rewrite $x'_{\rm BBN}=m_{\rm DM}/T_{\rm BBN}\times(T_{\rm BBN}/T'_{\rm BBN})$ such that one can solve Eq. \ref{eq:neff_non_rel_dec} for $T'_{\rm dec}/T_{\rm dec}$ as a function of $m_{\rm DM}$ if we fix $T_{\rm BBN}\simeq 1$ MeV. The result of this operation can be translated into the diagonal part of the green region of Figure \ref{fig:Neff_HS}.
\\

Now, we consider the possibility of extra degrees of freedom which may be contained in the HS. Contributions to $\Delta N_{\rm eff}$ from other HS particles is much more model dependent as it depends on how numerous they are at DM decoupling and if they are still relativistic or not. However, one can consider the simplest case where the DM annihilates into lighter particles which can decay or annihilate into SM particles. In this case, if they decay or annihilate into SM particles for example before the BBN occurs they will have no impact on BBN. Otherwise, their mass should be larger than $T'_{\rm BBN}$ in order to become non-relativistic before BBN and to have their abundance to be Boltzmann suppressed. However, if the mediator  mass is of the same scale than the DM mass, the $N_{\rm eff}$ constraint is still relevant, see Eqs. \ref{eq:neff_rel_dec} and \ref{eq:neff_non_rel_dec} and the green area in Figure \ref{fig:Neff_HS}. Instead, if the mediator is lighter than the DM and if the ratio of the masses is fixed, the diagonal part moves to the right by a factor of $m_{\rm DM}/m_{\rm med}$. Alternatively, if one fixes the mass of the mediator and not the mass ratio, one gets an unique upper bound on the hidden-to-visible temperature ratio: $T'/T < m_{\rm med}/T_{\rm BBN}$, in the same way that we obtained the horizontal part of the green area of Figure \ref{fig:Neff_HS}, see Eq. \ref{eq:neff_rel_dec}. In the massless mediator limit, we have

\myeq{
\frac{T'_{\rm dec}}{T_{\rm dec}}\leq 0.60\times\left(\frac{1}{g'^{\rm eff}_{\ast}}\right)^{1/4}\times\left(\frac{g'^{S}_{\ast}(T'_{\rm BBN})}{g'^{S}_{\ast}(T'_{\rm dec})}\right)^{1/3}\left(\frac{g'^{S}_{\ast}(T'_{\rm dec})}{g^{S}_{\ast}(T_{\rm BBN})}\right)^{1/3},\label{eq:neff_rel_dec_med}
}

\noindent where $g'^{\rm eff}_{\ast}$ includes all HS degrees of freedom which are still unsuppressed at the BBN time (see also \cite{Feng:2008mu}). This last constraint gives rise to an additional horizontal line depicted in light green in Figure \ref{fig:Neff_HS}. This extra line lies within the exclusion line from the contribution of DM for $m_{\rm DM}<$ MeV. This is due to the fact that we consider a Dirac DM ($g_{\rm DM}^{\rm eff}=3.5$) and a scalar mediator ($g_{\rm med}^{\rm eff}=1$). Thus, on the one hand, the region allowed by BBN extends to the white domain depicted in Figure \ref{fig:Neff_HS} if there are no extra degrees of freedom left at BBN epoch. On the other hand, values of the hidden-to-visible temperature ratio at DM decoupling larger than $\sim 1$ are excluded if there are still some extra degrees of freedom remaining at BBN time. We made this discussion as generic as possible, but of course, constraints from BBN can be much more restrictive in explicit models. For example, one could be worry if HS particles (DM or other) annihilate and/or decay during BBN. This could greatly impact the expansion rate and the energy transfer into the VS, see \cite{Berger:2016vxi,Hufnagel:2017dgo,Hufnagel:2018bjp}. However, this would require a concrete model and it goes out of the context of this chapter which is dedicated to model independent constraints on thermal DM candidate.

\subsection{Free-streaming constraints}\label{sec:streaming}
The second and last observational constraint we will consider in a model independent way is the so-called free-streaming (FS) constraint. This bound comes from the fact that a too light DM particle would have been relativistic for so long that it would have not permitted the formation of large scale structure which are observed today. The usual bound is given by data extracted from Lyman-$\alpha$ forest and gives $m_{\rm DM} > 5.3$ keV \cite{Irsic:2017ixq}. But, it is assuming a unique thermal bath in the early Universe, i.e. with $T'=T$. In this section, we will generalise this constraint for the case where $T'\neq T$ by converting the constraint on the average distance a collisionless DM particle travels after production, that is to say on the DM free-streaming horizon.
\\

The average momentum of a boson or a fermion population in thermal equilibrium with temperature $T'$ is given by 

\myeq{
\langle p \rangle \equiv \frac{\int d^3 p f(p) p}{\int d^3p f(p)} = T' \times 
\begin{cases}
3.15 \text{ for a fermion,} \\
2.70 \text{ for a boson.}
\end{cases}
}

\noindent Since a species can be characterised as non-relativistic when its average momentum lies around its mass, $\langle p \rangle\sim m$, the temperature for which a DM fermion candidate becomes non-relativistic (NR) is about $T'_{\rm NR}\simeq m_{\rm DM}/3.15$. This temperature can be converted in time unit thanks to the following relation: $t=1/2H(T')$ such that one can compute when the DM candidate stops to be relativistic. Depending on if this transition happens before of after the time of matter-radiation (MR) equality (i.e. $t_{\rm MR}= 1.9 \times 10^{11}$ s), the free-streaming horizon is given by $\lambda_{\rm FS}\equiv\int_{t_{\rm dec}}^{t_0} \langle v(t) \rangle/a(t) dt$ such that one has,

\myeq{
\lambda_{\rm FS} \simeq \frac{\sqrt{t_{\rm MR} t_{\rm NR}}}{a_{\rm MR}} \left( 5 + \log \frac{t_{\rm MR}}{t_{\rm NR}} \right) \left( \frac{g^{S}_{\ast}(T_{0}) + (\xi_{0})^{3}g'^{S}_{\ast}(T'_{0})}{g^{S}_{\ast}(T_{\rm dec}) + (\xi_{\rm dec})^{3}g'^{S}_{\ast}(T'_{\rm dec})} \right)^{1/3},\label{eq:fs_before}
}

\noindent if $t_{\rm NR}<t_{\rm MR}$ and where $a_{\rm MR} = 8.3 \times 10^{-5}$ is the scale factor at the time of matter-radiation equality. If $t_{\rm NR}>t_{\rm MR}$ we have instead, 

\myeq{
\lambda_{\rm FS} \simeq \left(\frac{6(t_{\rm MR}^2 t_{\rm NR})^{1/3} - t_{\rm MR}}{a_{\rm MR}} \right) \left( \frac{g^{S}_{\ast}(T_{0}) + (\xi_{0})^{3}g'^{S}_{\ast}(T'_{0})}{g^{S}_{\ast}(T_{\rm dec}) + (\xi_{\rm dec})^{3}g'^{S}_{\ast}(T'_{\rm dec})} \right)^{1/3}.\label{eq:fs_after}
}

\noindent In the last expression we used $\langle v (t) \rangle \simeq a_{\rm NR}/a \simeq (t_{\rm NR}/t)^{2/3}$ which is true only if $t>t_{\rm NR}$. Finally, one can verify that Eq. \ref{eq:fs_before} coincides with Eq. \ref{eq:fs_after} when $t_{\rm NR} = t_{\rm MR}$ as it should. 
\\

\begin{center}
\begin{figure}[h!]
\centering
\includegraphics[scale=0.825]{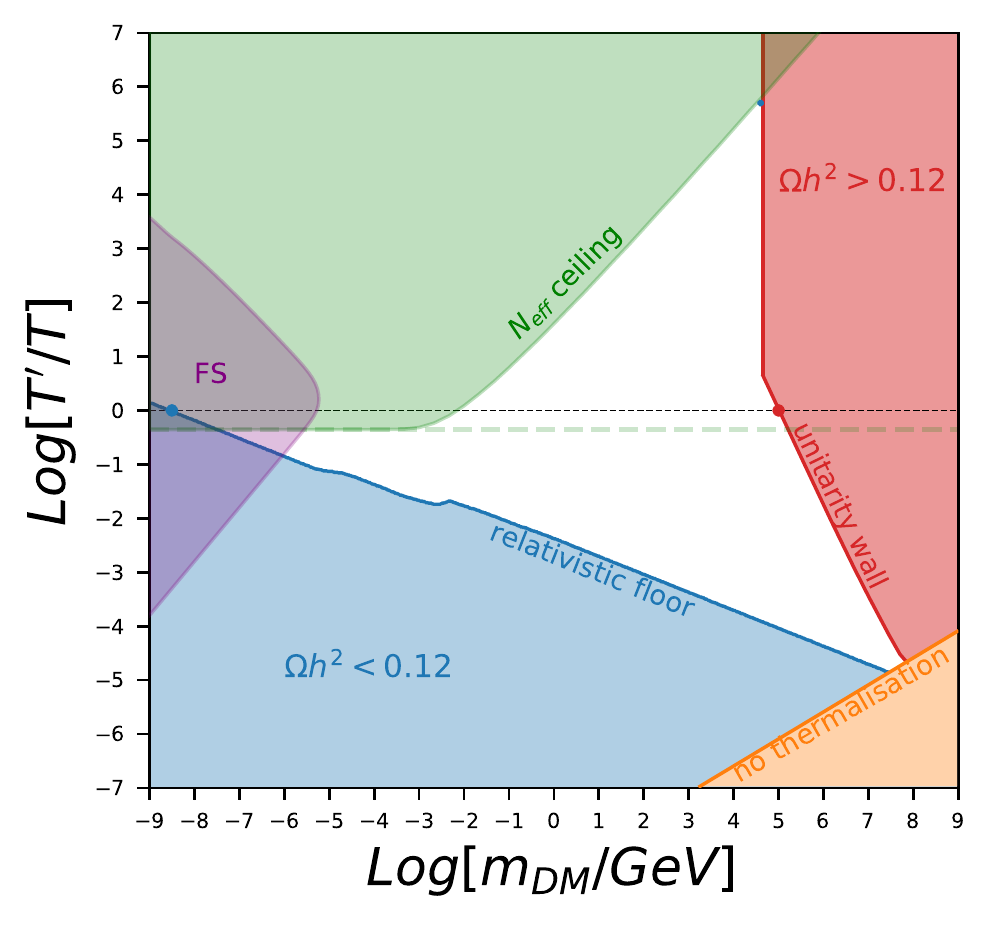}
\caption[Free-streaming exclusion area together with the $N_{\rm eff}$ ceiling, the relativistic floor, unitarity wall and thermalisation condition]{Free-streaming exclusion area (purple) together with the $N_{\rm eff}$ ceiling, the relativistic floor, unitarity wall and thermalisation condition.}
\label{fig:FS_HS}
\end{figure}
\end{center}

In order to obtain the free-streaming bound given in purple in Figure \ref{fig:FS_HS}, we imposed the usual free-streaming horizon upper bound given by $\lambda_{FS} \simeq 0.066$ Mpc, see \cite{Irsic:2017ixq}. The very specific behaviours of the purple exclusion area can be understood as following: if the hidden-to-visible temperature ratio is much smaller than unity ($T'\ll T$), the time when DM becomes non-relativistic goes like $t_{\rm NR}\sim (T'/T)^2/m_{\rm DM}^2$ such that the free-streaming length increases with the temperature ratio, $\lambda_{\rm FS} \propto (T'/T)/m_{\rm DM}$, up to the logarithm in Eq. \ref{eq:fs_before}. We have then that $T'/T<\left(m_{\rm DM}\times \text{factor}\right)$ if we impose that $\lambda_{\rm FS}<0.066$ Mpc. On the other hand, if the temperature ratio is larger than unity ($T'\gg T$), the time when DM becomes non-relativistic goes like $t_{\rm NR}\propto 1/T'^2$ such that the free-streaming length decreases with the temperature ratio, $\lambda_{\rm FS} \propto (T'/T)^{-4/3}/m_{\rm DM}$. The free-streaming horizon constraint gives then $T'/T> \left(\text{factor}/m_{\rm DM}^{3/4}\right)$.

\section{The 2D domain}
\begin{figure}[h!]
\centering
\includegraphics[scale=0.68]{FS_HS.pdf}
\includegraphics[scale=0.68]{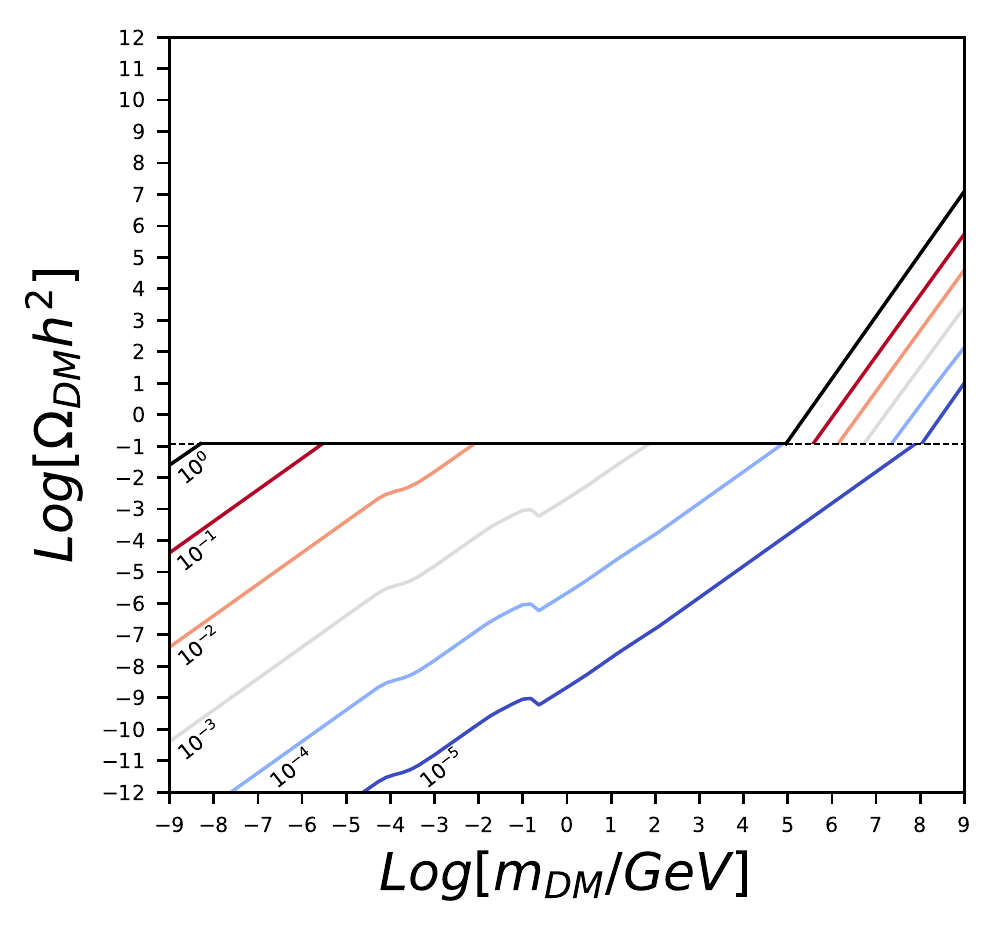}
\caption[Domain of thermal DM candidates]{Domain of thermal DM candidates (white). Left: Temperature ratio (at the time of DM decoupling) as function of the DM mass. Right: Relic density versus DM mass plane for various temperature ratios.}\label{fig:polygon}
\end{figure}

In this section, we will summarise the various theoretical and observational constraints discussed in Sections \ref{sec:2d_theoretical} and \ref{sec:2d_observational}. Left panel of Figure \ref{fig:polygon} shows bounds obtained from the relativistic floor of Eq. \ref{eq:TpTlowerbound} (blue), the unitarity wall of Eqs. \ref{eq:UB_VS} and \ref{eq:UB_HS} (red), the thermalisation line of Eq. \ref{eq:TH} (orange), the $N_{\rm eff}$ ceiling of Eqs. \ref{eq:neff_rel_dec} and \ref{eq:neff_non_rel_dec} (green) and, finally, the free-streaming constraint of Eqs. \ref{eq:fs_before} and \ref{eq:fs_after} (purple). Along the horizontal black dashed line for which $T'=T$, we recover the well known 1D mass range for a DM thermal candidate which was in equilibrium with the SM bath at decoupling and for which DM masses smaller than a few MeV are already excluded by BBN.
\\

This figure shows that the DM relic density lies below observations in the blue region (below the relativistic floor) and above observations in the red region (right of the unitarity wall). This means that suitable DM candidate should live within the white region of left panel of Figure \ref{fig:polygon}. That is to say that the DM mass and the hidden-to-visible temperature ratio at DM decoupling are bounded;

\myeq{
&m_{\rm DM} \in \left[1.1 \text{ keV}, 53 \text{PeV}\right],\\
&\frac{T'_{\rm dec}}{T_{\rm dec}} \in \left[1.6\times 10^{-5}, 5.0\times 10^{5}\right].
}

\noindent Note that the upper bound we get is as high as 53 PeV, which is about a factor 500 larger than the one we have for ordinary $T'=T$ freeze-out. The lower bound being as low as 1.1 keV, it is about a factor 1000 lower than for ordinary FO.
\\

Right panel of Figure \ref{fig:polygon} shows contour of the hidden-to-visible temperature ratio as a function of the DM mass and the DM relic density. The horizontal black dashed line indicates the observed DM relic density, i.e. $\Omega_{\rm DM} h^2 = 0.1188$. Below this line, the contours of the temperature ratio are obtained from a relativistic decoupling such that intersection of those contours with the horizontal dashed line correspond to the relativistic floor of left panel. Above the line $\Omega_{\rm DM} h^2 = 0.1188$, contours are the one obtained considering a non-relativistic freeze-out and assuming the maximal cross section allowed by unitarity (see Eq. \ref{eq:unitaritygeneralaverage}). Again, intersections of those lines with the horizontal dashed line correspond to the unitarity wall shown in the left panel of Fig. \ref{fig:polygon}.
\\

For $T'=T$, we recover the 1D bounds on the DM mass range given in Eq. \ref{eq:mass_range_1}. For $T' \leq T$, the allowed mass range shrinks, but also slowly shifts toward higher DM masses. As a consequence, the merging point of the relativistic floor and the unitarity wall gives a DM candidate around the PeV scale which corresponds to a hidden-to-visible temperature ratio at decoupling of order $T'/T\sim 10^{-5}$, see darkest blue line of bottom panel of Figure \ref{fig:polygon}.

\section{Explicit models in the 2D domain}\label{sec:models}
Now that we have drawn the 2D domain of all DM thermal candidate, we can study how concrete DM models can live in the domain depicted in Figure \ref{fig:polygon}. In all of these models, DM will annihilate into HS particles, and the way it will do so will distinguish the models.

\subsection{Scenario 1 : t-channel }\label{subsec:t-channel}
\begin{figure}[h!]
\centering
\includegraphics[scale=0.825]{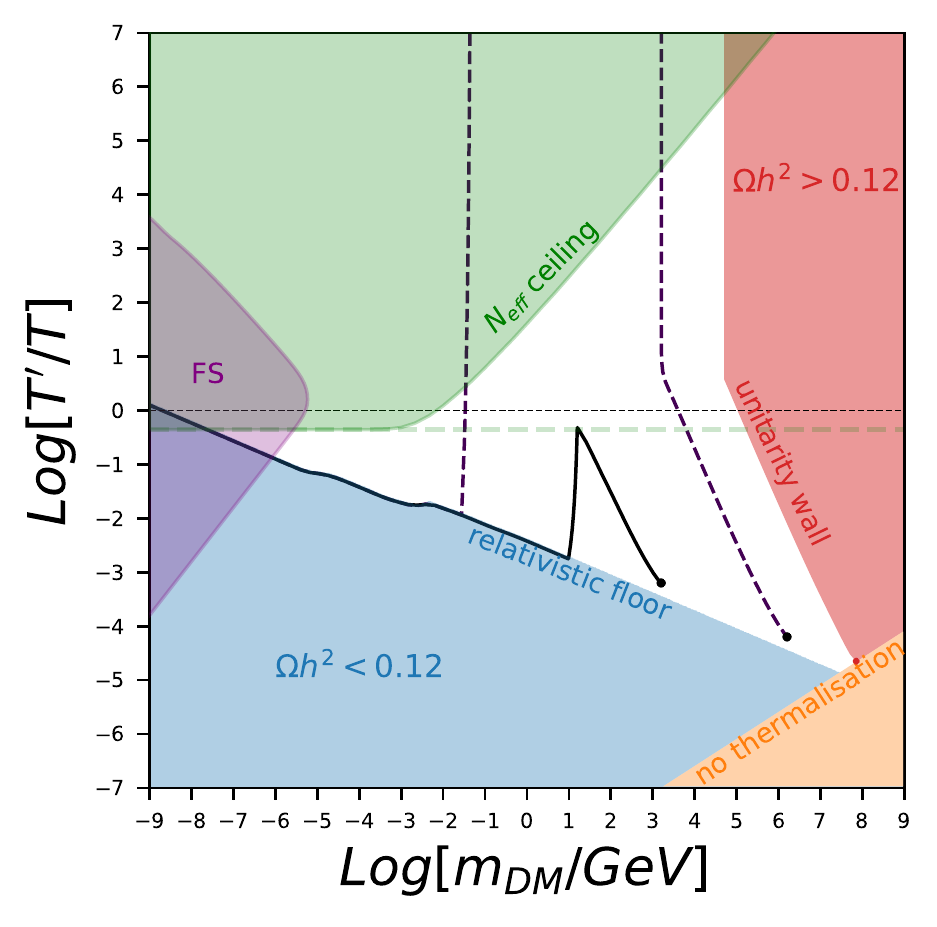}
\caption[Parameter space for DM freeze-out driven by DM annihilating into two dark photons in the t-channel]{Parameter space for DM freeze-out driven by DM annihilating into two dark photons in the t-channel. The parameters are $\alpha'=3\cdot 10^{-4}$ and $m_{\gamma'}=10$ GeV for the solid curve while $\alpha'=0.1$ and $m_{\gamma'}=30$ MeV for the black dashed curve.}\label{fig:polygon_t_ch_noDP}
\end{figure}

We first consider a very popular model that we have already studied in this thesis; The vector portal model (benchmark model B of Eq. \ref{eq:lag_km}) where the DM is a Dirac fermion charged under a new gauge group U(1)'. As above, the DM relic abundance is set by the freeze-out of the annihilation process of a pair of DM particles into a pair of massive vector bosons (see Figure \ref{fig:diag_DM-to-med} above), the new gauge boson. Here, in contrast to what we did in Eq. \ref{eq:lag_km}, we will not consider any kinetic mixing term as we are not interested by the fate of the mediator for the moment. Figure \ref{fig:polygon_t_ch_noDP} shows, in the 2D domain of the left panel of Figure \ref{fig:polygon}, the hidden-to-visible temperature ratio as a function of the DM mass for two choices of the free parameters which are the DM-to-med coupling $\alpha'$ and the vector boson mass $m_{\gamma'}$.
\\

The black solid line of Figure \ref{fig:polygon_t_ch_noDP} is for $\alpha' = 3\times 10^{-4}$ and $m_{\gamma'}=10$ GeV. Let us explain the several features visible of this solid line. First, in the $T'<T$ region of the parameter space, the HS content never plays a major role in the expansion of the Universe and the HS energy and entropy densities can be neglected when solving the Boltzmann equation for the DM yield. Second, as long as the DM is lighter than the vector boson, the only possible DM annihilation process in the theory is totally inefficient such that DM decouples while still relativistic, that is to say that $T'_{\rm dec}>m_{\rm DM}$. The DM relic abundance is then given by the relativistic floor of Eq. 4.2. Once the DM and the vector boson have the same mass, the DM annihilation process starts to be extremely efficient such that one needs that it induces a non-relativistic FO with a large Boltzmann suppression, which must be compensated with a larger value of the temperature ratio. For $m_{\rm DM}>m_{\gamma'}$, the DM annihilation cross section decreases with the DM mass $\langle\sigma v\rangle\sim 1/m_{\rm DM}^2$ such that the temperature ratio also has to decrease with the DM mass giving rise to the diagonal. Finally, at some point, the DM annihilation cross section will be so small that it could never thermalised the HS. At such an high DM mass, neither a non-relativistic decoupling nor a relativistic decoupling is possible and the curve stops (little black dot at the extreme right of the dashed black curve). This dot is the equivalent of the little red dot which lies at the intersection of the unitarity wall and the thermallization line, but for a smaller value of the DM-to-med coupling $\alpha'$.
\\

The black dashed line of Figure \ref{fig:polygon_t_ch_noDP} is for $\alpha' = 0.1$ and $m_{\gamma'}=30$ MeV. In this case, as the vector boson is lighter than in the previous case, the threshold where the DM annihilation process starts to be efficient happens for a lighter DM mass, at $m_{\rm DM}=30$ MeV. The increase in the annihilation cross section is so large than one need a HS temperature larger than the VS temperature, $T'>T$. In this part of the parameter space, the HS sector may dominate very quickly the Universe's expansion rate. Hence, the DM relic abundance is no longer impacted by the VS as the freeze-out occurs in the HS without feeling the presence of the VS through the expansion rate and the energy and entropy densities are dominated by the HS. As a consequence, the DM relic abundance does not depend on the VS temperature nor the hidden-to-visible temperature ratio. This behaviour is translated by vertical lines in the 2D domain of Figure \ref{fig:polygon_t_ch_noDP}. Finally, the fact that the right branch of the black dashed line of Figure \ref{fig:polygon_t_ch_noDP} is shifted to the right compared to the dashed black line is also easy to understand. This comes from the large increase in the DM-to-med coupling $\alpha'$. As the DM and the vector boson are more strongly coupled, they remain longer in chemical equilibrium and DM abundance is further suppressed. Thus, the DM mass can be larger before that the annihilation process stops to allow any thermalisation within the HS.

\subsection{Scenario 2: s-channel}\label{subsec:s-channel}
The second model we consider is none of the two benchmark models above. However, it is still built on a Dirac fermion DM, but which annihilates into lighter Dirac fermions through the production of a scalar state in the s-channel. This time the Lagrangian is given by

\myeq{
&\mathcal{L} = \mathcal{L}_{SM} + i \bar{\chi}\gamma_{\mu}\partial ^{\mu}\chi - m_{\rm DM}\bar{\chi}\chi + i \bar{\psi}\gamma_{\mu}\partial ^{\mu}\psi - m_{\psi}\bar{\psi}\psi\nonumber\\
&\hspace{1cm} + \partial ^{\mu}\phi\partial _{\mu}\phi - m_{\phi}^{2}\phi^{2} - y_{\chi}\phi\bar{\chi}\chi - y_{\psi}\phi\bar{\psi}\psi,
}

\noindent and the DM annihilation cross section, in the non-relativistic limit, by

\myeq{
\left\langle \sigma v\right\rangle = \frac{\pi\alpha_{x}^{2}}{m_{\rm DM}^{2}}\sqrt{1-\frac{m_{\psi}^{2}}{m_{\rm DM}^{2}}}\frac{1-\frac{m_{\psi}^{2}}{2m_{\rm DM}^{2}}}{\left(1-\frac{m_{\phi}^{2}}{4m_{\rm DM}^{2}}\right)^{2}+\frac{\Gamma_{\phi}^{2}m_{\phi}^{2}}{16m_{\rm DM}^{4}}},\label{eq:s_channel_sv}
}

\noindent where we defined $\alpha_x\equiv y_x y_\phi/4\pi$.
\\

\begin{figure}[h!]
\centering
\includegraphics[scale=0.825]{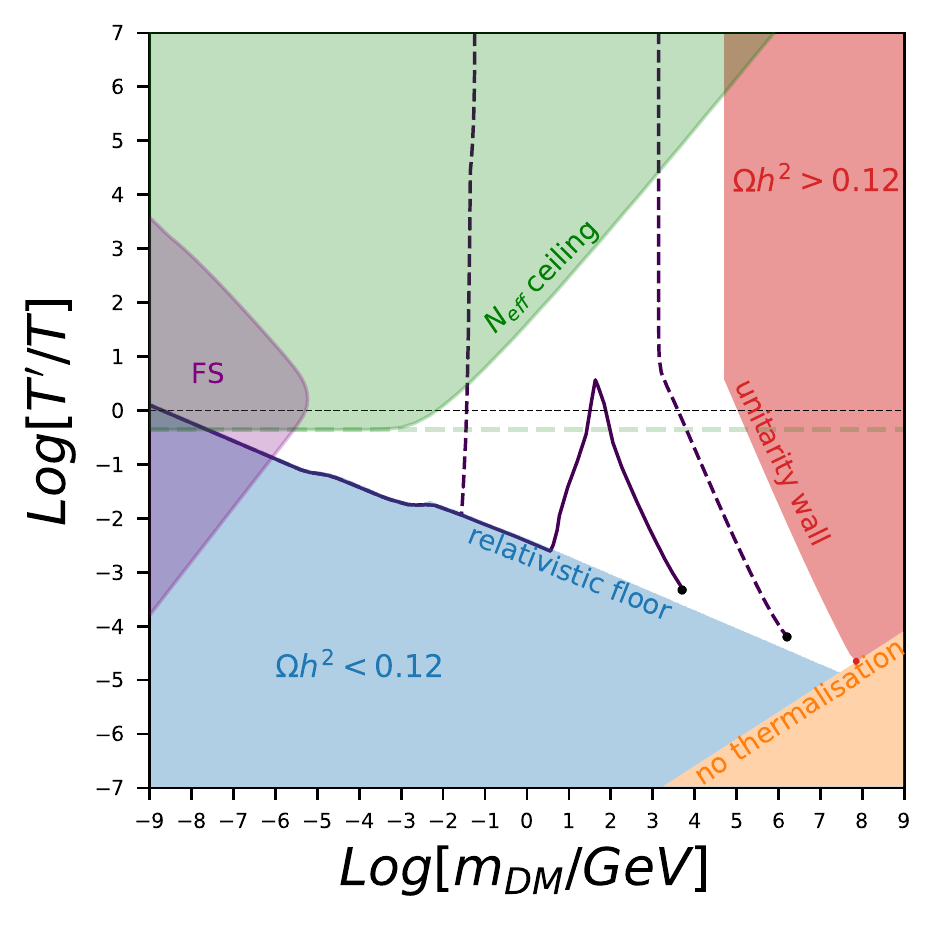}
\caption[Parameter space for DM freeze-out driven by DM annihilating into two Dirac fermions in the s-channel mediated by a scalar]{Parameter space for DM freeze-out driven by DM annihilating into two Dirac fermions in the s-channel mediated by a scalar. We used $\alpha_x=0.001$, $m_{\psi}=5 MeV$ and $m_{\phi}=100$ GeV for the dashed line and $\alpha_x=0.1$ and $m_{\psi}=30$ MeV with $m_{\phi}=1$ GeV for the solid line.}\label{fig:polygon_s_ch_noDP}
\end{figure}

We give in Figure \ref{fig:polygon_s_ch_noDP} the results obtained in this model. As one can see comparing Figures \ref{fig:polygon_t_ch_noDP} and \ref{fig:polygon_s_ch_noDP}, results are very similar in both models. The important difference with the previous model is the presence of a resonance when the mediator mass is twice the DM one. When this happens, the DM annihilation cross section is resonantly enhanced such that one need to largely increase the hidden-to-visible temperature ratio in order to not annihilate too much DM particles. As a result, around the resonance, the dependence of the temperature ratio as a function of the DM mass is different. This can be seen from Eq. \ref{eq:s_channel_sv} neglecting the DM mass ($m_{\rm DM}\ll m_\phi$) or neglecting the mediator mass ($m_\phi\ll m_{\rm DM}$). This gives a cross section going like $\langle\sigma v\rangle\sim m_{\rm DM}^2/m_\phi^4$ or like $\langle\sigma v\rangle\sim 1/m_{\rm DM}^2$ respectively. This behaviour is only visible on the black solid line ($\alpha_x=0.001$, $m_\psi = 10$ GeV and $m_\phi = 100$ GeV) as the choice of the parameters is such that the curve lies entirely in the region of the parameter space where the HS never dominates the energy nor the entropy density of the Universe. The dashed black line, on the other hand, does not show this behaviour as the resonance occurs when the expansion of the Universe is dominated by the HS.

\subsection{The role of the heavy mediator}
As seen in Section \ref{subsec:rel_floor}, DM can decouple in a relativistic way in two main ways, depending on whether the annihilation cross section is cut off at $T'>m_{\rm DM}$ by the mass of a heavy mediator it involves or by the mass of its (heavier) annihilation products. Here, we will look a bit more in details the scenario with heavy mediator. In particular, we will look at the interplay between the heavy mediator mass, the DM mass and the decoupling temperature $T'_{\rm dec}$. To this end, we keep considering the s-channel model of the previous subsection.
\\

To analyse in details what this relativistic decoupling mechanism implies for the parameter space of this explicit example of DM model, one can take the relativistic counterpart of the non-relativistic form of the DM annihilation cross section of Eq. \ref{eq:s_channel_sv}. Neglecting this time the width of the real scalar $\phi$ and the DM and finale states masses, we have,

\myeq{
\langle \sigma v\rangle\simeq \frac{8\pi\alpha_{x}^{2} T'^2}{m_{\phi}^4},\label{eq:s_channel_sv_rel}
}

\noindent in the range $m_{\rm DM}\lesssim T'\lesssim m_{\phi}$ which applies when the DM decouples while still relativistic. With an explicit annihilation cross section at hand, one can compute the absolute value of the temperature at DM decoupling solving the standard thermalisation condition $\Gamma/H=n_{\rm DM}^{eq}(T') \langle \sigma v\rangle/H=1$ along the relativistic floor (i.e. plugging \ref{eq:TpTlowerbound} and \ref{eq:s_channel_sv_rel} in the thermalisation condition). We get,

\myeq{
T'_{\rm dec}= 19.5 \text{ TeV} \times \left(\frac{1}{\alpha_{x}}\right)^{2/3}\left(\frac{m_{\phi}}{\text{PeV}}\right)^{4/3}\left(\frac{m_{\rm DM}}{\text{10 TeV}}\right)^{2/9}\left(\frac{100}{g_{\ast}^{S}(T_{\rm dec})}\right)^{1/18}\left(\frac{1}{g_{\rm DM}^{n}}\right)^{1/9}.\label{eq:Tprimedec}
}

\noindent We can still use the unitarity bound of Eq. \ref{eq:unitaritygeneralaverage} for this model as it was a model independent approach. Hence, the upper bound on the DM mass for this model goes down to

\myeq{
m_{\rm DM}\lesssim 30 \text{ PeV}\times \left(\mathcal{I}_{1}(x'_{\rm dec})\times x'_{\rm dec}\right)^{3/5}.\label{eq:mDMexplicitA}
}

\noindent where $\left(\mathcal{I}_{1}(x'_{\rm dec})\times x'_{\rm dec}\right)^{3/5}$ is of order unity such that we get back the absolute upper bound on the DM mass for a relativistic decoupling, see Eq. \ref{eq:m_max_rel}. As already said above, the fact that this upper bound is slightly smaller than the absolute upper bound on the mass of any DM thermal candidate (Eq. \ref{eq:GK_TpT_VS}) is to be expected as Eq. \ref{eq:mDMexplicitA} sits on the intersection of the relativistic floor and the thermalisation line while Eq. \ref{eq:GK_TpT_VS} sits on the intersection of the unitarity wall and the thermalisation line, see Figure \ref{fig:theroretical_bounds}. Actually, we get back to the 30 PeV upper bound of Eq. \ref{eq:m_max_rel} as it should for a relativistic decoupling
\\

The absolute upper bound on the DM mass coming from unitarity for this model given in Eq. \ref{eq:mDMexplicitA} also gives an upper bound on the heavy mediator versus DM mass ratio $m_{\phi}/m_{\rm DM}$ through the relationship between the decoupling temperature and the DM and heavy mediator masses, see Eq. \ref{eq:Tprimedec}. We have,

\myeq{
\frac{m_{\phi}}{m_{\rm DM}}< 680 \times \left(\alpha_{x}\right)^{1/2}\left(\frac{\text{PeV}}{m_{\rm DM}}\right)^{5/3}\left(\mathcal{I}_{1}(x'_{\rm dec})\right)^{3/4}.\label{eq:mSmDMupperbound}
}

\noindent Moreover, one can impose that the decoupling temperature is larger than the DM mass, ${T'_{\rm dec}\gtrsim m_{\rm DM}}$, in order to ensure a relativistic decoupling for the DM. We thus obtain a lower bound on the heavy mediator versus DM mass ratio,

\myeq{
\frac{m_{\phi}}{m_{\rm DM}}\gtrsim 9.78 \times \left(\alpha_{x}\right)^{1/2}\left(\frac{\text{PeV}}{m_{\rm DM}}\right)^{5/12}.\label{eq:mSmDMlowerbound}
}

\noindent Note that importantly, this last equation indicates that the smaller the DM mass the larger the heavy mediator-to-DM mass ratio has to be for DM to decouple relativistically. This stems from the fact that smaller the DM mass, larger the hidden-to-visible temperature ratio has to be in order to have enough DM abundance today (i.e. to keep sitting on the relativistic floor). Larger the temperature ratio, later the DM will decouple as its annihilation rate will stay longer more efficient than the Hubble rate. As we want the DM to decouple relativistically, we need to compensate this gain in the annihilation rate by decreasing the heavy mediator mass with respect to the DM one to make the DM to still decouple while relativistic.
\\

Finally, one can check that the unitarity bound also constraints the mass of the heavy mediator as a function of the decoupling temperature thanks to the explicit form of the annihilation cross section. The lower bound is thus given by,

\myeq{
\frac{m_{\phi}}{T'_{\rm dec}}> 2.38\times(\alpha_{x})^{1/2}\left(\mathcal{I}_{1}(x'_{\rm dec})\right)^{-1/4}.
}

\noindent This last bound is automatically satisfied for typical perturbative coupling, as expected.
\\

\begin{figure}[h!]
\centering
\includegraphics[scale=0.95]{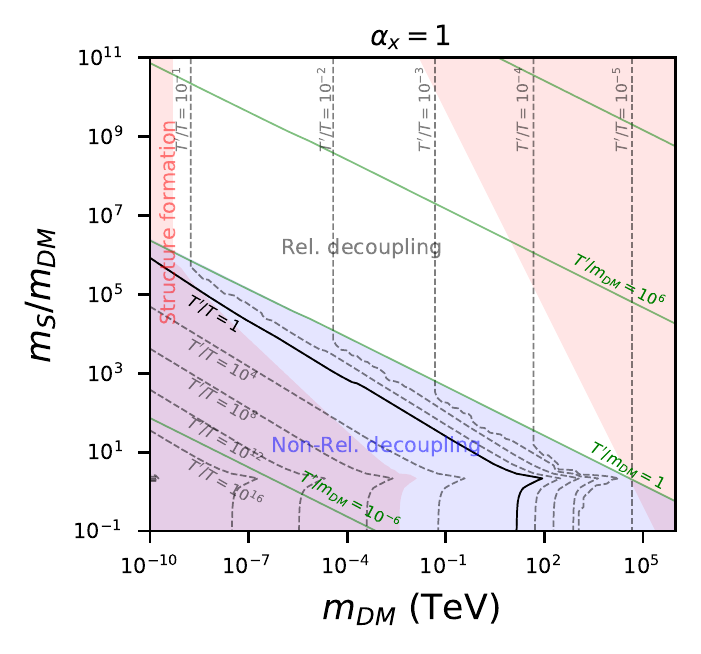}
\includegraphics[scale=0.95]{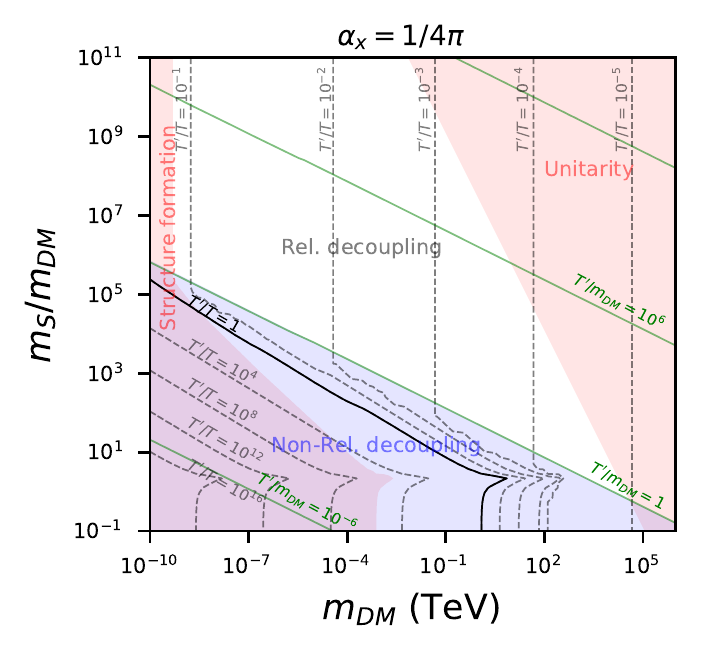}
\caption[Allowed parameter space for a relativistic decoupling in the s-channel]{Allowed parameter space for two choices of the couplings, $\alpha_{x}=1$ (left) and $\alpha_{x}=1/4\pi$ (right). The regions excluded by the structure formation and unitarity constraints are shown in red, while the blue region indicates the non-relativistic regime. We also show various contour of $T'_{\rm dec}/m_{\rm DM}$ and $T'_{\rm dec}/T_{\rm dec}$.}\label{fig:Mass_to_mass}
\end{figure}

Figure \ref{fig:Mass_to_mass} shows as a function of the DM mass and the heavy mediator-to-DM mass ratio, the value of the temperature ratio at decoupling one needs to account for the observed relic density in this model for two choices of the coupling $\alpha_{x}=1$ (left) and $\alpha_{x}=1/4\pi$ (right). Are also visible in Figure \ref{fig:Mass_to_mass}, the lower (blue) and upper (red) bounds on the mass ratio given by Eq. \ref{eq:mSmDMlowerbound} and Eq. \ref{eq:mSmDMupperbound} respectively. Constraint from large scale structure is also shown in red, as well as the isocontours of $T'_{\rm dec}/T_{\rm dec}$ and $T'_{\rm dec}/m_{\rm DM}$. Let us explain various features visible in Figure \ref{fig:Mass_to_mass}. When DM decouples while still relativistic (i.e. white region), the isocontour of the temperature ratio at decoupling, $T'_{\rm dec}/T_{\rm dec}$, are vertical because this ratio does not depend on the heavy mediator mass $m_{\phi}$ as the model saturates the relativistic floor of Eq. \ref{eq:TpTlowerbound}. The absolute unitarity bound given in Eq. \ref{eq:unitaritygeneralaverage} which corresponds to Eq. \ref{eq:mDMexplicitA} is saturated close to the floor ($T'_{\rm dec}/m_{\rm DM}\simeq 1$) as expected and for $m_{\phi}/m_{\rm DM}\sim 2$. 
\\

As said above, for smaller DM masses, the heavy mediator as to be much heavier than the DM $m_{\phi}\gg m_{\rm DM}$ so that the DM decouples while still relativistic, see Eq. \ref{eq:mSmDMlowerbound}.
\\

If the heavy mediator versus DM mass ratio is smaller, one enters in the non-relativistic decoupling regime shown in blue in Figure \ref{fig:Mass_to_mass}. This regime corresponds to a secluded freeze-out regime (see Subsection \ref{subsec:secluded} for more details) and results in this regime follow the analytical results of \cite{Chu:2011be} (see also \cite{Feng:2008mu,Hambye:2019dwd,Hambye_2020}). As we have seen in Chapter \ref{ch:prod}, in the secluded freeze-out (as in the ordinary freeze-out), the DM final relic abundance also depends on the annihilation cross section and not only on the DM mass and the hidden-to-visible temperature ratio as in the relativistic decoupling regime. Hence, since the annihilation cross section decreases quickly with the heavy mediator mass, $\langle \sigma v\rangle\propto 1/m_{\phi}^4$ when $m_{\phi}\gtrsim m_{\rm DM}$, the DM relic density will decrease quickly when $m_{\phi}/m_{\rm DM}$ decreases. This has to be compensated by a larger value of the hidden-to-visible temperature ratio in order to balance this fall of the annihilation cross section strength. This specific feature explains the behaviour of the temperature ratio isocontours in the non-relativistic regime for $m_{\phi}\gtrsim m_{\rm DM}$. One can also distinguish in Figure \ref{fig:Mass_to_mass} the effect of the resonance at $m_{\phi}\simeq 2 m_{\rm DM}$ for which the DM annihilation cross section is resonantly boosted and this requires a larger value of the temperature ratio.
\\

Finally, if the mediator $\phi$ is no longer heavier than the DM, $m_{\phi}<m_{\rm DM}$, the DM relic density becomes very quickly independent of the mediator $\phi$ mass as the model is now in a "light mediator" regime where the cross section is independent of the mediator mass $m_{\phi}$. This is why the contours of the temperature ratio are vertical in this region of the parameter space as it is the case in the relativistic regime. Note the large difference in the magnitude of the hidden-to-visible temperature ratio in those two regions. In fact, the $T'/T$ isocontours should be even more shifted towards the right than in Figure \ref{fig:Mass_to_mass} due to the fact that in the $m_{\phi}<m_{\rm DM}$ region, the DM can also annihilate into two $\phi$'s (${\rm DM} {\rm DM} \rightarrow \phi\phi$). This effect has not been taken into account as, here, we aimed to show how the results behave from a unique annihilation channel everywhere (i.e. from ${\rm DM} {\rm DM} \rightarrow \psi\psi$ mediated by a s-channel scalar $\phi$).

\part{Self-Interacting Dark Matter ways-out}
\chapter{SIDM in a colder Hidden Sector}\label{ch:TpT}
\yinipar{I}n the two first chapters, we have considered and motivated the possibility that DM undergoes self-interactions. One has seen in the framework of the two particularly simple explicit benchmark models A and B that many constraints apply to this setup and, actually, excluding that these models could have enough self-interactions as suggested by small scale structure anomalies. In the previous chapter, one has considered and motivated the possibility that the HS does not thermalise with the SM thermal bath, as a result of a feeble DM-to-SM connection. Having all of these chapters in mind, one will now ask the following question: What about self-interacting DM in a HS not thermally connected to the SM thermal bath? In particular, what become the constraints developed in Chapter \ref{ch:constr} in the hidden-to-visible temperature ratio versus DM mass plane? Indeed, constraints as considered in Chapter \ref{ch:constr} assumed a thermal connection between the dark sector and the SM, so that $T'=T$. But what happens for $T'\neq T$? To illustrate this, we consider here too the two portal models that we presented in Section \ref{subsec:portal} and defined by Eqs. \ref{eq:lag_hp} and \ref{eq:lag_km} for the Higgs portal and the kinetic mixing portal models respectively. The following discussion is mainly based on \cite{Hambye_2020} for the vector case, but also contain original unpublished work (scalar case) and use results from \cite{Hambye:2018dpi}.

\section{Hidden sector temperature}\label{sec:HS_temperature}
The first thing to do when considering a different temperature for the HS and the SM is to ensure that the HS bath has thermalised before DM decoupling such that a temperature for the HS can be defined at this moment. We have then to compute the minimal value of the HS sector coupling (corresponding to the DM-to-med coupling in the two benchmark models we consider) which allows a thermalisation of particles within the HS. That is to say that we look for the equivalent of the thermalisation condition in orange in Figure \ref{fig:polygon}. Since, we already discussed thermalisation conditions in Chapter \ref{ch:prod}, we simply recall here the condition which is of our interest in this section given first in Eq. \ref{eq:thermalisations},

\myeq{
\Gamma_{\rm DM\leftrightarrow med}\vert_{T'\simeq m_{\rm DM}}\gtrsim H\vert_{T'\simeq m_{\rm DM}}, \label{eq:th_HS}
}

\noindent with $H$ the Hubble rate as previously. Neglecting the portal connection to the SM there is no significant production of DM or mediator particles from the SM and the interaction rate in Eq. \ref{eq:th_HS} is simply given by

\myeq{
\Gamma_{\rm DM\leftrightarrow med} = \left\langle\sigma_{\rm DM \rightarrow med}v\right\rangle n_{\rm DM}^{eq}.
}

\begin{center}
\begin{figure}[h!]
\centering
\includegraphics[scale=1]{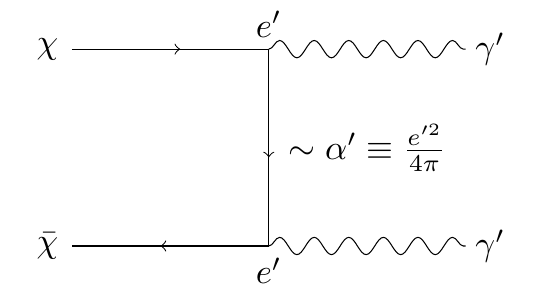}
\includegraphics[scale=1]{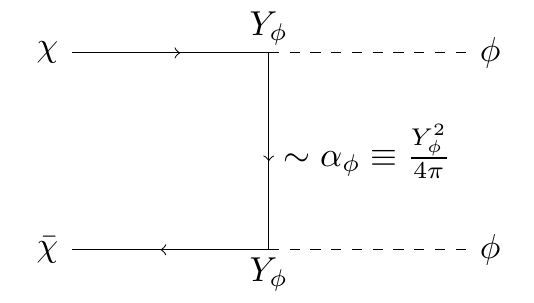}
\caption[Feynman diagram of DM annihilating into bosons]{Feynman diagrams corresponding to the DM annihilation processes responsible for the freeze-out mechanism for both the kinetic mixing portal (left) and the Higgs portal (right) models.}
\label{fig:diag_DM_annihilation_med}
\end{figure}
\end{center}

\noindent The thermally averaged DM annihilation cross sections at tree level (see corresponding Feynman diagrams in Figure \ref{fig:diag_DM_annihilation_s} above which we reproduce again in Figure \ref{fig:diag_DM_annihilation_med} for convenience) for the vector and the scalar portal models are given in Eqs. \ref{eq:sv_HP_SI} and \ref{eq:sv_KM_SI} and that we repeat here,

\myeq{
&\left\langle\sigma v\right\rangle_{\bar{\chi}\chi\rightarrow \gamma '\gamma '} = \frac{\pi \alpha '^{2}}{m_{\rm DM}^{2}}\sqrt{1-\frac{m_{\gamma'^{2}}}{m_{\rm DM}^{2}}},\label{eq:sv_KM}\\
&\left\langle\sigma v\right\rangle_{\bar{\chi}\chi\rightarrow \phi\phi} = \frac{3v^{2}}{4}\frac{\pi \alpha_{\phi}^{2}}{m_{\rm DM}^{2}}\sqrt{1-\frac{m_{\phi^{2}}}{m_{\rm DM}^{2}}},\label{eq:sv_HP}
}

\noindent where we kept same notations as previously, with $v$ the M\o ller velocity (see Chapter \ref{ch:som}).
\\

In the same way that we obtained, for a cross section which saturates the unitarity bound, the thermalisation line of Eq. \ref{eq:TH} and depicted in Figure \ref{fig:theroretical_bounds}, one can determine the minimal allowed value of the DM-to-med coupling in those specific benchmark models we consider. In order to do that, one must plug the expressions for the DM annihilation cross section  of Eqs. \ref{eq:sv_KM} and \ref{eq:sv_HP} and the DM relativistic number density into the thermalisation condition given in Eq. \ref{eq:th_HS}. These lower bounds are given as a function of the hidden-to-visible temperature ratio and the DM mass by,

\myeq{
&\alpha'\gtrsim 5.98\times 10^{-9}\frac{T }{T'}\left(\frac{\sqrt{g_{\star}^{eff}(T )}}{g_{DM}^{eff}(T')}\right)^{1/2}\left(\frac{m_{DM}}{100\text{~GeV}}\right)^{1/2},\label{eq:alphapth} \\
&\alpha _{\phi}\gtrsim 3.95\times 10^{-9}\frac{T }{T'}\left(\frac{\sqrt{g_{\star}^{eff}(T )}}{g_{DM}^{eff}(T')}\right)^{1/2}\left(\frac{m_{DM}}{100\text{~GeV}}\right)^{1/2}.\label{eq:alphaphith}
}

\noindent As long as the DM-to-med coupling is bigger than these lower bounds and if the connection to the SM is negligible, the dark sector will be totally decoupled from the SM and enters into a simple freeze-out phase within the HS. The source of the HS has not to be necessarily the SM as in the secluded freeze-out regime such as considered above (regime IV of Section \ref{subsec:secluded}). Indeed, as discussed in Chapter \ref{ch:prod}, the HS could have been produced just after the inflation in the same way that it is usually assumed that the SM is produced just after inflation. However, in practice, the dynamics is the same as in the secluded freeze-out mechanism because the source of the bath has no role other than setting the initial conditions. Thus, in order to have an analytic approximation of the DM relic density, one can use the formula obtained assuming an instantaneous FO (see Eq. \ref{eq:Oh2_FO_TpT}), but for the DM annihilation cross section of Eqs. \ref{eq:sv_KM} and \ref{eq:sv_HP} for the vector and scalar portal models respectively. For the former one we have,

\myeq{
&\hspace{-0.2cm}\frac{T'_{\rm dec}}{T_{\rm dec}} \simeq 1.73\times 10^{-2}\times\left(\frac{\alpha'}{10^{-3}}\right)^{2}\left(\frac{g_{\star}^{S}(T_{\rm dec})}{100}\right)\left(\frac{100}{g_{\star}^{eff}(T_{\rm dec})}\right)^{1/2}\left(\frac{23}{x'_{\rm dec}}\right)\left(\frac{\text{TeV}}{m_{\rm DM}}\right)^{2},\label{eq:TpT_KM}
}

\noindent in which we neglected the mediator mass as we are interested in the light mediator regime where $m_{\rm med}\ll m_{\rm DM}$. Similarly, for the scalar portal model using the cross section given in \ref{eq:sv_HP} we get,

\myeq{
&\hspace{-0.2cm}\frac{T'_{\rm dec}}{T_{\rm dec}} \simeq 1.70\times 10^{-3}\times\left(\frac{\alpha_{\phi}}{10^{-3}}\right)^{2}\left(\frac{g_{\star}^{S}(T_{\rm dec})}{100}\right)\left(\frac{100}{g_{\star}^{eff}(T_{\rm dec})}\right)^{1/2}\left(\frac{23}{x'_{\rm dec}}\right)^{2}\left(\frac{\text{TeV}}{m_{\rm DM}}\right)^{2}.\label{eq:TpT_HP}
}

\begin{figure}[h!]
\centering
\includegraphics[scale=0.7]{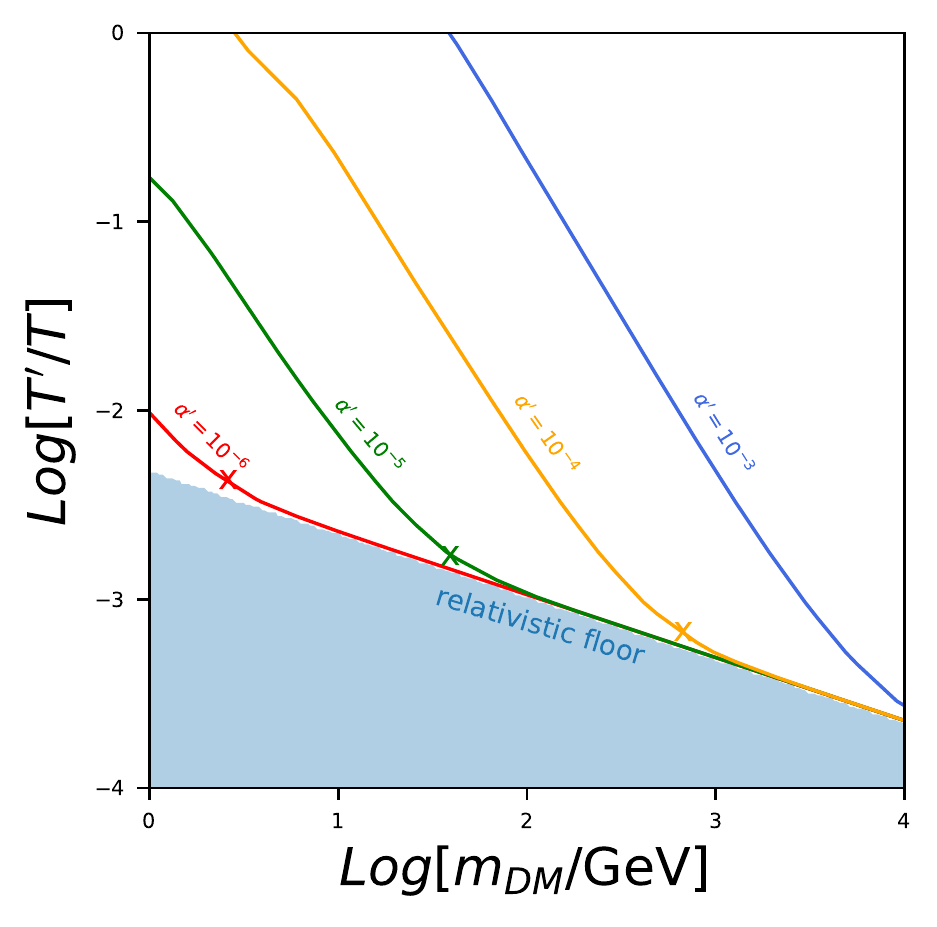}
\includegraphics[scale=0.7]{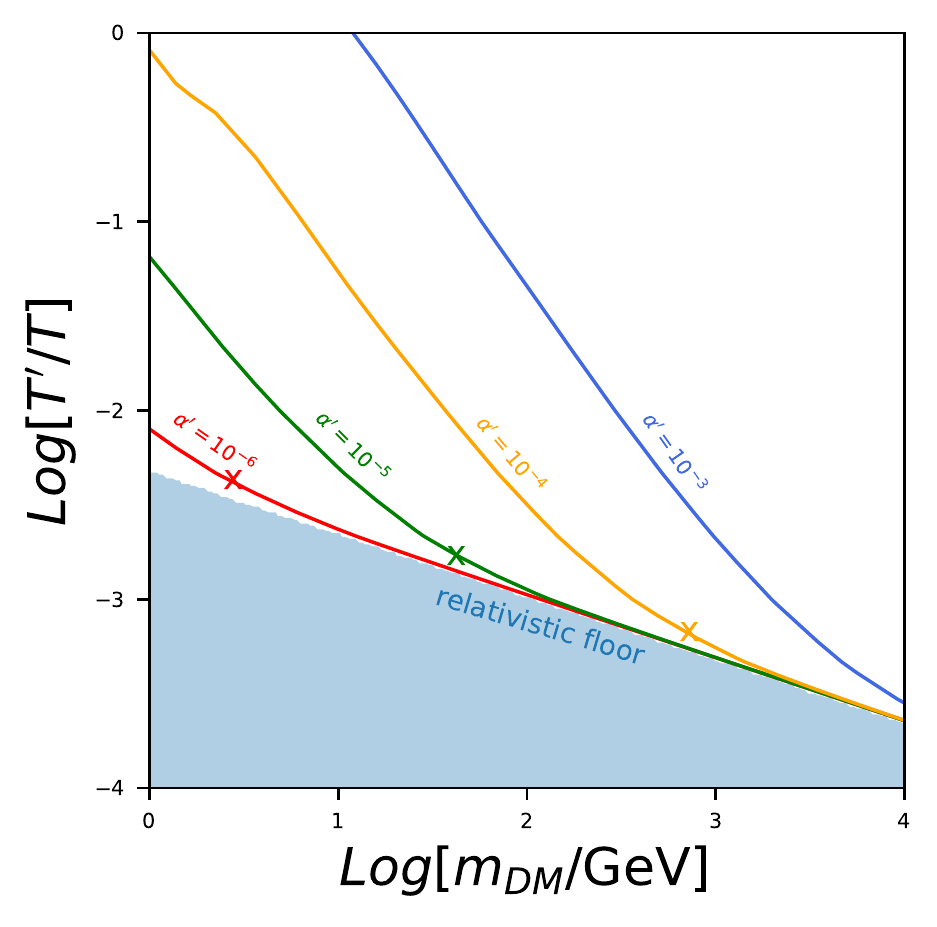}
\caption[Temperature ratio as a function of the DM mass for thermally disconnected DM portal models]{Needed value of $(T'/T)_{T=T_{dec}}$ as a function of the DM mass in order to satisfy the relic density constraint for four different values of the DM-to-med coupling. Results are obtained neglecting all connection to the SM for both the vector portal model (left) and the scalar portal model (right). The black solid line shows the model independent relativistic floor obtained in Eq. (\ref{eq:TpTlowerbound}). The crosses show the value of $m_{\rm DM}$  above which Eqs. (\ref{eq:alphapth}) and (\ref{eq:alphaphith}) are no longer satisfied, i.e.~above which $\Gamma/H |_{T'=m_{DM}}<1$.}
\label{fig:TpT_epsilon}
\end{figure}

\noindent Note the power of two on $x'_{\rm dec}$ in the scalar portal case which differs from the vector portal case. This is due to the velocity dependence of the annihilation cross section as $v^{2}\sim T'/m_{\rm DM}=1/x'$. Figure \ref{fig:TpT_epsilon} shows, within the domain of thermal DM candidates of Figure \ref{fig:polygon}, contours of the DM-to-med coupling which account for the DM relic density today. Note that in Figure \ref{fig:TpT_epsilon} we focus on the GeV to TeV DM mass range such that one cannot see the unitarity wall neither the thermalisation line. However, one can still recognise the relativistic floor in blue.
\\

Behaviours of curves shown in Figure \ref{fig:TpT_epsilon} are easy to understand with the help of Eqs. \ref{eq:TpT_KM} to \ref{eq:TpT_HP}. Indeed, from these equations one can see that the DM relic density scales as $\Omega_{DM}\propto (T'/T)/\langle \sigma v\rangle$ such that, for a fixed value of the DM-to-med coupling, the larger the DM mass is, the smaller the temperature ratio must be. As a consequence, the larger the DM mass is, the smaller the Boltzmann suppression must be in order to not deplete to much the DM abundance after it becomes non-relativistic. If one keeps increasing the DM mass, one will reach the point where the needed value of the temperature ratio is so small that the initial DM abundance is the same of what one would expect from a particle which decouples while still relativistic. In order words, the temperature ratio curve as a function of the DM mass will reach asymptotically the "relativistic floor"  of Eq \ref{eq:TpTlowerbound} (shown in blue in Figure \ref{fig:TpT_epsilon}). We also indicate with crosses in both panels of Figure \ref{fig:TpT_epsilon} where the approximate condition for the HS to thermalise ($\Gamma/H |_{T'=m_{DM}}=1$) stops to hold, in the same way we ended the curves with dots in Section \ref{sec:models}. Thus, for a fixed value of the DM-to-med coupling, on the one hand the HS thermalises if the DM mass is lighter than the mass shown by the cross. On the other hand, the HS does not thermalise for heavier DM masses. This is due to the fact that the annihilation rate scales as the cross section which decreases with the mass. Since the process responsible of the DM freeze-out cannot be responsible for the HS thermalisation for heavy DM masses, one needs to assume an additional interaction in the UV in order to satisfy the HS thermalisation hypothesis. The fact that the crosses lie close to the relativistic floor is to be expected. Indeed, looking at the heaviest DM mass (for a fixed value of the DM-to-med coupling) which allows the HS to thermalise, the thermalisation condition holds such that when the DM decouples, its number density has already been a little bit Boltzmann suppressed. This indicates that its associated temperature ratio has to be a little bit bigger than the one expected for relativistic decoupling. Conversely, looking at the lightest DM mass for which the thermalisation condition does not hold, one does not expect any sizeable Boltzmann suppression since in this case, the DM thermalises when $T'\sim m_{\rm DM}$ (and not before already). That is to say when the $\Gamma/H$ ratio is maximum. In this case, the needed temperature ratio lies very close to the relativistic floor.

\section{Portal strength to the visible sector}
Results shown in Figure \ref{fig:TpT_epsilon} were obtained assuming a negligible SM-to-med coupling and thus a negligible SM-to-DM coupling. The purpose of this section is to determinate the maximal allowed value of the mixing parameter $\epsilon/\lambda_{\Phi H}$ for previous results still being valid. Indeed, if one increases the mixing parameter, the SM could produce more DM and/or mediator particles such that the HS would be reheated. In that case, the hidden-to-visible temperature ratio would become bigger than the one assumed in Figure \ref{fig:TpT_epsilon}. The critera we used is to impose that the DM and mediator yields obtained by freeze-in from SM particles does not exceed the yields obtained at DM decoupling when neglecting the SM source terms. That is to say that we imposed that:

\myeq{
Y_{\rm DM}^{SM}(T'_{\rm dec})<Y_{\rm DM}^{HS}(T'_{\rm dec}),\\
Y_{\rm med}^{SM}(T'_{\rm dec})<Y_{\rm med}^{HS}(T'_{\rm dec}),
}

\noindent where $Y^{SM}(T'_{\rm dec})$ indicates the yield obtained through a freeze-in from SM particles at DM decoupling (given by Eq. \ref{nDMFIa}) while $Y^{HS}(T'_{\rm dec})$ stems for the yield obtained after a freeze-out in the HS at DM decoupling when neglecting the SM source terms (given by solution of Eq. \ref{BoltzEqIVa}).
\\

\begin{figure}[h!]
\centering
\includegraphics[scale=0.65]{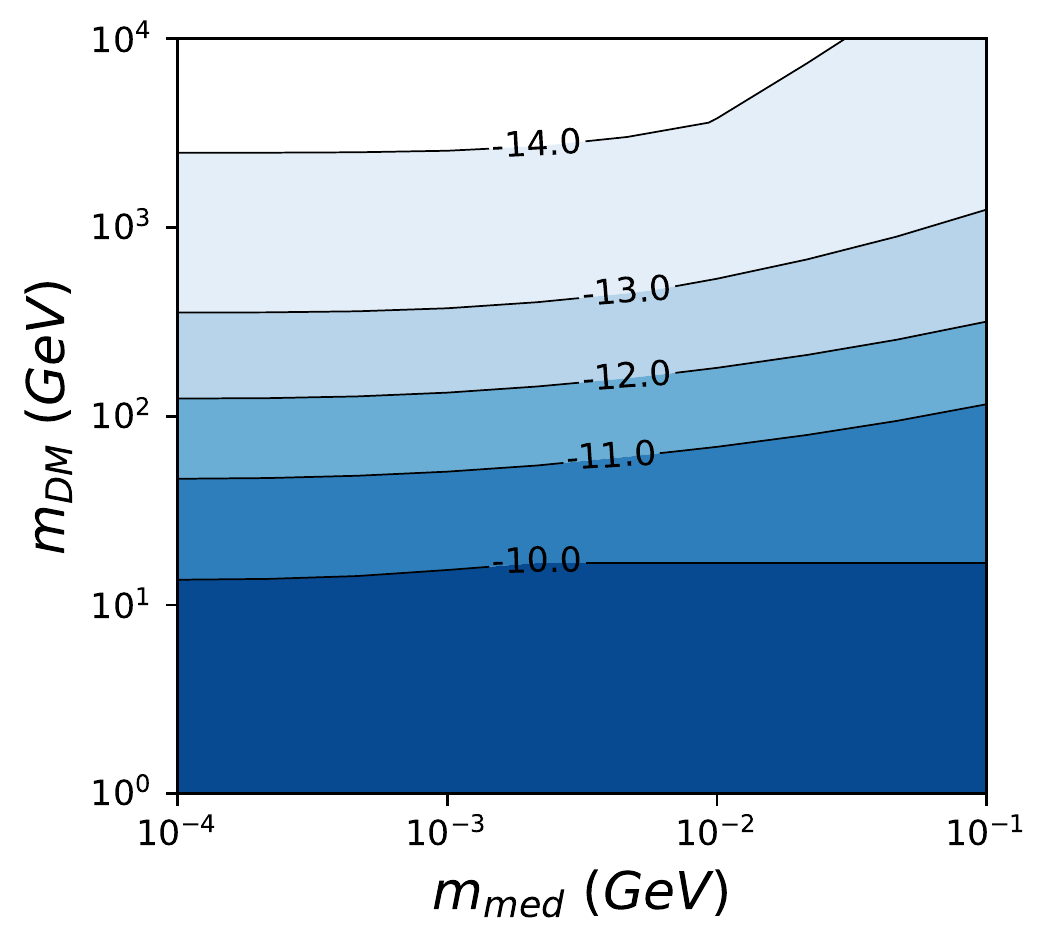}
\includegraphics[scale=0.65]{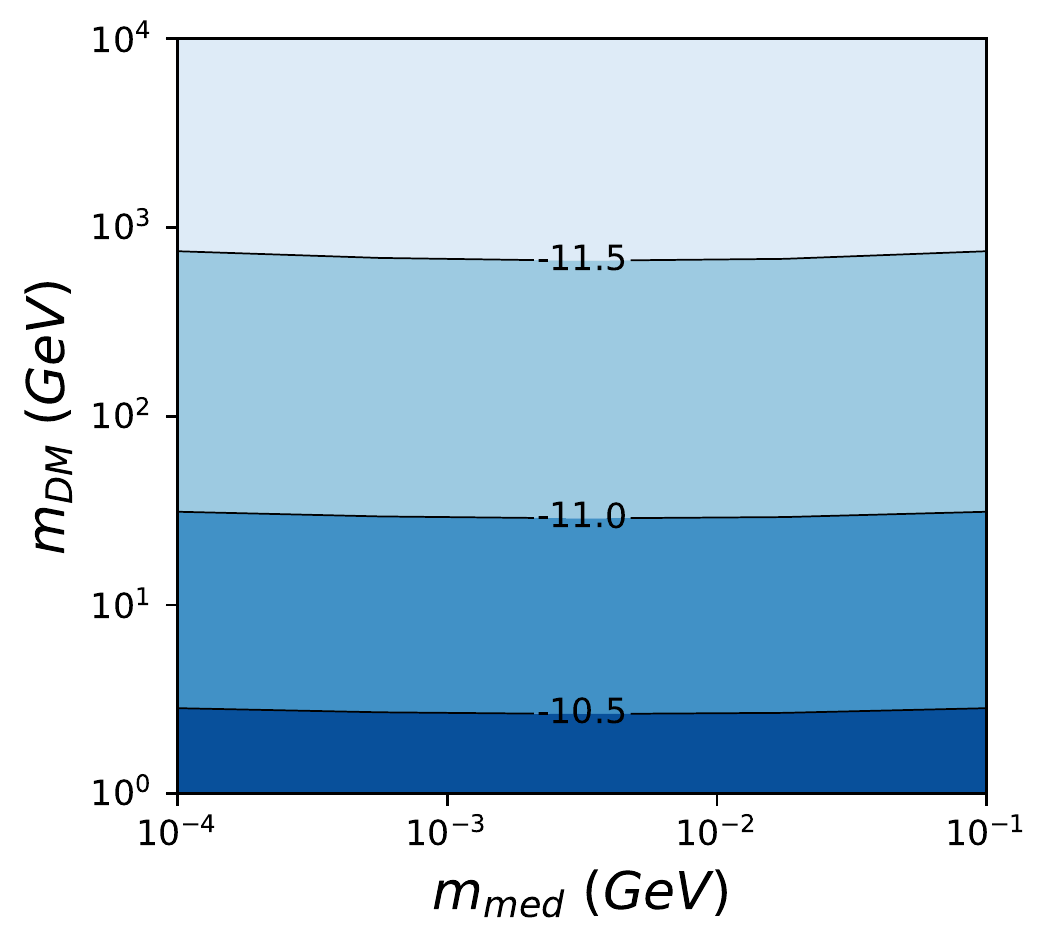}
\caption[Contour of the SM-to-med coupling for thermally disconnected DM portal models]{Maximal allowed value of the SM-to-med coupling such that the production of HS particles from SM never exceeds what one would have without the portal interaction. For the whole shown parameter space, we fixed $\alpha '=10^{-4}$ for the vector portal model (left) and for the scalar portal model $\alpha_\phi=10^{-4}$ (right) and used the temperature ratio given by Figure \ref{fig:TpT_epsilon}.}\label{fig:portalmax}
\end{figure}

Figure \ref{fig:portalmax} shows the upper bound on the SM-to-med coupling $\epsilon/\lambda_{\Phi H}$ in the DM versus mediator mass plane one has when assuming that the portal interaction never produce more DM and mediator that one would get without this connection to the SM. Every point in this plane is a DM candidate in the sense that, fixing the DM-to-med coupling to $\alpha'=10^{-4}$ for the vector portal model (left) and to $\alpha_{\phi}=10^{-4}$ as well as for the scalar portal model (right), we used the temperature ratio given by Figure \ref{fig:TpT_epsilon}\footnote{We recall that this temperature ratio were fixed in order to satisfy the DM relic abundance constraint.}. From Figure \ref{fig:portalmax}, one can conclude for example that for $m_{\rm DM} = 100$ GeV and $m_{med} = 0.1$ GeV, the maximal allowed value for results shown on Figure \ref{fig:TpT_epsilon} still being valid is $\epsilon\lesssim 10^{-11}$ in the vector portal model and $\lambda_{\Phi H}\lesssim 6.3\times10^{-12}$ in the scalar portal model.

\section{Constraints}
In Chapter \ref{ch:constr}, we have reviewed all relevant constraints one should take care of while considering SIDM models. But, during the whole chapter, we have considered a dark sector which is in thermal equilibrium with the SM or, at least, with the same temperature $T$. However, as we discussed already above, this requirement is not mandatory at all and one could have a HS with its own temperature. In such scenarios, some of the constraints reviewed in Chapter \ref{ch:constr} may change. In this section, we will go through all relevant constraints which are impacted if the HS thermal bath has a different temperature as the SM thermal bath.

\subsection{CMB}
\subsubsection{DM annihilation rate}
Let us start our analysis on changes of CMB constraints with the upper bound on the DM annihilation cross section into two mediators which could give rise to photons at recombination, see Eq. \ref{cmbconstraint}. In the case where both sector thermalise and if we only have one thermal bath and thus one temperature (i.e. $T'=T$), this constraint is already strong enough to exclude the whole vector portal model as soon as the dark photon is heavier than twice the electron mass, $m_{\gamma '}>2m_{e}$. Indeed, in this scenario, the new mediator decays into charged SM leptons and this would produce too many photons. This has already been studied at length, see \cite{Bringmann:2016din} for a deeper analysis. However, this bound does not entirely exclude the scalar portal model as the annihilation cross section of DM into two scalars proceeds in p-wave. Thus, the cross section is suppressed compared to a s-wave annihilation. In practice, both s-wave and p-wave cross sections are boosted thanks to the Sommerfeld effect, but, as we have seen in Chapter \ref{ch:som}, it turns out that the p-wave cross section is much less boosted than the s-wave one at recombination as a result of small velocities at this time, see Figure 1 of \cite{Bringmann:2016din}. As a consequence, the scalar portal model does really not suffer from this constraint.
\\

\begin{figure}[h!]
\centering
\includegraphics[scale=0.65]{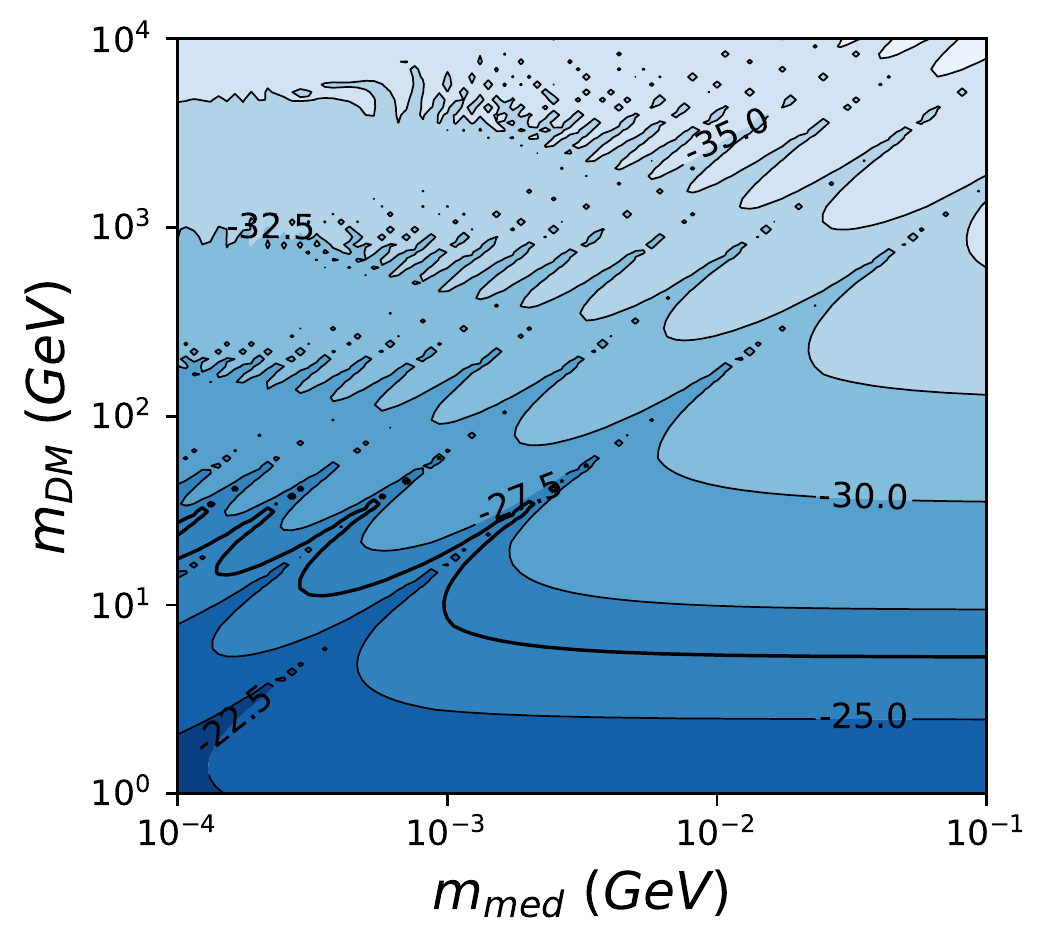}
\includegraphics[scale=0.65]{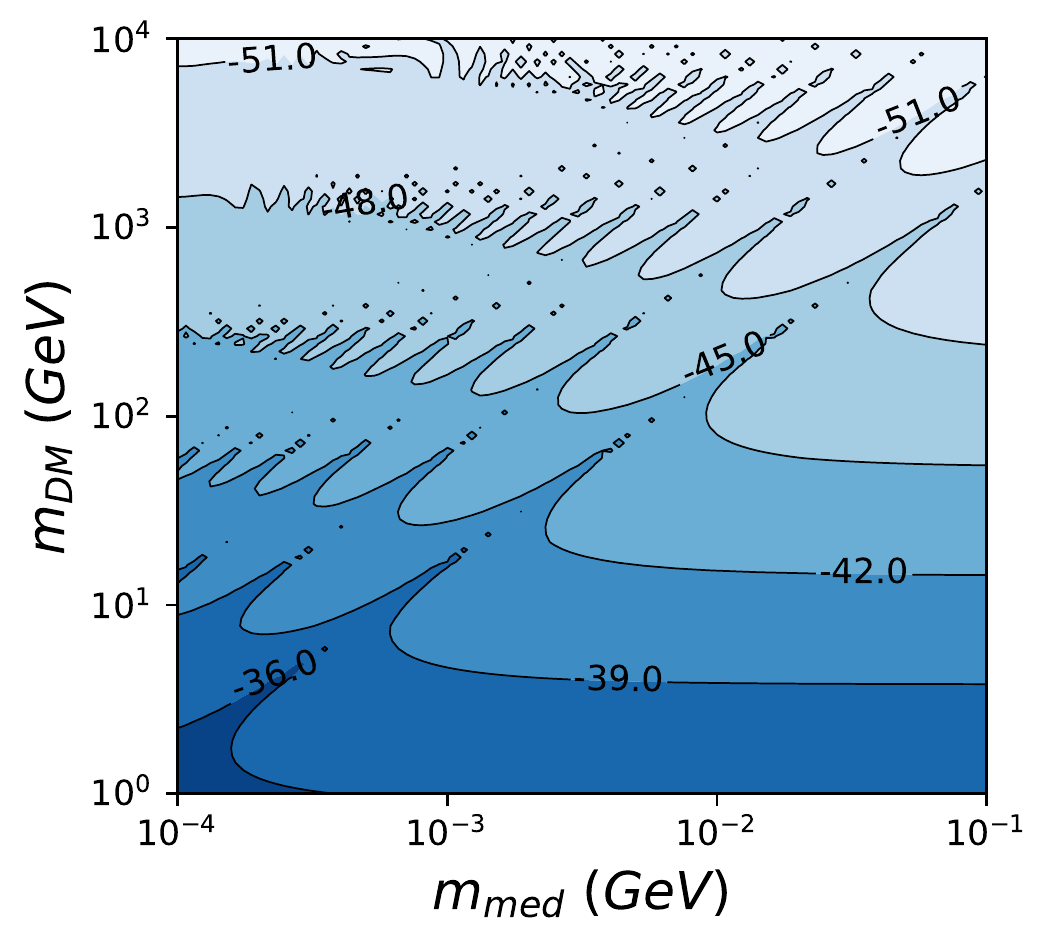}
\caption[Contour of the DM annihilation cross section at CMB for thermally disconnected DM portal models]{Contours of $\log_{10}\left[\frac{\left\langle \sigma v\right\rangle_{rec}}{\text{cm}^{3}\text{s}^{-1}}\frac{\text{GeV}}{m_{\rm DM}}\right]$ at recombination as a function of DM and mediator masses for the vector with $\alpha _{\gamma'}=10^{-4}$ (left) and scalar with $\alpha _{\phi}=10^{-4}$ (right) portal models. The solid black line gives the upper bound given by Eq. \ref{cmbconstraint}: $\left\langle \sigma v\right\rangle_{rec}/m_{\rm DM} = 4\times 10^{-27}$ $\text{cm}^{3}\text{s}^{-1}\text{GeV}^{-1}$ such that the region below this line is excluded.}\label{fig:CMB_BBN}
\end{figure}

On the other hand, in the case where the connection between the two sectors is not strong enough to allow thermalisation between the two baths, the HS has its own temperature, $T'\neq T$. As we have seen above, the required annihilation cross section in order to get the observed DM relic abundance is changed by a factor of $T'/T$, see Eq. \ref{eq:Oh2_FO_TpT}. Thus, if the HS thermal bath temperature is smaller than the SM thermal bath temperature, the annihilation cross section needed is suppressed with respect to the case $T'/T=1$, since it has been less Boltzmann suppressed in this case (its number density before Boltzmann suppression being smaller by a factor of $\left(T'/T\right)^{3}$). This suppression could potentially be strong enough to avoid the upper bound on the annihilation cross section given by the CMB. Figure \ref{fig:CMB_BBN} gives for the vector portal model (left) and the scalar portal model (right) contour of $\log_{10}\left[\frac{\left\langle \sigma v\right\rangle_{rec}}{\text{cm}^{3}\text{s}^{-1}}\frac{\text{GeV}}{m_{\rm DM}}\right]$ at recombination in the DM versus mediator mass plane. As previously, we used the temperature ratio given by Figure \ref{fig:TpT_epsilon} for all points in this parameter space. The key feature is given by the solid black line which indicates where the upper bound on this quantity sits in this plane. Every candidate below this line has an annihilation cross section into light mediators too strong at recombination and is excluded by CMB data.
\\

One can conclude from these results that the CMB constraints on the DM annihilation cross section at recombination is strongly relaxed as now it excludes DM masses below a few tens of GeV for the vector portal model. While the parameter space we consider for the scalar portal model is totally allowed by this constraint.

\subsubsection{Mediator decay}
The constraints from CMB on the mediator decay were summarised in Figure \ref{fig:CST-omega_med} and Eq. \ref{eq:CST-xray_med}. They both suggested two simple ways out: either one can reduce the mediator lifetime in order to make it decay long before recombination era, either one can reduce the mediator abundance when at the last scattering surface. The problem with the first way out is that if one shorten the mediator lifetime, one must increase its connection to the SM and this usually requires SM-to-med couplings above the upper limit given by the no thermalisation condition (see Figure \ref{fig:portalmax}). However, the second way out is automatically included in the thermally disconnected HS scenario. Indeed, if $T'<T$ the mediator number density left after having decoupled from the DM particles is suppressed by a factor of $(T'/T)^3$. This factor can be small enough to suppress the mediator number density and to make it below the upper bound given by CMB. Fixing the value of the SM-to-med coupling as given in Figure \ref{fig:portalmax} and the temperature ratio as given in Figure \ref{fig:TpT_epsilon}, one can check that the mediator lifetime is short enough in order to avoid the bound on the mediator decay from CMB. As an example,

\myeq{
m_{\rm med} = 10 \text{ MeV}\hspace{0.5cm}\text{and}\hspace{0.5cm} \frac{T'}{T}\lesssim 0.02 \hspace{0.5cm}\Rightarrow\hspace{0.5cm} m_{\rm DM}\gtrsim 117 \text{ GeV},\\
m_{\rm med} = 1.5 \text{ MeV}\hspace{0.5cm}\text{and}\hspace{0.5cm} \frac{T'}{T}\lesssim 0.04 \hspace{0.5cm}\Rightarrow\hspace{0.5cm} m_{\rm DM}\gtrsim 74 \text{ GeV}.
}

\noindent This shows how this CMB constraint on the mediator decay can be easily avoided once the hidden-to-visible temperature ratio is sizeably smaller than unity.

\subsubsection{$N_{eff}$}
Since the mediator number density is suppressed by a factor of $(T'/T)^3$ at recombination, this suppression also applies for the CMB constraint on the number of light degrees of freedom encoded in the value of $N_{eff}$. In practice, this constraint will also become quickly irrelevant for portal models like those we are considering. However, this suppression in the mediator number density is not the only effect that the thermally disconnected HS scenario brings with it. Indeed, if the new light mediator decays while still relativistic after neutrino decoupling, each decay process will inject less energy in the visible sector than in the case where the HS and SM baths are thermally connected. This is simply due to the fact that the mediator reservoir contains less energy since it is decreased by a suppression factor of $(T'/T)^4$ and that its decay is less boosted. As a consequence, there are much less decays which happen after neutrino decoupling such that constraints from CMB on $N_{eff}$ are even more weakened.
\\

The upper bound on the mediator number density as a function of its lifetime which actually applied when $T'<T$ requires a deeper analysis, but a conservative upper bound can be obtained by multiplying the upper bound got in the $T'=T$ case by the suppression factor $(T'/T)^3$. Thus, if one assumes that the mediator decouples while still relativistic (as it is the case in the light mediator scenario), one gets the upper bound on the lifetime given in Table \ref{table:Neff}.
\\

\begin{table}[h!]
\centering
%\fontsize{10}{12}
\begin{tabular}{|c|c|c|c|c|c|c|c|c|c|c|c|}
\hline
\multicolumn{2}{|c|}{\multirow{2}{*}{$\tau_{med}$ (sec)}} & \multicolumn{4}{c|}{$m_{\gamma '}$ (MeV)} & \multicolumn{4}{c|}{$m_{\phi}$ (MeV)} \\ \cline{3-10} 
\multicolumn{2}{|c|}{}                                    & 3     & 10     & 30    & 100    & 3     & 10     & 30    & 100   \\ \hline
\multirow{4}{*}{$\frac{T'}{T}$}          & 1              & $10^{1.1}$       & $10^{0.1}$      & $10^{-0.4}$     & $10^{-0.7}$      & $10^{2.1}$       & $10^{1.1}$      & $10^{0.1}$     & $10^{-0.5}$     \\ %\cline{2-10} 
                                         & 0.1            & $10^{7.3}$       & $10^{6.3}$      & $10^{5.3}$     & $10^{4.3}$      & $10^{8.3}$       & $10^{7.3}$      & $10^{6.3}$     & $10^{5.3}$     \\ %\cline{2-10} 
                                         & 0.01           & $>$       & $>$      & $>$     & $>$      & $>$     & $>$     & $>$    & $>$     \\ %\cline{2-10} 
                                         & 0.001          & $>$       & $>$      & $>$     & $>$      & $>$     & $>$     & $>$    & $>$     \\ \hline
\end{tabular}
%\normalsize
\caption[Upper bound on the light mediator lifetime from $N_{eff}$ at CMB for several values of the hidden-to-visible temperature ratio at DM decoupling]{Upper bound on the light mediator lifetime from $N_{eff}$ at CMB for several values of the hidden-to-visible temperature ratio at DM decoupling. This bounds are given for both the vector portal (left) and the scalar portal (right) models assuming a relativistic decoupling. The mention "$>$" indicates an upper bound greater than $10^{8}$ sec.}\label{table:Neff}
\end{table}

From the upper bounds given in Table \ref{table:Neff}, it is easy to conclude that the thermally disconnected HS scenario fulfill easily the $N_{eff}$ constraint at recombination. Indeed, as the mention of "$>$" indicates an upper bound greater than $10^{8}$ sec, a temperature ratio smaller than 0.01 is enough to guarantee that this constraint is avoided. 
\\

Note that for a stable mediator (or with a lifetime longer than the age of the Universe), as it is the case if the mediator is lighter than twice the electron mass or if there is no SM-to-med connection at all, the only relevant constraints are given by the requirement that the mediator relic abundance is negligible compared to the DM and by the Hubble constant. In both cases, the extra $(T'/T)^3$ suppression factor of the mediator number density makes those constraints easier to be fulfilled. 

\subsection{BBN}
As in the case of constraints coming from CMB, BBN constraints such that the Hubble constant, the entropy injection and the photodisintegration are all relaxed by the extra factor of $(T'/T)^3$ in the light mediator number density. One could then also consider a conservative upper bound by multiplying the light mediator number density by $(T'/T)^3$. This has been done in Table \ref{table:BBN} which gives the maximal value of the mediator lifetime allowed by the Hubble constant/entropy injection and photodisintegration constraints. As explained in Subsection \ref{subsec:BBN-CST}, photodisintegration constraint does not apply if the mediator mass is smaller than the photodisintegration threshold (i.e. if $m_{med}< 4.4$~MeV). In the same way, none of these constraints apply if the mediator is stable or if the mediator is already non-relativistic at BBN (its number density would have been Boltzmann suppressed).
\\

One can thus conclude that BBN constraints become quickly irrelevant in the case of a thermally disconnected HS. Typically, a hidden-to-visible temperature ratio below $T'/T<0.01$ is low enough to avoid all BBN constraints for most of the parameter space.

\begin{table}[h!]
\centering
\begin{tabular}{|c|c|c|c|c|c|c|c|c|c|c|c|}
\hline
\multicolumn{2}{|c|}{\multirow{2}{*}{$\tau_{med}$ (sec)}} & \multicolumn{4}{c|}{$m_{\gamma '}$ (MeV)} & \multicolumn{4}{c|}{$m_{\phi}$ (MeV)} \\ \cline{3-10} 
\multicolumn{2}{|c|}{}                                    & 3     & 10     & 30    & 100    & 3     & 10     & 30    & 100   \\ \hline
\multirow{4}{*}{$\frac{T'}{T}$}          & 1              & $10^{2.0}$       & $10^{1.0}$      & $10^{0.0}$     & $10^{-0.4}$      & $10^{2.4}$       & $10^{2.0}$      & $10^{1.0}$     & $10^{-0.1}$     \\ 
                                         & 0.1            & $10^{6.9}$       & $10^{4.1}$      & $10^{3.9}$     & $10^{3.8}$      & $10^{7.8}$       & $10^{4.2}$      & $10^{4.0}$     & $10^{3.9}$     \\ 
                                         & 0.01           & $>$       & $>$      & $10^{4.7}$     & $10^{4.5}$      & $>$     & $10^{7.6}$     & $10^{4.9}$    & $10^{4.7}$     \\ 
                                         & 0.001          & $>$       & $>$      & $>$     & $10^{7.0}$      & $>$     & $>$     & $>$    & $>$     \\ \hline
\end{tabular}
\caption[Upper bound on the light mediator lifetime from photodisintegration and Hubble constant/entropy injection during BBN for several values of the hidden-to-visible temperature ratio at DM decoupling]{Upper bound on the light mediator lifetime from photodisintegration and Hubble constant/entropy injection during BBN for several values of the hidden-to-visible temperature ratio at DM decoupling. This bounds are given for both the vector portal (left) and the scalar portal (right) models assuming a relativistic decoupling. The mention "$>$" indicates an upper bound greater than $10^{8}$ sec.}\label{table:BBN}
\end{table}

\subsection{Direct detection}
For the whole DM versus mediator mass plane we consider in Figure \ref{fig:portalmax}, the DM-to-SM connector $\kappa'/\kappa_{\phi}$ appears to be below or almost below current direct detection constraints given in Figure \ref{fig:DD_Constraints}. Indeed, we have $\kappa _{KM}\lesssim 10^{-10.5}$ for the vector portal model and $\kappa _{HP}\lesssim 10^{-13}\left(\frac{v_{\phi}}{\text{GeV}}\right)$ for the scalar portal model. This stems from the fact that, as we have already seen above (see Section Eqs. \ref{eq:alphapth} and \ref{eq:alphaphith}), if the temperature ratio is smaller than one, the DM-to-med coupling which is required by the relic density constraint, is smaller than the one one should expect in the thermally connected case (i.e. when $T'=T$). As a direct consequence, the DM scattering cross section on nucleon (given in Eqs. \ref{eq:cross-section-KM} and \ref{eq:cross-section-HP} for the vector and scalar portal models respectively) will be suppressed such that it is easier to satisfy the direct detection constraints in the thermally disconnected scenario. Note that it has already been suggested some time ago that considering a hidden-to-visible temperature ratio smaller than unity could relax the direct detection (see \cite{Chu:2016pew}). A value of $T'/T$ slightly below unity has also been considered in \cite{Foot:2014uba,Foot:2016wvj}, along another general (”dissipative”) framework.

\subsection{Indirect detection}
Once again, in the thermally disconnected scenario ($T'<T$), the required DM-to-med coupling is smaller than in the thermally connected scenario ($T'=T$). Thus, the DM annihilation cross section into two mediator particles is also smaller and constraints coming from indirect detection experiments becomes easy to be fulfilled. Such constraints are then relevant only for smaller DM masses (smaller than a few GeV), see Section \ref{sec:TpT_results} below.

\section{Summary plot}\label{sec:TpT_results}
Let us now present how all of these constraints look like in the DM versus mediator mass plane together with the other constraints developed in Chapter \ref{ch:constr}.

\subsection{Kinetic mixing portal}
\begin{figure}
\centering
\includegraphics[scale=0.6]{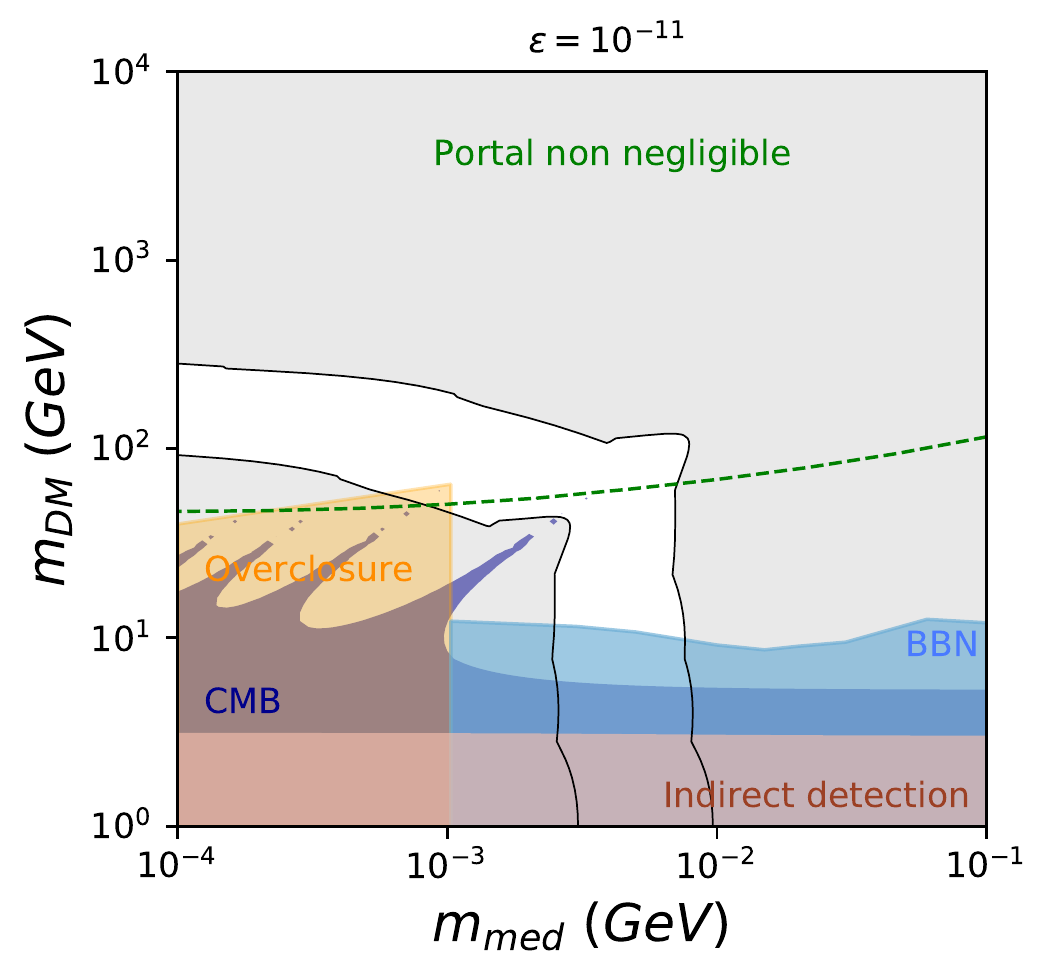}
\includegraphics[scale=0.6]{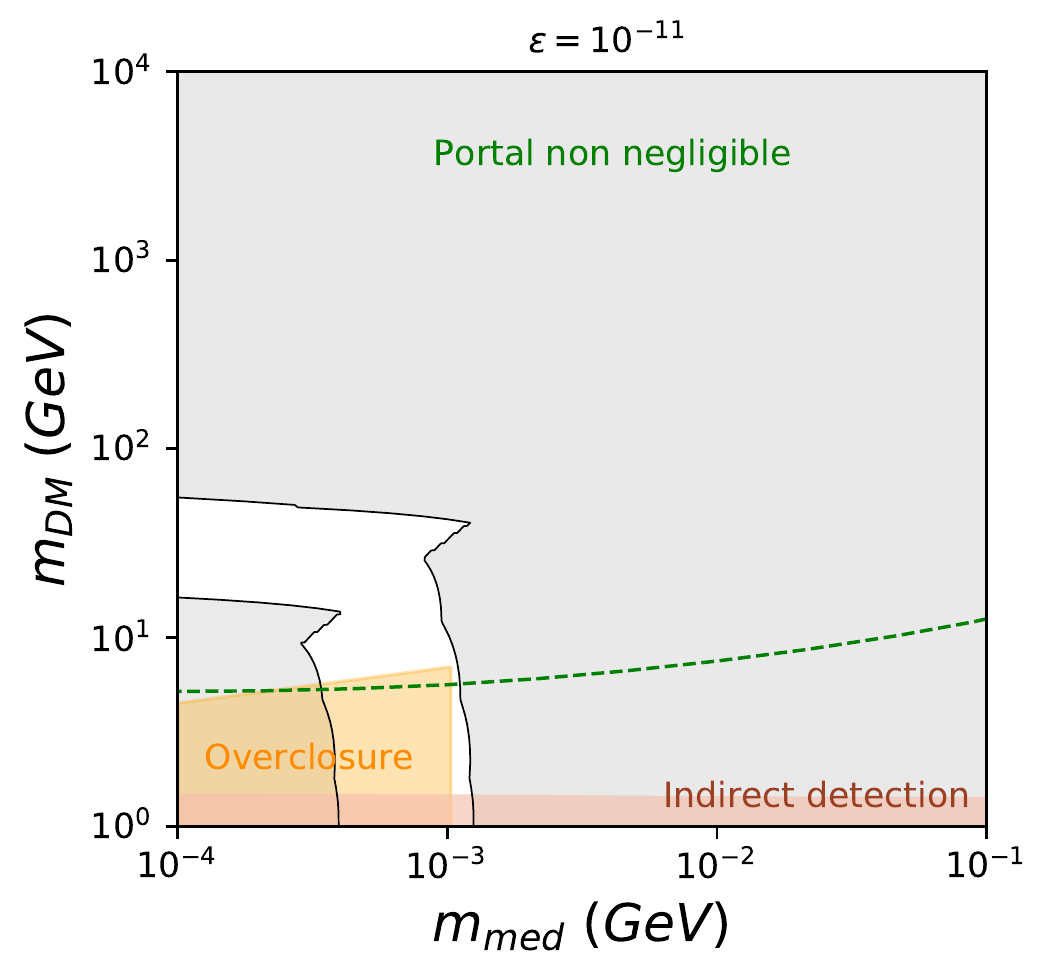}
\includegraphics[scale=0.6]{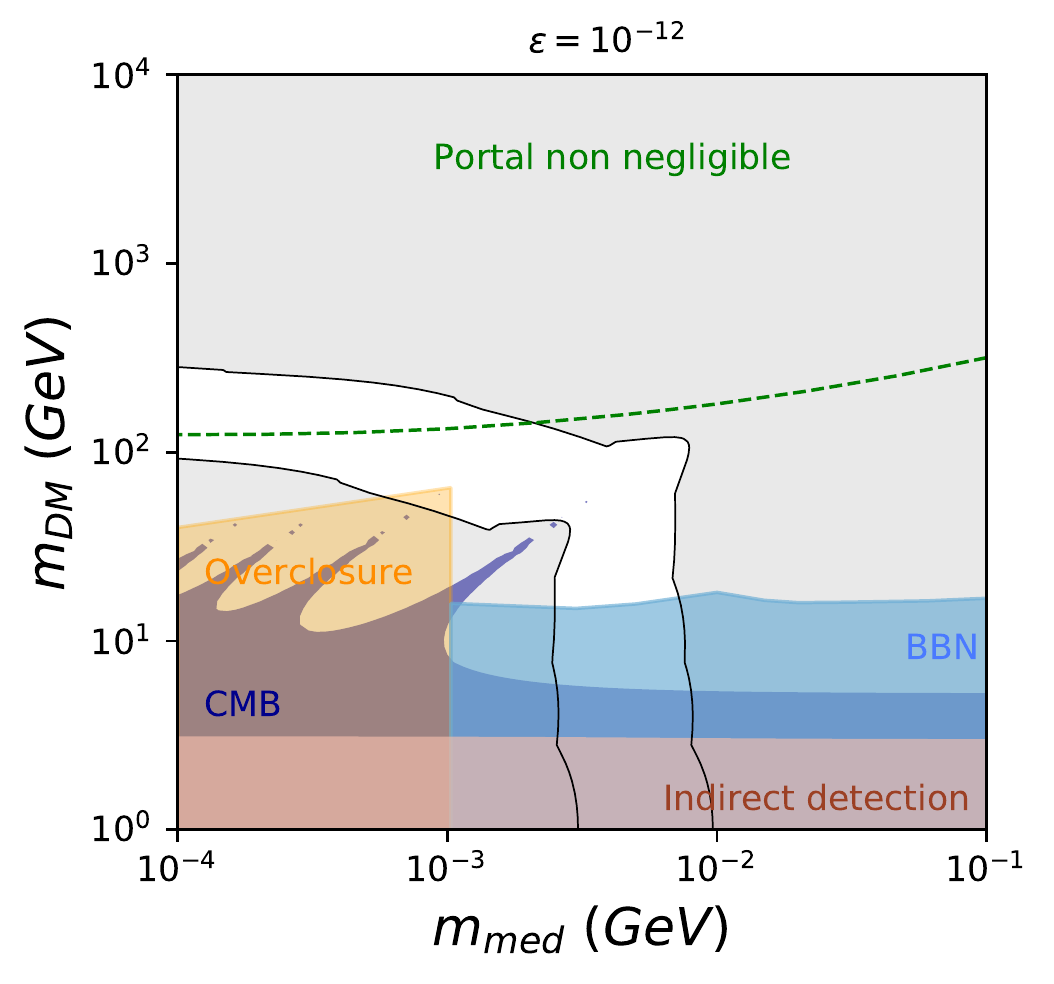}
\includegraphics[scale=0.6]{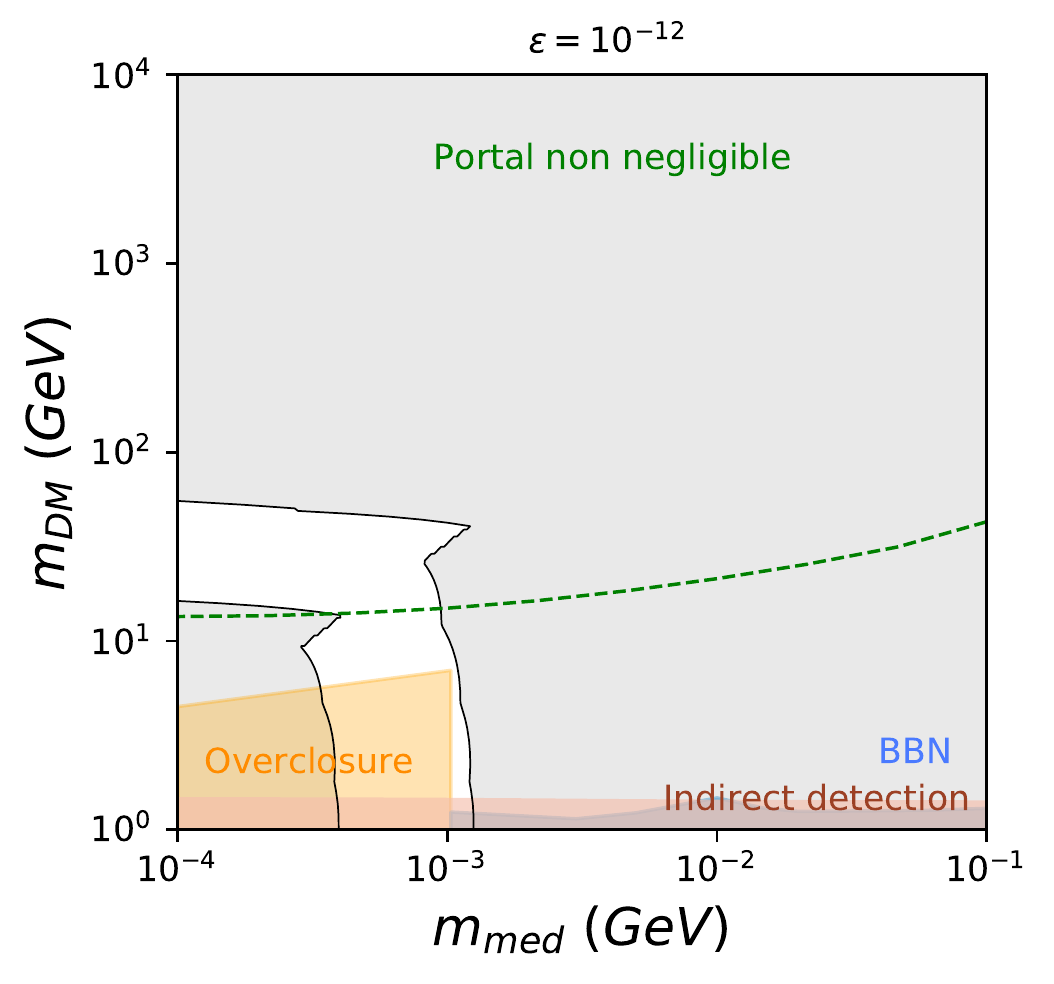}
\includegraphics[scale=0.6]{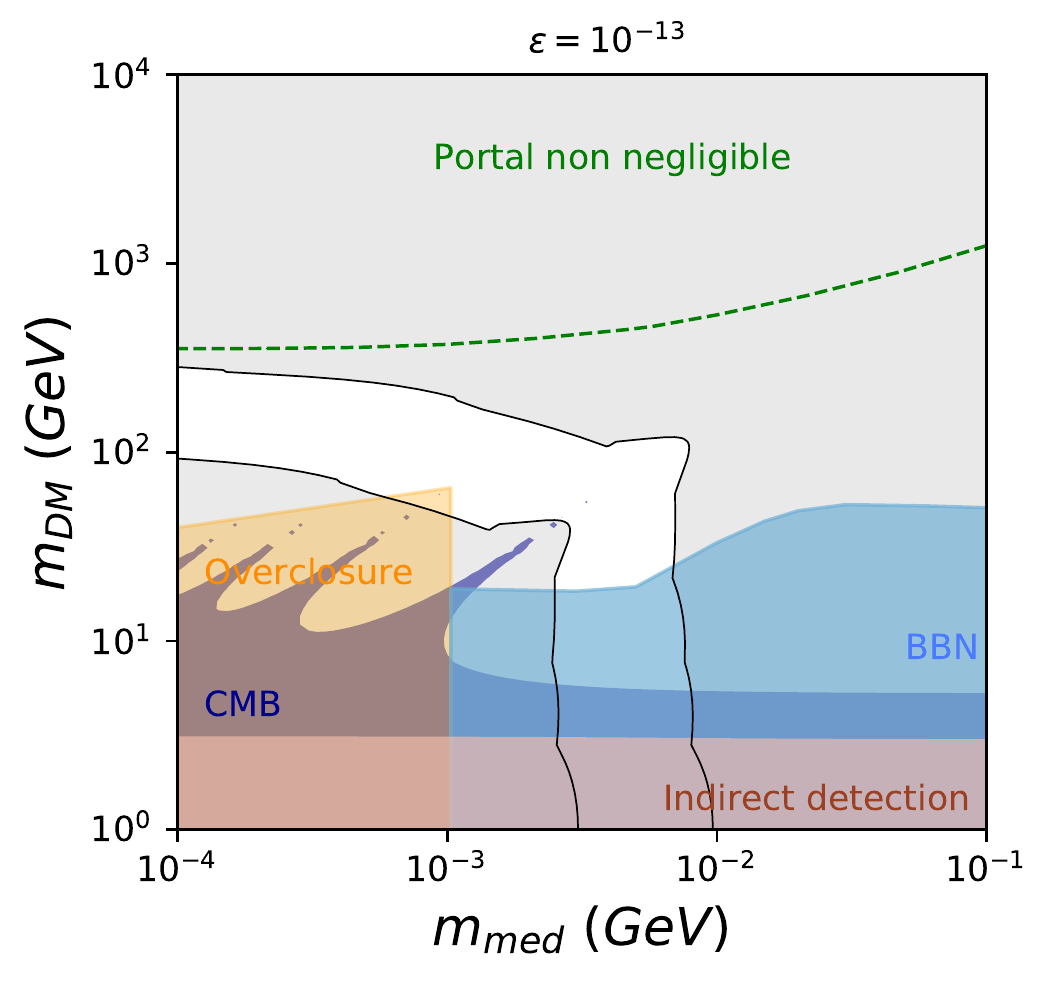}
\includegraphics[scale=0.6]{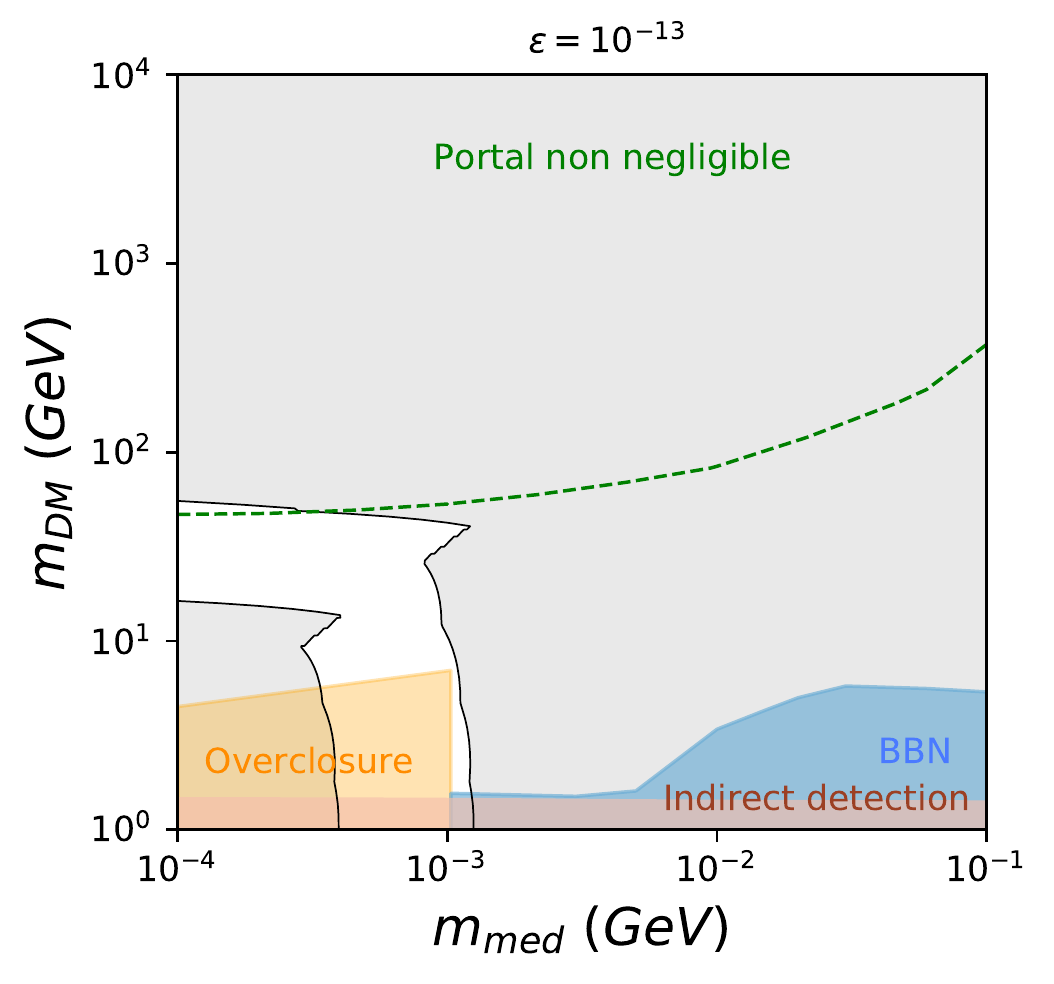}
\caption[Summary plot for thermally disconnected vector portal model]{Constraints from CMB, BBN, self-interaction, indirect and direct detection all together for the vector portal model with $\alpha '= 10^{-4}$  (left) and $\alpha '= 10^{-5}$ (right), for three different values of $\epsilon$.}\label{fig:CMB_SI_alphapV}
\end{figure}

Starting with the vector portal model, Figure \ref{fig:CMB_SI_alphapV} shows in the DM versus mediator mass plane which part of this parameter space all constraints previously discussed exclude for two values of the DM-to-med coupling: $\alpha '=10^{-4}$ (left) and $\alpha '=10^{-5}$ (right) and for three values of the SM-to-med connector: $\epsilon=10^{-11}$ (top), $\epsilon=10^{-12}$ (middle) and $\epsilon=10^{-13}$ (bottom).
\\

In each plot of Figure \ref{fig:CMB_SI_alphapV}, the light blue region is excluded by photodisintegration and/or Hubble constant/entropy injection at BBN combined with the $N_{eff}$ constraint. As explained above, these constraints have been determined by applying the $(T'/T)^3$ suppression factor to the dark photon number density of existing bounds in the thermally connected case (see \cite{Hufnagel:2018bjp}), as we believe constraints obtained in this way are conservative. The red region on the same plot shows the indirect detection constraints while the grey region indicates where the self-interactions constraints are not satisfied. As for direct detection constraints, they are not visible since they are satisfied for the whole DM versus mediator mass plane for these choices of the DM-to-med and SM-to-med couplings. Indeed, as we have seen in \ref{fig:DD_Constraints}, the direct detection constraints can be translated into an upper bound on the DM-to-SM coupling $\kappa'$ as a function of the DM. The maximal allowed value of this coupling is $\kappa'\lesssim 3\times 10^{-11}$ while this coupling is at most $1.13\times 10^{-12}$ (top left) and at least $3.58\times 10^{-15}$ (bottom right). The relic density constraint is automatically satisfied for the whole parameter space visible in these plots as we have fixed the value of the temperature ratio according to Figure \ref{fig:TpT_epsilon}. Since all this has been done assuming that the portal interaction has a negligible impact on both DM and mediator relic abundance, the green dashed line indicates the maximal value  of the DM mass for which this assumption holds (obtained as in Figure \ref{fig:portalmax}). Above this line the portal interaction starts to have more and more impact as we increase the DM mass. However this does not necessarily exclude this part of the parameter space, on the contrary, we expect that a sizeably region above the green dashed line is still allowed. To show that, one would have to compute the relic abundance including portal interactions, which is beyond the scope of this thesis. One would then enter a reannihilation (\ref{subsec:rean}) or a secluded freeze-out (\ref{subsec:secluded}) production regime (see \cite{Chu:2011be}).
\\

The differences between the left and right plots of Figure \ref{fig:CMB_SI_alphapV} can be easily understood. A lower DM-to-med coupling means a smaller DM annihilation rate and a smaller Boltzmann suppression of the DM number density at freeze-out. Once again, this imply a smaller hidden-to-visible temperature ratio (see Figure \ref{fig:TpT_epsilon}) and finally a smaller mediator density. This explain why the overclosure constraint as well as the BBN and CMB constraints are excluding a smaller part of the parameter space when $\alpha' =10^{-5}$ than when $\alpha' =10^{-4}$. The indirect detection constraint is also relevant for a smaller part of the parameter space; this is simply due to the smaller DM annihilation rate. The self-interaction constraints are more difficult to satisfy and require a smaller DM mass, but since the other constraints are not excluding this part of the parameter space anymore, we still have an order of magnitude in the DM mass and in the mediator mass which is allowed. Finally, the fact that the green dashed line appears to go down as we decrease the DM-to-med coupling is understood as following: as one decreases the DM connection to the mediator, the DM and mediator number densities become more sensitive to the small source term from SM.

\subsection{Higgs portal}
\begin{figure}
\centering
\includegraphics[scale=0.6]{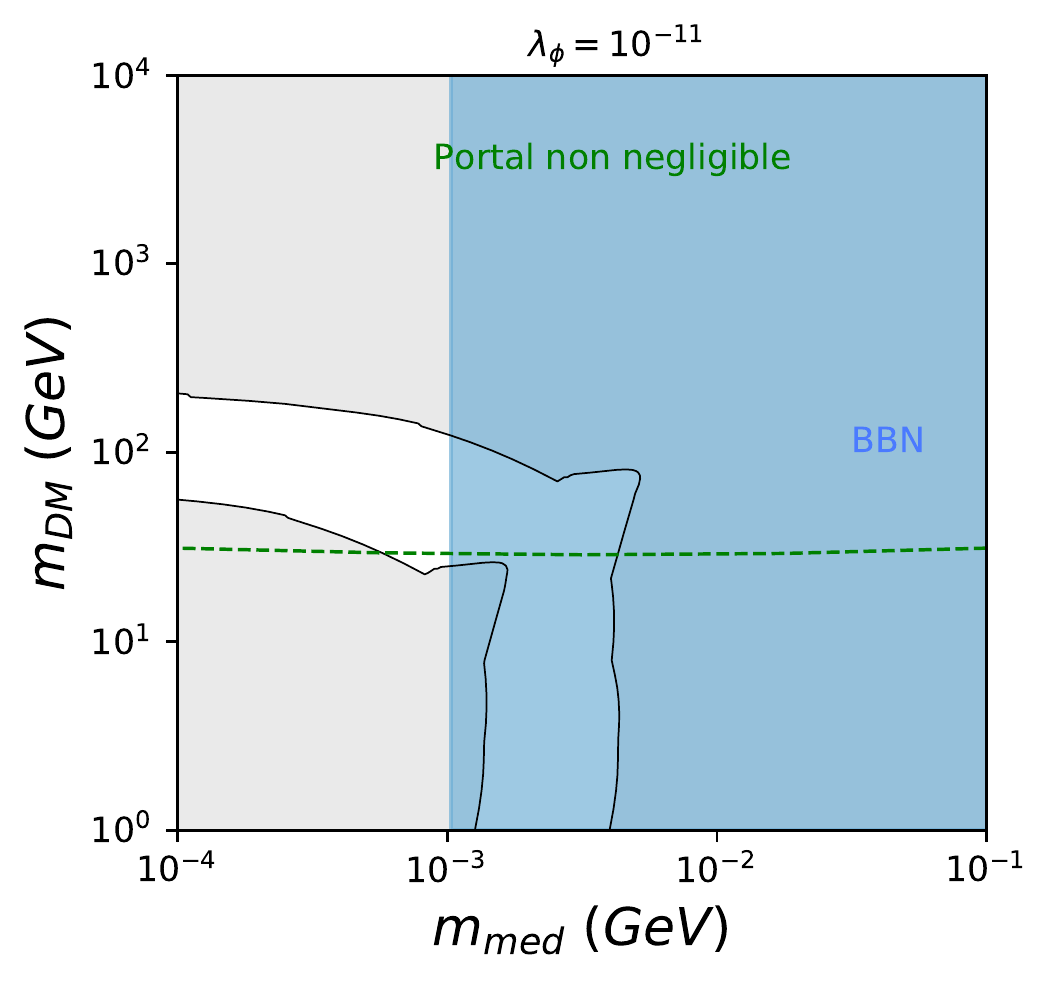}
\includegraphics[scale=0.6]{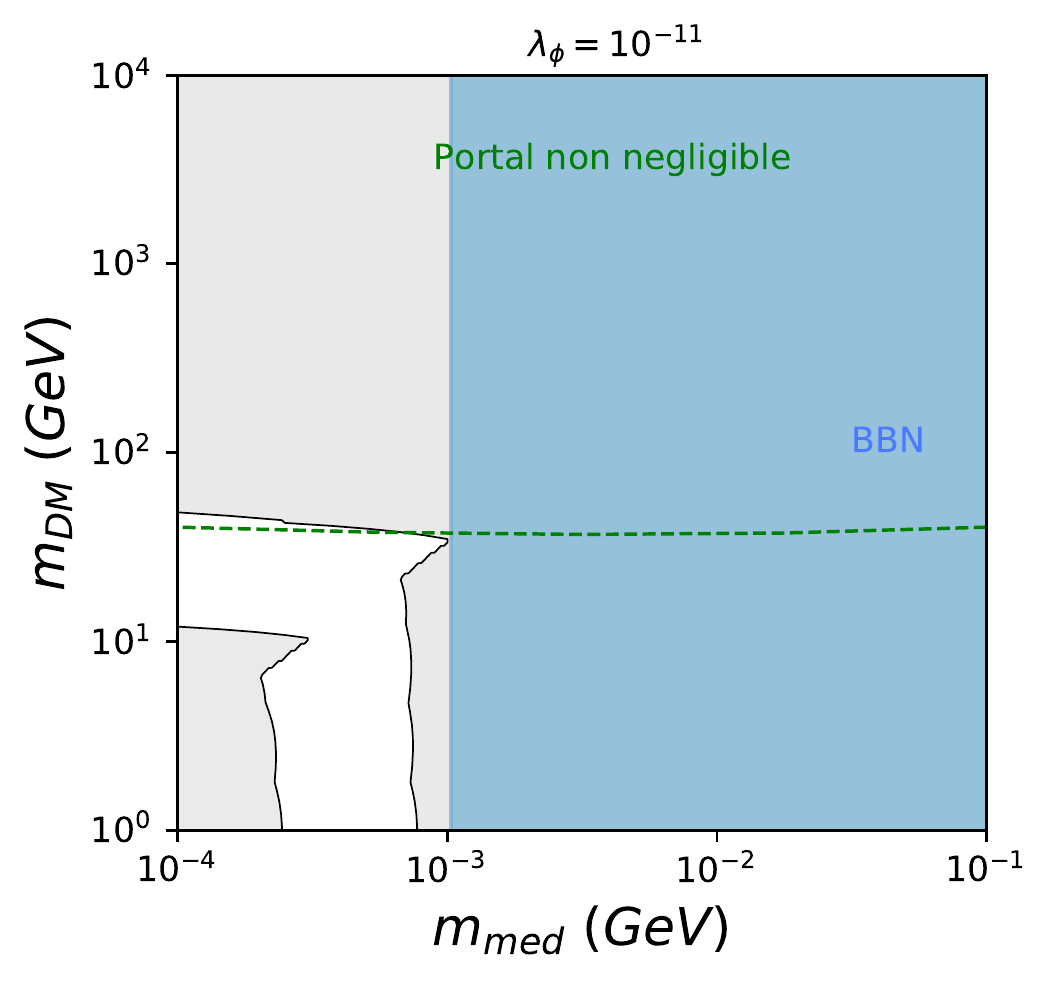}
\includegraphics[scale=0.6]{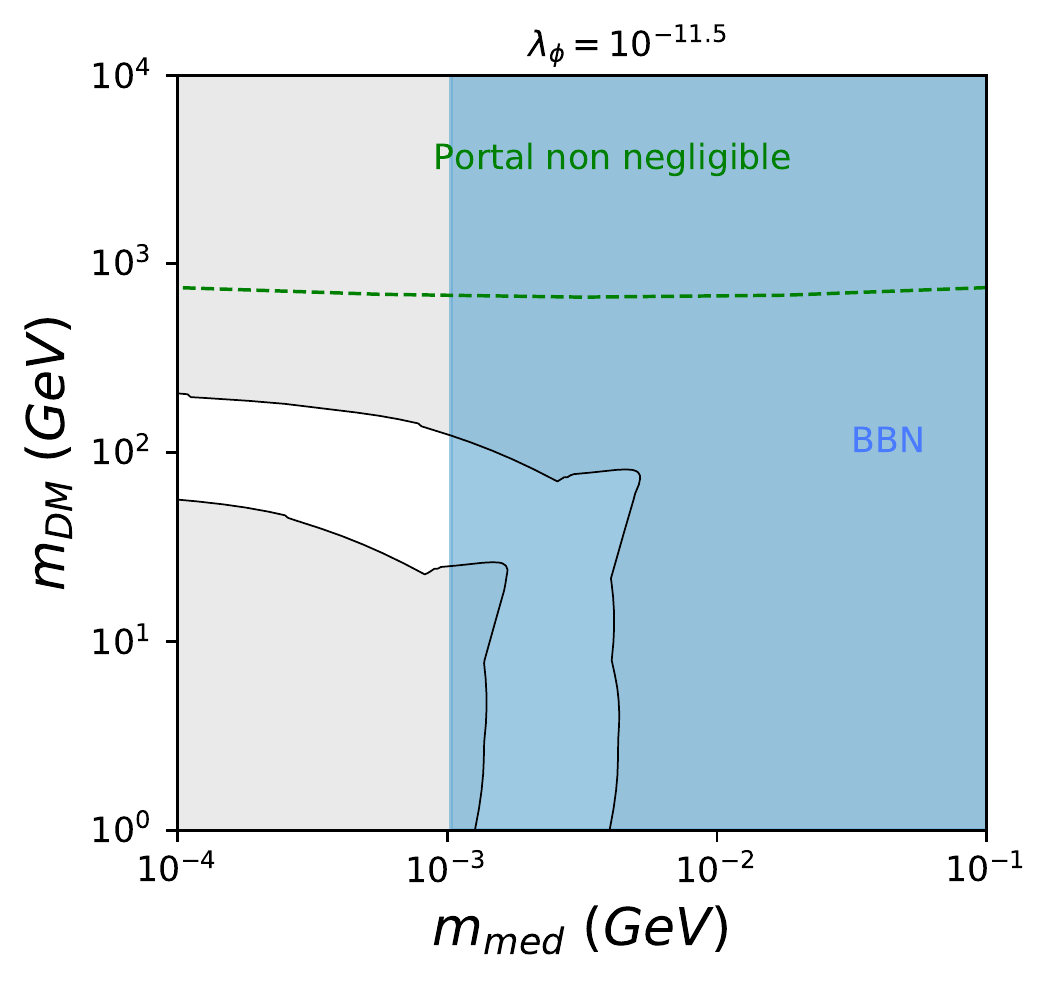}
\includegraphics[scale=0.6]{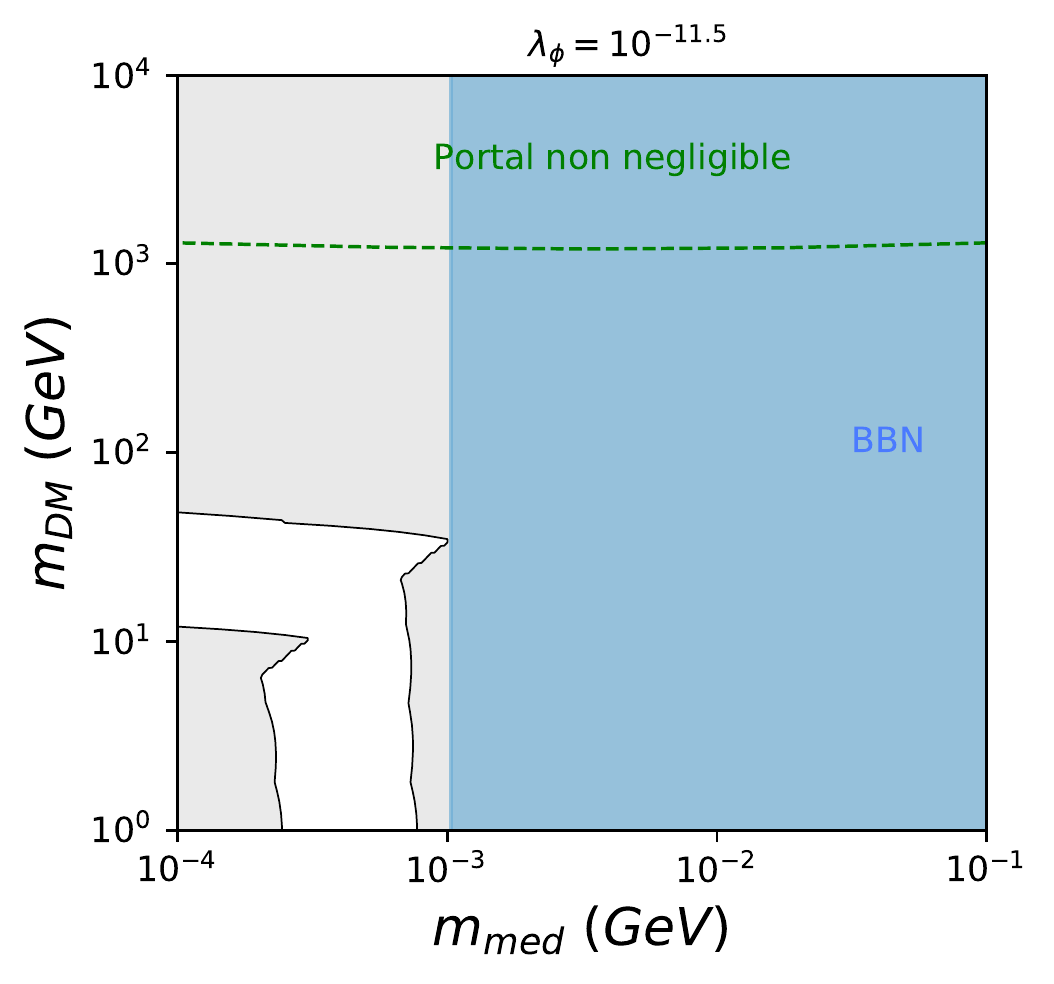}
\includegraphics[scale=0.6]{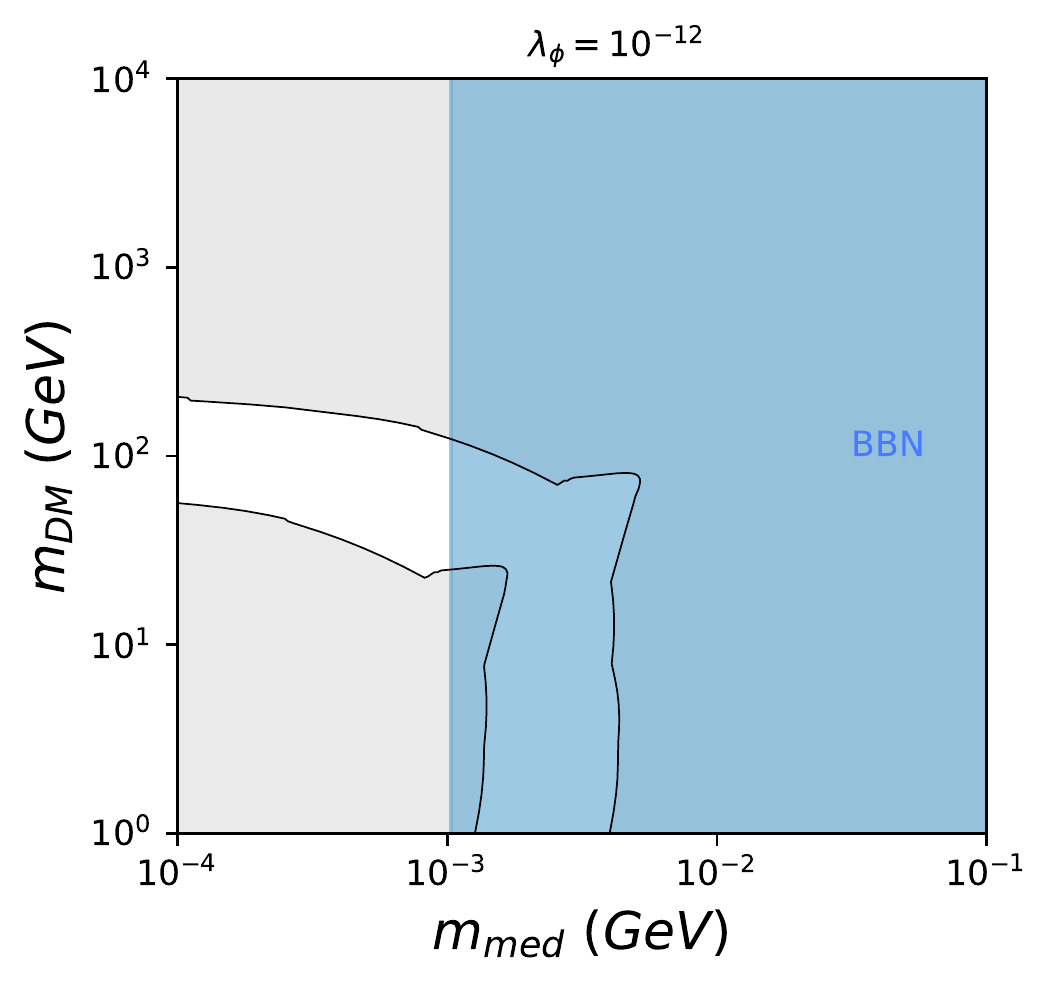}
\includegraphics[scale=0.6]{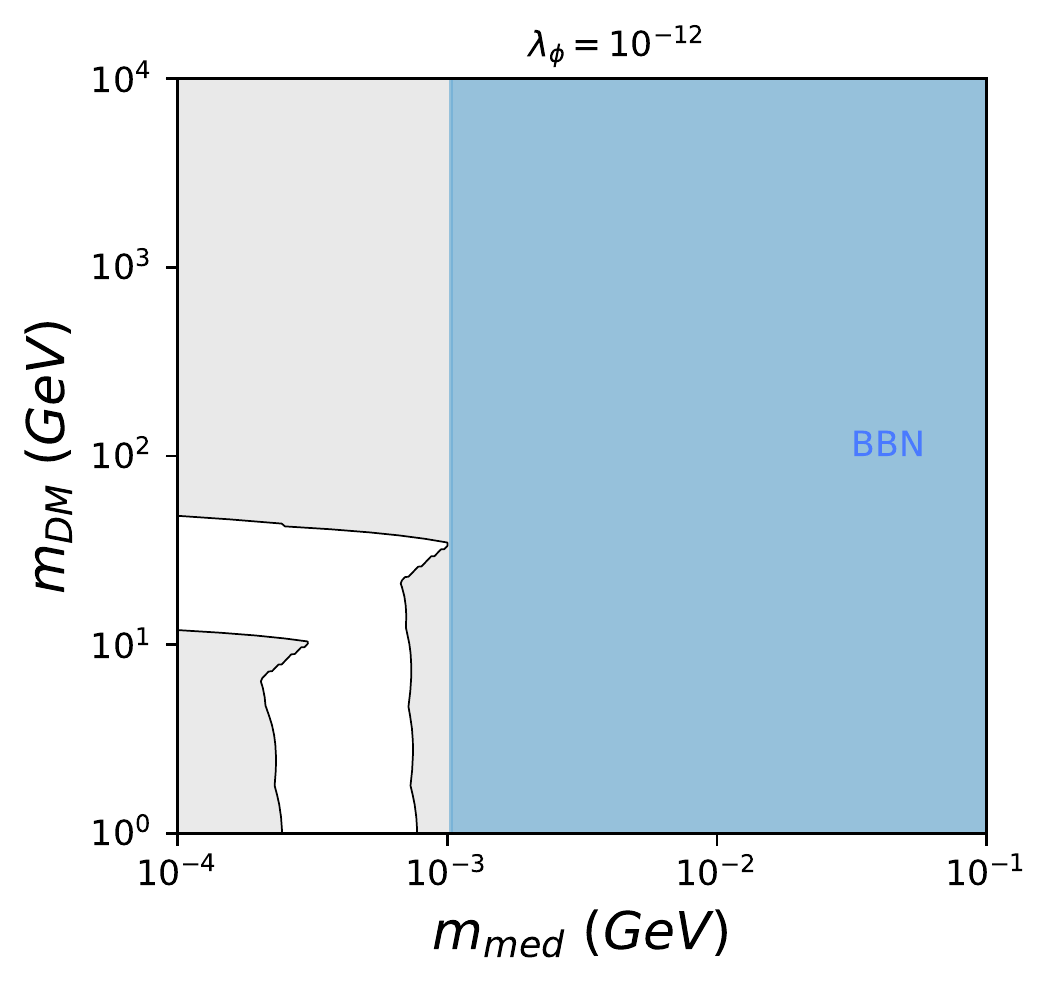}
\caption[Summary plot for thermally disconnected scalar portal model]{Constraints from CMB, BBN, self-interaction, indirect and direct detection all together for the scalar $A_{\phi}$ model with $\alpha _{\phi}= 10^{-4}$  (left) and $\alpha _{\phi}= 10^{-5}$ (right), for three different values of $\lambda _{\Phi H}$.}\label{fig:CMB_SI_alphapVphi}
\end{figure}

Switching now to the scalar portal model, Figure \ref{fig:CMB_SI_alphapVphi} shows the same parameter space as in Figure \ref{fig:CMB_SI_alphapV}, but for the Higgs portal. The DM-to-med coupling has also been fixed to $\alpha_\phi=10^{-4}$ for panels on the left while panels on the right are for $\alpha_\phi=10^{-5}$. Again, these plots are for three values of the SM-to-med connector: $\lambda_{\Phi H}=10^{-11}$ (top), $\lambda_{\Phi H}=10^{-11.5}$ (middle) and $\lambda_{\Phi H}=10^{-12}$ (bottom).
\\

As in the previous case, when the mediator is heavier than twice the electron mass (i.e. $m_{med}>2 m_e$), the mediator decay width\footnote{All relevant decay widths and cross sections can be found in Appendix \ref{app:cs}.} is suppressed by the mixing parameter. Furthermore, it is also suppressed due to the electron Yukawa coupling. As a consequence, the BBN constraints on the mediator lifetime impose to the mixing angle and thus the SM-to-med coupling to be large such that the mediator does not decay during or after BBN. However, hidden sector particles can still be actively produced from SM thanks to the quartic interaction ($H H\rightarrow \phi\phi$) which involves the SM-to-med coupling only. This can force thermalisation between the hidden and the visible sectors which contradicts the assumption of a thermally decoupled HS. The major consequence of this is that the whole parameter space is excluded by BBN constraints as soon as the mediator is heavier than twice the electron mass. This can be seen in Figure \ref{fig:CMB_SI_alphapVphi}. One can also see that the narrow region which in the case where both sector thermalise with each other is still not excluded (for $m_{DM}\sim 0.5$~GeV and $m_{med}\sim 1.1$~MeV \cite{Hufnagel:2018bjp}) disappears when the HS is colder than the visible sector. This is due to the fact that this region is viable thanks to a large value of the SM-to-med coupling (so that the light mediator can decay before BBN and avoid its constraints) but this large value of the Higgs portal deeply thermalises both sectors, such that $T'/T=1$.
\\

However, as one can conclude from Figure \ref{fig:CMB_SI_alphapVphi}, the scalar portal model is now widely open for mediator lighter than twice the electron mass, unlike for the case where both bath thermalise with each other. Below the electron threshold, the decay width is suppressed by loop processes and this leads to lifetime so large than they are forbidden in the $T'/T=1$ case but not anymore in the $T'/T<1$ case, i.e. from the suppression by a factor of $\left(T'/T\right)^{3}$. For $\alpha'=10^{-5}$ this even allows values of $m_{DM}$ below the GeV scale.

\section{What if no portal at all?}
As a final comment for this chapter, let us stress that from the discussion above, it appears that both the vector portal model and the scalar portal model are perfectly viable without any portal at all (i.e. with a stable mediator). In this case all constraints related to the decay of the light mediator, as well as direct and indirect detection, disappear and only the modification of the Hubble constant that the light mediator implies at BBN and CMB times, as well as the non-overclosure constraint are relevant. On the one hand, the former constraint can be easily satisfied by requiring that the mediator is non relativistic enough at these relevant times. This would then require a sufficiently high mediator mass. On the other hand, the latter constraint can be satisfied by requiring that the mediator constitutes a negligible fraction of the DM relic abundance today. This can be done imposing $\Omega_{\rm med}<\Omega_{\rm DM}$ which implies

\myeq{
\frac{T'_{\rm dec}}{T_{\rm dec}}\leq 1.14\times 10^{-2}\times \left(\frac{\text{MeV}}{m_{\rm med}}\right)^{1/3}\times  \left( \frac{g^{S}_{\ast}(T_{\rm dec})}{g^{\rm eff}_{\rm med}(T'_{\rm dec})} \right)^{1/3},
\label{lowerTprimestable}
}

\noindent where all quantities are taken at the DM decoupling. This upper bound lies well above the relativistic floor of Eq. \ref{eq:TpTlowerbound}, as a result of the fact that $m_{\rm med}\ll m_{\rm DM}$, so that this scenario is perfectly viable.

\chapter{Additional minimal solutions for SIDM}\label{ch:other_minimal}
\yinipar{I}n the previous chapter, the fact that the HS was thermally decoupled from the SM bath was used as a way to circumvent various constraints applying on SIDM with light mediator models. Indeed, we have seen at the end of Chapter \ref{ch:constr} (see more specifically Section \ref{sec:global_picture}) that particle physics experimental constraints were incompatible with small scale structure and cosmological constraints. Basically, a colder and thermally decoupled HS greatly weakens the cosmological constraints such that the tensions can be alleviated. Through this final chapter, we are going to see that even if there is no decoupled DM hidden sector, one can also alleviate the various tensions applying to SIDM in several other ways by introducing an additional particle on top of the DM and the light mediator particles. The way we will invoke an extra particle will differ in each scenario, but the general idea remains the same in all cases: to decouple the small scale structure constraint from the cosmological constraints (see below). We will end this chapter by seeing what could still be done if one does not introduce any extra degree of freedom and consider thermally coupled dark sector. This final chapter is based on \cite{Hambye_2020} and on results from \cite{Hambye:2018dpi} and unpublished results (see Chapter \ref{ch:som}).

\section{Subleading DM annihilation into light mediators}\label{sec:subdomannihwayout}
Let us start with the simplest way to decouple small scale structure constraints from cosmological constraints. As we said above, this makes the use of at least one extra new particle on top of the DM and the light mediator particles. This extra particle does not need necessarily to be light\footnote{In the sense that this extra particle can be heavier than the light mediator, but has still to be lighter than the DM.}, which allows this extra particle to easily fulfil all cosmological constraints in a much easier way than the light mediator. The simple idea is to assume that the coupling between the DM and the new heavy mediator is the one which fixes the DM relic density from a freeze-out of DM annihilation into heavy mediators, say $DM\,DM\rightarrow X\,X$. Thus in such a scenario, the DM annihilation rate into a pair of light mediators can be much smaller than the thermal value. This does not cause any problem for the small scale structure constraints as these can still be easily fulfilled for a large part of the DM versus mediator mass plane with couplings to the light mediator much smaller than the ones required to account for the DM relic density, see Figure \ref{fig:SI_Oh2}. Thus, even if it proceeds in a s-wave way, this allow to fulfil the CMB constraints on the DM annihilation rate into light mediators as seen in Figure \ref{fig:CST-CMB-1}. Alternatively, the annihilation channel responsible for the DM relic density, could be p-wave in order to satisfy this CMB constraint such that the heavy mediator should be a scalar boson. Moreover, the direct detection constraint is relaxed by the fact that the DM-to-med coupling is reduced. The simplest option of this kind we have found turns out to have as DM candidate a Dirac fermion, as a light mediator, a vector boson or a scalar boson and as a real or complex scalar $S$ as extra particle (we took it to be real and consider only interactions which contains an even number of them). Those two versions are simply the usual vector and scalar portal model we are used to, but with an additional scalar boson. Their Lagrangian are given by

\myeq{
&{\cal L}\owns -g \gamma '_\mu i\bar \psi \gamma^\mu \psi-\frac{\epsilon}{2} F^Y_{\mu\nu}F'^{\mu\nu}-y_S S \bar{\psi} \psi +\lambda_{H S} H^{\dagger}H S^{2},\label{eq:sub_KM}\\
&{\cal L}\owns- y_\phi \phi\overline{\chi} \chi +h.c. -\lambda \phi^\dagger \phi H^\dagger H-y_S S \bar{\psi} \psi +\lambda_{H S} H^{\dagger}HS^2 +\lambda_{\phi S} \phi^\dagger \phi S^2.\label{eq:sub_HP}
}

\begin{figure}
\centering
\includegraphics[scale=0.6]{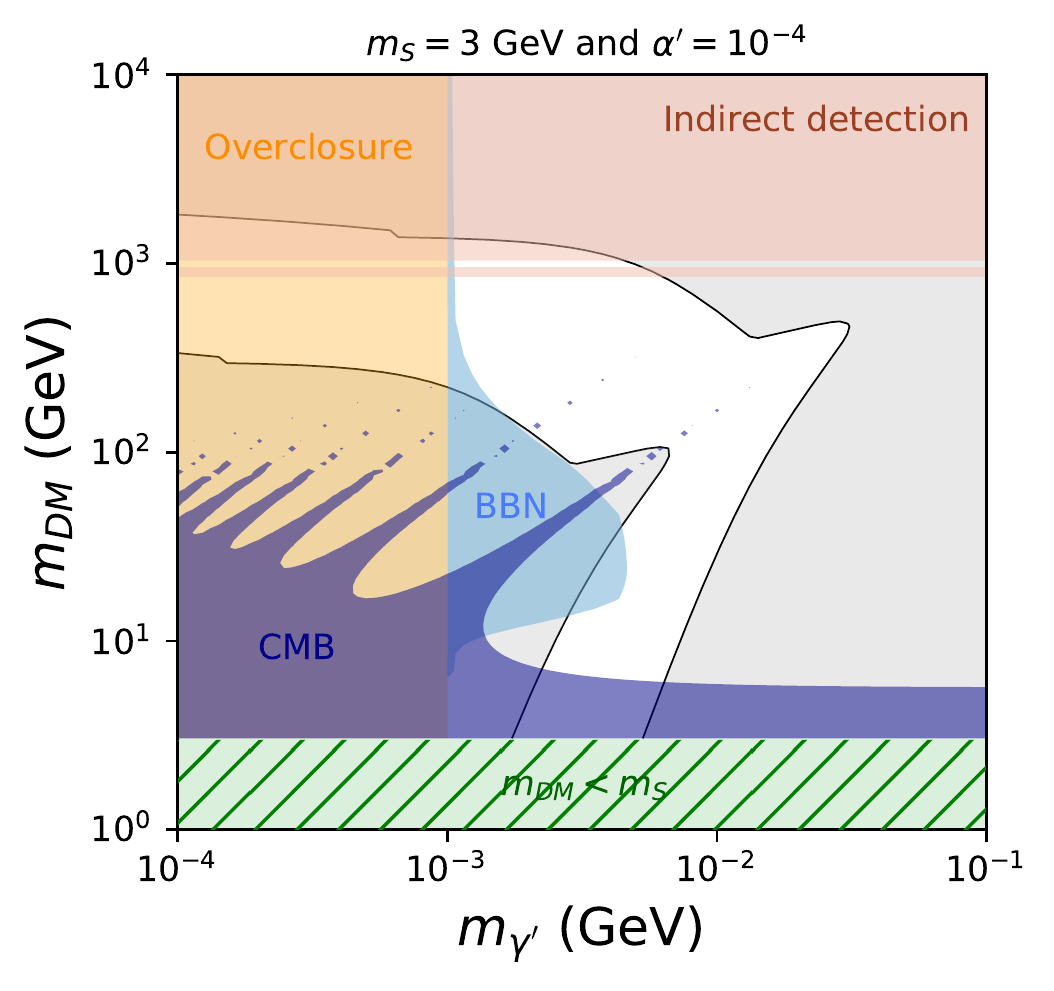}
\includegraphics[scale=0.6]{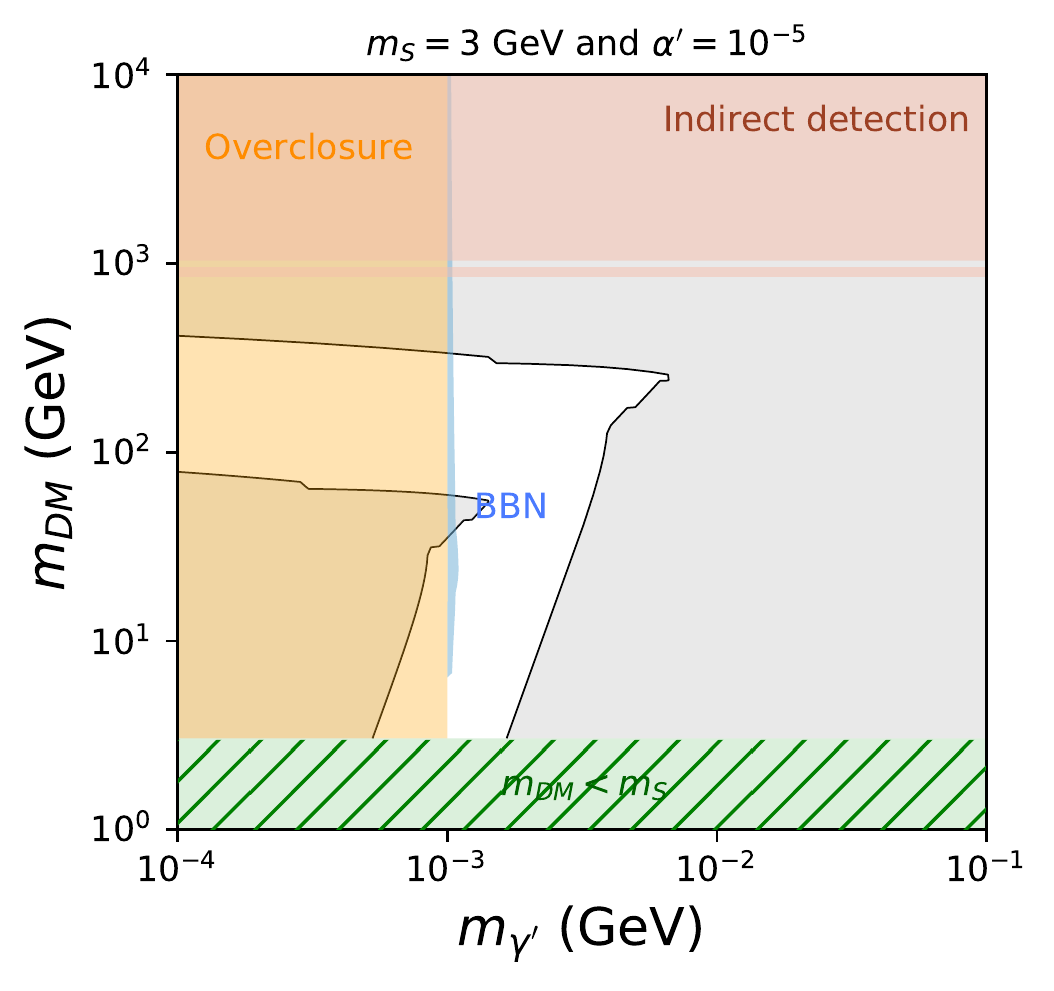}
\includegraphics[scale=0.6]{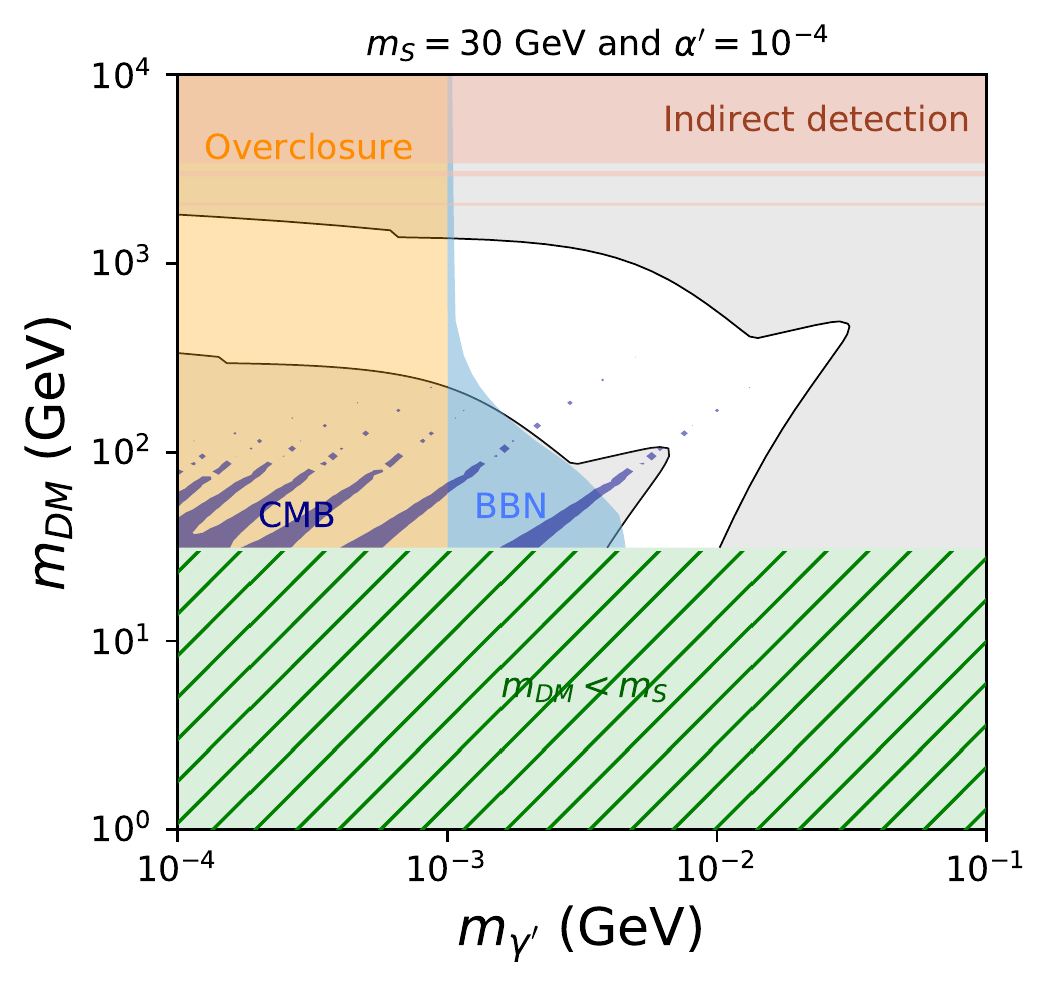}
\includegraphics[scale=0.6]{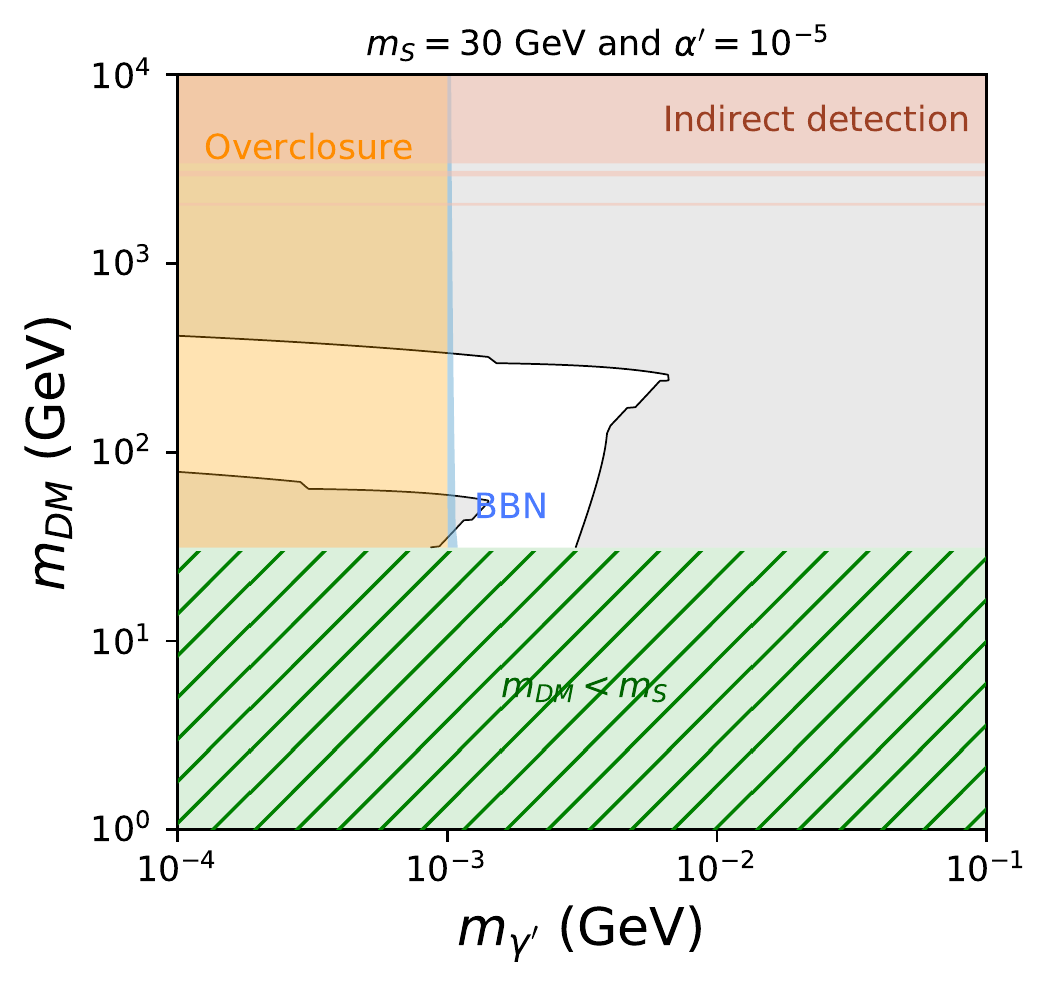}
\includegraphics[scale=0.6]{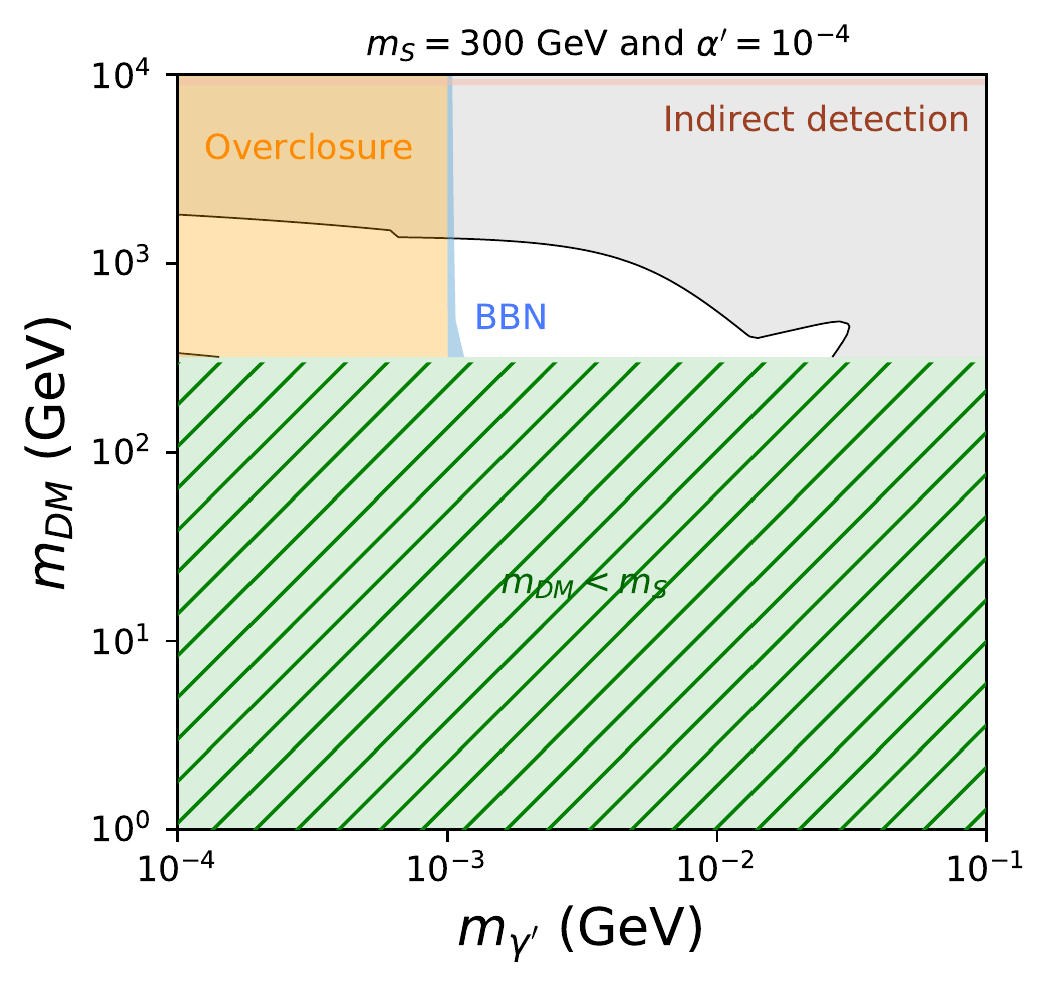}
\includegraphics[scale=0.6]{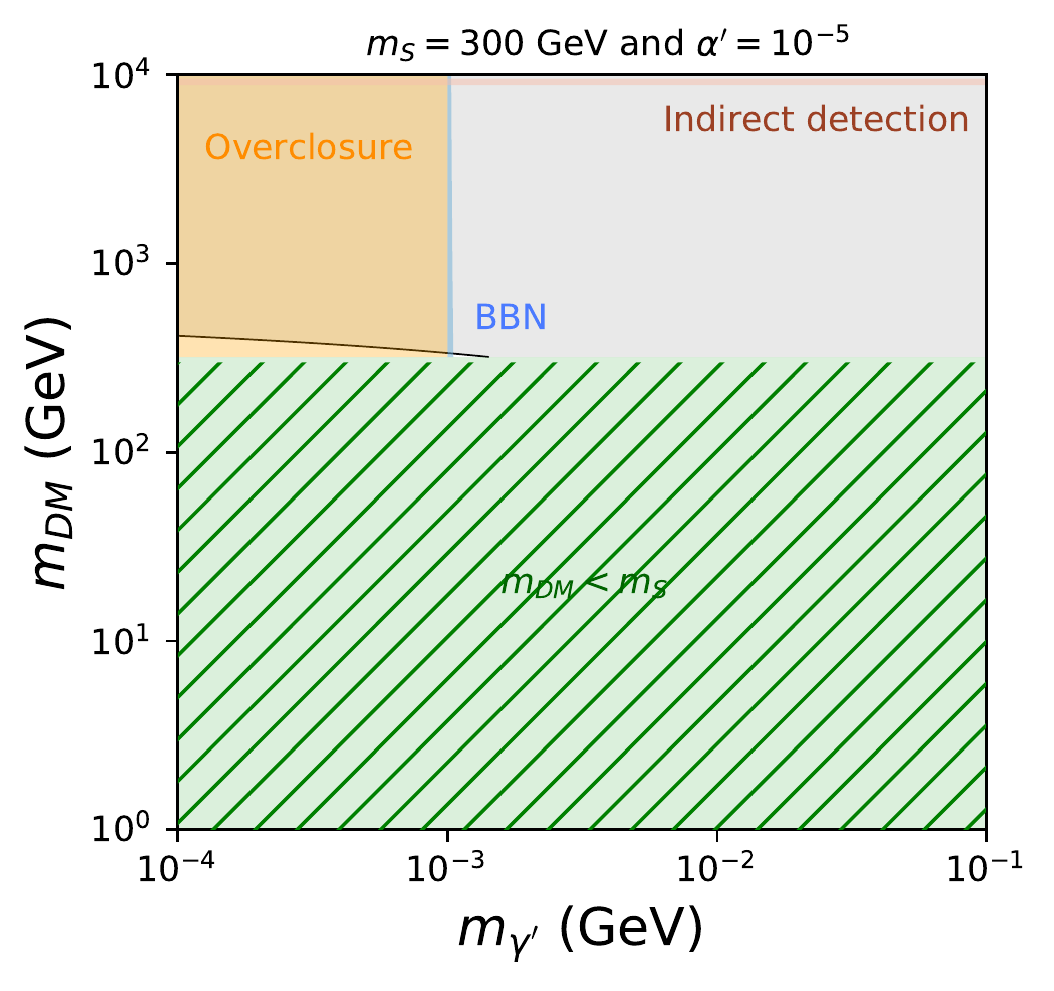}
\caption[Summary plot for the subleading DM annihilation scenario - vector portal model]{Constraints from CMB, BBN, self-interaction, indirect and direct detection all together for the subleading DM annihilation into light mediators with $\alpha '= 10^{-4}$  (left) and $\alpha '= 10^{-5}$ (right), for three different values of $m_S$.}\label{fig:BV}
\end{figure}

\begin{figure}
\centering
\includegraphics[scale=0.6]{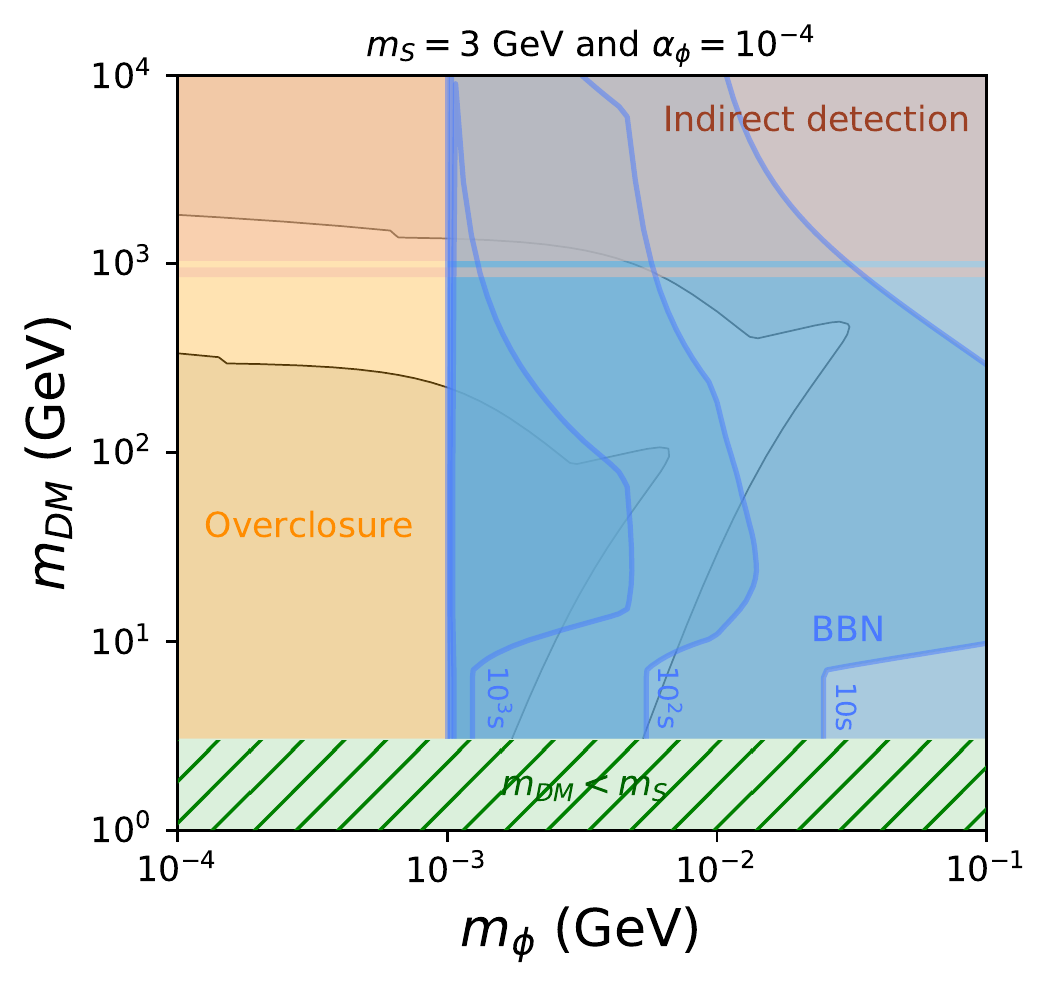}
\includegraphics[scale=0.6]{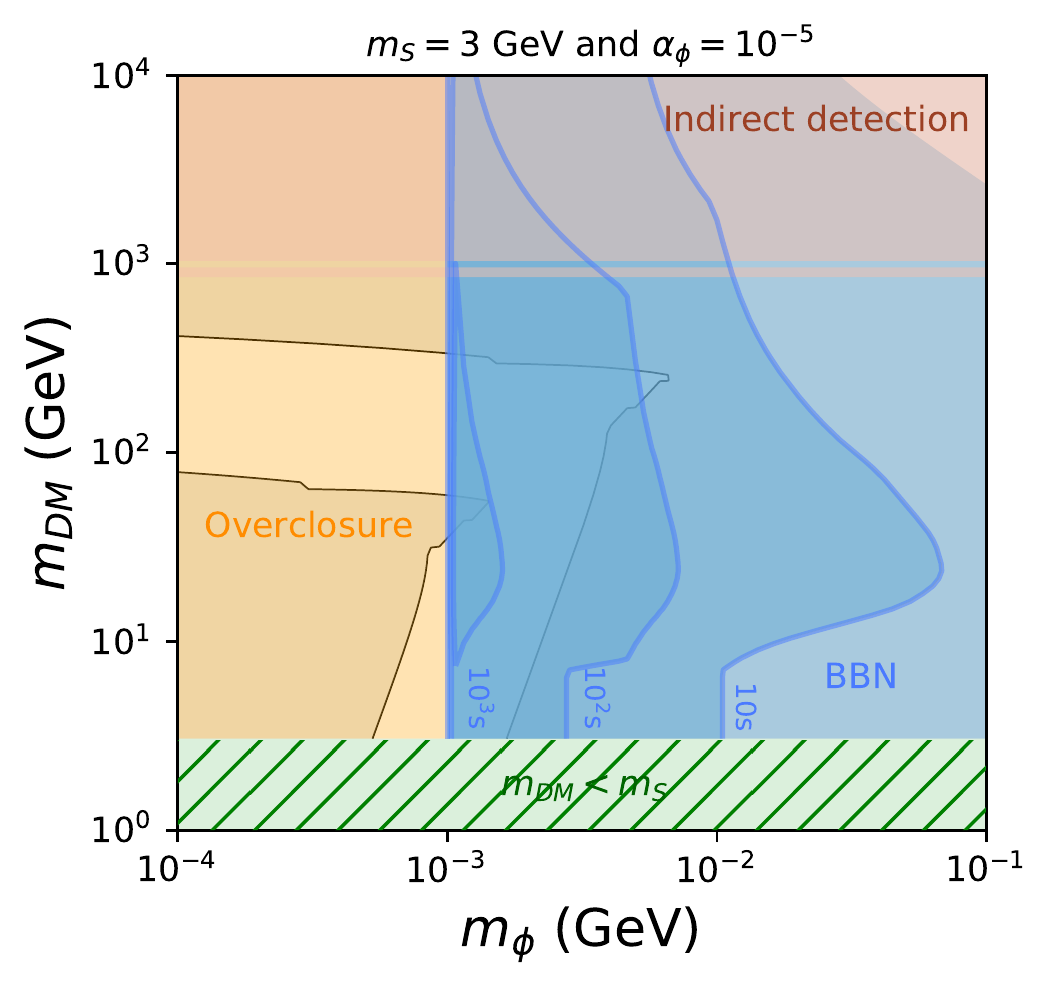}
\includegraphics[scale=0.6]{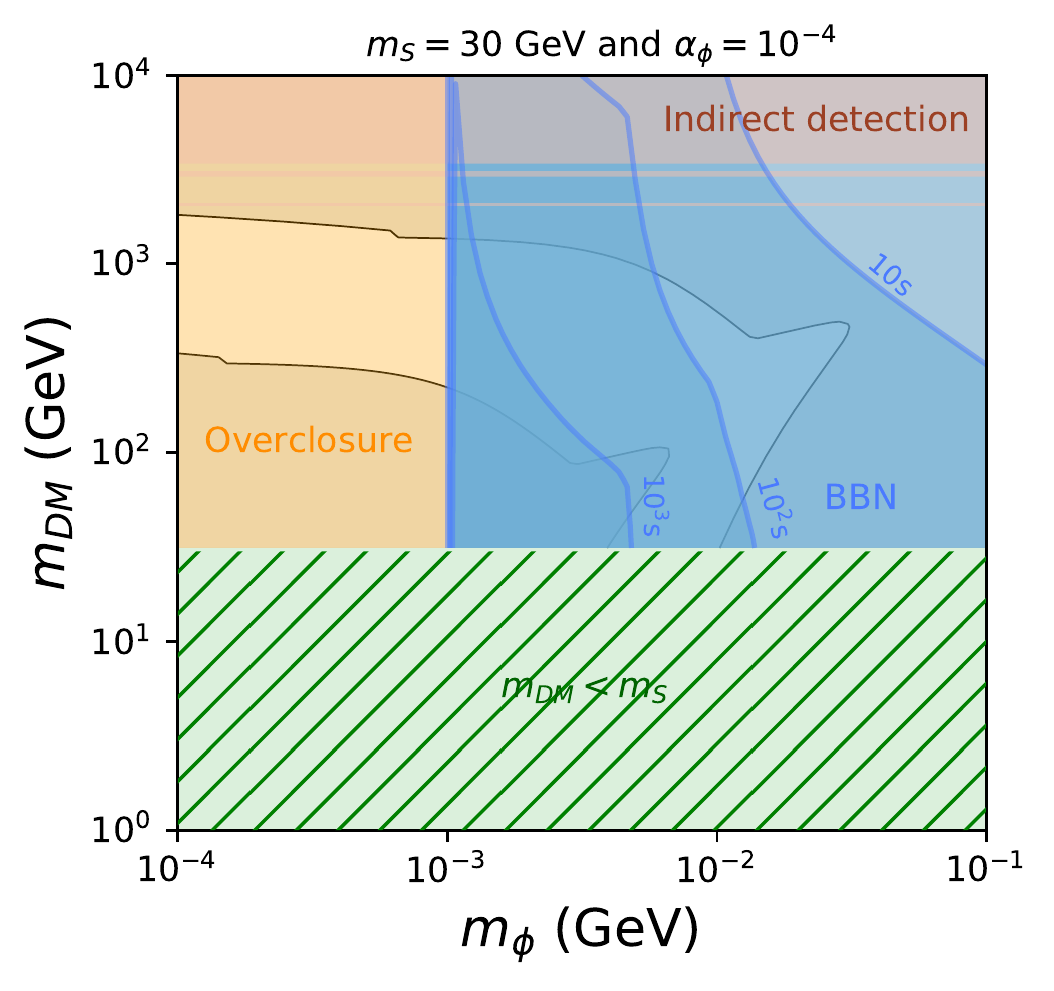}
\includegraphics[scale=0.6]{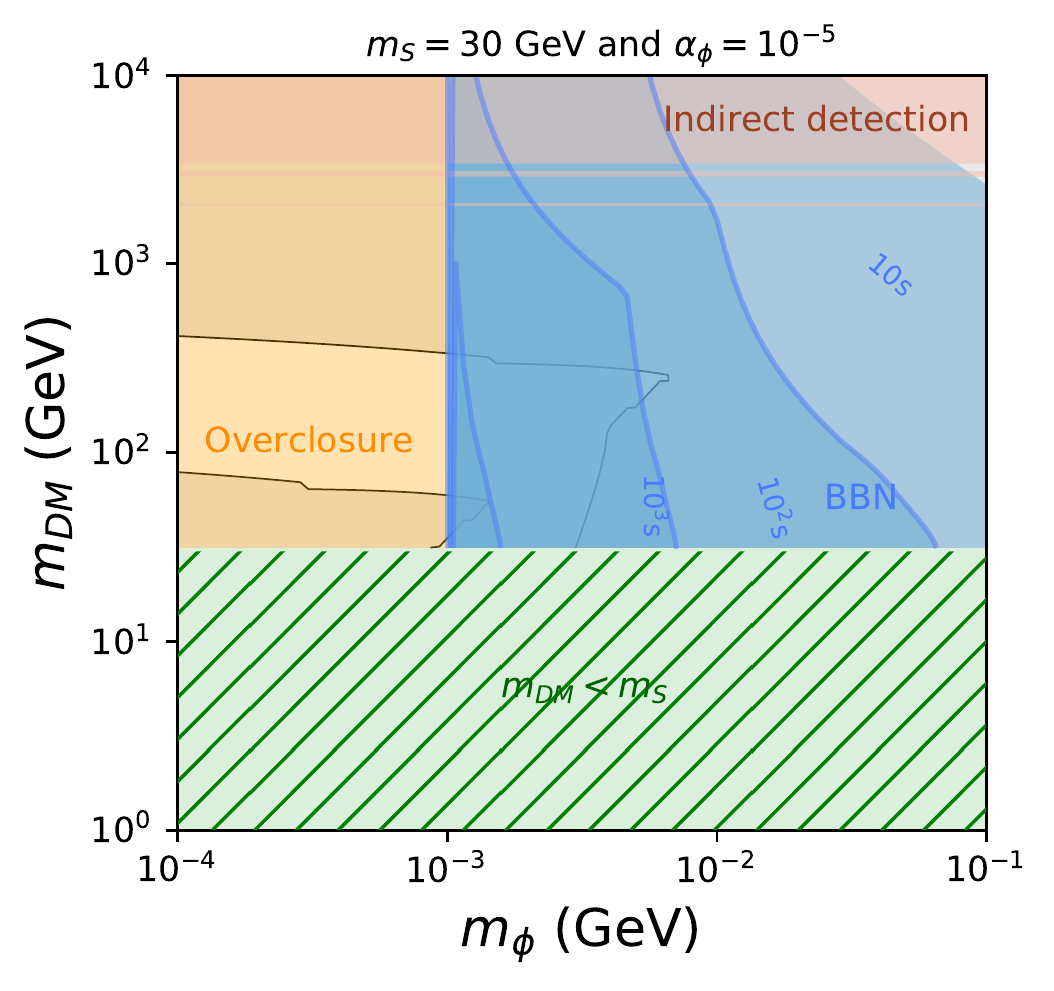}
\includegraphics[scale=0.6]{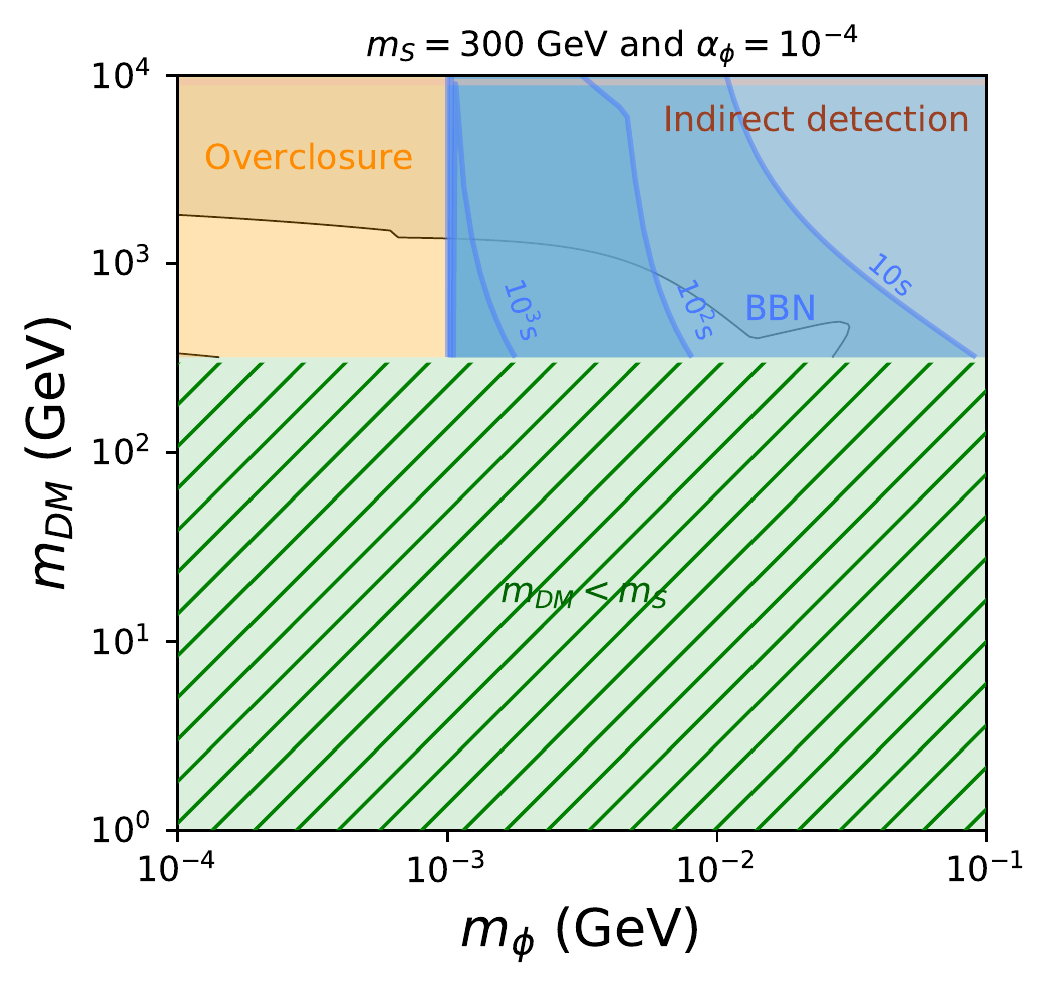}
\includegraphics[scale=0.6]{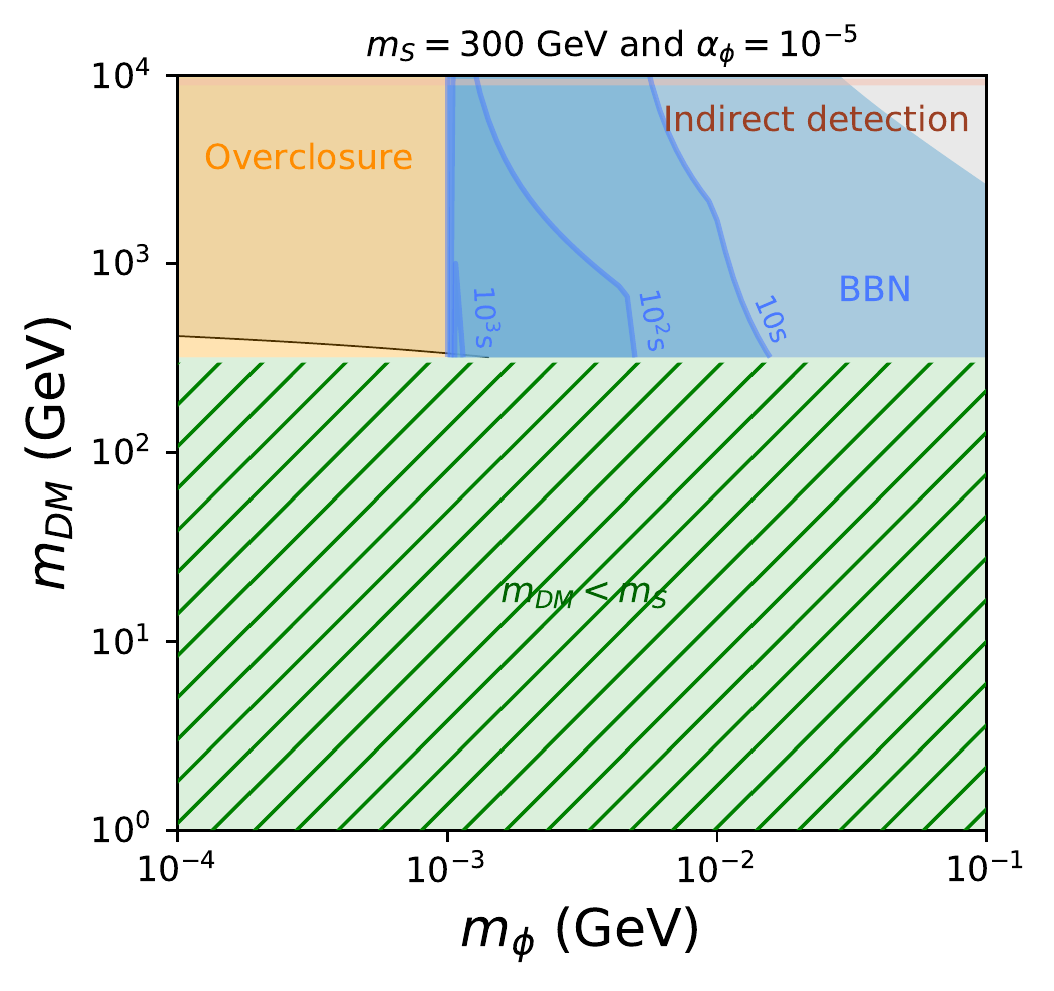}
\caption[Summary plot for the subleading DM annihilation scenario - scalar portal model]{Constraints from CMB, BBN, self-interaction, indirect and direct detection all together for the subleading DM annihilation into light mediators with $\alpha_{\phi}= 10^{-4}$  (left) and $\alpha_{\phi}= 10^{-5}$ (right), for three different values of $m_S$.}\label{fig:BS}
\end{figure}

\noindent Now, we will see how such a structure allows us to avoid each constraints. We present results obtained for the vector and the scalar models in Figures \ref{fig:BV} and \ref{fig:BS} respectively. For each figures, the left (resp. right) panels show results for the DM-to-med coupling fixed to $\alpha'/\alpha_{\phi}=10^{-4}$ (resp. $\alpha'/\alpha_{\phi}=10^{-5}$) and this for three values of the new heavy mediator mass: $m_{S}=3$ GeV (top), $m_{S}=30$ GeV (middle) and $m_{S}=300$ GeV (bottom). On these figures, the small scale structure exclusion areas are shown in shaded grey while the green hatched regions indicate the part of the parameter space which is kinetically forbidden as the heavy mediator is heavier than the DM. In this region, the DM freezes-out when the temperature reaches the heavy mediator mass scale (i.e. when $T\simeq m_{S}$) instead of the DM mass scale. Thus, the DM decouples while still relativistic and does not undergo through the usual Boltzmann suppression. As a consequence, its relic abundance is much bigger than the one required by observations. We also show the overclosure constraint in yellow which arises when the light mediator is stable which happens once its mass is smaller than twice the electron mass: $m_{\rm med}<2m_{e}$. This part of the parameter space is then also excluded.

\subsection{CMB}
As the new heavy mediator is a scalar boson in both the vector and the scalar models, the CMB does not suffer from any large distortion from DM annihilation into a pair of this heavy mediator. Indeed, we have seen in Eq. \ref{cmbconstraint} that p-wave annihilations do not produce enough charged particles at recombination in order to spoil the CMB spectra. Thus, it is the same than for the case of a DM annihilation into light scalar mediators. The scalar model is then not further constrained by the upper bound on the annihilation rate at recombination, this explains the absence of CMB constraint for all panels of Figure \ref{fig:BS}. The light mediator could still perturb the CMB in the vector model in which the DM annihilation into a pair of light mediators is of s-wave type. However, since this annihilation rate is subleading, its corresponding cross section is smaller than the thermal value such that it can easily fulfill the upper bound given Eq. \ref{cmbconstraint} and depicted in Figure \ref{fig:CST-CMB-1}. The CMB bound is shown in dark blue in Figure \ref{fig:BV}.
\\

Moreover, in both models the heavy and light mediators are connected to the SM through a kinetic mixing portal or a Higgs portal depending on if it is a vector or a scalar boson respectively. These mixing interactions do not have to be particularly small as one does not impose to the two baths to be thermally disconnected. As a consequence, all of these light and heavy mediators can decay fast enough into a pair of charged fermions to have any effect on the CMB or $N_{eff}$.

\subsection{BBN}
In both the vector and the scalar model, on the one hand we have that the mass of the new heavy mediator particle lies much above the MeV scale (from 3 GeV to 300 GeV in Figures \ref{fig:BV} and \ref{fig:BS}) such that it can easily decay before it would cause any problems for BBN at all. On the other hand, the light mediator mass still lies around the MeV scale and could be constrained. However, as in models presented in the previous chapter (see Section \ref{sec:HS_temperature}), the vector mediator can still decay fast enough to avoid BBN constraints in most of the parameter space, the forbidden part of the parameter space is depicted in light blue in \ref{fig:BV}. The scalar mediator model is a little bit more challenging if one keep requiring that its lifetime should not exceed one second ($\tau_{\phi}<1$ sec). If one relaxes slightly this bound, one can find some non-negligible part of the parameter space to be allowed. This can be seen in Figure \ref{fig:BS}, where we show in light blue several contours of the light mediator lifetime in seconds which excludes less and less of the parameter space if one relaxes more and more the $\tau_{\phi}<1$ sec constraint.

\subsection{Direct detection}
As for the previously defined vector and scalar portal models (see Subsection \ref{subsec:portal} and Section \ref{sec:CST-DD}), one can define a DM-to-SM coupling parameter which here, through the exchange of a $S$ scalar particle, is

\myeq{
\kappa_{S}\equiv \frac{y_{S}}{2}\sin (2\theta),
}

\noindent where $\theta$ is the $H-S$ mixing angle and where we follow the same procedure for the VEV's definitions as detailed in Subsection \ref{subsec:portal} for the portal model with a scalar $\phi$, see Eqs. \ref{eq:lag_hp} to \ref{eq:mixing_angle}. In terms of Lagrangian parameters (Eq. \ref{eq:sub_HP}), the mixing angle is thus given by

\myeq{
\tan(2\theta) = \frac{\lambda_{HS}v_{H}v_{S}}{\lambda_{S}v_{S}^{2}-\lambda_{H}v_{H}^{2}}.
}

\noindent Even if the new heavy mediator $S$ brings a new direct detection channel, since it is much heavier than the light mediator threshold, $m_{S}\gg 40$ MeV (see Section \ref{sec:CST-DD} and Eq. \ref{eq:DD-cross-section}), the DM scattering on nucleon process is strongly suppressed and direct detection experiments are not sensitive to this channel. There is thus no constraints coming from direct detection on the DM-to-SM coupling from this channel, $\kappa_{S}\ll \kappa_{HP}$ for considered mass ranges. Moreover, as the coupling between the DM and the light mediator particle is reduced in this scenario, the direct detection constraints for the light mediator exchanged channel are relaxed. Thus, tensions between direct detection constraints and BBN constraints are relaxed compared to models analysed in Chapter \ref{ch:secl}, which explains why the region allowed by BBN are wider in the present case. This behaviour is true for both vector and scalar portal, but since the constraints were much stronger in the scalar model than in the vector model, the resulting allowed region is still wider in the vector model than in the scalar one. These models illustrate well the fact that if one makes the light mediator to have a negligible impact on the DM production mechanism, it can be easy to fulfil all constraints, specially to alleviate the BBN versus direct detection tensions. For all plots shown in Figures \ref{fig:BV} and \ref{fig:BS}, the magnitude of the DM-to-SM connection occurring through the light mediator has been chosen such that we have the largest part of the parameter space which is allowed. In other words, the direct detection constraints are saturated for the whole parameter space, i.e. $\kappa'\sim\kappa_{KM}/(\text{a few})$ and $\kappa_{\phi}\sim(\text{a few})\times\kappa_{HP}$. However, we would like to emphasise that it is a choice in the way of presenting the results. One can decide to lower the DM-to-SM connection which would shrink a bit the parameter space allowed by BBN constraints.

\subsection{Indirect detection}
Since the DM annihilation into a pair of light mediators cross section at freeze-out is sub-dominant, the corresponding annihilation rate today is also reduced. As a consequence, indirect detection signal from DM annihilation into a pair of light mediators is reduced and there is no indirect detection constraint coming from this channel. Moreover, DM annihilation into a pair of heavy mediator is p-wave and so suppressed too. However, as explained in details in Subsection \ref{subsec:p_wave_ID}, when the heavy mediator $S$ becomes massless compared to the DM (that is to say when $m_{S}\ll m_{\rm DM}$), the Sommerfeld effect can be important, even for p-wave annihilation and one still finds some part of the parameter space that are excluded by indirect detection, depicted in red in Figures \ref{fig:BV} and \ref{fig:BS}.

\subsection{Detailed results}
Figures \ref{fig:BV} and \ref{fig:BS} are used to globally show what the allowed parameter space can look like, but does not give explicitly all parameters. Thus, we present in Table \ref{tab:sub} numerical examples of sets of parameters which satisfy all constraints for both the vector (top lines) and the scalar models (bottom lines).
\\

\begin{table*}[t]
\begin{center}
\resizebox{\textwidth}{!}{%
\begin{tabular}{|c|c|c|c|c|c|c|c|c|c|c|}
\hline
$m_{\rm DM}$                  & $m_{\gamma '}$                  & $m_S$                     & \multirow{2}{*}{$\alpha '$} & \multirow{2}{*}{$y_S$} & $\sigma _T/m_{\rm DM}$                  & \multirow{2}{*}{$\frac{\sigma_{{\rm DM DM} \rightarrow \gamma ' \gamma '}}{\sigma_{thermal}}$} & \multirow{2}{*}{$\kappa '$ $\left(\frac{\kappa '}{\kappa_{KM}^{DD}}\right)$} & \multirow{2}{*}{$\kappa_{S}$ $\left(\frac{\kappa_{S}}{\kappa_{HP}^{DD}}\right)$} & $\tau_{\gamma '}$               & $\tau_{S}$                \\
(GeV) & (MeV) & (GeV) &                             &                        & $(\text{cm}^2/\text{g})$ &                                                                              &                                                 &                                                 & (sec) & (sec) \\ \hline
83                        & 18                        & 31                        & $1.7\times 10^{-4}$         & 0.25                   & 0.18                                & $1.2\times 10^{-2}$                                                          & $1.8\times 10^{-11}$ (0.55)                                            & $1.1\times 10^{-10}$ ($\ll 1$)                                            & 0.30                      & 0.089                     \\
326                       & 12                        & 62                        & $6.5\times 10^{-5}$         & 0.51                   & 0.35                                & $1.2\times 10^{-4}$                                                          & $2.3\times 10^{-11}$ (0.35)                                            & $1.8\times 10^{-10}$ ($\ll 1$)                                            & 0.12                      & 0.006                     \\
617                       & 11                        & 12                        & $3.8\times 10^{-4}$         & 0.70                   & 0.13                                & $1.0\times 10^{-3}$                                                          & $4.4\times 10^{-11}$ (0.47)                                            & $4.4\times 10^{-10}$ ($\ll 1$)                                            & 0.22                      & 0.020                     \\ \hline
\hline
$m_{\rm DM}$                  & $m_{\phi}$                  & $m_S$                     & \multirow{2}{*}{$\alpha_{\phi}$} & \multirow{2}{*}{$y_S$} & $\sigma _T/m_{\rm DM}$                  & \multirow{2}{*}{$\frac{\sigma_{{\rm DM DM} \rightarrow \phi\phi}}{\sigma_{thermal}}$} & \multirow{2}{*}{$\kappa_{\phi}$ $\left(\frac{\kappa_{\phi}}{\kappa_{HP}^{DD}}\right)$} & \multirow{2}{*}{$\kappa_{S}$ $\left(\frac{\kappa_{S}}{\kappa_{HP}^{DD}}\right)$} & $\tau_{\phi}$               & $\tau_{S}$                \\
(GeV) & (MeV) & (GeV) &                             &                        & $(\text{cm}^2/\text{g})$ &                                                                              &                                                 &                                                 & (sec) & (sec) \\ \hline
0.5                       & 1.1                        & 0.01                        & $1.5\times 10^{-5}$         & 0.02                   & 0.19 & 0.23                                                          & $6.8\times 10^{-7}$ (0.60)                                            & $9.4\times 10^{-8}$ (0.09)                                            & 27                      & 1                     \\
2                       & 3                       & 0.01                        & $4.5\times 10^{-5}$         & 0.04                   & 0.13 & 0.13                                                          & $1.9\times 10^{-7}$ (0.99)                                            & $1.9\times 10^{-7}$ (0.99)                                            & 24                      & 1                     \\
326                       & 12                        & 62                        & $6.5\times 10^{-5}$         & 0.51                   & 0.35                                & $1.0\times 10^{-5}$                                                          & $1.4\times 10^{-8}$ (6.17)                                            & $1.8\times 10^{-10}$ ($\ll 1$)                                            & 38                    & 0.006                     \\ \hline
\end{tabular}
}
\end{center}
\caption[Examples of parameters values which satisfy the various constraints for the vector and the scalar models of the subleading DM scenario]{Examples of parameters values which satisfy the various constraints for the vector (top) and the scalar (bottom) models. $\kappa_{HP}^{DD}$ and $\kappa_{KM}^{DD}$ stand for the current experimental upper limit on $\kappa_{S}$ and $\kappa'$ respectively, see Figure \ref{fig:DD_Constraints}. That these sets of couplings satisfy the CMB constraint of Eq.~(\ref{cmbconstraint}) can be seen straightforwardly from comparing the value of $\sigma_{{\rm DM DM} \rightarrow \gamma ' \gamma '}/\sigma_{thermal}$ (or $\sigma_{{\rm DM DM} \rightarrow \phi\phi}/\sigma_{thermal}$) above with Figure \ref{fig:CST-CMB-1}. For all of these examples, the indirect detection signal is at least two orders of magnitudes below current experimental sensitivities (see Figure \ref{fig:ID_Constraints}).}
\label{tab:sub}
\end{table*}

Finally we give in Table \ref{tab:sub_sum} a schematic summary of how various constraint are evaded in this model.

\begin{table*}
\begin{center}
\resizebox{\textwidth}{!}{
\begin{tabular}{|c|c|}
\hline 
\multicolumn{2}{|c|}{Subleading DM annihilation into light mediators} \\ 
\hline 
Process diagram & Constraint \\ 
\hline 
\begin{minipage}{.37\textwidth}
\begin{center}
\includegraphics[scale=0.6]{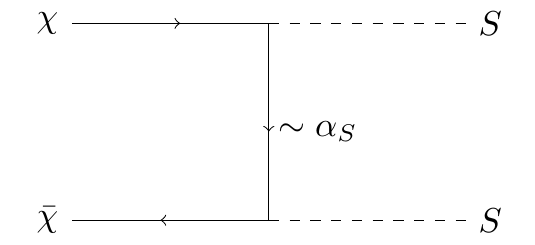}
\end{center}
\end{minipage}
&
\begin{minipage}{.5\textwidth}
\begin{center}
\begin{itemize}
\item[$\bullet$] $\alpha_{S}$ fixed by relic density
\item[$\bullet$] p-wave $\Rightarrow$ small enough for CMB
\item[$\bullet$] small enough for indirect detection
\end{itemize}
\end{center}
\end{minipage} \\ 
\hline 
\begin{minipage}{.37\textwidth}
\begin{center}
\includegraphics[scale=0.6]{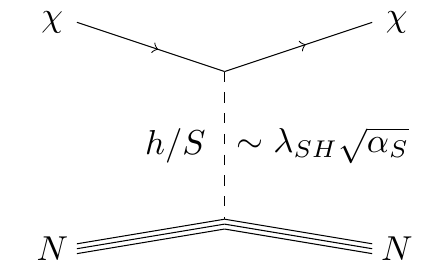}
\end{center}
\end{minipage}
&
\begin{minipage}{.5\textwidth}
\begin{center}
\begin{itemize}
\item[$\bullet$] $\lambda_{SH}$ fixed by direct detection
\end{itemize}
\end{center}
\end{minipage} \\ 
\hline 
\begin{minipage}{.37\textwidth}
\begin{center}
\includegraphics[scale=0.6]{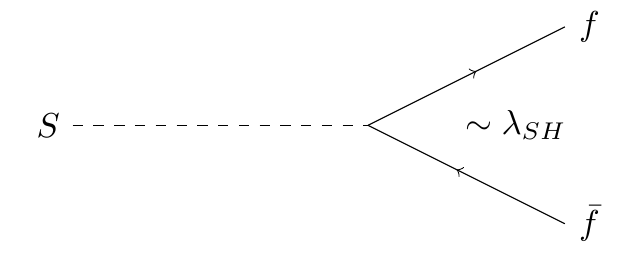}
\end{center}
\end{minipage}
&
\begin{minipage}{.5\textwidth}
\begin{center}
\begin{itemize}
\item[$\bullet$] $m_{S}> $ GeV $\Rightarrow$ fast enough for BBN
\end{itemize}
\end{center}
\end{minipage} \\ 
\hline 
\begin{minipage}{.37\textwidth}
\begin{center}
\includegraphics[scale=0.45]{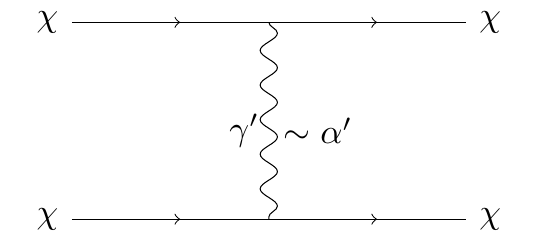}
\includegraphics[scale=0.45]{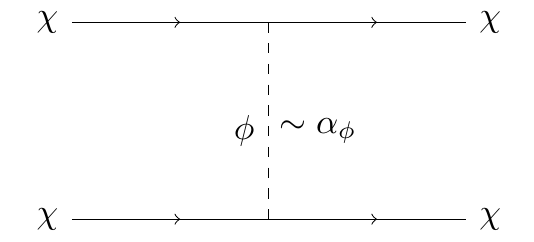}
\end{center}
\end{minipage}
&
\begin{minipage}{.5\textwidth}
\begin{center}
\begin{itemize}
\item[$\bullet$] $\alpha'/\alpha_{\phi}$ fixed by small scale structure
\end{itemize}
\end{center}
\end{minipage} \\ 
\hline 
\begin{minipage}{.37\textwidth}
\begin{center}
\includegraphics[scale=0.45]{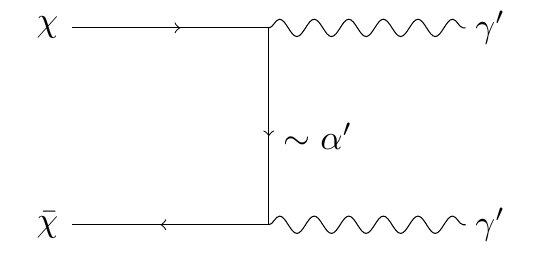}
\includegraphics[scale=0.45]{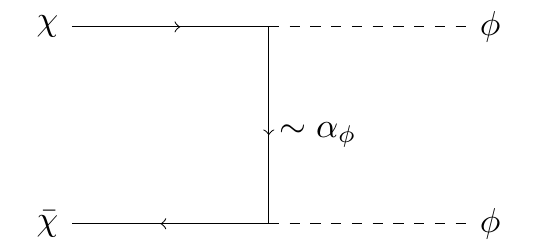}
\end{center}
\end{minipage}
&
\begin{minipage}{.5\textwidth}
\begin{center}
\begin{itemize}
\item[$\bullet$] $\alpha'/\alpha_{\phi}$ small enough for indirect detection
\item[$\bullet$] $\alpha'/\alpha_{\phi}$ small enough for CMB
\end{itemize}
\end{center}
\end{minipage} \\
\hline 
\begin{minipage}{.37\textwidth}
\begin{center}
\includegraphics[scale=0.45]{sub_XNApXN.pdf}
\includegraphics[scale=0.45]{sub_XNPXN.pdf}
\end{center}
\end{minipage}
&
\begin{minipage}{.5\textwidth}
\begin{center}
\begin{itemize}
\item[$\bullet$] $\epsilon/\lambda_{\Phi H}$ fixed by direct detection
\end{itemize}
\end{center}
\end{minipage} \\ 
\hline 
\begin{minipage}{.37\textwidth}
\begin{center}
\includegraphics[scale=0.38]{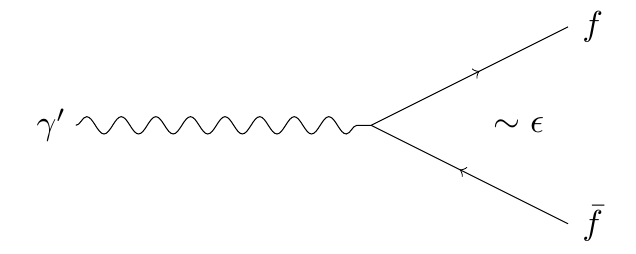}
\includegraphics[scale=0.38]{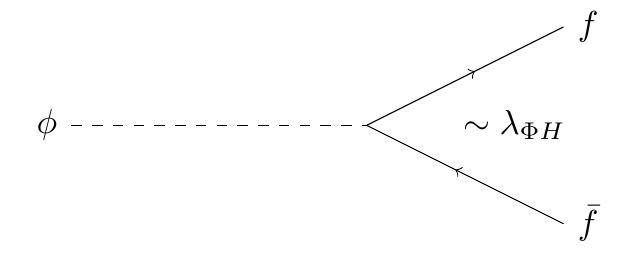}
\end{center}
\end{minipage}
&
\begin{minipage}{.5\textwidth}
\begin{center}
\begin{itemize}
\item[$\bullet$] $\epsilon/\lambda_{\Phi H}$ fast enough for BBN
\end{itemize}
\end{center}
\end{minipage} \\ 
\hline 
\end{tabular} 
}
\end{center}
\caption[Summary of how tensions are alleviated in the subleading annihilation option]{Summary of how tensions are alleviated in the subleading annihilation option.}
\label{tab:sub_sum}
\end{table*}

\section{The p-wave option}\label{sec:pwavemeddecay}
Another possibility is to consider models in which the DM annihilation into a pair of light mediators is p-wave, but still dominates the freeze-out. Thus for a Dirac DM candidate, the light mediator must be a scalar boson, $\phi$. Here, we do not decouple the channel responsible for the relic density and the one responsible for the strong self-interactions. As discussed above in Section \ref{sec:global_picture}, this could be problematic because these two constraints together generally impose a relatively strong connection between DM and the light mediator which would imply a relatively strong DM-to-SM connection. This would then be directly in opposition with cosmological constraints and particle physics experiments, see Figure \ref{fig:CST-SIDM_pic}. Thus here instead, the extra degree of freedom is assumed to be lighter (i.e. extra-light) than the already existing light mediator $\phi$ and will be used to reduce the light mediator number density. This reduction can be done through a decay of the light mediator into two extra light mediator particles ($\phi\rightarrow XX$) and/or through an annihilation process of two light mediators into two extra-light mediator particles ($\phi\phi\rightarrow XX$). The simplest model we could think of along these hypothesis is to assume that the extra-light mediator is a scalar boson $S$ (which we take here too to be real for definiteness, with an even number of them in all interactions), as following,

\myeq{
&{\cal L}\owns - (y_\phi \phi\overline{\chi^c} \chi +h.c.) -\lambda_{\phi H} \phi^\dagger \phi H^\dagger H-\lambda_{\phi S} \phi^\dagger \phi S^2-\lambda_{H S} H^\dagger H S^2,\label{eq:pwave}
}

\noindent with $\phi$ the light mediator and $S$ the extra-light scalar into which the light mediator decays and/or annihilates. For simplicity, we assume that the DM does not directly couple to the extra-light scalar (or that this coupling is negligible) which can be justified on the basis of a symmetry. Depending on the values of the coupling between the light and the extra-light scalars, the reduction of the number density of the light one will be dominated either by its decay or its annihilation. Note that, in order to decay, the light mediator $\phi$ must have a VEV.
\\

\begin{figure}[h!]
\centering
\includegraphics[scale=0.6]{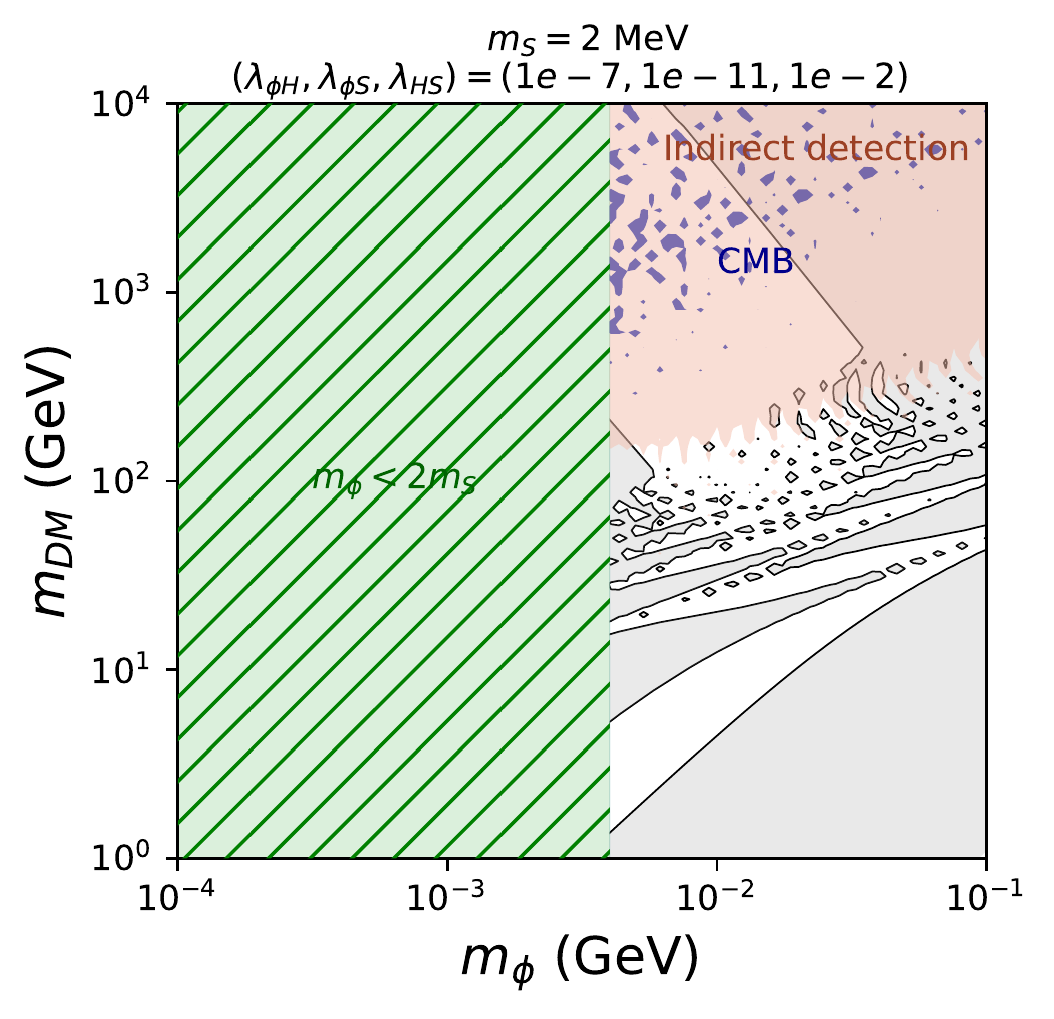}
\includegraphics[scale=0.6]{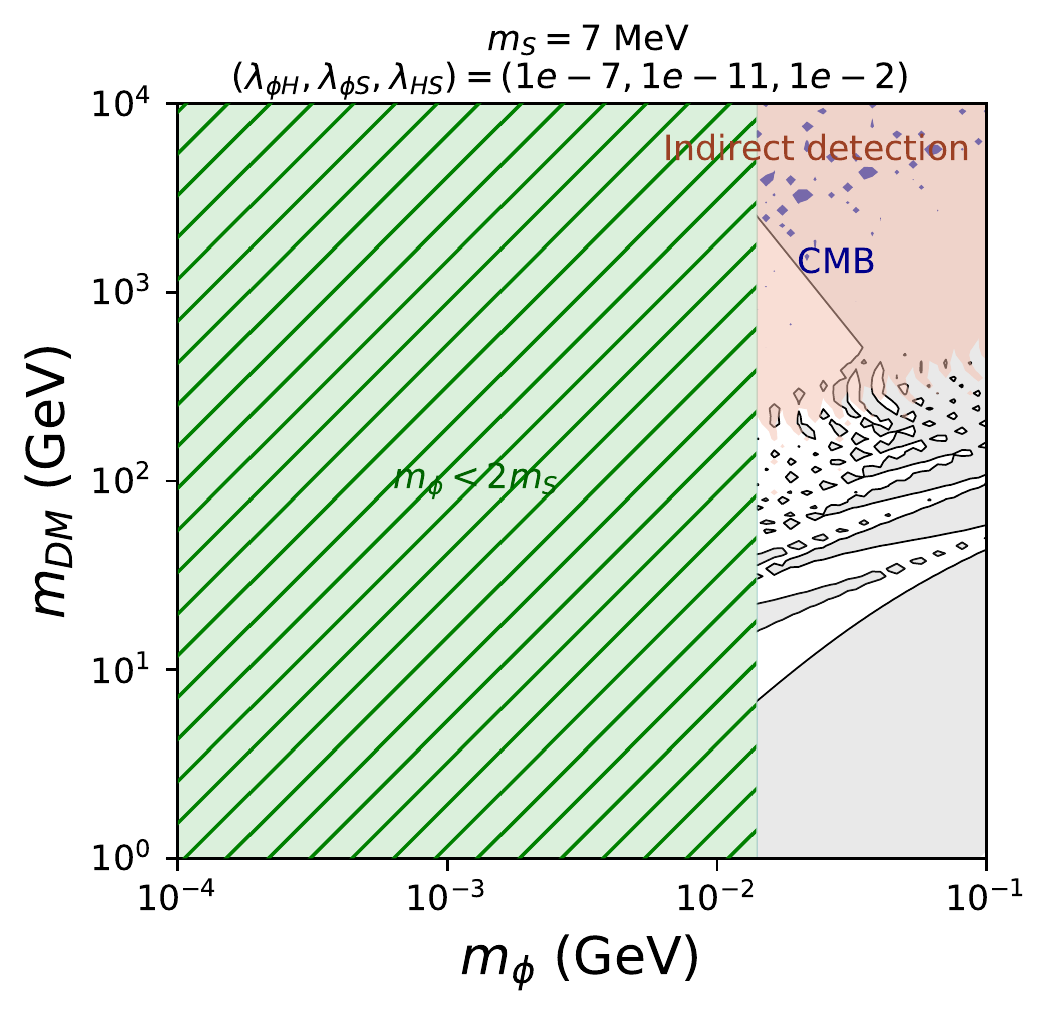}
\includegraphics[scale=0.6]{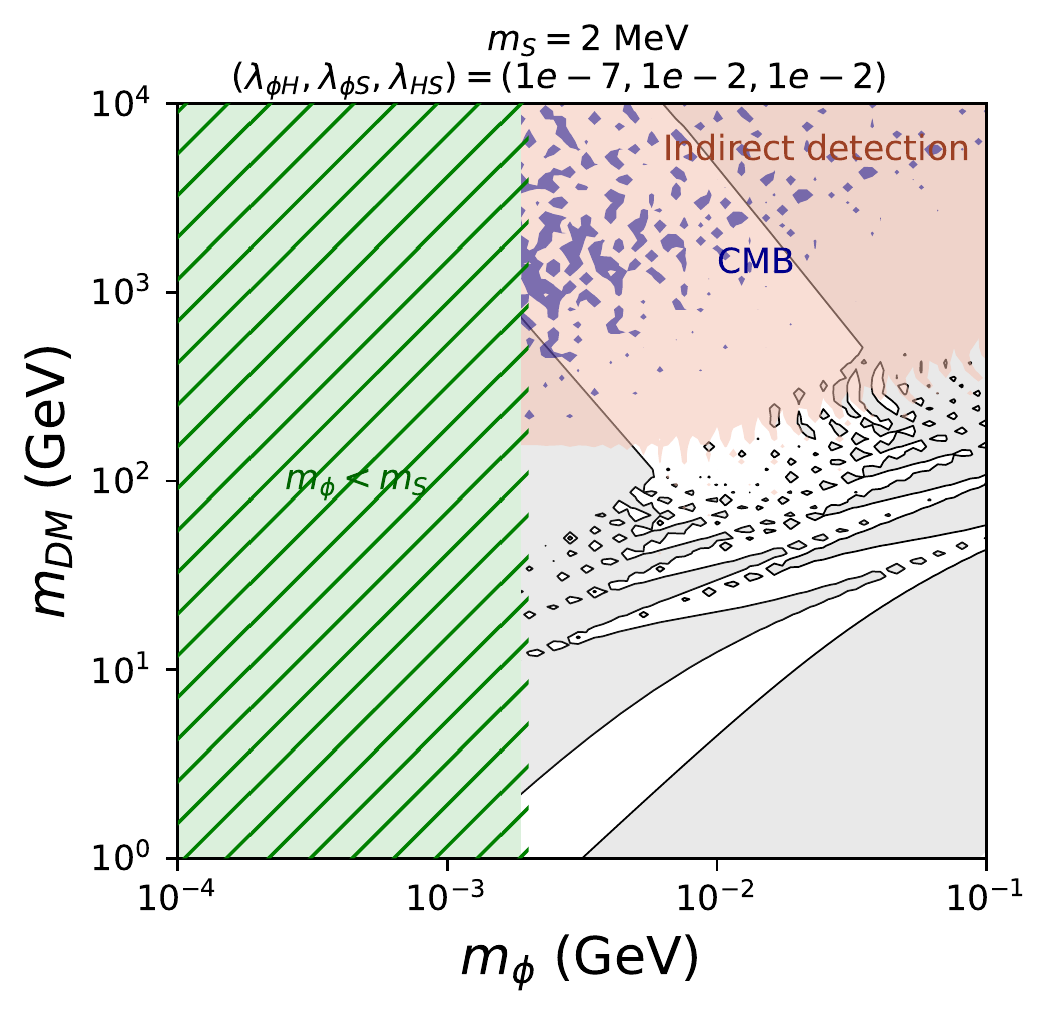}
\includegraphics[scale=0.6]{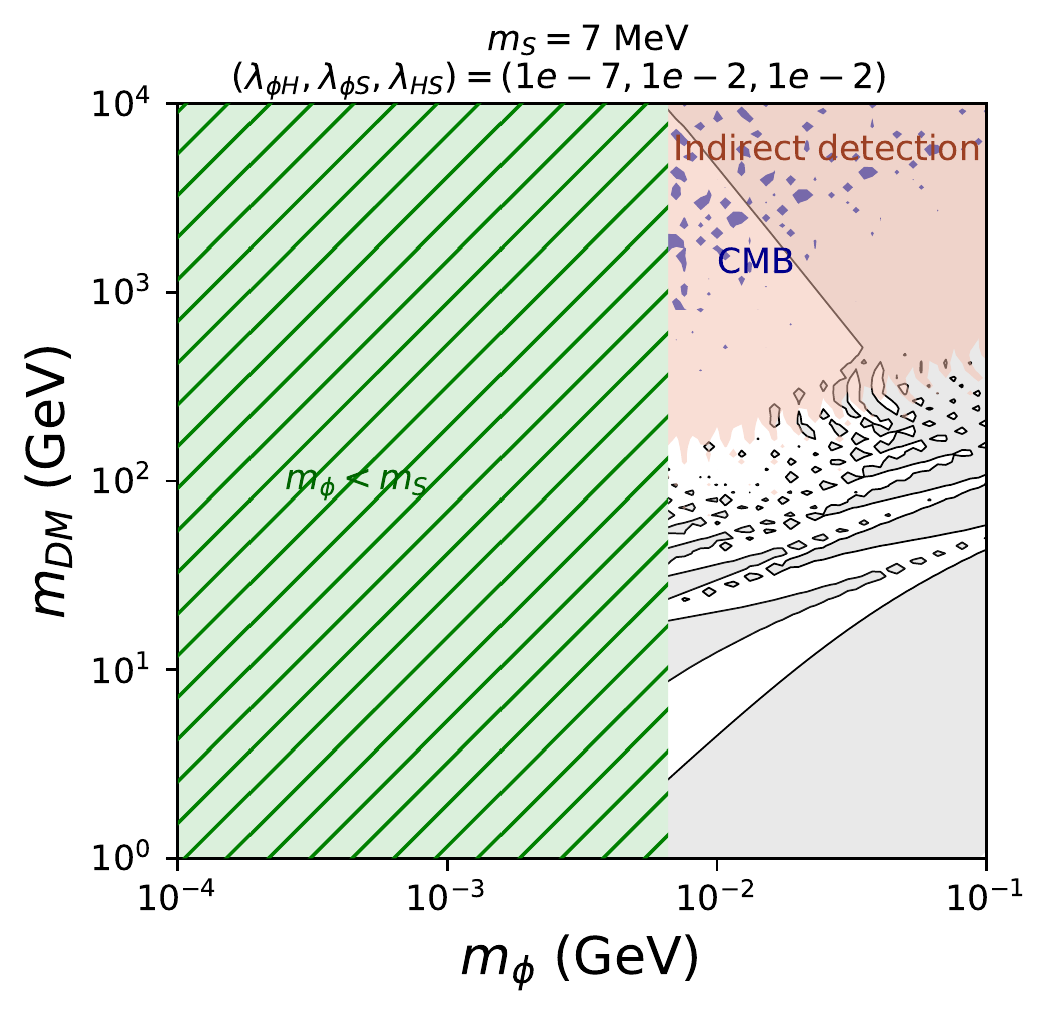}
\caption[Summary plot for the p-wave scenario]{Constraints from CMB, BBN, self-interaction, indirect and direct detection all together for the DM p-wave annihilation into light mediators with $m_{S}= 2$ MeV  (left) and $m_{S}= 7$ MeV (right), for the decaying (top panels) and the annihilating (bottom panels) regimes (see below).}\label{fig:CD}
\end{figure}

Figure \ref{fig:CD} gives the allowed parameter space for this model for two choices of the extra-light scalar mass and for two opposite cases: a feeble (top panels) and a strong (bottom panels) coupling between the light and the extra-light scalars. One can see that contrary to the previous model, here the different panels are very similar. This was to be expected as for a given choice of the couple ($m_{\rm DM}$, $m_{\phi}$), the self-interaction cross section is only a function of the DM-to-med coupling. Since this coupling is always fixed by the relic density constrained for all panels of Figure \ref{fig:CD}, the self-interaction cross section is the same whatever the choice of the other parameters is (i.e. $m_{S}$, $\lambda_{\phi H}$, $\lambda_{\phi S}$ and $\lambda_{HS}$).

\subsection{CMB}
Since the DM annihilation into a pair of scalar bosons is of the p-wave type and since the DM annihilates only into a pair of scalar bosons, the model cannot be constrained too strongly by CMB constraints on the DM annihilation cross section at recombination. However, the DM abundance is set by the freezing of the DM annihilation process into a pair of light mediator (i.e. a pair of $\phi$'s) such that the DM-to-med coupling can be big and that the annihilation at recombination can still be too high in some part of the parameter space, see darkblue contour in Figure \ref{fig:CD}. This also explains why the CMB constraint is the same for all panels of the very same figure: it only depends on the DM-to-med interaction which is the same as explain above. This is at the opposite than in the previous model where the DM abundance were fixed by the freezing of another process. Thus, on the one hand, the DM annihilation into $\phi$'s was negligible as seen in Table \ref{tab:CD} and could not be constrained by the upper bound on the DM annihilation cross section at recombination (see Eq. \ref{cmbconstraint}). On the other hand, the light and the extra-light scalars ($\phi$ and $S$ respectively) can both decay fast enough to have any sizable effect on the CMB. The $N_{eff}$ constraint is also satisfied from fast enough decay of the two new scalars of the theory.

\subsection{BBN}
The BBN and $N_{eff}$ constraints are solved in a similar way. Indeed, as said above, if the mediator $\phi$ has a relic density resulting from the DM annihilation into a pair of these mediators, this relic density must be decreased by the BBN time. A decay exclusively into SM particles through the Higgs portal is forbidden by direct detection (for $m_{\rm DM}\gtrsim 1$~GeV). A decay into extra hidden sector particles solves this problem. On the one hand the $\phi\rightarrow S S$ decay can be fast enough if $\lambda_{\phi S}$ is not too small. On the other hand, the $S\rightarrow SM SM$ decay can also be fast enough, through the mixing of the extra-light and the SM scalars ($S-h$) if the coupling between these particles (i.e. $\lambda_{H S}$) is large enough. The later feature can be realised without inducing a too large Higgs boson invisible decay width. For instance, if the extra-light scalar $S$ has a mass above the $e^+e^-$ threshold one has the following lifetime

\myeq{
&\tau_{\phi\rightarrow SS}\simeq  1\text{s}\cdot \left(\frac{m_{\phi}}{20\text{~MeV}}\right)\left(\frac{2.3\times 10^{-12}}{\lambda _{\phi S}}\right)^{2}\left(\frac{500\text{~MeV}}{v_{\phi}}\right)^{2},\label{eq:phidecay}\\
&\tau_{S\rightarrow e^+e^-}\simeq  1\text{s}\cdot \left(\frac{2\text{~MeV}}{m_{S}}\right)\left(\frac{7.8\times 10^{-3}}{\lambda _{H S}}\right)^{2}\left(\frac{500\text{~MeV}}{v_{S}}\right)^{2}.\label{eq:Sdecay}
}

\noindent From these expressions, one can see that it is easy to have the two scalars to decay before BBN such that BBN do not further constrain the model. Indeed, the mixing parameter between the light and the extra-light scalars $\lambda_{\phi S}$ does not have to be large (see Eq. \ref{eq:phidecay}). However, if the light mediator VEV is very small (that is to say if $v_{\phi}\ll 1$ MeV), the $\phi$ boson becomes stable and one would need a large mixing parameter between the light and the extra-light scalars in order to deplete the abundance of $\phi$'s through a two-to-two annihilation process, $\phi\phi\rightarrow SS$. We have thus here two types of solution emerging from this unique model. The solution were the light mediator abundance is depleted due to its decay into a pair of extra-light scalars and the solution were its abundance is depleted thanks to a fast annihilation into a pair of $S$'s.
\\

In both solutions, BBN constraints are easily avoided for the considered parameter space. This explain the total absence of constraint labeled as BBN in all panels of Figure \ref{fig:CD}.

\subsection{Direct detection}
Since the small scale structure constraints did not require any coupling between the DM and the extra-light scalar $S$, direct detection through DM-to-S coupling is small if this coupling is small or nonexistent. But, direct detection constraints could still be relevant for scattering of DM on nucleon through the mixing of the light mediator $\phi$ and the SM Higgs scalar. However, this interaction can also easily be suppressed, even if the mixing parameters $\lambda_{\phi S}$ and $\lambda_{HS}$ are sizable to account for BBN. Indeed, it would only requires that the transitions from light mediators $\phi$'s to SM scalars $H$'s are suppressed enough, i.e. that the $\lambda_{\phi H}$ coupling is small enough, as well as the product of $\lambda_{\phi S}$ and $\lambda_{H S}$.
\\

These requirements which allow us to avoid direct detection constraints are very easy to achieve in the "annihilating" solution because the mixing is suppressed by the very small VEV of $\phi$. They can also be fulfilled in the "decaying" solution, but a small tension remains between direct detection and the $\phi$ lifetime which tends to be too long for being in agreement with BBN constraints.

\subsection{Indirect detection}
As for the previous model, even if the DM annihilation cross section is p-wave, the Sommerfeld effect can considerably increase the annihilation rate in the same way that explained above in Subsection \ref{subsec:p_wave_ID}. It is even more true here, as the two light scalars can participate to the Sommerfeld effect and enhance the DM annihilation cross section and lead in this model to a large indirect detection rate which can be tested (and actually rules out a large part of the parameter space). Indirect detection constraints are shown in red in Figure \ref{fig:CD} and vary with the light mediator as the light mediator has to decay and/or annihilate into $S$'s for this model to be constrained. 
%So, even if for the whole parameter space the light mediator mass is irrelevant when compared to the DM mass, it becomes relevant when compared to the extra-light mediator mass as both masses can be very close.

\subsection{Results}
To sum up, all constraints are satisfied if both mixing parameter $\lambda_{\phi S}$ and $\lambda _{H S}$ are large enough (but not too large so that their product is small enough) and $\lambda_{\phi H}$ is small enough.
\\

As previously, a couple of numerical examples which satisfy all constraints are given in the two first lines of Table \ref{tab:CD}. For the second example the annihilation cross section of DM into $\phi$'s is of order $\left\langle\sigma v\right\rangle_{DM DM \rightarrow \phi\phi}\simeq 10^{-25} $ cm$^{3}$s$^{-1}$ in dwarf galaxies\footnote{This cross section has to be multiplied by four in order to get the DM annihilation cross section into SM charged fermions since each of the $\phi$'s will decay into two charged fermions.}. As can be seen from Fig. \ref{fig:ID_Constraints}, this can be constrained by indirect detection experiments. This is due to the fact, as discussed above (see Chapter \ref{ch:som}), that the Sommerfeld effect brings a $\sim 1/v^3$ enhancement for p-wave processes which compensates for the $v^2$ suppression which appears in p-wave cross section. For further details on this setup, note that with for example the first set of couplings in Table \ref{tab:CD}, the $S$ scalar thermalises with the SM thermal bath through $\lambda_{HS}$, whereas DM and $\phi$ thermalise through $y_\phi$. Both visible and hidden sector never thermalise through the $\lambda_{\phi S}$ interaction but does so through $\lambda_{\phi H}$ until $T\simeq 35$~MeV. Later on, the $T'/T$ ratio does not remain equal to unity but remains close to it.  For instance, at $t\simeq 1$~sec,  one has  $T'/T\simeq 0.95$.

\begin{table*}[t]
\begin{center}
\resizebox{\textwidth}{!}{%
\begin{tabular}{|c|c|c|c|c|c|c|c|c|c|c|c|c|c|c|}
\hline
$m_{\rm DM}$ & $m_{\phi}$ & $m_S$ & $v_\phi$ & $v_S$ & \multirow{2}{*}{$\lambda_{\phi H}$} & \multirow{2}{*}{$\lambda_{\phi S}$} & \multirow{2}{*}{$\lambda_{H S}$} & $\sigma_T/m_{\rm DM}$        & \multirow{2}{*}{$\frac{\kappa_{HP}}{\kappa_{HP}^{DD}}$} & $\tau_{\phi}$ & $\tau_{S}$ & $\Gamma_h^{inv}$ & $\langle \sigma_{\phi\phi\rightarrow S S} v \rangle$ & \multirow{2}{*}{$\Omega_{med}^{0} h^2$} \\
(GeV)    & (MeV)      & (MeV) & (MeV)    & (MeV) &                                     &                                     &                                  & $(\text{cm}^2/\text{g})$ &                                                 & (sec)         & (sec)      & (MeV)            & ($\text{GeV}^{-2}$)                                  &                                     \\ \hline
126      & 20         & 2     & 500      & 500   & $5.4\times 10^{-8}$                 & $6.2\times 10^{-12}$                & 0.016              & 0.28                     & 0.05                                            & 0.14          & 0.76       & 1.23             & $\ll \sigma_{thermal}$ & 0 \\ \hline
382      & 71         & 7     & 436      & 83    & $6.3\times 10^{-7}$                 & $3.2\times 10^{-11}$                & 0.014 & 0.21                     & 0.49                                            & 0.03          & 0.68       & 0.94             & $\ll \sigma_{thermal}$                                                    & 0                                   \\ \hline\hline
83        & 50          & 2     & $\ll 1$        & 500     & $6.3\times 10^{-8}$                                   & 0.010                                   & 0.015                                & 0.11                        & $\ll 1$                                               & $\gg 1$             & 0.87 & 1.08                & $2.5\times 10^{-5}$                                                    & $2.5\times 10^{-5}$                                   \\ \hline
173        & 300          & 10     & $\ll 1$        & 50     & $1.0\times 10^{-6}$                                   & 0.015 & 0.010 & 0.21 & $\ll 1$ & $\gg 1$             & 1.44          & 0.48                & $1.5\times 10^{-6}$                                                    & $4.0\times 10^{-4}$                                   \\ \hline
\end{tabular}
}
\end{center}
\caption[Examples of parameter values which satisfy the various constraints for a model with two light scalars for the p-wave option]{Examples of parameter values which satisfy the various constraints for a model with two light scalars. $\kappa_{HP}^{DD}$ stands for the current experimental upper limit on $\kappa_{HP}$ for the masses considered, see Fig.~\ref{fig:DD_Constraints}. $\Omega_{med}^{0} h^2$ refers to the relic density value of the mediator today.}
\label{tab:CD}
\end{table*}

Finally and as for previous solution, we give in Table \ref{tab:pwave_sum} a summary of how various constraints are evaded in this model.

\begin{table*}
\begin{center}
\resizebox{\textwidth}{!}{
\begin{tabular}{|c|c|}
\hline 
\multicolumn{2}{|c|}{P-wave option} \\ 
\hline 
Process diagram & Constraint \\ 
\hline 
\begin{minipage}{.37\textwidth}
\begin{center}
\includegraphics[scale=0.6]{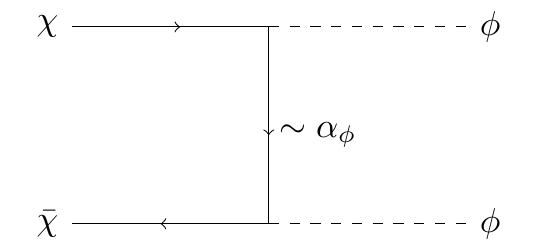}
\end{center}
\end{minipage}
&
\begin{minipage}{.5\textwidth}
\begin{center}
\begin{itemize}
\item[$\bullet$] $\alpha_{\phi}$ fixed by relic density
\item[$\bullet$] p-wave $\Rightarrow$ small enough for CMB
\item[$\bullet$] small enough for indirect detection
\end{itemize}
\end{center}
\end{minipage} \\ 
\hline 
\begin{minipage}{.37\textwidth}
\begin{center}
\includegraphics[scale=0.6]{sub_XXPXX.pdf}
\end{center}
\end{minipage}
&
\begin{minipage}{.5\textwidth}
\begin{center}
\begin{itemize}
\item[$\bullet$] $\alpha_{\phi}$ large enough for small scale structure
\end{itemize}
\end{center}
\end{minipage} \\ 
\hline 
\begin{minipage}{.37\textwidth}
\begin{center}
\includegraphics[scale=0.6]{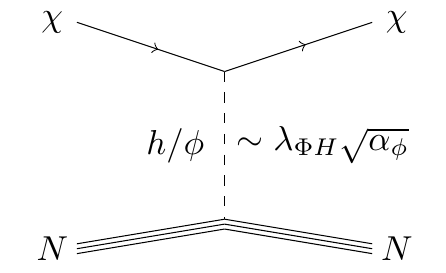}
\end{center}
\end{minipage}
&
\begin{minipage}{.5\textwidth}
\begin{center}
\begin{itemize}
\item[$\bullet$] $\lambda_{\Phi H}$ fixed by direct detection
\end{itemize}
\end{center}
\end{minipage} \\ 
\hline 
\begin{minipage}{.37\textwidth}
\begin{center}
\includegraphics[scale=0.4]{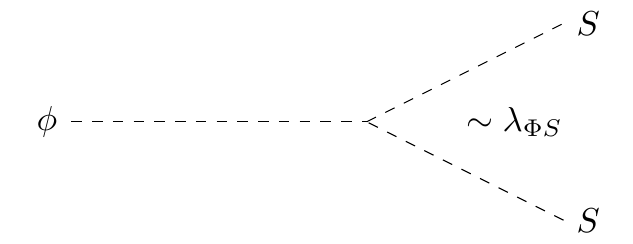}
\includegraphics[scale=0.4]{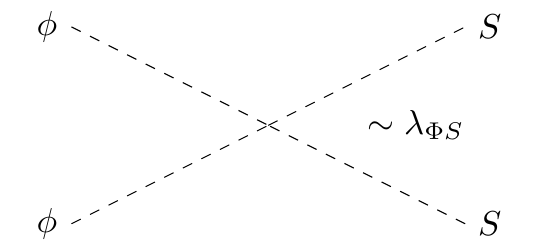}
\end{center}
\end{minipage}
&
\begin{minipage}{.5\textwidth}
\begin{center}
\begin{itemize}
\item[$\bullet$] $\lambda_{\Phi S}$ large enough to deplete $\phi$
\end{itemize}
\end{center}
\end{minipage} \\ 
%\hline 
%\begin{minipage}{.37\textwidth}
%\begin{center}
%\includegraphics[scale=0.6]{sub_XXSS.pdf}
%\end{center}
%\end{minipage}
%&
%\begin{minipage}{.5\textwidth}
%\begin{center}
%\begin{itemize}
%\item[$\bullet$] small enough for indirect detection
%\item[$\bullet$] p-wave $\Rightarrow$ small enough for CMB
%\end{itemize}
%\end{center}
%\end{minipage} \\
\hline 
\begin{minipage}{.37\textwidth}
\begin{center}
\includegraphics[scale=0.6]{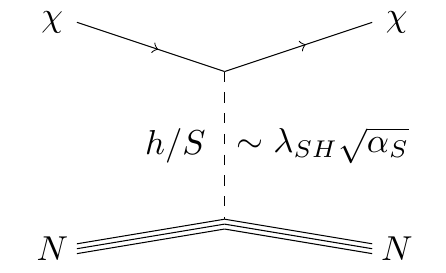}
\end{center}
\end{minipage}
&
\begin{minipage}{.5\textwidth}
\begin{center}
\begin{itemize}
\item[$\bullet$] $\lambda_{S H}$ fixed by direct detection
\end{itemize}
\end{center}
\end{minipage} \\ 
\hline 
\begin{minipage}{.37\textwidth}
\begin{center}
\includegraphics[scale=0.6]{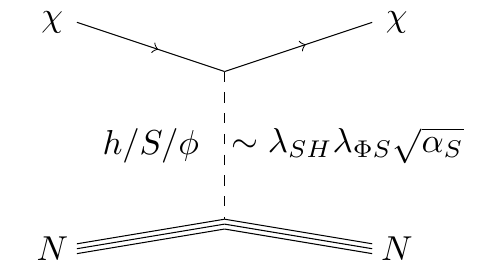}
\end{center}
\end{minipage}
&
\begin{minipage}{.5\textwidth}
\begin{center}
\begin{itemize}
\item[$\bullet$] $\lambda_{\Phi S}$ fixed by direct detection
\end{itemize}
\end{center}
\end{minipage} \\ 
\hline 
\begin{minipage}{.37\textwidth}
\begin{center}
\includegraphics[scale=0.6]{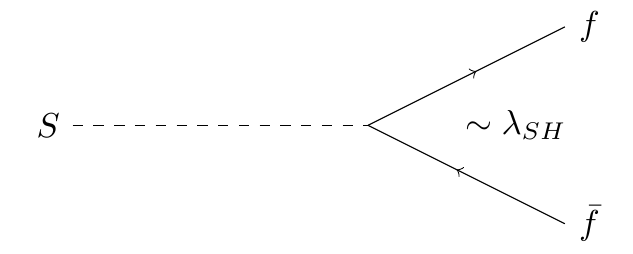}
\end{center}
\end{minipage}
&
\begin{minipage}{.5\textwidth}
\begin{center}
\begin{itemize}
\item[$\bullet$] $\lambda_{SH}$ fast enough for BBN
\end{itemize}
\end{center}
\end{minipage} \\ 
\hline 
\end{tabular} 
}
\end{center}
\caption[Summary of how tensions are alleviated in the p-wave option]{Summary of how tensions are alleviated in the p-wave option.}
\label{tab:pwave_sum}
\end{table*}

\section{Stable mediator option\label{sec:stablewayout}}
As discussed in the above section, if the mediator is absolutely stable, the main constraints are the non overclosure and modification of the Hubble constant once as all other constraints above do not apply. A simple way to avoid these constraints is to consider a scenario where the hidden-to-visible temperature ratio is sizably smaller than unity, as we have done in Chapter \ref{ch:secl}, see Eq. \ref{eq:TpTlowerbound} in particular. Instead, if one consider a HS with a temperature similar to the one of the visible sector (i.e. $T'/T\sim 1$), a more complicated possibility arises if the stable light mediator number density is reduced after its decoupling from an annihilation into extra lighter particles. This possibility would imply new constraints related to the existence of this extra light particle. One finds two minimal models realising this scenario. The first one has already been proposed in \cite{Duerr:2018mbd} (see also \cite{Ma:2017ucp}). It involves a s-wave annihilation of DM into a $\gamma '$ light mediator, followed by an annihilation of this $\gamma '$ into a lighter scalar $S$, followed by decay of this scalar $S$ into SM particles. Omitting less relevant scalar interactions, the Lagrangian of this model is:

\myeq{
&{\cal L}\owns -g \gamma '_\mu i\bar \psi \gamma^\mu \psi-g \gamma '_\mu i Q'_S (S^* \partial_\mu S - S \partial_\mu S^*) -\frac{\epsilon}{2} F^Y_{\mu\nu}F'^{\mu\nu} -\lambda_{H S} S^\dagger S H^\dagger H,\label{eq:lag_stable}
}

\noindent with $\psi$ the Dirac DM candidate and $S$ the additional extra-light scalar. The mass of the light mediator $\gamma '$ comes from spontaneous breaking of the $U(1)'$ gauge symmetry once the extra-light scalar $S$ acquires a VEV. The stability of the $\gamma '$ light mediator requires here an extra (charged conjugation) symmetry, in order that a kinetic mixing between the $\gamma '$ and the hypercharge gauge boson is forbidden. A nice feature of this model is that the extra-light scalar $S$ is the one one has anyway to introduce in the model if one assumes that the $U(1)'$ symmetry is spontaneously broken through the Brout-Englert-Higgs mechanism. The other minimal model where this turns out to be possible is nothing but the model of the previous section of Eq. \ref{eq:pwave}, in other regions of the parameter space as already discussed, assuming in particular now that the scalar mediator $\phi$ has no VEV, so that it doesn't decay.
\\

The phenomenology of the model described by the Lagrangian given in Eq. \ref{eq:lag_stable} has been analysed in detail in \cite{Duerr:2018mbd}. Once the non-overclosure constraint is satisfied for the light mediator $\gamma '$, from annihilation of this light mediator into a pair of extra-light scalars $S$, $\gamma ' \gamma '\rightarrow S S$, one is left essentially with constraints on the extra-light particles, the $S$'s. On the one hand the mass of this extra particle must be larger than twice the electron mass $m_{S}>2m_{e}$, otherwise its decay is loop suppressed and therefore too slow for CMB or BBN constraints to be satisfied. On the other hand, its mass must be below the 4.4 MeV threshold to avoid the BBN photodisintegration constraint. The $N_{eff}$, the Hubble constant and the entropy injection constraints are just enough satisfied from the fact that when the two mediators ($S$ and $\gamma '$) decouple, the hidden-to-visible temperature ratio $T'/T$ (which is equal to unity at DM decoupling) is already not anymore equal to unity, but slightly smaller (from decoupling of SM particles in between). The CMB constraint on the DM annihilation cross section given in Eq \ref{cmbconstraint} does not apply for DM annihilation into a pair of light mediator $DM DM\rightarrow \gamma' \gamma'$ because the $\gamma'$ is stable but applies for the $DM DM\rightarrow \gamma' S$ scattering, as well as for $\gamma' \gamma'\rightarrow S S$ scattering since it is the $S$'s particles which are connected to the SM. For the last scattering it is easily satisfied since this scattering is not enhanced by the Sommerfeld effect. For the former scattering, this constraint still leaves an allowed parameter space. Direct detection is satisfied from the fact that DM communicates with SM particles only through a DM-$\gamma '$-$S$-SM chain.
\\

\begin{figure}[h!]
\centering
\includegraphics[scale=0.65]{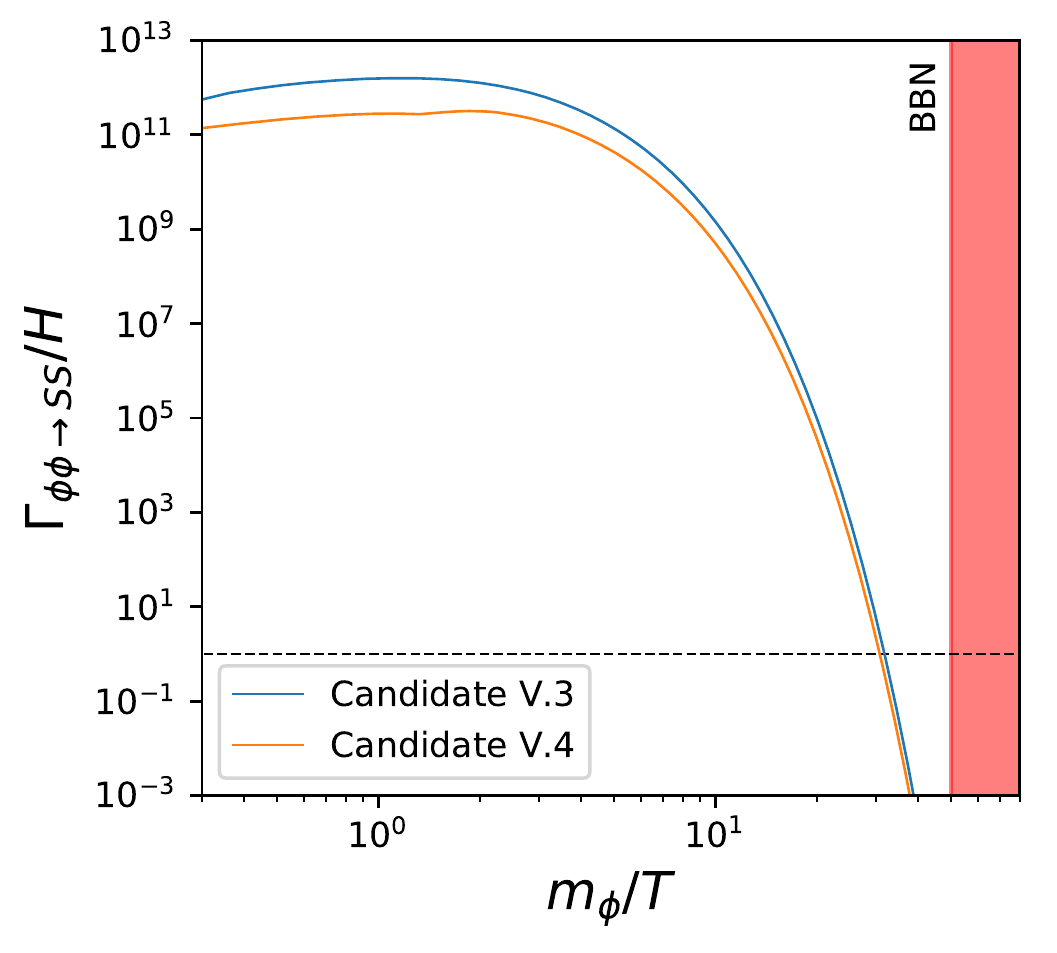}
\includegraphics[scale=0.65]{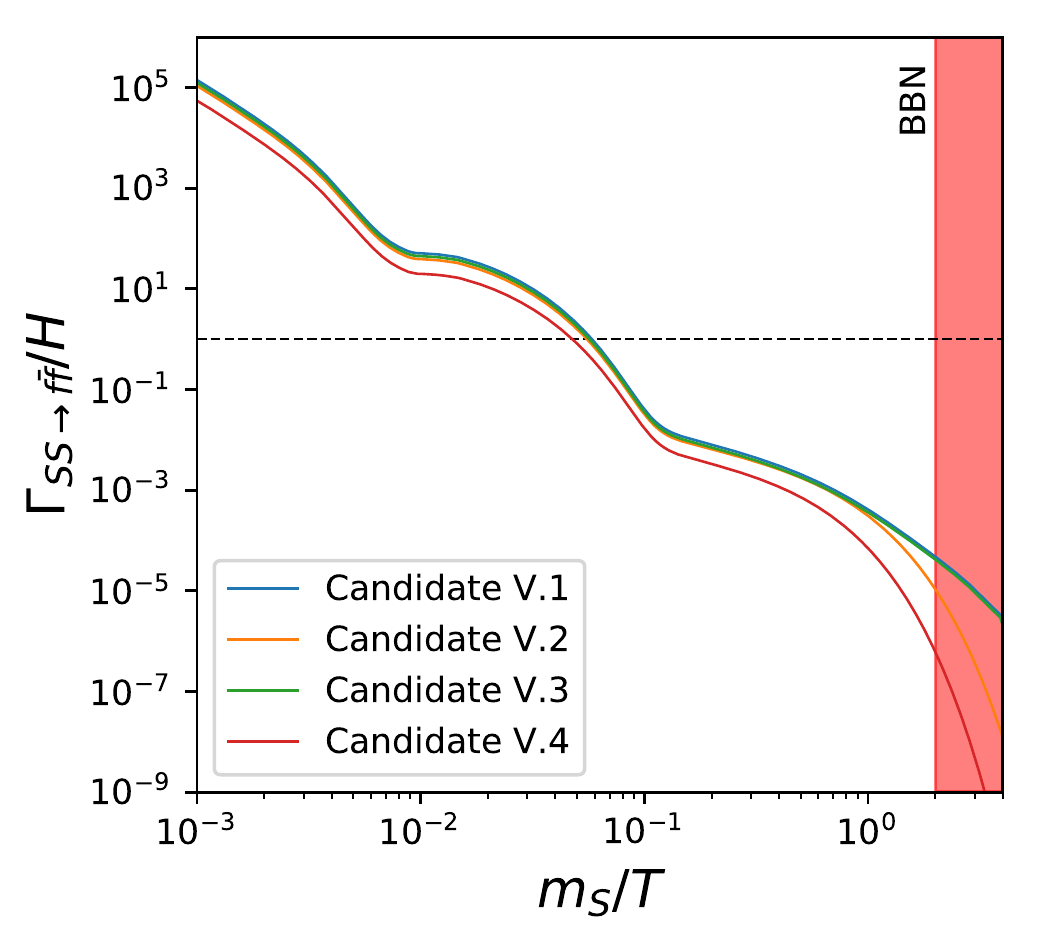}
\caption[Evolution of relevant annihilation rates in the stable mediator scenario]{Left: Evolution of the annihilation rate of a pair of $\phi$'s into a pair of $S$'s, normalised to the Hubble rate for the last two candidates of Table \ref{tab:CD}. The black dashed line represents the thermalisation line, ie: when $\Gamma _{\phi\phi\rightarrow SS}=H$. Right: Evolution of the annihilation rate of a pair of $S$'s into a pair of SM fermions normalised to the Hubble rate for all candidates of Table \ref{tab:CD}. The black dashed line represents the thermalisation line, ie: when $\Gamma _{SS\rightarrow f\bar{f}}=H$. One can also see when BBN starts in red in both plots.}\label{fig:C_Rate}
\end{figure}

As for the model of the previous section with a stable scalar mediator $\phi$, Eq. \ref{eq:pwave}, it can be successful in a perhaps less squeezed way. The first constraint it has to fulfill is that the annihilation process into a pair of extra-light scalars $\phi\phi\rightarrow S S$, which must be in thermal equilibrium at temperature of order $T\sim m_\phi$, such that the $\phi$ number density is later on Boltzmann suppressed. This will be the case for model described by Eq. \ref{eq:pwave} in another parameter regime than for the decay option of Section \ref{sec:pwavemeddecay}. As already mentioned above, one has to assume much larger values of the mixing between the two scalars $\lambda_{\phi S}$ than above in the decay option (i.e. for instance $\lambda_{\phi S}> \lambda _{\phi S}^{th}\simeq 6\times 10^{-9}$ for $m_{\rm med}=30$ MeV). Note that $S$ can annihilate into a pair of SM particles through the Higgs portal, but these $S S\rightarrow f \bar{f}$ annihilation processes decouple before the $\phi \phi\rightarrow S S$ process. The latter process decouples when $m_\phi/T\sim 30$.  This is shown in Figure \ref{fig:C_Rate} for two examples sets of parameters which satisfy all constraints and are given in Table \ref{tab:CD}. Thus, for these two examples (where $m_\phi/m_S\lesssim 30$), the $S$ scalar is a hot relic and its number density is not Boltzmann suppressed. Now, we are going to see how the various constraints are fulfilled which would make this stable mediator scenario easily viable.

\subsection{Non-overclosure}
The light mediator annihilation process into a pair of extra-light scalars (i.e. the $\phi\phi \rightarrow S S$ scattering through $\lambda_{\phi S}$ coupling of Eq. \ref{eq:lag_stable}) is fast enough in order to reduce the population of stable light mediator $\phi$ and to avoid the Universe to overclose. There is then no reduction of the parameter space coming from this constraint.

\subsection{CMB}
The DM annihilation into a pair of light mediators does not cause any large CMB distortion because the mediator is a scalar such that the annihilation is of p-wave type (see Eq. \ref{cmbconstraint}). Moreover, the light mediator annihilation process, from $\lambda_{\phi S}$ coupling, into a pair of extra-light scalars also reduces enough the light mediator population to fulfill the related $N_{eff}$ constraint. Finally, the decay of the extra-light scalar $S$ is fast enough to satisfy the $N_{eff}$ constraint associated to its decay. 

\subsection{BBN}
The decay rate of the extra-light particle $S$ decay, from $\lambda_{HS}$ coupling, is also fast enough to satisfy BBN Hubble constant and entropy injection constraints altogether with photodisintegration constraints. This requires to take a large enough mixing parameter between the extra-light scalar and the SM scalar boson $\lambda_{H S}$, see Eq. \ref{eq:Sdecay}. All this can be realised without inducing a too large Higgs boson invisible decay width. 

\subsection{Direct detection}
As in Section \ref{sec:pwavemeddecay}, the self-interaction constraints require a sizable DM-to-med coupling but does not require any connection between the DM and the extra-light scalar $S$. Thus, direct detection through DM-to-S coupling is small if the corresponding coupling, $\bar{\psi}\psi S$, is small or is irrelevant (i.e. if there is no coupling at all between these two particles). Moreover, the DM cannot scatter on nucleon at tree level through the DM-to-med coupling because the light mediator has no VEV.

\subsection{Indirect detection}
If the $\phi$ has no VEV and is stable, the annihilation of DM into $\phi$ does not produce any SM particles and DM annihilation produce SM particles only at the loop level. There is thus basically no relevant constraint coming from indirect detection experiments.

\section{The neutrino option \label{sec:neutrinowayout}}
If one does not introduce any other extra particle on top of the DM and the light mediator particles, the only possibility\footnote{Except for the $T'/T\neq 1$ solution discussed in Chapter \ref{ch:secl}.} for the light mediator to not spoil BBN observations while decaying is that it decays into neutrinos. For the decay to proceed into neutrinos exclusively without producing any electron/positron pairs, there are two simple possibilities. One needs either an \textit{eletronphobic} light mediator such that it does not couple to the electron lepton doublet and that the decay proceeds only into muon or tau neutrinos, or that the mediator is lighter than twice the electron mass. With a light scalar mediator it is not easy to have a dominant and fast enough decay into neutrinos. As for the vector light mediator option (i.e. the $\gamma '$ option), the possible portals which can make the vector mediator to decay into neutrinos are the kinetic mixing portal or the so-called "mass mixing portal". As already mentioned above, through kinetic mixing, and below the $e^+e^-$ threshold, the $\gamma '$ decays dominantly to three photons at a rate far too slow to proceed before BBN and even slower to neutrinos. Through the more involved mass mixing scenario, the decay can dominantly proceed into neutrinos at a rate fast enough to decay before the BBN occurs. The relevant interactions of the "mass mixing portal" are therefore

\myeq{
&{\cal L} \owns -g \gamma '_\mu J^\mu_{\rm DM} -\delta m^2 \gamma '_\mu Z_\mu,
}

\noindent where $J_\text{\rm DM}^\mu = \bar \psi \gamma^\mu \psi $, $J_\text{\rm DM}^\mu = \bar \chi \gamma^\mu \gamma^5 \chi $ and $J_\text{\rm DM}^\mu = i (S^* \partial_\mu S - S \partial_\mu S^*)$ for a Dirac, Majorana and scalar DM particle respectively. This model for the Dirac DM case has already been considered in Appendix B of \cite{Bringmann:2016din}, as a way out to the CMB constraints above summarised by Eq. \ref{cmbconstraint}, see the various constraints holding on it in Figure 4 of this reference.
\\

The kinetic mixing option does not allow the light mediator to decay into muon and/or tau neutrinos without producing electron neutrinos as well. However, the production of electron neutrinos can be avoided in models with an extra $U(1)'$ gauge symmetry, along which the corresponding gauge boson $Z'$ couples only to $\mu$ and $\tau$ flavours, as well as to DM. The most straightforward possibility is to assume a $L_\mu-L_\tau$ flavour symmetry so that gauge anomalies cancel,

\myeq{
&{\cal L} \owns -\frac{1}{4} F'_{\mu\nu} F'^{\mu\nu}- i g' \sum_{i} Q'_{i} \overline{\psi}_i \gamma^\mu Z'_\mu \psi_i  + i g'  Q'_{\rm DM} \overline{\chi} \gamma^\mu \chi Z'_\mu\chi\quad,\label{mutaumodel}
}

\noindent where the sum is over the muon and tau left-handed doublets, as well as over the muon and tau right-handed singlets (with $Q'=1$ for muon spinors and $Q'=-1$ for tau spinors). Here, for definiteness, $\chi$ is taken to be a Dirac fermion, but in principle the DM candidate could also be a scalar boson charged under the $U(1)'$. The gauge part of this model has been considered in many different contexts (see \cite{He:1990pn,Foot:1990mn,He:1991qd,Heeck:2011wj}), including as a possibility of explanation for the $(g-2)_\mu$ anomaly, \cite{Gninenko:2001hx,Baek:2001kca,Carone:2013uh,Altmannshofer:2014pba}. This requires a $Z'$ with mass 10-100 MeV and $g'\sim 5\times 10^{-4}$. With the adjunction of a DM particle, as in Eq. \ref{mutaumodel}, it has also been considered for various purposes \cite{Cirelli:2008pk,Baek:2008nz,Garani:2019fpa}. In \cite{Garani:2019fpa} it has been noted that the values of $m_{Z'}$ and $g'$ which fit well the $(g-2)_\mu$ anomaly, see above, can also lead to DM self-interactions with $\sigma_T/m_{\rm DM}\sim 1 \hbox{cm}^2/\hbox{g}$ (once $Q'_{\rm DM}$ has been fixed for the annihilation of DM into leptons to have the thermal value). Here we would like to point out that this model turns out to be good also to avoid the CMB and BBN constraints above. In particular, as already said, a decay of the light mediator into muon and/or tau neutrinos allow to avoid the CMB constraint of Eq. \ref{cmbconstraint}. The other CMB and the BBN constraints are avoided in a way similar to the way the first model we considered (see vector portal model of Chapter \ref{ch:secl}) avoids these constraints above (for $T'/T=1$). Note that all this has been analyzed in \cite{Kamada:2018zxi} in the framework of a model where, on top of the interactions of Eq. \ref{mutaumodel}, the scalar whose vev breaks spontaneously the $U(1)_{L_{\mu}-L_{\tau}}$ gauge symmetry is considered explicitly, assuming in addition that this scalar has Yukawa interactions with DM, so that for self-interactions the light mediator is this extra scalar, rather than the $Z'$.

\section{The asymmetric dark matter option\label{sec:asym}}
So far we have assumed everywhere that there is an equal number of DM particles and antiparticles. If instead we assume that DM is asymmetric, some of the constraints will change drastically. This possibility has been considered  in \cite{Baldes:2017gzu} in the context of a model where a dark proton and a dark electron couple to a dark photon. On the one hand, all constraints related to DM annihilation are trivially removed, since DM does not annihilate anymore. This concerns in particular the CMB constraint of Eq. \ref{cmbconstraint} and the indirect detection constraints. On the other hand, the constraints on the number of mediators and on its decay remain. In an asymmetric setup, still, DM thermalises with the light mediator at $T\gtrsim m_{\rm DM}$, and the resulting important symmetric DM component must be suppressed afterwards through an annihilation catastrophy. This annihilation catastrophy leaves the mediator as a hot relic, so one is left with as many light mediator as in the symmetric case. Thus, if the mediator is stable, the non-overclosure constraint remains fully relevant. Similarly, for an unstable mediator, remain relevant the $N_{eff}$ and mediator decay CMB constraints above, as well as the Hubble constant and entropy injection and photodisintegration BBN constraints, as well as direct detection constraints. In Refs. \cite{Baldes:2017gzu,Hufnagel:2018bjp}, authors have not considered in details the $N_{eff}$ CMB constraint, as well as Hubble constant and entropy injection BBN constraints. However, as already said above, if for the minimal vector model of Chapter \ref{ch:secl} (and similarly for the more involved vector model of Ref. \cite{Baldes:2017gzu}), one removes the CMB constraint given in Eq. \ref{cmbconstraint}, a proper incorporation of these constraints leaves allowed a relatively wide region of parameter space. However, for the scalar mediator model, to assume an asymmetric setup instead of a symmetric one does not change much the picture because in this model this CMB constraint was already avoided (from the fact that the annihilation is p-wave). In particular the strong tension between direct detection and BBN constraints remains.

\chapter*{Summary and outlooks}\label{ch:ccl}
\addcontentsline{toc}{chapter}{Summary and outlooks}
\markboth{Summary and outlooks}{Summary and outlooks}
\yinipar{A}ll along this thesis we have studied various aspects, and put forward several possible new properties, of models where the DM belongs to a HS that may be connected to the SM sector through portal interactions. These new properties and scenarios concern in particular the way models of this type can account for the DM relic abundance observed today in the Universe and/or the way they could allow large DM self-interactions. We have started by presenting the many relevant constraints which apply to scenarios where the large self-interactions needed to account for the small scale anomalies result from the exchange of a light mediator. To do so, we presented two very well known portal models which are the Higgs portal and the kinetic mixing portal models. Both of them have been used in this work as representatives of scalar and vector portal models respectively. The global study we have performed for these two portal models is also representative of what would happen in other portal-like models. At the end of the first chapter, it appeared that self-interacting DM models with light mediator are strongly constrained from three types of constraints which are in tensions with each other \cite{Hambye:2018dpi,Hambye_2020}. First, to alleviate small scale structure constraints, one requires a relatively strong connection between the DM and the light mediator. Second, cosmological and astrophysical observations suggest a short lifetime for the light mediator and thus a strong connection between the light mediator and SM particles. Third, particle physics experiments which try to detect or produce DM (or light mediator particles) have still not found any prove of the existence of such particles. This non-observation indicates that the DM should be not so strongly connected to the SM.
\\

The second chapter introduced in detail the so-called Sommerfeld effect and its consequences on DM portal models. This effect is at the origin of the enhancement of the DM self-interaction in presence of a light mediator and to look at it in more details, as we did in this chapter, allowed us to see how one could take advantage of this effect in order to reduce the tensions discussed in Chapter \ref{ch:constr}. In particular, we have shown (as a first main result) that the Sommerfeld enhancement can still be largely operative for a DM-to-med coupling quite smaller than the ones that we considered in Chapter \ref{ch:constr}, \cite{Hambye_2020}. This will be used later in Chapters \ref{ch:TpT} and \ref{ch:other_minimal} to build simple SIDM frameworks that account for all constraints. In this chapter \ref{ch:som}, we also stressed the fact that the Sommerfeld effect is important for the annihilation of DM particles in various contexts, at the time of the recombination (constrained by CMB data), as well as for production of cosmic rays today in the MW galactic centre or in DG (constrained by indirect detection). We showed here, in particular, that contrary to common belief, a DM p-wave annihilation could lead to observable signals in indirect detection experiments thanks to the Sommerfeld effect in the case of a large DM-to-med coupling, \cite{Hambye_2020}. This second main result will also be used in Chapter \ref{ch:other_minimal} to constraints models discussed within.
\\

The third chapter was dedicated to the way one could account for the amount of DM observed in the Universe in HS models. In these models, the HS does not contain the DM particles only, but it also contains other particles with which the DM can interact. Thus, in this chapter, we studied models containing, in full generality, three reservoirs (SM, DM and extra HS particles) with potentially three connections between them. The scalar and vector portal models considered in Chapter \ref{ch:constr} are perfect prototype models of this kind. Thus, to study how we account for the DM relic density in these models, as we did in Chapter \ref{ch:prod}, is interesting in itself as it studies a generic situation. It is also fully relevant for SIDM models based on a light mediator since these models have generically the very same three sector-three connections structure. We have then shown, as a third result, that with such a structure the DM relic abundance can be produced through no less than five distinct mechanisms which are the freeze-in, the sequential freeze-in (which is new), the reannihilation, the secluded freeze-out and finally the freeze-out \cite{Hambye:2019dwd,Vanderheyden:2021tih}. Four of those production mechanisms have two different realisations depending on if the DM is dominantly connected to the mediator or SM particles. This makes nine regimes among which four are new and for five of the nine regimes the DM is feebly connected to the SM.
\\

In Chapter \ref{ch:secl}, we looked in more details at the possibility that the DM relic density stems from a freeze-out in the HS which would occur when a possible portal between the HS and the VS has ceased to have any sizeable effect. This was already the case of one of the DM production mechanism that we saw in Chapter \ref{ch:prod}, the so-called secluded FO regime. Here, we developed more this possibility which holds also in general for models where there would be no connection at all between both the visible and hidden sectors, but where both sectors would be already there at the end of inflation. The purpose of this chapter was to determine, in a model independent way (i.e. beyond the two benchmark portal models) what is the domain of all DM thermal candidate. This domain stems from the two possibilities of FO that can hold in this case, the non-relativistic secluded FO and the relativistic secluded FO. The latter which is extremely simple and leads to a relic density which does not depend on the annihilation cross section, but only on the DM mass and on the hidden-to-visible temperature ratio, had, as mentioned above, strangely not be considered in any published paper before, which makes its study our fourth result and the general study of the domain to be our fifth result \cite{Hambye:2020lvy,Coy:2021ann}. The full $m_{\rm DM}-T'/T$ domain of possibilities we presented is bounded from below (for $T'/T$) by the "relativistic floor" (i.e. the relativistic FO case), from above (for $T'/T$) by the "unitarity wall" and the "$N_{eff}$ ceiling", from below (for $m_{\rm DM}$) by the generalisation of the Cowsik-McClelland bound and from above (for $m_{\rm DM}$) by generalisation of the Griest-Kamionkowski bound.
\\

From all these discussions in Chapters \ref{ch:constr} to \ref{ch:secl}, in Chapter \ref{ch:TpT} we come back to the problematic of conciliating all constraints that apply to the SIDM scenarios with light mediator. This necessarily requires to relax one of the assumptions made in Chapter \ref{ch:constr}. In this chapter, we considered what is probably the simplest possibility. This is to say that this possibility does not require to enlarge the Lagrangian of the two benchmark models with any extra new field. It only requires to give up the assumption that the HS is thermally connected to the SM thermal bath, just as we considered already in Chapters \ref{ch:prod} and \ref{ch:secl} for the DM relic density issue. So, we argue that this solution is minimal because it does not require any additional degree of freedom and moreover it is natural because it is what observations suggest: feebly coupled to the VS and strongly self-interacting. Indeed, we have seen that if the mixing parameter (the SM-to-med coupling) lies below the thermalisation threshold, such portal models can easily satisfy the DM relic abundance constraint, and avoid all kinds of other constraints still alleviating the tensions at small scale. We found, as a sixth result, that all free parameters can live in ranges of several orders of magnitude without difficulties which makes this solution totally viable and really interesting \cite{Hambye_2020}.
\\

In chapter \ref{ch:other_minimal}, other minimal solutions to the SIDM tensions have been studied at length. The general idea behind such scenarios is either to deplete the light mediator abundance prior to BBN by decay or further annihilations into lighter particles or to decrease its impact on the BBN and/or CMB using p-wave processes or neutrinos particles in the final state. These solutions are less minimal as they require extra degrees of freedom and in most cases their allowed parameter space is more squeezed, but remain simple solutions to the SIDM issue and constitute the final result of this thesis \cite{Hambye_2020}.
\\

The several aspects and new properties/scenarios of setups based on a VS-HS portal structure discussed here could be further studied in many ways, along several new avenues. Through this thesis, we have studied many aspects of self-interacting DM models with light mediator, from production mechanisms, to experimental constraints passing by a theoretical model building phase. A simple thermalised HS brings a lot of new and open questions. One could worry about what happens in more details in the transition regime between the relativistic and non-relativistic DM decoupling cases, close to the relativistic floor. One could also study further the fate of the light mediator in the hotter HS scenario. We hope that these works, and many others, could, step by step, ultimately lead to the elucidation of the fundamental question of what DM is made of, in terms of particles.

%----------------------------------------------------------------------------------------
%	THESIS CONTENT - APPENDICES
%----------------------------------------------------------------------------------------

\appendix % Cue to tell LaTeX that the following "chapters" are Appendices
\part{Appendices}
%\addcontentsline{toc}{part}{Appendices}
% Include the appendices of the thesis as separate files from the Appendices folder
% Uncomment the lines as you write the Appendices

\chapter{Cross sections}\label{app:cs}
In this first appendix, we give all relevant cross sections we may used in this thesis, for both the scalar and vector portal models.

\section{Higgs portal}
The scalar portal model detailed in Subsection \ref{subsec:HP} is described by the following Lagrangian,

\myeq{
\mathcal{L}&= \mathcal{L}_{SM}+ i\bar{\chi}\slashed{D}\chi - m_{\rm DM}\bar{\chi}\chi + Y_{\chi}\Phi\bar{\chi}\chi-\mu_{\Phi}^{2}\Phi^{2}+\lambda_{\Phi}\Phi^{4}-\mu_{H}^{2}H^{\dagger}H\nn\\
&\hspace{1cm}+\lambda_{H}\left(H^{\dagger}H\right)^{2}+\lambda_{3}\Phi H^{\dagger}H+\lambda_{\Phi H}\Phi^{2}H^{\dagger}H,
\label{eq:ap_lag_hp}
}

\noindent such that relevant processes in the symmetric, semi-symmetric and broken phases are as given in the following tables.
\\

%\begin{center}
%\begin{figure}[h!]
%\centering
%%\includegraphics[scale=0.8]{XXFF_v.pdf}
%\includegraphics[scale=0.8]{XXFF_s.pdf}
%\caption{Connection between the DM and SM baths goes through the production of the mediator in the s-channel for Higgs portal model. $f$ indicates a SM fermion.}
%\label{fig:ap_diag_DM-to-SM}
%\end{figure}
%\end{center}

%\begin{table*}
\begin{center}
%\resizebox{\textwidth}{!}{
\begin{tabular}{|>{\centering\arraybackslash}p{0.37\textwidth}|>{\centering\arraybackslash}p{0.53\textwidth}|}
\hline 
\multicolumn{2}{|c|}{$T_{EW}<T_{\Phi}<m_{\rm DM}$ (fully symmetric phase): $v_{\Phi}=0=v_{H}$} \\ 
\hline 
Process & Cross section \\ 
\hline
\begin{minipage}{.37\textwidth}
\begin{center}
\includegraphics[scale=0.6]{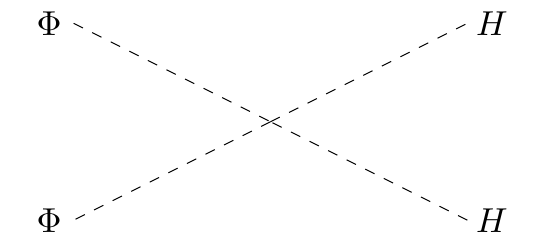}
\end{center}
\end{minipage}
& $\frac{\lambda_{\Phi H}^{2}}{\pi s}$ \\ 
\hline
\end{tabular}
%}
\end{center}\label{tab:annexe_HP_SYM}
%\end{table*}

%\begin{table*}
\begin{center}
%\resizebox{\textwidth}{!}{
\begin{tabular}{|>{\centering\arraybackslash}p{0.37\textwidth}|>{\centering\arraybackslash}p{0.53\textwidth}|}
\hline 
\multicolumn{2}{|c|}{$T_{EW}<m_{\rm DM}<T_{\Phi}$ (semi-symmetric phase): $v_{\Phi}\neq 0=v_{H}$} \\ 
\hline 
Process & Cross section \\ 
\hline
\begin{minipage}{.37\textwidth}
\begin{center}
\includegraphics[scale=0.6]{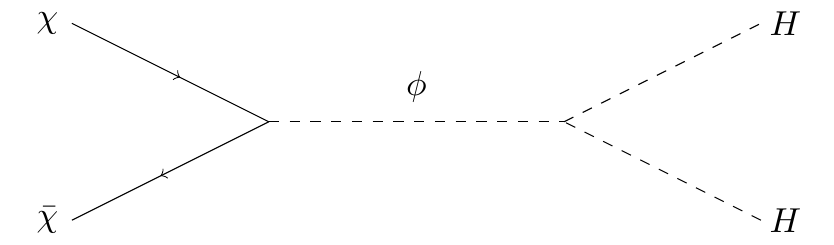}
\end{center}
\end{minipage}
& $\frac{\lambda ^{2}v_{\Phi}^{2}Y_{\chi}^{2}}{64\pi s^{2}}\left(1-\frac{4m_{\rm DM}^{2}}{s}\right)^{3/2}$ \\ 
\hline
\begin{minipage}{.37\textwidth}
\begin{center}
\includegraphics[scale=0.6]{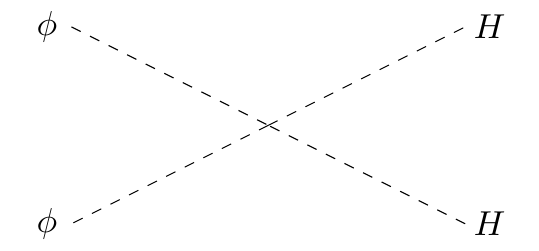}
\end{center}
\end{minipage}
& $\frac{\lambda_{\Phi H}^{2}}{4\pi s}\frac{1}{\sqrt{1-4\frac{m_{\phi}^{2}}{s}}}$ \\ 
\hline
\end{tabular}
%}
\end{center}\label{tab:annexe_HP_SEM}
%\end{table*}

%\begin{table*}
\begin{center}
%\resizebox{\textwidth}{!}{
\begin{tabular}{|>{\centering\arraybackslash}p{0.37\textwidth}|>{\centering\arraybackslash}p{0.53\textwidth}|}
\hline 
\multicolumn{2}{|c|}{$m_{\rm DM}<T_{EW}<T_{\Phi}$ (broken phase): $v_{\Phi}\neq 0\neq v_{H}$} \\ 
\hline 
Process & Cross section \\ 
\hline
\begin{minipage}{.37\textwidth}
\begin{center}
\includegraphics[scale=0.6]{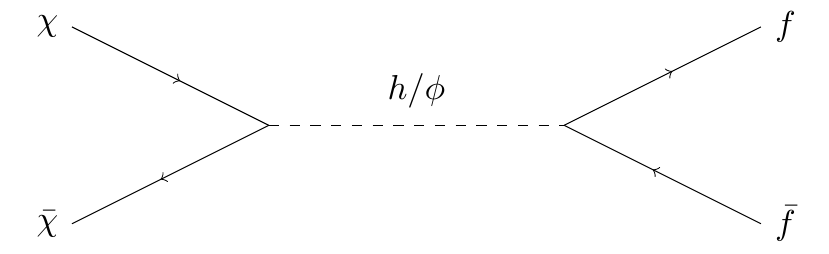}
\end{center}
\end{minipage}
& $N_{c}\frac{\kappa ^{2}}{8\pi s^{3}}\left(m_{f}^{2}m_{h}^{2}\right)\frac{\sqrt{1-\frac{4m_{f}^{2}}{s}}\left(1-\frac{4m_{\rm DM}^{2}}{s}\right)^{3/2}}{\left(1-\frac{m_{h}^{2}}{s}\right)^{2}+\frac{\Gamma _{h}^{2}m_{h}^{2}}{s^{2}}}$ \\ 
\hline
\begin{minipage}{.37\textwidth}
\begin{center}
\includegraphics[scale=0.6]{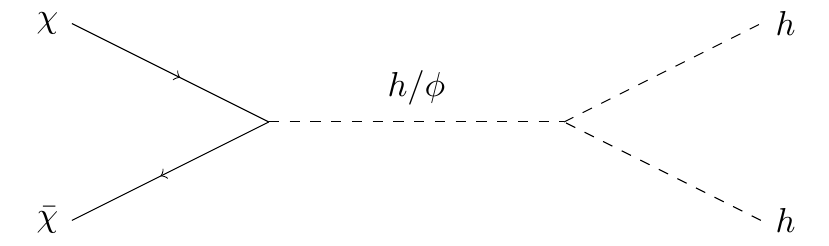}
\end{center}
\end{minipage}
& $\frac{\kappa ^{2}}{32\pi s^{4}}\left(\frac{m_{h}^{8}}{v_{H}^{2}}\right)\frac{\sqrt{1-\frac{4m_{h}^{2}}{s}}\left(1-\frac{4m_{\rm DM}^{2}}{s}\right)^{3/2}}{\left(1-\frac{m_{h}^{2}}{s}\right)^{2}+\frac{\Gamma _{h}^{2}m_{h}^{2}}{s^{2}}}$ \\ 
\hline
\multirow{2}{*}{
\begin{minipage}{.37\textwidth}
\begin{center}
\includegraphics[scale=0.6]{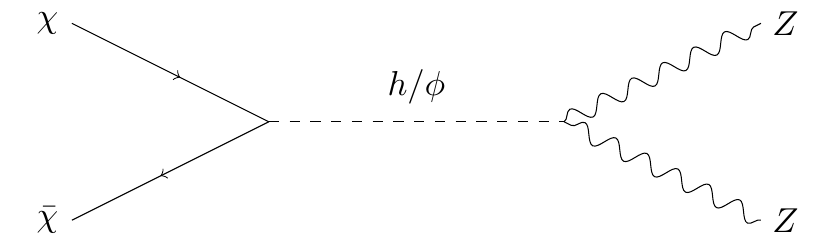}
\end{center}
\end{minipage}}
& $\frac{\kappa ^{2}}{288\pi s^{2}}\left(\frac{g^{2}}{m_{Z}^{2}c_{w}^{2}}\right)\frac{\sqrt{1-\frac{4m_{Z}^{2}}{s}}\left(1-\frac{4m_{\rm DM}^{2}}{s}\right)^{3/2}}{\left(1-\frac{m_{h}^{2}}{s}\right)^{2}+\frac{\Gamma _{h}^{2}m_{h}^{2}}{s^{2}}}$\\
& $\hspace{0.5cm}\times\left(1-4\frac{m_{Z}^{2}}{s}+12\frac{m_{Z}^{4}}{s^{2}}\right)$ \\ 
\hline
\multirow{2}{*}{
\begin{minipage}{.37\textwidth}
\begin{center}
\includegraphics[scale=0.6]{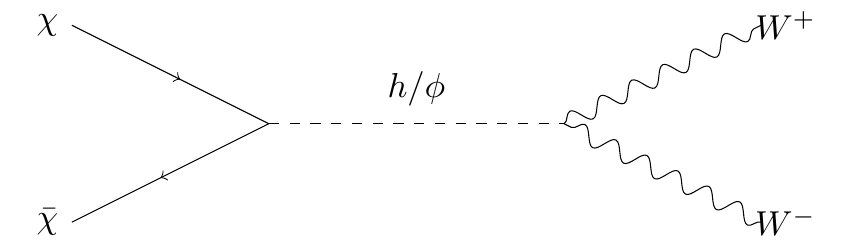}
\end{center}
\end{minipage}}
& $\frac{\kappa ^{2}}{288\pi s^{2}}\left(\frac{g^{2}}{m_{W}^{2}}\right)\frac{\sqrt{1-\frac{4m_{W}^{2}}{s}}\left(1-\frac{4m_{\rm DM}^{2}}{s}\right)^{3/2}}{\left(1-\frac{m_{h}^{2}}{s}\right)^{2}+\frac{\Gamma _{h}^{2}m_{h}^{2}}{s^{2}}}$ \\
& $\hspace{0.5cm}\times\left(1-4\frac{m_{W}^{2}}{s}+12\frac{m_{W}^{4}}{s^{2}}\right)$ \\
\hline
\begin{minipage}{.37\textwidth}
\begin{center}
\includegraphics[scale=0.6]{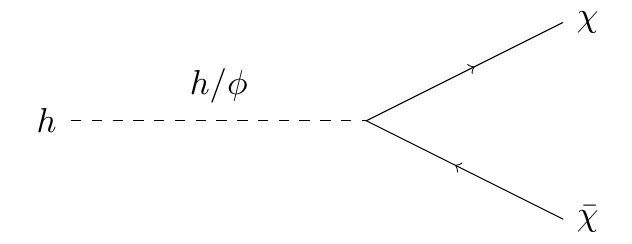}
\end{center}
\end{minipage}
& $\frac{\kappa ^{2}}{8\pi}m_{h}\left(1-\frac{4m_{\rm DM}^{2}}{m_{h}^{2}}\right)^{3/2}$ \\ 
\hline
\begin{minipage}{.37\textwidth}
\begin{center}
\includegraphics[scale=0.6]{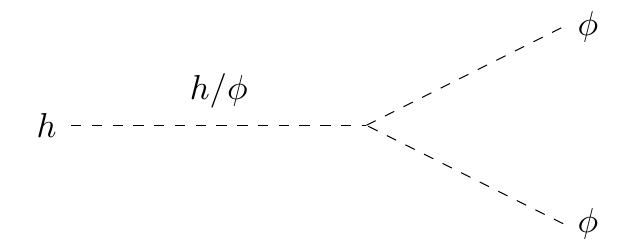}
\end{center}
\end{minipage}
& $\frac{\sin^{2}(2\beta_{\Phi H})}{32\pi}\frac{m_{h}^{2}}{v_{\Phi}}\sqrt{1-\frac{4m_{\phi}^{2}}{m_{h}^{2}}}\left(1-\frac{m_{\phi}^{2}}{m_{h}^{2}}\right)^{2}$ \\ 
\hline
\begin{minipage}{.37\textwidth}
\begin{center}
\includegraphics[scale=0.6]{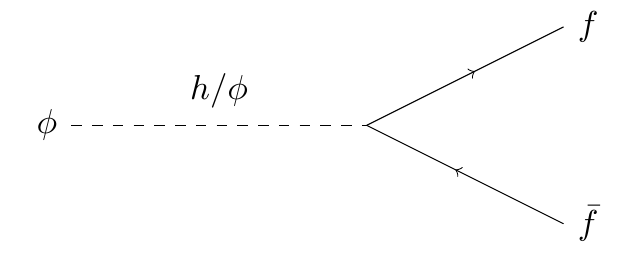}
\end{center}
\end{minipage}
& $\frac{Y_{f}^{2}\sin^{2}(\beta_{\Phi H})}{16\pi}m_{\phi}\left(1-\frac{4m_{f}^{2}}{m_{\phi}^{2}}\right)^{3/2}$ \\ 
\hline
\multirow{2}{*}{
\begin{minipage}{.37\textwidth}
\begin{center}
\includegraphics[scale=0.6]{hp_ap_BRO_XXPP.pdf}
\end{center}
\end{minipage}}
& $\hspace{-0.15cm}\frac{Y_{\chi}^{4}}{2\pi s}\left[\left(1+16\frac{m_{\rm DM}^{2}}{s}-32\frac{m_{\rm DM}^{4}}{s^{2}}\right)\tanh^{-1}\sqrt{1-\frac{4m_{\rm DM}^{2}}{s}}\right.$ \\
& $\hspace{0.5cm}\left.-\left(1+8\frac{m_{\rm DM}^{2}}{s}\right)\sqrt{1-4\frac{m_{\rm DM}^{2}}{s}}\right]$ \\
\hline
\begin{minipage}{.37\textwidth}
\begin{center}
\includegraphics[scale=0.6]{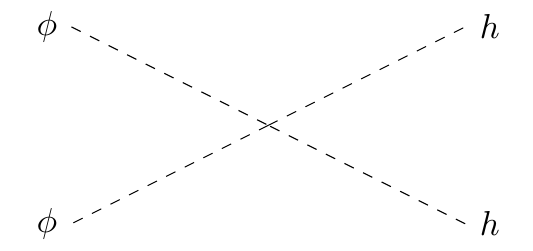}
\end{center}
\end{minipage}
& $\frac{\lambda_{\Phi H}^{2}}{16\pi s}\frac{\sqrt{1-4\frac{m_{h}^{2}}{s}}}{\sqrt{1-4\frac{m_{\phi}^{2}}{s}}}$ \\ 
\hline
\end{tabular}
%}
\end{center}\label{tab:annexe_HP_BRO}

\section{Kinetic mixing}
The vector portal model detailed in Subsection \ref{subsec:KM} is described by the following Lagrangian,

\myeq{
\mathcal{L}&\subset -\frac{1}{4} B'^{\mu\nu} B'_{\mu\nu} +\frac{1}{2}m_{\gamma'}^{2} B'^{\mu} B'_{\mu} - \hat \epsilon\,  m_{\gamma'}^{2} B^{\mu} B'_{\mu} - e' \bar{\chi}\gamma^\mu\chi  ( B'_\mu  - \hat \epsilon B_\mu),\label{eq:ap_lag_km_mass_mix}
}

\noindent such that relevant processes are as given in the following table for the massive dark photon case:
\\

\begin{center}
%\resizebox{\textwidth}{!}{
\begin{tabular}{|>{\centering\arraybackslash}p{0.37\textwidth}|>{\centering\arraybackslash}p{0.53\textwidth}|}
%\hline 
%\multicolumn{2}{|c|}{$m_{\rm DM}<T_{EW}<T_{\Phi}$ (broken phase)} \\ 
\hline 
Process & Cross section \\ 
\hline
\begin{minipage}{.37\textwidth}
\begin{center}
\includegraphics[scale=0.6]{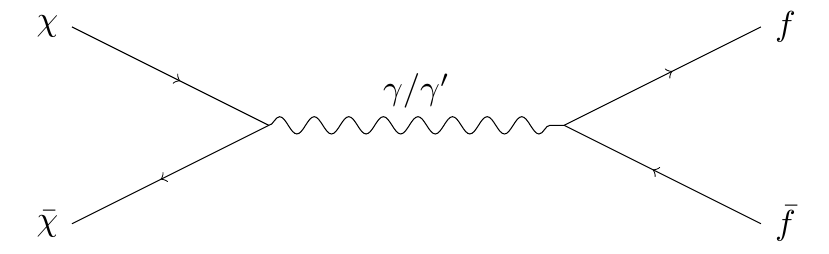}
\end{center}
\end{minipage}
& $16\pi N_{c}\frac{\alpha^{2}\kappa ^{2}}{s}\frac{\left(1-\frac{2m_{f}^{2}}{s}\right)\left(1-\frac{2m_{\rm DM}^{2}}{s}\right)}{\left(1-\frac{m_{\gamma'}^{2}}{s}\right)^{2}+\frac{\Gamma _{\gamma'}^{2}m_{\gamma'}^{2}}{s^{2}}}\sqrt{\frac{1-\frac{4m_{f}^{2}}{s}}{1-\frac{4m_{\rm DM}^{2}}{s}}}$ \\
\hline
\multirow{2}{*}{
\begin{minipage}{.37\textwidth}
\begin{center}
\includegraphics[scale=0.6]{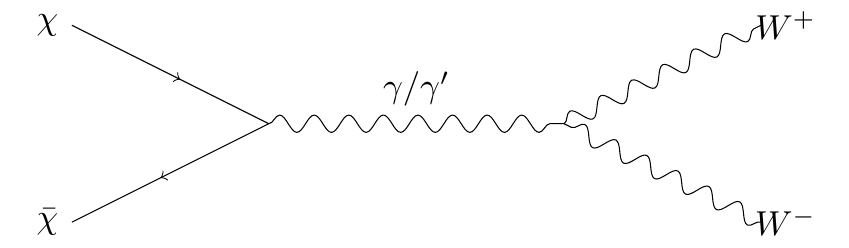}
\end{center}
\end{minipage}}
& $\frac{\pi\alpha^{2}\kappa ^{2}s}{m_{W}^{4}}\left(1-4\frac{m_{W}^{2}}{s}+12\frac{m_{W}^{4}}{s^{2}}\right)\left(1-\frac{m_{\rm DM}^{2}}{s}-6\frac{m_{W}^{4}}{s^{2}}\right.$ \\
& $\hspace{0.5cm}\left.+2\frac{m_{W}^{2}}{s}\left(1-4\frac{m_{\rm DM}^{2}}{s}\right)\right)\sqrt{\frac{1-\frac{4m_{W}^{2}}{s}}{1-\frac{4m_{\rm DM}^{2}}{s}}}$ \\
\hline
\begin{minipage}{.37\textwidth}
\begin{center}
\includegraphics[scale=0.6]{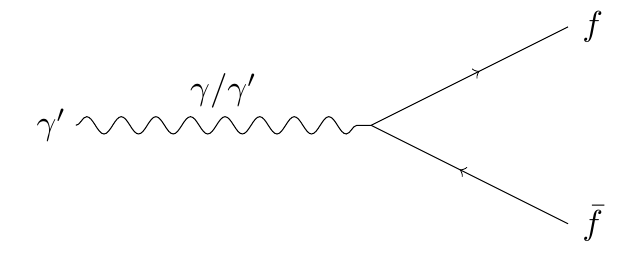}
\end{center}
\end{minipage}
& $N_{c}\cos^{2}\theta_{W}Q^{2}\frac{\alpha\epsilon^{2}}{3}m_{\gamma'}\left(1-\frac{4m_{f}^{2}}{m_{\gamma'}^{2}}\right)^{1/2}$ \\ 
\hline
\multirow{2}{*}{
\begin{minipage}{.37\textwidth}
\begin{center}
\includegraphics[scale=0.6]{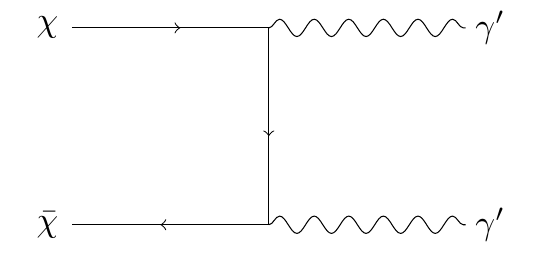}
\end{center}
\end{minipage}}
& $\frac{32\pi\alpha'^{2}}{s\left(1-\frac{4m_{\rm DM}^{2}}{s}\right)}\left[\left(1-8\frac{m_{\rm DM}^{2}}{s}\right)\tanh^{-1}\sqrt{1-4\frac{m_{\rm DM}^{2}}{s}}\right.$ \\ 
& $\hspace{0.5cm}\left.+\frac{m_{\rm DM}^{2}}{s}\sqrt{1-4\frac{m_{\rm DM}^{2}}{s}}\right]$ \\ 
\hline
\multirow{2}{*}{
\begin{minipage}{.37\textwidth}
\begin{center}
\includegraphics[scale=0.6]{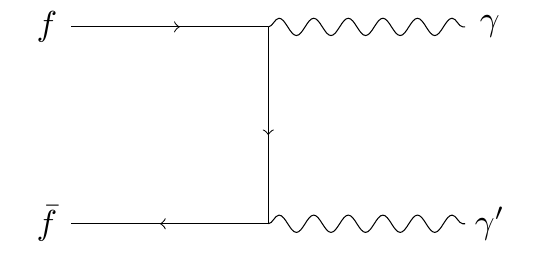}
\end{center}
\end{minipage}}
& $\frac{32\pi\alpha^{2}\epsilon^{2}}{s\left(1-\frac{4m_{f}^{2}}{s}\right)}\left[\left(1-8\frac{m_{f}^{2}}{s}\right)\tanh^{-1}\sqrt{1-4\frac{m_{f}^{2}}{s}}\right.$ \\ 
& $\hspace{0.5cm}\left.+\frac{m_{f}^{2}}{s}\sqrt{1-4\frac{m_{f}^{2}}{s}}\right]$ \\ 
\hline
\begin{minipage}{.37\textwidth}
\begin{center}
\includegraphics[scale=0.6]{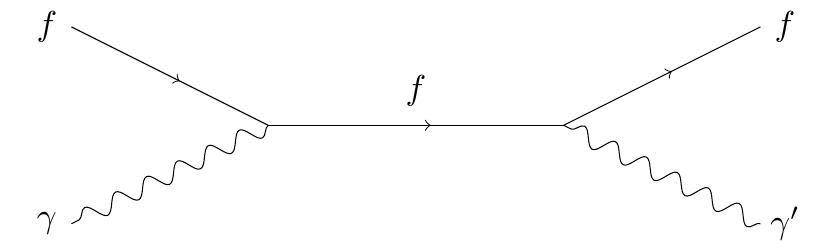}
\end{center}
\end{minipage}
& $\frac{4\pi\alpha^{2}\epsilon ^{2}}{s}\frac{\left(1+\frac{m_{f}^{2}}{s}\right)\left(1-5\frac{m_{f}^{2}}{s}+2\frac{m_{f}^{4}}{s^{2}}\right)}{1-\frac{m_{f}^{2}}{s}}$ \\
\hline
\end{tabular}
%}
\end{center}

In the case of a massless dark photon, one has instead:

\begin{center}
%\resizebox{\textwidth}{!}{
\begin{tabular}{|>{\centering\arraybackslash}p{0.37\textwidth}|>{\centering\arraybackslash}p{0.53\textwidth}|}
%\hline 
%\multicolumn{2}{|c|}{$m_{\rm DM}<T_{EW}<T_{\Phi}$ (broken phase)} \\ 
\hline 
Process & Cross section \\ 
\hline
\begin{minipage}{.37\textwidth}
\begin{center}
\includegraphics[scale=0.6]{km_ap_XXFF.pdf}
\end{center}
\end{minipage}
& $16\pi N_{c}\frac{\alpha^{2}\kappa ^{2}}{s}\left(1-\frac{2m_{f}^{2}}{s}\right)\left(1-\frac{2m_{\rm DM}^{2}}{s}\right)\sqrt{\frac{1-\frac{4m_{f}^{2}}{s}}{1-\frac{4m_{\rm DM}^{2}}{s}}}$ \\
\hline
\multirow{2}{*}{
\begin{minipage}{.37\textwidth}
\begin{center}
\includegraphics[scale=0.6]{km_ap_XXWW.pdf}
\end{center}
\end{minipage}}
& $\frac{\pi\alpha^{2}\kappa ^{2}s}{m_{W}^{4}}\left(1-4\frac{m_{W}^{2}}{s}+12\frac{m_{W}^{4}}{s^{2}}\right)\left(1-\frac{m_{\rm DM}^{2}}{s}-6\frac{m_{W}^{4}}{s^{2}}\right.$ \\
& $\hspace{0.5cm}\left.+2\frac{m_{W}^{2}}{s}\left(1-4\frac{m_{\rm DM}^{2}}{s}\right)\right)\sqrt{\frac{1-\frac{4m_{W}^{2}}{s}}{1-\frac{4m_{\rm DM}^{2}}{s}}}$ \\
\hline
\multirow{2}{*}{
\begin{minipage}{.37\textwidth}
\begin{center}
\includegraphics[scale=0.6]{km_ap_XXApAp.pdf}
\end{center}
\end{minipage}}
& $\frac{32\pi\alpha'^{2}}{s\left(1-\frac{4m_{\rm DM}^{2}}{s}\right)}\left[\left(1-8\frac{m_{\rm DM}^{2}}{s}\right)\tanh^{-1}\sqrt{1-4\frac{m_{\rm DM}^{2}}{s}}\right.$ \\ 
& $\hspace{0.5cm}\left.+\frac{m_{\rm DM}^{2}}{s}\sqrt{1-4\frac{m_{\rm DM}^{2}}{s}}\right]$ \\ 
\hline
\end{tabular}
%}
\end{center}
\chapter{Thermal effects}\label{app:th}
For comprehensiveness and completeness we will review in this appendix the main features of the dark photon production rate, extensively studied in e.g. \cite{Redondo:2008aa,Jaeckel:2008fi,Redondo:2008ec,An:2013yfc,Redondo:2013lna,Fradette:2014sza}.
\\

\begin{figure}[h!]
\centering
\includegraphics[scale=.2]{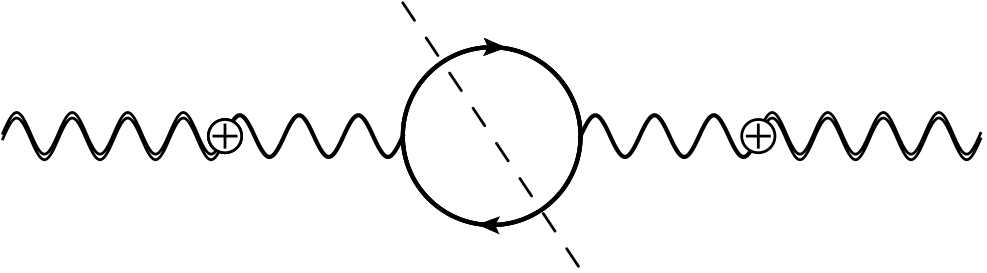}
\caption[The imaginary part of the dark photon propagator in a medium]{The imaginary part of the dark photon propagator (double wiggly lines) in a medium includes both its decay rate and creation rates.}
\label{fig:fg1_ap}
\end{figure}

As we did in Chapter \ref{ch:prod}, we will treat the kinetic mixing as a perturbation and we will work in the interaction basis. Thus, in such a framework, the dark photon self-energy of Figure \ref{fig:fg1_ap} captures thermal effects through those of the photons. We have,

\myeq{
\Pi_{\gamma'} = m^2_{\gamma'} + {\epsilon^2 m_{\gamma'}^4\over (K^2 - \Pi_{\gamma})},
}

\noindent \noindent where the momentum of the virtual dark photon $K$ is given by $K^2 \equiv \omega^2 - k^2$. The imaginary part of the dark photon propagator is given by, 

\myeq{
\operatorname{Im} \Pi_{\gamma'}= {\epsilon^2 m_{\gamma'}^4 \operatorname{Im}\Pi_\gamma \over (m_{\gamma'}^2- \operatorname{Re}\Pi_{\gamma})^2 +  \operatorname{Im}\Pi_{\gamma}^2},
}

\noindent where we used $K^2 = m_{\gamma'}^2$. This corresponds to a virtual photon which mixes into a dark photon on mass-shell. In vacuum, the imaginary part of the photon propagator $ \operatorname{Im}\Pi_\gamma$ is related to possible photon decay channels, but within a thermal bath, the imaginary part of the photon propagator also gets contributions from the emission and the absorption rates of photons from the medium. $\Gamma$ is the rate that rules the thermal equilibrium condition, 

\myeq{
f(\omega,t) - f_{\rm eq} \propto \exp(- \Gamma t),
}

\noindent with $f$ the distribution function of dark photons.

\myeq{
\operatorname{Im}\Pi_\gamma = - \omega \Gamma_\gamma = - \omega (\Gamma_{\gamma, {\rm em}} - \Gamma_{\gamma, {\rm abs}}),
}

\noindent where the minus sign between the rates of emission and of absorption stems for Bose-Einstein statistics while for Fermi-Dirac statistics one has \cite{Weldon:1983jn},

\myeq{
\operatorname{Im}\Pi_\gamma = - \omega \Gamma_\gamma = - \omega (\Gamma_{\gamma, {\rm em}} + \Gamma_{\gamma, {\rm abs}}).
}

\noindent Due to unitarity \cite{Weinberg:1979bt}, it is known that the amplitude squares for the emission and absorption process are equal. However, the emission rate of a photon of energy $\omega$ is Boltzmann suppressed compared to its corresponding absorption rate, 

\myeq{
\Gamma_{\rm em} = \exp{(- \omega/T)} \;\Gamma_{\rm abs},
}

\noindent such that one can write

\myeq{
\Gamma_{\gamma', {\rm em}} =   {\epsilon^2 m_{\gamma'}^4\Gamma_{\gamma, {\rm em}}  \over (m_{\gamma'}^2- \operatorname{Re}\Pi_{\gamma})^2 +  \omega^2(e^{\omega/T}-1)^2 \Gamma_{\gamma, {\rm em}}^2}.
\label{eq:emmissionrate_ap}
}

\noindent At finite temperature, the medium can also support the propagation of longitudinal modes or plasmon waves on top of transverse modes. These new modes describe the oscillations of charged particles which are contained in the thermal bath. Thus, one should definitively make the distinction between transverse (labeled with a T subscript) and longitudinal (labeled with a L subscript) self-energy such as emission and absorption rates,  $( \Gamma, \Pi) \rightarrow (\Gamma_{T,L}, \Pi_{T,L})$ \cite{An:2013yfc}. The simplest of the two is the transverse mode as transverse photons simply get a thermal mass with a small momentum dependence. In the relativistic regime and to leading order in the electroweak fine structure constant $\alpha$, we have

\myeq{
\operatorname{Re}\Pi_{\gamma,T} = \left\{
\begin{array}{ll}
\omega_P^2 = \sum_i q_i^2 {T^2/9} &\mbox{low $k$}\\
{3/ 2}\, \omega_P^2 & \mbox{large $k$}
\end{array}\right.,
}

\noindent where $\omega_P$ is the so-called plasma frequency and the sum is over relativistic charged particles \cite{Braaten:1993jw}.
\\

The longitudinal mode of the photon self-energy is more complicated due to the fact that longitudinal photons do no propagate in vacuum. Staying in the relativistic regime and at leading order in the electroweak fine structure constant $\alpha$, the photon self-energy takes the form \cite{Braaten:1993jw}

\myeq{
\Pi_{L}(\omega,k) = 3 \omega_P^2  {K^2\over k^2} \left({\omega\over 2 k} \log\left({\omega + k)\over (\omega - k)}\right) - 1\right),
\label{eq:PiL_ap}
}

\noindent where we used the definition of the longitudinal polarisation tensor given in \cite{An:2013yfc,Redondo:2013lna}: $\Pi_L \equiv \Pi_L^{\rm APP}$, which differs from that of \cite{Braaten:1993jw}, $\Pi_L^{\rm BS} = K^2/k^2 \Pi_L^{\rm APP}$. One can thus solve for $ \omega_L^2 - k^2 = \operatorname{Re}\Pi_L(\omega_L(k), k)$ which leads to

\myeq{
\operatorname{Re}\Pi_{\gamma,L} = \left\{
\begin{array}{ll}
\omega_P^2\,  K^2/\omega^2& k \sim 0\\
\sim 0 & k \gtrsim \omega_P
\end{array}\right. .
}

\begin{figure}[h!]
\centering
\includegraphics[scale=0.55]{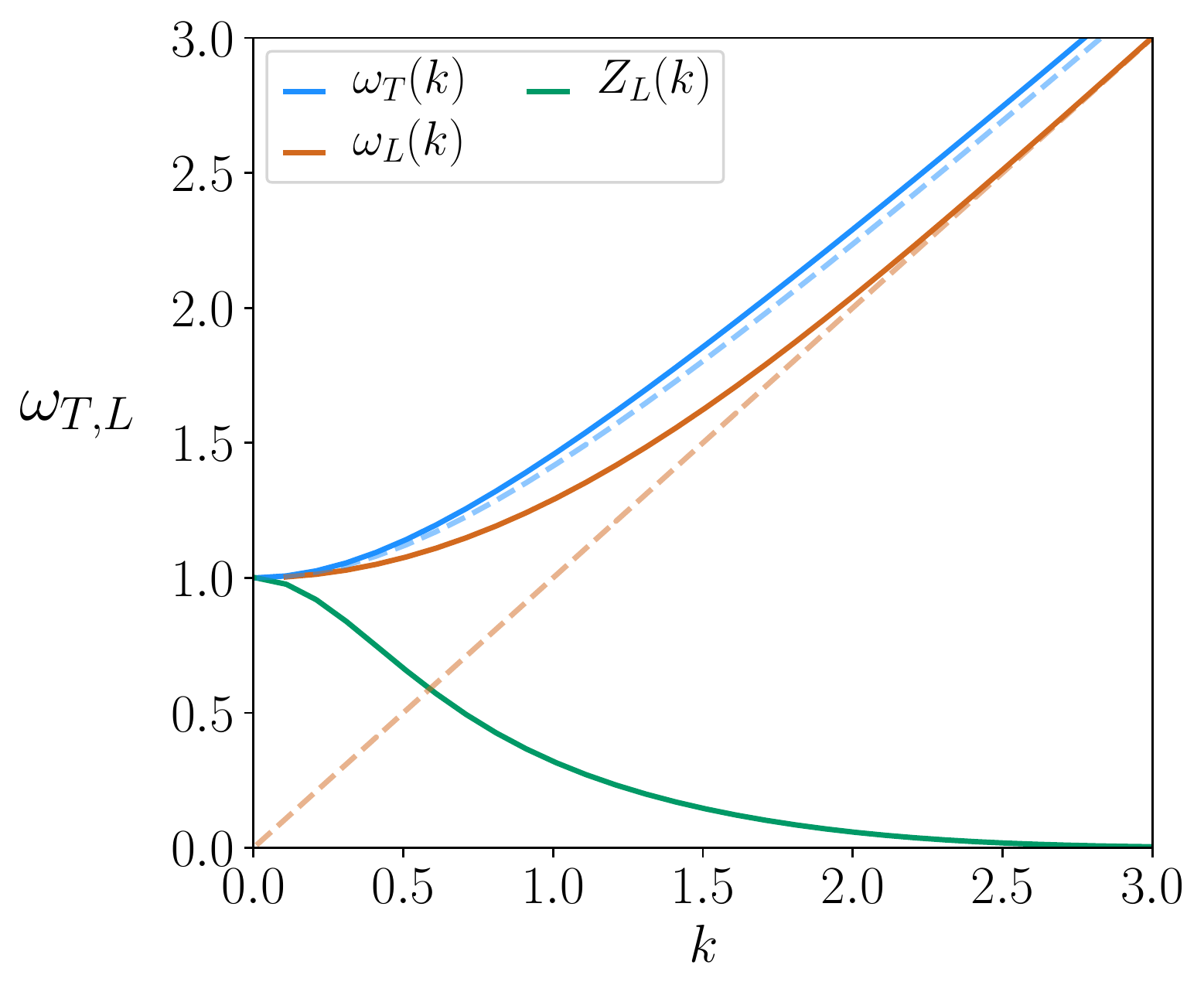}
\caption[Dispersion relations for transversal and longitudinal modes in the relativistic regime]{Dispersion relations for $\omega_T(k)$ (blue solid) and $\omega_L(k)$ (orange solid) in the relativistic regime $T\ll m_e$. They are normalised to $\omega_P=1$. The dashed orange line is the dispersion relation of a standard massive particle. The dispersion relation of the transverse mode has $\omega_T= \omega_P$ for small $k$ and $\omega_T \approx \sqrt{3/2 \omega_P^2 + k^2}$ for large $k$. That of the longitudinal mode has $\omega_L = \omega_P$ for small $k$ but asymptots to $\omega_L \approx k$ for large $k$. Also shown is the wave-function normalisation $Z_L$ (green solid) which goes to zero at large $k$, revealing that the longitudinal mode propagates only for small enough momenta. }\label{fig:omegaL_ap}
\end{figure}

\noindent The behaviours of the dispersion relations of the transverse (solid blue) and longitudinal (solid orange) modes as functions of the momentum $k$ are shown in Figure \ref{fig:omegaL_ap}.
\\

One can write for a longitudinal plasmon mode close to be on-shell (i.e. having $\omega \sim \omega_L$, $\operatorname{Re}\Pi_{\gamma,L}  \approx \omega_L^2\,  K^2/\omega^2$) \cite{Braaten:1993jw,An:2013yfc,Redondo:2013lna},

\myeq{
{1\over K^2 -\Pi_{L}}\approx{ \omega^2 \, Z_L\over K^2( \omega^2 - \omega_L^2) - i Z_L \omega^2 \operatorname{Im}\Pi_{L}},
\label{eq:plasmonOS_ap}
}

\noindent where, using Eq. \ref{eq:PiL_ap},  the wave-function normalisation $Z_L$ is given by

\myeq{
Z_L^{-1} = 1+ {k^2\over \omega_L}  {3 \omega_P^2 - \omega_L^2 + k^2\over  2 (\omega_L^2 - k^2)},
}

\noindent which satisfies  $Z_L \rightarrow 1$ as $k\rightarrow 0$ but tends to zero for $k\gtrsim \omega_P$, as one can see from the green solid line in Figure \ref{fig:omegaL_ap}. Note that, we considered the real part of the self-energy only at leading order in the electroweak fine structure constant $\alpha$, but the imaginary part $\operatorname{Im}\Pi_{L}$ may include higher order corrections in $\alpha$ (Compton emission, etc.). The fact that $Z_L$ goes to zero when $k\gtrsim \omega_P$ reflects the fact that the longitudinal mode mostly exists for moderate momenta. From this and from Eq. \ref{eq:emmissionrate_ap} we get

\myeq{
\Gamma_{\gamma',\, {\rm em}}^L =   {\epsilon^2 m_{\gamma'}^4 \tilde Z_L^2 \Gamma_{\gamma,\,{\rm em}}^L  \over  (\omega^2 - \omega_L^2)^2 + \omega^2 (e^{\omega/T}-1)^2 (\tilde Z_L \Gamma_{\gamma,\,{\rm em}}^L)^2},
\label{eq:GamLFinal_ap}
}

\noindent with $\tilde Z_L = \omega^2/m_{\gamma'}^2\,  Z_L$. Expression given in Eq. \ref{eq:GamLFinal_ap} agrees with the literature \cite{An:2013yfc,Redondo:2013lna,Rrapaj:2019eam}. However, it may be compared with Eq. 2.7 of \cite{Redondo:2013lna} using the fact that in Eq. \ref{eq:GamLFinal_ap} the longitudinal photon emission rates are to be calculated as in vacuum while in \cite{Redondo:2013lna} they include factors interpreted as wave-function normalisation: $\tilde Z_L \Gamma_{\gamma,\,{\rm em}}^L\vert_{\rm us}= \Gamma_{\gamma,\,{\rm em}}^L\vert_{\rm RR} $.
\\

Eq. \ref{eq:GamLFinal_ap} has to be compared to

\myeq{
\label{eq:GamRFinal}
\Gamma_{\gamma',\,{\rm em}}^T =   {\epsilon^2 m_{\gamma'}^4  \Gamma_{\gamma,\,{\rm em}}^T  \over  (m_{\gamma'}^2  - \omega_T^2)^2 + \omega^2 (e^{\omega/T}-1)^2 (\Gamma_{\gamma,\,{\rm em}}^T)^2}.
}

\noindent Comparing these expressions, one sees that at large $T$, $\omega_P \gg m_{\gamma'}$ and $\omega \sim \omega_P$,  such as the ratio of longitudinal and transversal modes of the rate scales as,

\myeq{
{\Gamma_{\gamma',\,{\rm em}}^L \over \Gamma_{\gamma',\,{\rm em}}^T} \approx {\omega^4\over m_{\gamma'}^4} {\Gamma_{\gamma,\,{\rm em}}^L \over \Gamma_{\gamma,\,{\rm em}}^T} \label{eq:highT}.
}

\noindent Furthermore, as the longitudinal production rate itself is proportional to $\propto m_{\gamma'}^2$\footnote{This is due to current conservation: $k_\mu J^\mu = 0\rightarrow\epsilon_\mu^L J^\mu \propto m_{\gamma'}$.}, one finds that the overall scaling of the ratio given in Eq. \ref{eq:highT} is $\omega^2/m_{\gamma'}^2 \gg 1$. Hence, at the end of the day, the dark photons production rate occurs dominantly through production of longitudinal photons at high temperature, $\omega_P \gtrsim m_{\gamma'}$ \cite{An:2013yfc,Redondo:2013lna}. On the other hand, at lower temperatures, which is the regime relevant for infra-red dominated freeze-in production, one has 

\myeq{
{\Gamma_{\gamma',\,{\rm em}}^L \over \Gamma_{\gamma',\,{\rm em}}^T} \approx Z_L^2 {\Gamma_{\gamma,\,{\rm em}}^L \over \Gamma_{\gamma,\,{\rm em}}^T} \sim Z_L^2 {m_{\gamma'}\over \omega} \ll 1\label{eq:lowT},
}

\noindent because $Z_L\gtrsim 1$. Thus the dark photon production is dominated by production of transverse photons as explained above, see also \cite{An:2013yfc,Redondo:2013lna}. One can then use the following replacement rule for the on-shell dark photon production as it proceeds essentially as in vacuum,

\myeq{
\epsilon \rightarrow \epsilon_{\rm eff}^2 = {\epsilon^2 m_{\gamma'}^4 \over (m_{\gamma'}^2- \operatorname{Re}\Pi_{\gamma})^2 +  \omega^2(e^{\omega/T}-1)^2 \Gamma_{\gamma}^2},
}

\noindent which basically explains the substitution rule stated in Eq.~(\ref{eq:substitution}).\\
\\

\begin{figure}
\centering
\includegraphics[scale=0.55]{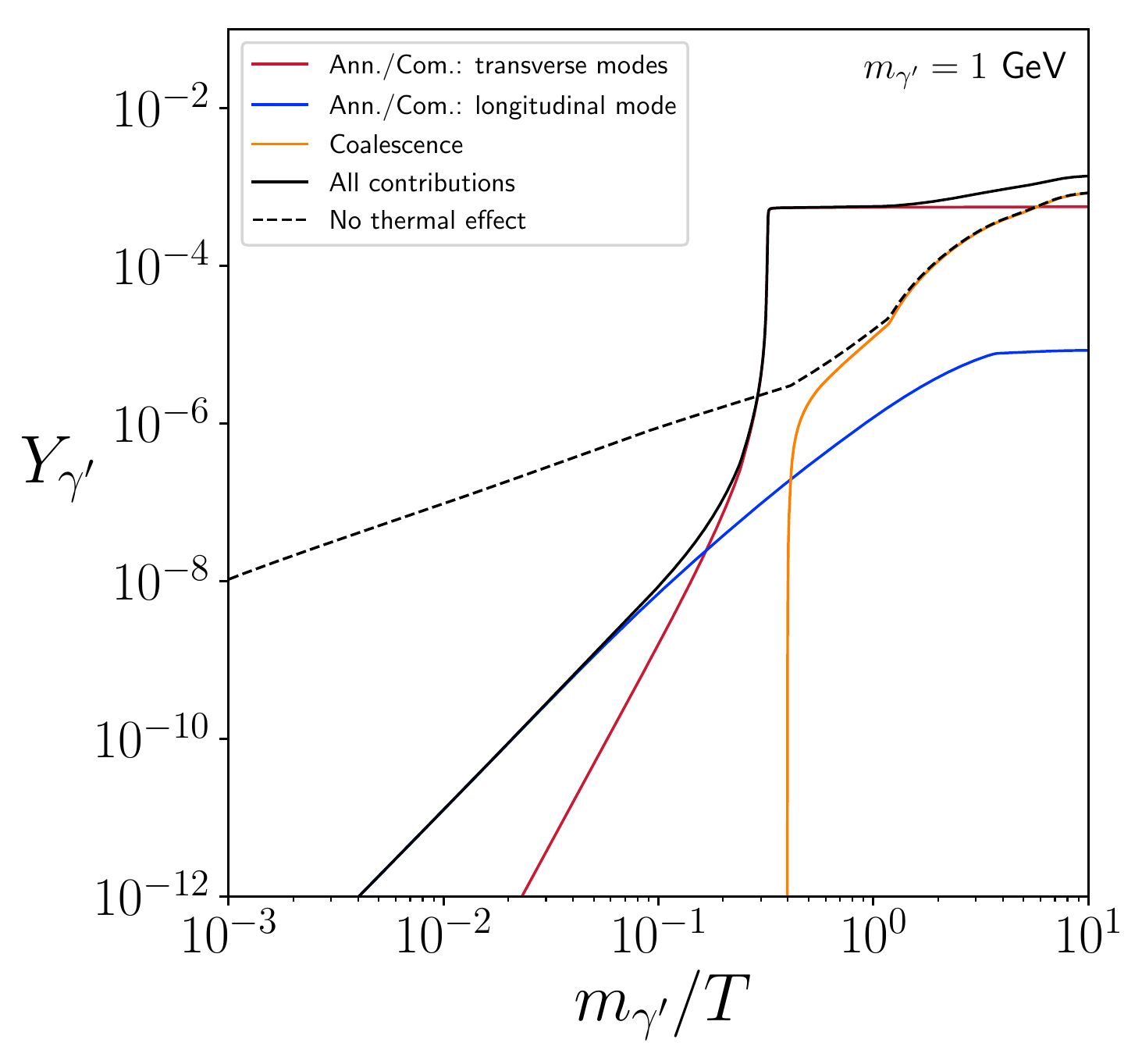}
\caption[Time evolution of the dark photon yield with all production contributions]{For $\epsilon=10^{-9}$, all contributions to the dark photon yield as a function of the inverse temperature. One can distinguish contributions from pair annihilation and Compton processes for both transverses (red dashed) and longitudinal modes (blue dashed), from the coalescence process (orange dashed) and from all contributions together (solid black). For the coalescence, we have taken into account the thermal corrections to the mass of the SM particles that annihilate into a dark photon \cite{Redondo:2008ec}.}\label{fig:DPProductionChannels}
\end{figure}

Finally, we give in Figure \ref{fig:DPProductionChannels} the time evolution of the dark photon yield highlighting all contributions to dark photon production. One can recognise the high temperature suppression of transversal modes and the low temperature suppression of the longitudinal mode of Eqs. \ref{eq:highT} and \ref{eq:lowT} respectively.

\chapter{Instantaneous freeze-out approximation}\label{app:inst_FO}
\section{The thermally connected HS case}
In this Appendix, we derive the DM relic abundance for a non-relativistic FO as a function of the DM annihilation cross section and the hidden-to-visible temperature ratio. Let us first review the very well known case (see e.g. \cite{Kolb:1990vq}) for which the DM is in thermal equilibrium with the SM bath.
\\

We assume that the DM FO occurs instantaneously such that the instant of DM decoupling is set when the Hubble expansion rate becomes larger than the DM annihilation rate: $\Gamma\vert_{T_{\rm dec}}\lesssim H\vert_{T_{\rm dec}}$. Defining $T_{\rm dec}$ as the exact instant where the two rates stop to be equal, on has

\myeq{
n^{eq}(T_{\rm dec})\langle\sigma v\rangle = \sqrt{\frac{8\pi}{3M_{\rm pl}^{2}}}\sqrt{\rho(x_{\rm dec})},\label{eq:n_sv_H_ap}
}

\noindent with $x_{\rm dec}\equiv m_{\rm DM}/T_{\rm dec}$, $M_{\rm pl}$ the Planck mass and where the DM number density is given by $n^{eq}(T_{\rm dec})\equiv g_{\rm DM}(m_{\rm DM}T_{\rm dec}/2\pi)^{3/2}e^{-m_{\rm DM}/T_{\rm dec}}$, using the non-relativistic approximation. Taking the logarithm of Eq. \ref{eq:n_sv_H_ap}, one can approximate the instant of DM decoupling $x_{\rm dec}$ by

\myeq{
&x_{\rm dec} \simeq \ln\left[0.038\left\langle\sigma v\right\rangle m_{\rm pl}m_{\rm DM}\left(\frac{g_{\rm DM}}{\sqrt{g^{\rm eff}_{\ast}(T_{\rm dec})}}\right)\right]\nonumber\\
&\hspace{0.5cm}+\frac{1}{2}\ln\ln\left[0.038\left\langle\sigma v\right\rangle m_{\rm pl}m_{\rm DM}\left(\frac{g_{\rm DM}}{\sqrt{g^{\rm eff}_{\ast}(T_{\rm dec})}}\right)\right].\label{eq:xp_dec_ap}
}

\noindent The DM relic density is defined by the DM energy density relative abundance today and can be related to the DM yield today \cite{Kolb:1990vq}: $\Omega_{\rm DM}h^{2}=2.5\times Y_{\rm DM}(T_{0})(m_{\rm DM}/\text{GeV})$. Then, assuming an instantaneous one has: $Y_{\rm DM}(T_{0})=Y_{\rm DM}(T_{\rm dec})=H(T_{\rm dec})/s(T_{\rm dec})\langle\sigma v\rangle$. Finally, plugging the expressions for the Hubble expansion rate as well as for the entropy energy density, we find the DM relic abundance in the instantaneous freeze-out approximation:

\myeq{
&\Omega _{\rm DM} h^{2} = 4.7\cdot10^8 \frac{g_{\rm DM}^{n}\sqrt{g^{\rm eff}_{\ast}(T_{\rm dec})}x_{\rm dec}}{g^{S}_{\ast}(T_{\rm dec})m_{\rm pl}\langle\sigma v\rangle\,{\rm GeV}},
\label{eq:Oh2_FO_ap}
}

\noindent where $x_{\rm dec}$ is given by Eq. \ref{eq:xp_dec_ap}.

\section{The thermally disconnected HS case}
The procedure to obtain the DM relic abundance in the instantaneous FO approximation when the DM particle is part of a HS which has never thermalised with the VS is exactly the same as in the previous section. The only difference is that one must take the HS contributions in the energy and entropy densities. We have,

\myeq{
& H(x') = \sqrt{\frac{8\pi}{3M_{\rm pl}^{2}}}\sqrt{\rho_{\rm VS}(x')+\rho_{\rm HS}(x')},\label{eq:H_universe_ap}\\
& s(x') = s_{\rm VS}(x')+s_{\rm HS}(x')\label{eq:s_universe_ap}.
}

\noindent In terms of the inverse hidden temperature $x'\equiv m_{\rm DM}/T'$ and the hidden-to-visible temperature ratio $\xi\equiv T'/T$, the energy and entropy densities of both sectors are given by,

\myeq{
&\rho_{\rm VS}(x') = \frac{\pi^{2}}{30}g_{\ast}^{\rm eff}\left(m_{\rm DM}/x'\xi\right)\left(\frac{m_{\rm DM}}{x'\xi}\right)^{4},\label{eq:rho_VS_ap}\\
&\rho_{\rm HS}(x') = \frac{\pi^{2}}{30}g'^{\rm eff}_{\ast}\left(m_{\rm DM}/x'\right)\left(\frac{m_{\rm DM}}{x'}\right)^{4},\label{eq:rho_HS_ap}\\
&s_{\rm VS}(x') = \frac{2\pi^{2}}{45}g_{\ast}^{S}\left(m_{\rm DM}/x'\xi\right)\left(\frac{m_{\rm DM}}{x'\xi}\right)^{3}\label{eq:s_VS_ap},\\
&s_{\rm HS}(x') = \frac{2\pi^{2}}{45}g'^{S}_{\ast}\left(m_{\rm DM}/x'\right)\left(\frac{m_{\rm DM}}{x'}\right)^{3}\label{eq:s_HS_ap}.
}

\noindent In that case, the instantaneous freeze-out approximation gives,

%\myeq{
%&\Omega _{\rm DM} h^{2} = 4.7\cdot10^8 \frac{g_{\rm DM}\sqrt{g^{\rm eff}_{\ast}(T_{\rm dec})}x'_{\rm dec}}{g^{S}_{\ast}(T_{\rm dec})m_{\rm pl}\langle\sigma v\rangle\,{\rm GeV}}\times\frac{T'_{\rm dec}}{T_{\rm dec}}\times\frac{\sqrt{1+\xi_{\rm dec}^{4}\frac{g'^{\rm eff}_{\ast}(T'_{\rm dec})}{g^{\rm eff}_{\ast}(T_{\rm dec})}}}{1+\xi_{\rm dec}^{3}\frac{g'^{S}_{\ast}(T'_{\rm dec})}{g^{S}_{\ast}(T_{\rm dec})}},
%\label{eq:Oh2_FO_TpT_ap}
%}

\myeq{
&\Omega _{\rm DM} h^{2} = 4.7\cdot10^8 \frac{g_{\rm DM}\sqrt{g^{\rm eff}_{\ast}(T_{\rm dec})+\xi_{\rm dec}^{4}g'^{\rm eff}_{\ast}(T'_{\rm dec})}x'_{\rm dec}\xi_{\rm dec}}{\left(g^{S}_{\ast}(T_{\rm dec})+\xi_{\rm dec}^{3}g'^{S}_{\ast}(T'_{\rm dec})\right)(T_{\rm dec})m_{\rm pl}\langle\sigma v\rangle\,{\rm GeV}},
\label{eq:Oh2_FO_TpT_ap}
}

\noindent with $x'_{\rm dec}$ given by,

%\myeq{
%&x'_{\rm dec} \simeq \ln\left[0.038\left(\frac{T'_{\rm dec}}{T_{\rm dec}}\right)^2\left\langle\sigma v\right\rangle m_{\rm pl}m_{\rm DM}\left(\frac{g_{\rm DM}}{\sqrt{g^{\rm eff}_{\ast}(T_{\rm dec})}}\right)\times\frac{1}{\sqrt{1+\xi_{\rm dec}^{4}\frac{g'^{\rm eff}_{\ast}(T'_{\rm dec})}{g^{\rm eff}_{\ast}(T_{\rm dec})}}}\right]\nonumber\\
%&\hspace{0.0cm}+\frac{1}{2}\ln\ln\left[0.038\left(\frac{T'_{\rm dec}}{T_{\rm dec}}\right)^2\left\langle\sigma v\right\rangle m_{\rm pl}m_{\rm DM}\left(\frac{g_{\rm DM}}{\sqrt{g^{\rm eff}_{\ast}(T_{\rm dec})}}\right)\times\frac{1}{\sqrt{1+\xi_{\rm dec}^{4}\frac{g'^{\rm eff}_{\ast}(T'_{\rm dec})}{g^{\rm eff}_{\ast}(T_{\rm dec})}}}\right]\label{eq:xp_dec}
%}

\myeq{
&x'_{\rm dec} \simeq \ln\left[0.038\xi_{\rm dec}^2\left\langle\sigma v\right\rangle m_{\rm pl}m_{\rm DM}\left(\frac{g_{\rm DM}}{\sqrt{g^{\rm eff}_{\ast}(T_{\rm dec})+\xi_{\rm dec}^{4}g'^{\rm eff}_{\ast}(T'_{\rm dec})}}\right)\right]\nonumber\\
&\hspace{0.0cm}+\frac{1}{2}\ln\ln\left[0.038\xi_{\rm dec}^2\left\langle\sigma v\right\rangle m_{\rm pl}m_{\rm DM}\left(\frac{g_{\rm DM}}{\sqrt{g^{\rm eff}_{\ast}(T_{\rm dec})+\xi_{\rm dec}^{4}g'^{\rm eff}_{\ast}(T'_{\rm dec})}}\right)\right].\label{eq:xp_dec}
}

\section{Further the instantaneous freeze-out approximation}
Let us now derive a slightly improved formula for the DM relic abundance. We no longer assume the instantaneous FO, but we will not have a fully analytical formula.
\\

Starting with the Boltzmann equation on the DM yield (Eq. \ref{eq:BE_yield}),

\myeq{
\frac{\diff Y}{\diff x} = - \frac{\left\langle\sigma v\right\rangle_{\bar{\chi}\chi\rightarrow XX}s}{xH}\left[Y_{\chi}^{2}-\left(Y^{eq}_{\chi}\right)^{2}\right].
}

\noindent One can use the HS temperature $x'\equiv m_{\rm DM}/T'$ instead of the visible one ($x=m_{\rm DM}/T$), we have

\myeq{
\frac{\diff Y}{\diff x'} = - \mathcal{G}(x')\left[Y_{\chi}^{2}-\left(Y^{eq}_{\chi}\right)^{2}\right],
}

\noindent where now all quantities are seen as a function of $x'$ and where we have defined $\mathcal{G}(x')\equiv \left\langle\sigma v\right\rangle s(x')/x'H(x')$. 
\\

%Let us now write down the Boltzmann equation for the offset to equilibrium defined as $\Delta \equiv Y-Y^{eq}$,
%
%\myeq{
%\frac{\diff \Delta}{\diff x'} = -\mathcal{G}(x')\Delta\left(\Delta+2 Y^{eq}\right)-\frac{\diff Y^{eq}}{\diff x'}.
%}

At late times, long after the freeze-out, the equilibrium Yield is totally negligible (Boltzmann suppressed) compared to the actual Yield. It is to say that $Y\gg Y^{eq}$ when $x'\gg x'_{\rm dec}$ such that the Boltzmann equation can be approximated as following,

\myeq{
\frac{\diff Y}{\diff x'} = - \mathcal{G}(x')Y_{\chi}^{2}.
}

\noindent This last equation can be integrated from the instant of decoupling $x'_{\rm dec}$ to today $x'_{0}$,

\myeq{
Y(x'_{0}) \simeq \left(\int_{x'_{\rm dec}}^{x'_{0}}\mathcal{G}(x')\diff x'\right)^{-1}.
}

\noindent One gets the DM relic abundance,

\myeq{
\Omega_{\rm DM}h^{2}=2.53\times 10^{8}\times\left(\frac{m_{DM}}{\text{GeV}}\right)\times\left(\int_{x'_{\rm dec}}^{x'_{0}}\frac{\left\langle\sigma v\right\rangle s(x')}{x'H(x')}\diff x'\right)^{-1},\label{eq:Oh2_integral_ap}
}

\noindent where an integral has still to be performed. As said above, the final formula obtained in this way is not fully analytical, but is numerically closer to what one would obtain solving the Boltzmann equation. However, one can check that it recovers the analytical formula obtained assuming an instantaneous FO by taking a constant cross section in \ref{eq:Oh2_integral_ap}. If one further assumes that the entropy and energy densities of the Universe are dominated by the VS, one gets

\myeq{
&H(x')\simeq 1.67\sqrt{g_{\star}^{eff}(m_{\rm DM}/x'\xi)}\frac{m_{\rm DM}^{2}}{M_{pl}x'^{2}\xi^{2}},\\
&s(x')\simeq \frac{2\pi^{2}}{45}g_{\star}^{S}(m_{\rm DM}/x'\xi)\frac{m_{\rm DM}^{3}}{x'^{3}\xi^{3}},
}

\noindent such that we finally have,

\myeq{
\Omega_{\rm DM}h^{2}=7.90\times 10^{-11}\times \frac{T'_{\rm dec}}{T_{\rm dec}}\times x'_{\rm dec}\times\left(\frac{\text{GeV}^{-2}}{\left\langle\sigma v\right\rangle}\right)\times\left(\frac{\sqrt{g_{\star}^{eff}(T_{\rm dec})}}{g_{\star}^{S}(T_{\rm dec})}\right).
}

%\include{Appendices/AppendixB}
%\include{Appendices/AppendixC}

%----------------------------------------------------------------------------------------
%	BIBLIOGRAPHY
%----------------------------------------------------------------------------------------

\printbibliography[heading=bibintoc]

@inproceedings{Vanderheyden:2021tih,
    author = "Vanderheyden, Laurent",
    title = "{Dark matter from dark photons}",
    eprint = "2105.07039",
    archivePrefix = "arXiv",
    primaryClass = "hep-ph",
    reportNumber = "ULB-TH/21-06",
    month = "5",
    year = "2021"
}

@article{Coy:2021ann,
    author = "Coy, Rupert and Hambye, Thomas and Tytgat, Michel H. G. and Vanderheyden, Laurent",
    title = "{The domain of thermal dark matter candidates}",
    eprint = "2105.01263",
    archivePrefix = "arXiv",
    primaryClass = "hep-ph",
    reportNumber = "ULB-TH/21-05",
    month = "5",
    year = "2021"
}

@ARTICLE{1932BAN.....6..249O,
       author = {{Oort}, J.~H.},
        title = "{The force exerted by the stellar system in the direction perpendicular to the galactic plane and some related problems}",
      journal = {Bull. Ast. Inst. Neth},
         year = 1932,
        month = aug,
       volume = {6},
        pages = {249},
       adsurl = {https://ui.adsabs.harvard.edu/abs/1932BAN.....6..249O}
}

@article{An:2009vq,
    author = "An, Haipeng and Chen, Shao-Long and Mohapatra, Rabindra N. and Zhang, Yue",
    title = "{Leptogenesis as a Common Origin for Matter and Dark Matter}",
    eprint = "0911.4463",
    archivePrefix = "arXiv",
    primaryClass = "hep-ph",
    reportNumber = "UMD-40762-471, UMD-PP-09-062, IC-2009-090",
    doi = "10.1007/JHEP03(2010)124",
    journal = "JHEP",
    volume = "03",
    pages = "124",
    year = "2010"
}

@article{PhysRevA.60.2118,
  title = {Consistent definitions for, and relationships among, cross sections for elastic scattering of hydrogen ions, atoms, and molecules},
  author = {Krstic, Predrag S. and Schultz, David R.},
  journal = {Phys. Rev. A},
  volume = {60},
  issue = {3},
  pages = {2118--2130},
  numpages = {0},
  year = {1999},
  month = {Sep},
  publisher = {American Physical Society},
  doi = {10.1103/PhysRevA.60.2118},
  url = {https://link.aps.org/doi/10.1103/PhysRevA.60.2118}
}

@article{Ibe:2009mk,
    author = "Ibe, Masahiro and Yu, Hai-bo",
    title = "{Distinguishing Dark Matter Annihilation Enhancement Scenarios via Halo Shapes}",
    eprint = "0912.5425",
    archivePrefix = "arXiv",
    primaryClass = "hep-ph",
    doi = "10.1016/j.physletb.2010.07.026",
    journal = "Phys. Lett. B",
    volume = "692",
    pages = "70--73",
    year = "2010"
}

@article{Kaul:1981uk,
    author = "Kaul, Romesh K.",
    title = "{Technicolor}",
    reportNumber = "PRINT-81-0258 (BANGALORE)",
    doi = "10.1103/RevModPhys.55.449",
    journal = "Rev. Mod. Phys.",
    volume = "55",
    pages = "449",
    year = "1983"
}

@article{Montull:2012ik,
    author = "Montull, Marc and Riva, Francesco",
    title = "{Higgs discovery: the beginning or the end of natural EWSB?}",
    eprint = "1207.1716",
    archivePrefix = "arXiv",
    primaryClass = "hep-ph",
    doi = "10.1007/JHEP11(2012)018",
    journal = "JHEP",
    volume = "11",
    pages = "018",
    year = "2012"
}

@article{Carmi:2012in,
    author = "Carmi, Dean and Falkowski, Adam and Kuflik, Eric and Volansky, Tomer and Zupan, Jure",
    title = "{Higgs After the Discovery: A Status Report}",
    eprint = "1207.1718",
    archivePrefix = "arXiv",
    primaryClass = "hep-ph",
    reportNumber = "LPT-ORSAY-12-77",
    doi = "10.1007/JHEP10(2012)196",
    journal = "JHEP",
    volume = "10",
    pages = "196",
    year = "2012"
}

@article{Barbieri:2015lqa,
    author = "Barbieri, Riccardo and Greco, Davide and Rattazzi, Riccardo and Wulzer, Andrea",
    title = "{The Composite Twin Higgs scenario}",
    eprint = "1501.07803",
    archivePrefix = "arXiv",
    primaryClass = "hep-ph",
    doi = "10.1007/JHEP08(2015)161",
    journal = "JHEP",
    volume = "08",
    pages = "161",
    year = "2015"
}

@article{Hooper:2008im,
    author = "Hooper, Dan and Zurek, Kathryn M.",
    title = "{A Natural Supersymmetric Model with MeV Dark Matter}",
    eprint = "0801.3686",
    archivePrefix = "arXiv",
    primaryClass = "hep-ph",
    reportNumber = "FERMILAB-PUB-07-587-A",
    doi = "10.1103/PhysRevD.77.087302",
    journal = "Phys. Rev. D",
    volume = "77",
    pages = "087302",
    year = "2008"
}

@article{Feng:2008ya,
    author = "Feng, Jonathan L. and Kumar, Jason",
    title = "{The WIMPless Miracle: Dark-Matter Particles without Weak-Scale Masses or Weak Interactions}",
    eprint = "0803.4196",
    archivePrefix = "arXiv",
    primaryClass = "hep-ph",
    reportNumber = "UCI-TR-2008-10",
    doi = "10.1103/PhysRevLett.101.231301",
    journal = "Phys. Rev. Lett.",
    volume = "101",
    pages = "231301",
    year = "2008"
}

@article{Akerib:2016vxi,
    author = "Akerib, D. S. and others",
    collaboration = "LUX",
    title = "{Results from a search for dark matter in the complete LUX exposure}",
    eprint = "1608.07648",
    archivePrefix = "arXiv",
    primaryClass = "astro-ph.CO",
    doi = "10.1103/PhysRevLett.118.021303",
    journal = "Phys. Rev. Lett.",
    volume = "118",
    number = "2",
    pages = "021303",
    year = "2017"
}

@article{Sirunyan:2018dub,
    author = "Sirunyan, Albert M. and others",
    collaboration = "CMS",
    title = "{Search for dark matter particles produced in association with a top quark pair at $\sqrt{s} =$ 13 TeV}",
    eprint = "1807.06522",
    archivePrefix = "arXiv",
    primaryClass = "hep-ex",
    reportNumber = "CMS-EXO-16-049, CERN-EP-2018-183",
    doi = "10.1103/PhysRevLett.122.011803",
    journal = "Phys. Rev. Lett.",
    volume = "122",
    number = "1",
    pages = "011803",
    year = "2019"
}

@article{Sirunyan:2018xlo,
    author = "Sirunyan, Albert M and others",
    collaboration = "CMS",
    title = "{Search for narrow and broad dijet resonances in proton-proton collisions at $ \sqrt{s}=13 $ TeV and constraints on dark matter mediators and other new particles}",
    eprint = "1806.00843",
    archivePrefix = "arXiv",
    primaryClass = "hep-ex",
    reportNumber = "CMS-EXO-16-056, CERN-EP-2018-123",
    doi = "10.1007/JHEP08(2018)130",
    journal = "JHEP",
    volume = "08",
    pages = "130",
    year = "2018"
}

@article{Sirunyan:2018fpy,
    author = "Sirunyan, Albert M. and others",
    collaboration = "CMS",
    title = "{Search for dark matter produced in association with a Higgs boson decaying to $\gamma\gamma$ or $\tau^+\tau^-$ at $\sqrt{s} =$ 13 TeV}",
    eprint = "1806.04771",
    archivePrefix = "arXiv",
    primaryClass = "hep-ex",
    reportNumber = "CMS-EXO-16-055, CERN-EP-2018-129",
    doi = "10.1007/JHEP09(2018)046",
    journal = "JHEP",
    volume = "09",
    pages = "046",
    year = "2018"
}

@article{Sirunyan:2018wcm,
    author = "Sirunyan, Albert M and others",
    collaboration = "CMS",
    title = "{Search for new physics in dijet angular distributions using proton\textendash{}proton collisions at $\sqrt{s}=$ 13 TeV and constraints on dark matter and other models}",
    eprint = "1803.08030",
    archivePrefix = "arXiv",
    primaryClass = "hep-ex",
    reportNumber = "CMS-EXO-16-046, CERN-EP-2018-036",
    doi = "10.1140/epjc/s10052-018-6242-x",
    journal = "Eur. Phys. J. C",
    volume = "78",
    number = "9",
    pages = "789",
    year = "2018"
}

@article{Sirunyan:2018gka,
    author = "Sirunyan, Albert M and others",
    collaboration = "CMS",
    title = "{Search for dark matter in events with energetic, hadronically decaying top quarks and missing transverse momentum at $ \sqrt{s}=13 $ TeV}",
    eprint = "1801.08427",
    archivePrefix = "arXiv",
    primaryClass = "hep-ex",
    reportNumber = "CMS-EXO-16-051, CERN-EP-2017-299",
    doi = "10.1007/JHEP06(2018)027",
    journal = "JHEP",
    volume = "06",
    pages = "027",
    year = "2018"
}

@article{Aaboud:2018xdl,
    author = "Aaboud, M. and others",
    collaboration = "ATLAS",
    title = "{Search for dark matter in events with a hadronically decaying vector boson and missing transverse momentum in $pp$ collisions at $\sqrt{s} = 13$ TeV with the ATLAS detector}",
    eprint = "1807.11471",
    archivePrefix = "arXiv",
    primaryClass = "hep-ex",
    reportNumber = "CERN-EP-2018-083",
    doi = "10.1007/JHEP10(2018)180",
    journal = "JHEP",
    volume = "10",
    pages = "180",
    year = "2018"
}

@article{Sirunyan:2017leh,
    author = "Sirunyan, Albert M and others",
    collaboration = "CMS",
    title = "{Search for top squarks and dark matter particles in opposite-charge dilepton final states at $\sqrt{s}=$ 13 TeV}",
    eprint = "1711.00752",
    archivePrefix = "arXiv",
    primaryClass = "hep-ex",
    reportNumber = "CMS-SUS-17-001, CERN-EP-2017-252",
    doi = "10.1103/PhysRevD.97.032009",
    journal = "Phys. Rev. D",
    volume = "97",
    number = "3",
    pages = "032009",
    year = "2018"
}

@article{Aaboud:2017phn,
    author = "Aaboud, Morad and others",
    collaboration = "ATLAS",
    title = "{Search for dark matter and other new phenomena in events with an energetic jet and large missing transverse momentum using the ATLAS detector}",
    eprint = "1711.03301",
    archivePrefix = "arXiv",
    primaryClass = "hep-ex",
    reportNumber = "CERN-EP-2017-230",
    doi = "10.1007/JHEP01(2018)126",
    journal = "JHEP",
    volume = "01",
    pages = "126",
    year = "2018"
}

@article{Aaboud:2017rzf,
    author = "Aaboud, Morad and others",
    collaboration = "ATLAS",
    title = "{Search for dark matter produced in association with bottom or top quarks in $\sqrt{s}=13$ TeV pp collisions with the ATLAS detector}",
    eprint = "1710.11412",
    archivePrefix = "arXiv",
    primaryClass = "hep-ex",
    reportNumber = "CERN-EP-2017-229",
    doi = "10.1140/epjc/s10052-017-5486-1",
    journal = "Eur. Phys. J. C",
    volume = "78",
    number = "1",
    pages = "18",
    year = "2018"
}

@article{Aaboud:2017bja,
    author = "Aaboud, M. and others",
    collaboration = "ATLAS",
    title = "{Search for an invisibly decaying Higgs boson or dark matter candidates produced in association with a $Z$ boson in $pp$ collisions at $\sqrt{s} =$ 13 TeV with the ATLAS detector}",
    eprint = "1708.09624",
    archivePrefix = "arXiv",
    primaryClass = "hep-ex",
    reportNumber = "CERN-EP-2017-166",
    doi = "10.1016/j.physletb.2017.11.049",
    journal = "Phys. Lett. B",
    volume = "776",
    pages = "318--337",
    year = "2018"
}

@article{Hooper:2019xss,
    author = "Hooper, Dan and Leane, Rebecca K. and Tsai, Yu-Dai and Wegsman, Shalma and Witte, Samuel J.",
    title = "{A systematic study of hidden sector dark matter:application to the gamma-ray and antiproton excesses}",
    eprint = "1912.08821",
    archivePrefix = "arXiv",
    primaryClass = "hep-ph",
    reportNumber = "MIT-CTP/5157, FERMILAB-PUB-19-628-A",
    doi = "10.1007/JHEP07(2020)163",
    journal = "JHEP",
    volume = "07",
    number = "07",
    pages = "163",
    year = "2020"
}

@article{ArkaniHamed:2005yv,
    author = "Arkani-Hamed, Nima and Dimopoulos, Savas and Kachru, Shamit",
    title = "{Predictive landscapes and new physics at a TeV}",
    eprint = "hep-th/0501082",
    archivePrefix = "arXiv",
    reportNumber = "SLAC-PUB-10928, HUTP-05-A0001, SU-ITP-04-44",
    month = "1",
    year = "2005"
}

@article{Strassler:2006im,
    author = "Strassler, Matthew J. and Zurek, Kathryn M.",
    title = "{Echoes of a hidden valley at hadron colliders}",
    eprint = "hep-ph/0604261",
    archivePrefix = "arXiv",
    doi = "10.1016/j.physletb.2007.06.055",
    journal = "Phys. Lett. B",
    volume = "651",
    pages = "374--379",
    year = "2007"
}

@article{Cvetic:2012kj,
    author = "Cvetic, Mirjam and Halverson, James and Piragua, Hernan",
    title = "{Stringy Hidden Valleys}",
    eprint = "1210.5245",
    archivePrefix = "arXiv",
    primaryClass = "hep-ph",
    reportNumber = "UPR-1241-T, NSF-KITP-12-187",
    doi = "10.1007/JHEP02(2013)005",
    journal = "JHEP",
    volume = "02",
    pages = "005",
    year = "2013"
}

@article{Blinnikov:1982eh,
    author = "Blinnikov, S. I. and Khlopov, M. Yu.",
    title = "{ON POSSIBLE EFFECTS OF 'MIRROR' PARTICLES}",
    reportNumber = "ITEP-11-1982",
    journal = "Sov. J. Nucl. Phys.",
    volume = "36",
    pages = "472",
    year = "1982"
}

@article{Foot:1991bp,
    author = "Foot, Robert and Lew, H. and Volkas, R. R.",
    title = "{A Model with fundamental improper space-time symmetries}",
    reportNumber = "UM-P-91-54, OZ-91-10",
    doi = "10.1016/0370-2693(91)91013-L",
    journal = "Phys. Lett. B",
    volume = "272",
    pages = "67--70",
    year = "1991"
}

@article{Cowsik:1972gh,
    author = "Cowsik, R. and McClelland, J.",
    title = "{An Upper Limit on the Neutrino Rest Mass}",
    doi = "10.1103/PhysRevLett.29.669",
    journal = "Phys. Rev. Lett.",
    volume = "29",
    pages = "669--670",
    year = "1972"
}

@article{Hodges:1993yb,
    author = "Hodges, H.M.",
    title = "{Mirror baryons as the dark matter}",
    doi = "10.1103/PhysRevD.47.456",
    journal = "Phys. Rev. D",
    volume = "47",
    pages = "456--459",
    year = "1993"
}

@article{Berezhiani:1995am,
    author = "Berezhiani, Z.G. and Dolgov, A.D. and Mohapatra, R.N.",
    title = "{Asymmetric inflationary reheating and the nature of mirror universe}",
    eprint = "hep-ph/9511221",
    archivePrefix = "arXiv",
    reportNumber = "INFN-FE-16-95, UMD-PP-96-15, IFIC-95-64, FTUV-95-61",
    doi = "10.1016/0370-2693(96)00219-5",
    journal = "Phys. Lett. B",
    volume = "375",
    pages = "26--36",
    year = "1996"
}

@article{Hambye:2020lvy,
    author = "Hambye, Thomas and Lucca, Matteo and Vanderheyden, Laurent",
    title = "{Dark matter as a heavy thermal hot relic}",
    eprint = "2003.04936",
    archivePrefix = "arXiv",
    primaryClass = "hep-ph",
    reportNumber = "ULB-TH/20-02, ULB-TH/20-02",
    doi = "10.1016/j.physletb.2020.135553",
    journal = "Phys. Lett. B",
    volume = "807",
    pages = "135553",
    year = "2020"
}

@article{Bennett:2019ewm,
    author = "Bennett, Jack J. and Buldgen, Gilles and Drewes, Marco and Wong, Yvonne Y.Y.",
    title = "{Towards a precision calculation of the effective number of neutrinos $N_{\rm eff}$ in the Standard Model I: The QED equation of state}",
    eprint = "1911.04504",
    archivePrefix = "arXiv",
    primaryClass = "hep-ph",
    doi = "10.1088/1475-7516/2020/03/003",
    journal = "JCAP",
    volume = "03",
    pages = "003",
    year = "2020"
}

@article{Fields:2019pfx,
    author = "Fields, Brian D. and Olive, Keith A. and Yeh, Tsung-Han and Young, Charles",
    title = "{Big-Bang Nucleosynthesis after Planck}",
    eprint = "1912.01132",
    archivePrefix = "arXiv",
    primaryClass = "astro-ph.CO",
    reportNumber = "UMN--TH--3902/19, FTPI--MINN--19/25",
    doi = "10.1088/1475-7516/2020/03/010",
    journal = "JCAP",
    volume = "03",
    pages = "010",
    year = "2020",
    note = "[Erratum: JCAP 11, E02 (2020)]"
}

@article{Hufnagel:2017dgo,
    author = "Hufnagel, Marco and Schmidt-Hoberg, Kai and Wild, Sebastian",
    title = "{BBN constraints on MeV-scale dark sectors. Part I. Sterile decays}",
    eprint = "1712.03972",
    archivePrefix = "arXiv",
    primaryClass = "hep-ph",
    reportNumber = "DESY-17-211",
    doi = "10.1088/1475-7516/2018/02/044",
    journal = "JCAP",
    volume = "02",
    pages = "044",
    year = "2018"
}

@article{Irsic:2017ixq,
    author = "Ir\v{s}i\v{c}, Vid and others",
    title = "{New Constraints on the free-streaming of warm dark matter from intermediate and small scale Lyman-$\alpha$ forest data}",
    eprint = "1702.01764",
    archivePrefix = "arXiv",
    primaryClass = "astro-ph.CO",
    doi = "10.1103/PhysRevD.96.023522",
    journal = "Phys. Rev. D",
    volume = "96",
    number = "2",
    pages = "023522",
    year = "2017"
}

@article{Slatyer_2010,
	doi = {10.1088/1475-7516/2010/02/028},
	url = {https://doi.org/10.1088/1475-7516/2010/02/028},
	year = 2010,
	month = {feb},
	publisher = {{IOP} Publishing},
	volume = {2010},
	number = {02},
	pages = {028--028},
	author = {Tracy R Slatyer},
	title = {The Sommerfeld enhancement for dark matter with an excited state},
	journal = {Journal of Cosmology and Astroparticle Physics}
}

@article{PhysRevLett.90.225002,
  title = {Scattering in the Attractive Yukawa Potential in the Limit of Strong Interaction},
  author = {Khrapak, S. A. and Ivlev, A. V. and Morfill, G. E. and Zhdanov, S. K.},
  journal = {Phys. Rev. Lett.},
  volume = {90},
  issue = {22},
  pages = {225002},
  numpages = {4},
  year = {2003},
  month = {Jun},
  publisher = {American Physical Society},
  doi = {10.1103/PhysRevLett.90.225002},
  url = {https://link.aps.org/doi/10.1103/PhysRevLett.90.225002}
}

@article{PhysRevE.70.056405,
  title = {Momentum transfer in complex plasmas},
  author = {Khrapak, Sergey A. and Ivlev, Alexey V. and Morfill, Gregor E.},
  journal = {Phys. Rev. E},
  volume = {70},
  issue = {5},
  pages = {056405},
  numpages = {9},
  year = {2004},
  month = {Nov},
  publisher = {American Physical Society},
  doi = {10.1103/PhysRevE.70.056405},
  url = {https://link.aps.org/doi/10.1103/PhysRevE.70.056405}
}

@article{Cassel_2010,
	doi = {10.1088/0954-3899/37/10/105009},
	url = {https://doi.org/10.1088/0954-3899/37/10/105009},
	year = 2010,
	month = {aug},
	publisher = {{IOP} Publishing},
	volume = {37},
	number = {10},
	pages = {105009},
	author = {S Cassel},
	title = {Sommerfeld factor for arbitrary partial wave processes},
	journal = {Journal of Physics G: Nuclear and Particle Physics}
}

@misc{PlanckPicture,
  title = {ESA - Planck Collaboration},
  howpublished = {\url{https://sci.esa.int/s/WLGmGdw}},
  note = {Accessed: 2021-01-15}
}

@article{Foot:2016wvj,
    author = "Foot, Robert and Vagnozzi, Sunny",
    title = "{Solving the small-scale structure puzzles with dissipative dark matter}",
    eprint = "1602.02467",
    archivePrefix = "arXiv",
    primaryClass = "astro-ph.CO",
    doi = "10.1088/1475-7516/2016/07/013",
    journal = "JCAP",
    volume = "07",
    pages = "013",
    year = "2016"
}

@article{Essig:2010gu,
    author = "Essig, Rouven and Harnik, Roni and Kaplan, Jared and Toro, Natalia",
    title = "{Discovering New Light States at Neutrino Experiments}",
    eprint = "1008.0636",
    archivePrefix = "arXiv",
    primaryClass = "hep-ph",
    reportNumber = "SLAC-PUB-14197, FERMILAB-PUB-10-274-T",
    doi = "10.1103/PhysRevD.82.113008",
    journal = "Phys. Rev. D",
    volume = "82",
    pages = "113008",
    year = "2010"
}

@article{Salucci:1996bf,
    author = "Salucci, P. and Persic, M.",
    title = "{Dark matter halos around galaxies}",
    eprint = "astro-ph/9703027",
    archivePrefix = "arXiv",
    reportNumber = "SISSA-86-97-A",
    journal = "ASP Conf. Ser.",
    volume = "117",
    pages = "1",
    year = "1997"
}

@article{Kirkman:2003uv,
    author = "Kirkman, David and Tytler, David and Suzuki, Nao and O'Meara, John M. and Lubin, Dan",
    title = "{The Cosmological baryon density from the deuterium to hydrogen ratio towards QSO absorption systems: D/H towards Q1243+3047}",
    eprint = "astro-ph/0302006",
    archivePrefix = "arXiv",
    doi = "10.1086/378152",
    journal = "Astrophys. J. Suppl.",
    volume = "149",
    pages = "1",
    year = "2003"
}

@inproceedings{Sarkar:1998gx,
    author = "Sarkar, Subir",
    title = "{Big bang nucleosynthesis: Reprise}",
    booktitle = "{2nd International Heidelberg Conference on Dark Matter in Astro and Particle Physics}",
    eprint = "astro-ph/9903183",
    archivePrefix = "arXiv",
    reportNumber = "OUTP-99-18-P",
    pages = "108--130",
    month = "7",
    year = "1998"
}

@article{Bennett:2003bz,
    author = "Bennett, C.L. and others",
    collaboration = "WMAP",
    title = "{First year Wilkinson Microwave Anisotropy Probe (WMAP) observations: Preliminary maps and basic results}",
    eprint = "astro-ph/0302207",
    archivePrefix = "arXiv",
    doi = "10.1086/377253",
    journal = "Astrophys. J. Suppl.",
    volume = "148",
    pages = "1--27",
    year = "2003"
}

@article{PhysRevLett.60.1797,
  title = {Axions from SN1987A},
  author = {Turner, Michael S.},
  journal = {Phys. Rev. Lett.},
  volume = {60},
  issue = {18},
  pages = {1797--1800},
  numpages = {0},
  year = {1988},
  month = {May},
  publisher = {American Physical Society},
  doi = {10.1103/PhysRevLett.60.1797},
  url = {https://link.aps.org/doi/10.1103/PhysRevLett.60.1797}
}

@article{PhysRevA.40.1185,
  title = {Generalized Laguerre representation: Application to relativistic two-photon decay rates},
  author = {Goldman, S. P.},
  journal = {Phys. Rev. A},
  volume = {40},
  issue = {3},
  pages = {1185--1193},
  numpages = {0},
  year = {1989},
  month = {Aug},
  publisher = {American Physical Society},
  doi = {10.1103/PhysRevA.40.1185},
  url = {https://link.aps.org/doi/10.1103/PhysRevA.40.1185}
}

@article{PhysRevD.36.2201,
  title = {Axions and stars},
  author = {Frieman, Joshua A. and Dimopoulos, Savas and Turner, Michael S.},
  journal = {Phys. Rev. D},
  volume = {36},
  issue = {8},
  pages = {2201--2210},
  numpages = {0},
  year = {1987},
  month = {Oct},
  publisher = {American Physical Society},
  doi = {10.1103/PhysRevD.36.2201},
  url = {https://link.aps.org/doi/10.1103/PhysRevD.36.2201}
}

@article{PhysRevD.39.1020,
  title = {Axions and SN 1987A},
  author = {Burrows, Adam and Turner, Michael S. and Brinkmann, R. P.},
  journal = {Phys. Rev. D},
  volume = {39},
  issue = {4},
  pages = {1020--1028},
  numpages = {0},
  year = {1989},
  month = {Feb},
  publisher = {American Physical Society},
  doi = {10.1103/PhysRevD.39.1020},
  url = {https://link.aps.org/doi/10.1103/PhysRevD.39.1020}
}

@article{Pospelov:2010hj,
    author = "Pospelov, Maxim and Pradler, Josef",
    title = "{Big Bang Nucleosynthesis as a Probe of New Physics}",
    eprint = "1011.1054",
    archivePrefix = "arXiv",
    primaryClass = "hep-ph",
    doi = "10.1146/annurev.nucl.012809.104521",
    journal = "Ann. Rev. Nucl. Part. Sci.",
    volume = "60",
    pages = "539--568",
    year = "2010"
}

@article{Trimble:1987ee,
    author = "Trimble, Virginia",
    title = "{Existence and Nature of Dark Matter in the Universe}",
    doi = "10.1146/annurev.aa.25.090187.002233",
    journal = "Ann. Rev. Astron. Astrophys.",
    volume = "25",
    pages = "425--472",
    year = "1987"
}

@article{Ma:2017ucp,
    author = "Ma, Ernest",
    title = "{Inception of Self-Interacting Dark Matter with Dark Charge Conjugation Symmetry}",
    eprint = "1704.04666",
    archivePrefix = "arXiv",
    primaryClass = "hep-ph",
    reportNumber = "UCRHEP-T576, UCRHEP-T576-(APR-2017)",
    doi = "10.1016/j.physletb.2017.06.067",
    journal = "Phys. Lett. B",
    volume = "772",
    pages = "442--445",
    year = "2017"
}

@article{He:1990pn,
    author = "He, X.G. and Joshi, Girish C. and Lew, H. and Volkas, R.R.",
    title = "{NEW Z-prime PHENOMENOLOGY}",
    reportNumber = "UM-P-90/42, OZ-P-90/16",
    doi = "10.1103/PhysRevD.43.R22",
    journal = "Phys. Rev. D",
    volume = "43",
    pages = "22--24",
    year = "1991"
}

@article{Foot:1990mn,
    author = "Foot, Robert",
    title = "{New Physics From Electric Charge Quantization?}",
    reportNumber = "MAD/TH/90-14",
    doi = "10.1142/S0217732391000543",
    journal = "Mod. Phys. Lett. A",
    volume = "6",
    pages = "527--530",
    year = "1991"
}

@article{He:1991qd,
    author = "He, Xiao-Gang and Joshi, Girish C. and Lew, H. and Volkas, R.R.",
    title = "{Simplest Z-prime model}",
    reportNumber = "CERN-TH-6084-91, UM-P-91-32, OZ-91-07",
    doi = "10.1103/PhysRevD.44.2118",
    journal = "Phys. Rev. D",
    volume = "44",
    pages = "2118--2132",
    year = "1991"
}

@article{Heeck:2011wj,
    author = "Heeck, Julian and Rodejohann, Werner",
    title = "{Gauged~ $L_\mu  -  L_\tau$~ Symmetry~ at~ the~ Electroweak~ Scale}",
    eprint = "1107.5238",
    archivePrefix = "arXiv",
    primaryClass = "hep-ph",
    doi = "10.1103/PhysRevD.84.075007",
    journal = "Phys. Rev. D",
    volume = "84",
    pages = "075007",
    year = "2011"
}

@article{Gninenko:2001hx,
    author = "Gninenko, S.N. and Krasnikov, N.V.",
    title = "{The Muon anomalous magnetic moment and a new light gauge boson}",
    eprint = "hep-ph/0102222",
    archivePrefix = "arXiv",
    doi = "10.1016/S0370-2693(01)00693-1",
    journal = "Phys. Lett. B",
    volume = "513",
    pages = "119",
    year = "2001"
}

@article{Baek:2001kca,
    author = "Baek, Seungwon and Deshpande, N.G. and He, X.G. and Ko, P.",
    title = "{Muon anomalous g-2 and gauged L(muon) - L(tau) models}",
    eprint = "hep-ph/0104141",
    archivePrefix = "arXiv",
    reportNumber = "KAIST-TH-2001-08",
    doi = "10.1103/PhysRevD.64.055006",
    journal = "Phys. Rev. D",
    volume = "64",
    pages = "055006",
    year = "2001"
}

@article{Carone:2013uh,
    author = "Carone, Christopher D.",
    title = "{Flavor-Nonuniversal Dark Gauge Bosons and the Muon g-2}",
    eprint = "1301.2027",
    archivePrefix = "arXiv",
    primaryClass = "hep-ph",
    doi = "10.1016/j.physletb.2013.03.011",
    journal = "Phys. Lett. B",
    volume = "721",
    pages = "118--122",
    year = "2013"
}

@article{Altmannshofer:2014pba,
    author = "Altmannshofer, Wolfgang and Gori, Stefania and Pospelov, Maxim and Yavin, Itay",
    title = "{Neutrino Trident Production: A Powerful Probe of New Physics with Neutrino Beams}",
    eprint = "1406.2332",
    archivePrefix = "arXiv",
    primaryClass = "hep-ph",
    doi = "10.1103/PhysRevLett.113.091801",
    journal = "Phys. Rev. Lett.",
    volume = "113",
    pages = "091801",
    year = "2014"
}

@article{Cirelli:2008pk,
    author = "Cirelli, Marco and Kadastik, Mario and Raidal, Martti and Strumia, Alessandro",
    title = "{Model-independent implications of the e+-, anti-proton cosmic ray spectra on properties of Dark Matter}",
    eprint = "0809.2409",
    archivePrefix = "arXiv",
    primaryClass = "hep-ph",
    reportNumber = "IFUP-TH-2008-27, SACLAY-T08-139",
    doi = "10.1016/j.nuclphysb.2008.11.031",
    journal = "Nucl. Phys. B",
    volume = "813",
    pages = "1--21",
    year = "2009",
    note = "[Addendum: Nucl.Phys.B 873, 530--533 (2013)]"
}

@article{Baek:2008nz,
    author = "Baek, Seungwon and Ko, Pyungwon",
    title = "{Phenomenology of U(1)(L(mu)-L(tau)) charged dark matter at PAMELA and colliders}",
    eprint = "0811.1646",
    archivePrefix = "arXiv",
    primaryClass = "hep-ph",
    doi = "10.1088/1475-7516/2009/10/011",
    journal = "JCAP",
    volume = "10",
    pages = "011",
    year = "2009"
}

@article{Baldes:2017gzu,
    author = "Baldes, Iason and Cirelli, Marco and Panci, Paolo and Petraki, Kalliopi and Sala, Filippo and Taoso, Marco",
    title = "{Asymmetric dark matter: residual annihilations and self-interactions}",
    eprint = "1712.07489",
    archivePrefix = "arXiv",
    primaryClass = "hep-ph",
    reportNumber = "CERN-TH-2017-272, DESY-17-232, NIKHEF-2017-070",
    doi = "10.21468/SciPostPhys.4.6.041",
    journal = "SciPost Phys.",
    volume = "4",
    number = "6",
    pages = "041",
    year = "2018"
}

@article{Garani:2019fpa,
    author = "Garani, Raghuveer and Heeck, Julian",
    title = "{Dark matter interactions with muons in neutron stars}",
    eprint = "1906.10145",
    archivePrefix = "arXiv",
    primaryClass = "hep-ph",
    reportNumber = "ULB-TH/19-05, UCI-TR-2019-17",
    doi = "10.1103/PhysRevD.100.035039",
    journal = "Phys. Rev. D",
    volume = "100",
    number = "3",
    pages = "035039",
    year = "2019"
}

@article{Kamada:2018zxi,
    author = "Kamada, Ayuki and Kaneta, Kunio and Yanagi, Keisuke and Yu, Hai-Bo",
    title = "{Self-interacting dark matter and muon $g-2$ in a gauged U$(1)_{L_{\mu} - L_{\tau}}$ model}",
    eprint = "1805.00651",
    archivePrefix = "arXiv",
    primaryClass = "hep-ph",
    reportNumber = "CTPU-PTC-18-10, UMN-TH-3716/18, FTPI-MINN-18/07, UT-18-10, UMN-TH-3716-18, FTPI-MINN-18-07",
    doi = "10.1007/JHEP06(2018)117",
    journal = "JHEP",
    volume = "06",
    pages = "117",
    year = "2018"
}

@article{Dubinski:1991bm,
    author = "Dubinski, John and Carlberg, R.G.",
    title = "{The Structure of cold dark matter halos}",
    doi = "10.1086/170451",
    journal = "Astrophys. J.",
    volume = "378",
    pages = "496",
    year = "1991"
}

@article{Navarro:1995iw,
    author = "Navarro, Julio F. and Frenk, Carlos S. and White, Simon D.M.",
    title = "{The Structure of cold dark matter halos}",
    eprint = "astro-ph/9508025",
    archivePrefix = "arXiv",
    doi = "10.1086/177173",
    journal = "Astrophys. J.",
    volume = "462",
    pages = "563--575",
    year = "1996"
}

@article{Navarro:1996gj,
    author = "Navarro, Julio F. and Frenk, Carlos S. and White, Simon D.M.",
    title = "{A Universal density profile from hierarchical clustering}",
    eprint = "astro-ph/9611107",
    archivePrefix = "arXiv",
    doi = "10.1086/304888",
    journal = "Astrophys. J.",
    volume = "490",
    pages = "493--508",
    year = "1997"
}

@article{Flores:1994gz,
    author = "Flores, Ricardo A. and Primack, Joel R.",
    title = "{Observational and theoretical constraints on singular dark matter halos}",
    eprint = "astro-ph/9402004",
    archivePrefix = "arXiv",
    reportNumber = "SCIPP-93-01-REV, SCIPP-93-01",
    doi = "10.1086/187350",
    journal = "Astrophys. J. Lett.",
    volume = "427",
    pages = "L1--4",
    year = "1994"
}

@article{Moore:1994yx,
    author = "Moore, B.",
    title = "{Evidence against dissipationless dark matter from observations of galaxy haloes}",
    doi = "10.1038/370629a0",
    journal = "Nature",
    volume = "370",
    pages = "629",
    year = "1994"
}

@article{Moore:1999gc,
    author = "Moore, Ben and Quinn, Thomas R. and Governato, Fabio and Stadel, Joachim and Lake, George",
    title = "{Cold collapse and the core catastrophe}",
    eprint = "astro-ph/9903164",
    archivePrefix = "arXiv",
    doi = "10.1046/j.1365-8711.1999.03039.x",
    journal = "Mon. Not. Roy. Astron. Soc.",
    volume = "310",
    pages = "1147--1152",
    year = "1999"
}

@article{Burkert:1995yz,
    author = "Burkert, A.",
    title = "{The Structure of dark matter halos in dwarf galaxies}",
    eprint = "astro-ph/9504041",
    archivePrefix = "arXiv",
    doi = "10.1086/309560",
    journal = "IAU Symp.",
    volume = "171",
    pages = "175",
    year = "1996"
}

@article{McGaugh:1998tq,
    author = "McGaugh, Stacy S. and de Blok, W.J.G.",
    title = "{Testing the dark matter hypothesis with low surface brightness galaxies and other evidence}",
    eprint = "astro-ph/9801123",
    archivePrefix = "arXiv",
    doi = "10.1086/305612",
    journal = "Astrophys. J.",
    volume = "499",
    pages = "41",
    year = "1998"
}

@article{vandenBosch:2000rza,
    author = "van den Bosch, Frank C. and Swaters, Rob A.",
    title = "{Dwarf galaxy rotation curves and the core problem of dark matter halos}",
    eprint = "astro-ph/0006048",
    archivePrefix = "arXiv",
    doi = "10.1046/j.1365-8711.2001.04456.x",
    journal = "Mon. Not. Roy. Astron. Soc.",
    volume = "325",
    pages = "1017",
    year = "2001"
}

@article{Borriello:2000rv,
    author = "Borriello, Annamaria and Salucci, Paolo",
    title = "{The Dark matter distribution in disk galaxies}",
    eprint = "astro-ph/0001082",
    archivePrefix = "arXiv",
    doi = "10.1046/j.1365-8711.2001.04077.x",
    journal = "Mon. Not. Roy. Astron. Soc.",
    volume = "323",
    pages = "285",
    year = "2001"
}

@article{deBlok:2001rgg,
    author = "de Blok, W.J.G. and McGaugh, Stacy S. and Rubin, Vera C.",
    title = "{High-Resolution Rotation Curves of Low Surface Brightness Galaxies. II. Mass Models}",
    doi = "10.1086/323450",
    journal = "Astron. J.",
    volume = "122",
    pages = "2396--2427",
    year = "2001"
}

@article{deBlok:2001hbg,
    author = "de Blok, W.J.G. and McGaugh, Stacy S. and Bosma, Albert and Rubin, Vera C.",
    title = "{Mass density profiles of LSB galaxies}",
    eprint = "astro-ph/0103102",
    archivePrefix = "arXiv",
    doi = "10.1086/320262",
    journal = "Astrophys. J. Lett.",
    volume = "552",
    pages = "L23--L26",
    year = "2001"
}

@article{Marchesini:2002vm,
    author = "Marchesini, D. and D'Onghia, E. and Chincarini, G. and Firmani, C. and Conconi, P. and Molinari, E. and Zacchei, A.",
    title = "{Halpha rotation curves: the soft core question}",
    eprint = "astro-ph/0202075",
    archivePrefix = "arXiv",
    doi = "10.1086/341475",
    journal = "Astrophys. J.",
    volume = "575",
    pages = "801--813",
    year = "2002"
}

@article{Gentile:2005de,
    author = "Gentile, Gianfranco and Burkert, A. and Salucci, P. and Klein, U. and Walter, F.",
    title = "{The dwarf galaxy DDO 47 as a dark matter laboratory: testing cusps hiding in triaxial halos}",
    eprint = "astro-ph/0506538",
    archivePrefix = "arXiv",
    doi = "10.1086/498939",
    journal = "Astrophys. J. Lett.",
    volume = "634",
    pages = "L145--L148",
    year = "2005"
}

@article{Gentile:2006hv,
    author = "Gentile, Gianfranco and Salucci, Paolo and Klein, Uli and Granato, Gian Luigi",
    title = "{NGC 3741: Dark halo profile from the most extended rotation curve}",
    eprint = "astro-ph/0611355",
    archivePrefix = "arXiv",
    doi = "10.1111/j.1365-2966.2006.11283.x",
    journal = "Mon. Not. Roy. Astron. Soc.",
    volume = "375",
    pages = "199--212",
    year = "2007"
}

@article{KuziodeNaray:2006wh,
    author = "Kuzio de Naray, Rachel and McGaugh, Stacy S. and de Blok, W.J.G. and Bosma, A.",
    title = "{High Resolution Optical Velocity Fields of 11 Low Surface Brightness Galaxies}",
    eprint = "astro-ph/0604576",
    archivePrefix = "arXiv",
    doi = "10.1086/505345",
    journal = "Astrophys. J. Suppl.",
    volume = "165",
    pages = "461--479",
    year = "2006"
}

@article{KuziodeNaray:2007qi,
    author = "Kuzio de Naray, Rachel and McGaugh, Stacy S. and de Blok, W.J.G.",
    title = "{Mass Models for Low Surface Brightness Galaxies with High Resolution Optical Velocity Fields}",
    eprint = "0712.0860",
    archivePrefix = "arXiv",
    primaryClass = "astro-ph",
    doi = "10.1086/527543",
    journal = "Astrophys. J.",
    volume = "676",
    pages = "920--943",
    year = "2008"
}

@article{Salucci:2007tm,
    author = "Salucci, Paolo and Lapi, A. and Tonini, C. and Gentile, G. and Yegorova, I. and Klein, U.",
    title = "{The Universal Rotation Curve of Spiral Galaxies. 2. The Dark Matter Distribution out to the Virial Radius}",
    eprint = "astro-ph/0703115",
    archivePrefix = "arXiv",
    doi = "10.1111/j.1365-2966.2007.11696.x",
    journal = "Mon. Not. Roy. Astron. Soc.",
    volume = "378",
    pages = "41--47",
    year = "2007"
}

@article{Bullock:1999he,
    author = "Bullock, James S. and Kolatt, Tsafrir S. and Sigad, Yair and Somerville, Rachel S. and Kravtsov, Andrey V. and Klypin, Anatoly A. and Primack, Joel R. and Dekel, Avishai",
    title = "{Profiles of dark haloes. Evolution, scatter, and environment}",
    eprint = "astro-ph/9908159",
    archivePrefix = "arXiv",
    doi = "10.1046/j.1365-8711.2001.04068.x",
    journal = "Mon. Not. Roy. Astron. Soc.",
    volume = "321",
    pages = "559--575",
    year = "2001"
}

@article{Oman:2015xda,
    author = "Oman, Kyle A. and others",
    title = "{The unexpected diversity of dwarf galaxy rotation curves}",
    eprint = "1504.01437",
    archivePrefix = "arXiv",
    primaryClass = "astro-ph.GA",
    doi = "10.1093/mnras/stv1504",
    journal = "Mon. Not. Roy. Astron. Soc.",
    volume = "452",
    number = "4",
    pages = "3650--3665",
    year = "2015"
}

@article{deNaray:2009xj,
    author = "Kuzio de Naray, Rachel and Martinez, Gregory D. and Bullock, James S. and Kaplinghat, Manoj",
    title = "{The Case Against Warm or Self-Interacting Dark Matter as Explanations for Cores in Low Surface Brightness Galaxies}",
    eprint = "0912.3518",
    archivePrefix = "arXiv",
    primaryClass = "astro-ph.CO",
    doi = "10.1088/2041-8205/710/2/L161",
    journal = "Astrophys. J. Lett.",
    volume = "710",
    pages = "L161",
    year = "2010"
}

@article{Kauffmann:1993gv,
    author = "Kauffmann, G and White, Simon D.M. and Guiderdoni, B.",
    title = "{The Formation and Evolution of Galaxies Within Merging Dark Matter Haloes}",
    journal = "Mon. Not. Roy. Astron. Soc.",
    volume = "264",
    pages = "201",
    year = "1993"
}

@article{Moore:1999nt,
    author = "Moore, B. and Ghigna, S. and Governato, F. and Lake, G. and Quinn, Thomas R. and Stadel, J. and Tozzi, P.",
    title = "{Dark matter substructure within galactic halos}",
    eprint = "astro-ph/9907411",
    archivePrefix = "arXiv",
    doi = "10.1086/312287",
    journal = "Astrophys. J. Lett.",
    volume = "524",
    pages = "L19--L22",
    year = "1999"
}

@article{Klypin:1999uc,
    author = "Klypin, Anatoly A. and Kravtsov, Andrey V. and Valenzuela, Octavio and Prada, Francisco",
    title = "{Where are the missing Galactic satellites?}",
    eprint = "astro-ph/9901240",
    archivePrefix = "arXiv",
    doi = "10.1086/307643",
    journal = "Astrophys. J.",
    volume = "522",
    pages = "82--92",
    year = "1999"
}

@article{Zavala:2009ms,
    author = "Zavala, J. and Jing, Y.P. and Faltenbacher, A. and Yepes, G. and Hoffman, Y. and Gottlober, S. and Catinella, B.",
    title = "{The velocity function in the local environment from LCDM and LWDM constrained simulations}",
    eprint = "0906.0585",
    archivePrefix = "arXiv",
    primaryClass = "astro-ph.CO",
    doi = "10.1088/0004-637X/700/2/1779",
    journal = "Astrophys. J.",
    volume = "700",
    pages = "1779--1793",
    year = "2009"
}

@article{Zwaan:2009dz,
    author = "Zwaan, Martin A. and Meyer, Martin J. and Staveley-Smith, Lister",
    title = "{The velocity function of gas-rich galaxies}",
    eprint = "0912.1754",
    archivePrefix = "arXiv",
    primaryClass = "astro-ph.CO",
    doi = "10.1111/j.1365-2966.2009.16188.x",
    journal = "Mon. Not. Roy. Astron. Soc.",
    volume = "403",
    pages = "1969",
    year = "2010"
}

@article{BoylanKolchin:2011de,
    author = "Boylan-Kolchin, Michael and Bullock, James S. and Kaplinghat, Manoj",
    title = "{Too big to fail? The puzzling darkness of massive Milky Way subhaloes}",
    eprint = "1103.0007",
    archivePrefix = "arXiv",
    primaryClass = "astro-ph.CO",
    doi = "10.1111/j.1745-3933.2011.01074.x",
    journal = "Mon. Not. Roy. Astron. Soc.",
    volume = "415",
    pages = "L40",
    year = "2011"
}

@article{BoylanKolchin:2011dk,
    author = "Boylan-Kolchin, Michael and Bullock, James S. and Kaplinghat, Manoj",
    title = "{The Milky Way's bright satellites as an apparent failure of LCDM}",
    eprint = "1111.2048",
    archivePrefix = "arXiv",
    primaryClass = "astro-ph.CO",
    doi = "10.1111/j.1365-2966.2012.20695.x",
    journal = "Mon. Not. Roy. Astron. Soc.",
    volume = "422",
    pages = "1203--1218",
    year = "2012"
}

@article{Tollerud:2014zha,
    author = "Tollerud, Erik J. and Boylan-Kolchin, Michael and Bullock, James S.",
    title = "{M31 Satellite Masses Compared to LCDM Subhaloes}",
    eprint = "1403.6469",
    archivePrefix = "arXiv",
    primaryClass = "astro-ph.GA",
    doi = "10.1093/mnras/stu474",
    journal = "Mon. Not. Roy. Astron. Soc.",
    volume = "440",
    number = "4",
    pages = "3511--3519",
    year = "2014"
}

@article{Garrison-Kimmel:2014vqa,
    author = "Garrison-Kimmel, Shea and Boylan-Kolchin, Michael and Bullock, James S. and Kirby, Evan N.",
    title = "{Too Big to Fail in the Local Group}",
    eprint = "1404.5313",
    archivePrefix = "arXiv",
    primaryClass = "astro-ph.GA",
    doi = "10.1093/mnras/stu1477",
    journal = "Mon. Not. Roy. Astron. Soc.",
    volume = "444",
    number = "1",
    pages = "222--236",
    year = "2014"
}

@article{PhysRev.104.1466,
  title = {Inelastic and Elastic Scattering of 187-Mev Electrons from Selected Even-Even Nuclei},
  author = {Helm, Richard H.},
  journal = {Phys. Rev.},
  volume = {104},
  issue = {5},
  pages = {1466--1475},
  numpages = {0},
  year = {1956},
  month = {Dec},
  publisher = {American Physical Society},
  doi = {10.1103/PhysRev.104.1466},
  url = {https://link.aps.org/doi/10.1103/PhysRev.104.1466}
}

@article{Bondarenko:2019vrb,
    author = "Bondarenko, Kyrylo and Boyarsky, Alexey and Bringmann, Torsten and Hufnagel, Marco and Schmidt-Hoberg, Kai and Sokolenko, Anastasia",
    title = "{Direct detection and complementary constraints for sub-GeV dark matter}",
    eprint = "1909.08632",
    archivePrefix = "arXiv",
    primaryClass = "hep-ph",
    reportNumber = "DESY-19-140",
    doi = "10.1007/JHEP03(2020)118",
    journal = "JHEP",
    volume = "03",
    pages = "118",
    year = "2020"
}

@article{Khachatryan:2016vau,
    author = "Aad, Georges and others",
    collaboration = "ATLAS, CMS",
    title = "{Measurements of the Higgs boson production and decay rates and constraints on its couplings from a combined ATLAS and CMS analysis of the LHC pp collision data at $ \sqrt{s}=7 $ and 8 TeV}",
    eprint = "1606.02266",
    archivePrefix = "arXiv",
    primaryClass = "hep-ex",
    reportNumber = "CERN-EP-2016-100, ATLAS-HIGG-2015-07, CMS-HIG-15-002",
    doi = "10.1007/JHEP08(2016)045",
    journal = "JHEP",
    volume = "08",
    pages = "045",
    year = "2016"
}

@article{Sirunyan:2018owy,
    author = "Sirunyan, Albert M and others",
    collaboration = "CMS",
    title = "{Search for invisible decays of a Higgs boson produced through vector boson fusion in proton-proton collisions at $\sqrt{s} =$ 13 TeV}",
    eprint = "1809.05937",
    archivePrefix = "arXiv",
    primaryClass = "hep-ex",
    reportNumber = "CMS-HIG-17-023, CERN-EP-2018-139",
    doi = "10.1016/j.physletb.2019.04.025",
    journal = "Phys. Lett. B",
    volume = "793",
    pages = "520--551",
    year = "2019"
}

@article{Kolb:2017jvz,
      author         = "Kolb, Edward W. and Long, Andrew J.",
      title          = "{Superheavy dark matter through Higgs portal operators}",
      journal        = "Phys. Rev.",
      volume         = "D96",
      year           = "2017",
      number         = "10",
      pages          = "103540",
      doi            = "10.1103/PhysRevD.96.103540",
      eprint         = "1708.04293",
      archivePrefix  = "arXiv",
      primaryClass   = "astro-ph.CO",
      SLACcitation   = "%%CITATION = ARXIV:1708.04293;%%"
}

@article{Kolb:1998ki,
      author         = "Kolb, Edward W. and Chung, Daniel J. H. and Riotto,
                        Antonio",
      title          = "{WIMPzillas!}",
      booktitle      = "{Trends in theoretical physics II. Proceedings, 2nd La
                        Plata Meeting, Buenos Aires, Argentina, November
                        29-December 4, 1998}",
      journal        = "AIP Conf. Proc.",
      volume         = "484",
      year           = "1999",
      number         = "1",
      pages          = "91-105",
      doi            = "10.1063/1.59655",
      note           = "[,592(1999)]",
      eprint         = "hep-ph/9810361",
      archivePrefix  = "arXiv",
      primaryClass   = "hep-ph",
      reportNumber   = "FERMILAB-CONF-98-325-A",
      SLACcitation   = "%%CITATION = HEP-PH/9810361;%%"
}

@article{Chung:1998zb,
      author         = "Chung, Daniel J. H. and Kolb, Edward W. and Riotto,
                        Antonio",
      title          = "{Superheavy dark matter}",
      journal        = "Phys. Rev.",
      volume         = "D59",
      year           = "1998",
      pages          = "023501",
      doi            = "10.1103/PhysRevD.59.023501",
      eprint         = "hep-ph/9802238",
      archivePrefix  = "arXiv",
      primaryClass   = "hep-ph",
      reportNumber   = "FERMILAB-PUB-98-021-A, CERN-TH-98-37, OUTP-98-02-P,
                        FERMILAB-FERMILAB-PUB-98-021-A, CERN-CERN-TH-98-37,
                        OXFORD-U. --OUTP-98-02-P",
      SLACcitation   = "%%CITATION = HEP-PH/9802238;%%"
}

@article{Chway:2019kft,
      author         = "Chway, Dongjin and Jung, Tae Hyun and Shin, Chang Sub",
      title          = "{Dark matter filtering-out effect during a first-order
                        phase transition}",
      year           = "2019",
      eprint         = "1912.04238",
      archivePrefix  = "arXiv",
      primaryClass   = "hep-ph",
      reportNumber   = "CTPU-19-37",
      SLACcitation   = "%%CITATION = ARXIV:1912.04238;%%"
}

@article{Sigurdson:2009uz,
      author         = "Sigurdson, Kris",
      title          = "{Hidden Hot Dark Matter as Cold Dark Matter}",
      year           = "2009",
      eprint         = "0912.2346",
      archivePrefix  = "arXiv",
      primaryClass   = "astro-ph.CO",
      SLACcitation   = "%%CITATION = ARXIV:0912.2346;%%"
}

@article{Baker:2019ndr,
      author         = "Baker, Michael J. and Kopp, Joachim and Long, Andrew J.",
      title          = "{Filtered Dark Matter at a First Order Phase Transition}",
      year           = "2019",
      eprint         = "1912.02830",
      archivePrefix  = "arXiv",
      primaryClass   = "hep-ph",
      SLACcitation   = "%%CITATION = ARXIV:1912.02830;%%"
}

@article{Randall:2015xza,
      author         = "Randall, Lisa and Scholtz, Jakub and Unwin, James",
      title          = "{Flooded Dark Matter and S Level Rise}",
      journal        = "JHEP",
      volume         = "03",
      year           = "2016",
      pages          = "011",
      doi            = "10.1007/JHEP03(2016)011",
      eprint         = "1509.08477",
      archivePrefix  = "arXiv",
      primaryClass   = "hep-ph",
      SLACcitation   = "%%CITATION = ARXIV:1509.08477;%%"
}

@article{Tenkanen:2016jic,
      author         = "Tenkanen, Tommi and Vaskonen, Ville",
      title          = "{Reheating the Standard Model from a hidden sector}",
      journal        = "Phys. Rev.",
      volume         = "D94",
      year           = "2016",
      number         = "8",
      pages          = "083516",
      doi            = "10.1103/PhysRevD.94.083516",
      eprint         = "1606.00192",
      archivePrefix  = "arXiv",
      primaryClass   = "astro-ph.CO",
      reportNumber   = "HIP-2016-18-TH",
      SLACcitation   = "%%CITATION = ARXIV:1606.00192;%%"
}

@article{Harigaya:2016nlg,
      author         = "Harigaya, Keisuke and Ibe, Masahiro and Kaneta, Kunio and
                        Nakano, Wakutaka and Suzuki, Motoo",
      title          = "{Thermal Relic Dark Matter Beyond the Unitarity Limit}",
      journal        = "JHEP",
      volume         = "08",
      year           = "2016",
      pages          = "151",
      doi            = "10.1007/JHEP08(2016)151",
      eprint         = "1606.00159",
      archivePrefix  = "arXiv",
      primaryClass   = "hep-ph",
      reportNumber   = "IPMU16-0071, CTPU-16-14",
      SLACcitation   = "%%CITATION = ARXIV:1606.00159;%%"
}

@article{Bramante:2017obj,
      author         = "Bramante, Joseph and Unwin, James",
      title          = "{Superheavy Thermal Dark Matter and Primordial
                        Asymmetries}",
      journal        = "JHEP",
      volume         = "02",
      year           = "2017",
      pages          = "119",
      doi            = "10.1007/JHEP02(2017)119",
      eprint         = "1701.05859",
      archivePrefix  = "arXiv",
      primaryClass   = "hep-ph",
      SLACcitation   = "%%CITATION = ARXIV:1701.05859;%%"
}

@article{Blanco:2017sbc,
      author         = "Blanco, Carlos and Harding, J. Patrick and Hooper, Dan",
      title          = "{Novel Gamma-Ray Signatures of PeV-Scale Dark Matter}",
      journal        = "JCAP",
      volume         = "1804",
      year           = "2018",
      number         = "04",
      pages          = "060",
      doi            = "10.1088/1475-7516/2018/04/060",
      eprint         = "1712.02805",
      archivePrefix  = "arXiv",
      primaryClass   = "hep-ph",
      reportNumber   = "FERMILAB-PUB-17-568-A, LA-UR-17-31003",
      SLACcitation   = "%%CITATION = ARXIV:1712.02805;%%"
}

@article{Cirelli:2018iax,
      author         = "Cirelli, Marco and Gouttenoire, Yann and Petraki,
                        Kalliopi and Sala, Filippo",
      title          = "{Homeopathic Dark Matter, or how diluted heavy substances
                        produce high energy cosmic rays}",
      journal        = "JCAP",
      volume         = "1902",
      year           = "2019",
      pages          = "014",
      doi            = "10.1088/1475-7516/2019/02/014",
      eprint         = "1811.03608",
      archivePrefix  = "arXiv",
      primaryClass   = "hep-ph",
      SLACcitation   = "%%CITATION = ARXIV:1811.03608;%%"
}

@article{Heurtier:2019beu,
      author         = "Heurtier, Lucien and Partouche, Hervé",
      title          = "{Spontaneous Freeze Out of Dark Matter From an Early
                        Thermal Phase Transition}",
      journal        = "Phys. Rev.",
      volume         = "D101",
      year           = "2020",
      number         = "4",
      pages          = "043527",
      doi            = "10.1103/PhysRevD.101.043527",
      eprint         = "1912.02828",
      archivePrefix  = "arXiv",
      primaryClass   = "hep-ph",
      reportNumber   = "CPHT-RR065.112019",
      SLACcitation   = "%%CITATION = ARXIV:1912.02828;%%"
}

@article{Heurtier:2019eou,
      author         = "Heurtier, Lucien and Huang, Fei",
      title          = "{Inflaton portal to a highly decoupled EeV dark matter
                        particle}",
      journal        = "Phys. Rev.",
      volume         = "D100",
      year           = "2019",
      number         = "4",
      pages          = "043507",
      doi            = "10.1103/PhysRevD.100.043507",
      eprint         = "1905.05191",
      archivePrefix  = "arXiv",
      primaryClass   = "hep-ph",
      SLACcitation   = "%%CITATION = ARXIV:1905.05191;%%"
}

@article{Davoudiasl:2019xeb,
      author         = "Davoudiasl, Hooman and Mohlabeng, Gopolang",
      title          = "{Getting a THUMP from a WIMP}",
      year           = "2019",
      eprint         = "1912.05572",
      archivePrefix  = "arXiv",
      primaryClass   = "hep-ph",
      SLACcitation   = "%%CITATION = ARXIV:1912.05572;%%"
}

@article{Berlin:2017ife,
      author         = "Berlin, Asher",
      title          = "{WIMPs with GUTs: Dark Matter Coannihilation with a
                        Lighter Species}",
      journal        = "Phys. Rev. Lett.",
      volume         = "119",
      year           = "2017",
      pages          = "121801",
      doi            = "10.1103/PhysRevLett.119.121801",
      eprint         = "1704.08256",
      archivePrefix  = "arXiv",
      primaryClass   = "hep-ph",
      reportNumber   = "SLAC-PUB-16953",
      SLACcitation   = "%%CITATION = ARXIV:1704.08256;%%"
}

@article{Kim:2019udq,
      author         = "Kim, Hyungjin and Kuflik, Eric",
      title          = "{Superheavy Thermal Dark Matter}",
      journal        = "Phys. Rev. Lett.",
      volume         = "123",
      year           = "2019",
      number         = "19",
      pages          = "191801",
      doi            = "10.1103/PhysRevLett.123.191801",
      eprint         = "1906.00981",
      archivePrefix  = "arXiv",
      primaryClass   = "hep-ph",
      SLACcitation   = "%%CITATION = ARXIV:1906.00981;%%"
}

@article{Berlin:2016vnh,
      author         = "Berlin, Asher and Hooper, Dan and Krnjaic, Gordan",
      title          = "{PeV-Scale Dark Matter as a Thermal Relic of a Decoupled
                        Sector}",
      journal        = "Phys. Lett.",
      volume         = "B760",
      year           = "2016",
      pages          = "106-111",
      doi            = "10.1016/j.physletb.2016.06.037",
      eprint         = "1602.08490",
      archivePrefix  = "arXiv",
      primaryClass   = "hep-ph",
      reportNumber   = "FERMILAB-PUB-16-040-A",
      SLACcitation   = "%%CITATION = ARXIV:1602.08490;%%"
}

@article{Berlin:2016gtr,
      author         = "Berlin, Asher and Hooper, Dan and Krnjaic, Gordan",
      title          = "{Thermal Dark Matter From A Highly Decoupled Sector}",
      journal        = "Phys. Rev.",
      volume         = "D94",
      year           = "2016",
      number         = "9",
      pages          = "095019",
      doi            = "10.1103/PhysRevD.94.095019",
      eprint         = "1609.02555",
      archivePrefix  = "arXiv",
      primaryClass   = "hep-ph",
      reportNumber   = "FERMILAB-PUB-16-318-A",
      SLACcitation   = "%%CITATION = ARXIV:1609.02555;%%"
}

@article{Hambye_2020,
	doi = {10.1088/1475-7516/2020/05/001},
	url = {https://doi.org/10.1088/1475-7516/2020/05/001},
	year = 2020,
	month = {may},
	publisher = {{IOP} Publishing},
	volume = {2020},
	number = {05},
	pages = {001--001},
	author = {Thomas Hambye and Laurent Vanderheyden},
	title = {Minimal self-interacting dark matter models with light mediator},
	journal = {Journal of Cosmology and Astroparticle Physics}
}

@article{Vikhlinin:2005mp,
    author = "Vikhlinin, Alexey and Kravtsov, A. and Forman, W. and Jones, C. and Markevitch, M. and Murray, S.S. and Van Speybroeck, L.",
    title = "{Chandra sample of nearby relaxed galaxy clusters: Mass, gas fraction, and mass-temperature relation}",
    eprint = "astro-ph/0507092",
    archivePrefix = "arXiv",
    doi = "10.1086/500288",
    journal = "Astrophys. J.",
    volume = "640",
    pages = "691--709",
    year = "2006"
}

@article{Albert:2016emp,
    author = "Albert, A. and others",
    title = "{Results from the search for dark matter in the Milky Way with 9 years of data of the ANTARES neutrino telescope}",
    eprint = "1612.04595",
    archivePrefix = "arXiv",
    primaryClass = "astro-ph.HE",
    doi = "10.1016/j.physletb.2017.03.063",
    journal = "Phys. Lett. B",
    volume = "769",
    pages = "249--254",
    year = "2017",
    note = "[Erratum: Phys.Lett.B 796, 253--255 (2019)]"
}

@article{Clowe:2006eq,
    author = "Clowe, Douglas and Bradac, Marusa and Gonzalez, Anthony H. and Markevitch, Maxim and Randall, Scott W. and Jones, Christine and Zaritsky, Dennis",
    title = "{A direct empirical proof of the existence of dark matter}",
    eprint = "astro-ph/0608407",
    archivePrefix = "arXiv",
    reportNumber = "SLAC-PUB-12078",
    doi = "10.1086/508162",
    journal = "Astrophys. J. Lett.",
    volume = "648",
    pages = "L109--L113",
    year = "2006"
}

@article{1991MNRAS.249..523B,
       author = {{Begeman}, K.~G. and {Broeils}, A.~H. and {Sanders}, R.~H.},
        title = "{Extended rotation curves of spiral galaxies : dark haloes and modified dynamics.}",
      journal = {mnras},
         year = 1991,
        month = apr,
       volume = {249},
        pages = {523},
          doi = {10.1093/mnras/249.3.523},
       adsurl = {https://ui.adsabs.harvard.edu/abs/1991MNRAS.249..523B}
}

@article{Zwicky:1933gu,
    author = "Zwicky, F.",
    title = "{Die Rotverschiebung von extragalaktischen Nebeln}",
    doi = "10.1007/s10714-008-0707-4",
    journal = "Helv. Phys. Acta",
    volume = "6",
    pages = "110--127",
    year = "1933"
}

@article{1970ApJ...159..379R,
       author = {{Rubin}, Vera C. and {Ford}, W. Kent, Jr.},
        title = "{Rotation of the Andromeda Nebula from a Spectroscopic Survey of Emission Regions}",
      journal = {apj},
         year = 1970,
        month = feb,
       volume = {159},
        pages = {379},
          doi = {10.1086/150317},
       adsurl = {https://ui.adsabs.harvard.edu/abs/1970ApJ...159..379R}
}

@article{Aghanim:2018eyx,
      author         = "Aghanim, N. and others",
      title          = "{Planck 2018 results. VI. Cosmological parameters}",
      collaboration  = "Planck",
      year           = "2018",
      eprint         = "1807.06209",
      archivePrefix  = "arXiv",
      primaryClass   = "astro-ph.CO",
      SLACcitation   = "%%CITATION = ARXIV:1807.06209;%%"
}

@article{Abdallah_2016,
archivePrefix = {arXiv},
arxivId = {1607.08142},
author = {Abdallah, H. and Abramowski, A. and Aharonian, F. and {Ait Benkhali}, F. and Akhperjanian, A. G. and Ang{\"{u}}ner, E. and Arrieta, M. and Aubert, P. and Backes, M. and Balzer, A. and Barnard, M. and Becherini, Y. and {Becker Tjus}, J. and Berge, D. and Bernhard, S. and Bernl{\"{o}}hr, K. and Birsin, E. and Blackwell, R. and B{\"{o}}ttcher, M. and Boisson, C. and Bolmont, J. and Bordas, P. and Bregeon, J. and Brun, F. and Brun, P. and Bryan, M. and Bulik, T. and Capasso, M. and Carr, J. and Casanova, S. and Chakraborty, N. and Chalme-Calvet, R. and Chaves, R. C.G. and Chen, A. and Chevalier, J. and Chr{\'{e}}tien, M. and Colafrancesco, S. and Cologna, G. and Condon, B. and Conrad, J. and Couturier, C. and Cui, Y. and Davids, I. D. and Degrange, B. and Deil, C. and Dewilt, P. and Djannati-Ata{\"{i}}, A. and Domainko, W. and Donath, A. and Drury, L. O. and Dubus, G. and Dutson, K. and Dyks, J. and Dyrda, M. and Edwards, T. and Egberts, K. and Eger, P. and Ernenwein, J. P. and Eschbach, S. and Farnier, C. and Fegan, S. and Fernandes, M. V. and Fiasson, A. and Fontaine, G. and F{\"{o}}rster, A. and Funk, S. and F{\"{u}}{\ss}ling, M. and Gabici, S. and Gajdus, M. and Gallant, Y. A. and Garrigoux, T. and Giavitto, G. and Giebels, B. and Glicenstein, J. F. and Gottschall, D. and Goyal, A. and Grondin, M. H. and Grudzi{\'{n}}ska, M. and Hadasch, D. and Hahn, J. and Hawkes, J. and Heinzelmann, G. and Henri, G. and Hermann, G. and Hervet, O. and Hillert, A. and Hinton, J. A. and Hofmann, W. and Hoischen, C. and Holler, M. and Horns, D. and Ivascenko, A. and Jacholkowska, A. and Jamrozy, M. and Janiak, M. and Jankowsky, D. and Jankowsky, F. and Jingo, M. and Jogler, T. and Jouvin, L. and Jung-Richardt, I. and Kastendieck, M. A. and Katarzy{\'{n}}ski, K. and Katz, U. and Kerszberg, D. and Kh{\'{e}}lifi, B. and Kieffer, M. and King, J. and Klepser, S. and Klochkov, D. and Klu{\'{z}}niak, W. and Kolitzus, D. and Komin, Nu and Kosack, K. and Krakau, S. and Kraus, M. and Krayzel, F. and Kr{\"{u}}ger, P. P. and Laffon, H. and Lamanna, G. and Lau, J. and Lees, J. P. and Lefaucheur, J. and Lefranc, V. and Lemi{\`{e}}re, A. and Lemoine-Goumard, M. and Lenain, J. P. and Leser, E. and Lohse, T. and Lorentz, M. and Lui, R. and Lypova, I. and Marandon, V. and Marcowith, A. and Mariaud, C. and Marx, R. and Maurin, G. and Maxted, N. and Mayer, M. and Meintjes, P. J. and Menzler, U. and Meyer, M. and Mitchell, A. M.W. and Moderski, R. and Mohamed, M. and Mor{\aa}, K. and Moulin, E. and Murach, T. and {De Naurois}, M. and Niederwanger, F. and Niemiec, J. and Oakes, L. and Odaka, H. and Ohm, S. and {\"{O}}ttl, S. and Ostrowski, M. and Oya, I. and Padovani, M. and Panter, M. and Parsons, R. D. and {Paz Arribas}, M. and Pekeur, N. W. and Pelletier, G. and Petrucci, P. O. and Peyaud, B. and Pita, S. and Poon, H. and Prokhorov, D. and Prokoph, H. and P{\"{u}}hlhofer, G. and Punch, M. and Quirrenbach, A. and Raab, S. and Reimer, A. and Reimer, O. and Renaud, M. and {De Los Reyes}, R. and Rieger, F. and Romoli, C. and Rosier-Lees, S. and Rowell, G. and Rudak, B. and Rulten, C. B. and Sahakian, V. and Salek, D. and Sanchez, D. A. and Santangelo, A. and Sasaki, M. and Schlickeiser, R. and Sch{\"{u}}ssler, F. and Schulz, A. and Schwanke, U. and Schwemmer, S. and Seyffert, A. S. and Shafi, N. and Simoni, R. and Sol, H. and Spanier, F. and Spengler, G. and Spie{\ss}, F. and Stawarz, L. and Steenkamp, R. and Stegmann, C. and Stinzing, F. and Stycz, K. and Sushch, I. and Tavernet, J. P. and Tavernier, T. and Taylor, A. M. and Terrier, R. and Tluczykont, M. and Trichard, C. and Tuffs, R. and {Van Der Walt}, J. and {Van Eldik}, C. and {Van Soelen}, B. and Vasileiadis, G. and Veh, J. and Venter, C. and Viana, A. and Vincent, P. and Vink, J. and Voisin, F. and V{\"{o}}lk, H. J. and Vuillaume, T. and Wadiasingh, Z. and Wagner, S. J. and Wagner, P. and Wagner, R. M. and White, R. and Wierzcholska, A. and Willmann, P. and W{\"{o}}rnlein, A. and Wouters, D. and Yang, R. and Zabalza, V. and Zaborov, D. and Zacharias, M. and Zdziarski, A. A. and Zech, A. and Zefi, F. and Ziegler, A. and {\.{Z}}ywucka, N.},
doi = {10.1103/PhysRevLett.117.111301},
eprint = {1607.08142},
issn = {10797114},
journal = {Physical Review Letters},
month = {sep},
number = {11},
publisher = {American Physical Society (APS)},
title = {{Search for Dark Matter Annihilations towards the Inner Galactic Halo from 10 Years of Observations with H.E.S.S.}},
url = {http://dx.doi.org/10.1103/PhysRevLett.117.111301},
volume = {117},
year = {2016}
}

@article{Abramowski_2013,
archivePrefix = {arXiv},
arxivId = {1301.1173},
author = {Abramowski, A. and Acero, F. and Aharonian, F. and Akhperjanian, A. G. and Anton, G. and Balenderan, S. and Balzer, A. and Barnacka, A. and Becherini, Y. and {Becker Tjus}, J. and Bernl{\"{o}}hr, K. and Birsin, E. and Biteau, J. and Bochow, A. and Boisson, C. and Bolmont, J. and Bordas, P. and Brucker, J. and Brun, F. and Brun, P. and Bulik, T. and Carrigan, S. and Casanova, S. and Cerruti, M. and Chadwick, P. M. and Chaves, R. C.G. and Cheesebrough, A. and Colafrancesco, S. and Cologna, G. and Conrad, J. and Couturier, C. and Dalton, M. and Daniel, M. K. and Davids, I. D. and Degrange, B. and Deil, C. and Dewilt, P. and Dickinson, H. J. and Djannati-Ata{\"{i}}, A. and Domainko, W. and Drury, L. O.C. and Dubus, G. and Dutson, K. and Dyks, J. and Dyrda, M. and Egberts, K. and Eger, P. and Espigat, P. and Fallon, L. and Farnier, C. and Fegan, S. and Feinstein, F. and Fernandes, M. V. and Fernandez, D. and Fiasson, A. and Fontaine, G. and F{\"{o}}rster, A. and F{\"{u}}{\ss}ling, M. and Gajdus, M. and Gallant, Y. A. and Garrigoux, T. and Gast, H. and Giebels, B. and Glicenstein, J. F. and Gl{\"{u}}ck, B. and G{\"{o}}ring, D. and Grondin, M. H. and H{\"{a}}ffner, S. and Hague, J. D. and Hahn, J. and Hampf, D. and Harris, J. and Heinz, S. and Heinzelmann, G. and Henri, G. and Hermann, G. and Hillert, A. and Hinton, J. A. and Hofmann, W. and Hofverberg, P. and Holler, M. and Horns, D. and Jacholkowska, A. and Jahn, C. and Jamrozy, M. and Jung, I. and Kastendieck, M. A. and Katarzy{\'{n}}ski, K. and Katz, U. and Kaufmann, S. and Kh{\'{e}}lifi, B. and Klepser, S. and Klochkov, D. and Klu{\aa}niak, W. and Kneiske, T. and Komin, Nu and Kosack, K. and Kossakowski, R. and Krayzel, F. and Kr{\"{u}}ger, P. P. and Laffon, H. and Lamanna, G. and Lefaucheur, J. and Lemoine-Goumard, M. and Lenain, J. P. and Lennarz, D. and Lohse, T. and Lopatin, A. and Lu, C. C. and Marandon, V. and Marcowith, A. and Masbou, J. and Maurin, G. and Maxted, N. and Mayer, M. and McComb, T. J.L. and Medina, M. C. and M{\'{e}}hault, J. and Menzler, U. and Moderski, R. and Mohamed, M. and Moulin, E. and Naumann, C. L. and Naumann-Godo, M. and {De Naurois}, M. and Nedbal, D. and Nekrassov, D. and Nguyen, N. and Niemiec, J. and Nolan, S. J. and Ohm, S. and {De O{\~{n}}a Wilhelmi}, E. and Opitz, B. and Ostrowski, M. and Oya, I. and Panter, M. and Parsons, R. D. and {Paz Arribas}, M. and Pekeur, N. W. and Pelletier, G. and Perez, J. and Petrucci, P. O. and Peyaud, B. and Pita, S. and P{\"{u}}hlhofer, G. and Punch, M. and Quirrenbach, A. and Raue, M. and Reimer, A. and Reimer, O. and Renaud, M. and {De Los Reyes}, R. and Rieger, F. and Ripken, J. and Rob, L. and Rosier-Lees, S. and Rowell, G. and Rudak, B. and Rulten, C. B. and Sahakian, V. and Sanchez, D. A. and Santangelo, A. and Schlickeiser, R. and Schulz, A. and Schwanke, U. and Schwarzburg, S. and Schwemmer, S. and Sheidaei, F. and Skilton, J. L. and Sol, H. and Spengler, G. and Stawarz{\L}. and Steenkamp, R. and Stegmann, C. and Stinzing, F. and Stycz, K. and Sushch, I. and Szostek, A. and Tavernet, J. P. and Terrier, R. and Tluczykont, M. and Trichard, C. and Valerius, K. and {Van Eldik}, C. and Vasileiadis, G. and Venter, C. and Viana, A. and Vincent, P. and V{\"{o}}lk, H. J. and Volpe, F. and Vorobiov, S. and Vorster, M. and Wagner, S. J. and Ward, M. and White, R. and Wierzcholska, A. and Wouters, D. and Zacharias, M. and Zajczyk, A. and Zdziarski, A. A. and Zech, A. and Zechlin, H. S.},
doi = {10.1103/PhysRevLett.110.041301},
eprint = {1301.1173},
issn = {00319007},
journal = {Physical Review Letters},
month = {jan},
number = {4},
publisher = {American Physical Society (APS)},
title = {{Search for photon-linelike signatures from dark matter annihilations with H.E.S.S.}},
url = {http://dx.doi.org/10.1103/PhysRevLett.110.041301},
volume = {110},
year = {2013}
}

@article{Ackermann_2015,
archivePrefix = {arXiv},
arxivId = {1506.00013},
author = {Ackermann, M. and Ajello, M. and Albert, A. and Anderson, B. and Atwood, W. B. and Baldini, L. and Barbiellini, G. and Bastieri, D. and Bellazzini, R. and Bissaldi, E. and Blandford, R. D. and Bloom, E. D. and Bonino, R. and Bottacini, E. and Brandt, T. J. and Bregeon, J. and Bruel, P. and Buehler, R. and Buson, S. and Caliandro, G. A. and Cameron, R. A. and Caputo, R. and Caragiulo, M. and Caraveo, P. A. and Cecchi, C. and Charles, E. and Chekhtman, A. and Chiang, J. and Chiaro, G. and Ciprini, S. and Claus, R. and Cohen-Tanugi, J. and Conrad, J. and Cuoco, A. and Cutini, S. and D'Ammando, F. and {De Angelis}, A. and {De Palma}, F. and Desiante, R. and Digel, S. W. and {Di Venere}, L. and Drell, P. S. and Drlica-Wagner, A. and Favuzzi, C. and Fegan, S. J. and Franckowiak, A. and Fukazawa, Y. and Funk, S. and Fusco, P. and Gargano, F. and Gasparrini, D. and Giglietto, N. and Giordano, F. and Giroletti, M. and Godfrey, G. and Gomez-Vargas, G. A. and Grenier, I. A. and Grove, J. E. and Guiriec, S. and Gustafsson, M. and Hewitt, J. W. and Hill, A. B. and Horan, D. and J{\'{o}}hannesson, G. and Johnson, R. P. and Kuss, M. and Larsson, S. and Latronico, L. and Li, J. and Li, L. and Longo, F. and Loparco, F. and Lovellette, M. N. and Lubrano, P. and Malyshev, D. and Mayer, M. and Mazziotta, M. N. and McEnery, J. E. and Michelson, P. F. and Mizuno, T. and Moiseev, A. A. and Monzani, M. E. and Morselli, A. and Murgia, S. and Nuss, E. and Ohsugi, T. and Orienti, M. and Orlando, E. and Ormes, J. F. and Paneque, D. and Pesce-Rollins, M. and Piron, F. and Pivato, G. and Rain{\`{o}}, S. and Rando, R. and Razzano, M. and Reimer, A. and Reposeur, T. and Ritz, S. and S{\'{a}}nchez-Conde, M. and Schulz, A. and Sgr{\`{o}}, C. and Siskind, E. J. and Spada, F. and Spandre, G. and Spinelli, P. and Tajima, H. and Takahashi, H. and Thayer, J. B. and Tibaldo, L. and Torres, D. F. and Tosti, G. and Troja, E. and Vianello, G. and Werner, M. and Winer, B. L. and Wood, K. S. and Wood, M. and Zaharijas, G. and Zimmer, S.},
doi = {10.1103/PhysRevD.91.122002},
eprint = {1506.00013},
issn = {15502368},
journal = {Physical Review D - Particles, Fields, Gravitation and Cosmology},
number = {12},
publisher = {American Physical Society (APS)},
title = {{Updated search for spectral lines from Galactic dark matter interactions with pass 8 data from the Fermi Large Area Telescope}},
url = {http://dx.doi.org/10.1103/PhysRevD.91.122002},
volume = {91},
year = {2015}
}

@article{Ackermann_2012,
archivePrefix = {arXiv},
arxivId = {1205.2739},
author = {Ackermann, M. and Ajello, M. and Albert, A. and Baldini, L. and Barbiellini, G. and Bechtol, K. and Bellazzini, R. and Berenji, B. and Blandford, R. D. and Bloom, E. D. and Bonamente, E. and Borgland, A. W. and Brigida, M. and Buehler, R. and Buson, S. and Caliandro, G. A. and Cameron, R. A. and Caraveo, P. A. and Casandjian, J. M. and Cecchi, C. and Charles, E. and Chekhtman, A. and Chiang, J. and Ciprini, S. and Claus, R. and Cohen-Tanugi, J. and Conrad, J. and D'Ammando, F. and {De Palma}, F. and Dermer, C. D. and {Do Couto E Silva}, E. and Drell, P. S. and Drlica-Wagner, A. and Edmonds, Y. and Essig, R. and Favuzzi, C. and Fegan, S. J. and Focke, W. B. and Fukazawa, Y. and Funk, S. and Fusco, P. and Gargano, F. and Gasparrini, D. and Germani, S. and Giglietto, N. and Giordano, F. and Giroletti, M. and Glanzman, T. and Godfrey, G. and Grenier, I. A. and Guiriec, S. and Gustafsson, M. and Hadasch, D. and Hayashida, M. and Horan, D. and Hughes, R. E. and Kamae, T. and Kn{\"{o}}dlseder, J. and Kuss, M. and Lande, J. and Lionetto, A. M. and {Llena Garde}, M. and Longo, F. and Loparco, F. and Lovellette, M. N. and Lubrano, P. and Mazziotta, M. N. and Michelson, P. F. and Mitthumsiri, W. and Mizuno, T. and Moiseev, A. A. and Monte, C. and Monzani, M. E. and Morselli, A. and Moskalenko, I. V. and Murgia, S. and Naumann-Godo, M. and Norris, J. P. and Nuss, E. and Ohsugi, T. and Okumura, A. and Orlando, E. and Ormes, J. F. and Paneque, D. and Panetta, J. H. and Pesce-Rollins, M. and Piron, F. and Pivato, G. and Porter, T. A. and Prokhorov, D. and Rain{\`{o}}, S. and Rando, R. and Razzano, M. and Reimer, O. and Roth, M. and Sbarra, C. and Scargle, J. D. and Sgr{\`{o}}, C. and Siskind, E. J. and Snyder, A. and Spinelli, P. and Suson, D. J. and Takahashi, H. and Tanaka, T. and Thayer, J. G. and Thayer, J. B. and Tibaldo, L. and Tinivella, M. and Torres, D. F. and Tosti, G. and Troja, E. and Vandenbroucke, J. and Vasileiou, V. and Vianello, G. and Vitale, V. and Waite, A. P. and Winer, B. L. and Wood, K. S. and Yang, Z. and Zimmer, S.},
doi = {10.1103/PhysRevD.86.022002},
eprint = {1205.2739},
issn = {15502368},
journal = {Physical Review D - Particles, Fields, Gravitation and Cosmology},
number = {2},
publisher = {American Physical Society (APS)},
title = {{Fermi LAT search for dark matter in gamma-ray lines and the inclusive photon spectrum}},
url = {http://dx.doi.org/10.1103/PhysRevD.86.022002},
volume = {86},
year = {2012}
}

@inproceedings{Alexander:2016aln,
archivePrefix = {arXiv},
arxivId = {hep-ph/1608.08632},
author = {Alexander, Jim and Others},
eprint = {1608.08632},
primaryClass = {hep-ph},
title = {{Dark Sectors 2016 Workshop: Community Report}},
url = {http://lss.fnal.gov/archive/2016/conf/fermilab-conf-16-421.pdf},
year = {2016}
}

@article{An:2013yfc,
archivePrefix = {arXiv},
arxivId = {1302.3884},
author = {An, Haipeng and Pospelov, Maxim and Pradler, Josef},
doi = {10.1016/j.physletb.2013.07.008},
eprint = {1302.3884},
issn = {03702693},
journal = {Physics Letters, Section B: Nuclear, Elementary Particle and High-Energy Physics},
number = {4-5},
pages = {190--195},
title = {{New stellar constraints on dark photons}},
volume = {725},
year = {2013}
}

@article{Aprile:2018dbl,
archivePrefix = {arXiv},
arxivId = {1805.12562},
author = {Aprile, E. and Aalbers, J. and Agostini, F. and Alfonsi, M. and Althueser, L. and Amaro, F. D. and Anthony, M. and Arneodo, F. and Baudis, L. and Bauermeister, B. and Benabderrahmane, M. L. and Berger, T. and Breur, P. A. and Brown, A. and Brown, A. and Brown, E. and Bruenner, S. and Bruno, G. and Budnik, R. and Capelli, C. and Cardoso, J. M.R. and Cichon, D. and Coderre, D. and Colijn, A. P. and Conrad, J. and Cussonneau, J. P. and Decowski, M. P. and {De Perio}, P. and {Di Gangi}, P. and {Di Giovanni}, A. and Diglio, S. and Elykov, A. and Eurin, G. and Fei, J. and Ferella, A. D. and Fieguth, A. and Fulgione, W. and {Gallo Rosso}, A. and Galloway, M. and Gao, F. and Garbini, M. and Geis, C. and Grandi, L. and Greene, Z. and Qiu, H. and Hasterok, C. and Hogenbirk, E. and Howlett, J. and Itay, R. and Joerg, F. and Kaminsky, B. and Kazama, S. and Kish, A. and Koltman, G. and Landsman, H. and Lang, R. F. and Levinson, L. and Lin, Q. and Lindemann, S. and Lindner, M. and Lombardi, F. and Lopes, J. A.M. and Mahlstedt, J. and Manfredini, A. and {Marrod{\'{a}}n Undagoitia}, T. and Masbou, J. and Masson, D. and Messina, M. and Micheneau, K. and Miller, K. and Molinario, A. and Mor{\aa}, K. and Murra, M. and Naganoma, J. and Ni, K. and Oberlack, U. and Pelssers, B. and Piastra, F. and Pienaar, J. and Pizzella, V. and Plante, G. and Podviianiuk, R. and Priel, N. and {Ram{\'{i}}rez Garc{\'{i}}a}, D. and Rauch, L. and Reichard, S. and Reuter, C. and Riedel, B. and Rizzo, A. and Rocchetti, A. and Rupp, N. and {Dos Santos}, J. M.F. and Sartorelli, G. and Scheibelhut, M. and Schindler, S. and Schreiner, J. and Schulte, D. and Schumann, M. and {Scotto Lavina}, L. and Selvi, M. and Shagin, P. and Shockley, E. and Silva, M. and Simgen, H. and Thers, D. and Toschi, F. and Trinchero, G. and Tunnell, C. and Upole, N. and Vargas, M. and Wack, O. and Wang, H. and Wang, Z. and Wei, Y. and Weinheimer, C. and Wittweg, C. and Wulf, J. and Ye, J. and Zhang, Y. and Zhu, T.},
doi = {10.1103/PhysRevLett.121.111302},
eprint = {1805.12562},
issn = {10797114},
journal = {Physical Review Letters},
number = {11},
pages = {111302},
title = {{Dark Matter Search Results from a One Ton-Year Exposure of XENON1T}},
volume = {121},
year = {2018}
}

@article{Aprile:2017iyp,
archivePrefix = {arXiv},
arxivId = {1705.06655},
author = {Aprile, E. and Aalbers, J. and Agostini, F. and Alfonsi, M. and Amaro, F. D. and Anthony, M. and Arneodo, F. and Barrow, P. and Baudis, L. and Bauermeister, B. and Benabderrahmane, M. L. and Berger, T. and Breur, P. A. and Brown, A. and Brown, A. and Brown, E. and Bruenner, S. and Bruno, G. and Budnik, R. and B{\"{u}}tikofer, L. and Calv{\'{e}}n, J. and Cardoso, J. M.R. and Cervantes, M. and Cichon, D. and Coderre, D. and Colijn, A. P. and Conrad, J. and Cussonneau, J. P. and Decowski, M. P. and {De Perio}, P. and {Di Gangi}, P. and {Di Giovanni}, A. and Diglio, S. and Eurin, G. and Fei, J. and Ferella, A. D. and Fieguth, A. and Fulgione, W. and {Gallo Rosso}, A. and Galloway, M. and Gao, F. and Garbini, M. and Gardner, R. and Geis, C. and Goetzke, L. W. and Grandi, L. and Greene, Z. and Grignon, C. and Hasterok, C. and Hogenbirk, E. and Howlett, J. and Itay, R. and Kaminsky, B. and Kazama, S. and Kessler, G. and Kish, A. and Landsman, H. and Lang, R. F. and Lellouch, D. and Levinson, L. and Lin, Q. and Lindemann, S. and Lindner, M. and Lombardi, F. and Lopes, J. A.M. and Manfredini, A. and Mariş, I. and {Marrod{\'{a}}n Undagoitia}, T. and Masbou, J. and Massoli, F. V. and Masson, D. and Mayani, D. and Messina, M. and Micheneau, K. and Molinario, A. and Mor{\aa}, K. and Murra, M. and Naganoma, J. and Ni, K. and Oberlack, U. and Pakarha, P. and Pelssers, B. and Persiani, R. and Piastra, F. and Pienaar, J. and Pizzella, V. and Piro, M. C. and Plante, G. and Priel, N. and Rauch, L. and Reichard, S. and Reuter, C. and Riedel, B. and Rizzo, A. and Rosendahl, S. and Rupp, N. and Saldanha, R. and {Dos Santos}, J. M.F. and Sartorelli, G. and Scheibelhut, M. and Schindler, S. and Schreiner, J. and Schumann, M. and {Scotto Lavina}, L. and Selvi, M. and Shagin, P. and Shockley, E. and Silva, M. and Simgen, H. and Sivers, M. V. and Stein, A. and Thapa, S. and Thers, D. and Tiseni, A. and Trinchero, G. and Tunnell, C. and Vargas, M. and Upole, N. and Wang, H. and Wang, Z. and Wei, Y. and Weinheimer, C. and Wulf, J. and Ye, J. and Zhang, Y. and Zhu, T.},
doi = {10.1103/PhysRevLett.119.181301},
eprint = {1705.06655},
issn = {10797114},
journal = {Physical Review Letters},
number = {18},
pages = {181301},
title = {{First Dark Matter Search Results from the XENON1T Experiment}},
volume = {119},
year = {2017}
}

@article{Kobzarev:1966qya,
author = {Batyghin, V. V.},
doi = {10.1016/0375-9601(68)90603-8},
issn = {03759601},
journal = {Physics Letters A},
number = {1},
pages = {64--65},
title = {{On the possibility of experimental observation of Hard Vavilov-{\v{C}}erenkov radiation}},
volume = {28},
year = {1968}
}

@article{Berezhiani:2000gw,
archivePrefix = {arXiv},
arxivId = {hep-ph/0008105},
author = {Berezhiani, Zurab and Comelli, Denis and Villante, Francesco L.},
doi = {10.1016/S0370-2693(01)00217-9},
eprint = {0008105},
issn = {03702693},
journal = {Physics Letters, Section B: Nuclear, Elementary Particle and High-Energy Physics},
number = {3-4},
pages = {362--375},
primaryClass = {hep-ph},
title = {{The early mirror universe: Inflation, baryogenesis, nucleosynthesis and dark matter}},
volume = {503},
year = {2001}
}

@article{Berezhiani:2008gi,
archivePrefix = {arXiv},
arxivId = {hep-ph/0810.1317},
author = {Berezhiani, Zurab and Lepidi, Angela},
doi = {10.1016/j.physletb.2009.10.023},
eprint = {0810.1317},
issn = {03702693},
journal = {Physics Letters, Section B: Nuclear, Elementary Particle and High-Energy Physics},
number = {3},
pages = {276--281},
primaryClass = {hep-ph},
title = {{Cosmological bounds on the "millicharges" of mirror particles}},
volume = {681},
year = {2009}
}

@article{Berger:2016vxi,
archivePrefix = {arXiv},
arxivId = {1605.07195},
author = {Berger, Joshua and Jedamzik, Karsten and Walker, Devin G.E.},
doi = {10.1088/1475-7516/2016/11/032},
eprint = {1605.07195},
issn = {14757516},
journal = {Journal of Cosmology and Astroparticle Physics},
number = {11},
pages = {32},
title = {{Cosmological constraints on decoupled dark photons and dark Higgs}},
volume = {2016},
year = {2016}
}

@article{Boddy:2015efa,
archivePrefix = {arXiv},
arxivId = {astro-ph.CO/1504.04024},
author = {Boddy, Kimberly K. and Kumar, Jason},
doi = {10.1103/PhysRevD.92.023533},
eprint = {1504.04024},
issn = {15502368},
journal = {Physical Review D - Particles, Fields, Gravitation and Cosmology},
number = {2},
pages = {23533},
primaryClass = {astro-ph.CO},
title = {{Indirect detection of dark matter using MeV-range gamma-ray telescopes}},
volume = {92},
year = {2015}
}

@article{Bondarenko_2018,
archivePrefix = {arXiv},
arxivId = {1712.06602},
author = {Bondarenko, Kyrylo and Boyarsky, Alexey and Bringmann, Torsten and Sokolenko, Anastasia},
doi = {10.1088/1475-7516/2018/04/049},
eprint = {1712.06602},
issn = {14757516},
journal = {Journal of Cosmology and Astroparticle Physics},
number = {4},
pages = {49},
publisher = {{\{}IOP{\}} Publishing},
title = {{Constraining self-interacting dark matter with scaling laws of observed halo surface densities}},
url = {https://doi.org/10.1088{\%}2F1475-7516{\%}2F2018{\%}2F04{\%}2F049},
volume = {2018},
year = {2018}
}

@article{Braaten:1993jw,
archivePrefix = {arXiv},
arxivId = {hep-ph/9302213},
author = {Braaten, Eric and Segel, Daniel},
doi = {10.1103/PhysRevD.48.1478},
eprint = {9302213},
issn = {05562821},
journal = {Physical Review D},
number = {4},
pages = {1478--1491},
primaryClass = {hep-ph},
title = {{Neutrino energy loss from the plasma process at all temperatures and densities}},
volume = {48},
year = {1993}
}

@article{Bringmann:2016din,
archivePrefix = {arXiv},
arxivId = {1612.00845},
author = {Bringmann, Torsten and Kahlhoefer, Felix and Schmidt-Hoberg, Kai and Walia, Parampreet},
doi = {10.1103/PhysRevLett.118.141802},
eprint = {1612.00845},
issn = {10797114},
journal = {Physical Review Letters},
number = {14},
pages = {141802},
title = {{Strong Constraints on Self-Interacting Dark Matter with Light Mediators}},
volume = {118},
year = {2017}
}

@article{Chang:2016ntp,
archivePrefix = {arXiv},
arxivId = {1611.03864},
author = {Chang, Jae Hyeok and Essig, Rouven and McDermott, Samuel D.},
doi = {10.1007/JHEP01(2017)107},
eprint = {1611.03864},
issn = {10298479},
journal = {Journal of High Energy Physics},
number = {1},
pages = {107},
title = {{Revisiting Supernova 1987A constraints on dark photons}},
volume = {2017},
year = {2017}
}

@article{Chang:2018rso,
archivePrefix = {arXiv},
arxivId = {1803.00993},
author = {Chang, Jae Hyeok and Essig, Rouven and McDermott, Samuel D.},
doi = {10.1007/JHEP09(2018)051},
eprint = {1803.00993},
issn = {10298479},
journal = {Journal of High Energy Physics},
number = {9},
pages = {51},
title = {{Supernova 1987A constraints on sub-GeV dark sectors, millicharged particles, the QCD axion, and an axion-like particle}},
volume = {2018},
year = {2018}
}

@article{Chu:2016pew,
archivePrefix = {arXiv},
arxivId = {1609.00399},
author = {Chu, Xiaoyong and Garcia-Cely, Camilo and Hambye, Thomas},
doi = {10.1007/JHEP11(2016)048},
eprint = {1609.00399},
issn = {10298479},
journal = {Journal of High Energy Physics},
number = {11},
pages = {48},
title = {{Can the relic density of self-interacting dark matter be due to annihilations into Standard Model particles?}},
volume = {2016},
year = {2016}
}

@article{Chu:2011be,
archivePrefix = {arXiv},
arxivId = {1112.0493},
author = {Chu, Xiaoyong and Hambye, Thomas and Tytgat, Michel H.G.},
doi = {10.1088/1475-7516/2012/05/034},
eprint = {1112.0493},
issn = {14757516},
journal = {Journal of Cosmology and Astroparticle Physics},
number = {5},
pages = {34},
title = {{The four basic ways of creating dark matter through a portal}},
volume = {2012},
year = {2012}
}

@article{Cirelli:2016rnw,
archivePrefix = {arXiv},
arxivId = {hep-ph/1612.07295},
author = {Cirelli, Marco and Panci, Paolo and Petraki, Kalliopi and Sala, Filippo and Taoso, Marco},
doi = {10.1088/1475-7516/2017/05/036},
eprint = {1612.07295},
issn = {14757516},
journal = {Journal of Cosmology and Astroparticle Physics},
number = {5},
pages = {36},
primaryClass = {hep-ph},
title = {{Dark Matter's secret liaisons: Phenomenology of a dark U(1) sector with bound states}},
volume = {2017},
year = {2017}
}

@article{Clowe:2003tk,
    author = "Clowe, Douglas and Gonzalez, Anthony and Markevitch, Maxim",
    archivePrefix = "arXiv",
    doi = "10.1086/381970",
    eprint = "astro-ph/0312273",
    journal = "Astrophys.\ J.",
    pages = "596--603",
    title = "{Weak lensing mass reconstruction of the interacting cluster 1E0657-558: Direct evidence for the existence of dark matter}",
    volume = "604",
    year = "2004"
}

@book{cohen1973mecanique2,
author = {Cohen-Tannoudji, C and Diu, B and Lalo{\"{e}}, F},
isbn = {9782705657338},
number = {vol.{\~{}}1},
publisher = {Masson},
series = {Collection Enseignement des sciences, 16},
title = {{Mecanique Quantique, vol.II}},
url = {https://books.google.fr/books?id=RnQfAQAAMAAJ https://books.google.fr/books?id=dnQfAQAAMAAJ},
year = {1995}
}

@article{Cui:2017nnn,
archivePrefix = {arXiv},
arxivId = {1708.06917},
author = {Cui, Xiangyi and Abdukerim, Abdusalam and Chen, Wei and Chen, Xun and Chen, Yunhua and Dong, Binbin and Fang, Deqing and Fu, Changbo and Giboni, Karl and Giuliani, Franco and Gu, Linhui and Gu, Yikun and Guo, Xuyuan and Guo, Zhifan and Han, Ke and He, Changda and Huang, Di and He, Shengming and Huang, Xingtao and Huang, Zhou and Ji, Xiangdong and Ju, Yonglin and Li, Shaoli and Li, Yao and Lin, Heng and Liu, Huaxuan and Liu, Jianglai and Ma, Yugang and Mao, Yajun and Ni, Kaixiang and Ning, Jinhua and Ren, Xiangxiang and Shi, Fang and Tan, Andi and Wang, Cheng and Wang, Hongwei and Wang, Meng and Wang, Qiuhong and Wang, Siguang and Wang, Xiuli and Wang, Xuming and Wu, Qinyu and Wu, Shiyong and Xiao, Mengjiao and Xie, Pengwei and Yan, Binbin and Yang, Yong and Yue, Jianfeng and Zhang, Dan and Zhang, Hongguang and Zhang, Tao and Zhang, Tianqi and Zhao, Li and Zhou, Jifang and Zhou, Ning and Zhou, Xiaopeng},
doi = {10.1103/PhysRevLett.119.181302},
eprint = {1708.06917},
issn = {10797114},
journal = {Physical Review Letters},
number = {18},
pages = {181302},
title = {{Dark Matter Results from 54-Ton-Day Exposure of PandaX-II Experiment}},
volume = {119},
year = {2017}
}

@article{Das:2016ced,
archivePrefix = {arXiv},
arxivId = {hep-ph/1611.04606},
author = {Das, Anirban and Dasgupta, Basudeb},
doi = {10.1103/PhysRevLett.118.251101},
eprint = {1611.04606},
issn = {10797114},
journal = {Physical Review Letters},
number = {25},
primaryClass = {hep-ph},
title = {{Selection Rule for Enhanced Dark Matter Annihilation}},
volume = {118},
year = {2017}
}

@article{Duerr:2018mbd,
archivePrefix = {arXiv},
arxivId = {1804.10385},
author = {Duerr, Michael and Schmidt-Hoberg, Kai and Wild, Sebastian},
doi = {10.1088/1475-7516/2018/09/033},
eprint = {1804.10385},
issn = {14757516},
journal = {Journal of Cosmology and Astroparticle Physics},
number = {9},
pages = {33},
title = {{Self-interacting dark matter with a stable vector mediator}},
volume = {2018},
year = {2018}
}

@article{Englert:1964et,
author = {Englert, F. and Brout, R.},
doi = {10.1103/PhysRevLett.13.321},
issn = {00319007},
journal = {Physical Review Letters},
number = {9},
pages = {321--323},
title = {{Broken symmetry and the mass of gauge vector mesons}},
volume = {13},
year = {1964}
}

@inproceedings{Essig:2013lka,
archivePrefix = {arXiv},
arxivId = {1311.0029},
author = {Essig, Rouven and Jaros, JA John A. and Wester, William and Others},
booktitle = {Proceedings, 2013 Community Summer Study on the Future of U.S. Particle Physics: Snowmass on the Mississippi (CSS2013): Minneapolis, MN, USA, July 29-August 6, 2013},
eprint = {1311.0029},
title = {{Working Group Report: New Light Weakly Coupled Particles}},
url = {http://www.slac.stanford.edu/econf/C1307292/docs/IntensityFrontier/NewLight-17.pdf},
year = {2013}
}

@article{Essig:2013goa,
archivePrefix = {arXiv},
arxivId = {1309.4091},
author = {Essig, Rouven and Kuflik, Eric and McDermott, Samuel D. and Volansky, Tomer and Zurek, Kathryn M.},
doi = {10.1007/JHEP11(2013)193},
eprint = {1309.4091},
issn = {10298479},
journal = {Journal of High Energy Physics},
number = {11},
pages = {193},
title = {{Constraining light dark matter with diffuse X-ray and gamma-ray observations}},
volume = {2013},
year = {2013}
}

@article{Feng:2009mn,
archivePrefix = {arXiv},
arxivId = {0905.3039},
author = {Feng, Jonathan L. and Kaplinghat, Manoj and Tu, Huitzu and Yu, Hai Bo},
doi = {10.1088/1475-7516/2009/07/004},
eprint = {0905.3039},
issn = {14757516},
journal = {Journal of Cosmology and Astroparticle Physics},
number = {7},
pages = {4},
title = {{Hidden charged dark matter}},
volume = {2009},
year = {2009}
}

@article{Feng:2009hw,
archivePrefix = {arXiv},
arxivId = {0911.0422},
author = {Feng, Jonathan L. and Kaplinghat, Manoj and Yu, Hai Bo},
doi = {10.1103/PhysRevLett.104.151301},
eprint = {0911.0422},
issn = {00319007},
journal = {Physical Review Letters},
number = {15},
pages = {151301},
title = {{Halo-shape and relic-density exclusions of sommerfeld-enhanced dark matter explanations of cosmic ray excesses}},
volume = {104},
year = {2010}
}

@article{Feng:2008mu,
archivePrefix = {arXiv},
arxivId = {0808.2318},
author = {Feng, Jonathan L. and Tu, Huitzu and Yu, Hai Bo},
doi = {10.1088/1475-7516/2008/10/043},
eprint = {0808.2318},
issn = {14757516},
journal = {Journal of Cosmology and Astroparticle Physics},
number = {10},
pages = {43},
title = {{Thermal relics in hidden sectors}},
volume = {2008},
year = {2008}
}

@article{Foot:2014uba,
archivePrefix = {arXiv},
arxivId = {1409.7174},
author = {Foot, R. and Vagnozzi, S.},
doi = {10.1103/PhysRevD.91.023512},
eprint = {1409.7174},
issn = {15502368},
journal = {Physical Review D - Particles, Fields, Gravitation and Cosmology},
number = {2},
pages = {23512},
title = {{Dissipative hidden sector dark matter}},
volume = {91},
year = {2015}
}

@article{Fornengo:2011sz,
archivePrefix = {arXiv},
arxivId = {hep-ph/1108.4661},
author = {Fornengo, N. and Panci, P. and Regis, M.},
doi = {10.1103/PhysRevD.84.115002},
eprint = {1108.4661},
issn = {15507998},
journal = {Physical Review D - Particles, Fields, Gravitation and Cosmology},
number = {11},
pages = {115002},
primaryClass = {hep-ph},
title = {{Long-range forces in direct dark matter searches}},
volume = {84},
year = {2011}
}

@article{Fradette:2014sza,
archivePrefix = {arXiv},
arxivId = {1407.0993},
author = {Fradette, Anthony and Pospelov, Maxim and Pradler, Josef and Ritz, Adam},
doi = {10.1103/PhysRevD.90.035022},
eprint = {1407.0993},
issn = {15502368},
journal = {Physical Review D - Particles, Fields, Gravitation and Cosmology},
number = {3},
pages = {35022},
title = {{Cosmological constraints on very dark photons}},
volume = {90},
year = {2014}
}

@article{Gondolo:1990dk,
author = {Gondolo, Paolo and Gelmini, Graciela},
doi = {10.1016/0550-3213(91)90438-4},
issn = {05503213},
journal = {Nuclear Physics, Section B},
number = {1},
pages = {145--179},
title = {{Cosmic abundances of stable particles: Improved analysis}},
volume = {360},
year = {1991}
}

@article{Goodsell:2009xc,
archivePrefix = {arXiv},
arxivId = {0909.0515},
author = {Goodsell, Mark and Jaeckel, Joerg and Redondo, Javier and Ringwald, Andreas},
doi = {10.1088/1126-6708/2009/11/027},
eprint = {0909.0515},
issn = {11266708},
journal = {Journal of High Energy Physics},
number = {11},
pages = {27},
title = {{Naturally light hidden photons in LARGE volume string compactifications}},
volume = {2009},
year = {2009}
}

@article{Griest:1989wd,
author = {Griest, Kim and Kamionkowski, Marc},
doi = {10.1103/PhysRevLett.64.615},
issn = {00319007},
journal = {Physical Review Letters},
number = {6},
pages = {615--618},
title = {{Unitarity limits on the mass and radius of dark-matter particles}},
volume = {64},
year = {1990}
}

@article{Hambye:2010zb,
archivePrefix = {arXiv},
arxivId = {1012.4587},
author = {Hambye, Thomas},
doi = {10.22323/1.110.0098},
eprint = {1012.4587},
issn = {18248039},
journal = {Proceedings of Science},
pages = {98},
title = {{On the stability of particle dark matter}},
volume = {IDM2010},
year = {2010}
}

@article{Hambye:2019dwd,
archivePrefix = {arXiv},
arxivId = {1908.09864},
author = {Hambye, Thomas and Tytgat, Michel H. G. and Vandecasteele, J{\'{e}}r{\^{o}}me and Vanderheyden, Laurent},
doi = {10.1103/PhysRevD.100.095018},
eprint = {1908.09864},
journal = {Phys. Rev.},
number = {9},
pages = {95018},
title = {{Dark matter from dark photons: a taxonomy of dark matter production}},
url = {http://arxiv.org/abs/1908.09864},
volume = {D100},
year = {2019}
}

@article{Hambye:2018dpi,
archivePrefix = {arXiv},
arxivId = {1807.05022},
author = {Hambye, Thomas and Tytgat, Michel H.G. and Vandecasteele, J{\'{e}}r{\^{o}}me and Vanderheyden, Laurent},
doi = {10.1103/PhysRevD.98.075017},
eprint = {1807.05022},
issn = {24700029},
journal = {Physical Review D},
number = {7},
pages = {75017},
title = {{Dark matter direct detection is testing freeze-in}},
volume = {98},
year = {2018}
}

@article{Harvey:2015hha,
archivePrefix = {arXiv},
arxivId = {1503.07675},
author = {Harvey, David and Massey, Richard and Kitching, Thomas and Taylor, Andy and Tittley, Eric},
doi = {10.1126/science.1261381},
eprint = {1503.07675},
issn = {10959203},
journal = {Science},
number = {6229},
pages = {1462--1465},
title = {{The nongravitational interactions of dark matter in colliding galaxy clusters}},
volume = {347},
year = {2015}
}

@article{Heeba:2019jho,
archivePrefix = {arXiv},
arxivId = {1908.09834},
author = {Heeba, Saniya and Kahlhoefer, Felix},
eprint = {1908.09834},
title = {{Probing the freeze-in mechanism in dark matter models with U(1)' gauge extensions}},
url = {http://arxiv.org/abs/1908.09834},
year = {2019}
}

@article{Ackerman:mha,
archivePrefix = {arXiv},
arxivId = {1504.00773},
author = {Heo, Jae Ho and Kim, C. S.},
doi = {10.3938/jkps.68.715},
eprint = {1504.00773},
issn = {19768524},
journal = {Journal of the Korean Physical Society},
number = {5},
pages = {715--721},
title = {{Light dark matter and dark radiation}},
volume = {68},
year = {2016}
}

@article{Higgs:1964pj,
author = {Higgs, Peter W.},
doi = {10.1103/PhysRevLett.13.508},
issn = {00319007},
journal = {Physical Review Letters},
number = {16},
pages = {508--509},
title = {{Broken symmetries and the masses of gauge bosons}},
volume = {13},
year = {1964}
}

@article{Hufnagel:2018bjp,
archivePrefix = {arXiv},
arxivId = {1808.09324},
author = {Hufnagel, Marco and Schmidt-Hoberg, Kai and Wild, Sebastian},
doi = {10.1088/1475-7516/2018/11/032},
eprint = {1808.09324},
issn = {14757516},
journal = {Journal of Cosmology and Astroparticle Physics},
number = {11},
pages = {32},
title = {{BBN constraints on MeV-scale dark sectors. Part II: Electromagnetic decays}},
volume = {2018},
year = {2018}
}

@article{Jaeckel:2008fi,
archivePrefix = {arXiv},
arxivId = {0804.4157},
author = {Jaeckel, Joerg and Redondo, Javier and Ringwald, Andreas},
doi = {10.1103/PhysRevLett.101.131801},
eprint = {0804.4157},
issn = {00319007},
journal = {Physical Review Letters},
number = {13},
pages = {131801},
title = {{Signatures of a hidden cosmic microwave background}},
volume = {101},
year = {2008}
}

@article{Jaeckel:2010ni,
archivePrefix = {arXiv},
arxivId = {1002.0329},
author = {Jaeckel, Joerg and Ringwald, Andreas},
doi = {10.1146/annurev.nucl.012809.104433},
eprint = {1002.0329},
issn = {0163-8998},
journal = {Annual Review of Nuclear and Particle Science},
number = {1},
pages = {405--437},
title = {{The Low-Energy Frontier of Particle Physics}},
volume = {60},
year = {2010}
}

@article{Kahlhoefer:2017ddj,
archivePrefix = {arXiv},
arxivId = {1707.08571},
author = {Kahlhoefer, Felix and Kulkarni, Suchita and Wild, Sebastian},
doi = {10.1088/1475-7516/2017/11/016},
eprint = {1707.08571},
issn = {14757516},
journal = {Journal of Cosmology and Astroparticle Physics},
number = {11},
pages = {16},
title = {{Exploring light mediators with low-threshold direct detection experiments}},
volume = {2017},
year = {2017}
}

@article{Oh:2015xoa,
    author = "Oh, Se-Heon and others",
    title = "{High-resolution mass models of dwarf galaxies from LITTLE THINGS}",
    eprint = "1502.01281",
    archivePrefix = "arXiv",
    primaryClass = "astro-ph.GA",
    doi = "10.1088/0004-6256/149/6/180",
    journal = "Astron. J.",
    volume = "149",
    pages = "180",
    year = "2015"
}

@article{Kamada:2016euw,
archivePrefix = {arXiv},
arxivId = {1611.02716},
author = {Kamada, Ayuki and Kaplinghat, Manoj and Pace, Andrew B. and Yu, Hai Bo},
doi = {10.1103/PhysRevLett.119.111102},
eprint = {1611.02716},
issn = {10797114},
journal = {Physical Review Letters},
number = {11},
pages = {111102},
title = {{Self-Interacting Dark Matter Can Explain Diverse Galactic Rotation Curves}},
volume = {119},
year = {2017}
}

@article{Kane:2015qea,
archivePrefix = {arXiv},
arxivId = {1502.05406},
author = {Kane, Gordon L. and Kumar, Piyush and Nelson, Brent D. and Zheng, Bob},
doi = {10.1103/PhysRevD.93.063527},
eprint = {1502.05406},
issn = {24700029},
journal = {Physical Review D},
number = {6},
pages = {63527},
title = {{Dark matter production mechanisms with a nonthermal cosmological history: A classification}},
volume = {93},
year = {2016}
}

@article{Kaplinghat:2015aga,
archivePrefix = {arXiv},
arxivId = {1508.03339},
author = {Kaplinghat, Manoj and Tulin, Sean and Yu, Hai Bo},
doi = {10.1103/PhysRevLett.116.041302},
eprint = {1508.03339},
issn = {10797114},
journal = {Physical Review Letters},
number = {4},
pages = {41302},
title = {{Dark Matter Halos as Particle Colliders: Unified Solution to Small-Scale Structure Puzzles from Dwarfs to Clusters}},
volume = {116},
year = {2016}
}

@article{Kazanas:2014mca,
archivePrefix = {arXiv},
arxivId = {1410.0221},
author = {Kazanas, Demos and Mohapatra, Rabindra N. and Nussinov, Shmuel and Teplitz, Vigdor L. and Zhang, Yongchao},
doi = {10.1016/j.nuclphysb.2014.11.009},
eprint = {1410.0221},
issn = {05503213},
journal = {Nuclear Physics B},
pages = {17--29},
title = {{Supernova bounds on the dark photon using its electromagnetic decay}},
volume = {890},
year = {2015}
}

@article{Klasen:2013ypa,
archivePrefix = {arXiv},
arxivId = {hep-ph/1309.2777},
author = {Klasen, Michael and Yaguna, Carlos E.},
doi = {10.1088/1475-7516/2013/11/039},
eprint = {1309.2777},
issn = {14757516},
journal = {Journal of Cosmology and Astroparticle Physics},
number = {11},
pages = {39},
primaryClass = {hep-ph},
title = {{Warm and cold fermionic dark matter via freeze-in}},
volume = {2013},
year = {2013}
}

@book{Kolb:1990vq,
author = {Kolb, Edward W. and Turner, Michael S.},
booktitle = {Nature},
doi = {10.1038/294521a0},
issn = {00280836},
number = {5841},
pages = {521--526},
publisher = {Addison-Wesley},
title = {{The early Universe}},
volume = {294},
year = {1981}
}

@article{Aaij:2019bvg,
    author = "Aaij, Roel and others",
    collaboration = "LHCb",
    title = "{Search for $A'\to\mu^+\mu^-$ Decays}",
    eprint = "1910.06926",
    archivePrefix = "arXiv",
    primaryClass = "hep-ex",
    reportNumber = "LHCb-PAPER-2019-031, CERN-EP-2019-212",
    doi = "10.1103/PhysRevLett.124.041801",
    journal = "Phys. Rev. Lett.",
    volume = "124",
    number = "4",
    pages = "041801",
    year = "2020"
}

@article{Krnjaic:2015mbs,
archivePrefix = {arXiv},
arxivId = {1512.04119},
author = {Krnjaic, Gordan},
doi = {10.1103/PhysRevD.94.073009},
eprint = {1512.04119},
issn = {24700029},
journal = {Physical Review D},
number = {7},
pages = {73009},
title = {{Probing light thermal dark matter with a Higgs portal mediator}},
volume = {94},
year = {2016}
}

@book{Bellac:2011kqa,
author = {{Le Bellac}, Michel},
booktitle = {Thermal Field Theory},
doi = {10.1017/cbo9780511721700},
isbn = {9780511885068, 9780521654777},
publisher = {Cambridge University Press},
series = {Cambridge Monographs on Mathematical Physics},
title = {{Thermal Field Theory}},
url = {http://www.cambridge.org/mw/academic/subjects/physics/theoretical-physics-and-mathematical-physics/thermal-field-theory?format=AR},
year = {1996}
}

@article{Loeb:2010gj,
archivePrefix = {arXiv},
arxivId = {astro-ph.CO/1011.6374},
author = {Loeb, Abraham and Weiner, Neal},
doi = {10.1103/PhysRevLett.106.171302},
eprint = {1011.6374},
issn = {00319007},
journal = {Physical Review Letters},
number = {17},
pages = {171302},
primaryClass = {astro-ph.CO},
title = {{Cores in dwarf galaxies from dark matter with a Yukawa potential}},
volume = {106},
year = {2011}
}

@article{2016,
archivePrefix = {arXiv},
arxivId = {1601.06590},
author = {{MAGIC Collaboration} and Ahnen, M. L. and Ansoldi, S. and Antonelli, L. A. and Antoranz, P. and Babic, A. and Banerjee, B. and Bangale, P. and {Barres De Almeida}, U. and Barrio, J. A. and {Becerra Gonz{\'{a}}lez}, J. and Bednarek, W. and Bernardini, E. and Biasuzzi, B. and Biland, A. and Blanch, O. and Bonnefoy, S. and Bonnoli, G. and Borracci, F. and Bretz, T. and Carmona, E. and Carosi, A. and Chatterjee, A. and Clavero, R. and Colin, P. and Colombo, E. and Contreras, J. L. and Cortina, J. and Covino, S. and {Da Vela}, P. and Dazzi, F. and {De Angelis}, A. and {De Lotto}, B. and {De O{\~{n}}a Wilhelmi}, E. and {Delgado Mendez}, C. and {Di Pierro}, F. and {Dominis Prester}, D. and Dorner, D. and Doro, M. and Einecke, S. and {Eisenacher Glawion}, D. and Elsaesser, D. and Fern{\'{a}}ndez-Barral, A. and Fidalgo, D. and Fonseca, M. V. and Font, L. and Frantzen, K. and Fruck, C. and Galindo, D. and {Garc{\'{i}}a L{\'{o}}pez}, R. J. and Garczarczyk, M. and {Garrido Terrats}, D. and Gaug, M. and Giammaria, P. and Godinovi{\'{c}}, N. and {Gonz{\'{a}}lez Mu{\~{n}}oz}, A. and Guberman, D. and Hahn, A. and Hanabata, Y. and Hayashida, M. and Herrera, J. and Hose, J. and Hrupec, D. and Hughes, G. and Idec, W. and Kodani, K. and Konno, Y. and Kubo, H. and Kushida, J. and {La Barbera}, A. and Lelas, D. and Lindfors, E. and Lombardi, S. and Longo, F. and L{\'{o}}pez, M. and L{\'{o}}pez-Coto, R. and L{\'{o}}pez-Oramas, A. and Lorenz, E. and Majumdar, P. and Makariev, M. and Mallot, K. and Maneva, G. and Manganaro, M. and Mannheim, K. and Maraschi, L. and Marcote, B. and Mariotti, M. and Mart{\'{i}}nez, M. and Mazin, D. and Menzel, U. and Miranda, J. M. and Mirzoyan, R. and Moralejo, A. and Moretti, E. and Nakajima, D. and Neustroev, V. and Niedzwiecki, A. and {Nievas Rosillo}, M. and Nilsson, K. and Nishijima, K. and Noda, K. and Orito, R. and Overkemping, A. and Paiano, S. and Palacio, J. and Palatiello, M. and Paneque, D. and Paoletti, R. and Paredes, J. M. and Paredes-Fortuny, X. and Persic, M. and Poutanen, J. and {Prada Moroni}, P. G. and Prandini, E. and Puljak, I. and Rhode, W. and Rib{\'{o}}, M. and Rico, J. and {Rodriguez Garcia}, J. and Saito, T. and Satalecka, K. and Schultz, C. and Schweizer, T. and Shore, S. N. and Sillanp{\"{a}}{\"{a}}, A. and Sitarek, J. and Snidaric, I. and Sobczynska, D. and Stamerra, A. and Steinbring, T. and Strzys, M. and Takalo, L. and Takami, H. and Tavecchio, F. and Temnikov, P. and Terzi{\'{c}}, T. and Tescaro, D. and Teshima, M. and Thaele, J. and Torres, D. F. and Toyama, T. and Treves, A. and Verguilov, V. and Vovk, I. and Ward, J. E. and Will, M. and Wu, M. H. and Zanin, R. and Aleksi{\'{c}}, J. and Wood, M. and Anderson, B. and Bloom, E. D. and Cohen-Tanugi, J. and Drlica-Wagner, A. and Mazziotta, M. N. and S{\'{a}}nchez-Conde, M. and Strigari, L.},
doi = {10.1088/1475-7516/2016/02/039},
eprint = {1601.06590},
issn = {14757516},
journal = {Journal of Cosmology and Astroparticle Physics},
number = {2},
pages = {039--039},
publisher = {IOP Publishing},
title = {{Limits to dark matter annihilation cross-section from a combined analysis of MAGIC and Fermi-LAT observations of dwarf satellite galaxies}},
url = {http://dx.doi.org/10.1088/1475-7516/2016/02/039},
volume = {2016},
year = {2016}
}

@article{Mahoney:2017jqk,
archivePrefix = {arXiv},
arxivId = {hep-ph/1706.08871},
author = {Mahoney, Cameron and Leibovich, Adam K. and Zentner, Andrew R.},
doi = {10.1103/PhysRevD.96.043018},
eprint = {1706.08871},
issn = {24700029},
journal = {Physical Review D},
number = {4},
pages = {43018},
primaryClass = {hep-ph},
title = {{Updated constraints on self-interacting dark matter from Supernova 1987A}},
volume = {96},
year = {2017}
}

@article{Markevitch:2003at,
      author         = "Markevitch, Maxim and Gonzalez, A. H. and Clowe, D. and
                        Vikhlinin, A. and David, L. and Forman, W. and Jones, C.
                        and Murray, S. and Tucker, W.",
      title          = "{Direct constraints on the dark matter self-interaction
                        cross-section from the merging galaxy cluster 1E0657-56}",
      journal        = "Astrophys. J.",
      volume         = "606",
      year           = "2004",
      pages          = "819-824",
      doi            = "10.1086/383178",
      eprint         = "astro-ph/0309303",
      archivePrefix  = "arXiv",
      primaryClass   = "astro-ph",
      SLACcitation   = "%%CITATION = ASTRO-PH/0309303;%%"
}

@article{McDermott:2010pa,
archivePrefix = {arXiv},
arxivId = {1011.2907},
author = {McDermott, Samuel D. and Yu, Hai Bo and Zurek, Kathryn M.},
doi = {10.1103/PhysRevD.83.063509},
eprint = {1011.2907},
issn = {15507998},
journal = {Physical Review D - Particles, Fields, Gravitation and Cosmology},
number = {6},
pages = {63509},
title = {{Turning off the lights: How dark is dark matter?}},
volume = {83},
year = {2011}
}

@article{Peter:2012jh,
archivePrefix = {arXiv},
arxivId = {1208.3026},
author = {Peter, Annika H.G. and Rocha, Miguel and Bullock, James S. and Kaplinghat, Manoj},
doi = {10.1093/mnras/sts535},
eprint = {1208.3026},
issn = {00358711},
journal = {Monthly Notices of the Royal Astronomical Society},
number = {1},
pages = {105--120},
title = {{Cosmological simulations with self-interacting dark matter - II. Halo shapes versus observations}},
volume = {430},
year = {2013}
}

@article{Poulin:2016anj,
archivePrefix = {arXiv},
arxivId = {1610.10051},
author = {Poulin, Vivian and Lesgourgues, Julien and Serpico, Pasquale D.},
doi = {10.1088/1475-7516/2017/03/043},
eprint = {1610.10051},
issn = {14757516},
journal = {Journal of Cosmology and Astroparticle Physics},
number = {3},
pages = {43},
title = {{Cosmological constraints on exotic injection of electromagnetic energy}},
volume = {2017},
year = {2017}
}

@article{Randall:2007ph,
    author = "Randall, Scott W. and Markevitch, Maxim and Clowe, Douglas and Gonzalez, Anthony H. and Bradac, Marusa",
    archivePrefix = "arXiv",
    doi = "10.1086/587859",
    eprint = "0704.0261",
    journal = "Astrophys.\ J.",
    pages = "1173--1180",
    primaryClass = "astro-ph",
    title = "{Constraints on the Self-Interaction Cross-Section of Dark Matter from Numerical Simulations of the Merging Galaxy Cluster 1E 0657-56}",
    volume = "679",
    year = "2008"
}

@article{Redondo:2008aa,
archivePrefix = {arXiv},
arxivId = {0801.1527},
author = {Redondo, Javier},
doi = {10.1088/1475-7516/2008/07/008},
eprint = {0801.1527},
issn = {14757516},
journal = {Journal of Cosmology and Astroparticle Physics},
number = {7},
pages = {8},
title = {{Helioscope bounds on hidden sector photons}},
volume = {2008},
year = {2008}
}

@article{Redondo:2008ec,
archivePrefix = {arXiv},
arxivId = {0811.0326},
author = {Redondo, Javier and Postma, Marieke},
doi = {10.1088/1475-7516/2009/02/005},
eprint = {0811.0326},
issn = {14757516},
journal = {Journal of Cosmology and Astroparticle Physics},
number = {2},
pages = {5},
title = {{Massive hidden photons as lukewarm dark matter}},
volume = {2009},
year = {2009}
}

@article{Redondo:2013lna,
archivePrefix = {arXiv},
arxivId = {1305.2920},
author = {Redondo, Javier and Raffelt, Georg},
doi = {10.1088/1475-7516/2013/08/034},
eprint = {1305.2920},
issn = {14757516},
journal = {Journal of Cosmology and Astroparticle Physics},
number = {8},
pages = {34},
title = {{Solar constraints on hidden photons re-visited}},
volume = {2013},
year = {2013}
}

@article{Riemer-Sorensen:2015kqa,
archivePrefix = {arXiv},
arxivId = {astro-ph.CO/1507.01378},
author = {Riemer-S{\o}rensen, S. and Wik, D. and Madejski, G. and Molendi, S. and Gastaldello, F. and Harrison, F. A. and Craig, W. W. and Hailey, C. J. and Boggs, S. E. and Christensen, F. E. and Stern, D. and Zhang, W. W. and Hornstrup, A.},
doi = {10.1088/0004-637X/810/1/48},
eprint = {1507.01378},
issn = {15384357},
journal = {Astrophysical Journal},
number = {1},
pages = {48},
primaryClass = {astro-ph.CO},
title = {{Dark matter line emission constraints from NuSTAR observations of the bullet cluster}},
volume = {810},
year = {2015}
}

@article{Rocha:2012jg,
archivePrefix = {arXiv},
arxivId = {1208.3025},
author = {Rocha, Miguel and Peter, Annika H.G. and Bullock, James S. and Kaplinghat, Manoj and Garrison-kimmel, Shea and O{\~{n}}orbe, Jose and Moustakas, Leonidas A.},
doi = {10.1093/mnras/sts514},
eprint = {1208.3025},
issn = {00358711},
journal = {Monthly Notices of the Royal Astronomical Society},
number = {1},
pages = {81--104},
title = {{Cosmological simulations with self-interacting dark matter - I. Constant-density cores and substructure}},
volume = {430},
year = {2013}
}

@article{Rrapaj:2019eam,
archivePrefix = {arXiv},
arxivId = {hep-ph/1904.10567},
author = {Rrapaj, Ermal and Sieverding, Andre and Qian, Yong-Zhong},
doi = {10.1103/physrevd.100.023009},
eprint = {1904.10567},
issn = {2470-0010},
journal = {Physical Review D},
number = {2},
primaryClass = {hep-ph},
title = {{Rate of dark photon emission from electron positron annihilation in massive stars}},
volume = {100},
year = {2019}
}

@article{Cline:2013gha,
archivePrefix = {arXiv},
arxivId = {hep-ph/1306.4710},
author = {Scribano, A and Chung, YS and Renton, P and Pukhov, O and Kato, Y and Clark, AG and Kirk, M and Niu, H and Berryhill, J and Anikeev, K and Rossin, R and Smith, C and Moore, R and Glenzinski, D and Velev, G and Korn, A and Roy, A and Introzzi, G and Carlsmith, D and Wu, X and Liss, TM and Lyons, L and LeCompte, T and Scott, A and Demers, S and Worm, S and Rakitine, A and Joshi, U and Loreti, M and Wester, WC and Flaugher, B and Vaiciulis, T and Tannenbaum, B and Grim, G and Rosenson, L and Lobban, O and Shimojima, M and McIntyre, P and DeJongh, F and Klimenko, S and Garfinkel, AF and Fan, Q and Yeh, GP and Barnes, VE and Savard, P and Yoshida, T and Kartal, S and Kaneko, T and Gorelov, I and Conway, J and Hao, W and Miscetti, S and Dominguez, A and Bromberg, C and Erbacher, R and Amaral, P and Robertson, WJ and Chang, PS and Artikov, A and Suzuki, T and Nagaslaev, V and Bolla, G and Gao, T and Newman-Holmes, C and Hollebeek, R and Tipton, P and Chiarelli, G and Heiss, A and Pellett, D and Shapiro, MD and Donati, S and Demortier, L and Lath, A and Tkaczyk, S and Liu, JB and Martin, A and Holloway, L and Deninno, M and Kim, MJ and Nelson, C and Hartmann, F and Einsweiler, K and Berge, JP and Kim, SB and Feild, RG and Thomson, E and Pagliarone, C and Paoletti, R and Gatti, P and Sedov, A and Bedeschi, F and Kim, BJ and Kim, HS and Toyoda, H and Gallinaro, M and Pescara, L and Khazins, D and Ukegawa, F and Haas, RM and Paus, C and Campbell, M and Takashima, R and Huth, J and Lander, R and Merkel, P and Colijn, AP and Ristori, L and Shepard, PF and van den Brink, S and Kambara, H and Orejudos, W and Azfar, F and Kovacs, E and Affolder, T and D'Onofrio, M and Zetti, F and Ashmanskas, W and Phillips, TJ and Moore, E and Shah, T and Wang, C and Ptohos, F and Ratnikov, F and Furic, I and Wicklund, E and Nakano, I and Solodsky, A and Pauletta, G and Pompos, A and Segler, S and Meyer, A and Mishina, M and Vejcik, S and Heinrich, J and Lai, N and Barbaro-Galtieri, A and Kasha, H and Sumorok, K and Ward, B and Sansoni, A and Lammel, S and Sanchez, C and Nahn, S and Vucinic, D and Brubaker, E and Tsybychev, D and Nakada, H and Kirsch, L and Wolinski, J and Bloom, K and Turini, N and Derwent, PF and Schmidt, MP and Liu, YC and Mitselmakher, G and Loken, J and Sill, A and Rolli, S and Kirby, M and Morita, Y and Schlabach, P and Wilkes, T and Siegrist, J and Connolly, A and Ngan, CYP and Culbertson, R and Matthews, JAJ and Watts, T and Ambrose, D and Tamburello, P and Madrak, R and Karr, K and Green, C and Waters, D and Bauer, G and Cheng, MT and Gris, P and Belforte, S and Stuart, D and Bensinger, J and Wagner, RG and Albrow, MG and Wang, MJ and Savoy-Navarro, A and Punzi, G and Hill, C and Wolinski, S and Spiegel, L and Proudfoot, J and Ruiz, A and Ferretti, C and Brandl, A and Nelson, T and Okusawa, T and Antos, J and Cauz, D and Zanetti, A and Blusk, SR and Handler, R and Hardman, AD and Schmitt, M and Fukui, Y and Gallas, A and Schmidt, EE and Gotra, Y and Spinella, F and Castro, A and Arisawa, T and Kuwabara, T and Byrum, KL and Miller, JS and Bocci, A and Foster, GW and Happacher, F and Handa, T and Kurino, K and Errede, S and Mukherjee, A and Chan, AW and Giannetti, P and Stefanini, A and Ohmoto, T and Oh, YD and Yoh, J and Semenov, A and Roser, R and Calafiura, P and Mulhearn, M and Bachacou, H and Webb, R and Pope, G and Leone, S and Minato, H and Hara, K and James, E and Hughes, R and Thompson, AS and Busetto, G and Munar, A and Kim, DH and Field, RD and Baroiant, S and Fang, HC and Tanaka, M and Gold, M and Tecchio, M and Sato, H and Bellinger, J and Papadimitriou, V and Blumenfeld, B and Carithers, W and Maeshima, K and Paulini, M and Friedman, J and Seiya, Y and {St Denis}, R and Yao, W and Moulik, T and Neu, C and Lukens, P and Bruner, N and Wang, SM and Thurman-Keup, R and Caskey, W and Patrick, J and Riegler, W and Wallace, NB and Dagenhart, D and Yeh, P and Acosta, D and Nachtman, J and Palmonari, F and Tollestrup, A and Takano, T and Semeria, F and Pitts, KT and Martignon, G and Slaughter, AJ and Kroll, J and Bishai, M and Oishi, R and Hocker, A and Bodek, A and Lockyer, NS and Wan, Z and Murat, P and Menguzzato, M and Winn, D and Moggi, N and Jones, M and Lucchesi, D and Apollinari, G and Yu, Z and Miyazaki, Y and Piacentino, G and Koehn, P and Kruse, M and Fernandez, JP and Nodulman, L and da Costa, JG and Waschke, S and Chu, ML and Latino, G and Brozovic, M and Errede, D and Yu, S and Teng, PK and Cranshaw, J and Ikeda, H and Neuberger, D and Veramendi, G and Wyss, J and Spalding, J and Herndon, M and Gay, C and Garcia-Sciveres, M and Budd, HS and Riveline, M and Takikawa, K and Sidoti, A and Vidal, R and Bortoletto, D and Ivanov, A and Rimondi, F and Yosef, C and Zucchelli, S and Scodellaro, L and Litvintsev, DO and Muller, T and de Troconiz, JF and Hahn, SR and Winer, BL and Singh, P and Dittmann, JR and Bailey, S and Miller, R and von der Mey, M and Dell'Orso, M and Done, J and Tether, S and Unel, MK and Franklin, M and Shibayama, T and Sliwa, K and Kelley, K and Christofek, L and Menzione, A and Dell'Agnello, S and Goldstein, J and Trischuk, W and Wilson, P and Kondo, K and Ohsugi, T and Tseng, J},
doi = {10.1103/PhysRevD},
eprint = {1306.4710},
issn = {0556-2821},
journal = {Physical Review D},
number = {5},
pages = {55025},
primaryClass = {hep-ph},
title = {{Measurement of the B+ total cross section and B+ differential cross section d sigma/dp(T) in p(p)over-bar collisions at root s=1.8 TeV}},
volume = {65},
year = {2002}
}

@article{Slatyer:2012yq,
archivePrefix = {arXiv},
arxivId = {1211.0283},
author = {Slatyer, Tracy R.},
doi = {10.1103/PhysRevD.87.123513},
eprint = {1211.0283},
issn = {15507998},
journal = {Physical Review D - Particles, Fields, Gravitation and Cosmology},
number = {12},
pages = {123513},
title = {{Energy injection and absorption in the cosmic dark ages}},
volume = {87},
year = {2013}
}

@article{Slatyer:2015jla,
archivePrefix = {arXiv},
arxivId = {1506.03811},
author = {Slatyer, Tracy R.},
doi = {10.1103/PhysRevD.93.023527},
eprint = {1506.03811},
issn = {24700029},
journal = {Physical Review D},
number = {2},
pages = {23527},
title = {{Indirect dark matter signatures in the cosmic dark ages. I. Generalizing the bound on s -wave dark matter annihilation from Planck results}},
volume = {93},
year = {2016}
}

@article{ANDP:ANDP19314030302,
author = {Sommerfeld, A.},
doi = {10.1002/andp.19314030302},
issn = {15213889},
journal = {Annalen der Physik},
number = {3},
pages = {257--330},
publisher = {WILEY-VCH Verlag},
title = {{{\"{U}}ber die Beugung und Bremsung der Elektronen}},
url = {http://dx.doi.org/10.1002/andp.19314030302},
volume = {403},
year = {1931}
}

@article{Stueckelberg:1900zz,
author = {Stueckelberg, E C G},
doi = {10.5169/seals-110852},
journal = {Helv.Phys.Acta},
pages = {299--328},
title = {{Interaction forces in electrodynamics and in the field theory of nuclear forces}},
volume = {11},
year = {1938}
}

@article{Tulin:2017ara,
archivePrefix = {arXiv},
arxivId = {1705.02358},
author = {Tulin, Sean and Yu, Hai Bo},
doi = {10.1016/j.physrep.2017.11.004},
eprint = {1705.02358},
issn = {03701573},
journal = {Physics Reports},
pages = {1--57},
title = {{Dark matter self-interactions and small scale structure}},
volume = {730},
year = {2018}
}

@article{Tulin:2012wi,
archivePrefix = {arXiv},
arxivId = {1210.0900},
author = {Tulin, Sean and Yu, Hai Bo and Zurek, Kathryn M.},
doi = {10.1103/PhysRevLett.110.111301},
eprint = {1210.0900},
issn = {00319007},
journal = {Physical Review Letters},
number = {11},
pages = {111301},
title = {{Resonant dark forces and small-scale structure}},
volume = {110},
year = {2013}
}

@article{Tulin:2013teo,
archivePrefix = {arXiv},
arxivId = {1302.3898},
author = {Tulin, Sean and Yu, Hai Bo and Zurek, Kathryn M.},
doi = {10.1103/PhysRevD.87.115007},
eprint = {1302.3898},
issn = {15507998},
journal = {Physical Review D - Particles, Fields, Gravitation and Cosmology},
number = {11},
pages = {115007},
title = {{Beyond collisionless dark matter: Particle physics dynamics for dark matter halo structure}},
volume = {87},
year = {2013}
}

@article{Vogelsberger:2012ku,
archivePrefix = {arXiv},
arxivId = {1201.5892},
author = {Vogelsberger, Mark and Zavala, Jesus and Loeb, Abraham},
doi = {10.1111/j.1365-2966.2012.21182.x},
eprint = {1201.5892},
issn = {00358711},
journal = {Monthly Notices of the Royal Astronomical Society},
number = {4},
pages = {3740--3752},
title = {{Subhaloes in self-interacting galactic dark matter haloes}},
volume = {423},
year = {2012}
}

@article{Weinberg:1979bt,
author = {Weinberg, Steven},
doi = {10.1103/PhysRevLett.42.850},
issn = {00319007},
journal = {Physical Review Letters},
number = {13},
pages = {850--853},
title = {{Cosmological production of baryons}},
volume = {42},
year = {1979}
}

@article{Weldon:1983jn,
author = {Weldon, H. Arthur},
doi = {10.1103/PhysRevD.28.2007},
issn = {05562821},
journal = {Physical Review D},
number = {8},
pages = {2007--2015},
title = {{Simple rules for discontinuities in finite-temperature field theory}},
volume = {28},
year = {1983}
}

@article{Zavala:2012us,
archivePrefix = {arXiv},
arxivId = {1211.6426},
author = {Zavala, Jes{\'{u}}s and Vogelsberger, Mark and Walker, Matthew G.},
doi = {10.1093/mnrasl/sls053},
eprint = {1211.6426},
issn = {17453925},
journal = {Monthly Notices of the Royal Astronomical Society: Letters},
number = {1},
pages = {L20--L24},
title = {{Constraining self-interacting dark matter with the milky way's dwarf spheroidals}},
volume = {431},
year = {2013}
}
%\pagebreak
%%\bibliographystyle{abbrv}
%\bibliographystyle{unsrt}
%\bibliography{All}

\newpage\hbox{}\thispagestyle{empty}\newpage
\newpage\hbox{}\thispagestyle{empty}\newpage

\pagestyle{plain}
%\includepdf[page={1},noautoscale = true,scale=3]{Cover_page/DM_SM_v2_b5.pdf}
\includepdf[page={1}]{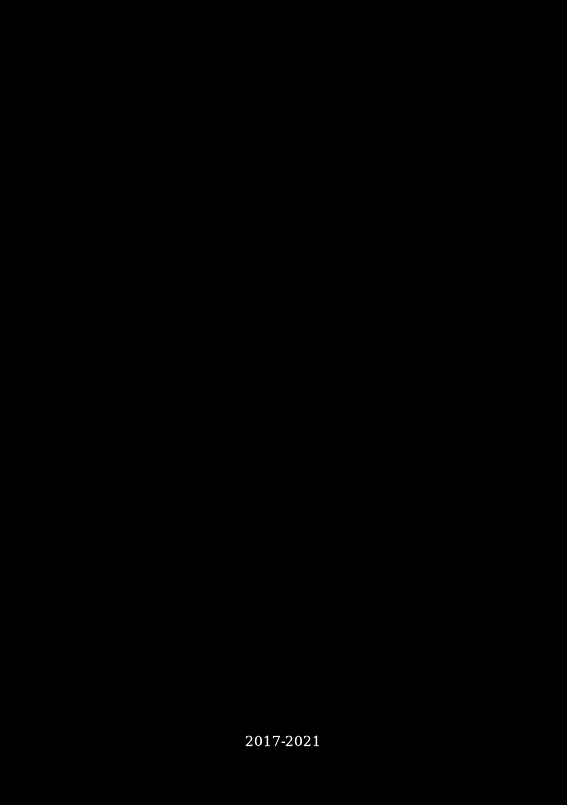}
\vfill
\newpage
\thispagestyle{empty}

\end{document}